\documentclass[aps,prl,groupedaddress,showpacs]{revtex4-1}
\usepackage{graphicx,amsmath}
\begin{document}

% Use the \preprint command to place your local institutional report
% number in the upper righthand corner of the title page in preprint mode.
% Multiple \preprint commands are allowed.
% Use the 'preprintnumbers' class option to override journal defaults
% to display numbers if necessary
%\preprint{}

%Title of paper
\title{Rayleigh-Benard convection in a hard disk system}

% repeat the \author .. \affiliation  etc. as needed
% \email, \thanks, \homepage, \altaffiliation all apply to the current
% author. Explanatory text should go in the []'s, actual e-mail
% address or url should go in the {}'s for \email and \homepage.
% Please use the appropriate macro foreach each type of information

% \affiliation command applies to all authors since the last
% \affiliation command. The \affiliation command should follow the
% other information
% \affiliation can be followed by \email, \homepage, \thanks as well.
\author{P.L. Garrido}
\email[]{garrido@onsager.ugr.es}
\affiliation{Instituto Carlos I de F{\'\i}sica Te{\'o}rica y Computacional. Universidad de Granada. E-18071 Granada. Spain }

\date{\today}
\begin{abstract}
We do a generic study of the behavior of a hard disk system under the action of a thermal gradient in presence of an uniform gravity field. We observe the conduction-convection transition and measure the main system observables and fields as the thermal current, global pressure, velocity field, temperature field,...
We can highlight two of the main results of this overall work: (1) for large enough thermal gradients and a given  gravity, we show that the hydrodynamic fields (density, temperature and velocity) have a natural scaling form with the gradient. And (2) we show that local equilibrium holds if the mechanical pressure and the thermodynamic one are not equal, that is, the Stoke's assumption does not hold in this case. Moreover we observe that the best fit to the data is obtained when the bulk viscosity  depends on the mechanical pressure.
\end{abstract}
\pacs{18-3e}
\maketitle

\section{Introduction}

Fluids are one of the subjects in physics that have been studied more intensively during the last centuries \cite{Batchelor}\cite{Chandra} \cite{Gallavotti}. Navier-Stokes (NS) equations have been the key ingredient in the understanding of observed fluid's phenomena in nature. Their analytic and numerical solutions are, nowadays, essential to many different scenarios, from the weather prediction up to the aircrafts design. In spite of the overall success obtained by its application to practical problems, there is a lack of a  rigorous connection between the underlying microscopic (typically hamiltonian) dynamics and the macroscopic NS-equations. Therefore there are still several important open issues connected to the NS main hypothesis and key ingredients used to build them. Let us mention a few of them as the {\it local equilibrium hypothesis} from which we assume that thermodynamics may be applied (up to some extend) at each macroscopic point of the fluid, the mathematical form of the transport coefficients (heat conduction and viscosities) as a function of the local fields, or the goodness of considering that the local heat conduction follows the Fourier's law or that the stress tensor is linear on the velocity gradients (newtonian fluid). The numerous efforts that have been done so far in the NS derivation show the enormous difficulties on such task. To have some flavour on it, let us mention  the nice review by Esposito et al. about the derivation of NS equations from the Boltzmann equation \cite{Esposito}. 
However, there is another way to study the connection between microscopic and macroscopic descriptions: the computer simulation tool. 

We know that fluids are typically in nonequilibrium states due to the action of external agents as temperature gradients or driving fields and there appear long range correlations in space and time. Therefore it is expected that their behavior depends strongly on the boundary conditions we impose. That makes a computer simulation of a fluid a very complicated task. It is necessary a priori a very large number of particles to have a reasonable collective description of the fluid and to limit (at some extent) the effects of the details of the boundaries in the bulk system behavior. Also, the long range correlations makes very difficult to make a finite size study of the simulation by trying to extrapolate any macroscopic observable value from a sequence of measurements of it in systems with increasing number of particles.
Finally, the time correlation functions are also power like which implies the need of very long runs to get a large enough set of almost independent configurations to do any observable averaging. In practice we immediately are limited by the finite power of our nowadays computers and we can use a very limited number of particles in our simulations (very far from the Avogrado's number) from which  we can get  some useful data in a computation time that does not exceeds a human life span. Obviously, with such small amount of particles it is difficult to expect any reasonable connection between those simulations and the behavior described by the Navier Stokes equations. In any case, there were some attempts during the eighties with some pioneering computer simulations trying to see macroscopic hydrodynamic behavior on particle systems. Let us mention for instance the work of D.C. Rapaport and E. Clementi \cite{Rapaport_1986} in which they simulate a fluid flow past a cylindrical obstacle by using a set of particles with a short range pair interaction potential. 

We have recently discovered \cite{delPozo} that a bidimensional hard disks system under the action of a temperature gradient, presents a nice property that we call {\it boundary decoupling} when studying its stationary state properties. That is, the system rearranges itself in such a way that there is a bulk part of it in which the local hydrodynamic properties hold: the local equilibrium hypothesis and the Fourier's law. The finite size dependences are then restricted to the values of the average heat current crossing the system and the local constant pressure. In other words, for a given stationary state with measured values of the heat current and pressure, the temperature profile in the system bulk is the one corresponding to the solution of the Fourier's law and the local temperature and density are related by the hard-disk equilibrium equation of state. We have checked this property also for a hard disk system under a shear forcing at the boundaries \cite{delPozo2}. We want to push further the possibility to get some insight on the hydrodynamic behavior of a system by doing computer simulations. The {\it B{\'e}nard problem} is, in our opinion, the natural next problem to be studied by a computer simulation of a hard disk particle system. 

The {\it B{\'e}nard Problem} consists in a fluid that is heated from below under the presence of gravity. It is observed that above some critical value of the temperature gradient the convection starts and the fluid move in rolls occupying the full system. This is the {\it Rayleigh-B{\'e}nard instability} that was observed experimentally by B{\'e}nard and afterwards there was interpreted theoretically by Rayleigh.
That is, this nonequilibrium problem includes a transition from conducting to convecting states and a nonzero hydrodynamic velocity field in the convecting stationary state (see for instance some recent books and reviews in \cite{Mutabazi}, \cite{Lappa} and \cite{Boden}). 

There are several papers studying the B{\'e}nard problem by using simulations of hard disks. Let us mention, for instance, the work from M. Marechal et al. \cite{Mar87}\cite{Mar88} and A. Puhl et al. \cite{Puhl}. Mareschal and co-workers observed the transition conducting-convecting for a system of hard disks  even for a very small number of particles ($N=5040$). However, the very large velocity fluctuations prevented, in that time, to get accurate measurements of local magnitudes. Nevertheless they managed to get some hydrodynamics velocity profiles that they could compare with a numerical resolution of NS equations getting a very reasonable accordance. This system was later studied also by D.C. Rapaport \cite{Rapaport_1988}\cite{Rapaport_1992} showing the influence of the boundary conditions and the initial configurations on the stationary behavior of the hard disk system. Finally, let us also mention the work from D. Risso and P. Cordero in which the study the onset of the transition \cite{Risso}.

In this paper we show a complete set of computer simulations results whose goal is two fold: first we want to get some insights on the stationary behavior of the system and its dependence on the external parameters (gravity and temperature gradient) by observing the heat current, internal energy, pressure,... and second, we would like to check some of the basic Navier Stokes assumptions like the local equilibrium hypothesis, the Fourier's law, the difference between the mechanical pressure versus the thermodynamic one or the efects of our system compressibility  in the transition \cite{Manela}.

  This paper is divided in several parts. Let us make a summary of the main items and results obtain on each part.
  
\begin{itemize}
\item {\it I. The physical properties of a hard system:} First we introduce a catalog of known properties of a hard disks system. We do not know the equilibrium equation of state but we express all the thermodynamic equilibrium quantities as a function of it. We also derive the hydrostatic formula in equilibrium. Finally we write down the  Navier Stokes equations,  the general structure of the transport coefficients and their particular analytical form obtained in the Enskog approximation of the Boltzmann equation for hard disks.

\item {\it II. The model:}  We describe the model, the  dynamics of particles and the boundary conditions used in the simulation. We also discuss the election of the parameter values for the simulation. We explain how we do the local measurements of the magnitudes in order to get the hydrodynamic fields. In a computer simulation like this where the ratio between noise and signal is expected to be not small is important to do a careful analysis of errors. We explain the way we have done such analysis. Let us remark that along the paper we use an strict $3\sigma$ error interval criterion. Finally we study the relation between the averages of the magnitudes on a local region and its continuous value. That is important to have some clean scheme to define the spatial derivatives of any field.

\item {\it III. Computer simulation results of equilibrium states:} We check the program and the way we measure the global and local observables by doing a set of simulations at equilibrium. That is, when the temperatures on both plates are equal but maintaining the gravity field.  We show that the local densities and temperatures follow accurately the theoretical results. Moreover, the local pressure measured by using the virial expression is found to give the correct results. Finally we observe that it is meaningless  to measure local fluctuations of the magnitudes because they are very sensitive to the size of the local cells that in our case is certainly very small.

\item {\it IV. Computer simulation results of nonequilibrium stationary states: global magnitudes.} In this part we show the behavior of the global observables as a function of the temperature gradient $\Delta T$ and the intensity of the gravity field $g$. The kinetic energy per particle increases with $g$ in contrast to what happens in equilibrium and for large $\Delta T$ it grows like $\simeq \Delta T/3$ for all $g$'s. The fluctuations of the kinetic energy increase with $\Delta T$  for all $g$'s and for a fixed $\Delta T$ it increases with $g$. The interesting thing is that the fluctuations divided by the average kinetic energy to the square presents a non-monotonous behavior for a given $g$ value: starting from the equilibrium, it decreases with $\Delta T$, reaches a minimum and then it increases. As we show later in the paper, this minimum is consistent with the transition point between conducting and convecting regimes. Moreover, the critical value of the gradient  at which the transition occurs increases with $g$. That behavior contrast with the one derived in the Boussinesq approximation of the Navier Stokes equations in which the critical value decreases with $g$. We explain there that the reason is that our system is compressible and then one should correct the Boussinesq assumptions accordingly.

We also measure the hydrodynamic kinetic energy, that is, the average particle velocity on a cell, squared and summed over all cells. Obviously this magnitude should be zero in the conducting regime and non-zero in the convecting one. We obtain the transition point from the data that is consistent with other methods used in the paper. We also study the hydrodynamic kinetic energy associated with each velocity component. We observe a kind of equipartition independently of the $g$'s for large values of $\Delta T$: $e_{u,2}\simeq 3.4 e_{u,1}$ where $e_{u,i}$ is the hydrodynamic kinetic energy associated to the $i$th component of the velocity.

The Pressure is computed at the top and bottom boundaries as the moment variation of the particles when colliding with it. We observe that the global barometric formula holds always, that is the difference between the pressures at bottom and top of the system is proprotional to the system weight. This confirms the correctness of the computer simulation. Nevertheless the pressure at each boundary has a non trivial dependece on $\Delta T$ and $g$ which we analyze.

Finally we study the energy current that cross each thermal bath boundary. We first check that the current coming into the system is exactly the same to the outgoing one. For large values of $\Delta T$ the current goes like $\Delta T^{3/2}$ for all $g$'s. In general we see that  the gravity hinders the heat current for small values of $\Delta T$ while it favours it for large values of it where it correspond to the fully developed convective regime. We define a second critical thermal gradient as the one in which the thermal current for a given $g$ value equals the one when $g=0$ (pure conducting case). That is, beyond such value, convection becomes ``more efficient'' that conduction in transporting heat through the system for a given temperature gradient. This second critical gradient is necessary to explain some observed system behavior later in the paper.
 
\item {\it V. Computer simulation results of nonequilibrium stationary states: spatial structures.} First we study the distribution of the current line lengths for a set of stationary velocity fields. For conducting states we reasonably fit a gamma distribution to the data. In the convecting regime there is a distribution that depends on the system geometry and we analyze its form. We also analyze  the fraction of velocity vectors whose modulus is larger than their standard deviation, $pr$. That is, the amount of ``real'' (or nonfluctuating) hydrodynamic velocity vector. We observe that this magnitude is directly correlated with the average length of the current lines in each case. Moreover, that relation is independent on the value of $g$. We see that for all $g$'s we obtain $pr\simeq 0.68$ when we are at the second critical gradient. This value coincides with the critical fraction of the covered surface at the percolation threshold of a system of overlapping disks of radius $r$. That is, the overall picture maybe that when convection begins, the current lines are small. Their length grow with the value of the thermal gradient up to a critical value in which the current lines extends and connect the system boundaries. We conclude that this phenomena is related with a percolation like transition that occurs precisely at the second critical temperature gradient. 

We study the spatial structure of the hydrodynamic velocity components. For large enough temperature gradients it seems that there is  well define limiting velocity profile. Moreover we show that each hydrodynamic velocity component has an universal scaled  field (universal in the sense that, once the field is scaled, it does not depend on $\Delta T$) for large enough values of $\Delta T$. The scaled field, $\tilde u_{1,2}(x,y)$ is obtained by
\begin{equation}
\tilde u_{1,2}=\frac{u_{1,2}(x,y)}{\sigma(u_{1,2})}
\end{equation}
where $u_{1,2}(x,y)$ is the components $1,2$ of the measured hydrodynamic field and $\sigma(u_{1,2})$ are their stardard deviation (the spatial average of the hydrodynamic velocity field is zero). In the process we observe the nice property: $\sigma(u_2)=2\sigma(u_1)$ for all fields with any $\Delta T$ and $g$ values. We  study  the Inertial Tensor and the fourth moment of scaled fields with different values of $\Delta T$ to look for any systematic dependence on the temperature gradient. Moreover, we compute the average mutual euclidean distance between scaled configurations to  confirm that, for large enough values of $\Delta T$ such distance tends to zero. The overall analysis is lengthly because the initial data fields have non-negligible fluctuating spatial behavior and intrinsic error bars that should be taken into account to finally obtain a consistent set of results.

We also study the Temperature field. The average $y$-profile, $\tau(y)$ is non-linear and it is convex for $g=0$ and $5$ and it has infection points for $g=10$ and $15$. We see thermal gap in the boundaries, characterize its behavior  and compute the effective thermal gradient that is $\simeq 0.82 \Delta T$ for all $g$-values. Finally we show also the existence of a universal scaled Temperature field that is defined by
\begin{equation}
\tilde T(x,y)=\frac{T(x,y)-\tau(y)}{\sigma(T)}
\end{equation}
where $\sigma(t)$ is the standard deviation with respect $\tau(y)$. 
In the same spirit we also analyze  the density and pressure fields. In both cases we observe the existence of universal fields.

\item {\it VI. Connecting with hydrodynamics.} In this section we try to check some of the Navier Stokes ansatz and properties at the stationary state. Dynamic properties are beyond the scope of this article. The first thing we check is the continuity equation that express the mass conservation at the hydrodynamic level:
$\nabla(\rho u)=0$ from the measured local density and velocity fields. This give us some confidence on the use of the field spatial derivatives in spite that the relative error after derivatives increases up to an average of $10\%$, large but enough to check some equation consistencies. Second we see if local equilibrium holds at each cell. We observe that it is the case for systems at the conducting regime. However we see a small and noisy systematic deviation from local equilibrium for systems at the convecting regime. We argue that it is so because of the difference between the mechanical pressure (the one we measure) and the thermodyncamic one. Their difference it is known to be proportional to the divergence of the hydrodynamic velocity. The hydrodynamic velocity divergence is small and the data is noisy but nevertheless we manage to discard the Stoke's assumption ($\nabla u=0$) and we find that the best fit occurs when we assume that the bulk viscosity is itself dependent on the mechanical pressure. This is a weak-like result but indicates a way to go deep inside of the properties of the NS equations.
Finally we check stationary NS equations for the non-convective states obtaining a reasonable good coincidence with the computer simulation results.

\end{itemize}

The overall work has been an effort to extract a coherent and systematic description of the Rayleigh-Benard problem from a microscopic computer simulation of a hard disk system. Of course there are many interesting items and observables we could study. Many of them we honestly tried to analyze (for instance local fluctuations) but we finally applied the strict criterium that anything whose error analysis obscure the average structure should be discarded. Maybe, one may focus on a particular item to design a simulation that gets much better statistics necessary to obtain with good confidence its behavior. We feel that it can be done and it is possible to study hydrodynamic behaviors from microscopic models.

\section{The physical properties of a hard disk system}

Hard disks is one of the paradigmatic models in many body physics. Having a very simple dynamics it contains many nontrivial  statistical properties that come from collective effects of its individuals. For instance, it has a  phase transition \cite{Alder62} or its time correlation functions have the long time tails \cite{Wain71}. Moreover its exact Equation of State (EOS) is not yet
derived in the context of the equilibrium statistical physics \cite{Mulero}.  Let us state some basic properties that are known and that we are going to use during the computer simulation data analysis.

\begin{itemize}
\item {\it Equilibrium properties:}
The canonical partition function for a system with $N$ disks with mass $m$ and radius $r$ enclosed  in a box of sides $L_x$ an $L_y$ at a temperature $T$ is given by
 \begin{equation}
 Z(N,S,T)=\frac{1}{h^{2N}N!}\left(2\pi mk_B T \right)^N Z_c(N,S)
 \end{equation} 
 where $S$ is the box surface ($S=L_xL_y$) and $Z_c$ is the configurational partition function that, in the hard disk case, it doesn't depend on the temperature. $Z_c$ is not known but it can be connected with the equation of state that has been extensively studied numerically and very good analytical expressions have been derived from the extensive computer simulations \cite{Mulero}. The equation of state for a hard disk system has the general form:
 \begin{equation}
PS=Nk_BT H(\rho)  \quad ,  \quad \rho=\frac{N\pi r^2}{S}
\end{equation}
A well known simple form for $H$ was proposed by Henderson \cite{Hen75} that fits  very well the data for de areal density interval $\rho\in[0,0.5]$:
 \begin{equation}
 H(\rho)=\frac{1+\rho^2/8}{(1-\rho)^2}
\end{equation} 
The equation of state is connected with $Z_c$ by the expression:
\begin{equation}
P=k_BT\frac{\partial Z_c}{\partial S}\quad\Rightarrow\quad NH(\rho)=-\rho\frac{\partial Z_c}{\partial \rho}
\end{equation}
If we assume that  $Z_c\simeq \exp[N\bar f(\rho)]$ for $N$ large enough, we obtain:
\begin{equation}
\bar f(\rho)=\bar f(\rho^*)-\int_{\rho^*}^{\rho}d\tilde\rho\,\,\frac{H(\tilde\rho)}{\tilde\rho}
\end{equation}
where $\rho^*$ is any arbitrary value. We know that in the limit $\rho\rightarrow 0$ the hard disk system behaves like an ideal gas. In such limit $f(0)=\ln S$, substituting  $\rho^*=0$  above we get
\begin{equation}
\bar f(\rho)=\frac{1}{N}ln Z_c=ln\left(\frac{\pi r^2N}{\rho}\right)-\int_{0}^{\rho}d\tilde\rho\,\,\frac{H(\tilde\rho)-1}{\tilde\rho}
\end{equation}
and the hard disk partition function can be written for $N$ large:
\begin{equation}
Z\simeq e^{Nf(\rho)}\quad\quad,\quad f(\rho)=ln\frac{T}{\rho}-\int_{0}^{\rho}d\tilde\rho\,\,\frac{H(\tilde\rho)-1}{\tilde\rho}+C
\end{equation}
where $C$ is the constant
\begin{equation}
C=ln\left(\frac{2mk_Be\pi^2 r^2}{h^2}\right)
\end{equation}
 From it we can compute all thermodynamic magnitudes, for example, the entropy is given by
 \begin{equation}
\frac{s}{k_B}=f(\rho)+1
\end{equation}
the specific heats at constant pressure and volume are respectively 
\begin{equation}
\frac{c_P}{k_B}=1+\frac{H(\rho)^2 }{H(\rho)+\rho \frac{dH(\rho)}{d\rho}}\quad , \quad c_V=k_B
\end{equation}
and the coefficients $\alpha$ and $\kappa_T$ are 
\begin{equation}
\alpha=\frac{1}{T}\frac{H(\rho)}{H(\rho)+\rho \frac{dH(\rho)}{d\rho}}\quad , \quad \kappa_T=\frac{\pi r^2}{k_B T\rho}\frac{1}{H(\rho)+\rho \frac{dH(\rho)}{d\rho}}
\end{equation}

\item{\it Hard disks under the action of an external field at equilibrium: the hydrostatic formula}

Let us assume now that over the hard disk system, there is an external field acting on the particles. That is, we should add to the Hamiltonian a term of the form:
\begin{equation}
H_{ext}=\sum_{i=1}^NU(\vec r_i)
\end{equation} 
Then, to obtain the equilibrium properties it is convenient to use the grand canonical ensemble. The computation simplifies if we consider a very slow varying external field (we follow here the arguments in Martin-Lof's book \cite{MartinLof}). That is, $U(\vec r)=\tilde U(\vec r/L)$ being $\tilde U$ a smooth function in $R^2$. We can define  a box with a volume that grows with $L^2$ at a given position $\vec x$. When $L\rightarrow\infty$ the particles of the bulk of the box only see an external constant field of value $\tilde U(\vec x)$ and then $H_{ext}=N\tilde U(\vec x)$. The local thermodynamic properties of the system can be  computed using the grand canonical ensemble just substituting the chemical potential $\mu\rightarrow \mu-\tilde U(\vec x)$. For instance, the equation of state is given in a parametric form:
\begin{eqnarray}
P&=&k_BTa(T,\mu-\tilde U(\vec x))\nonumber\\
\tilde\rho&=&k_BT\frac{d}{d\mu}a(T,\mu-\tilde U(\vec x))\nonumber
\end{eqnarray}
where $\tilde\rho$ is the particle density, $a=a(T,\mu)=\lim_{S\rightarrow\infty}S^{-1}\log {\mathcal Z}$ and ${\mathcal Z}$ is the gran canonical partition function. We can eliminate the parameter $\mu$ by applying the gradient to the pressure and using the definition of $\tilde\rho$. Thus, we obtain the hydrostatic formula:
\begin{equation}
\nabla P=-\tilde\rho(\vec x)\nabla\tilde U
\end{equation}
Notice that the full argument is general and it is not restricted to a hard disk system.

\item{Navier-Stokes Equations for hard disk systems:}

Navier Stokes equations describe the dynamics of a macroscopic fluid \cite{Gallavotti} and they are constructed assuming: (1) the local conservation of density, linear momentum and  energy, (2) the local equilibrium property and (3) some local constitutive relations (Fourier's law, Newtonian fluid). The macroscopic evolving fields are: the local mass density, $\tilde\rho(x,t)$, the local velocity components, $u_i(x,t)\, ,\, i=1,\ldots, d$, and the local temperature, $T(x,t)$. Their evolution is determined by:
\begin{eqnarray}
\partial_t\tilde\rho+\sum_i\partial_i(\tilde\rho u_i)&=&0\nonumber\\
\partial_t u_i+\sum_j u_j\partial_j u_i&=&\frac{1}{\tilde\rho}\sum_j \partial_j\tau_{ij}+g_i\quad (i=1,\ldots,d)\nonumber\\
\tilde\rho\tilde c_v\left(\partial_t T+\sum_i u_i\partial_i T \right)&=&-T\frac{\partial P}{\partial T}\biggr\vert_{\tilde\rho}\sum_i\partial_i u_i+\sum_{ij}\tau'_{ij}\partial_i u_j+\sum_{ij}\partial_i\left(\kappa_{ij}\partial_j T \right) \label{NS}
\end{eqnarray}
 Assuming that we study the so called {\it newtonian fluids} then: 
\begin{eqnarray}
\tau_{ij}&=&-P\delta_{ij}+\tau'_{ij}\nonumber\\
\tau'_{ij}&=&\eta\left(\partial_i u_j+\partial_j u_i \right)+\eta' \delta_{ij}\sum_k \partial_k u_k 
\end{eqnarray}
Finally, the local equilibrium property also implies that  the equilibrium equation of state applies to the local fields: $P=P(\tilde\rho,T)$.  Observe that  we also need to know the equilibrium mass specific heat at constant volume $\tilde c_v=\tilde c_v(\tilde\rho,T)$ and  the transport coefficient functions: the thermal conductivity $\kappa=\kappa(\tilde\rho,T)$, the shear viscosity $\eta=\eta(\tilde\rho,T)$ and the second viscosity $\eta'=\eta'(\tilde\rho,T)$ . All these equations plus the value of the external field $g_i$ and the boundary conditions for the fields give us  a complete description of the evolution of a fluid at the macroscopic level. 

For hard disks system we know the EOS and then 
\begin{equation}
\frac{\partial P}{\partial T}\biggr\vert_{\tilde\rho}=\frac{P}{T}
\end{equation}
and $c_v=k_B$. Notice that  $\rho=N\pi r^2/L_xL_y$  and it is related to the mass density by: $\tilde\rho=m\rho/\pi r^2$ where $m$ is the mass of one disk and $r$ its radius.

Moreover one can show that the transport coefficients have also a simple dependence on the temperature:  
\begin{eqnarray}
\eta&=&\frac{\sqrt{k_Bm}}{r}\sqrt{T}E(\rho) \nonumber\\
\eta'&=&\frac{\sqrt{k_Bm}}{r}\sqrt{T}E'(\rho)\nonumber\\
\kappa&=&\frac{\sqrt{k_B^3}}{r\sqrt{m}}\sqrt{T}K(\rho) \label{tc}
\end{eqnarray}

We are going to be interested on the stationary solutions when the external field is $g_j=-g\delta_{j,2}$. The resulting equations after we substitute these formulas into eqs. (\ref{NS}) are: 
\begin{eqnarray}
\partial_1(\rho u_1)+\partial_2(\rho u_2)&=&0 \label{s1}\\
\rho\left(u_1\partial_1 u_1 +u_2\partial_2 u_1 \right)&=&-\partial_1Q+\pi r\partial_1\left[\sqrt{T}\left(2E\partial_1u_1+E'(\partial_1u_1+\partial_2u_2) \right)\right]\nonumber\\
&+&\pi r\partial_2\left[\sqrt{T}E\left(\partial_1u_2+\partial_2u_1\right)\right]\label{s2}\\
\rho\left(u_1\partial_1 u_2 +u_2\partial_2 u_2 \right)&=&-\partial_2Q+\pi r\partial_2\left[\sqrt{T}\left(2E\partial_2u_2+E'(\partial_1u_1+\partial_2u_2) \right)\right]\nonumber\\
&+&\pi r\partial_1\left[\sqrt{T}E\left(\partial_1u_2+\partial_2u_1\right)\right]-\rho g\label{s3}\\
\rho\left(u_1\partial_1 T +u_2\partial_2 T \right)&=&-Q(\partial_1u_1+\partial_2u_2)+2\pi r\sqrt{T}E\left((\partial_1 u_1)^2+(\partial_2 u_2)^2\right)+\pi r\sqrt{T}E'\left(\partial_1u_1+\partial_2u_2\right)^2\nonumber\\
&+&\pi r\sqrt{T}E\left(\partial_1u_2+\partial_2u_1\right)^2+\pi r\left[\partial_1\left(\sqrt{T}K\partial_1T\right)+\partial_2\left(\sqrt{T}K\partial_2T\right)\right]\label{s4}
\end{eqnarray}
where we have defined the {\it reduced pressure}, $Q=\pi r^2 P$, and  let us remind that $E$, $E'$ and $K$ are functions of $\rho$. We have also done, after the computation, the following mapping:
\begin{equation}
k_BT\rightarrow T\quad ,\quad mg\rightarrow g \quad,\quad \sqrt{m}u_{1,2}\rightarrow u_{1,2}
\end{equation}
With this mapping the equation of state is:
\begin{equation}
Q=TF(\rho)=T\rho H(\rho)
\end{equation}
Observe that we use in all the equation the areal density $\rho$ instead the mass density $\tilde\rho$.

The NS equations for hard disks simplify when looking for non-convective  stationary solutions:
\begin{equation}
\rho=\rho^{NC}(y)\quad, \quad u_x=u_y=0\quad,\quad T=T^{NC}(y)
\end{equation} 
that are solutions of:
\begin{eqnarray}
\frac{dQ^{NC}}{dy}&=&-g\rho^{NC}\label{nc1}\\
\sqrt{T}K(\rho^{NC})\frac{dT^{NC}}{dy}&=&-J\label{nc2}\\
Q^{NC}&=&T^{NC}F(\rho^{NC})\label{nc3}
\end{eqnarray}
with the boundary conditions:
\begin{equation}
T(0)=T_0\quad,\quad T(1)=T_1=1\quad,\quad \bar\rho=\int_{0}^1dy \rho^{NC}(y)
\end{equation}
where $T_0$, $g$ and $\bar\rho$ are the control parameters. Observe that $J$ is a constant that it is be obtained as a function of the external parameters.

\item{\it{Transport coefficients (in the Enskog approximation):}}
Gass  computed the transport coefficients for the hard disk system in the Enskog approximation \cite{Gass71}. The shear viscosity $\eta$, the bulk viscosity $\xi$ and the thermal conductivity $\kappa$ are:
\begin{eqnarray}
\eta&=&\frac{\sqrt{k_Bm}}{r}\sqrt{T}E(\rho) \quad , \quad E(\rho)=\frac{0.255}{\sqrt{\pi}\chi(\rho)}\left[1+2\rho\chi(\rho)+3.4197\rho^2 \chi(\rho)^2\right]\nonumber\\
\xi&=&\frac{\sqrt{k_Bm}}{r}\sqrt{T}E_B(\rho) \quad , \quad E_B(\rho)=1.2734 \rho^2\chi(\rho)\nonumber\\
\kappa&=&\frac{\sqrt{k_B^3}}{r\sqrt{m}}\sqrt{T}K(\rho) \quad , \quad K(\rho)=\frac{1.029}{\chi(\rho)}\left[1+3\rho\chi(\rho)+3.4874\rho^2 \chi(\rho)^2\right]\label{ensk}
\end{eqnarray}
where
\begin{equation}
\chi(\rho)=\frac{H(\rho)-1}{2\rho}
\end{equation}
and the second viscosity $\eta'=\xi-\eta$.

\end{itemize}

\section{The model}

Our system is composed of a set of $N$ hard disk particles that move in a square box of side $L=1$. Each disk have unit mass ($m=1$) and its radius ($r$) is chosen such that the areal density is $\bar\rho=N\pi r^2/L^2$, that is $r=(\bar\rho L^2/(N\pi))^{1/2}$. The dynamics of a given disk has two parts: (1) free flight and (2) collisions. 

(1) Free fligth: Between collisions each disk $i$ is under the unique action of a constant external field $\vec a=(0,-g)$. Thus, its equations of motion are given by $$\frac{d\vec v_i}{dt}=\vec a $$ whose trivial solutions are 
\begin{equation}
\vec r_i(t)=\vec r_i(0)+\vec v_i(0)t+\frac{1}{2}\vec a t^2\quad,\quad \vec v_i(t)=\vec v_i(0)+\vec a t
\end{equation} 
where $\vec r_i(0)$ and $\vec v_i(0)$ are the values of the disk position and velocity {\it after} a collision of the disk has taken place.

(2) Collisions: A disk have three types of collisions, with another disk, with the box vertical boundaries and with the box horizontal boundaries. (a) Two  disks collide when their distance equals $2r$. When it happens, the linear moment and the kinetic energy is conserved and it is assumed that the velocity components perpendicular to the vector that joints the centers of the disks keep unaltered during the collision. That is, we do not consider the existence of disk internal rotation around its center. (b) When the disk collides with a vertical side of the box at $x=0$ or $x=1$ it gets perfectly reflected, that is $v_1\rightarrow -v_1$ and $v_2\rightarrow v_2$. (c) when the disk hits a horizontal boundary at $y=0$ or $y=1$ the disks changes its second component of its velocity by getting a random value chosen from a Maxwellian distribution with temperature $T_0$ or $T_1$ when hitting at $y=0$ or $y=1$ respectively. That is $v_1\rightarrow v_1$ and $v_2\rightarrow v_2=s_{0,1}(-2T_{0,1}\log(1-\xi))$ being $\xi$ an uniform random variable between $[0,1]$ and $s_0=1$ and $s_1=-1$.

\begin{figure}[h]
\begin{center}
\includegraphics[height=6cm,clip]{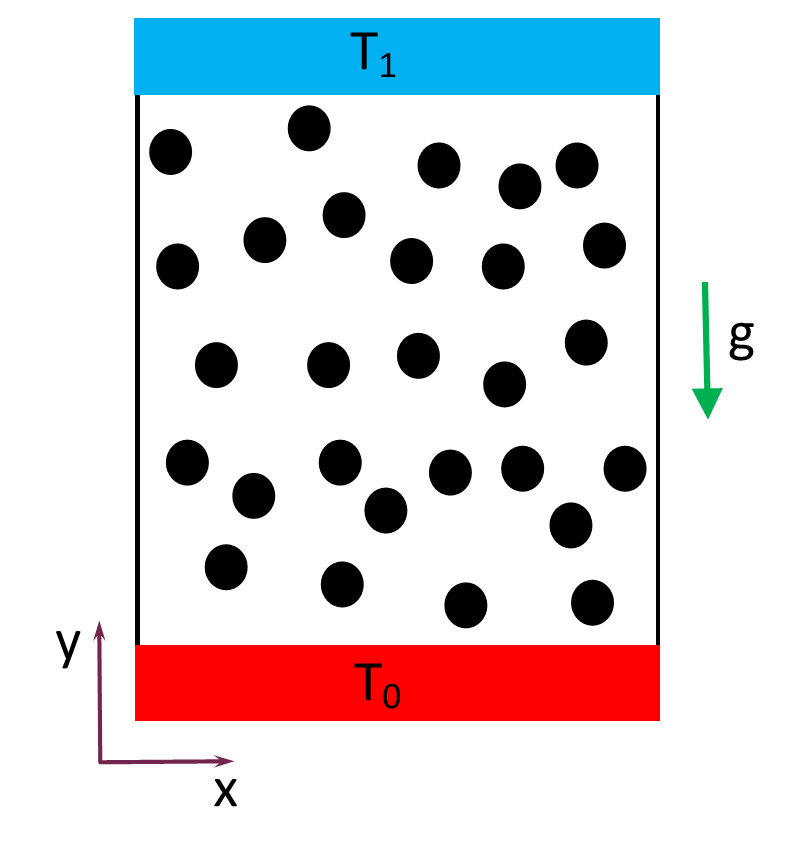} %basicconf.pub%
\end{center}
\kern -0.5truecm
\caption{Schematic view of the system simulated in this paper.}
\end{figure}

Initially the disks are placed regularly on the box with a random initial velocity vector with a modulus that is the square root of the sum of the two thermal bath temperatures. We evolve the system during $5\times 10^4$ collisions per particle ($\simeq 5\times 10^7$ collisions) when we check that the system is in the stationary state. Then we begin to take measures of different observables each $100$ collisions per particle. The simulation extends during $10^7$ collisions per particle ($\simeq 10^{10}$ total number of collisions) and we obtain  $M=10^5$ data for averaging. The observables have typical errors of $3\sigma/\sqrt{M}\simeq 0.01 \sigma$ that is sufficient to analyze with some detail many of their interesting behavior.

Computer simulations of hard disks systems have many advantages. First, the intrinsic dynamics is very symple and it can be implemented in the computer with very high efficiency without any loss of precision (see for instance \cite{Cordero95}\cite{Rapa09}). Moreover, its purely kinetic structure makes possible to fix one of the system external parameters $(T_0,T_1,g)$ by just  a time rescaling. In other words: If we rescale time, $t=\alpha t'$, the particle velocities rescale by $v=v'/\alpha$ and then, choosing the temperature of the thermal baths being $T'_{0,1}=\alpha^2 T_{0,1}$ and the gravity field $g'=\alpha^2 g$, the dynamic evolution of the system with parameters $(T_0,T_1,g)$ is indistinguishable from the one with parameters $(T_0',T_1',g')$. We can arbitrarily choose $\alpha=1/\sqrt T_1$ in order to fix to one the temperature of the the thermal bath at top. Therefore we can do the simulations varying only two parameters  $(T_0',T_1'=1,g')$. In order to obtain the behavior of any observable for an arbitrary value of $T_1$ we just should undo the corresponding time rescale on the variables defining the magnitude. 

In order to choose the set of values for the parameters $\bar\rho$ (the average areal density), $T_0$ and $g$ we should pay attention to the phenomena we want to describe. In this case we would like to study the transient from non-convective regime to the convective one. Then it is reasonable to use the Rayleigh coefficient ($Ra$) that  is a non-dimensional magnitude that localizes the parameter regions where the fluid is in a convective/non-convective state. It is defined by
\begin{equation}
Ra=\frac{\alpha g\Delta T L_y^3}{\nu\bar\kappa}
\end{equation}
where $g$ is the intensity of the external field, $\Delta T=T_0-T_1$, $\nu=\eta/\tilde\rho$ is the kinematic viscosity, $\bar\kappa=\kappa/\tilde\rho\tilde C_P$ and $\tilde\rho=m\rho/\pi r^2$ is the mass density and $C_P=c_p/m$ is the specific heat capacity per unit mass.

For a hard disks case it is known that linearizing NS equations under the Boussinesq approximation there is a critical Rayleigh number $Ra_c=27\pi^4/4\simeq 657.51$ for the stress free boundary condition case (see for instance S. Chandrasekhar \cite{Chandra}). Convection appears when the system has a Rayleigh number above the  critical value.

We have computed $Ra$ by using the Henderson equation of state and the Enskog transport coefficients. We choose the number of particles to be small enough to have a very fast system evolution to get the largest quantity of data to average. One may think that $N$ small would implies very large finite size effects. However,  we already shown in previous studies with hard disks \cite{delPozo} that the system have a {\it boundary decoupling} property in which the system re adapts its bulk configuration to behave as it was an infinite system with effective boundary conditions composed by the ones we define in the simulation plus a small region with disks around them. For the computer simulation in this paper we chose $N=957$.
 
\begin{figure}
\begin{center}
\includegraphics[width=6cm,clip]{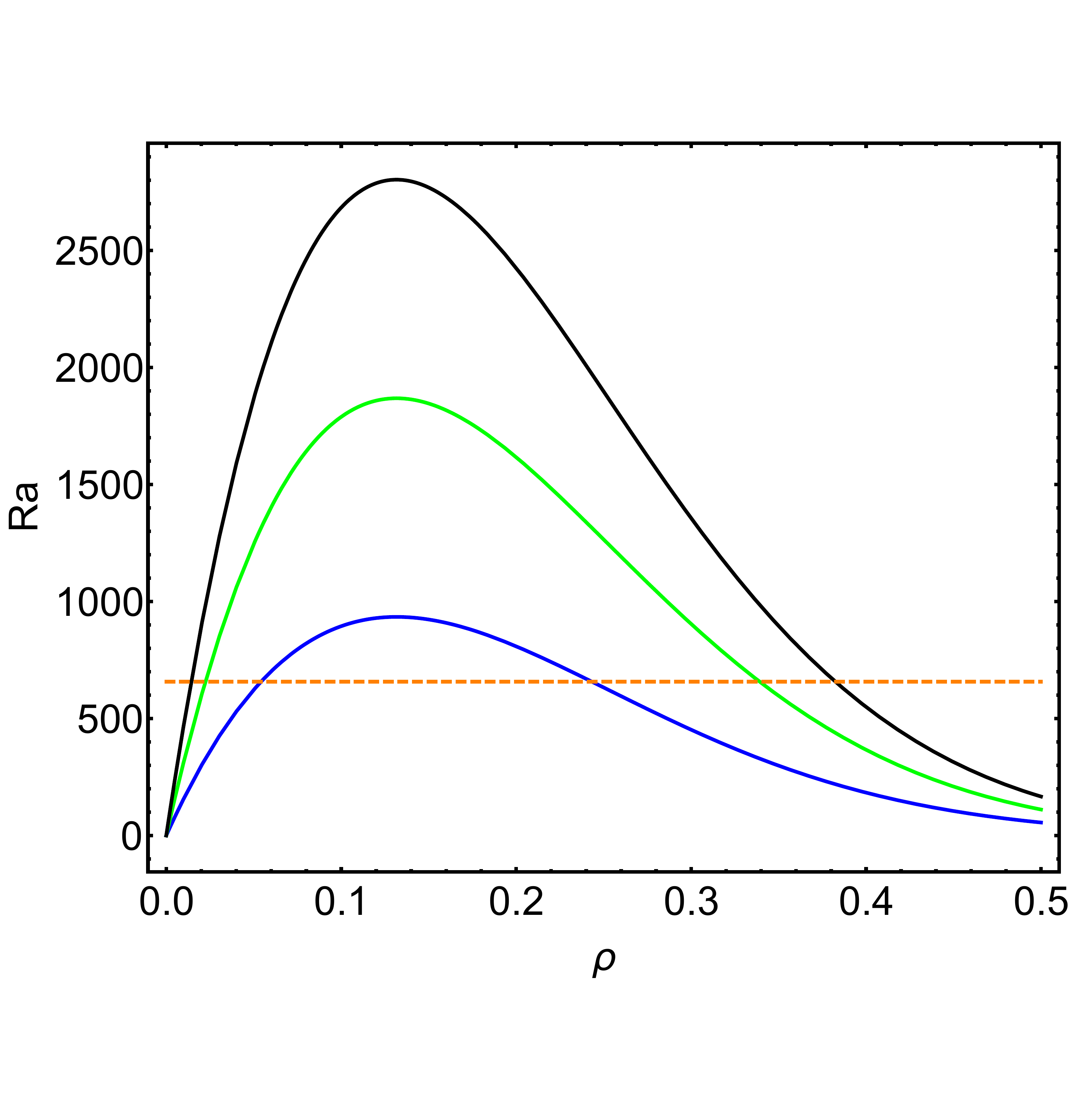}
\includegraphics[width=6cm,clip]{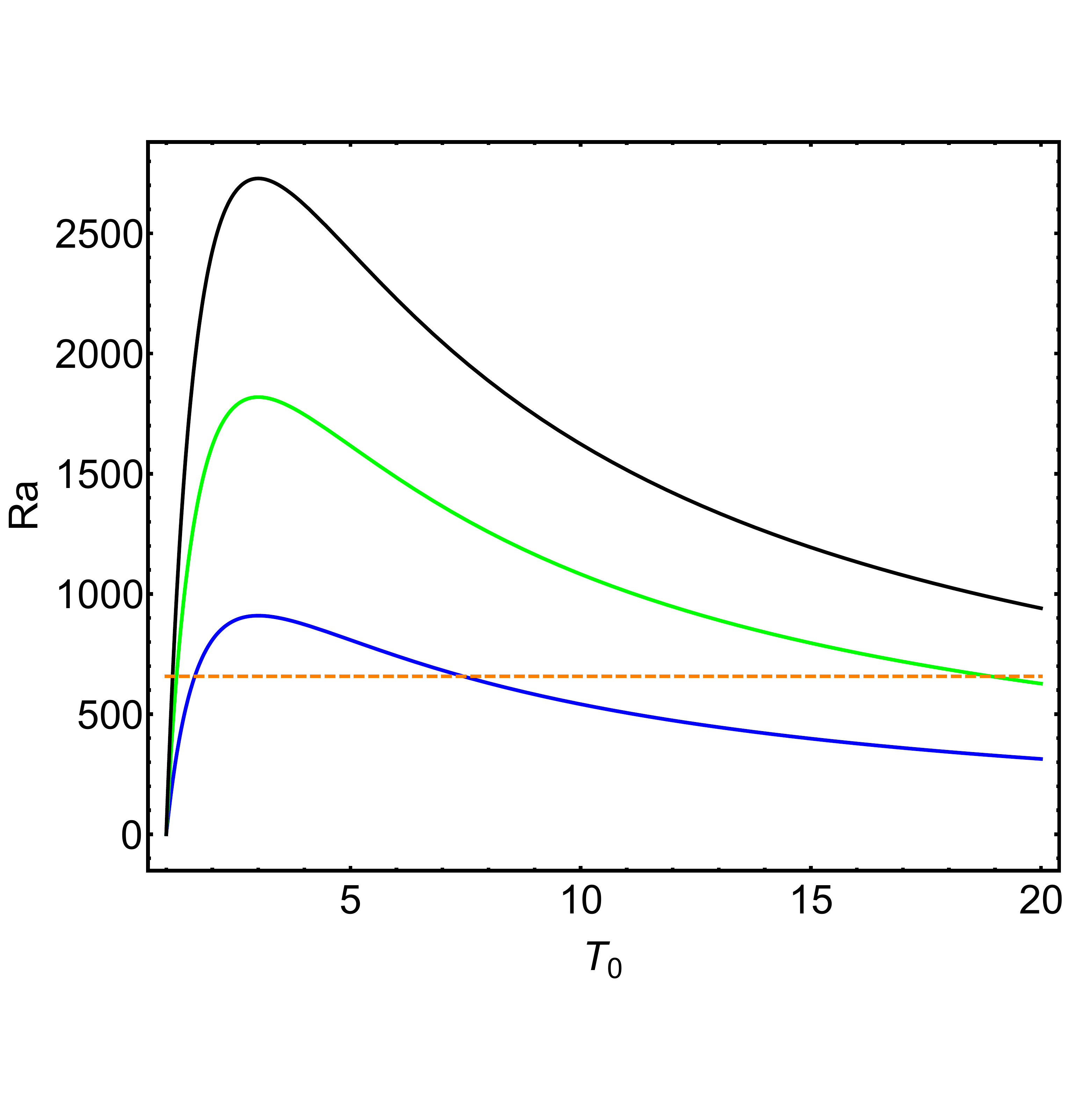}
\end{center}
\kern -1.2cm
\caption{Left: Rayleigh number in the Enskog approximation as a function of density for $T_0=5$, $T_1=0$ and $g=5$ (blue curve), $g=10$ (green curve) and  $g=15$ (black curve). Dashed orange line is the critical Rayleigh number $R_c=27\pi^4/4\simeq 657.51$ obtained using Boussinesq approximation on the Navier Stokes equations. Right: Rayleigh number as a function fo $T_0$ for a fixed value of $\rho=0.2$ for the same values of $g$ on the left figure}\label{R1}
\end{figure}

 In figure \ref{R1} we show $Ra$ computed using the Enskog transpot coefficients and the Henderson equation of state. In figure \ref{R1} left we see the behavior of $Ra$ for $T_0=5$ for different values of $g=5$, $10$ and $15$. We see that the maximum value of $Ra$ apears for low densities and it is above the critical Rayleigh number. Then, we have chosen that the average system density to be $\bar\rho=0.2$.  Figure \ref{R1} right we plot $Ra$ as a function of $T_0$ for fix $\bar\rho=0.2$ to have some idea of the critical value of $T_0$ that separates non-convective to convective regimes. We get the following critical temperatures:
 $T_0^c(g=5)=1.6205$,  $T_0^c(g=10)=1.2233$ and  $T_0^c(g=15)=1.1376$ . That is, when we increase the external field, the convective regime appears for lower temperatures.
 
From now on we will call to the simulated parameters $(T_0, T_1=1,g)$ whose chosen values are: $g=0$, $5$, $10$ and $15$, and  $T_0=1,1.2,1.4,1.6,1.8, 2.0, 2.2, 2.4, 2.6, 2.8, 3, 4, \ldots, 19, 20$. That is, we have done $112$ computer simulations. 

Before begining the analysis of the data we have obtained let us make some comments about notation, how the measurements have been done and the error bars criteria we have used.

\begin{itemize}

\item{Notation:} In general a vector $u$ have two components $u=(u_1,u_2)$. In order to do local measurements, we divide the unit box into $30\times 30$ (virtual) square cells of side $1/30$. A given cell has the position $(n,l)$ with $n,l=1,\ldots,30$. We express hydrodynamic magnitudes  at a given {\it macroscopic point} $(x,y)$ with $x,y\in[0,1]$ that correspond to the values measured of some microscopic observable at the cell $(n,l)$ such that   $x=(n-1)/30+1/60$ and $y=(l-1)/30+1/60$. That is, we use the center of each cell as the macroscopic position of the hydrodynamic fields. Finally, a macroscopic hydrodynamic field is written as $u=(u_1(x,y),u_2(x,y))$

\item{Local Measurements:}
The local measurements have been done in the following way. Let $a(r,v)$ be a magnitude that depend on a set of particle positions and velocities $r_i$ and $v_i$ respectively. Say for instance the kinetic energy of a particle: $a(r_i,v_i)=v_i^2/2=(v_{i,1}^2+v_{i,2}^2)/2$ or the local potential energy of a particle: $a(r_i,v_i)=g r_{i,2}$ where $r_i=(r_{i,1},r_{i,2})$. The extensive value of $a(r,v)$ on the cell $(n,l)$  ($n,l=1,\ldots, 30$) at time $t$ is given by: 
\begin{equation}
A(n,l;t)=\sum_{i:r_i(t)\in B(n,l)}a(r_i(t),v_i(t))
\end{equation}
where $B(n,l)$ is the spatial domain corresponding to the cell $(n,l)$. The total number of particles involved in such sum is given by:
\begin{equation}
N(n,l;t)=\sum_{i:r_i(t)\in B(n,l)} 1
\end{equation}
We do $M$ measurement of $A(n,l;t)$ during the system evolution at the stationary state. Then the average value per particle of $A$ is given by:
\begin{equation}
a(n,l)=\frac{\sum_{t=1}^M A(n,l;t)}{\sum_{t=1}^M N(n,l;t)}\label{aver}
\end{equation} 
We use this method because it has a better convergence to the limiting value when $M\rightarrow\infty$ and its fluctuations are smaller than in case we averaged $A(n,l;t)/N(n,l;t)$ over time.

\item{Analysis of errors:}
Let $\{A(t)\}_{t=1}^M$ the set of $M$ measurements of a given observable (local or global one). We assume that the data sequence is time decorrelated and then the law of large numbers apply. That is, the averaged value of $A$ behaves for large $M$ values as:
\begin{equation}
A(M)\simeq A(\infty)+\sigma(A;M)\xi
\end{equation}
where
\begin{equation}
A(M)=\frac{1}{M}\sum_{t=1}^M A(t)\quad\quad,\quad \sigma(A;M)=\frac{1}{M}\sqrt{\sum_{t=1}^M(A_t-A(M))^2}
\end{equation}
and $\xi$ is a Gaussian random variable with zero mean and variance one. In this paper the error in getting the $M\rightarrow\infty$ value is given by
\begin{equation}
A(\infty)=A(M)\pm 3\sigma(A;M)
\end{equation}
that is, we are assuming that the $99.7\%$ of the data is in the error interval we show in the analysis.

Sometimes, once we have computed the average value of the observable $A$ and its error, we need to study a function of it, say $f(A)$. Let us assume that the error is small and It can be considered as a perturbation of $A(\infty)$, then we can do a Taylor expansion:
\begin{equation}
f(A(M))=f(A(\infty)+\sigma(A;M)\xi)=f(A(\infty))+f'(A(\infty))\sigma(A;M)\xi+\frac{1}{2}f''(A(\infty))\sigma(A;M)^2\xi^2+\ldots
\end{equation}
If we average this expression over the random values $\xi$ we obtain that $\langle f(A(M))\rangle=f(A(\infty))+\frac{1}{2}f''(A(\infty))\sigma(A;M)^2$. There is a small but systematic shift on the observed value $f(A(M))$ with respect the desired $f(A(\infty))$. Let us define the observable:
\begin{equation}
B=f(A(M))-\frac{1}{2}f''(A(\infty))\sigma(A;M)^2
\end{equation}
Obviously the average value of $B$ is $f(A(\infty))$  (by construction) up to order $\sigma^4$ and the variance of $B$ is:
\begin{eqnarray}
\sigma(B)^2&=&\langle(B-f(A(\infty)))^2\rangle=\langle\left(f'(A(\infty))\sigma(A;M)\xi+\frac{1}{2}f''(A(\infty))\sigma(A;M)^2(\xi^2-1) \right)^2\rangle\nonumber\\
&=&\sigma(A;M)^2\left[f'(A(\infty))^2+\frac{1}{2}f''(A(\infty))^2\sigma(A;M)^2\right]\label{error2}
\end{eqnarray}
The error in computing $B$ is then  $B=f(A(\infty))\pm 3\sigma(B)$ and we conclude that
\begin{equation}
f(A(\infty))=f(A(M))-\frac{1}{2}f''(A(\infty))\sigma(A;M)^2\pm 3\sigma(B)\label{error1}
\end{equation}
Typically the systematic shift is smaller than the error interval and therefore not relevant for the analysis. However, sometimes we obtain small but positive data values for an observable that one knows that it should be zero in average and we should introduce this correction in order to do a correct data analysis. 

Let us consider another typical case: when we compute an observable as an spatial average of local observables with their own error. That is, let $A(x,y)$ a local observable with errors $\epsilon(x,y)$. Let
\begin{equation}
B=\frac{1}{N_C}\sum_{(x,y)}A(x,y)
\end{equation}
If we assume that the local random variables $\xi(x,y)$  associated to the errors $\epsilon(x,y)$ are {\it spatially independent}, then we can apply the Lyapunov Central Limit Theorem (see Appendix I) and  the error of $B$ is given by:
\begin{equation}
\epsilon(B)=\frac{1}{N_C}\left[ \sum_{(x,y)}\epsilon(x,y)^2\right]^{1/2}
\end{equation}
Notice that the assumption of spatial independence is not so obvious because the fluid has typically long range correlations that should affect to the data noise behavior. 

However, let us assume that the data error is {\it totally correlated} in the sense that the local random variable can be written as $\xi(x,y)=\epsilon(x,y)\xi$  where $\xi$ is now a Gaussian noise with zero mean and variance one. In this case the error of $B$ is given by:
\begin{equation}
\epsilon(B)=\frac{1}{N_C}\sum_{(x,y)}\epsilon(x,y)
\end{equation}
Observe that the totally correlated case has an error much larger that the uncorrelated one. It is not clear a priori what scheme to choose to define errors in this case without computing the data correlation function that it is very costly from a computational point of view. In any case the true error on $B$ would have a value in between of the two limiting cases.

\item{Differences between Local values and cell averages:}

To obtain local values of an observable we have build a set of virtual cells where we have
measured the desired magnitudes. That is an arbitrary choice and we could define other partition and associate a value of the magnitude on each of its elements. Therefore we can assume that at the stationary state, for each microscopic observable, there exists a function defined in ${\mathcal R}^2$ such that its averages over any partition is the value measured on the cell. In other words:
 \begin{equation}
 \tilde F(m,n)=\frac{1}{\Delta^2}\int_{\tilde x(m,n)-\Delta/2}^{\tilde x(m,n)+\Delta/2}dx \int_{\tilde y(m,n)-\Delta/2}^{\tilde y(m,n)+\Delta/2}dy\, F(x,y)
  \end{equation}
  where $\tilde F(m,n)$ is the average value of the observable that we obtain in the computer simulation and $F(x,y)$ its, assumed existing, underlying continuum density. We consider that 
 $\tilde x(m,n)=(m-1/2)\Delta$, $\tilde y(m,n)=(n-1/2)\Delta$ where $\Delta=1/N_C$ and $m,n=1,\ldots,N_C$ ($N_C=30$ in our computer simulations). The question that arises is what is the relation between this local averages and the value of  $F$ and its derivatives at the point $(\tilde x(m,n),\tilde y(m,n))$. In order to find such relations  let us write the value of $\tilde F$ for neighboring cells:
 \begin{equation}
 \tilde F(m+s,n+t)=\frac{1}{\Delta^2}\int_{\tilde x(m,n)+s\Delta-\Delta/2}^{\tilde x(m,n)+s\Delta+\Delta/2}dx \int_{\tilde y(m,n)+t\Delta-\Delta/2}^{\tilde y(m,n)+t\Delta+\Delta/2}dy\, F(x,y)
 \end{equation} 
 We can expand the last expression assuming that $F$ is analytic and $\Delta$ is small enough, then:
 \begin{eqnarray}
  \tilde F(m+s,n+t)&=&\sum_{n'=0}^{\infty}\Delta^{2n'}\sum_{k'=0}^{n'} F_{(2k',2n'-2k')}(m,n) a(s,k)a(t,n'-k')\nonumber\\
  &+&\sum_{n'=0}^{\infty}\Delta^{2n'+2}\sum_{k'=0}^{n'} F_{(2k'+1,2n'-2k'+1)}(m,n) b(s,k)b(t,n'-k')\nonumber\\
   &+&\sum_{n'=0}^{\infty}\Delta^{2n'+1}\sum_{k'=0}^{n'} F_{(2k'+1,2n'-2k')}(m,n) b(s,k)a(t,n'-k')\nonumber\\
    &+&\sum_{n'=0}^{\infty}\Delta^{2n'+1}\sum_{k'=0}^{n'} F_{(2k',2n'-2k'+1)}(m,n) a(s,k)b(t,n'-k')\label{des}
 \end{eqnarray}
 where $F_{(n',l')}(m,n)=\partial_x^{n'}\partial_y^{l'} F(x,y)\vert_{\tilde x(m,n),\tilde y(m,n))}$ and
 \begin{eqnarray}
 a(s,l)&=&\sum_{k=0}^l\frac{1}{(2k)!(2l-2k+1)!}\frac{s^{2k}}{2^{2l-2k}}\nonumber\\
  b(s,l)&=&\sum_{k=0}^l\frac{1}{(2k+1)!(2l-2k+1)!}\frac{s^{2k+1}}{2^{2l-2k}}\nonumber\label{des2}
 \end{eqnarray}
 Observe that $a(-s,l)=a(s,l)$ and $b(-s,l)=-b(s,l)$. These symmetry properties allows us to break eq. (\ref{des}) in a set of four equations:
 \begin{eqnarray}
  G_1(m,n;s,t)&\equiv&\frac{1}{4}\left[ \tilde F(m+s,n+t)+ \tilde F(m+s,n-t)+ \tilde F(m-s,n+t)+ \tilde F(m-s,n-t)\right]\nonumber\\
  &=&\sum_{n'=0}^{\infty}\Delta^{2n'}\sum_{k'=0}^{n'} F_{(2k',2n'-2k')}(m,n) a(s,k)a(t,n'-k')\nonumber\\
   G_2(m,n;s,t)&\equiv&\frac{1}{4}\left[ \tilde F(m+s,n+t)- \tilde F(m+s,n-t)-\tilde F(m-s,n+t)+ \tilde F(m-s,n-t)\right]\nonumber\\
  &=&\sum_{n'=0}^{\infty}\Delta^{2n'+2}\sum_{k'=0}^{n'} F_{(2k'+1,2n'-2k'+1)}(m,n) b(s,k)b(t,n'-k')\nonumber\\
  G_3(m,n;s,t)&\equiv&\frac{1}{4}\left[ \tilde F(m+s,n+t)+ \tilde F(m+s,n-t)- \tilde F(m-s,n+t)- \tilde F(m-s,n-t)\right]\nonumber\\
  &=&\sum_{n'=0}^{\infty}\Delta^{2n'+1}\sum_{k'=0}^{n'} F_{(2k'+1,2n'-2k')}(m,n) b(s,k)a(t,n'-k')\nonumber\\
  G_4(m,n;s,t)&\equiv&\frac{1}{4}\left[ \tilde F(m+s,n+t)- \tilde F(m+s,n-t)+ \tilde F(m-s,n+t)- \tilde F(m-s,n-t)\right]\nonumber\\ 
  &=&\sum_{n'=0}^{\infty}\Delta^{2n'+1}\sum_{k'=0}^{n'} F_{(2k',2n'-2k'+1)}(m,n) a(s,k)b(t,n'-k')\label{des3}
 \end{eqnarray}
 Let us study the $G_1$ case (the others follow straight forward). First, let us cut the infinite sum on the right hand side by keeping only up to order $\Delta^{2N_0}$. That is, 
 \begin{equation}
 G_1(m,n;s,t)=\sum_{n'=0}^{N_0}\Delta^{2n'}\sum_{k'=0}^{n'} F_{(2k',2n'-2k')}(m,n) a(s,k)a(t,n'-k')+ O(\Delta^{2N_0+2})\label{des5}
 \end{equation}
 The unknowns are the set of derivatives $F_{(n,l)}$ and the data are $G_1(m,n;s,t)$ by varying the values of $s$ and $t$ (being always integers). We have $L=(N_0+1)(N_0+2)/2$ unknowns and then we should give a set of $L$-values of pairs $(s,t)$ in order to have a linear set of equations to be solved. We choose the set of natural values $(s,t)$ such that $s+t=m$ with $m=0,\ldots,N_0$ and $s=0,\ldots,m$. 
 The tricky issue is to codify the index $(s,t)$ and/or $(k',n'-k')$ in order to create a new index $\alpha$ running from $1,\ldots,L$. It is easy to define $\alpha$ as a function of $(s,t)$:
 \begin{equation}
 \alpha=\frac{m(m+1)}{2}+s+1\quad\quad, m=t+s\quad\quad, m=0,1,\ldots,N_0\quad\quad, s=0,1,\ldots,m\label{des4}
 \end{equation}
 where $\alpha=1,\ldots,L$. We also need the inverse relation, that is, $(s,t)$ as a function of $\alpha$. Let us assume that $s=0$ in eq.(\ref{des4}). In this case $m=(-1+\sqrt{1+8(\alpha-1)})/2$. For $s>0$ the value of $m$ should be the same, therefore
 \begin{equation}
 m(\alpha)=\text{Int}\left(\frac{-1+\sqrt{1+8(\alpha-1)}}{2} \right)\quad \quad s(\alpha)=\alpha-1-\frac{m(\alpha)(m(\alpha)+1)}{2}
 \end{equation}
 Therefore we can write eq.(\ref{des5}):
 \begin{equation}
 G_1(m,n;s(\alpha),t(\alpha))=\sum_{\beta=1}^L  a(s(\alpha),s(\beta))a(t(\alpha),t(\beta)) \Delta^{2n(\beta)} F_{(2s(\beta),2t(\beta))}(m,n) +O(\Delta^{2N_0+2})
 \end{equation}
 where $t(\alpha)=m(\alpha)-s(\alpha)$. With this codification we manage to keep all terms in a given perturbation order. By inverting these  set of linear equations we can express $F$'s as a function of $G_1$ up to order $\Delta^{2N_0}$.
 Similar strategy can be followed for $G_2$, $G_3$ and $G_4$. The only change is that $b(0,t)=0$ and we should use a little different set of index $(s,t)$. In particular
 \begin{eqnarray}
 G_2&=&G_2(s(\alpha)+1,m(\alpha)-s(\alpha)+1)\nonumber\\
 G_3&=&G_3(s(\alpha)+1,m(\alpha)-s(\alpha))\nonumber\\
 G_4&=&G_4(s(\alpha),m(\alpha)-s(\alpha)+1)
 \end{eqnarray}
 For example, if we choose $N_0=1$ we get the first derivatives:
 \begin{eqnarray}
 F(\tilde x(m),\tilde y(n))&=&\frac{1}{24}\left[-\tilde F(m-1,n)-\tilde F(m,n-1)+28 \tilde F(m,n)-\tilde F(m,n+1)-\tilde F(m+1,n)\right]\nonumber\\
 &+& O(\Delta^4)\nonumber\\
 \partial_x F(x,y)\vert_{\tilde x(m,n),\tilde y(m,n)}&=&\frac{1}{48\Delta}\biggl[5 \tilde F(m-2,n)+\tilde F(m-1,n-1)-36 \tilde F(m-1,n)+\tilde F(m-1,n+1)\nonumber\\
 &-&\tilde F(m+1,n-1)+36 \tilde F(m+1,n)-\tilde F(m+1,n+1)-5 \tilde F(m+2,n)\biggr]+O(\Delta^4)\nonumber\\
 \partial_y F(x,y)\vert_{\tilde x(m,n),\tilde y(m,n)}&=&\frac{1}{48\Delta}\biggl[5 \tilde F(m,n-2)+\tilde F(m-1,n-1)-36 \tilde F(m,n-1)+\tilde F(m+1,n-1)\nonumber\\
 &-&\tilde F(m-1,n+1)+36 \tilde F(m,n+1)-\tilde F(m+1,n+1)-5 \tilde F(m,n+2)\biggr]+O(\Delta^4)\nonumber\\
 \partial_x^{2} F(x,y)\vert_{\tilde x(m,n),\tilde y(m,n)}&=&\frac{1}{\Delta^2}\left[\tilde F(m-1,n)-2 \tilde F(m,n)+\tilde F(m+1,n)\right]+O(\Delta^2)\nonumber\\
\partial_y^{2} F(x,y)\vert_{\tilde x(m,n),\tilde y(m,n)}&=&\frac{1}{\Delta^2}\left[\tilde F(m,n-1)-2 \tilde F(m,n)+\tilde F(m,n+1)\right]+O(\Delta^2)\nonumber\\
\partial_x\partial_y F(x,y)\vert_{\tilde x(m,n),\tilde y(m,n)}&=&\frac{1}{96\Delta^2}\biggl[44\left (\tilde F(m-1,n-1)-\tilde F(m-1,n+1)-\tilde F(m+1,n-1)+\tilde F(m+1,n+1)\right)\nonumber\\
&-&5 \left(\tilde F(m-1,n-2)-\tilde F(m-1,n+2)-\tilde F(m+1,n-2)+\tilde F(m+1,n+2)\right)\nonumber\\
&-&5\left(\tilde F(m-2,n-1)-\tilde F(m-2,n+1)-\tilde F(m+2,n-1)+\tilde F(m+2,n+1)\right)\biggr]\nonumber\\
&+&O(\Delta^4)\nonumber\\
 \end{eqnarray}
 
 We have checked the effect of using $\tilde F(m,n)$ or $F(\tilde x(m),\tilde y(n))$ for different observables. In figure \ref{des6} we show as examples the differences between the cell-values and the point values (obtained by the above expressions) for the  temperature and the x-component of the hydrodynamic velocity fields (defined below) for $T_0=17$ and $g=10$. We observe how the average value (over the cells) of the differences are $-0.0004 (0.0009)$ and $0.0000 (0.0003)$ and its standard deviations are $0.008$ and $0.003$ for $T$ and $u_1$ respectively. The statistical errors are of order $0.1$ and $0.02$ in these cases. Moreover, the distribution of the differences are quite random and the normalized distribution (subtracting the average and dividing by the standard deviation) has a third central moment of $-0.056$ and $-0.023$ and a kurtosis of $0.667$ and  $0.589$ for $T$ and $u_1$ cases respectively. That is, we can use $\tilde F(m,n)$ with the confidence that the 
 corrections do not introduce systematic deviations to the average values because it is just a small noise to be added to the already larger statistical error.  Observe that in \cite{delPozo} we measured the global hydrodynamic velocity moments with errors smaller than $10^{-3}$ and it was necessary there to use this corrections to do a correct analysis. 
 
 Nevertheless we have build a well defined scheme to obtain the derivatives of an observable  at any given central cell point. That is going to be very useful when comparing with the Navier-Stokes equations.
 
 \begin{figure}
\begin{center}
\includegraphics[width=6cm,clip]{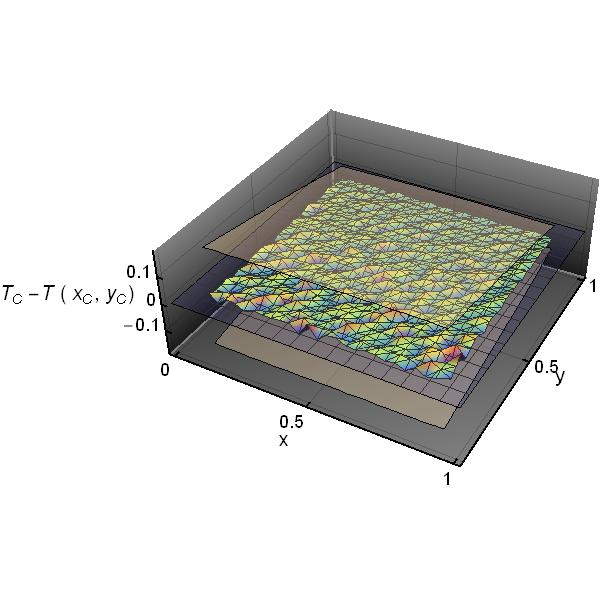}%temp_E10_prueba.nb
\includegraphics[width=6cm,clip]{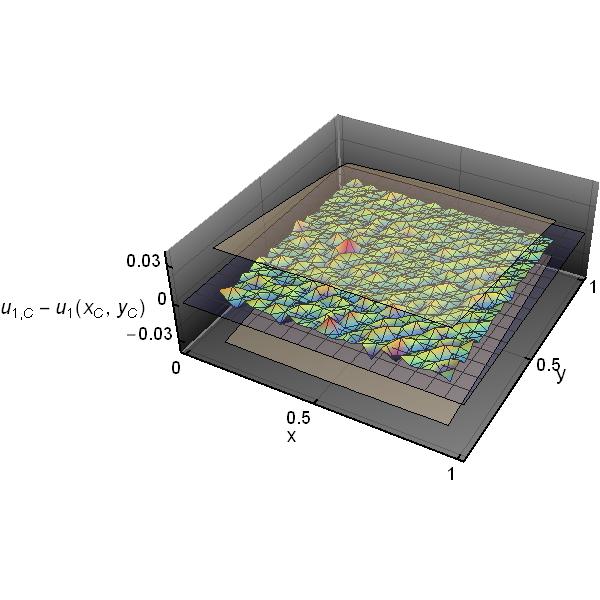} %vx_E10_prueba.nb
\includegraphics[width=6cm,clip]{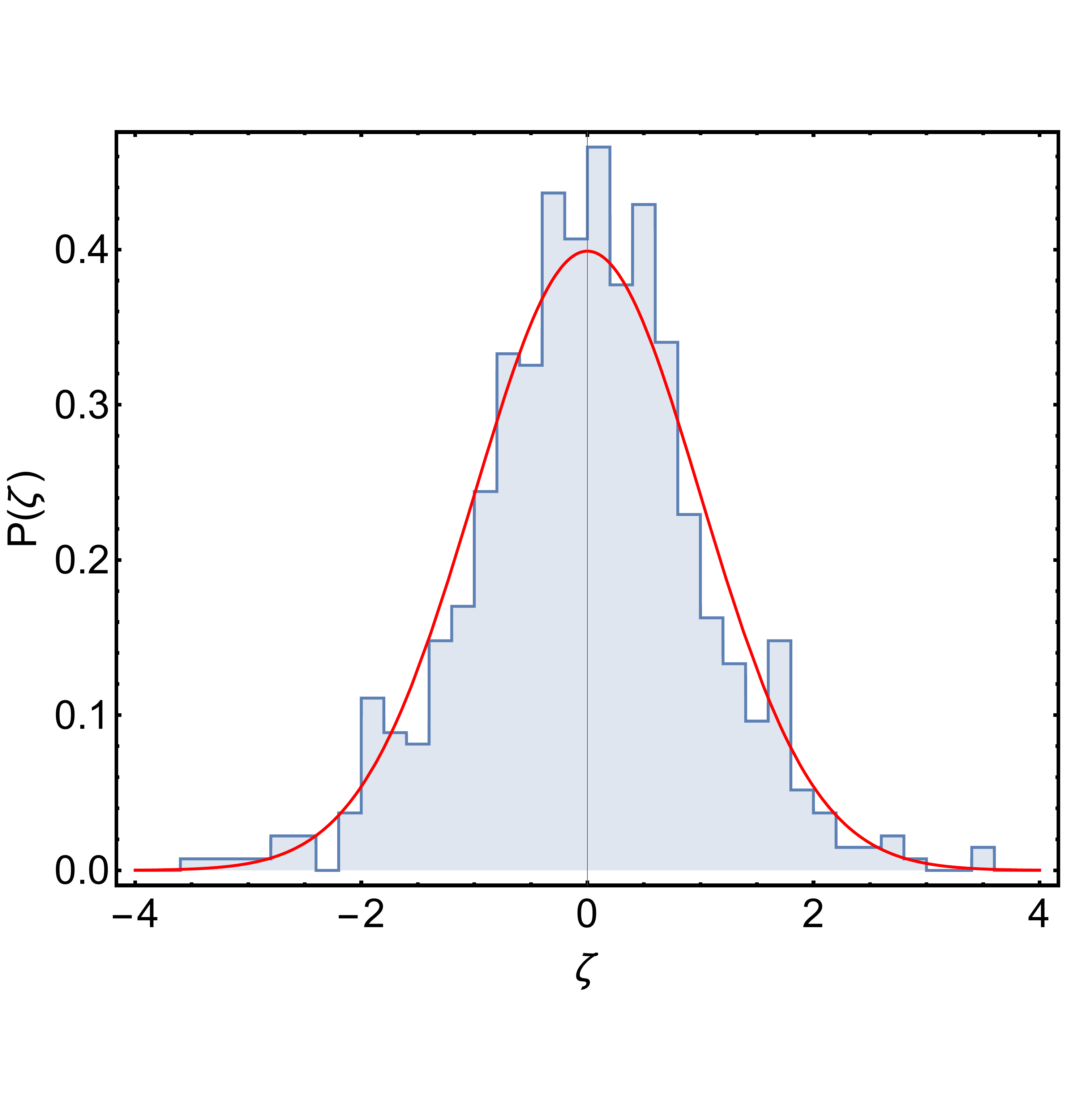}%temp_E10_prueba.nb
\includegraphics[width=6cm,clip]{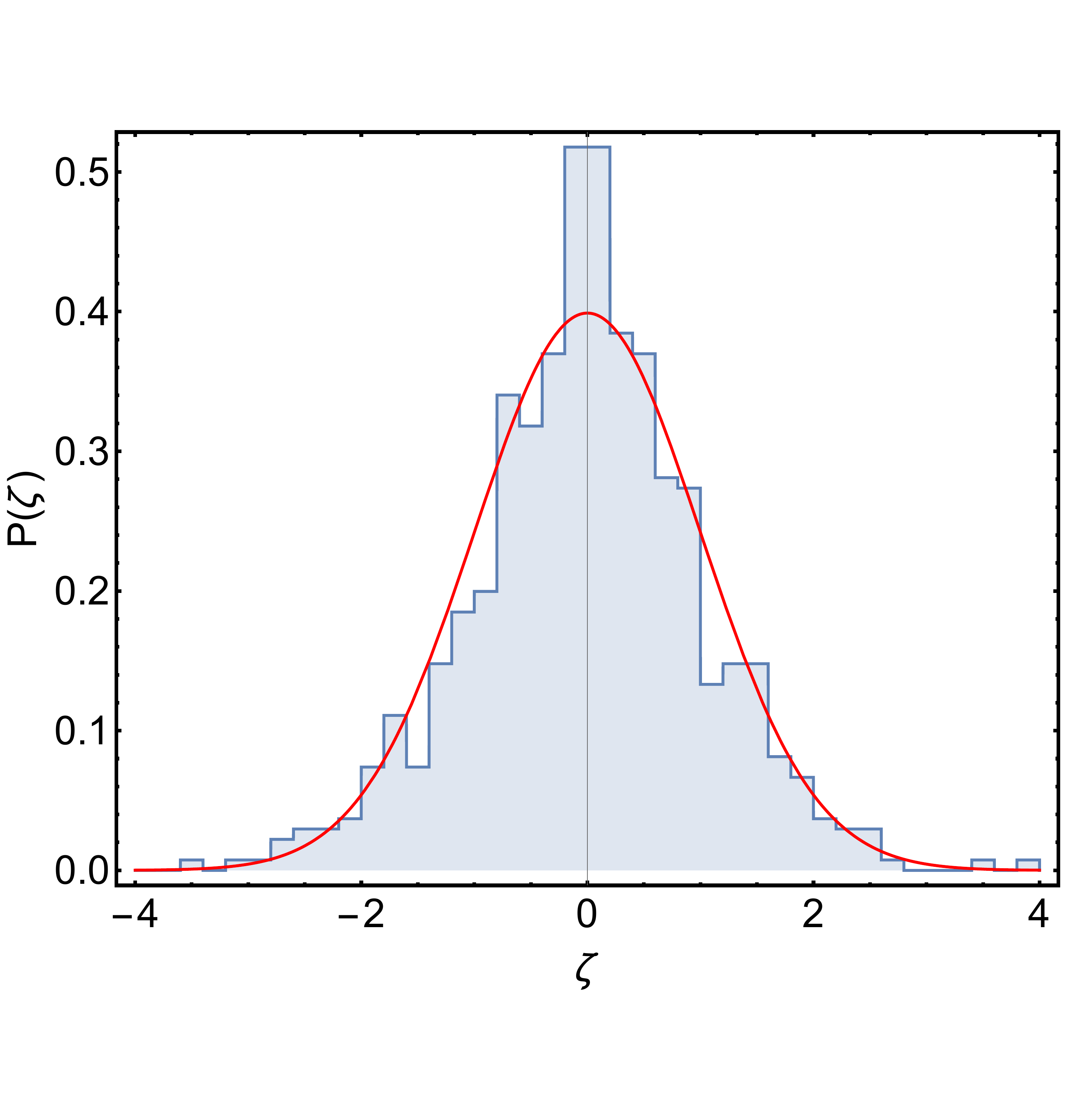} %vx_E10_prueba.nb
\end{center}
\kern -1.2cm
\caption{Differences between $\tilde F(m,n)\equiv F_C$ and $F(\tilde x(m,n), \tilde y(m,n))\equiv F(x_C,y_C)$ for the temperature field, $T$, (upper left) and the x-component of the velocity field $u_1$ (upper right) for $T_0=17$ and $g=10$. The Gray smooth surfaces are the statistical error interval. Below each figure there is the corresponding normalized probability distribution for the differences. Solid red lines are the Gaussian $N(0,1)$.
}\label{des6}
\end{figure} 
 \end{itemize}

\section{Computer Simulation Results: Equilibrium}

At equilibrium, we know that a system at temperature $T$,  the pressure and the density are related by the {\it equation of state} and the {\it hydrostatic formula}, that in our system can be written:
\begin{eqnarray}
Q(y)&=&T \rho(y) H(\rho(y))\\
\frac{dQ(y)}{dy}&=&-g\rho(y) \label{eqpress0}
\end{eqnarray}
where $Q(y)=P(y)\pi r^2$ and we have considered the mass of the disk to be one: $m=1$. This expressions  and the boundary conditions, determine the behavior of $Q$ and $\rho$ as a function of the height $y$. 

The analytic expression of $H(\rho)$ for the hard disk system is unknown. However it is well known that the Henderson expression for $H(\rho)=(1+\rho^2/8)/(1-\rho)^2$ is a very good approximation of the equation of state (EOS) in the range of $\rho\in[0,0.5]$ with an relative error smaller than the $1\%$. We  used Henderson's EOS  to solve the differential equation and we got:
\begin{eqnarray}
Q&=&T\rho\frac{1+\rho^2/8}{(1-\rho)^2}\nonumber\\
y&=&-\frac{1}{g}\left[\log\rho -\frac{7}{8}\log(1-\rho)+\frac{7}{8(1-\rho)}+\frac{9}{8(1-\rho)^2}+C\right] \label{eqpress}
\end{eqnarray}
where the constant $C$ is fixed by giving a point: $(y_0,\rho_0)$.
We can check the behavior of our computer program  by measuring the stationary density and pressure profiles for the equilibrium case $T_0=T_1=1$ and to compare with the exact results.

The local areal density, $\rho(n,l)$,  at cell  $(n,l)$ is computed as:
\begin{equation}
\rho(n,l)=\frac{N(n,l)\pi r^2}{\Delta^2}\label{ro}
\end{equation}
where $N(n,l)$ is the average number of particles with its center at the cell $(n,l)$ at the stationary state, $r$ is the radius of the particles and $\Delta$ is the side length of the cell ($\Delta=1/30$ in all the simulations). This definition is computationally very fast and convenient by its simplicity but it has a drawback when computing the density at the boundary cells: they present a systematic deviation due to the surface exclusion around the walls and therefore, the density is underestimated at these cells. Therefore in our analysis we should exclude the boundary cells and even their neighbor cells to minimize these effects.

\begin{figure}[h!]
\begin{center}
\includegraphics[height=5cm,clip]{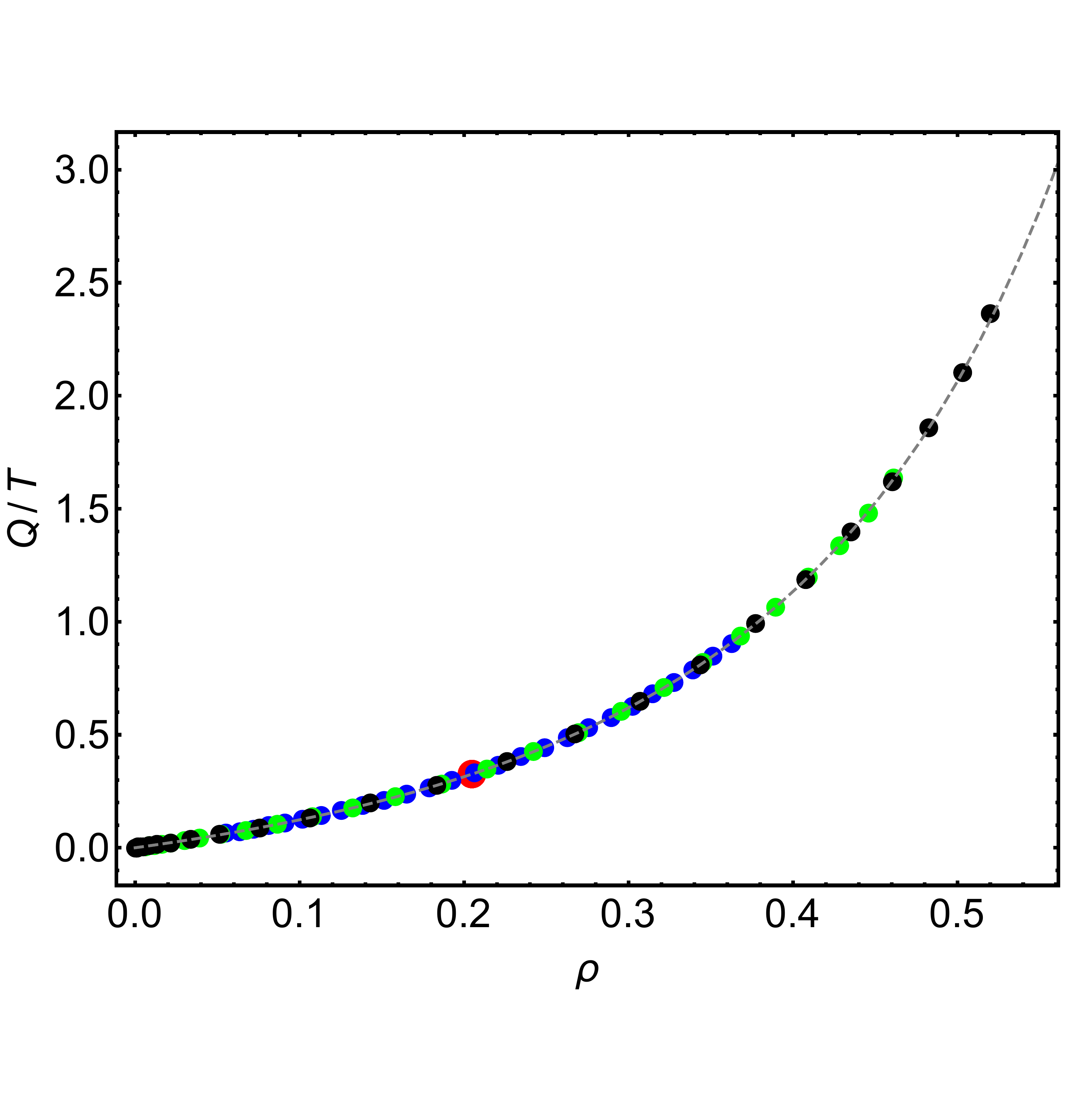} %press_prof_y_eq0.nb
\includegraphics[height=5cm,clip]{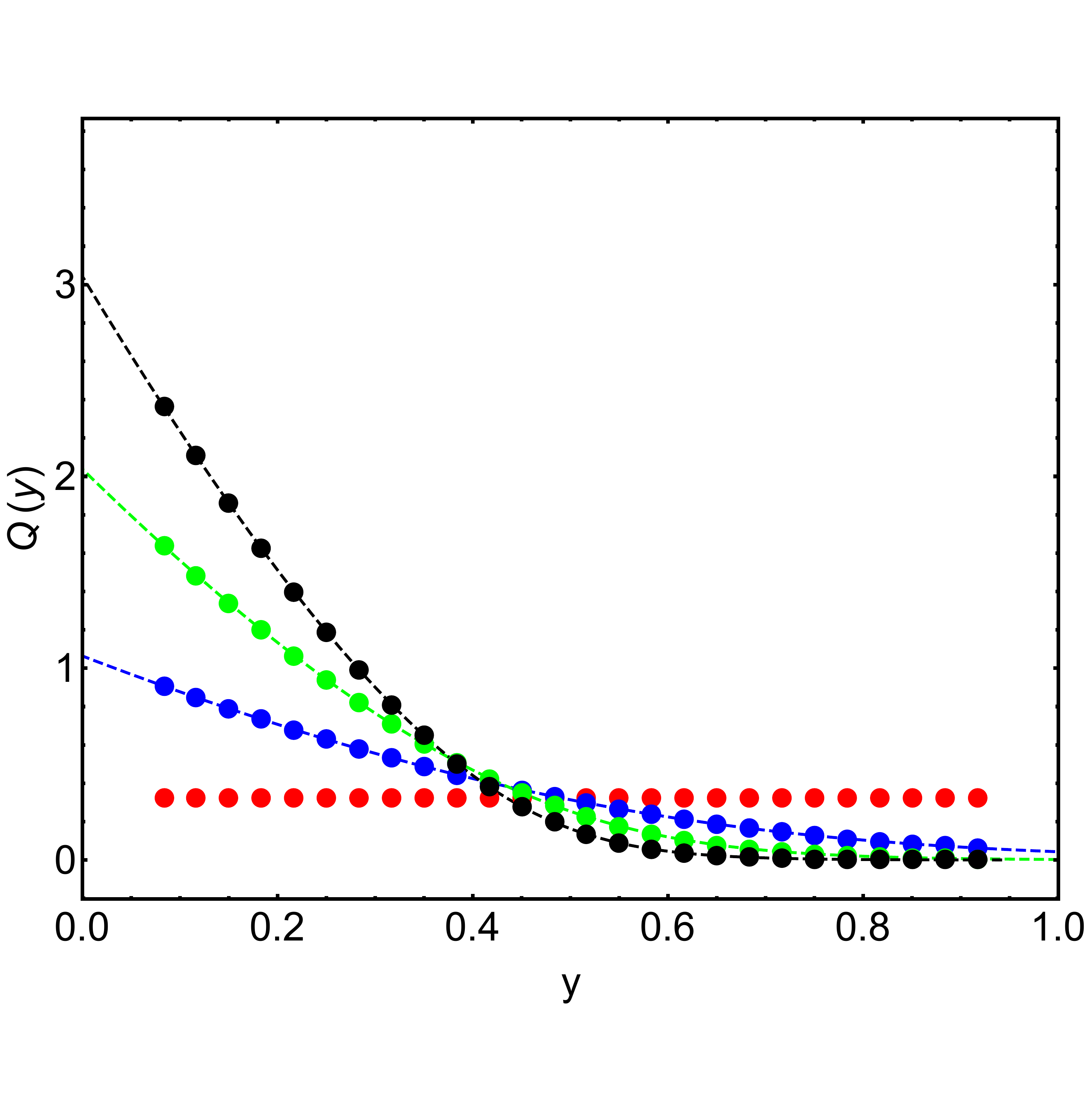}  
\includegraphics[height=5cm,clip]{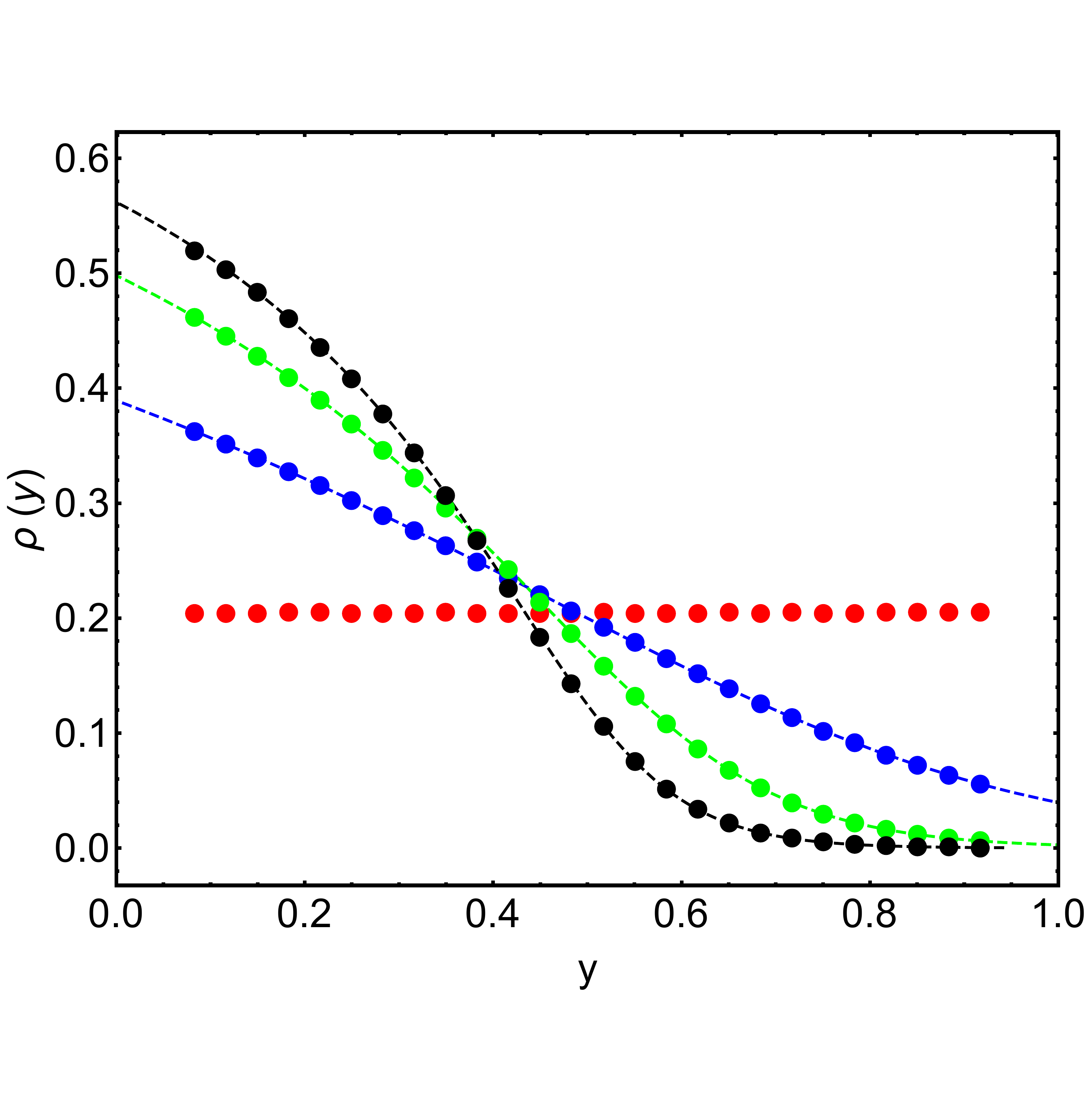}
\end{center}
\kern -1.cm
\caption{Equilibrium behavior ($T_0=T_1=1$). Left: Equation of State (EOS). The measured reduced pressure $Q(y)/T$ versus $\rho(y)$ for $g=0$ (red big dot), $g=5$ (blue dots), $g=10$ (green dots) and $g=15$ (black dots).  Center: Density height profile. Right: Virial pressure height profile. Dashed lines are the predictions using the hydrostatic formula and the Henderson's EOS (see text).  \label{eq_prof}}
\end{figure}

We use the virial theorem to compute the pressure field $P(x,y)$. Let us first derive the expression of $P$ for the hard disk case at equilibrium with no external field. Let $\alpha$ be a square spatial region of side $L$ where there are $N$ interacting particles of mass $m$. The {\it virial theorem} states:
 \begin{equation}
 \langle \sum_{i=1}^N r_i\cdot F_i\rangle=-2\langle E_c\rangle
 \end{equation}
 where $r_i$ is the vector position of particle $i$, $F_i$ is the total force acting on particle $i$, $E_c$ is the system kinetic energy observable and $\langle\cdot\rangle$ is the equilibrium ensemble average. A simple proof can be given using the ergodic theorem:
 \begin{eqnarray}
 \langle \sum_{i=1}^N r_i\cdot F_i\rangle&=&\lim_{\tau\rightarrow\infty}\frac{1}{\tau}\int_0^\tau dt \sum_{i=1}^N r_i(t)\cdot F_i(t)\nonumber\\
 &=&\lim_{\tau\rightarrow\infty}\frac{m}{\tau}\int_0^\tau dt \sum_{i=1}^N r_i(t)\cdot \frac{d^2 r_i(t)}{dt^2}\nonumber\\
 &=&-\lim_{\tau\rightarrow\infty}\frac{m}{\tau}\int_0^\tau dt \sum_{i=1}^N \left(\frac{dr_i}{dt}\right)^2=-2\langle E_c\rangle
 \end{eqnarray}
 The forces acting over each particle are the sum of two terms: internal and external, that is
 \begin{equation}
 F_i=\sum_{j\neq i}F_{ij}+F^{ext}_i
 \end{equation}
 Let us include this fact into the left hand side of the virial theorem. First, let us assume that the external force only acts at the system boundaries and its force is perpendicular to the box side with constant magnitude $F$. Then:
 \begin{equation}
\langle \sum_{i=1}^N r_i\cdot F^{ext}_i\rangle=\left(x_R-x_L-y_U+y_D\right)\tilde F=-2L\tilde F=-2SP
  \end{equation}
  where $x_{L,D}$ and $y_{U,D}$ are the coordinates of the box sides and $\tilde F$ is the average total force applied on each side. This average force is assumed to be constant in magnitude because it is assumed that the system is in mechanical equilibrium. Finally we have defined the {\it mechanical pressure} $P=\tilde F/L$.

 The inter-particle forces give the following contribution:
 \begin{equation}
 \langle \sum_{i=1}^N r_i\cdot \sum_{j\neq i}F_{ij}\rangle=\frac{1}{2}\sum_{i,j}\langle(r_i-r_j)\cdot F_{ij}\rangle=\sum_{\langle i,j\rangle}\langle(r_i-r_j)\cdot F_{ij}\rangle
 \end{equation}
  where we have use the fact that $F_{ij}=-F_{ji}$ ad $\langle i,j\rangle$ is the set of different pair of particles. For hard disks, the interaction only occurs when a pair of particles collide. Let us assume that the collision occurs in a time interval $2\epsilon$ very small. Then we can write $F_{ij}$
as the variation of the linear moment on such time interval:
  \begin{equation}
  F_{ij}(t_n)\simeq\frac{1}{2\epsilon}\left(p_i(t_n+\epsilon)-p_i(t_n-\epsilon)\right)
  \end{equation}
  where $t_n$ is the time where the collision occurs. Because the collision is elastic, the moment along the vector that connects the center of the disks are exchanged between the particles:
  \begin{equation}
 r_{ij}\cdot p_i(t_n-\epsilon)=r_{ij}\cdot p_j(t_n+\epsilon)
 \end{equation}
 where $r_{ij}=r_i-r_j$. Therefore,
 \begin{equation}
 r_{ij}\cdot F_{ij}=\frac{1}{2\epsilon}\left(r_{ij}p_i(t_n+\epsilon)-r_{ij}p_i(t_n-\epsilon)\right)=\frac{1}{2\epsilon}r_{ij}\cdot p_{ij}
 \end{equation}
 where $p_{ij}=p_i(t_n+\epsilon)-p_j(t_n+\epsilon)$. Finally, we can write:
 \begin{equation}
  \langle \sum_{i=1}^N r_i\cdot \sum_{j\neq i}F_{ij}\rangle=\lim_{\tau\rightarrow\infty}\frac{1}{\tau}\sum_{n:t_n\in[0,\tau]}r_{ij}\cdot p_{ij}
 \end{equation}
 where the sum is over all the pair collisions occurring in the time interval $[0,\tau]$.
 
 The virial theorem for hard disks can be written:
 \begin{equation}
 -2SP+\lim_{\tau\rightarrow\infty}\frac{1}{\tau}\sum_{n:t_n\in[0,\tau]}r_{ij}\cdot p_{ij}=-2NT
 \end{equation}
 where at equilibrium $\langle E_c\rangle=NT$ being $T$ the system temperature. Finally:
 \begin{equation}
 P=\frac{\rho T}{\pi r^2}+\lim_{\tau\rightarrow\infty}\frac{1}{2 S \tau}\sum_{n:t_n\in[0,\tau]}r_{ij}\cdot p_{ij}
 \end{equation}
 where $\rho=N\pi r^2/S$.

When the external field is turned on, and assuming that the conditions to derive the hydrostatic formula hold, we can use the virial expression {\it locally}. That is so because we showed that the system is in a kind of {\it local equilibrium} with a given temperature and a local mean density and the mechanical pressure obtained with the virial theorem is, in such conditions, equal to the thermodynamic pressure. That is

 \begin{equation}
 P(x,y)=\frac{\rho(x,y) e_c(x,y)}{\pi r^2}+\lim_{\tau\rightarrow\infty}\frac{1}{2 \Delta^2 \tau}\sum_{n:t_n\in[0,\tau]}r_{ij}\cdot p_{ij}\label{pre0}
 \end{equation}
where $\Delta=1/30$ is the side length of a cell, $e_c(x,y)=\langle E_c(x,y)\rangle/\langle N(x,y)\rangle$ is the average total kinetic energy in the cell and the sum runs over all particle-particle collisions that occur at the cell $(x,y)$ at the time interval $[0,\tau]$ assuming that at time $0$ the system is at the stationary state.  
\begin{figure}[h!]
\begin{center}
\includegraphics[height=6cm,clip]{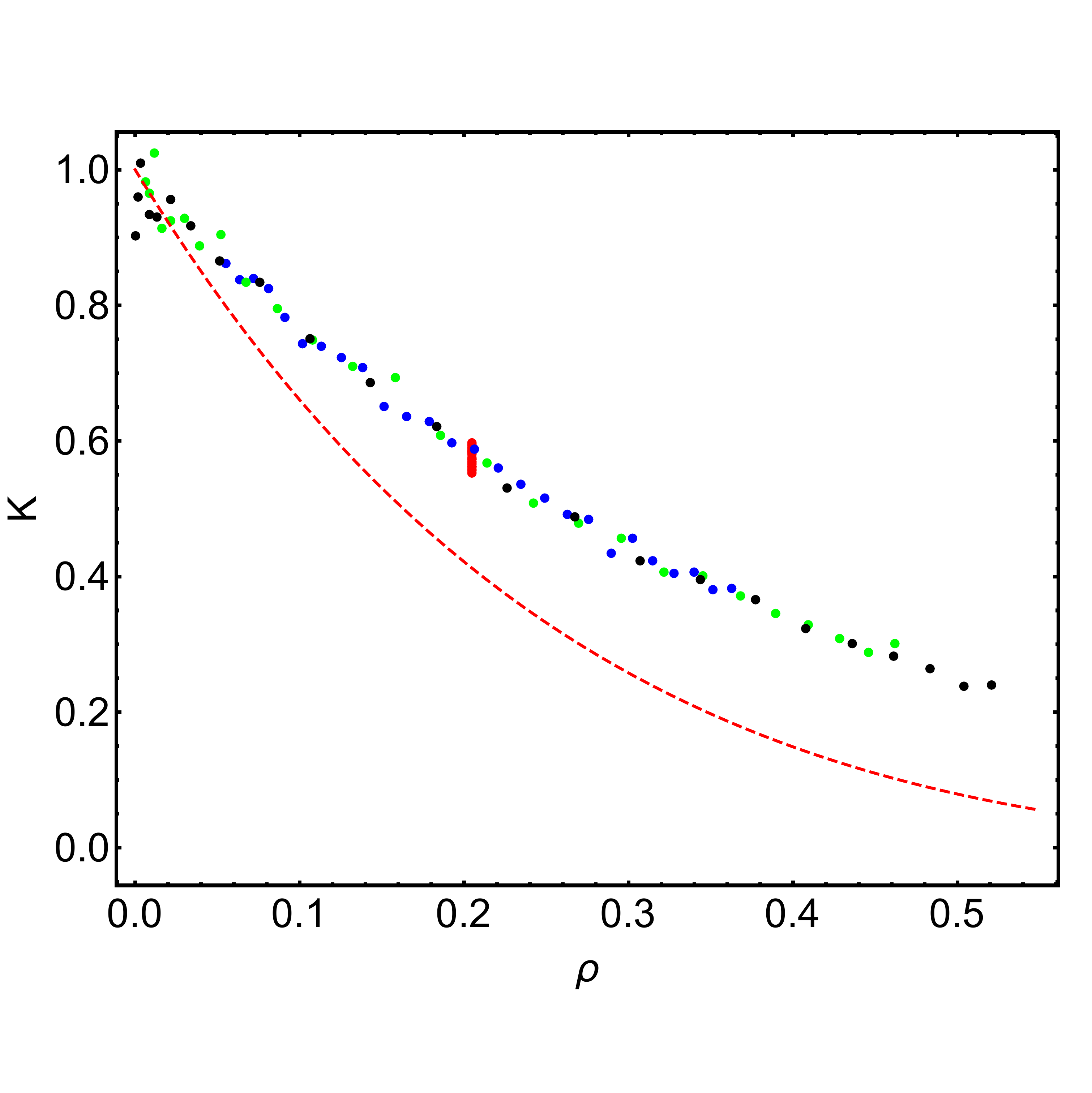} %compresi.nb
\end{center}
\kern -1.cm
\caption{$K$ equilibrium behavior ($T_0=T_1=1$) as a function of $\rho$  for $g=0$ (red  dots), $g=5$ (blue dots), $g=10$ (green dots) and $g=15$ (black dots).  The red dashed line is the theoretical curve assuming Henderson's EOS \label{denf1}}
\end{figure}

We have measured the cell values for $\rho$ and $P$ for the equilibrium case  ($T_0=T_1=1$). Then, we have done averages along the $x$-rows to build the $\rho(y)$ and $P(y)$ functions. We discarded the two first boundary cells in $x$ and $y$. We show in figure \ref{eq_prof} the obtained results and the comparison with the theory. First we plotted $Q(y)/T=\pi r^2P(y)/T$ vs $\rho(y)$ for the $26$ different values of $y$ and $g=0$, $5$, $10$ and $15$. We observe how all the data scale in a universal curve. Moreover, the Henderson EOS fits very well such curve. That means that the {\it local equilibrium} used in the derivation of the hydrostatic formula is correct and we are capable to capture such property even we are using very small sized cells. That could be due to the strong chaotic disk behavior where, in practice, the spatial and temporal correlations decay very fast. Second, we see how the profiles of $\rho(y)$ and $Q(y)$ for different values of $g$ also follow the solutions obtained with the hydrostatic formula with the Henderson's EOS. 

We also study the relative local fluctuations of the number of particles:
\begin{equation}
K(x,y)=\frac{\langle (N(x,y)-\langle N(x,y)\rangle)^2\rangle}{\langle N(x,y)\rangle}\label{denf0}
\end{equation}
where $N(x,y)$ is the observable number of particles at cell $(x,y)$. We know from equilibrium statistical mechanics that 
\begin{equation}
K(x,y)=\frac{\langle N(x,y)\rangle}{\Delta^2}T(x,y)\kappa_T(x,y)
\end{equation}
where $\Delta$ is the cell side length ($\Delta=1/30$ in our case). $\kappa_T(x,y)$ is the isothermal compressibility that it may be derive from the EOS and for hard disks case it can be written
\begin{equation}
\kappa_T(x,y)=\frac{\pi r^2}{T\rho(x,y)}\frac{1}{H(\rho(x,y))+\rho(x,y) H'(\rho(x,y))}
\end{equation}
with $H'(\rho)$ being the derivative of $H$. As in the pressure case above, we have done averages along the $x$-rows and we have discarded the two first boundary cells. We show in figure \ref{denf1} the data obtained for the equilibrium case for different values of $g$.  We also compare with the theoretical case using the Henderson's EOS. We observe (1) all the data collapse over a common curve which implies that the local equilibrium is again correct and (2) it deviates significantly from the theoretical value. We thing that this is due to the size of the cell in which we measure the fluctuations. In our case $\Delta\simeq 4.09 r$ which is too small to consistently capture the local pair correlation behavior as it were an infinite system. Therefore we are seeing strong size effects due to the size of our virtual cells. A complete discussion about the behavior of $K$ for hard disks at equilibrium when measuring it on small cells can be found on \cite{Roman}. In order to get reasonable results we would need to simulate systems with at least $10^7$ particles which it is far from our actual computer capabilities. We  conclude that there is no reason to study local fluctuations of any magnitudes due to such large finite size effects. In any case our simulation is able to describe with precision the local equilibrium behavior of the hard disk gas and the averages of the main magnitudes as the density, temperature and pressure.

\begin{figure}[h]
\begin{center}
\includegraphics[height=6cm,clip]{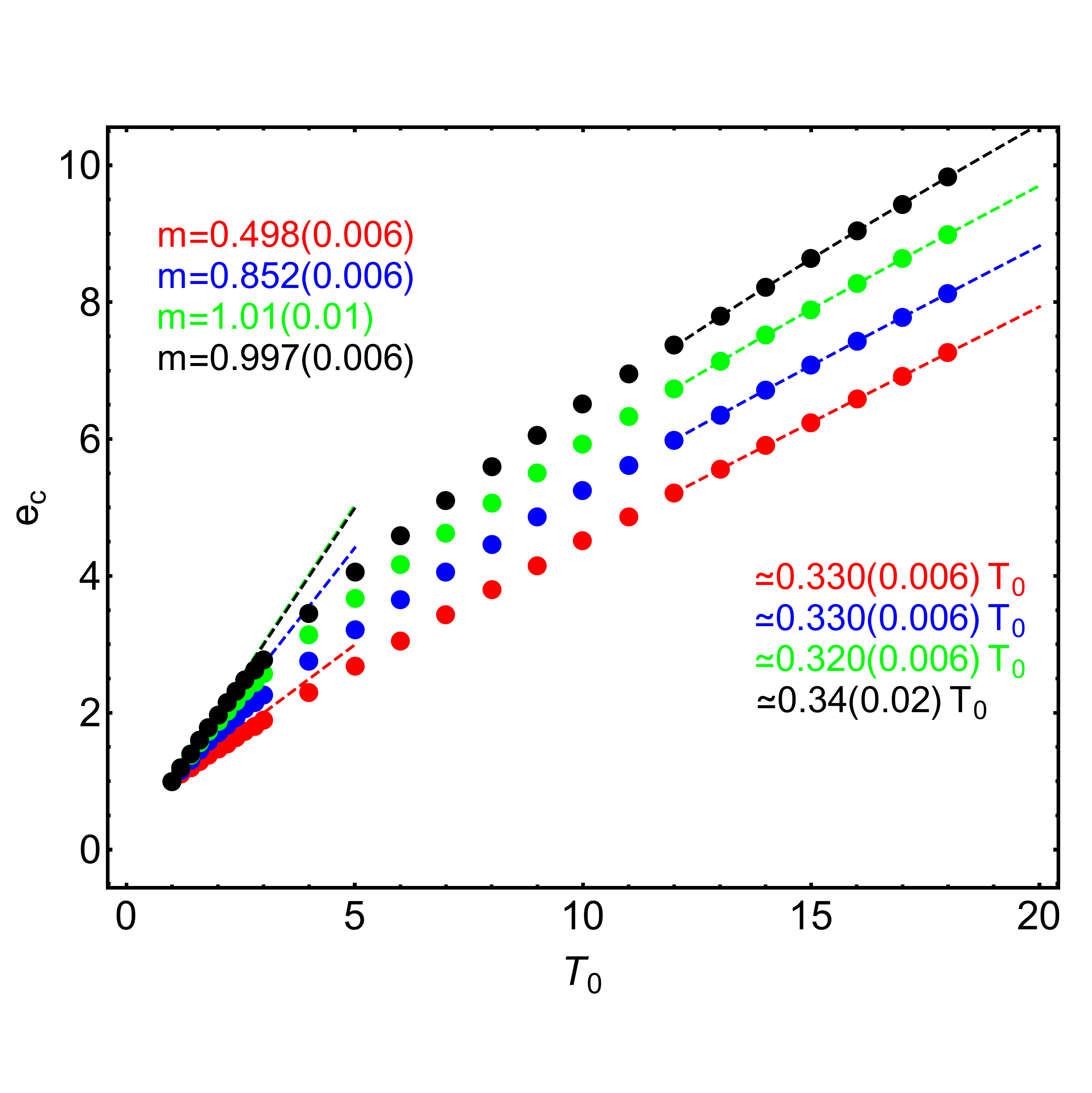} %global_temp.nb%
\includegraphics[height=6cm]{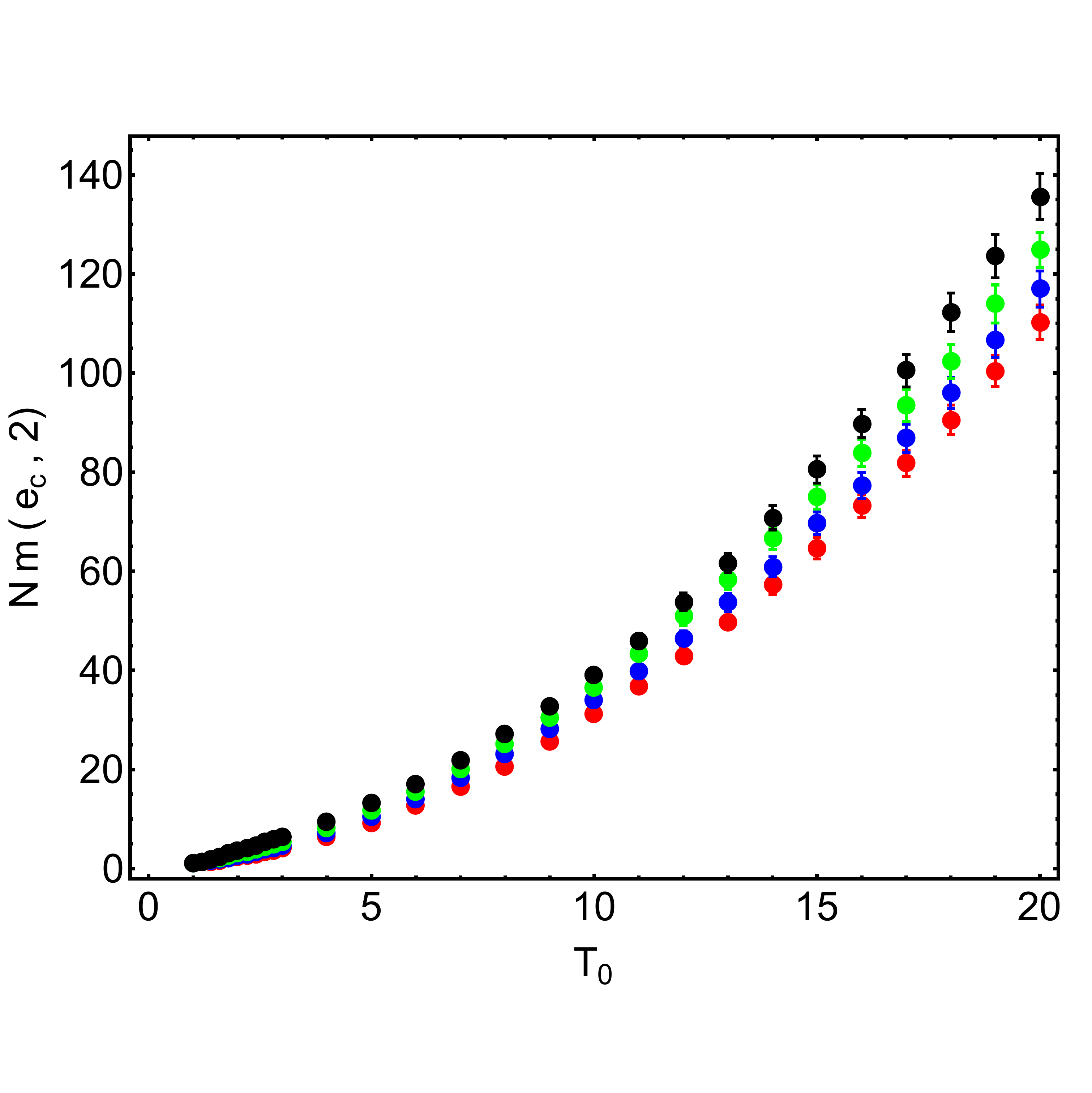}  %global_temp2.nb%
\includegraphics[height=6cm]{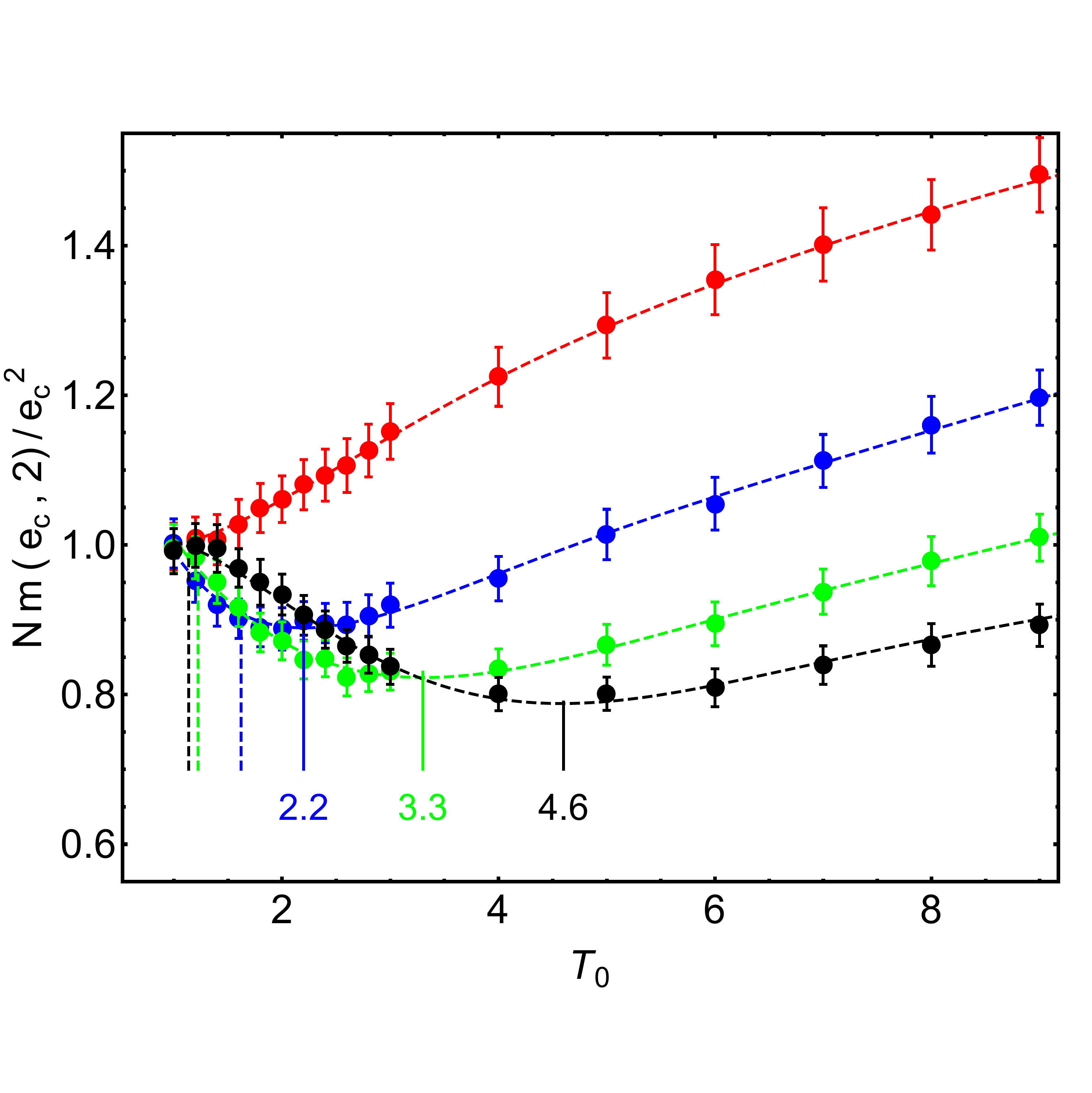}  %global_temp2.nb%
\end{center}
\kern -1.2cm
\caption{Top-Left: Averaged kinetic energy, $e_c$, as a function of $T_0=1,1.2,\ldots, 20$ for $g=0$ (red dots), $g=5$ (blue dots), $g=10$ (green dots) and $g=15$ (black dots). The error bars are included. Dotted lines show the tangents of the curves at $T_0=1$ and the asymptotic linear behavior for large $T_0$ values. Top-Right: Second momenta of $e_c$  multiplied by $N$ as a function of $T_0$. Bottom: Relative variance of $e_c$: $m(e_c,2)/e_c^2$. Dotted curves are phenomenological fits to compute the location of the data minimum value. They are $T_0^c=2.2$, $3.3$ and $4.6$
for $g=5$, $10$ and $15$ respectively. Dashed vertical lines are the corresponding predicted values obtained from the Enskog approximation of the transport coefficients.\label{ener}}
\end{figure}

\section{Computer Simulation Results: Global Magnitudes}

We have measured several global magnitudes like  the {\it averaged kinetic energy} ($e_c$),  the {\it potential energy} ($e_p$), the {\it hydrodynamic kinetic energy} ($e_u$), the {\it pressure} ($P$) and the {\it energy current} across the boundaries ($J$). All of them will give us a general overview of the system behavior. Moreover, we are particularly interested in knowing if the trasition non-convective to convective  appears or not as an abrupt change on their functional behavior and they can give us some clue to determine with precission the transition point. 

\begin{itemize}
\item {\it The averaged kinetic energy:} The averaged kinetic energy for a given disk configuration at time $t$ is given by:
\begin{equation}
e_c(p,t)=\frac{1}{N}\sum_{i=1}^N \frac{p_i(t)^2}{2m}
\end{equation}
We have computed its time average and its second moment:
\begin{equation}
e_c=\frac{1}{M}\sum_{t=1}^M e_c(p,t)\equiv \langle e_c(p,t)\rangle \quad, \quad m(e_c,2)= \langle e_c(p,t)^2\rangle- \langle e_c(p,t)\rangle^2
\end{equation}
We observe at figure \ref{ener} (left) that the averaged kinetic energy grows monotonically with $T_0$ in a smooth nonlinear way. The kinetic energy per particle grows with $g$ for any fixed $T_0$ value. That is, for a fixed external gradient (non zero) any positive variation of the external field increments the system kinetic energy and this hypothetical process  does a net work over the system (speaking in terms of classical equilibrium thermodynamics). Let us remark that variations of $g$ does not affect the value of the averaged kinetic energy at equilibrium  ($T_0=T_1=1$) that is $e_c=1$. Therefore, the dependence on $g$ of $e_c$  is just a pure nonequilibrium property.
The slope of $e_c$ at $T_0=1$ grows with $g$:$m=0.498 (0.006)$, $0.852 (0.006)$, $1.01 (0.01)$ and $0.997 (0.006)$ for $g=0$, $5$, $10$ and $g=15$ respectively. 
 For values  $T_0>15$  $e_c$  tends to an asymptotic linear behavior with a almost common slope of $m'=0.33$ for any $g$ value. In fact we get $m'=0.330 (0.006)$, $0.330 (0.006)$, $0.320 (0.006)$ and $0.34 (0.02)$ for the $g=0$, $5$, $10$ and $15$  cases respectively. That is far from the limiting slope associated to a pure linear temperature profile: $e=(T_0+T_1)/2$ that has a slope equal to $0.5$. It could be though that the thermal resistance at the boundaries makes the system to behave in the bulk with  effective temperatures $T_0^*$ and $T_1^*$. If that was so we could identify: $m' (T_0)=(T_0^*(T_0)+T_1^*(T_0))/2$. We'll see later that $T_0^*(T_0)$ grows linearly with $T_0$: $T_0^*\simeq 0.17 T_0$ and $T_1^*(T_0)$ tends to a constant value. Therefore $m'$ does not match with the linear profile assumption and we can conclude that for large $T_0$ values there is not a linear profile of temperatures.  

Finally let us to remark that there is no trace of any discontinuity and/or singularity in the data.  Therefore this macroscopic observable does not reflects any singular property on the convective-non-convective transition. 

The second moment relative to $e_c$: $m(e_c,2)/e_c^2$  behaves in a more interested manner. At $T_0=1$ (equilibrium) its value corresponds to $1/N=1/957=0.00104..$ as the law of large number predicts. We see at figure \ref{ener} (bottom) how the relative fluctuations, for a fixed $g>0$ value, decrease as we increment $T_0$, it reaches a minimum and then it grows. This is in contrast with the equilibrium case ($g=0$) in which the relative fluctuations grow monotonically with $T_0$. 
We get an idea on where it is located the minimum by fitting a $9$th degree polynomial to the data (plotted by a dotted line in the figure) and equating to zero its derivative. We find that the minimum values are at  $T_0^c=2.2$, $3.3$  and $4.6$ for $g=5$, $10$ and $15$ respectively. We will see that this singular point where the relative kinetic energy fluctuations  has the smallest value seems to be related to the critical temperature that defines the transition from the non-convective to a convective regime. However these critical temperatures does not coincide with the ones we already computed in the Enskog approximation (dashed vertical lines on figure \ref{ener} (right)). Moreover our critical values increase with $g$ while the ones from the Rayleigh parameter decrease with $g$. One explanation to this discrepancy is that the Rayleigh critical value is obtained in the Boussinesq approximation that is assuming that the fluid is non-compressible. Some numerical analysis of the linearized Navier Stokes equations found that the critical temperatures for the onset of convection, increase with $g$ \cite{Bormann}. In fact, the increment is argumented to be given by $g\alpha T L_y^4/c_p$. Using the values with the Henderson's EOS and $\bar\rho=0.2$, we find that the critical temperatures are: $T_0^{c (comp)}=3.2$, $4.6$ and $5.9$ for $g=5$, $10$ and $15$ respectively. These values are not too far from the ones we have found and the linear dependence on $g$ is similar to ours which is quite remarkable.

Let us finish these arguments by pointing out what would happens it the parameter $A(T_0,g)=m(e_c,2)/e_c^2$ that localizes the critical value for the temperature were correct. Then one could conclude that:  
\begin{eqnarray}
\frac{dA}{dT_0}&<&0\quad\quad\text{if}\quad T<T_0^c\nonumber\\
\frac{dA}{dT_0}&>&0\quad\quad\text{if}\quad T>T_0^c\nonumber
\end{eqnarray}
and, at the critical temperature  the following relation should be true:
\begin{equation}
\frac{d m(e_c,2)}{dT_0}\biggr\vert_{T_0^c}=2\frac{m(e_c,2)}{e_c}\frac{de_c}{dT_0}\biggr\vert_{T_0^c}
\end{equation}
It could be interesting to check in some other scenarios this relation to exclude the possibility that we are just seeing a numerical mirage. 

\begin{figure}[h]
\begin{center}
\includegraphics[height=6cm,clip]{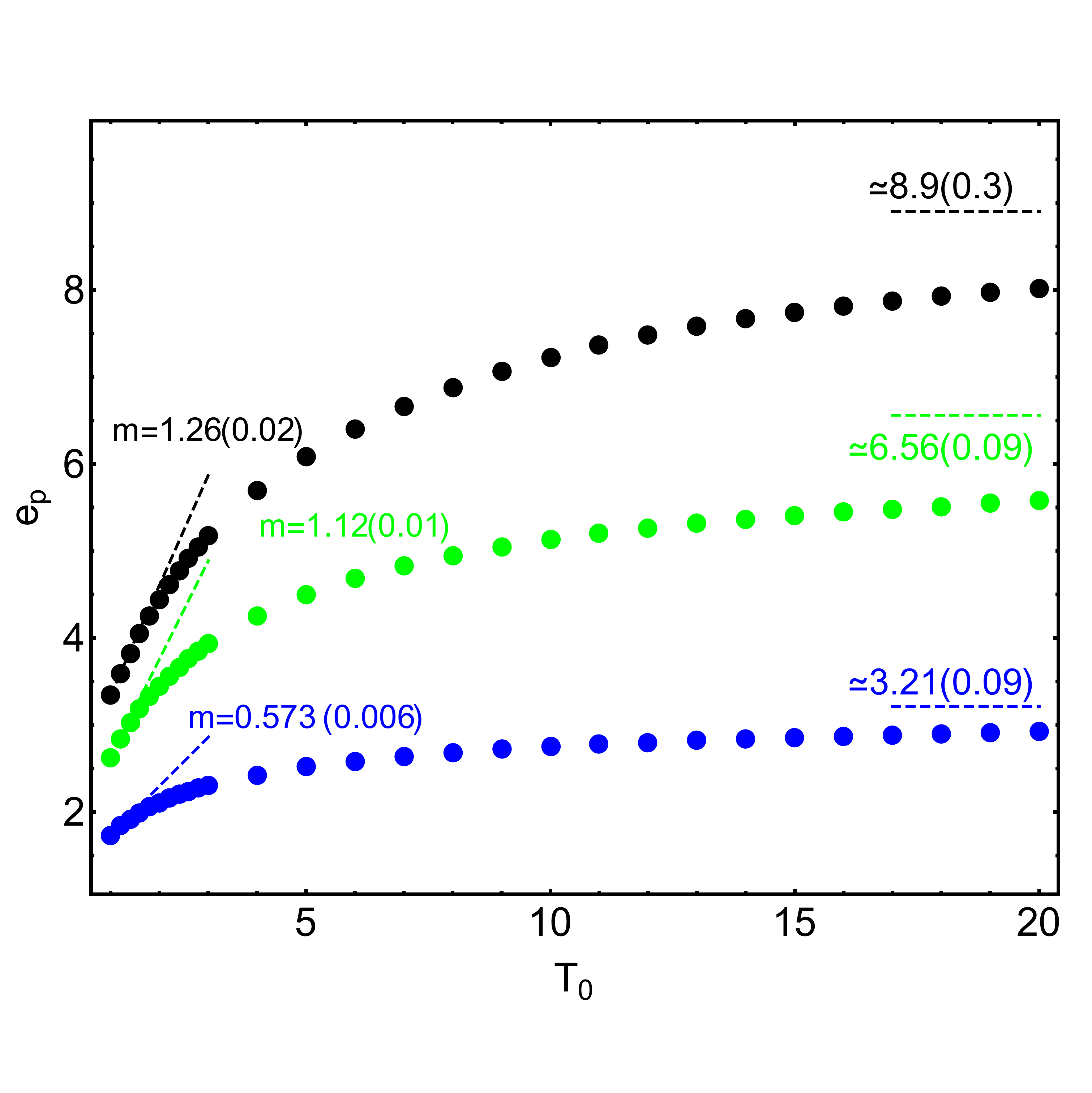} %global_potential_energy.nb %
\end{center}
\kern -1.cm
\caption{Averaged potential energy, $e_p$, as a function of $T_0=1,1.2,\ldots, 20$ for $g=5$ (blue dots),  $g=10$ (green dots) and $g=15$ (black dots).  \label{enerp}}
\end{figure}

\item {\it The potential energy:} It is the energy associated to the constant external acceleration  that is applied to the particles ($\vec a=-g\vec j$).The potential energy per particle is then defined as:
\begin{equation}
e_p=\frac{1}{N}\sum_{i=1}^N mgr_{i,2}
\end{equation}
where $r_{i,2}$ is the vertical coordinate associated to the particle $i$ assuming that at $r_{i,2}=0$ there is located the thermal bath at temperature $T_0$. We observe on fig. \ref{enerp} that $e_p$ monotonically grows with $T_0$. That is, the center of mass of the system increases its height when we increase the temperature at the bottom. From the data we can get the asymptotic value fo $e_p$ when $T_0\rightarrow\infty$ by fitting a polynomial of the form: $e_p=e_p^a+a_1/T_0+\ldots+a_5/T_0^5$, obtaining: $e_p^a=3.21 (0.09)$, $6.56 (0.09)$ and $8.9 (0.3)$ for $g=5$, $10$ and $15$ respectively. We cannot exclude a linear dependence of ${e_p}^a$ on $g$. One may easely check that the potential energy per particle of a uniform disk system in which all the particles were at the top is $e_p^t=g(1-\rho/2\rho^t)$ and with our data $\rho^t\simeq 0.3$ which is far from the liquid-hexatic phase transition that is at $\rho_c\simeq 0.70$ \cite{Engel} . That is, even in the infinite $T_0$ limit, the system would still be at a liquid-like phase.

Finally we have found that the slope of the fitted function of the data near the equilibrium point  is found to be $m=0.573 (0.006)$,  $1.12 (0.01)$ and $1.26 (0.02)$ for $g=5$, $10$ and $15$ respectively. Observe that in this case that the slopes are not compatible with a linear dependence on $g$. Just for completeness, the values at $T_0=1$ are $e_p=1.7365 (0.0003)$, $2.6263 (0.0005)$ and $3.343 (0.001)$.

 \begin{figure}[h]
\begin{center}
\includegraphics[height=6cm,clip]{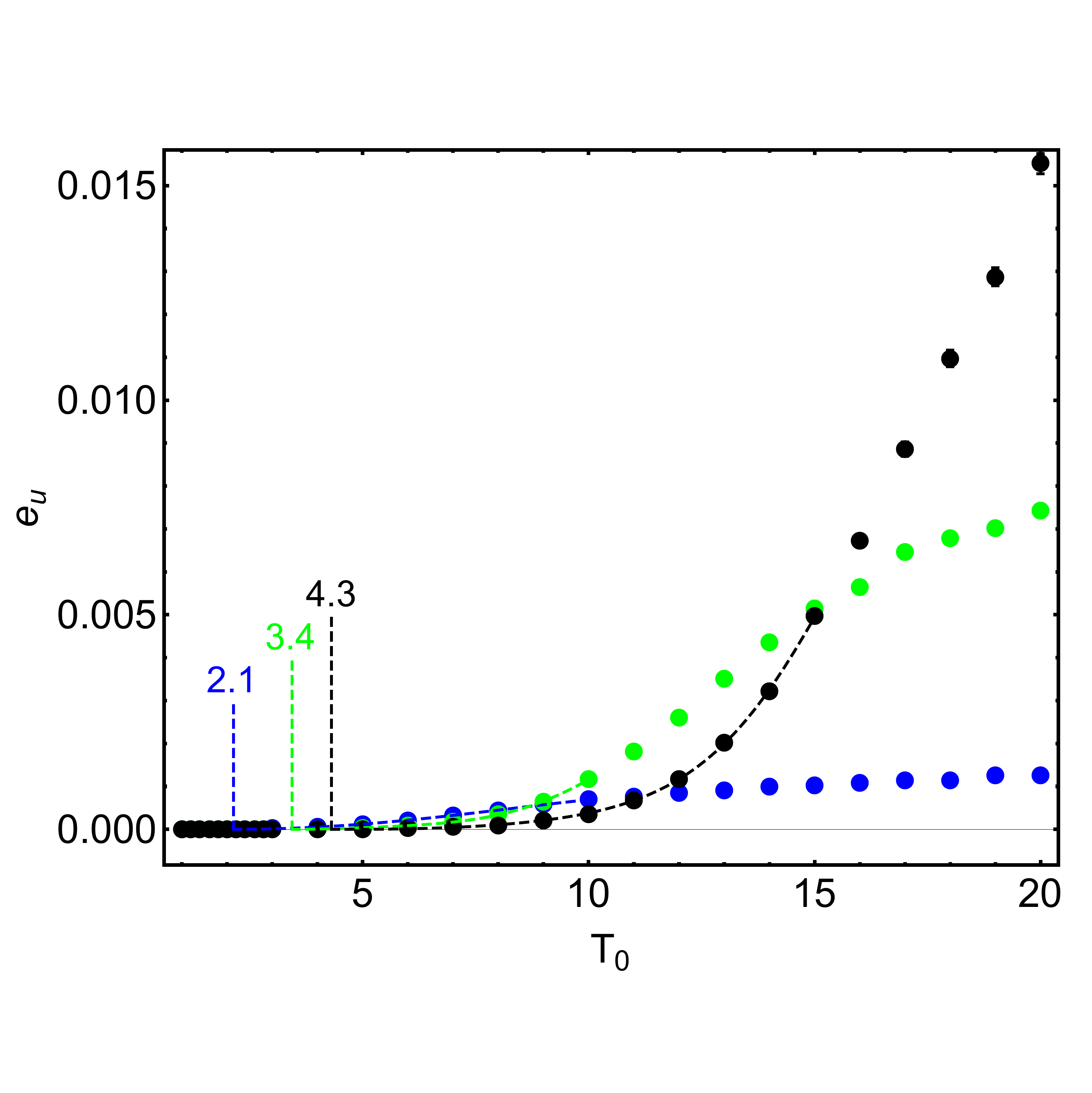} %global_kinetic_energy.nb %
\end{center}
\kern -1.2cm
\caption{Averaged hydrodynamic energy, $e_u$, as a function of $T_0=1,1.2,\ldots, 20$ for $g=5$ (blue dots), $g=10$ (green dots) and $g=15$ (black dots). The curves are polynomial fits to the points (see text). Error bars are included. \label{enerk}}
\end{figure}

\item  {\it The hydrodynamic kinetic energy:} It is defined as the kinetic energy associated to the velocity of the center of mass at each cell. That is, first lets us define  the average velocity measured at cell $(n,l)$:
\begin{equation}
u(n,l;M)=\frac{1}{N(n,l)M}\sum_{t=1}^M \sum_{i:r(i,t)\in B(n,l)} v_i(t) \quad , \quad N(n,l)=\frac{1}{M}\sum_{t=1}^M \sum_{i:r(i,t)\in B(n,l)}1\label{velocell}
\end{equation}
where we have follow the local average definition in eq. \ref{aver}.  Then, we define the hydrodynamic kinetic energy:
\begin{equation}
e_u(M)=\frac{1}{2N_C}\sum_{(n,l)}\frac{\rho(n,l)}{\rho}u(n,l;M)^2
\end{equation}
where $\rho(n,l)$ is the mean areal density at cell $(n,l)$, $N_C$ is the total number of cells and the $M$ argument indicates that we have done a large but finite number of measurements.

First observe that the measured values are three orders of magnitude smaller compared to the total kinetic energy or the potential energy ones. This is due to the hydrodynamic separation of scales. This makes the numerical analysis of the convective structures very hard and that's why we need very large time averages to get some clear picture of the system behavior.

In general we expect that $u(n,l;M)=0$, $\forall (n,l)$ for the non-convective states, that is for $T_0<T_0^c$ and  $u(n,l;M)\neq 0$ for the convective states  ($T_0>T_0^c$).  As we have commented in the error analysis section above, the finite number of measurements makes that $u(n,l;M)$ is typically nonzero due to the small but systematic effect of the noisy data behavior. We should correct this fact using the expressions \ref{error1} and \ref{error2} that in our case are given by:
\begin{equation}
e_u\equiv e_u(M\rightarrow \infty)\simeq e_u(M)- \frac{1}{18\rho N_C}\sum_{(n,l)}\rho(n,l)\epsilon(u(n,l))^2 \pm \epsilon(e_u)\label{eu}
\end{equation}
with
\begin{equation}
\epsilon(e_u)=\frac{1}{2\rho N_C}\left[ \sum_{(n,l)}u(n,l)^2\left(4\rho(n,l)^2\epsilon(u(n,l))^2+\epsilon(\rho(n,l))^2u(n,l)^2\right)  \right]^{1/2}\label{eu2}
\end{equation}
where
$\epsilon(u(n,l))$ is the measured error of $u(n,l)$  and $\epsilon(\rho)$ is the measured error fo $\rho(n,l)$ (both are three times their standard deviation).

Observe that we subtract to the measured value of $e_u(M)$ a small but monotonous increasing  function of $T_0$. One checks that the values so obtained have a nice average of zero near $T_0=1$ and we can try now to fit an appropriate function. We choose  $e(u)_{fit}=0$ for $T<T_0^c$ and $=(T-T_0^c)^2 (a_0+a_1T_0^2+\ldots a_7 T_0^8)$ for $T_0>T_0^c$ where the fitting parameters are $T_0^c$ and $a$'s.  We choose $(T-T_0^c)^2$ as a minimal assumption of asking continuity and first derivative zero both at $T_0^c$. We fit this function to the points in the intervals $T_0\in[2.6,12]$, $T_0\in[4,13]$  and $T_0\in[5,16]$ and we obtain: $T_0^c=2.1$, $3.4$ and $T_0^c=4.3$ for $g=5$, $10$ and $15$ respectively. These values change a bit when choosing other polynomials degree and/or the interval of fitting data. However they are consistent with the results obtained from the energy fluctuations and we think that the critical temperature should be around such values. 
 \begin{figure}[h]
\begin{center}
\includegraphics[height=6cm,clip]{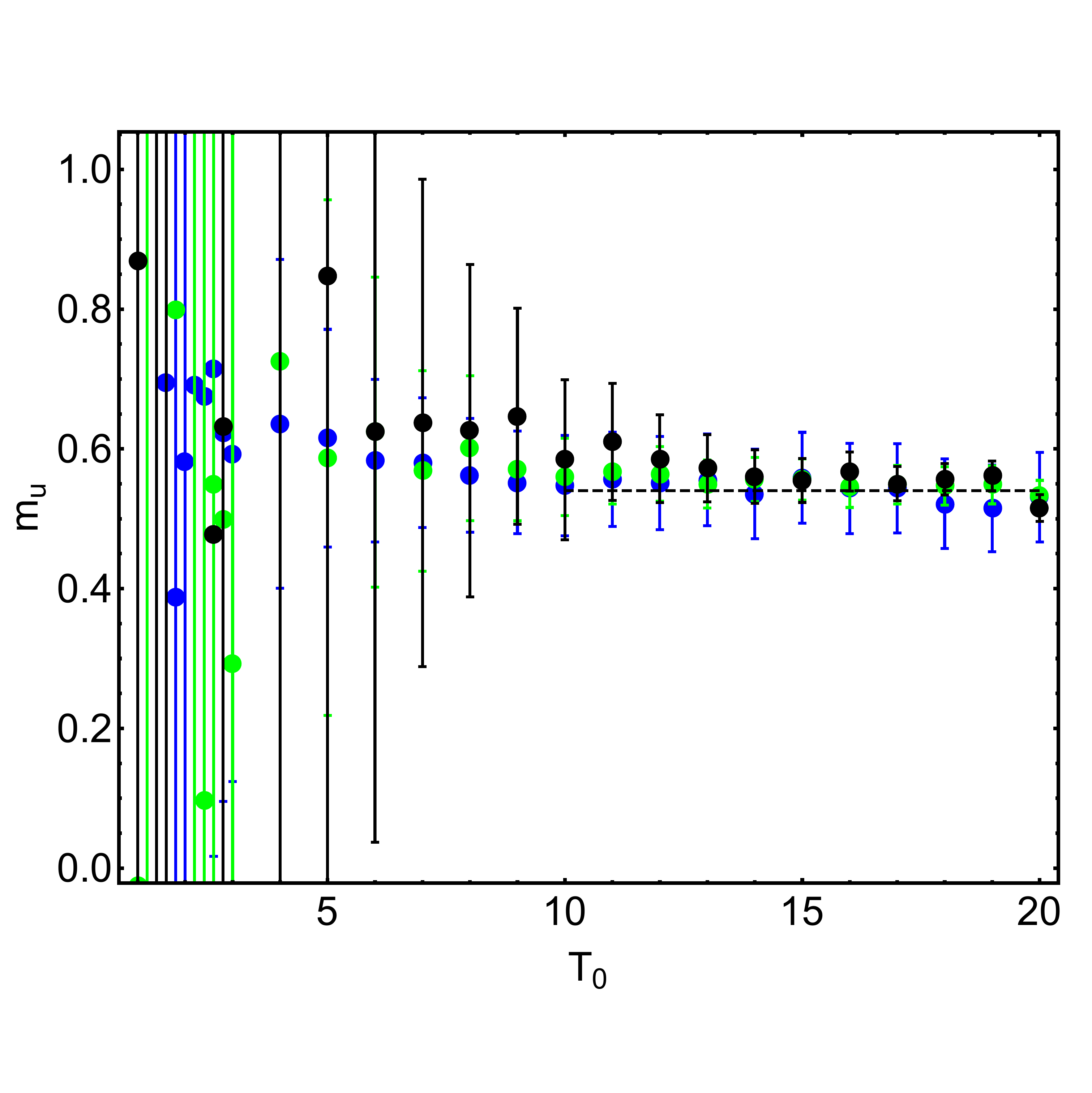} %global_vx_vy.nb %%
\end{center}
\kern -1.2cm
\caption{Anisotropy parameter, $m_u$, as a function of $T_0=1,1.2,\ldots, 20$ for $g=5$, (blue dots), $g=10$ (green dots) and $g=15$ (black dots).  \label{op}}
\end{figure}

Finally it is natural to define a kind of anisotropy parameter that measures the relative difference on the $x$ and $y$ components of the hydrodynamic kinetic energy.
We define
\begin{equation}
m_u=\frac{1}{2 e_uN_C}\sum_{(n,l)}\frac{\rho(n,l)}{\rho}\left[u(n,l;M)_2^2-u(n,l;M)_1^2\right]
\end{equation}
First observe that in the non-convective region this observable is singular because the hydrodynamic velocities are zero and this is reflected with a very large set of error bars (that are computed in a similar way as in the $e_u$ case (see \ref{eu} and \ref{eu2})  for the data in that region. The curious result is that this parameter seems to stabilize to a constant value $\simeq 0.55$  independently (up to error bars) of the $T_0$ and $g$ parameters. That is, the relative anisotropic behavior seems to be universal with respect the boundary conditions ($T_0$) and the external forcing ($g$). This  implies that there is a kind of energy equipartition:
\begin{equation}
e_{u,2}-e_{u,1}=0.54(e_{u_1}+e_{u,2})\quad\implies\quad e_{u,2}\simeq 3.4\, e_{u,1}
\end{equation}
independently on $g$ and for large enough values of $T_0$.

 \begin{figure}[h]
\begin{center}
\includegraphics[height=6cm,clip]{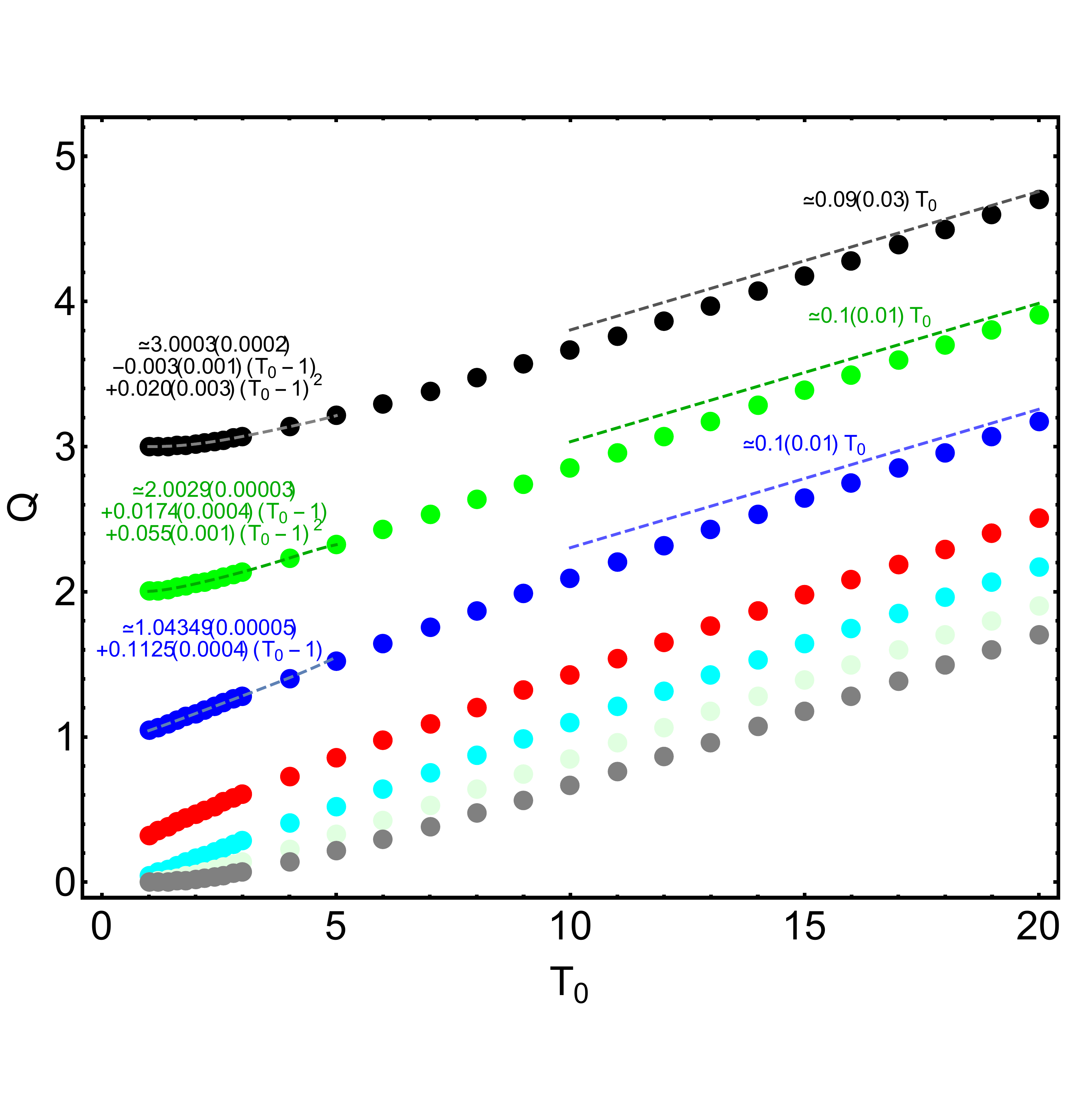}	    %global_pressure.nb %
\includegraphics[height=6.5cm,clip]{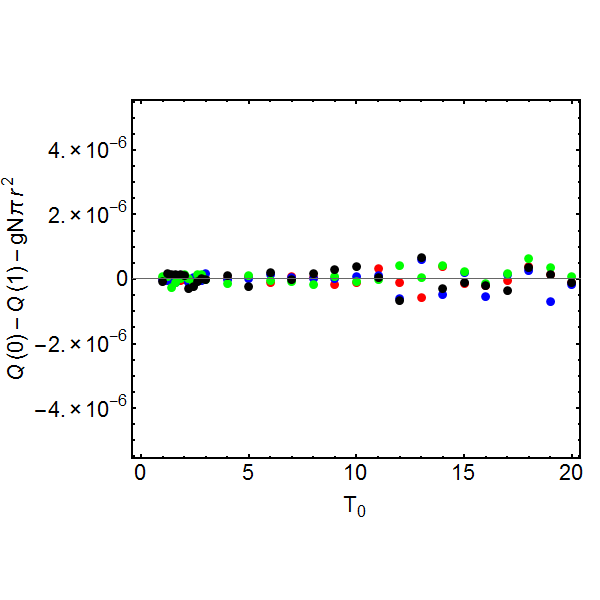}
\end{center}
\kern -1.2cm
\caption{Right: Reduced pressure at the thermal baths ($Q=P\pi r^2$) ($Q(0)$ at $T_0$ bath and $Q(1)$ at $T_1=1$ bath) as a function of $T_0=1,1.2,\ldots, 20$. Red dots: $(g=0,Q(0))$, Pink dots: $(g=0,Q(1))$, Blue dots: $(g=5,Q(0))$, Cyan dots: $(g=5,Q(1))$,
Green dots: $(g=10,Q(0))$, Light Green dots: $(g=10,Q(1))$, Black dots: $(g=15,Q(0))$ and Gray dots: $(g=15,Q(1))$. The right dotted curves are polynomial fits to the points (see text) and the left dotted lines are the asymptotic predicted behavior. Left: Difference of hot and cold bath reduced pressures minus the barometric contribution $gN/\pi r^2$ for $g=0$ (red dots), $g=5$ (blue dots), $g=10$ (green dots) and $g=15$ (black dots) \label{press}}
\end{figure}
\item {\it The Pressure:} We have computed  the system pressure on each of the thermal bath boundaries as the sum of the moment variation when particles collide with  the bath divided by the absolute time measuring interval times the boundary length:
\begin{equation}
P=\frac{m}{\Delta t L_x}\sum_{t_{col}\in\Delta t}\left(v_{i,2}'(t_{col})-v_{i,2}(t_{col})\right)
\end{equation}
where $v'$ and $v$ are the particle velocity right after and right before the collision respectively. We first observe that the global barometric formula: 
\begin{equation}
Q(0)-Q(1)=g N\pi r^2\quad , \quad Q=P\pi r^2
\end{equation}
holds in all cases as figure \ref{press} shows with high precision. This property is independent on the state of the fluid: non-convective or convective. However, the pressure has a non-trivial dependence with $T_0$. When $g=5$ the data is concave over all the $T_0$ interval. The slope at $T_0=1$ is $0.1125 (0.0004)$ and it gradually diminishes until its asymptotic value $0.10 (0.01)$. However, for $g=10$, the data has a small slope of $0.1740 (0.0004)$ at  $T_0=1$ and it is convex near it, it changes convexity at around $T_0=10$ and it finally tends to the same asymptotic behavior as in the $g=5$ case: $Q\simeq 0.10 (0.01)T_0$. Finally, for $g=15$ the slope at $T_0=1$ is zero and, again, it tends to the same asymptotic value. Observe that the $Q$ at $T_0=1$ tend to have the value $g\bar\rho$ because the pressure at the top goes to zero. We do not appreciate any singularity in the curve near the transition between non-convective and convective regimes.  
 \begin{figure}[h]
\begin{center}
\includegraphics[height=6cm,clip]{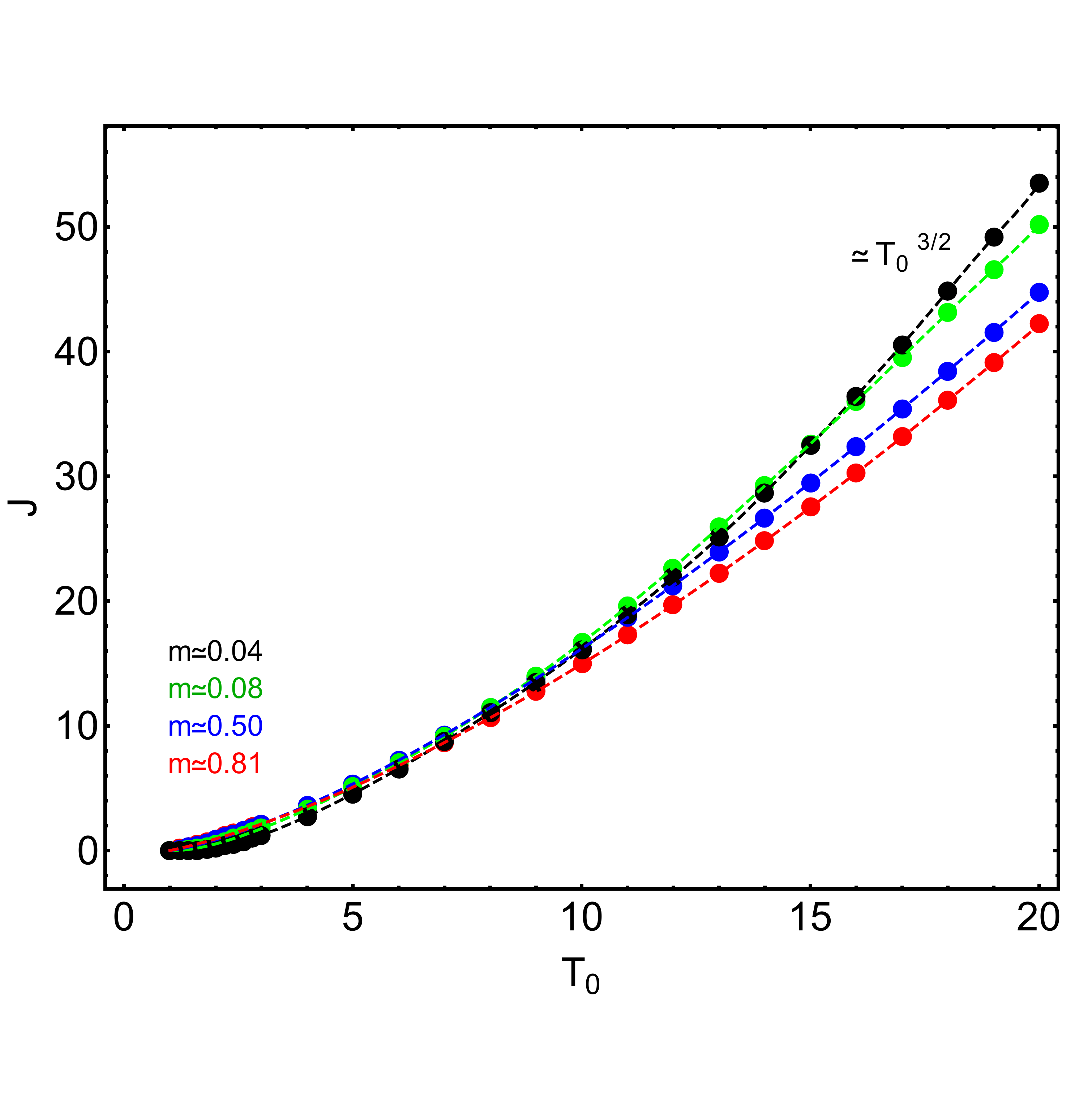}         %global_current.nb %
\includegraphics[height=6cm,clip]{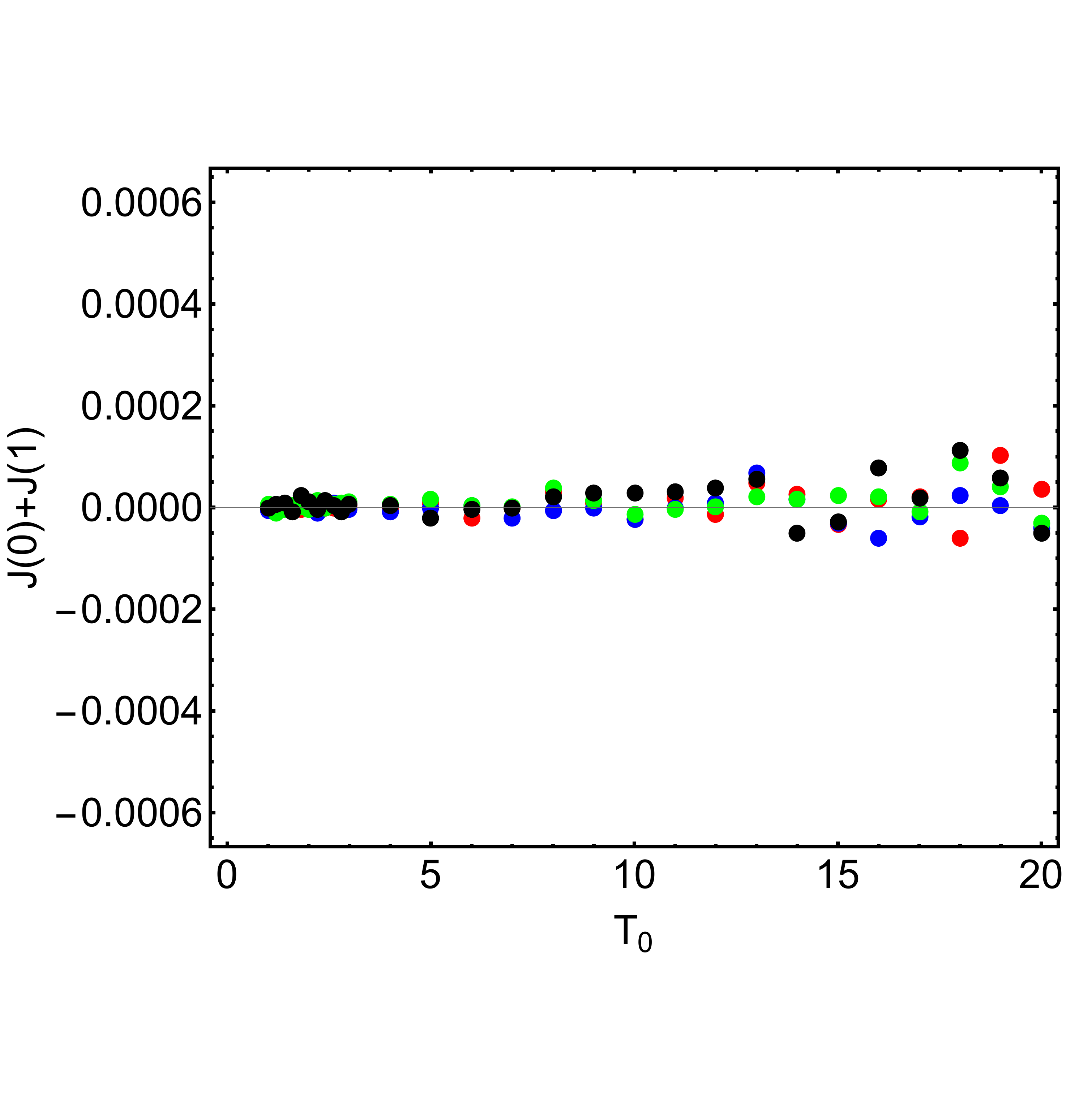}
\end{center}
\kern -1.2cm
\caption{Left: Reduced heat current at the bottom thermal bath, $J=\bar J r m^{1/2}$, as a function of $T_0$ for $g=0$ (red dots), $5$ (blue dots),  $10$ (green dots) and $15$ (black dots). The dotted curves are phenomenological  fits to the points (see text). $m$ is the slope of the curves at $T_0=1$ and all the curves behave $J\simeq T_0^{3/2}$ for large values of $T_0$. Right: $J(0)+J(1)$  ($J(y)$ is the energy current at boundary $y$) for $g=0$ (red dots), $g=5$ (blue dots), $10$ (green dots) and $15$ (black dots). \label{curr}}
\end{figure}
\item {\it The current of energy:} 
We have computed  the energy current that cross each of the thermal bath boundaries as the sum of the kinetic energy variation when particles collide with  the bath divided by the absolute time measuring interval and the boundary length:
\begin{equation}
\bar J=\frac{1}{2\Delta t L_x}\sum_{t_{col}\in\Delta t}\left(v_{i,2}'^2(t_{col})-v_{i,2}^2(t_{col})\right)
\end{equation}
We see on figure \ref{curr} the averaged behavior of $J=\bar J r m^{1/2}$  as a function of $T_0$ for $g=0$, $5$, $10$ and $15$. First we observe that $J(0)=-J(1)$ as expected because there isn't any internal energy dissipation mechanism. Second, its behavior is not linear in $T_0$ implying that one should assume that the thermal conductivity cannot be a constant if we assume that the Fourier's Law applies. Third, for large values of $T_0$ we see that the data seem to follow the same asymptotic behavior  independently of $g$:
\begin{equation}
J\simeq   m_1 T_0^{3/2}
\end{equation}
where $m_1=0.4726 (0.0003)$, $0.4870 (0.0009)$, $0.559 (0.005)$ and $0.672 (0.002)$ for $g=0$, $5$, $10$ and $15$ respectively. 

The current presents a slope at $T_0=1$ that dependes on $g$: $m=0.81$, $0.58$, $0.08$ and $0.04$ for $g=0$, $5$, $10$ and $15$ respectively.

\begin{figure}[h]
\begin{center}
\includegraphics[height=6cm,clip]{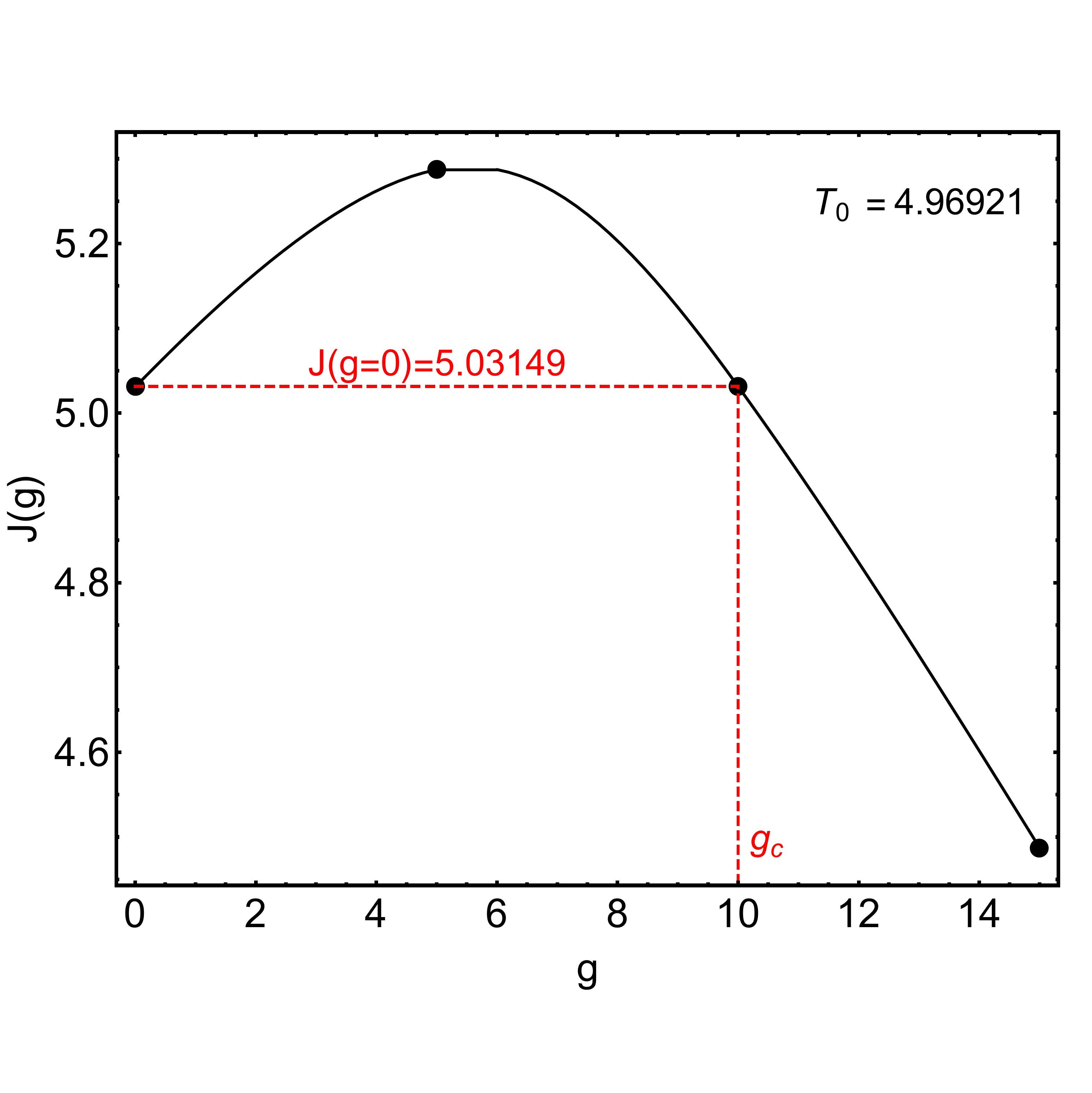}         %global_current.nb %
\includegraphics[height=6cm,clip]{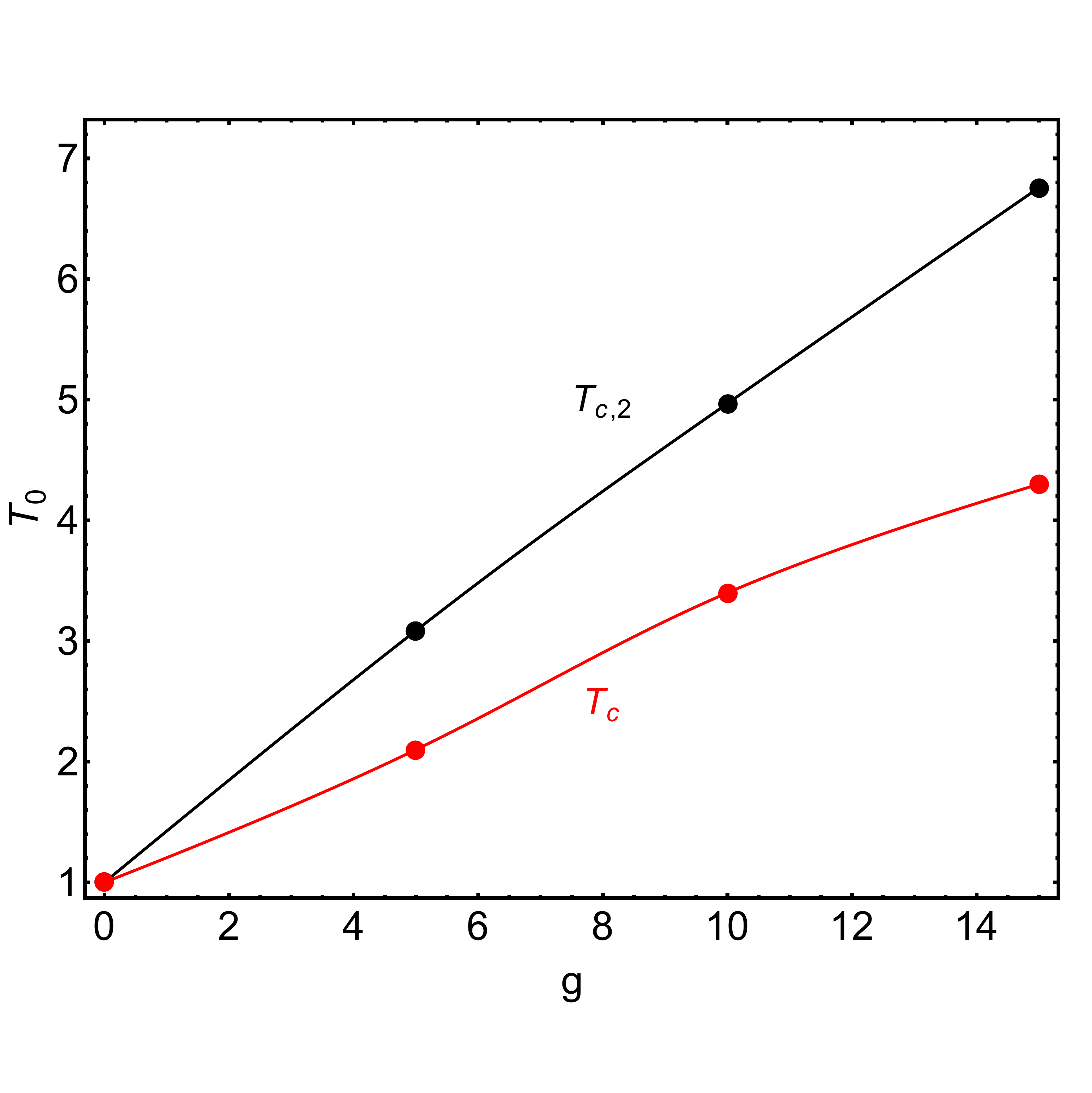}
\end{center}
\kern -1.2cm
\caption{Left: Heat current values as a function of $g$ for the temperature $T_0=4.96921$ at which the current at $g=10$ equals to the one at $g=0$. They are obtained from the global fits done in figure \ref{curr}. Solid line is just a help for the eyes. Right: Black dots are the $T_0$ values at which the heat current of a system with fix $g$ equals the one with $g=0$. Red dots are the computed critical temperature values such that for $T_0$ above them, the  hydrodynamic kinetic energy is not zero. 
 \label{curr2}}
\end{figure}
\begin{table}[h!]
\begin{center}
\resizebox*{!}{3cm}{ 
\begin{tabular}{|c|c|c|}
\hline
g&$T_c$&$T_{c,2}$\\ \hline
\hline
5&2.1&3.1\\ \hline
10&3.4&5.0\\ \hline
15&4.3&6.8\\ \hline
\end{tabular}}
\end{center}
\caption{Computed Temperature critical values. $T_c$ separates non-convecting from convecting states for a given $g$. $T_{c,2}$ separates {\it bad conducting} from {\it fully convecting state} (see text for more explanation). \label{tab1}}
\end{table}
From the behavior of the heat currents in figure \ref{curr} we observe some interesting physical properties:

\begin{itemize}
\item (1) Any curve $J(T_0;g)$ with a fixed $g$ value have a (non trivial) crossing point with the curve $J(T_0;g=0)$. We call such point $T_{c,2}$ and it is solution of the equation $J(T_{c,2};g)=J(T_{c,2};0)$ (see Table \ref{tab1} for computed values). We can conclude this property by seeing the behavior of the slopes of $J(T_0;g)$ near $T_0=1$: $m(g)<m(g=0)$ for all $g$'s. $T_{c,2}(g)$ separates neatly two regimes for any fixed $g$: (1) the temperatures at which the heat current is smaller than the one without gravity, $T_0<T_{c,2}(g)$. We will call this a {\it bad conducting state} because gravity is penalizing heat conduction and (2) the temperatures in which the sytem has a larger current than the corresponding without gravity,  $T_0>T_{c,2}(g)$. We call it {\it fully convecting state} because the action of $g$ through convection is increasing the energy transmission with respect the $g=0$ case.

\item (2) For any fixed value of $T_0$, there exists a $g_c$ in which its heat current equals the one of a system with $g=0$: that is $J(T_0;g_c)=J(T_0;0)$. Again we can observe two different regimes: (1) For $g<g_c$  $J(T_0;g)>J(T_0;0)$ and (2) $g>g_c$  where $J(T_0;g)<J(T_0;0)$. Observe this behavior in figure \ref{curr2} (left) in a particular case. First we found that the crossing point between the currents with $g=10$ with the ones with $g=0$ is at $T_{c,2}=4.96921$ with a common heat current of $J=5.0315$. Then we use this temperature to know the currents for the cases $g=5$ and $15$. Finally we can also conclude that for any fixed value of $T_0$ there exists a  $g^*<g_c$  at which the heat current reachs a {\it maximum}.

\item (3) We plot at figure \ref{curr2} (right) the temperatures at which the hydrodynamic kinetic energy begin to be nonzero ($T_c$) together with the temperatures at which the heat current equals the one without gravity ($T_{c,2}$). We observe an interesting pattern. It seems that for any fixed $g>0$ there are two critical temperature values that separates three regions: (A) $T_0<T_c$ where there is no macroscopic velocity field and there is only heat conduction (there are {\it bad conducting states}). (B) $T_0\in[T_c,T_{c,2}]$ where there is a small hydrodynamic field but still the heat current is less than the correponding with $g=0$ and (C) we have hydrodynamic velocity fields and the states are {\it fully convecting} ones (in the sense we defined above). If this is the general picture we would like to know if, for instance, there is a coherent convective behavior in region B.
\end{itemize}

 \end{itemize}

 \begin{figure}
\begin{center}
\includegraphics[height=4.5cm,clip]{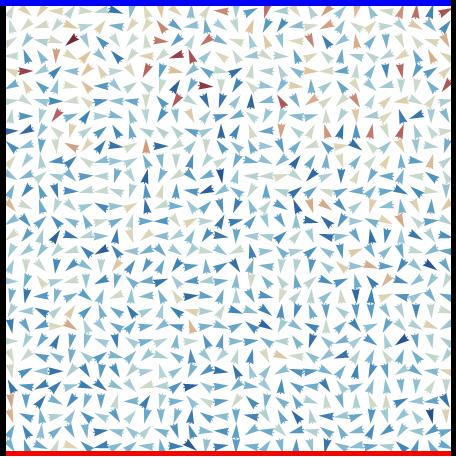}   %velo_field_1_g5.nb
\includegraphics[height=4.5cm,clip]{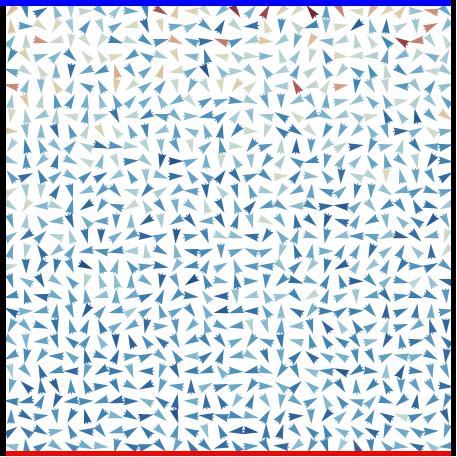}   %velo_field_1_g10.nb
\includegraphics[height=4.5cm,clip]{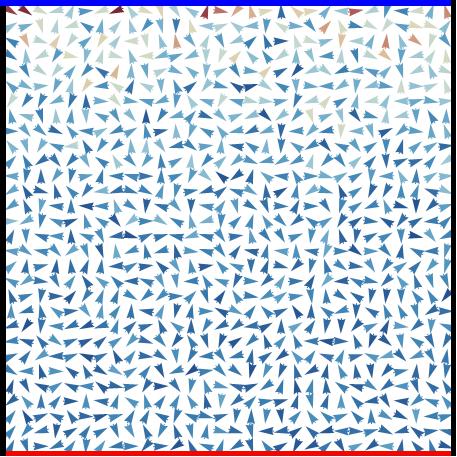}  %velo_field_1_g15.nb
\includegraphics[height=4.5cm,clip]{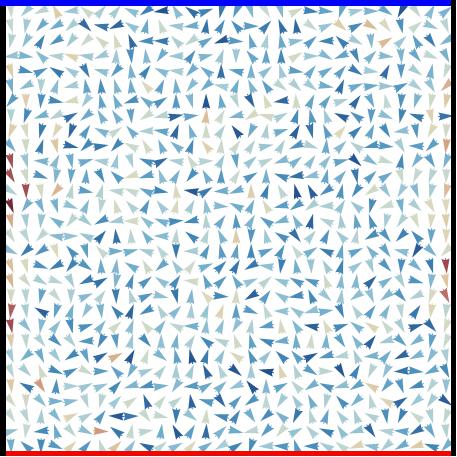}  
\includegraphics[height=4.5cm,clip]{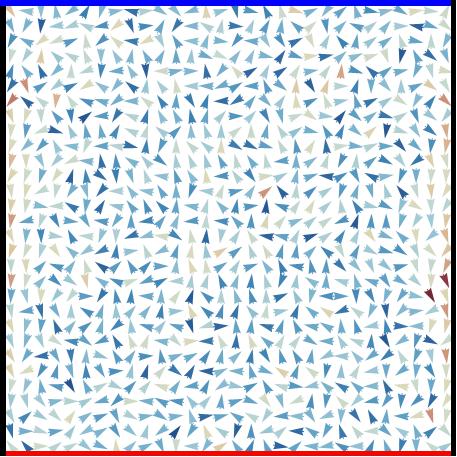}
\includegraphics[height=4.5cm,clip]{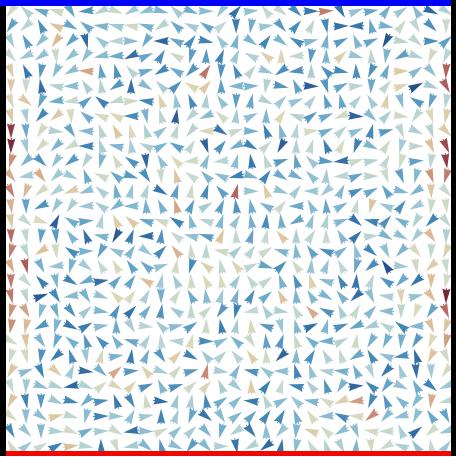}
\includegraphics[height=4.5cm,clip]{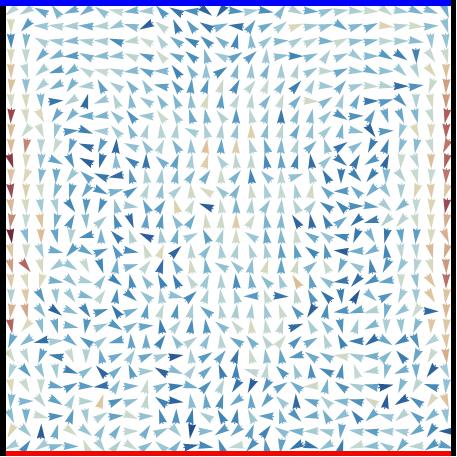}
\includegraphics[height=4.5cm,clip]{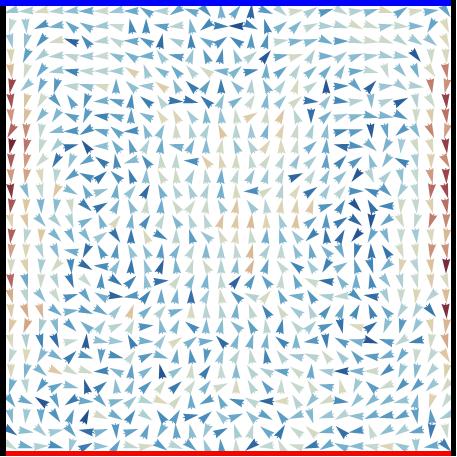}
\includegraphics[height=4.5cm,clip]{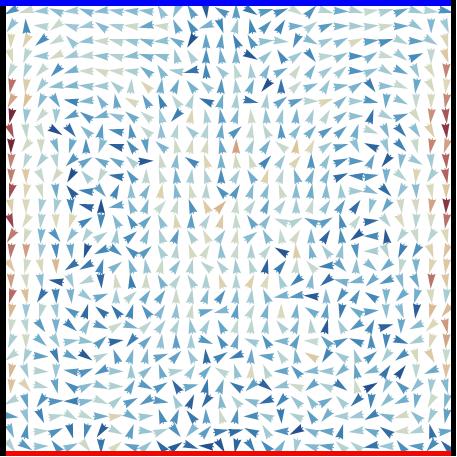}
\includegraphics[height=4.5cm,clip]{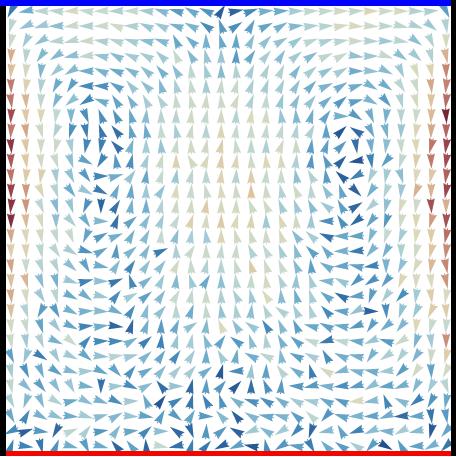}
\includegraphics[height=4.5cm,clip]{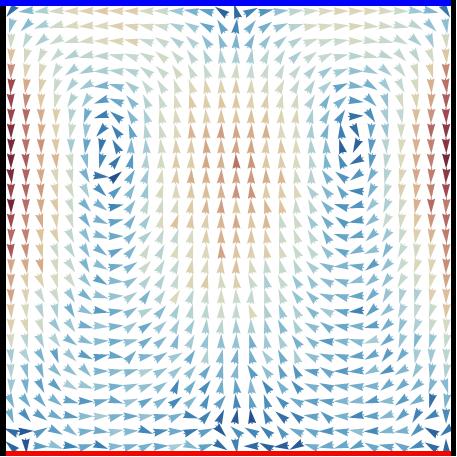}
\includegraphics[height=4.5cm,clip]{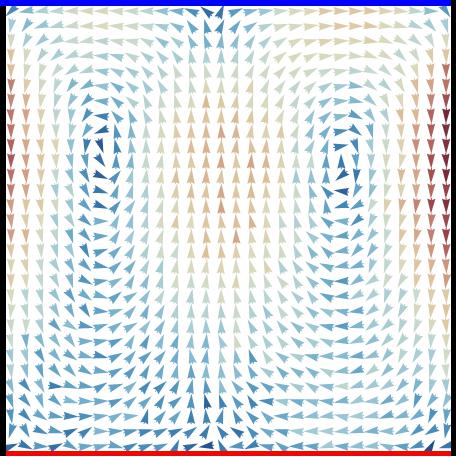}
\includegraphics[height=4.5cm,clip]{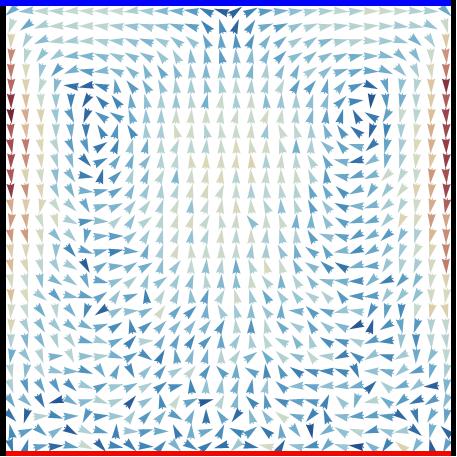}
\includegraphics[height=4.5cm,clip]{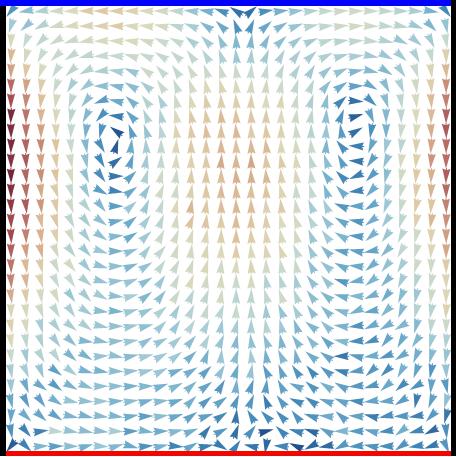}
\includegraphics[height=4.5cm,clip]{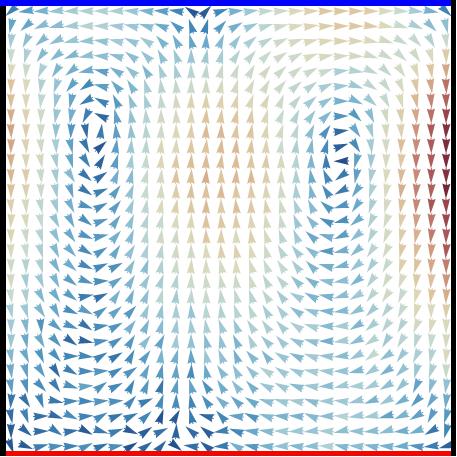}
\end{center}
\caption{Hydrodynamic velocity field. First column: $g=5$ and $T_0=1.4$, $T_0=2.6$, $T_0=4$, $T_0=10$ and $T_0=20$ from top to bottom.  Second column: $g=10$ and $T_0=1.8$, $T_0=4$, $T_0=6$, $T_0=15$ and $T_0=20$ from top to bottom. Third column: $g=15$ and $T_0=2$, $T_0=6$, $T_0=8$, $T_0=15$ and $T_0=20$ from top to bottom. \label{velofield}}
\end{figure}

\section{Spatial structures}

Once clarified (more or less) the existence of two well defined regimes: non-convective and convective, we can study the spatial structures of the local magnitudes. In general, as Navier-Sokes equations predicts,  we expect that a given local observable $A$  should be only function of $y$ for the non-convective states, $A=A(y)$, and  $A=A(x,y)$ in the convective states. We are going to analyze among others the behavior of the local  {\it hydrodynamic velocity}, {\it temperature}, {\it areal density} and the {\it virial pressure}.

\subsection{\it THE HYDRODYNAMIC VELOCITY FIELD} 

The velocity field is measured as the average center of mass velocity, eq.(\ref{velocell}), at each of the $900$ ($30\times 30$) virtual cells in which we have divided our system. The modulus of the hydrodynamic velocity is typically very small compared with the averaged mean particle velocity (for instance, for the case $g=10$ and $T_0=20$, the largest cell averaged value is $0.11$ compared with the mean particle velocity of around $3$) Therefore we needed very large time averages to resolve the structure beyond the underlying chaotic like behavior of the particles. Moreover, as we will see, the field structure can only be analyzed with care far from the critical temperature where we can discriminate the field from fluctuations.  

We see in figure \ref{velofield} some snapshots of the configurations for $g=5$, $10$ and $15$ cases and five different temperatures.They have been chosen with the following criteria: the first one is below $T_c$, the second between $T_c$ and $T_{c,2}$, and the rest beyond $T_{c,2}$. It is clear that the disordered behavior of the configurations below $T_c$ (first row). For temperatures between $T_c$ and $T_{c,2}$ some incipient local order is devised but fluctuations seems to dominate along the system. For large temperatures two rolls of convective fluid develops.  For the $g=5$ the rolls are noisier than the ones of the $g=10$ or $g=15$ cases,  in particular near the hot thermal bath. The color of the arrows indicate the velocity magnitude: from the smallest value in the given configuration (dark blue) to the largest value (dark red). We observe how the fluid goes up in the middle of the system with a intermediate velocity and the  fluid goes down following the vertical boundaries with higher velocity. Note that the hot thermal bath gives to the particles a lot of energy but it does not directly contribute to the coherence of the macroscopic movement of the fluid. The external field $g$ is the necessary mechanism that activates such movement: it seems to us that it is essential the acceleration that suffers the particles that goes the way down that, together with the neighbor interactions, tends to align the hydrodynamic velocity vector on the y-direction. This ordered fluid lose part of it acquired momenta when interacting with the disordered hot region but maintains most part of its coherence. Moreover, the created ordered structure is capable to trap and align (in average) the very hot particles from the hot reservoir and the fluid accelerates even going against the external field.   

\begin{table}
\begin{center}
\begin{tabular}{c c}
 \begin{minipage}{.3\textwidth}
 \includegraphics[height=3.5cm,clip]{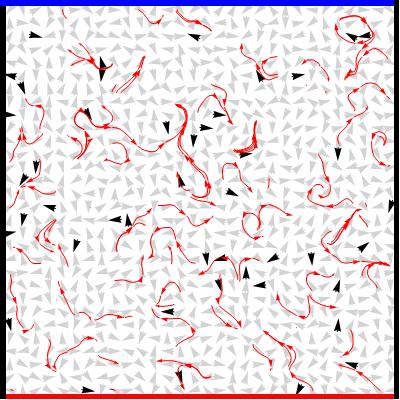}
    \end{minipage}
&
 \begin{minipage}{.3\textwidth}
 \kern 0.3cm
 \includegraphics[height=4.5cm,clip]{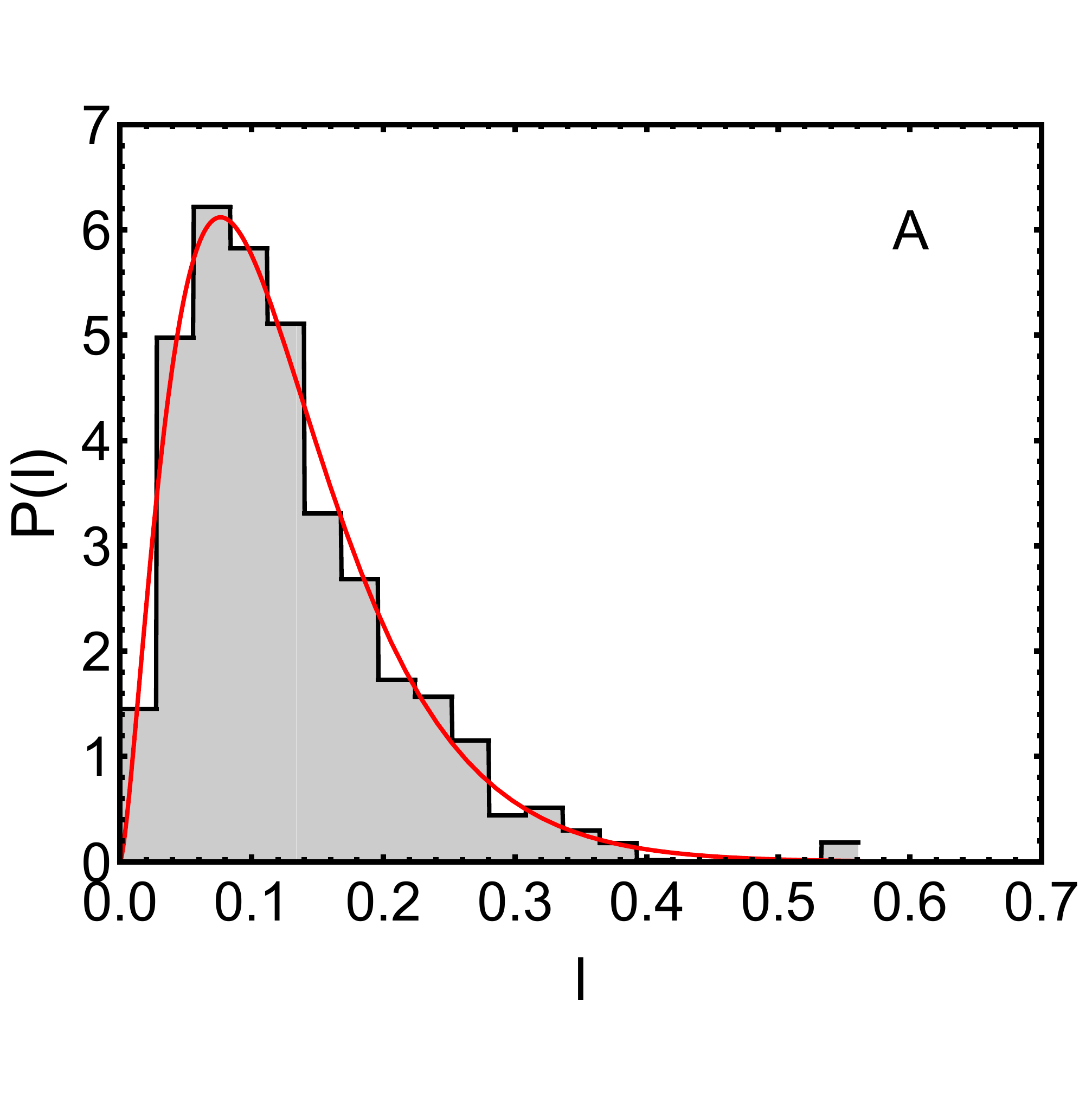}
    \end{minipage}\\[-15pt]
  \begin{minipage}{.3\textwidth}
   \includegraphics[height=3.5cm,clip]{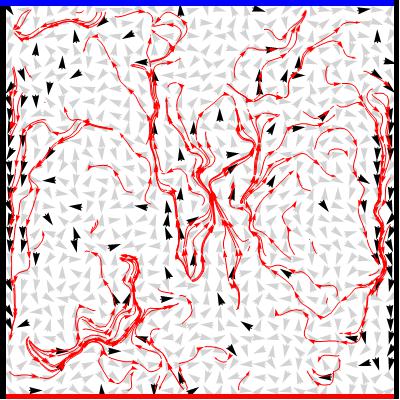}
    \end{minipage}
&
 \begin{minipage}{.3\textwidth}
 \kern 0.3cm
   \includegraphics[height=4.5cm,clip]{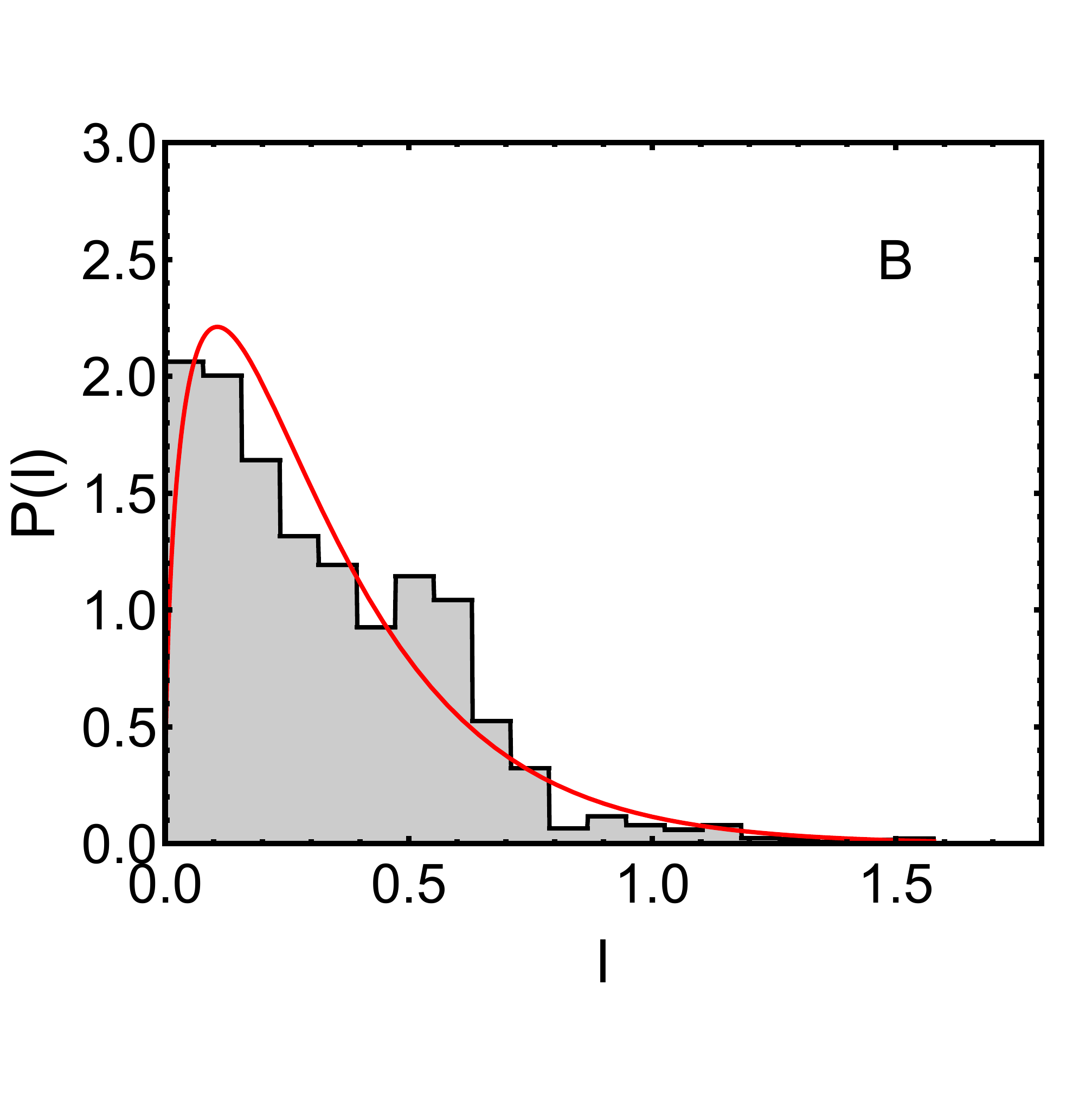}
    \end{minipage}  \\[-15pt]
 \begin{minipage}{.3\textwidth}
   \includegraphics[height=3.5cm,clip]{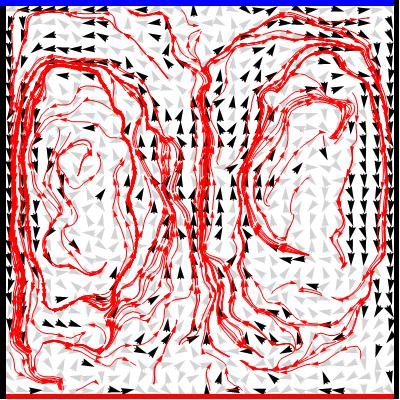}
    \end{minipage}
&
 \begin{minipage}{.3\textwidth}
 \kern 0.3cm
   \includegraphics[height=4.5cm,clip]{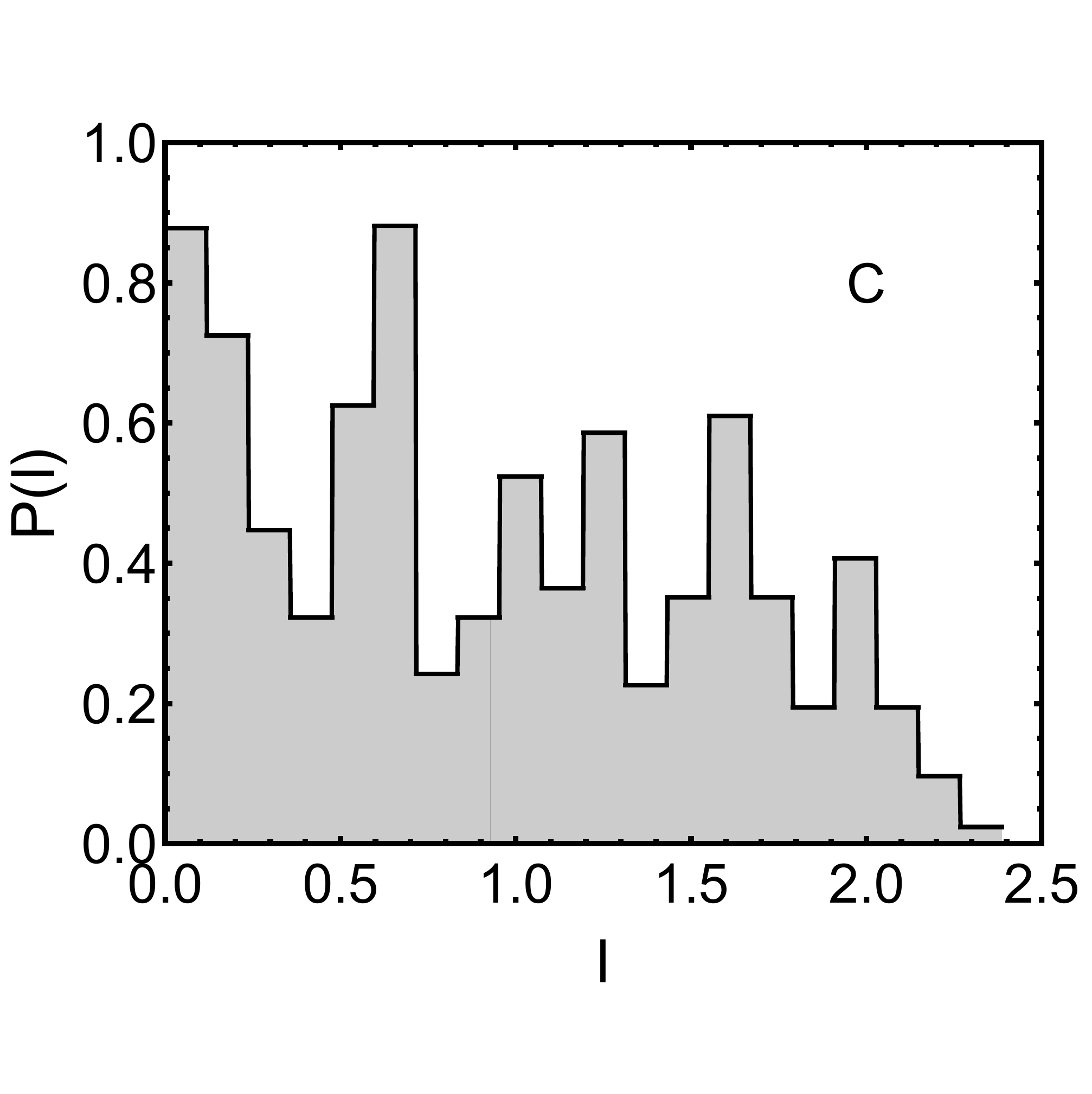}
    \end{minipage}  \\[-15pt]     
     \begin{minipage}{.3\textwidth}
   \includegraphics[height=3.5cm,clip]{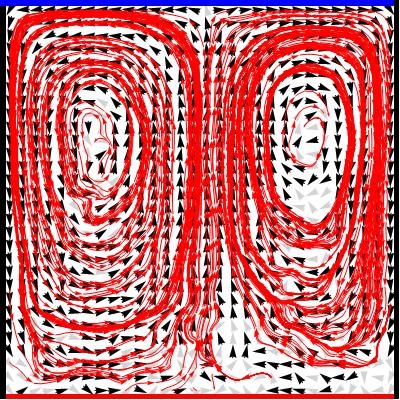}
    \end{minipage}
&
 \begin{minipage}{.3\textwidth}
 \kern 0.3cm
   \includegraphics[height=4.5cm,clip]{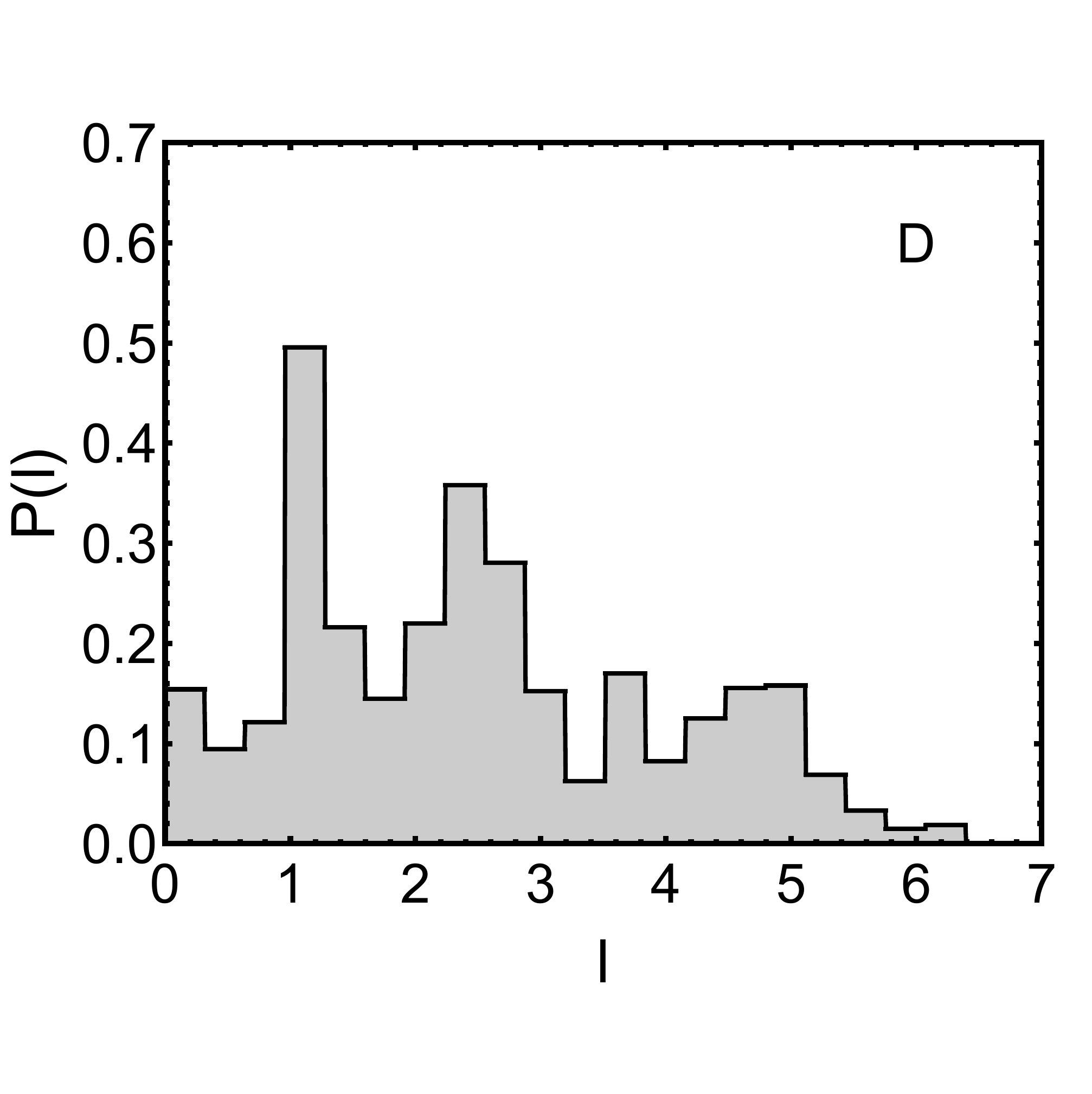}
    \end{minipage}  \\[-15pt]
     \begin{minipage}{.3\textwidth}
   \includegraphics[height=3.5cm,clip]{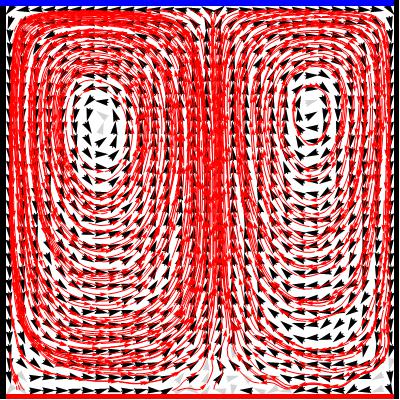}
    \end{minipage}
&
 \begin{minipage}{.3\textwidth}
 \kern 0.3cm
   \includegraphics[height=4.5cm,clip]{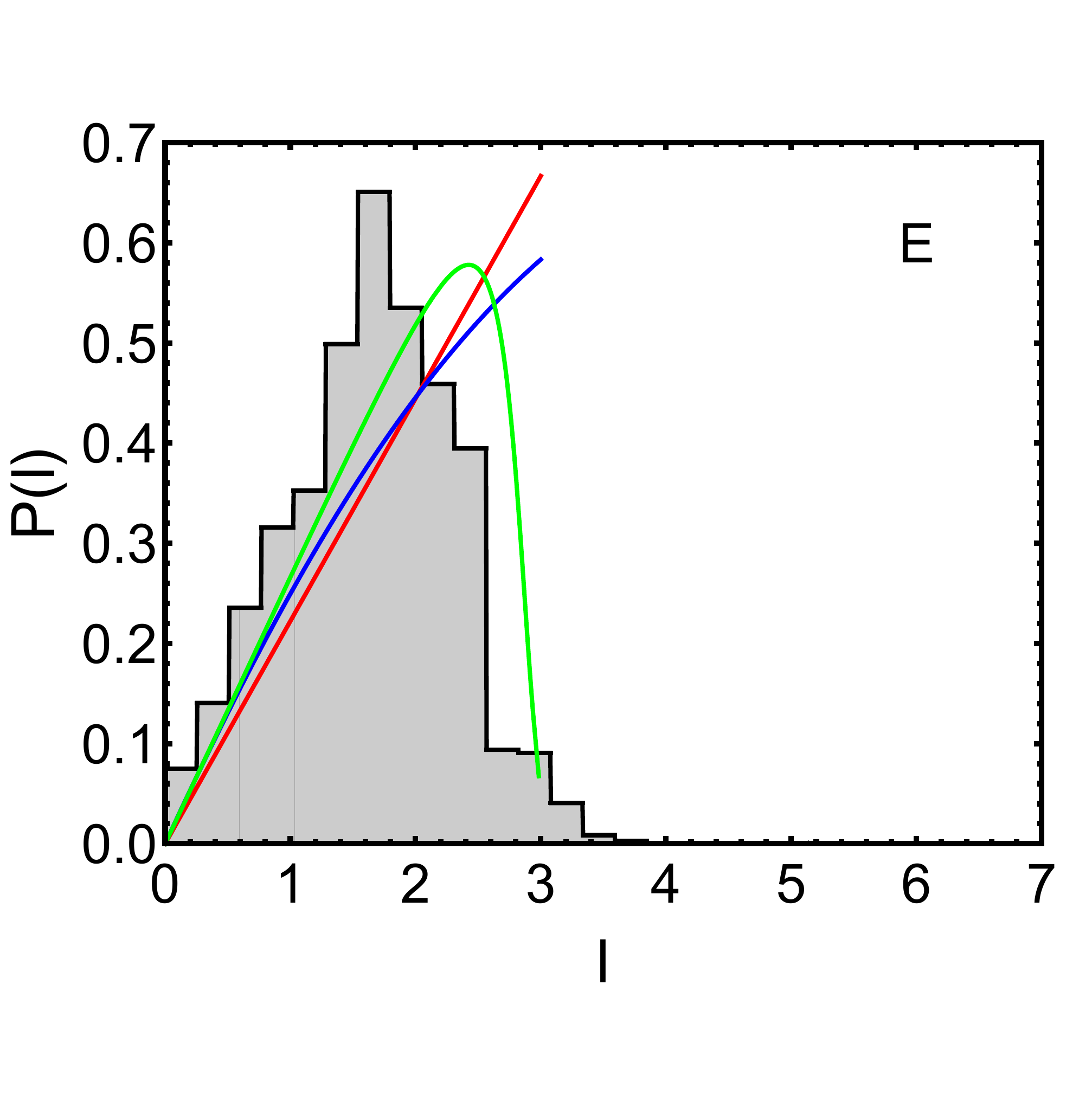}
    \end{minipage} 
\end{tabular}
\end{center}
\caption{Left column: Stream lines sample for the hydrodynamic velocity field in the case of $g=10$ and $T_0=1.8$, $T_0=4$, $T_0=6$, $T_0=10$ and $T_0=20$ from top to bottom. Right column: corresponding probability distribution of stream line lengths for $10000$ random initial points. Color curves are fitted distributions (see text)  \label{stream}}
\end{table}

\subsubsection{Stream lines}

In order to analyze the onset of convection from the spatial structure of the velocity vector field we have computed the distribution of its stream lines for each case. We define {\it stream line} as the trajectory build from an initial point whose point tangents are the given fixed vector field. That is, let $\vec u=(u_1(x,y),u_2(x,y))$ a given vector field and $(x_0,y_0)$ an arbitrary initial point. Then, the stream line is solution of the differential equation
\begin{equation}
\frac{dy}{dx}=\frac{u_2(x,y)}{u_1(x,y)}
\end{equation}
or in parametric form:
\begin{equation}
\frac{dx}{ds}=u_1(x(s),y(s))\quad ,\quad \frac{dy}{ds}=u_2(x(s),y(s))
\end{equation}
 The numerical solutions can be found just by using a Runge-Kutta integrator forward and backwards in time from the initial point. In our case we have a vector field defined over a grid. Then, we can reconstruct the vector field at each integration step and in a given point by linear interpolating it from neighboring grid sites. Because the vector field is not generated by an analytic function and there are noise effects, we should include a stop condition on the integration: $\vec u(n+1)\cdot\vec u(n)<0$ where $\vec u(n)$ is the vector used in the $n$-th step integration (from point $\vec r(n)$ to the next one $\vec r(n+1)$) and $\vec u(n+1)$ is the interpolation vector computed at the point $\vec r(n+1)$ (integration of stream curves from a discrete vector field are used in imaging processing, see for instance ref. \cite{Stalling} for some other technicalities). For any given vector field we generate $10000$ stream lines with random chosen initial point uniformly distributed inside the vector grid and we compute the statistics of the path-lengths. In Table \ref{stream} (left column) we show a small sample of $100$ stream lines corresponding to the $g=10$ case  and temperatures $T_0=1.8$ (A), $T_0=4$ (B), $T_0=6$ (C), $T_0=10$ (D) and $T_0=20$ (E). We choose these points on purpose to show the typical behavior at each of the regions.
 
\begin{figure}[h!]
\begin{center}
 \includegraphics[height=5cm,clip]{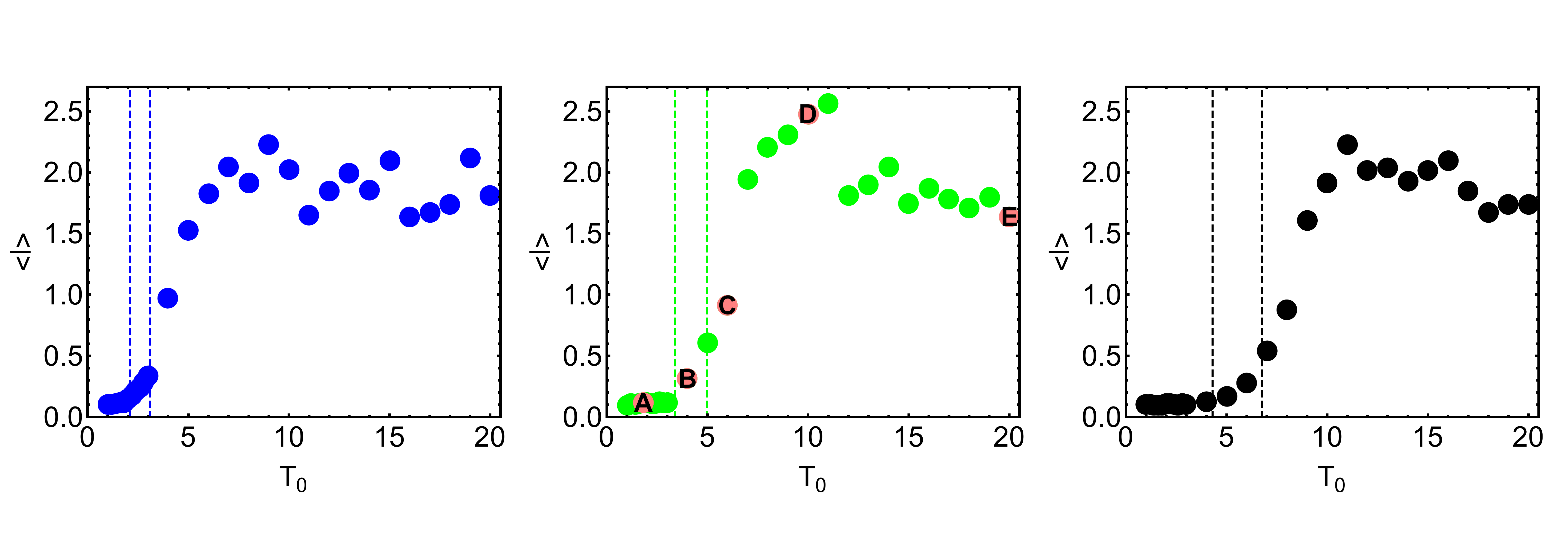}%velo_field_an4.nb
 \end{center}
 \kern -0.75cm
 \caption{Average stream length versus $T_0$ for $g=5$ (left) $g=10$ (center) and $g=15$ (right). The labels shown at the $g=10$ case correspond to the temperatures $T_0=1.8$ (A), $T_0=4$ (B), $T_0=6$ (C), $T_0=10$ (D) and $T_0=20$ (E) whose explicit length distribution and stream lines plot are shown in Table \ref{stream}. Vertical dashed lines correspond to $T_c$ and $T_{c,2}$ for each $g$ value. \label{str1}}
\end{figure}

\begin{figure}[h!]
\begin{center}
 \includegraphics[height=5cm,clip]{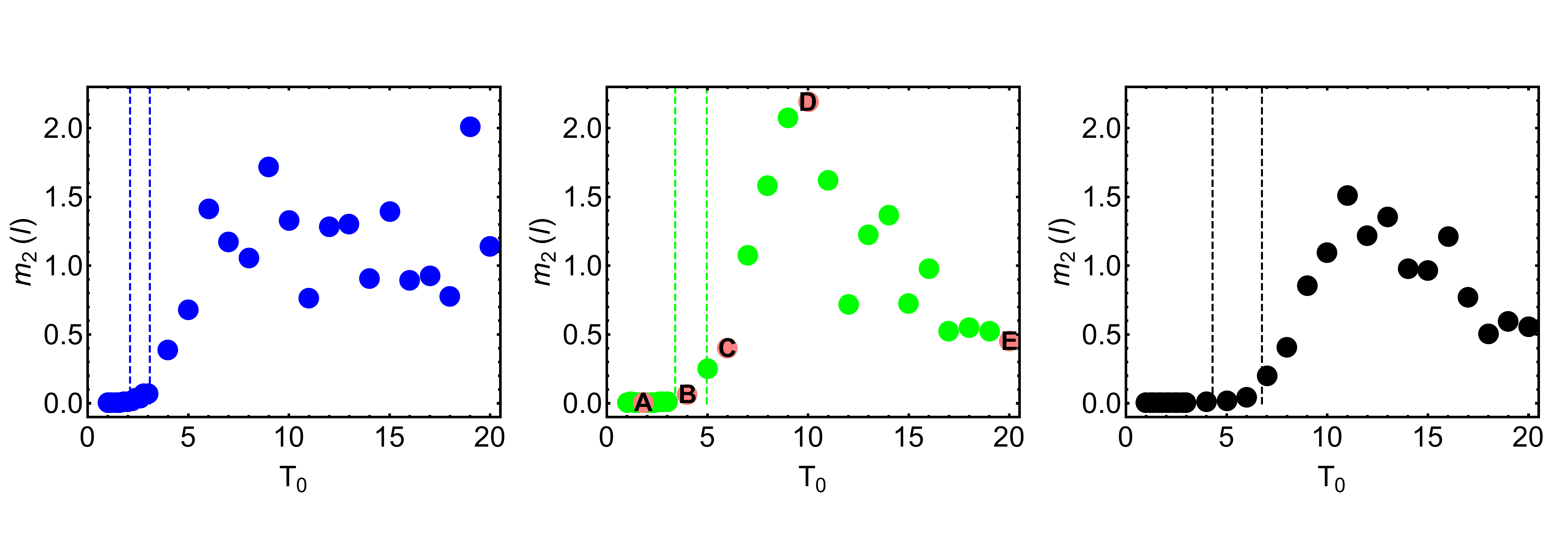}%velo_field_an4.nb
 \end{center}
 \kern -0.75cm
 \caption{Variance of the stream length distribution versus $T_0$ for $g=5$ (left) $g=10$ (center) and $g=15$ (right). The labels shown at the $g=10$ case correspond to the temperatures $T_0=1.8$ (A), $T_0=4$ (B), $T_0=6$ (C), $T_0=10$ (D) and $T_0=20$ (E) whose explicit length distribution and stream lines plot are shown in Table \ref{stream}. Vertical dashed lines correspond to $T_c$ and $T_{c,2}$ for each $g$ value. \label{str2}}
\end{figure}

The averaged stream line length versus $T_0$ for $g=5$, $g=10$ and $g=15$ is shown in figure \ref{str1}. In the non-convecting region it gets values of about $0.1$ and it grows softly until it reaches $T_c$ when it grows fast until it reaches the value $2$ at a $T_0\simeq 10$ from which it fluctuates around. Observe that $3$ is the perimeter of a rectangle of sides $1\times 1/2$. The big peak around $T_0=10$ for $g=10$ is just due to the discrete structure of our vector field that favors the appearance of large spirals when our algorithm is applied  (think in a closed trajectory with all vectors aligned to it except for one  pointing out). We can explicitly see this effect in the stream trajectories  sample corresponding to the point $D$ in figure \ref{stream}. For large $T_0$ values (point E) we have mainly closed stream lines and the spirals tend to disappear. We are pretty sure that if we do averaging over other stationary vector field realizations (assuming that there is a fluctuating part on it due to the finite time averaging used) and we simulate much larger systems would erase such peaked behavior. The variance of the stream length distribution is shown in figure \ref{str2} where  we again observe the large variance around $T_0=10$ for $g=10$ indicating the strong sensibility on the length to the initial chosen random point.

\begin{figure}[h!]
\begin{center}
 \includegraphics[height=5cm,clip]{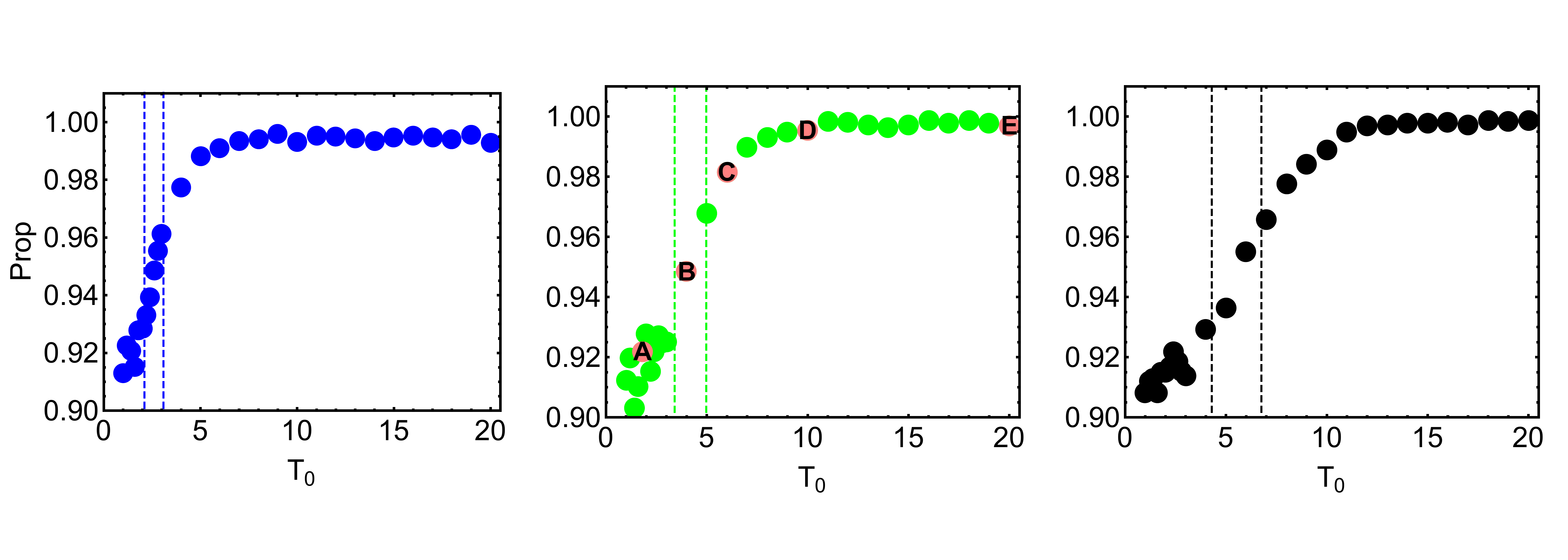}%velo_field_an4.nb
 \end{center}
 \kern -0.75cm
 \caption{Fraction of initial points (Prop) for which the stream lines lengths are larger than $1/200$ as a function of $T_0$ for $g=5$ (left) $g=10$ (center) and $g=15$ (right). The labels shown at the $g=10$ case correspond to the temperatures $T_0=1.8$ (A), $T_0=4$ (B), $T_0=6$ (C), $T_0=10$ (D) and $T_0=20$ (E) whose explicit length distribution and stream lines plot are shown in Table \ref{stream}. Vertical dashed lines correspond to $T_c$ and $T_{c,2}$ for each $g$ value. \label{str0}}
\end{figure}

The distribution of lengths changes its nature from the non-convective to convective regions as we see in Table \ref{stream}. First observe that there are a set of initial seeds whose stream length is less that $1/200$. That is, the algorithm stops just at the first steps. Such points correspond typically to strong noisy regions where local hydrodynamic velocity fields are dominated by microscopic fluctuations. We show at figure \ref{str0} the proportion of points with length larger than $1/200$. In all cases, the noisy regions tend to diminish when we increase $T_0$. For $T_0<T_c$ (non-convecting region) there are always about $8\%$ of such points. The proportion diminishes to zero for $T_0>>T_c$ with a sigmoid like of behavior. It is curious that $T_{c,2}$ seem to show, in all cases, the typical intermediate point where the proportion of strong noisy points is of about $3\%$ 
\begin{figure}[h!]
\begin{center}
 \includegraphics[height=5cm,clip]{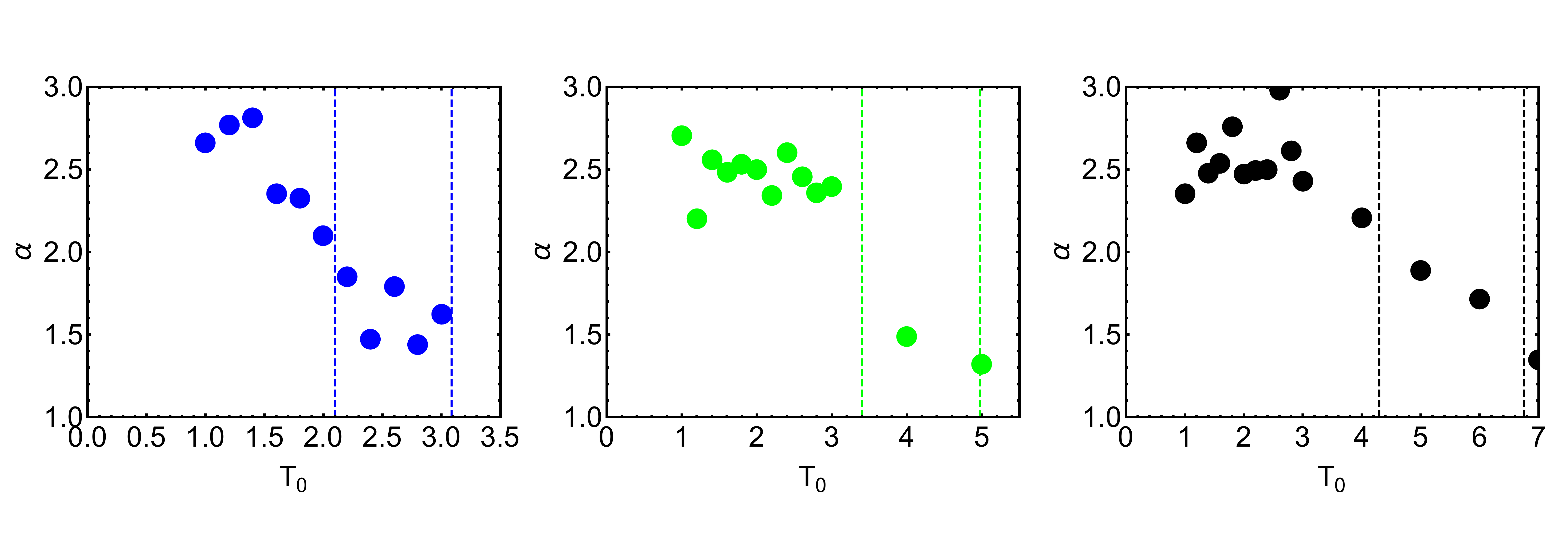}%velo_field_an4.nb
 \end{center}
 \kern -0.75cm
 \caption{Fitted value of exponent $\alpha$ versus $T_0<T_{c,2}$ for $g=5$ (left) $g=10$ (center) and $g=15$ (right). Vertical dashed lines correspond to $T_c$ and $T_{c,2}$ for each $g$ value. \label{stra}}
\end{figure}

\begin{figure}[h!]
\begin{center}
 \includegraphics[height=5cm,clip]{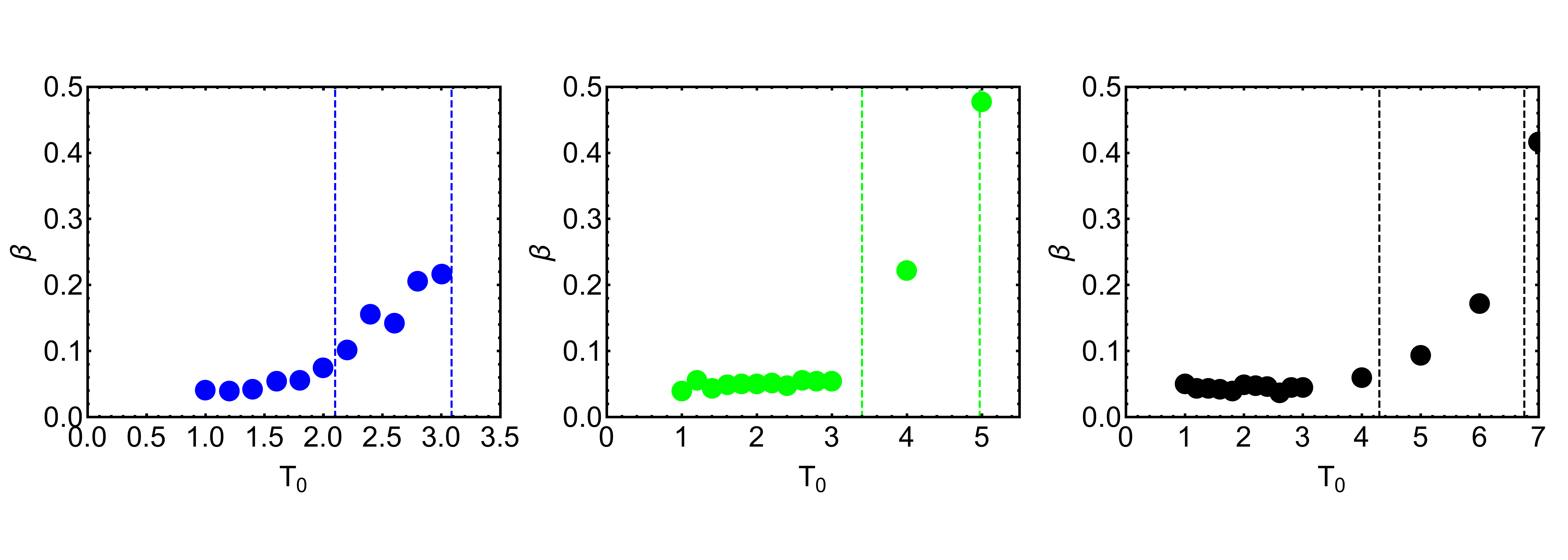}%velo_field_an4.nb
 \end{center}
 \kern -0.75cm
 \caption{Fitted value of exponent $\beta$ versus $T_0<T_{c,2}$ for $g=5$ (left) $g=10$ (center) and $g=15$ (right). Vertical dashed lines correspond to $T_c$ and $T_{c,2}$ for each $g$ value. \label{strb}}
\end{figure}
For the non-convective region ($T_0<T_c$) and once discarded the strong noisy data points set, the remainder distribution is reasonable well fitted by a gamma-distribution:
\begin{equation}
P(l;\alpha,\beta)=\frac{1}{\beta^{\alpha}\Gamma(\alpha)}l^{\alpha-1}e^{-l/\beta}
\end{equation}
 as we show with the red curve in Table \ref{stream} (A). The $\alpha$-fitted values are about $2.5$ (see figure \ref{stra}). The $\beta$ fitted values grows with $T_0$ similarly as $\langle l\rangle$ does (in fact it is just a natural length scale). The Gamma distribution fit  fails when we analyze the data corresponding to convecting states ($T_0>T_c$). The point (B) with $T_c<T_0<T_{c,2}$ has a distribution that looks like a shifted-Gamma distribution with a bump that reflects the existence of large stream lines (see for instance the red curve in Table \ref{stream} (B)). For $T_0>T_{c,2}$ the Gamma distribution is totally lost and there appears a extended distribution with some nontrivial peaked structure. We see in Table \ref{stream} (C) how there are a collection of ``channels''  where the long stream lines focus. They are responsible of the distribution expansion to higher $l$-values and to the existence of local peaks on it. The distribution at the point (D) confirms our idea about the existence of spirals. We see there how the probability of having stream lines larger than $3$ is non-zero. Even there are lines with length around $6$  which implies the existence of trajectories wrapping around half of the system several times. We recover a more physical behavior for a fluid  at the point $(E)$ ($T_0=20$) where the distribution drops fast at $l\simeq 3$. We can construct a simple model to understand, in some way, this limiting stream line distribution.

\begin{figure}[h!]
\begin{center}
 \includegraphics[height=5cm,clip]{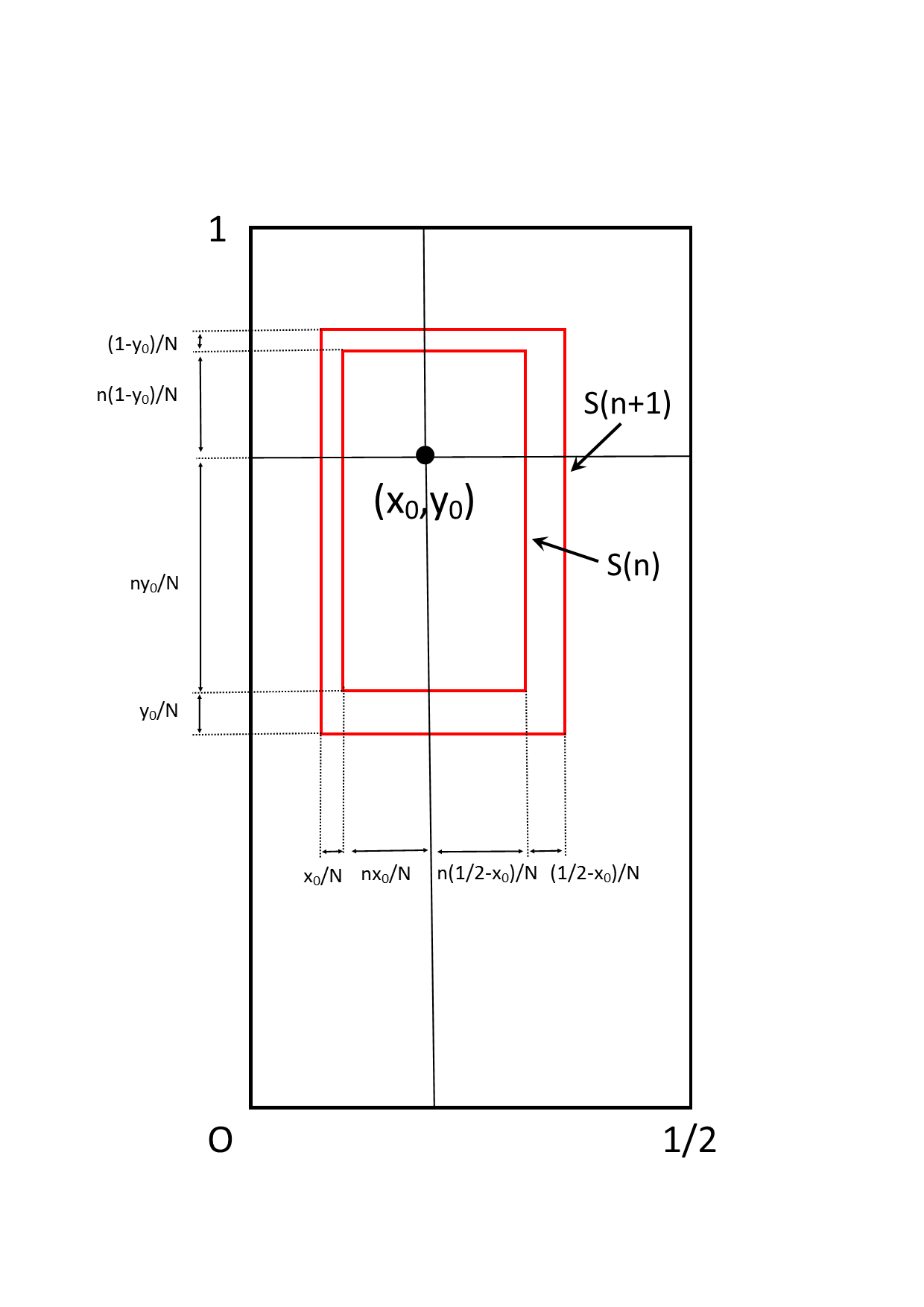}%model1.pub
 \end{center}
 \kern -0.75cm
 \caption{Schematic construction of $N$ rectangular stream lines around the point $(x_0,y_0)$. \label{strlines1}}
\end{figure}

Let us assume that there are $N$ closed stream lines turning around the point $P:(x_0,y_0)$. We also assume that along the principal axis with center at $P$, the lines are uniformly distributed. This implies that their crossing point at the axis with origin at $P$ are mutually separated by $(1/2-x_0)/N$ along the $PX$ positive axis, $x_0/N$ along the $PX$ negative axis, $(1-y0)/N$ along the $PY$ positive axis and $y0/N$ along the $PY$ negative axis. Stream lines should also be perpendicular to the axis at the crossing points. With this set of assumptions we just have to build a family of self avoiding curves connecting the crossing points. The idea is to build them with some structural parameter in such a way that the limiting value of such curves should be the containing rectangle. 

The simplest example is to assume that the stream lines are just rectangles (see figure \ref{strlines1}). In this case the $n$-th stream line is built of $8$ pieces, two for each of the four quadrants:
\begin{eqnarray}
y=y_0+n (1-y_0)/N\quad x\in[x_0,x_0+n(1/2-x_0)/N] &\quad ;\quad& x=x_0+n(1/2-x_0)/N \quad y\in[y_0,y_0+n(1-y_0)/N]\\ \nonumber
y=y_0+n (1-y_0)/N\quad x\in[x_0-n x_0/N,x_0] &\quad ;\quad& x=x_0-n x_0/N \quad y\in[y_0,y_0+n(1-y_0)/N]\\
\nonumber
y=y_0-n y_0/N \quad x\in[x_0,x_0+n(1/2-x_0)/N] &\quad ;\quad&x=x_0+n(1/2-x_0)/N \quad y\in[y_0-n y_0/N,y_0]\\
\nonumber
y=y_0-n y_0/N \quad x\in[x_0-n x_0/N,x_0] &\quad ;\quad&x=x_0-n x_0/N \quad y\in[y_0-n y_0/N,y_0]
\end{eqnarray}

\begin{figure}[h!]
\begin{center}
 \includegraphics[height=4cm,clip]{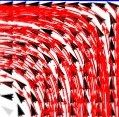}%spirals.nb
\vskip 0.2cm
  \includegraphics[height=4cm,clip]{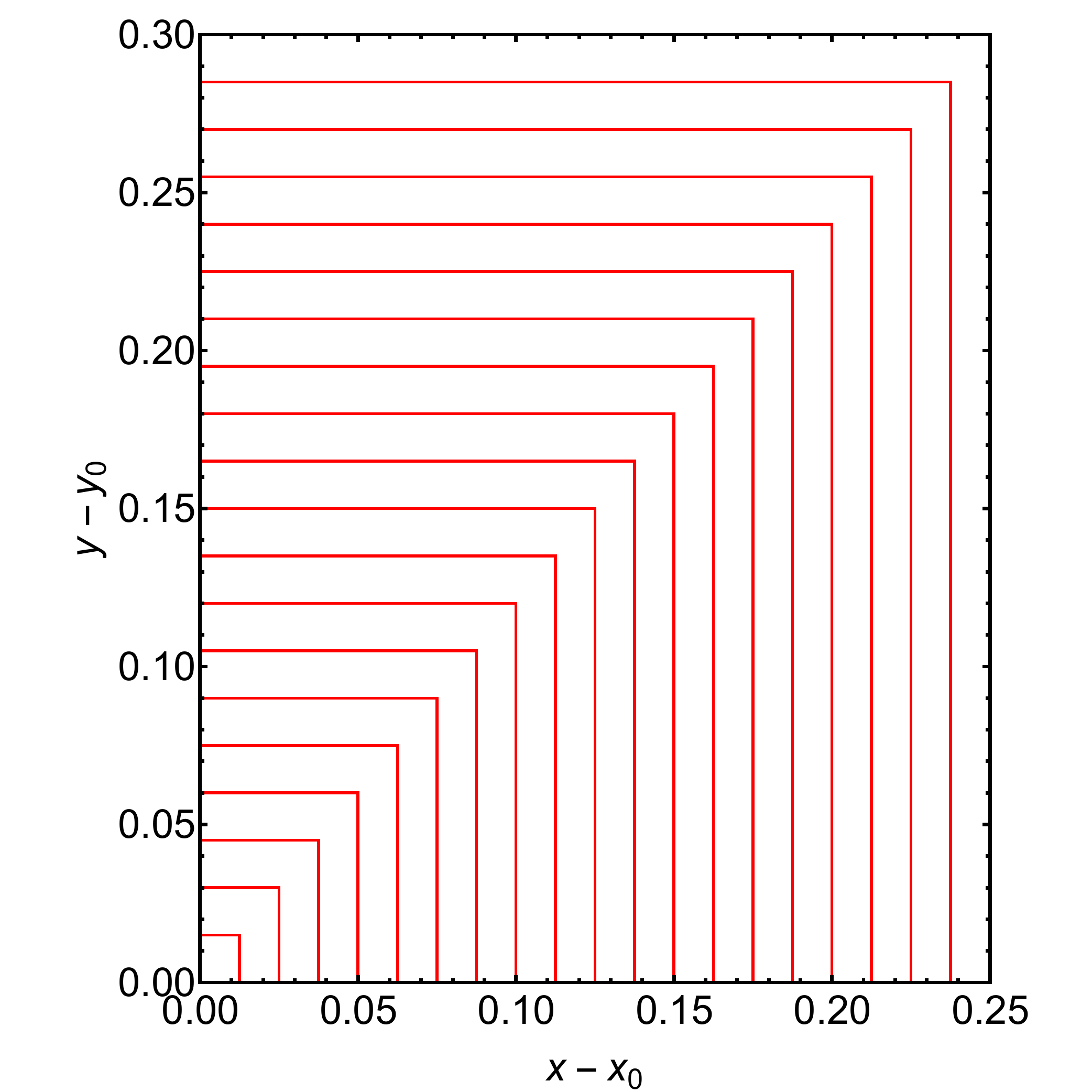}
   \includegraphics[height=4cm,clip]{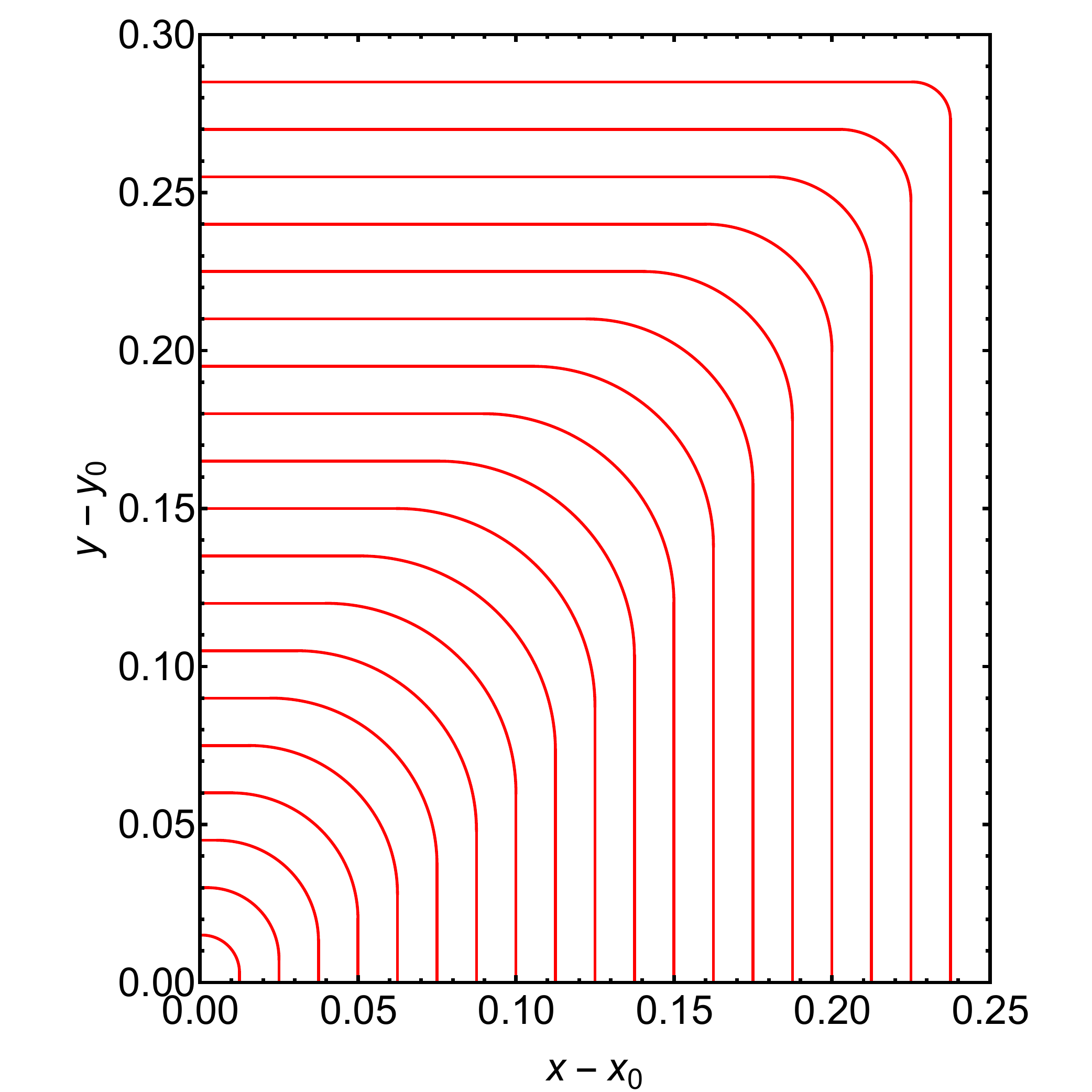}
    \includegraphics[height=4cm,clip]{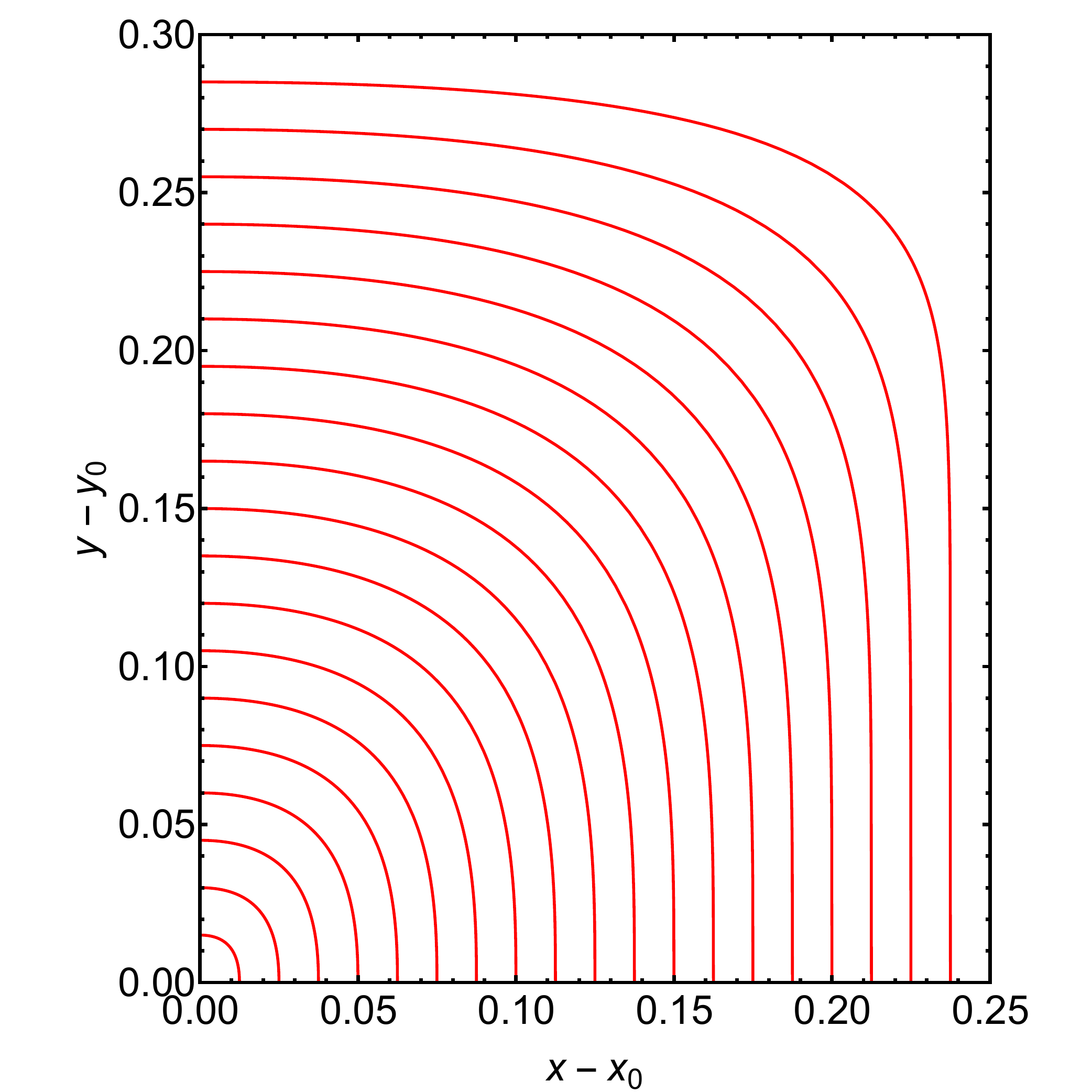}
 \end{center}
 \kern -0.25cm
 \caption{Top figure: Detail of the stream lines obtained in the case $g=10$ and $T_0=20$ (see figure \ref{stream}). We plot just the first quadrant of the left  roll whose center is at $(x_0,y_0)=(0.25,0.7)$. Figures Bottom: Equidistant rectangular stream lines (left), including a circular corner (center) and a continuous function (see text). \label{stream_types}}
\end{figure}

The $n$-th rectangle surface is $S(n)=n^2/2N^2$ with $n=1,\ldots, N$. The length of the $n$-th stream line is just the perimeter of the rectangle: $l(n)=3n/N$. Then, the probability of having a stream line of length $l(n)$ can be built as the surface diference between rectangles $n$ and $n-1$:
\begin{equation}
Q(l(n))=2(S(n)-S(n-1))=\frac{2n-1}{N^2}
\end{equation}
and the probability distribution: 
\begin{equation}
P(l)=Q(l)/\Delta l=\frac{2n-1}{3N}=\frac{2}{9}l-\frac{1}{3N}\quad l=\frac{3}{N}, \frac{6}{N},\ldots, 3
\end{equation}
we can do the limit $N\rightarrow\infty$ and $P(l)=2l/9$ with $l\in[0,3]$.  It is trivial to compute the average lenth and the variance:
\begin{equation}
\langle l\rangle= 2\qquad , \qquad m_2(l)=\frac{1}{2}
\end{equation}
 We can compare these results with the distribution in the case $g=10$ and $T_0=20$ (see right figure E in \ref{stream}).
There the red line is the distribution we have obtained with our model. It is clear that we have captured some of the observed behavior: increasing probability with $l$, cutoff probability at $l=3$, average $l$ of around $2$ and variance of  around $0.5$ (see figure \ref{str2}). Obviously, the rounding structure should be due to the curvy form of the stream lines. 
We tried two other types of stream lines in order to check this point (see figure \ref{stream_types})

First we introduced a circular corner of radius $R(n)$ to the above rectangular model of the stream lines (see the figure \ref{stream_types} bottom-center):
\begin{equation}
R(n)=f\left(\frac{n}{N}\right) min (L_1,L_2)\quad ,\quad f(u)=u(1-u)
\end{equation}
where $L_1$ and $L_2$ are the length of each quadrant with origin $(x_0,y_0)$: $L_1=1/2-x_0$, $L_2=1-y_0$ (first quadrant), $L_1=1/2-x_0$, $L_2=y_0$ (second quadrant), $L_1=x_0$, $L_2=y_0$ (third quadrant) and $L_1=x_0$, $L_2=1-y_0$ (fourth quadrant).
In this case the surface surrounded by the $n$-th stream line is given by:
\begin{equation}
S(n)=\frac{n^2}{2N^2}-(1-\frac{\pi}{4})\theta(x_0,y_0)f\left(\frac{n}{N}\right)^2 
\end{equation} 
with
\begin{equation}
\theta(x_0,y_0)= \left[min (x_0,y_0)\right]^2+\left[min (\frac{1}{2}-x_0,y_0)\right]^2+ \left[min (x_0,1-y_0)\right]^2+\left[min (\frac{1}{2}-x_0,1-y_0)\right]^2
\end{equation}
Following the same steps we did above for the rectangular stream lines case we obtain in the $N\rightarrow\infty$ limit
\begin{equation}
P(l)=2\left[u-2\theta(x_0,y_0)(1-\frac{\pi}{4})f(u)\frac{df}{du} \right]\left(\frac{dl}{du}\right)^{-1}
\end{equation}
where $l\in[0,3]$, $u\in[0,1]$,
\begin{equation}
l=3u-(2-\pi/2)w(x_0,y_0)f(u)
\end{equation}
and
\begin{equation}
w(x_0,y_0)=min (x_0,y_0)+min (\frac{1}{2}-x_0,y_0)+min (x_0,1-y_0)+min (\frac{1}{2}-x_0,1-y_0)
\end{equation}
This distribution applied to the case $g=10$ and $T_0=20$ give us $\langle l\rangle=1.9392$ and $m_2(l)=0.5190$. Moreover, the distribution form is plotted as a blue line in Table \ref{stream} (E) and there we observe that there is a curvature but still it is far to the observed one.

Finally we tried with a continuous curve at each quadrant. To connect them we ask to the curves be perpendicular to the crossing points at the quadrant axis with lengths $(L_1,L_2)$. Let $x_1=nL_1/N$  the crossing point of the n-th stream line on the x-axis (with center at $(x_0,y_0)$) and $y_1=nL_2/N$ the corresponding one with the y-axis. Then we have chosen the function 
\begin{equation}
y(x;n/N)=y_1 \tilde y(x/x_1;n/N)\equiv\frac{y_1}{1-\beta}(1-\frac{x}{x_1})^\beta-\frac{\beta y_1}{1-\beta}(1-\frac{x}{x_1})
\end{equation}
with $\beta=\beta_0(1-n/N)^{\beta_1}$ to compute $P(l)$. Observe that the function is tunned in such a form that $y(0)=y_1$, $y'(0)=0$, $y(x_1)=0$ and $y'(x_1)=\infty$. The distribution obtained in the limit $N\rightarrow\infty$ is, in this case,
\begin{equation}
P(l)=\left(u^2 \frac{dg}{du}+2ug(u)\right)\left(\frac{dl}{du}\right)^{-1}
\end{equation}
where
\begin{equation}
l(u)=u h(u)
\end{equation}
and
\begin{eqnarray}
g(u)&=&\int_0^1 dz\tilde y(z;u)\nonumber\\
h(u)&=&(\frac{1}{2}-x_0)\int_0^1 dz \left(1+\left(\frac{1-y_0}{1/2-x_0}\frac{d\tilde y}{dz}\right)^2\right)^{1/2}+
(\frac{1}{2}-x_0)\int_0^1 dz \left(1+\left(\frac{y_0}{1/2-x_0}\frac{d\tilde y}{dz}\right)^2\right)^{1/2}\\ \nonumber
&+&(\frac{1}{2}-x_0)\int_0^1 dz \left(1+\left(\frac{y_0}{x_0}\frac{d\tilde y}{dz}\right)^2\right)^{1/2}+
(\frac{1}{2}-x_0)\int_0^1 dz \left(1+\left(\frac{1-y_0}{x_0}\frac{d\tilde y}{dz}\right)^2\right)^{1/2}
\end{eqnarray}

\begin{figure}[h!]
\begin{center}
 \includegraphics[height=5cm,clip]{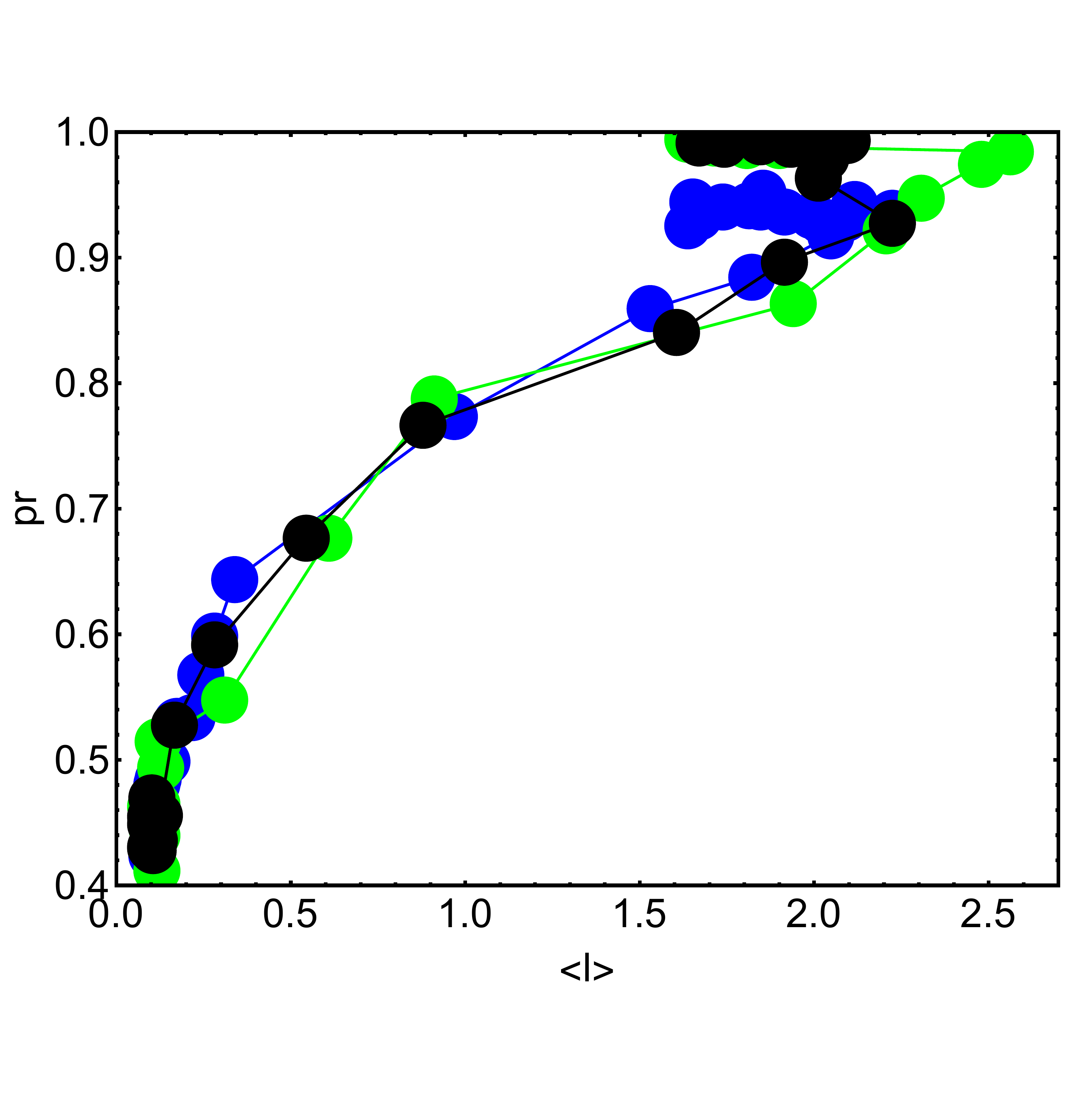}%velo_field_an4.nb
 \end{center}
 \kern -0.75cm
 \caption{Ratio of vectors whose modulus are larger than its standard deviation (pr) versus the average stream length ($\langle l\rangle$) for $g=5$ (blue dots) $g=10$ (green dots) and $g=15$ (black dots). \label{strc}}
\end{figure}
This distribution for $\beta_0=0.5$ and $\beta_1=0.5$ has the form plotted in figure \ref{stream_types} (bottom-left). In this case we find that $\langle l\rangle=1.8495$ and $m_2(l)=0.4496$. The overall distribution has the form shown by the green curve in Table \ref{stream} (E-green). Observe that the distribution has a similar form compared to the measured one but shifted to the right. This is probably due to the simplified assumption of equidistant crossing points through the axis centered at $(x_0,y_0)$. We cannot go further because the data we have obtained has a limited precision and playing with other distributions would just be a game. We wanted only to show that it is possible to do a systematic study of stream lines statistics  and we can unveil some interesting properties that characterize, from another point of view, the transition between the non-convective regime to the convective one, from a gamma distribution to a peaked distribution build from closed curved stream lines. It is an open issue to understand the transition between those extreme cases but it is beyond the computer effort done in this work. In order to do that study, one should need to focus the simulation in creating many vector configurations at the stationary state, for each one generating stream lines and, finally, do and overall precise study of the resulting distribution.
Nevertheless it could be very interesting to do such hard work. We plot in figure \ref{strc} the ratio of vectors whose modulus are larger than its standard deviation versus the measured stream length for all $g$ and $T_0$ values. We see how all the points corresponding to different $g$-values scale in a unique curve. That is, the average length depends on the proportion of ``ordered'' vector points (the ones whose modulus value is larger to the typical fluctuation interval and then it can be discriminate from the underlying noise). It seems that the stream line distribution maybe related to a kind of percolation phenomena. The overall picture from this point of view could be:
\begin{itemize}
\item  Non-convective regime ($T_0<T_c$): The proportion of non-noisy vector points ($pr$) are less than $\simeq 0.5$ and the averaged length seems to to be almost constant with $pr$.
\item Bad conducting regime ($T_c<T_0<T_{c,2}$): We observe that the values for which $T_0\simeq T_{c,2}$ correspond to  $\langle l\rangle\simeq 0.6$ and $pr\simeq 0.68$ in all cases. Notice that the $2$-d percolation of overlapping random disks of radius $r$ has a critical covered surface at the percolation threshold of $\simeq 0.67$ that is very near to our observed value. We can naturally connect both phenomena and we can conclude that when the set of ``ordered'' vectors percolates, the stream lines are long enough to connect both thermal baths and the {\it fully convection regime} appears.
\item Fully convecting regime ($T_0>T_{c,2}$): the noisy points tend to zero.
\end{itemize}
We are observing a typical fluctuating phenomena that it could disappear when the number of disks, or lattice points tends to infinity. However there are some open questions that we cannot resolve at this point: There exists the region $[T_c,T_{c,2}]$ at the hydrodynamic limit? If the answer is positive it would imply the existence of disordered closed stream lines of relative short length. That is, there would be solutions of Navier Stokes equations different from the well known regular convective ones. If  the answer is no, it would imply that $T_c=T_{c,2}$ in the hydrodynamic limit and the regular convective solutions would be the unique non-zero ones beyond the pure conducting regime.

\begin{figure}
\begin{center}
\includegraphics[height=5cm,clip]{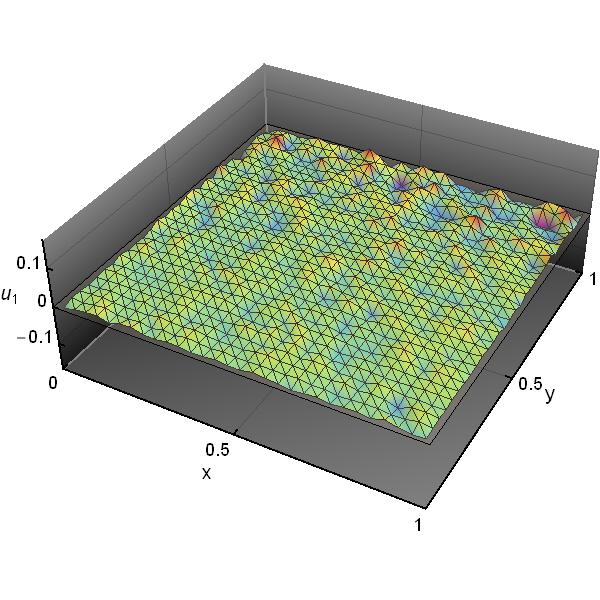}   %velo_field_show_vx.nb
\includegraphics[height=5cm,clip]{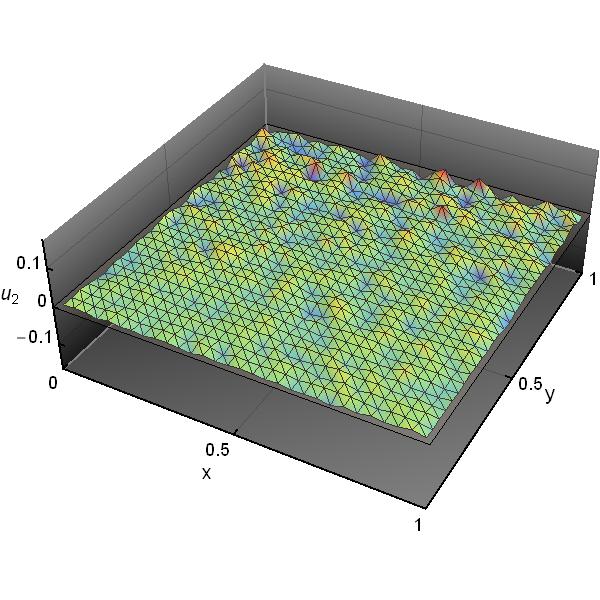}   %velo_field_show_vy.nb
\newline
\vglue -1cm
\includegraphics[height=5cm,clip]{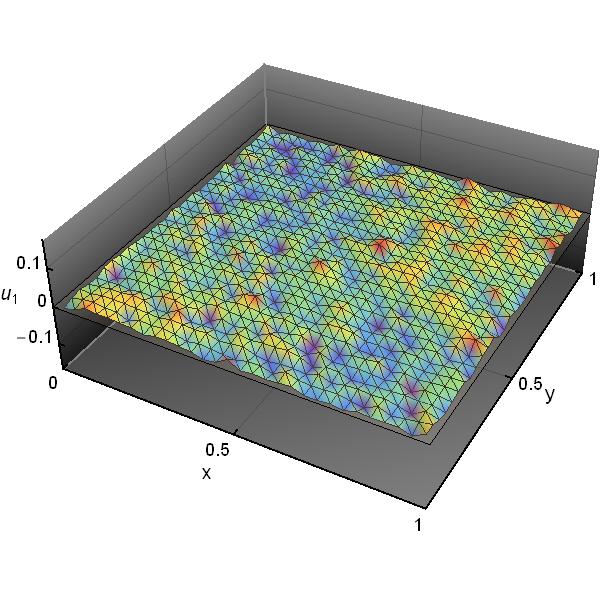}   %velo_field_show_vx.nb
\includegraphics[height=5cm,clip]{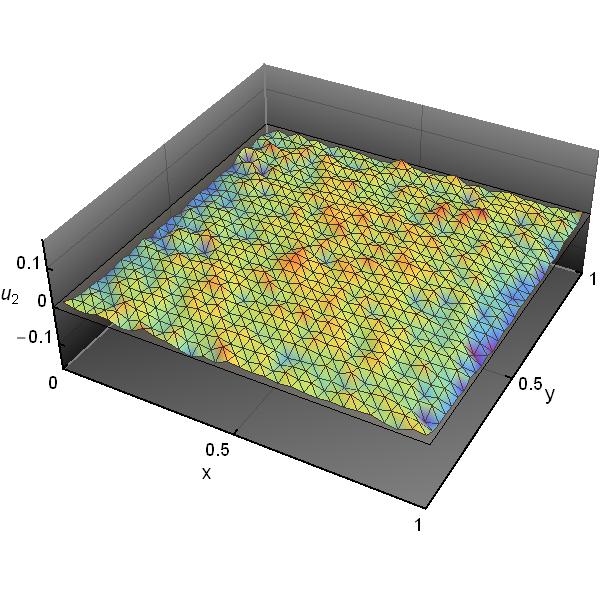}   %velo_field_show_vy.nb
\newline
\vglue -1cm
\includegraphics[height=5cm,clip]{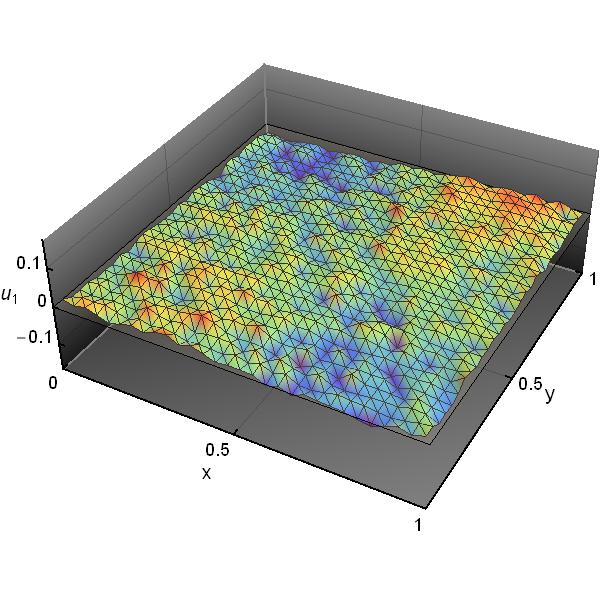}   %velo_field_show_vx.nb
\includegraphics[height=5cm,clip]{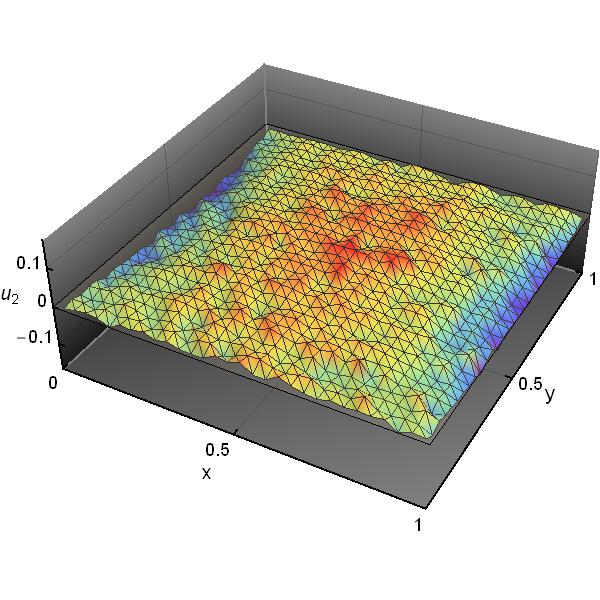}   %velo_field_show_vy.nb
\newline
\vglue -1cm
\includegraphics[height=5cm,clip]{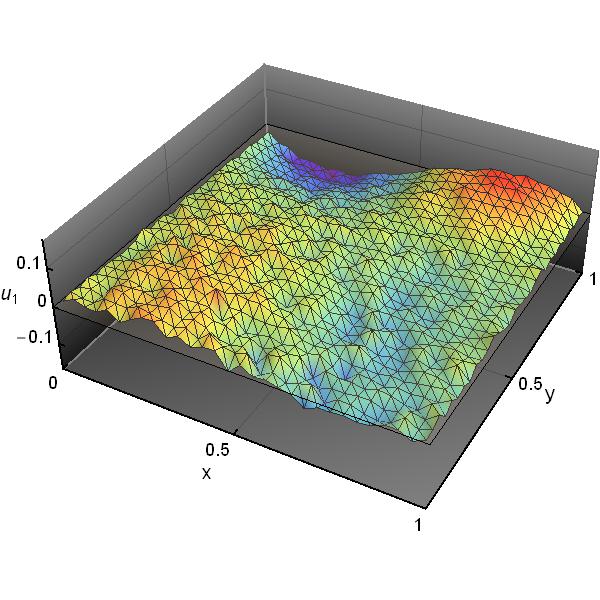}   %velo_field_show_vx.nb
\includegraphics[height=5cm,clip]{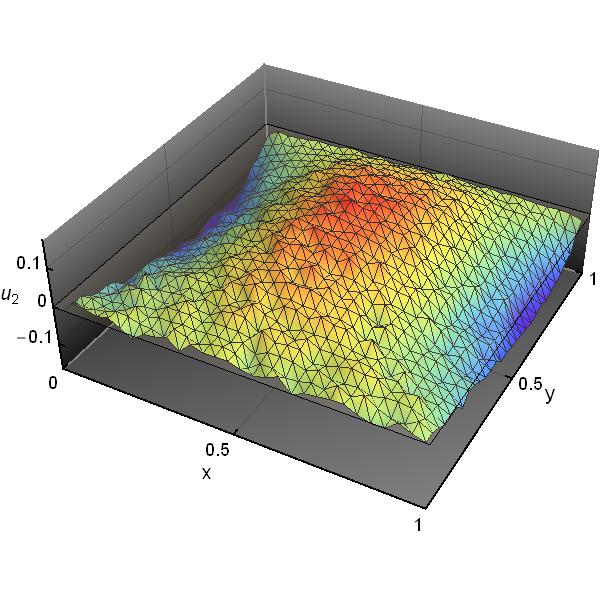}   %velo_field_show_vy.nb
\newline
\vglue -1cm
\includegraphics[height=5cm,clip]{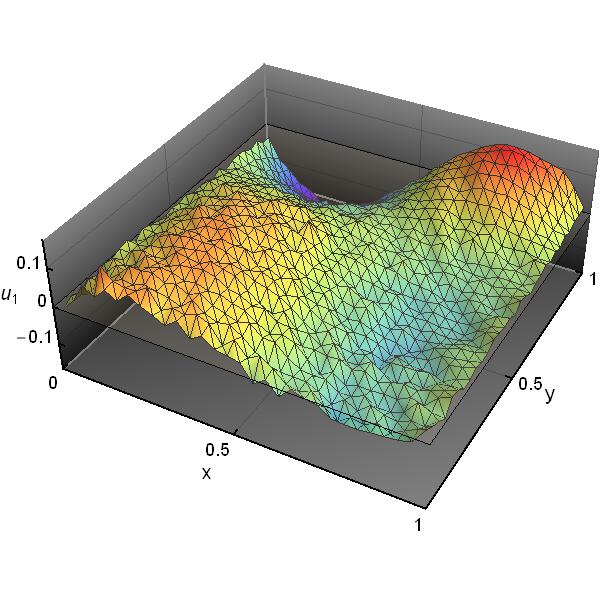}   %velo_field_show_vx.nb
\includegraphics[height=5cm,clip]{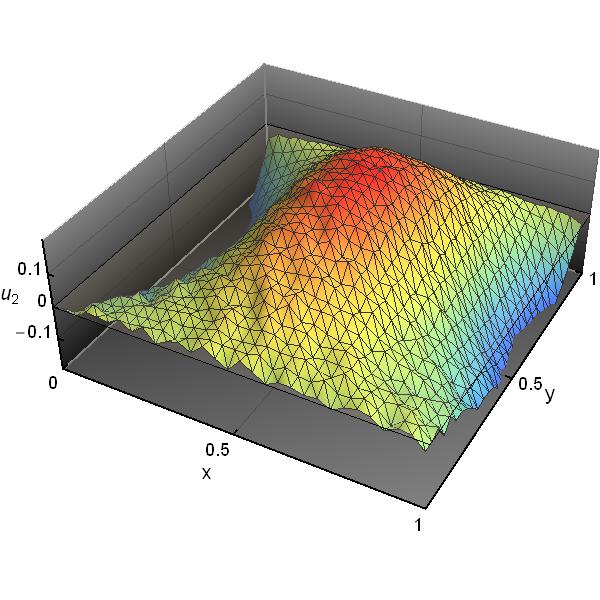}   %velo_field_show_vy.nb
\newline
\end{center}
\caption{Hydrodynamic velocity field components $u_1$ and $u_2$ (left and right columns respectively) for $g=10$ and, from top to bottom: $T_0=1.8$, $4$, $6$, $10$ and $20$.   \label{velofieldxy}}
\end{figure}

\subsubsection{The velocity vector field $u=(u_1(x,y),u_2(x,y))$ at the fully convective regime}

We show in figure \ref{velofieldxy} the averaged values obtained for the components of the hydrodynamic velocity field $u=(u_1(x,y),u_2(x,y))$ for $g=10$ and $T_0$ values corresponding to the particular cases studied in figure \ref{stream}. Observe how above $T_c$ there is a well defined spatial ordering for each velocity component that gets clearer as we increment the value of $T_0$. This favors the idea that the percolating phenomena we described in the above section appears only at the fluctuating level of description and it may disappear in the  hydrodynamic limit.

\begin{figure}[h!]
\begin{center}
\includegraphics[height=6cm,clip]{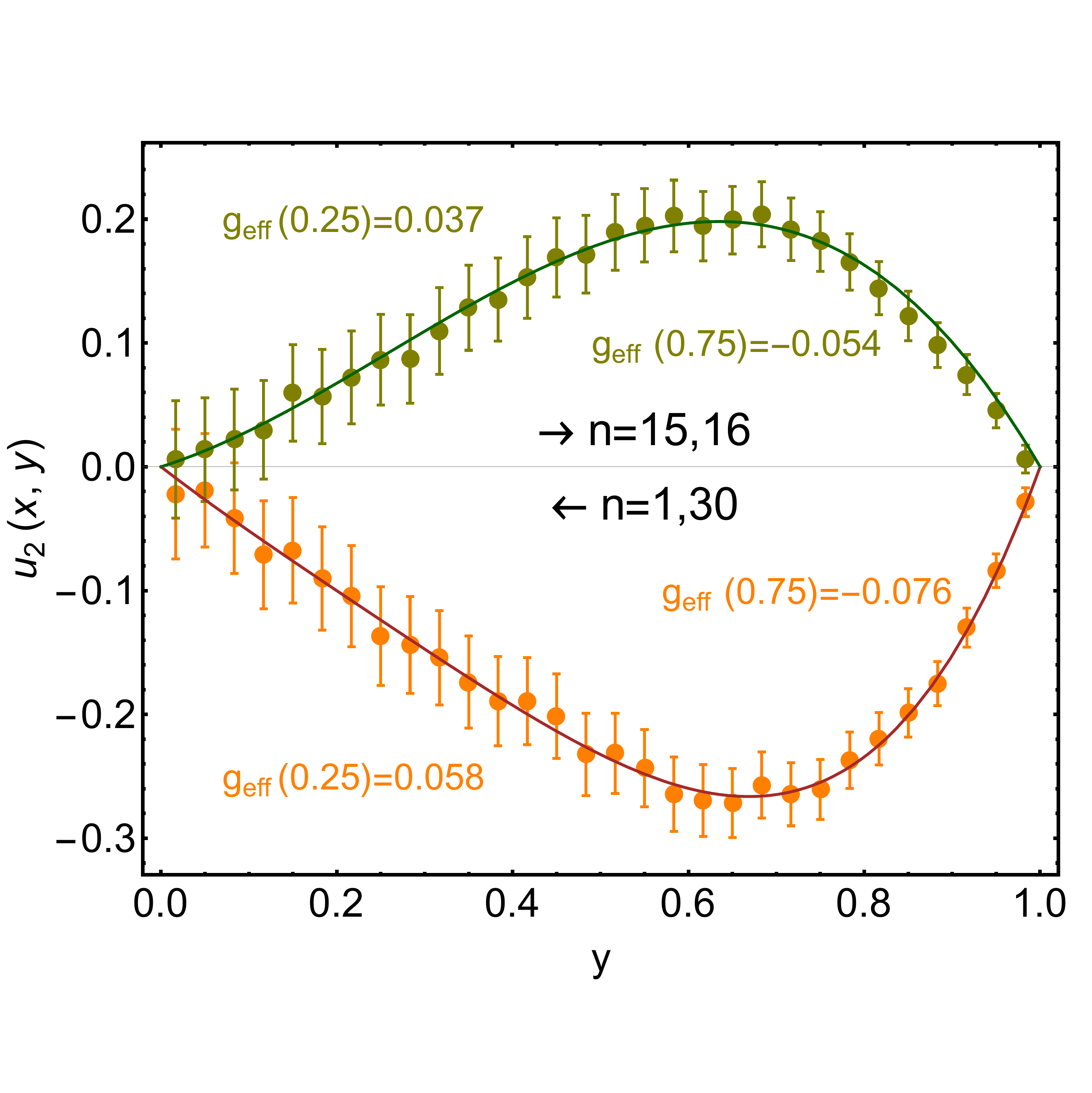}        %velo_profiles.nb
\includegraphics[height=6cm,clip]{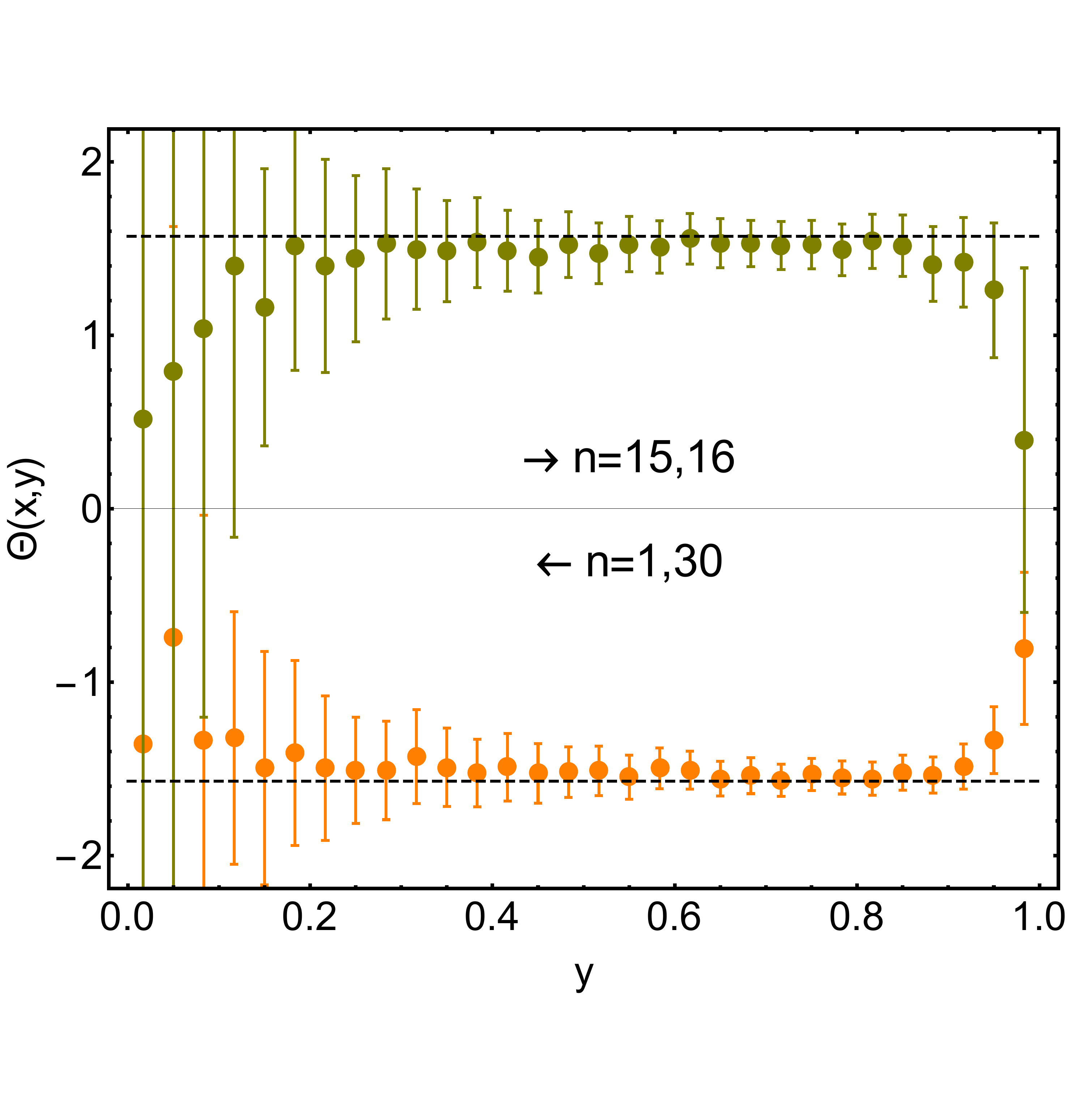}
\end{center}
\kern -1.cm
\caption{Left: Hydrodynamic velocity field component $u_2$ as a function of the vertical coordinate $y\in[0,1]$ for the $g=10$ and $T_0=19$ case. Orange points are the average of columns $n=1$ and $n=30$ and green points are the average of columns $n=15$ and $n=16$. Solid lines are polynomial fits to the data (see text). Right: $\theta=\arctan(u_2/u_1)\in [-\pi/2,\pi/2]$. Dotted lines are $\theta=\pi/2$ (top) and $\theta=-\pi/2$ (bottom)\label{veloprofile}}
\end{figure}
\begin{figure}
\begin{center}
\includegraphics[height=6cm,clip]{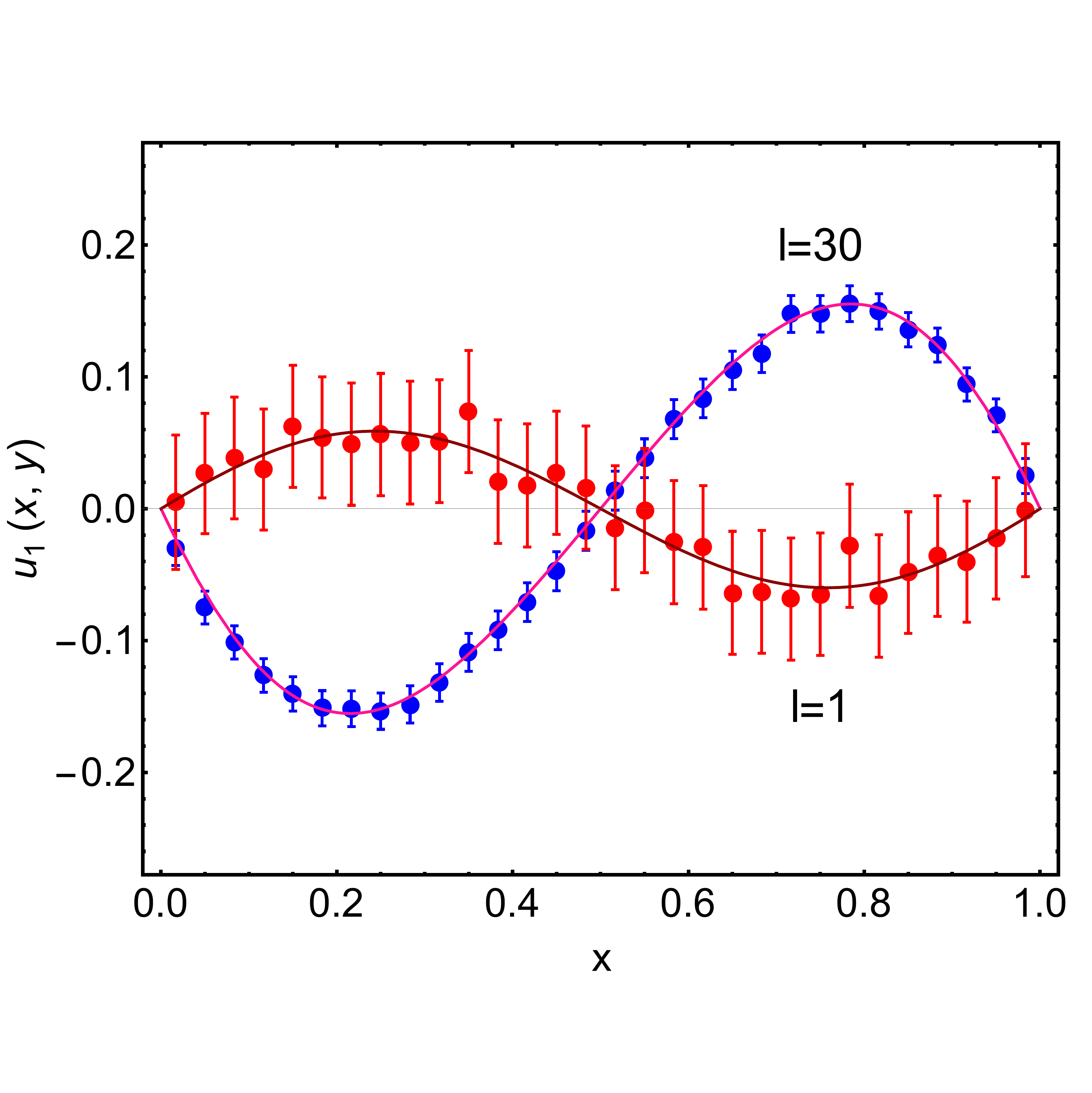}      %velo_profiles.nb
\includegraphics[height=6cm,clip]{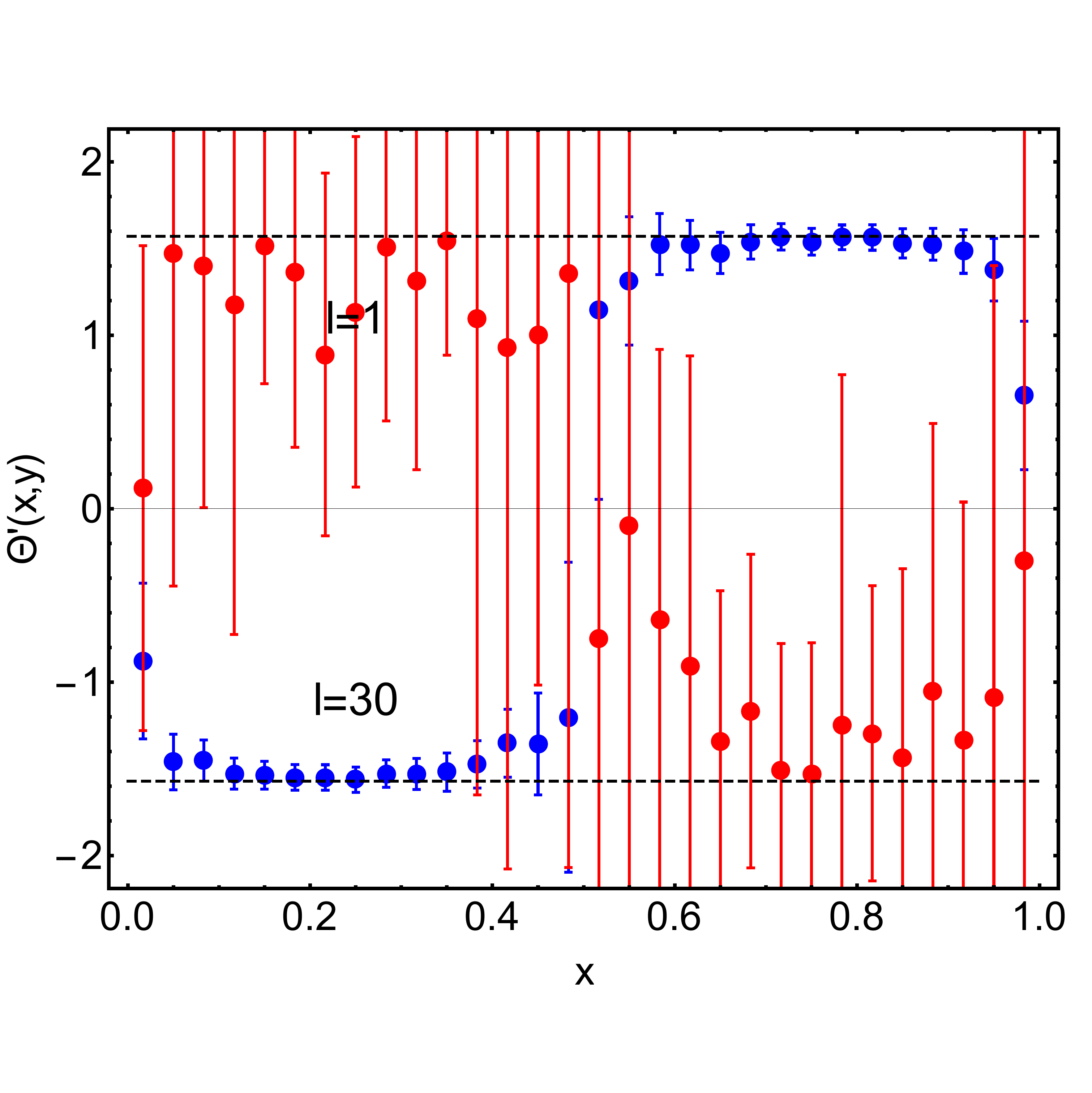}
\end{center}
\kern -1.cm
\caption{Left: Hydrodynamic velocity field component $u_1$ as a function of the horizontal coordinate $x\in[0,1]$ for the $g=10$ and $T_0=19$ case. Red  and blue points are the values of rows $l=1$ and  $l=30$ respectively.   Solid lines are polynomial fits to the data (see text). Right: $\theta'=\arctan(u_1/\vert u_2\vert)\in [-\pi/2,\pi/2]$. Dotted lines as previous figure. \label{veloprofile2}}
\end{figure}

We analyze with more detail the behavior of the components of the hydrodynamic velocity in some particular case.
Let's first study $u_2$ as a function of $l$ for the extreme columns $n=1$ and $n=30$ (averaged by symmetry)  and the central columns $n=15$ and $n=16$  (averaged by symmetry) for the case $g=10$ and $T_0=19$. In figures \ref{velofield} we see that the fluid goes down from $l=30$ to $l=1$ in the columns $n=1$ and $n=30$ and goes up from $l=1$ to $l=30$ in the columns $n=15$ and $n=16$. Let us remind that the center of the cell $(n,l)$ are located at $x=(n-1)/30+1/60$ and $y=(l-1)/30+1/60$.

First observe that for such columns the orientation of the hydrodynamic velocity vectors is practically on the $y$-direction (see right figure on \ref{velofield}). That is, these columns  maybe considered as stream lines of the fluid.
The velocity  is almost zero at $y=0$ and it starts to increase as we move towards smaller $y$-values. It reaches its maximum value and  decrease the velocity up to zero (orange data points). Just the contrary happens with the central columns: from $y=0$  it first accelerates and latter it decelerates (green data points). We can fit  to such profiles  the functions:
\begin{eqnarray} 
u_2(x,y)&=&-y (1 - y) (0.55 + 0.12 y +1.28 y^2)\quad,\quad x=1/60, 59/60 \quad (n=1,30)\nonumber\\
u_1(x,y)&=&y (1 - y) (0.22 + 1.00 y) \qquad,\quad  x=29/60, 31/60\quad (n=15,16)
\end{eqnarray}
 We can define a kind of {\it vertical effective  fluid acceleration} as 
\begin{equation}
g_{eff}=\frac{1}{2} \frac{\partial u_2^2}{\partial y}
\end{equation}
and from the fitted functions we can get an idea of the acceleration values. For instance: $g_{eff}=0.058$ for $y=0.25$ and $g_{eff}=-0.076$ for $y=0.75$ for columns $x=1$ and $30$ and  $g_{eff}=0.037$ for $y=0.25$ and $g_{eff}=-0.054$ for $y=0.75$ for columns $n=15$ and $16$. All these typical values are far from the external one: $g=10$. That is, the fluid hydrodynamic velocity field  is having a lot of  friction  from the underlying chaotic movement of particles being at local equilibrium. From Navier-Stokes equations we see that the effective acceleration on the y-direction is the competition of the external forcing, $g$, the pressure gradient and the viscous terms. We'll try later to study  the role of the last two on the value of $g_{eff}$ once clarified the behavior of the pressure profile.

We see in figure \ref{veloprofile2} the behavior of $u_1$ as a function of $x$ for the rows $l=1$ (bottom) and $l=30$ (top). The data near the hot bath ($y=1$) is very noisy compared with the data measured on the top. Again the fluid is aligned flowing along the $x$-direction. The profiles seem to follow a periodic function with modulated amplitude:
\begin{equation}
u_1(x,y)= A(x)\sin(a_1x+a_2))\quad,\quad y=1
\end{equation}
where $A(x)$ is a second order polynomial and $a$'2 are fitted parameters.

\begin{figure}
\begin{center}
\includegraphics[height=6cm,clip]{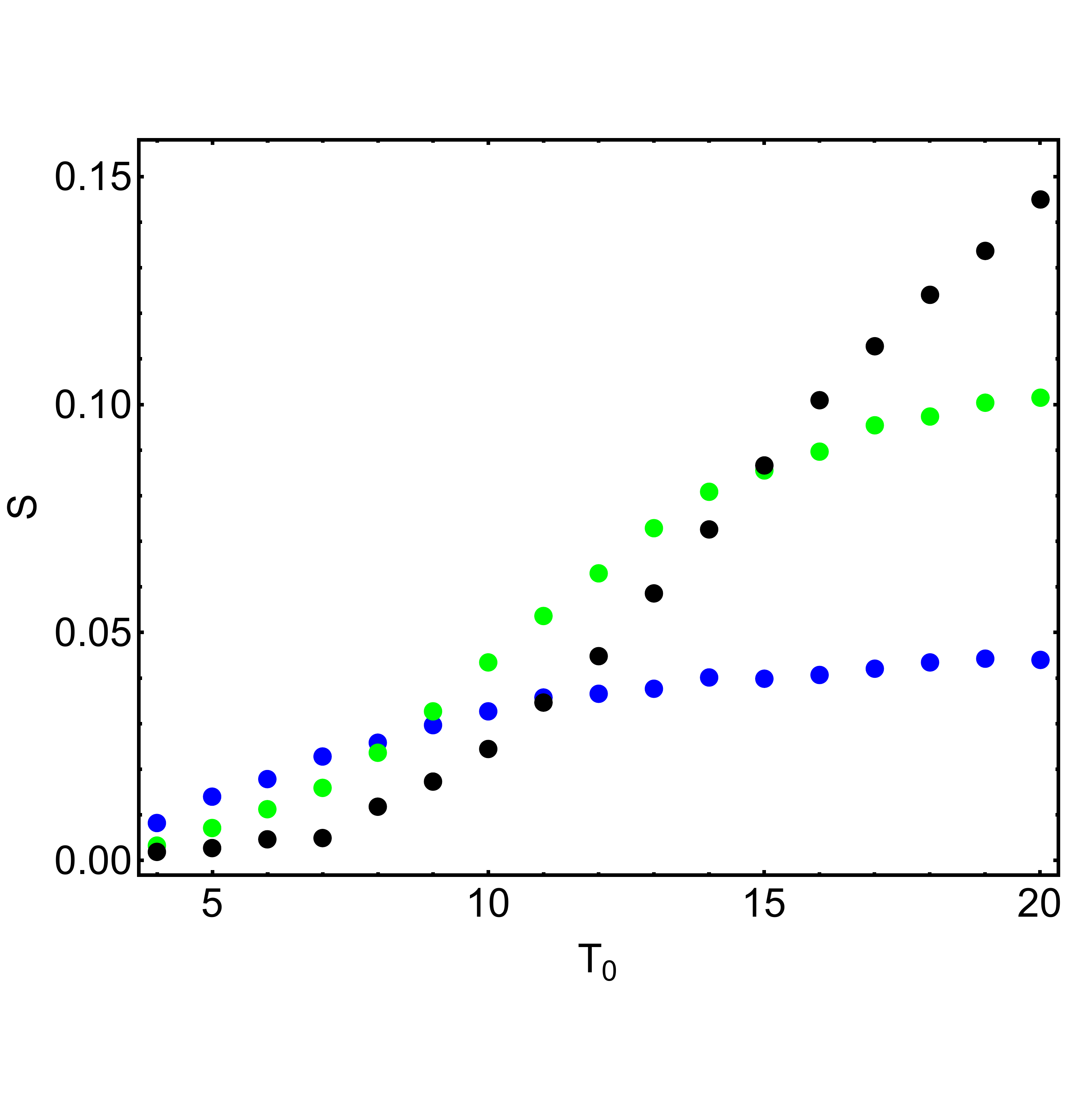} %velo_profiles3.nb
\end{center}
\kern -1.cm
\caption{Area associated to the profile of the hydrodynamic velocity field $u_2$ for row $n=30$ as a function of $T_0$ for $g=5$ (blue dots), $g=10$ (green dots) and $g=15$ (black dots). 
 \label{veloprofile3}}
\end{figure}

Qualitatively, the general form of the $u_{1,2}$ profiles is similiar independently of the values of $g$ or $T_0$.  However the amplitudes of the profiles increase with $T_0$ for a fixed $g$ and with $g$ for a fixed $T_0$ . This is shown in  figure \ref{veloprofile3} where we plot the area contained on the $u_1$ profiles for the top row $30$. We observe how the area increase as $T_0$ increase to a limit value (at least for the $g=5$ and $g=10$ cases). Moreover, the values are increased more than twice as we double $g$ which implies a nonlinear behavior as a function of $g$ that we cannot study with the actual set of computer simulations.  The existence of a limiting area implies that it should exists a limiting profile for a fixed $g$ and $T_0$ large enough. That is, the convective fluid absorbs always a maximum quantity of energy (kinetic). It also indicates that the $T_0$ values controls a kind of resistance to the macroscopic fluid movement that always exists even for large values of $T_0$ where its value is minimum. The behavior with $g$ reflects again that the power engine of the convective fluid is the external acceleration.

The idea of the existence of a limiting profile for $T_0\rightarrow\infty$ inspire us about the existence of a scaling profile at least in the asymptotic regime, that is, when $T_0$ is large enough. In fact, we are going to try to show below that, up to our computational precision level, there is a hydrodynamic universal profile for $T_0$ large enough values. We see in figure \ref{velofieldxy} that the spatial structure of the velocity field changes with $T_0$: they have peaks and valleys more and more pronounced when we increase it. The most naive form to see if there is an scaled common field  is to look for a natural measure of the peak-valley differences, rescale each configuration by it and compare them. As scaling parameter we have chosen the configuration standard deviation:
\begin{equation}
\sigma(u_{1,2})=\left[\frac{1}{N_C}\sum_{(x,y)} (u_{1,2}(x,y)-\bar u_{1,2})^2\right]^{1/2}\quad,\quad \bar u_{1,2}=\frac{1}{N_C}\sum_{(x,y)}u_{1,2}(x,y)
\end{equation}
where $N_C$ is the number of cell points and the sum runs over all $(x,y)$-cells. We define a {\it scaled configuration} as:
\begin{equation}
u_{1,2}^{(s)}(x,y)=\frac{u_{1,2}(x,y)-\bar u_{1,2}}{\sigma(u_{1,2})}
\end{equation}
\begin{figure}
\begin{center}
\includegraphics[height=5cm,clip]{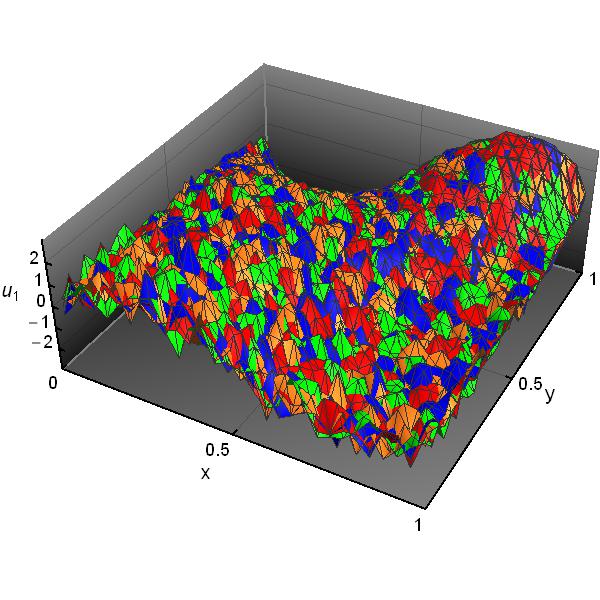}   %velo_field_show_vx2.nb
\includegraphics[height=5cm,clip]{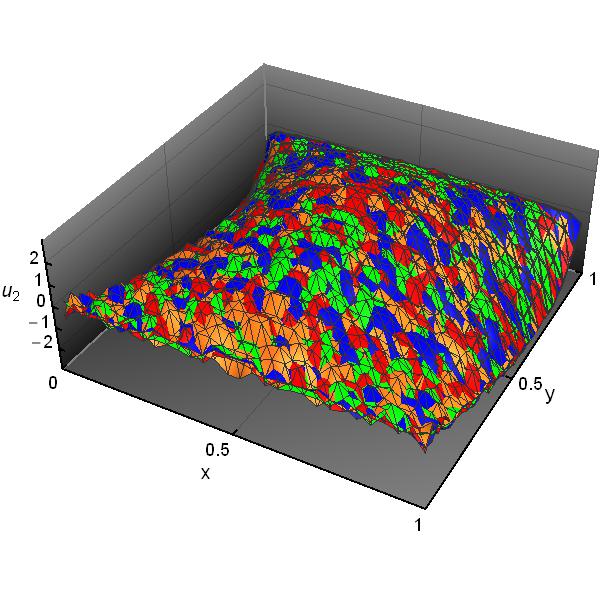}   %velo_field_show_vy2.nb
\newline
\vglue -1cm
\includegraphics[height=5cm,clip]{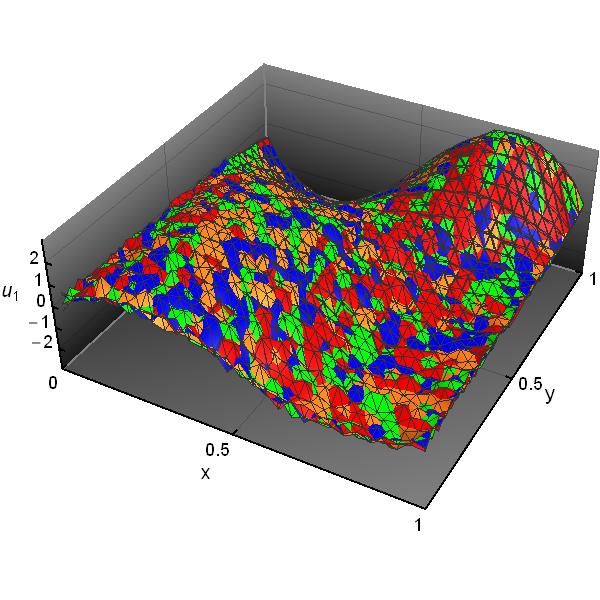}   %velo_field_show_vx2.nb
\includegraphics[height=5cm,clip]{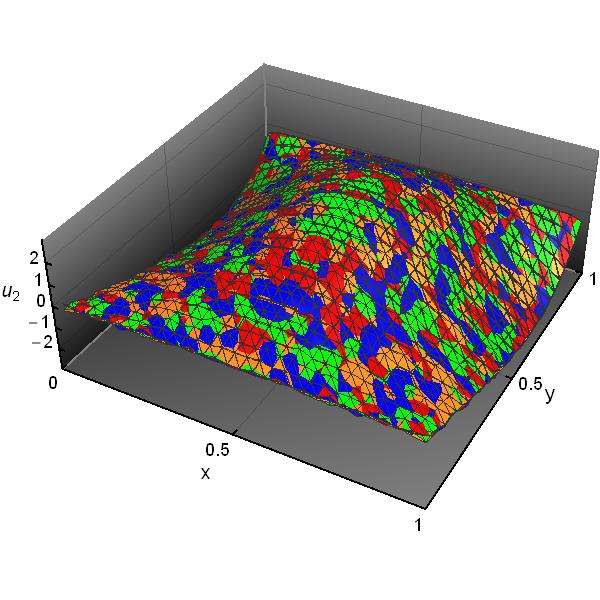}   %velo_field_show_vy2.nb
\newline
\vglue -1cm
\includegraphics[height=5cm,clip]{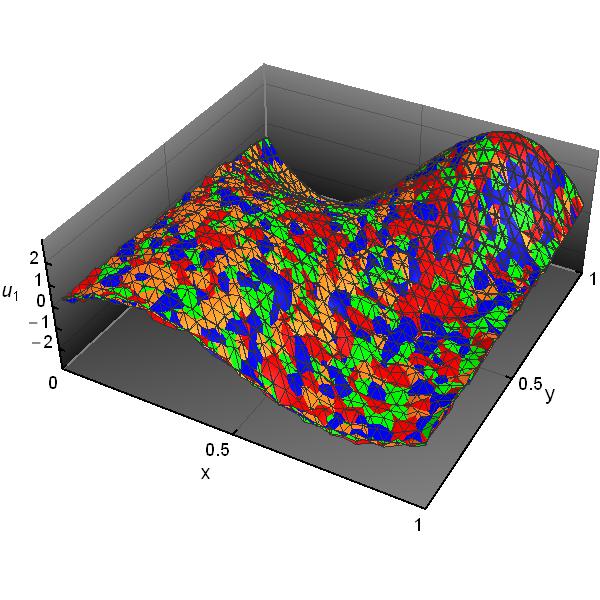}   %velo_field_show_vx2.nb
\includegraphics[height=5cm,clip]{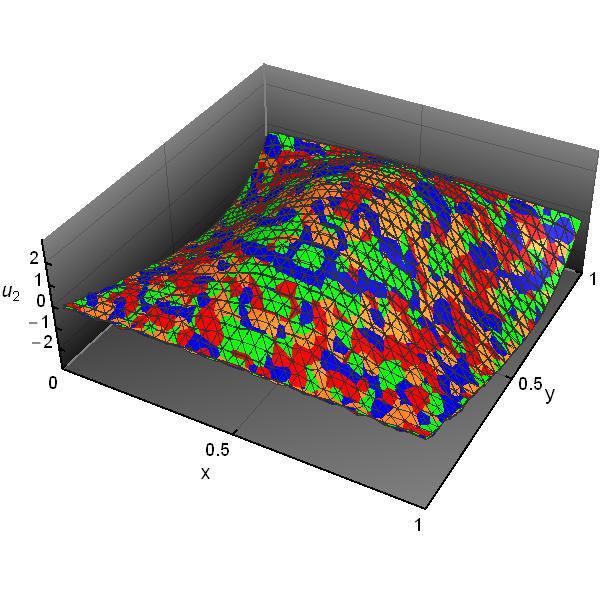}   %velo_field_show_vy2.nb
\newline
\end{center}
\caption{Superposition of scaled hydrodynamic velocity field components $u_1$ and $u_2$ (left and right columns respectively) and, from top to bottom: $g=5$, $10$ and $15$. Colors are for $T_0=17$, $18$, $19$ and $20$.  \label{velofieldscaled}}
\end{figure}

In figure \ref{velofieldscaled} we show the superposition of four different $T_0$ scaled configurations for $g=5$, $10$ and $15$. We see, at a glance, that except for noise effects the configurations seems to be the same. Obviously, these pictures  by themselves don't prove anything but they are a good starting point. The hard work now is to attempt to show that there is not a systematic $T_0$ dependence on such scaled configurations. From the numerical analysis point of view, we should be very careful. We have already seen that just a variable fluctuation introduces systematic deviations on any function of it. Then we should painfully handle all these effects. The scheme we have followed for this analysis is:
\begin{itemize}
\item {\it Preparation of the scaled fields:} We'll take into account the effect of the configuration error bars in the values of $\sigma(u_{1,2})$. From it we'll find the corrections on the scaled fields coming from the fluctuations. 
\item {\it Analysis of the scaled configuration spatial structure and mutual comparison:} For each configuration we'll measure the eigenvalues of the inertial moments and the fourth momenta. Finally we also study the mutual distance between configurations. All of this just for looking any systematic $T_0$ dependence.
\item {\it Obtaining the universal field:} We'll average the set of configurations that we assume numerically indistinguishable, Fourier transforms it, we'll discard the noisy modes and we'll do the inverse Fourier transform to get a smooth scaled configuration.
 Finally, we'll study the differences between the universal configuration and each of the original scaled ones. 
\end{itemize}

\begin{enumerate}
\item  {\it Preparation of the scaled fields}

Let $v(x,y;\xi)$ and $\sigma(x,y)$ the averaged measured field and its standard deviation respectively. We assume that the $v$-field depends on an random noise field that prevents it to have a smooth value. 
Then, by assuming that the central limit theorem applies, we can write
\begin{equation}
v(x,y;\xi)=v(x,y)+\sigma(x,y)\xi(x,y)
\end{equation}
where $v$ is the field we would like to know and each $\xi(x,y)$ is a Gaussian random variable with zero average and unit variance. Let us mention that one should expect a spatially correlated noise because fluctuations are based on several dynamically conserved quantities like density, momenta and energy, for instance. However, the assumption of independence maybe enough to catch the key ingredients of the errors and the corrections to the measured averaged magnitudes. 

We know that the fields $v(x,y)$ and $\sigma(x,y)$ have a natural symmetry: $v(x,y)=\lambda v(1-x,y)$ and $\sigma(x,y)=\sigma(1-x,y)$ with $\lambda=\pm 1$. That's because the physics we observe in the system: the magnitudes behave in a mirror like way with respect the axis $x=1/2$ even in the convective state when two symmetric rolls appear. Therefore, the observables depending on  the $u_1$ component of the hydrodynamic velocity field should have $\lambda=-1$ and any other observable $\lambda=1$. Obviously our measured configurations do not strictly respect this symmetry due to the fluctuations. We symmetrize our configurations and their errors to impose such important property: $v(x,y;\xi)\rightarrow (v(x,y;\xi)+\lambda v(1-x,y;\xi))/2$ and $\sigma(x,y)\rightarrow (\sigma(x,y;\xi)^2+\sigma(1-x,y;\xi)^2)^{1/2}/2$. With this averaging we also  improve the statistics and therefore the observable error bars.

We can get the averaged field value and its second momenta:
\begin{eqnarray}
\bar v(\xi)&=&\frac{1}{N_C}\sum_{(x,y)}v(x,y;\xi)=\bar v+\frac{1}{N_C}\sum_{(x,y)}\sigma(x,y)\xi(x,y)\nonumber\\
m_2(v;\xi)&=&\frac{1}{N_C}\sum_{(x,y)}(v(x,y;\xi)-\bar v(\xi))^2=m_2(v)+\frac{2}{N_C}\sum_{(x,y)}(v(x,y)-\bar v)\sigma(x,y)\xi(x,y)+\frac{1}{N_C}\sum_{(x,y)}\sigma(x,y)^2\xi(x,y)^2\nonumber\\
&+&\frac{1}{N_C^2}\sum_{(x,y)}\sum_{(x',y')}\sigma(x,y)\sigma(x',y')\xi(x,y)\xi(x',y')\label{sca}
\end{eqnarray}
where 
\begin{equation}
\bar v=\frac{1}{N_C}\sum_{(x,y)}v(x,y)\quad,\quad m_2(v)=\frac{1}{N_C}\sum_{(x,y)}(v(x,y)-\bar v)^2
\end{equation}

we follow now the same strategy we explained in the introduction about the error analysis. We define the random variable
\begin{equation}
B=m_2(v;\xi)-(1-\frac{1}{N_C})\frac{1}{N_C}\sum_{(x,y)}\sigma(x,y)^2
\end{equation}
we see that
\begin{eqnarray}
\langle B\rangle&=&m_2(v)\nonumber\\
\langle m_2(B)\rangle&=&\frac{4}{N_C^2}\sum_{(x,y)}(v(x,y)-\bar v)^2\sigma(x,y)^2
\end{eqnarray}
where $\langle\cdot\rangle$ is the average over the random fields values $\xi$. Then we can made the strong asumption that the central limit theorem applies to $B$ in the sense that the square root of its second moment give us  the error on the computation of its average. Then
\begin{equation}
B=m_2(v)\pm 3\langle m_2(B)\rangle^{1/2}\Rightarrow m_2(v)=m_2(v;\xi)-(1-\frac{1}{N_C})\frac{1}{N_C}\sum_{(x,y)}\sigma(x,y)^2 \pm 3\langle m_2(B)\rangle^{1/2}
\end{equation}
In the computation we have assumed that $\sigma(x,y)$ are small and we have disregard orders $\sigma^4$ in $m_2(B)$ because it is used only on the error bar definition. There, the higher order corrections are negligible because our $3\sigma$ criteria is generous enough and it gives us already an idea of the precision level in our computations. 

We are ready to derive the corrections to the scaled configuration that we define as
\begin{equation}
v^{(s)}(x,y;\xi)=\frac{v(x,y;\xi)-\bar v(\xi)}{\sigma(v;\xi)}
\end{equation}
where $\sigma(v;\xi)^2=m_2(v;\xi)$. Then
\begin{equation}
v^{(s)}(x,y)=\frac{v(x,y)-\bar v}{\sigma(v)}=v^{(s)}(x,y;\xi)\left[ 1+\frac{1}{2}(1-\frac{1}{N_C})\beta(0)-\frac{3}{2N_C}\beta(2)+\frac{\sigma(x,y)^2}{N_C\sigma(v)^2}\right]-\frac{1}{N_C}\beta(1)\pm 3\frac{\sigma(x,y)}{\sigma(v)}
\end{equation}
with
\begin{equation}
\beta(n)=\frac{1}{N_C}\sum_{(x,y)}v^{(s)}(x,y;\xi)^{n}\frac{\sigma(x,y)^2}{\sigma(v)^2}
\end{equation}

\begin{figure}[h!]
\begin{center}
\includegraphics[height=5cm,clip]{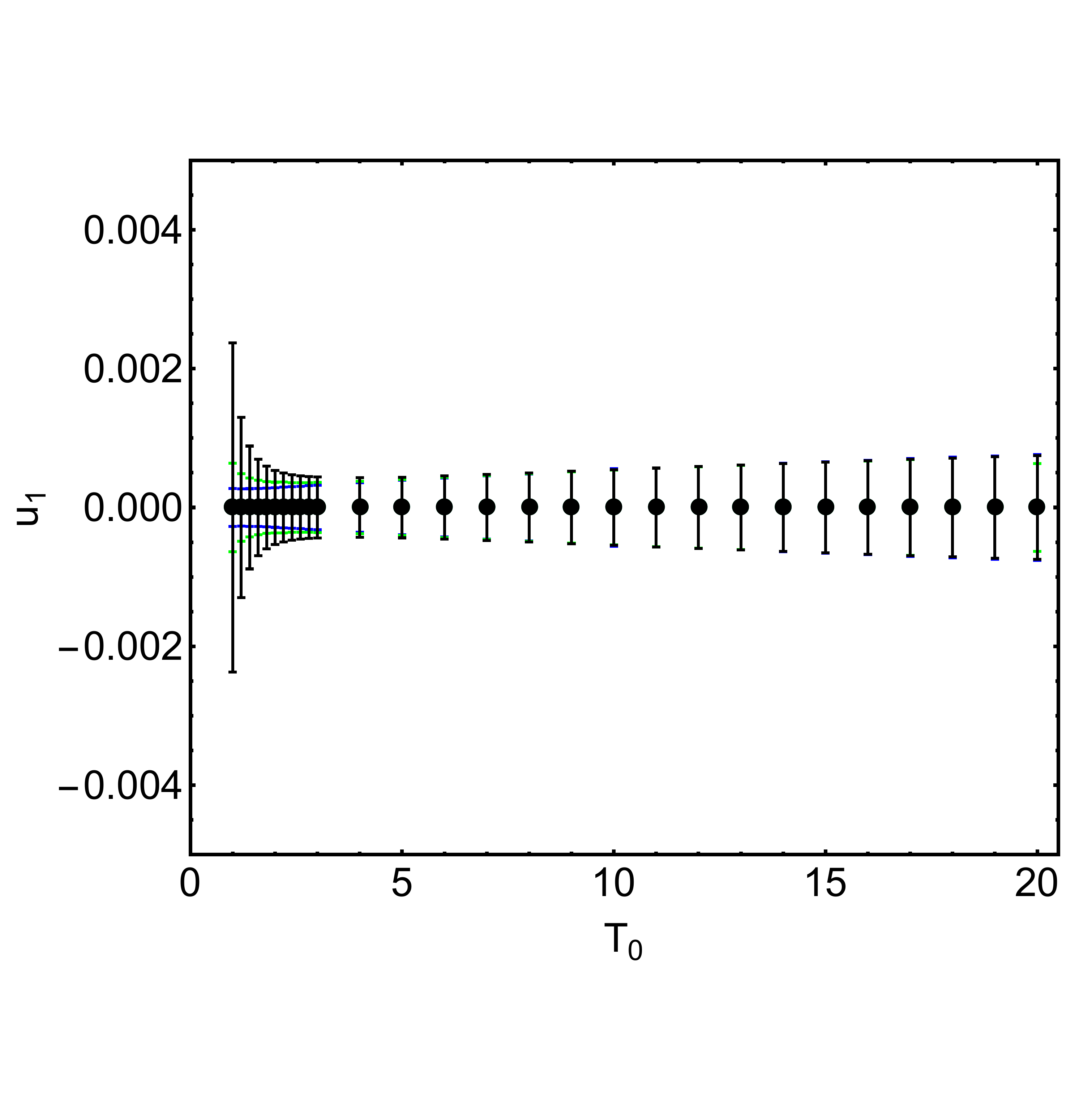}  %velo_moments_3_2_vx.nb
\includegraphics[height=5cm,clip]{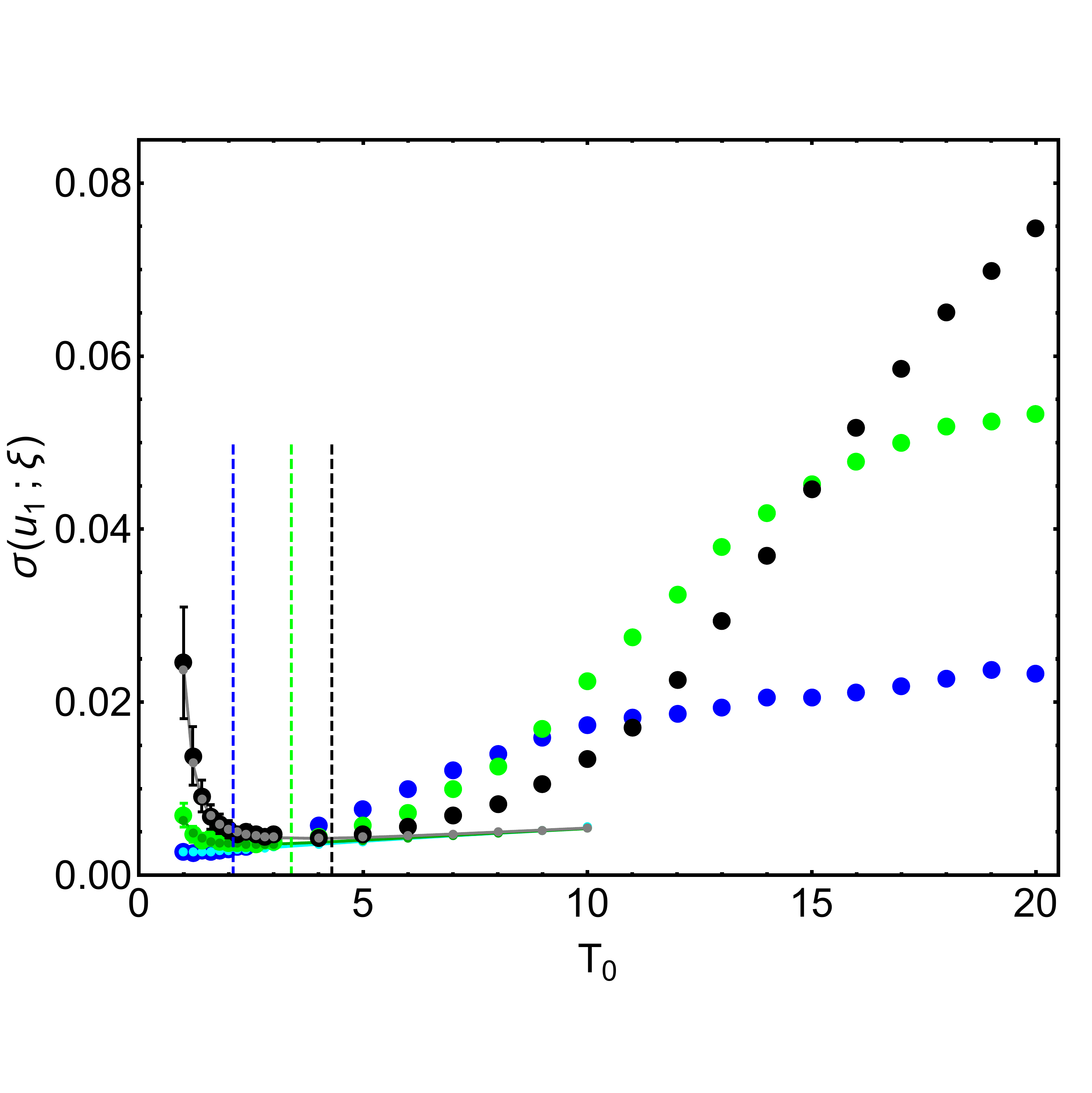}  
\includegraphics[height=5cm,clip]{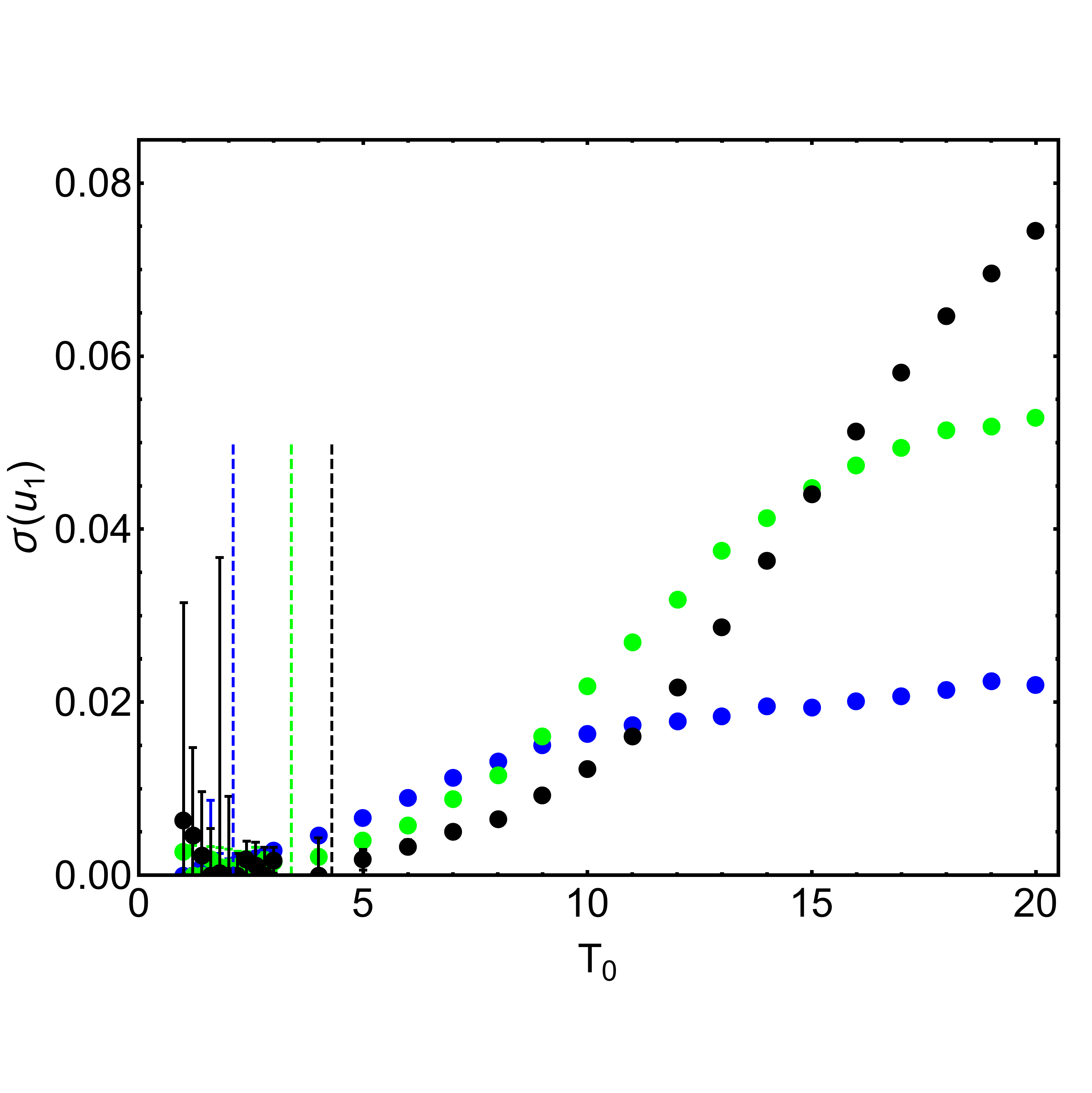}  
\end{center}
\kern -1.cm
\caption{Averaged hydrodynamic velocity field x-component, $u_1$ (left figure), its raw standard deviation $\sigma(u_1;\xi)$  (center figure) defined by eq. \ref{sca}, and the standard deviation after corrections due to fluctuations $\sigma(u_1)$ (left figure)  as a functions of $T_0$.  Blue, green and black points are for $g=5$, $g=10$ and $g=15$ respectively. Dotted vertical lines show the $T_c$ values corresponding to each $g$. Continuous lines (gray,dark green and cyan) are the computed $\sigma(u_1)$ assuming that it comes from only from the local equilibrium distribution (see text). The small dots are the corrections to $\sigma(u_1;\xi)$ assuming a white noise fluctuation. \label{vx_sigma}}
\end{figure}
\begin{figure}[h!]
\begin{center}
\includegraphics[height=5cm,clip]{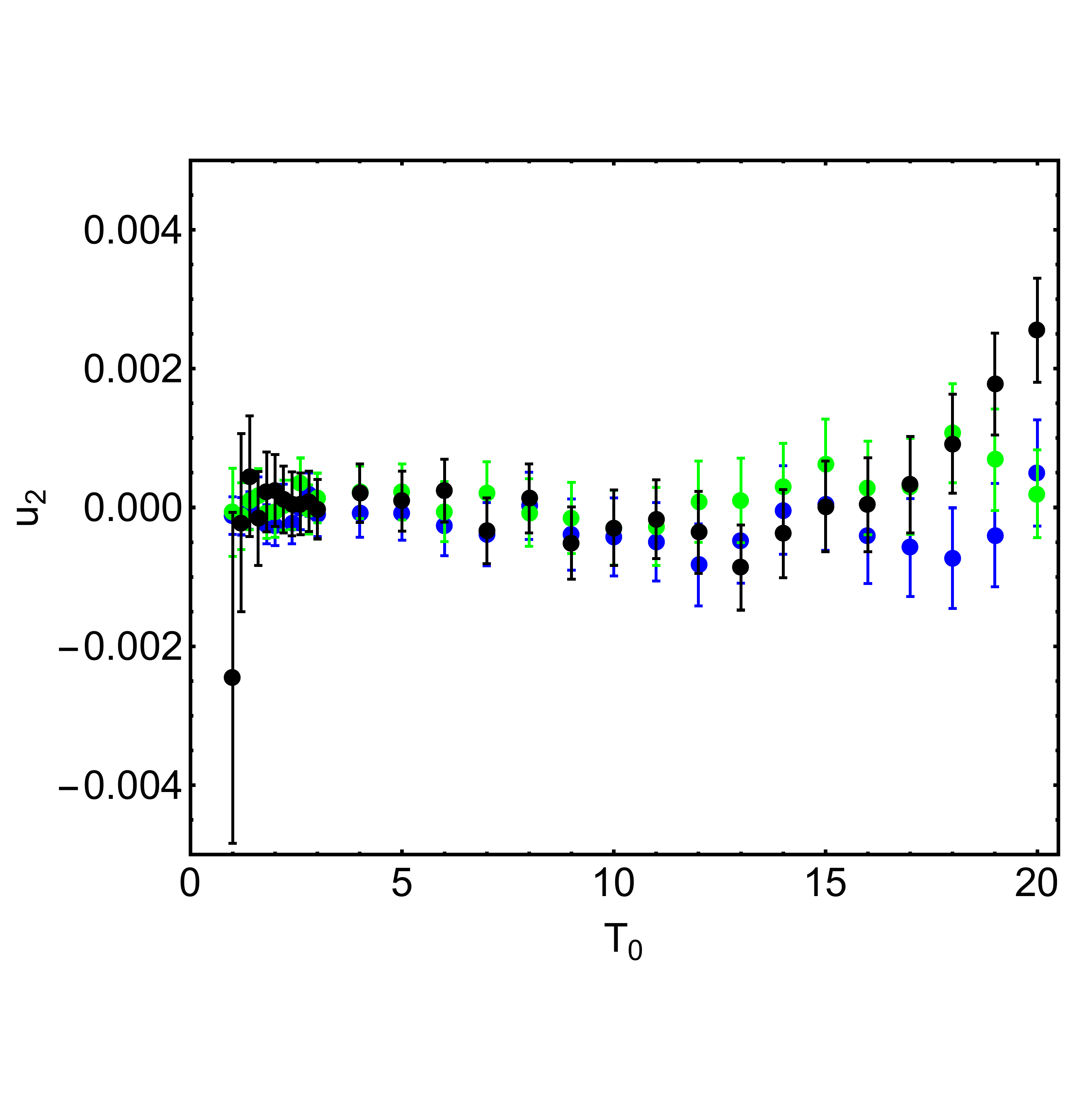}  %velo_moments_3_2_vy.nb
\includegraphics[height=5cm,clip]{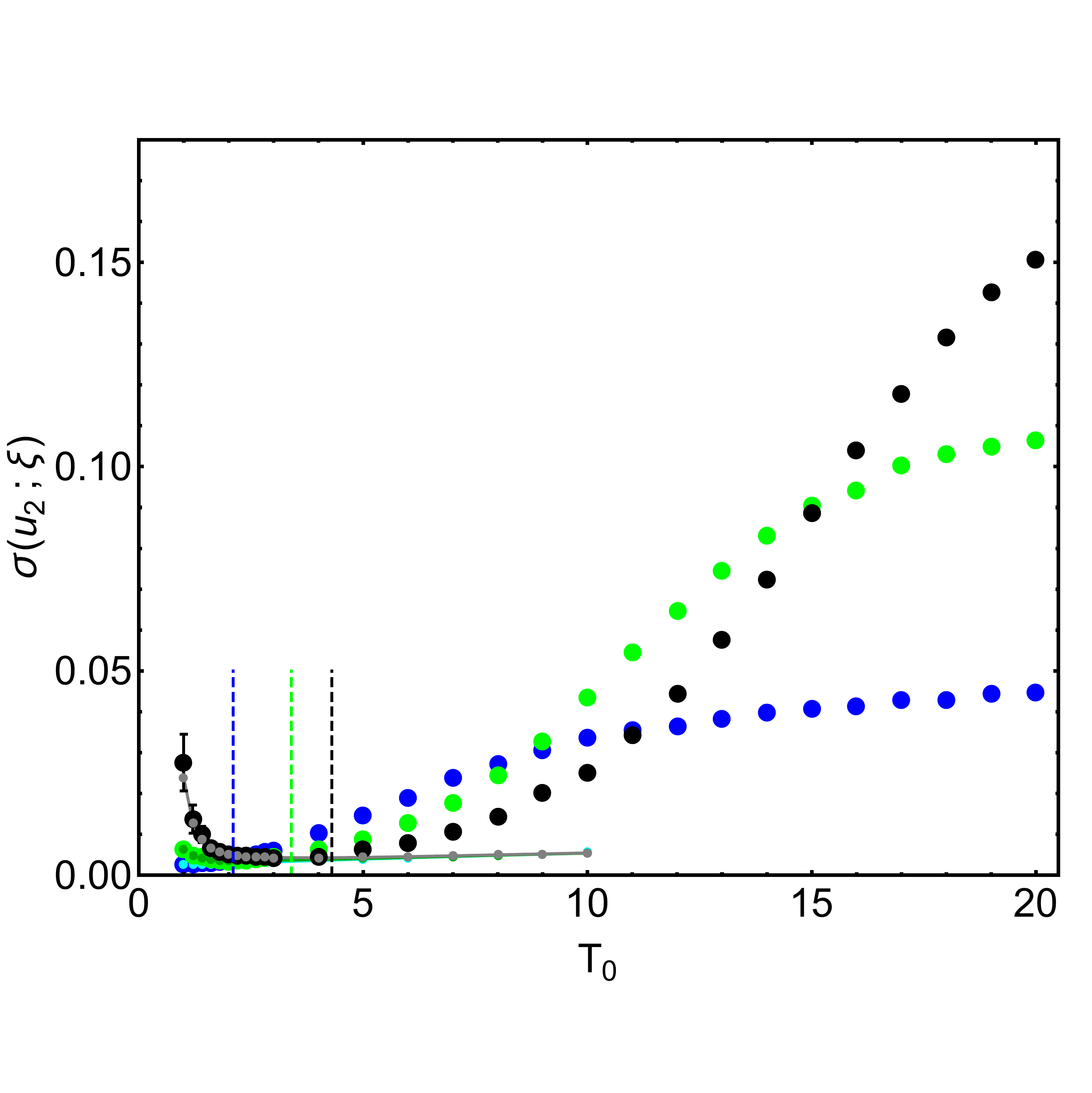}  
\includegraphics[height=5cm,clip]{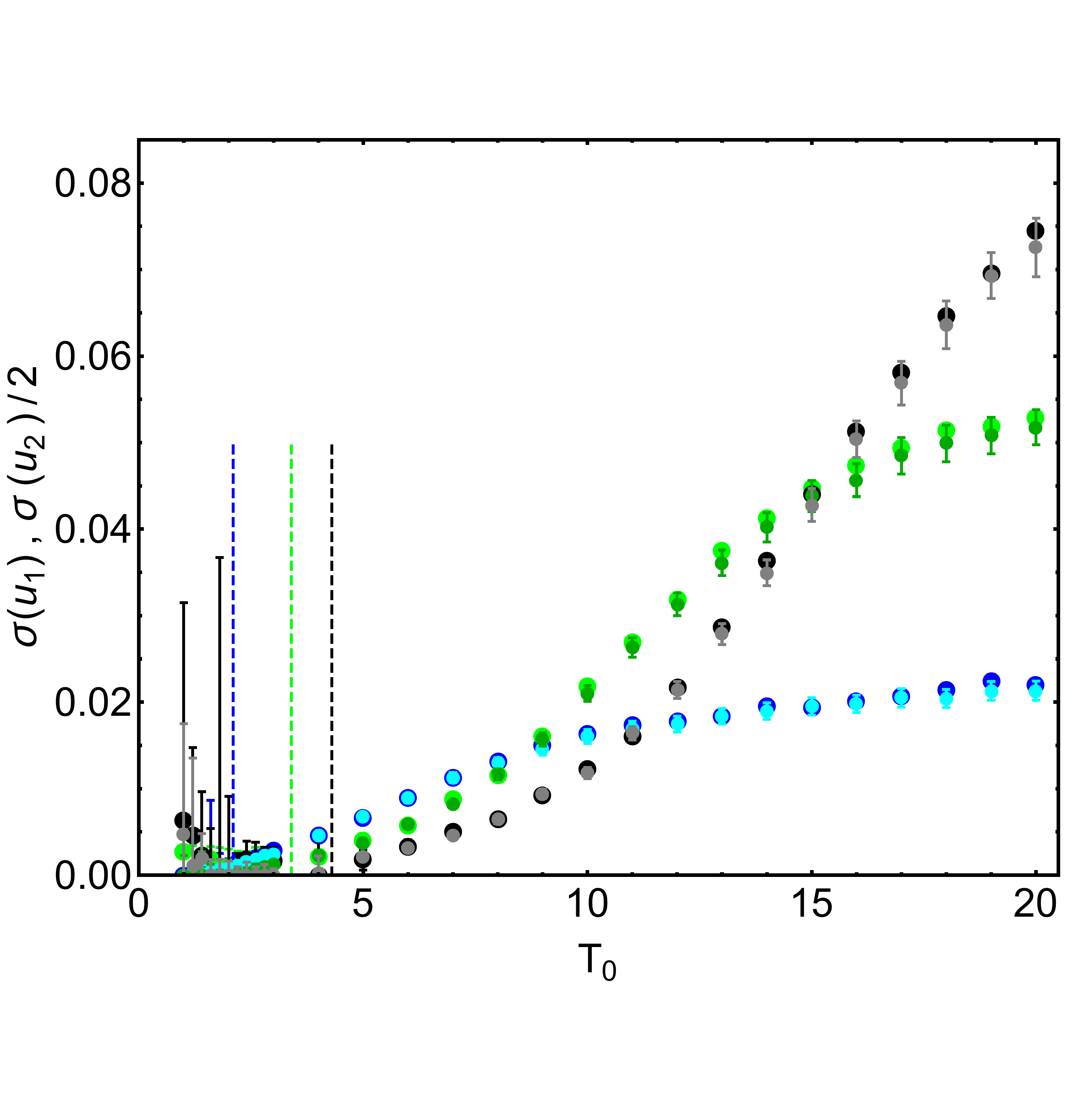} %velo_moments_3_2_vx.nb  
\end{center}
\kern -1.cm
\caption{Averaged hydrodynamic velocity field y-component, $u_2$ (left figure), its raw standard deviation $\sigma(u_2;\xi)$  (center figure) defined by eq. \ref{sca}  as a functions of $T_0$.  Blue, green and black points are for $g=5$, $g=10$ and $g=15$ respectively. Dotted vertical lines show the $T_c$ values corresponding to each $g$. Continuous lines (gray,dark green and cyan) are the computed $\sigma(u_1)$ assuming that it comes from only from the local equilibrium distribution (see text). The small dots are the corrections to $\sigma(u_1;\xi)$ assuming a white noise fluctuation. Right figure: is a comparison between the standard deviation after corrections due to fluctuations $\sigma(u_1)$ (blue, green and black for $g=5$, $10$ and $15$ respectively) and $\sigma(u_2)/2$ (cyan, dark green and gray for $g=5$, $10$ and $15$ respectively) as a functions of $T_0$.   \label{vy_sigma}}
\end{figure}

We apply the above analysis to the components of the hydrodynamic velocity field. We show in figures \ref{vx_sigma} and \ref{vy_sigma} how they  average value, $\bar u_{1,2}(\xi)$, and standard deviation, $\sigma(u_{1,2};\xi)$, behave. $\bar u_1(\xi)$ is strictly zero because of the symmetrization procedure.  It is interesting to see the systematic queue that appear on $\sigma(u_{1,2};\xi)$ for small $T_0$ values. It is only due to the local equilibrium fluctuations of the hydrodynamic velocity. If we assume that below the transition the average hydrodynamic velocity at any cell is zero we only see the fluctuations due to the law of large numbers. That is, the modulus of each component is given by $u_{1,2}(x,y)\simeq \sqrt{T(x,y)}/\sqrt{N_D(x,y)}$ being $T(x,y)$ the local temperature and $N_D(x,y)$ the total number of data used in the time averaging, both at the cell $(x,y)$. In our case $N_D(x,y)=N t \rho(x,y)\Delta^2/\rho$ where $N=957$ is the number of disks in the simulation, $t=100000$ is the total amount of measurements, $\rho(x,y)$ is the averaged areal density at the cell, $\Delta=1/30$ is the cell side and $\rho$ is the system areal density. Then
\begin{equation}
\sigma(u_{1,2})_{le}^2=\frac{1}{N_C}\sum_{(x,y)}u_{1,2}^2(x,y)=\frac{\rho}{\Delta^2NtN_C}\sum_{(x,y)}\frac{T(x,y)}{\rho(x,y)}
\end{equation}
We plot this result as continuum lines in the center of figures \ref{vx_sigma} and \ref{vy_sigma} after symmetrizing the data. Observe how the line follows almost perfectly the measured averages. Moreover, we also plot the corrections we have computed above due to the fluctuating character of the data (small dots) and they follow the line and the big dots. That is
\begin{equation}
\sigma(u_{1,2})_{le}^2=(1-\frac{1}{N_C})\frac{1}{N_C}\sum_{(x,y)}\sigma(x,y)^2 
\end{equation}
in the non-convecting region.
We see again in this example the importance to subtract such noise contribution  (even though it is well understood and it  should disappear in the limit for infinite larger averaging or number of disks) because it introduces a systematic deviation in our analysis obscuring, sometimes, the hydrodynamic bulk behavior of the system. 
Finally, the figures in \ref{vx_sigma} and \ref{vy_sigma}  right show the value of $\sigma(u_{1,2})$ after subtracting the noise term contribution. We couldn't find a simple scaling behavior with $g$. Nevertheless, we found a striking simple relation:
\begin{equation}
\sigma(u_2)=2\sigma(u_1) 
\end{equation}
for every $T_0$ value of any given $g$. See for instance in figure \ref{vy_sigma} how the points overlap all over the $T_0$ range for the three $g$ values.
At this moment, we do not know why a parameter that measures the average variation of the field should follow such relation. 

Once we have prepared with care the scaled fields we may start its analysis to see it they are similar or there is a small but systematic dependence on $T_0$.

\item {\it Analysis of the scaled configuration spatial structure and mutual comparison:}

We have measured some structural field parameters. First we have computed the Inertial Tensor with respect the field center of mass. That is,
\begin{equation}
I_{\alpha,\beta}(u_{1,2}^{(s)})=\frac{1}{N_C}\sum_{(x,y)}\left[r'(x,y;u_{1,2}^{(s)})^2\delta_{\alpha,\beta}-r'_\alpha(x,y;u_{1,2}^{(s)})r'_\beta(x,y;u_{1,2}^{(s)})\right]
\end{equation}
where
\begin{eqnarray}
r'(x,y;u_{1,2}^{(s)})&=&r(x,y;u_{1,2}^{(s)})-R(u_{1,2}^{(s)})\nonumber\\
r(x,y;u_{1,2}^{(s)})&=&(x,y,u_{1,2}^{(s)}(x,y))\nonumber\\
R(u_{1,2}^{(s)})&=&\frac{1}{N_C}\sum_{(x,y)}r(x,y;u_{1,2}^{(s)})
\end{eqnarray}
The Inertial Tensor has the following form:
\begin{equation}
I(u_{1,2}^{(s)})=
\begin{pmatrix}
13/12&0&I_1\\
0&13/12&I_2\\
I_1&I_2&1/6
\end{pmatrix}
\end{equation}
where
\begin{equation}
I_1=-\frac{1}{N_C}\sum_{(x,y)}(x-\frac{1}{2})u_{1,2}^{(s)}(x,y)\quad,\quad I_2=-\frac{1}{N_C}\sum_{(x,y)}(y-\frac{1}{2})u_{1,2}^{(s)}(x,y)
\end{equation}
Observe that $I_2=0$ for $u_1^{(s)}$ and $I_1=0$ for $u_2^{(s)}$. From it, we can compute its principal axis and the semi principal diameters that are just the eigenvectors and one over the square root of the eigenvalues. We are mainly interested on the set eigenvalues because they contain an averaged information of the topological form of the field. In fact they inverse square root are the main axis of the Poinsot's ellipsoid. The eigenvalues can be analytically computed and are:
\begin{equation}
\lambda_1=\frac{13}{12}\quad,\quad \lambda_\pm = \frac{1}{2}\left[\frac{5}{4}\pm\sqrt{\frac{121}{144}+4(I_1^2+I_2^2)}\right]
\end{equation}
We compute $I_1$ and $I_2$ using $u_{1,2}(x,y;\xi)$ and therefore the eigenvalues are noise-dependent: $\lambda(\xi)$. Following the same technique above explained we can get the noise-corrected eigenvalues that they are given by
\begin{equation}
\lambda_\pm=\lambda_\pm (\xi)\mp \frac{4\beta_2 (I_1^2+I_2^2)}{\sqrt{\frac{121}{144}+4(I_1^2+I_2^2)}}
\end{equation}
where we show here only the $N_C\rightarrow\infty$ correction. Nevertheless we use the full corrected form in the results we show in the figures. We observe in the figure \ref{eigen_v} (left) how $\lambda$'s have a nontrivial dependence on $T_0$ and the noise term correction doesn't change too much $\lambda(\xi)$. By other hand for the field $u_2^{(2)}$ the values of $I_2$ are so small that the eigenvalues follow the value corresponding to $I_2=0$, that is, $\lambda_1=13/12=1.08333...$. Finally we observe that the scale of variation is of $10^{-3}$ which is of the order of the hydrodynamic velocity resolution. That is, the Inertial Tensor is not giving us a clear answer to our question about the existence of a universal scaled field over a region or, at least, an asymptotic universal one.  

\begin{figure}[h!]
\begin{center}
\includegraphics[height=5cm,clip]{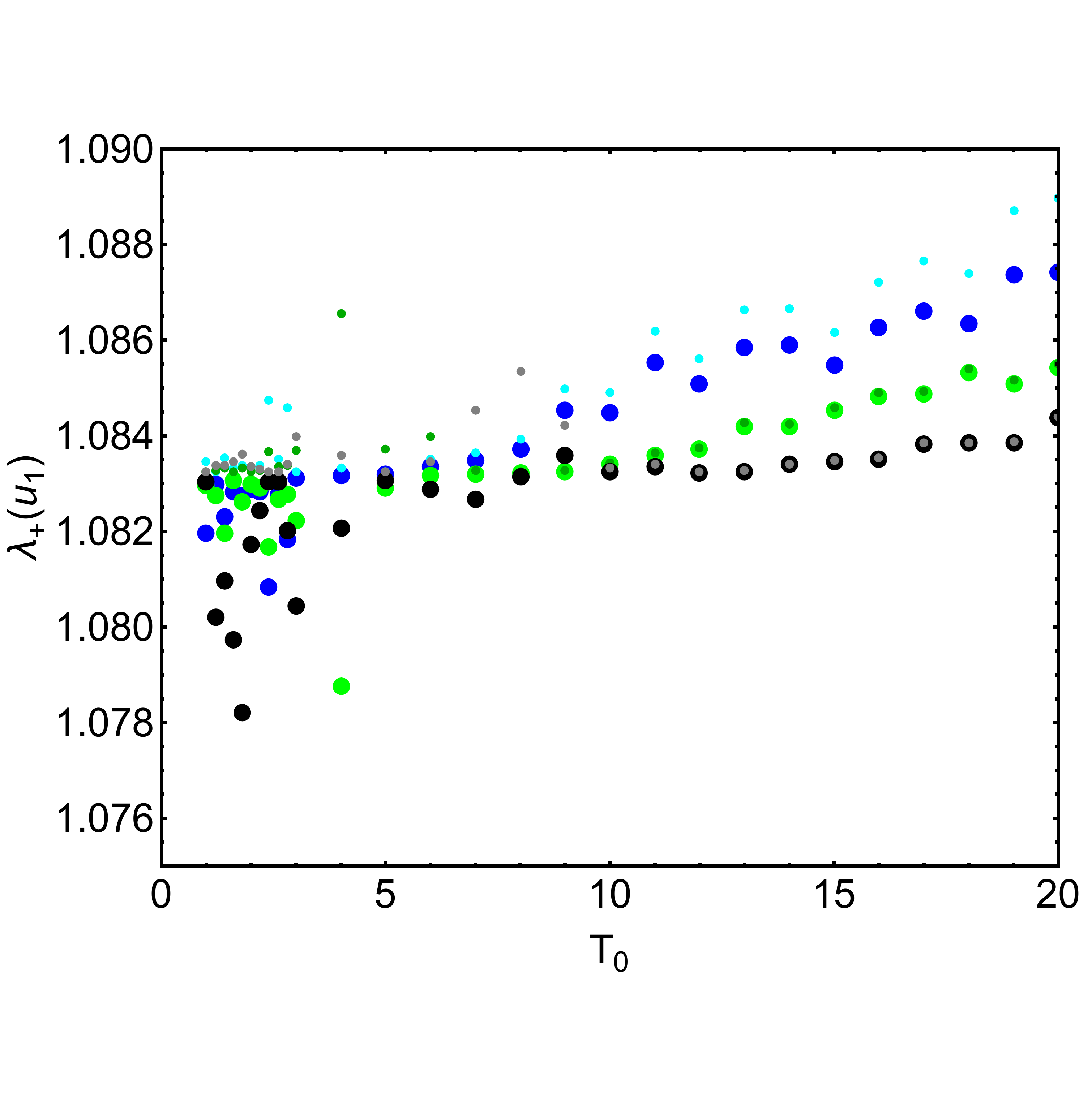}  %velo_moments_3_2_vx.nb
\includegraphics[height=5cm,clip]{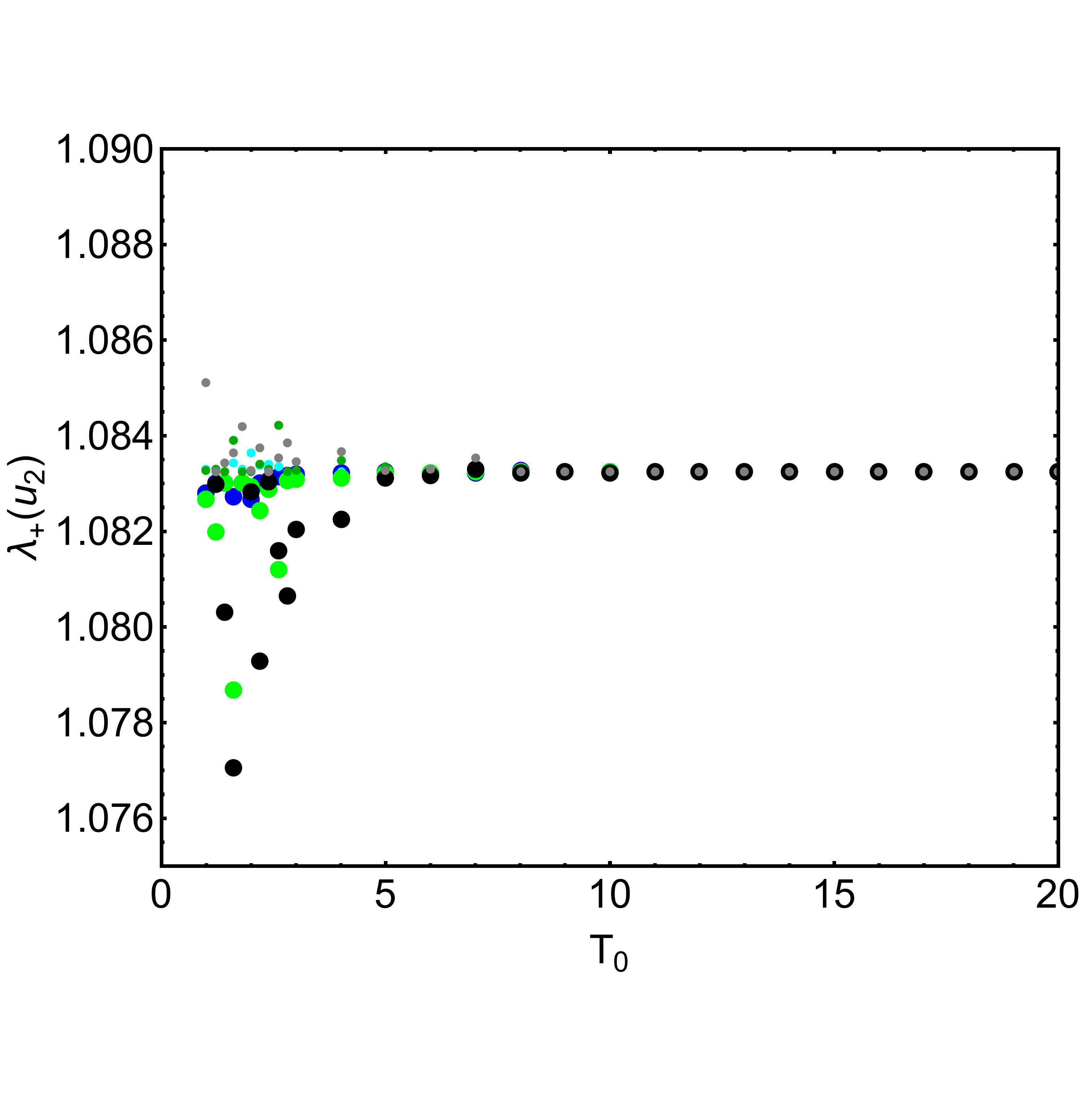}  %velo_moments_3_2_vy.nb 
\end{center}
\kern -1.cm
\caption{Inertial Tensor eigenvalue $\lambda_+$ for $u_1^{(s)}$-field (left figure) and  for $u_2^{(s)}$-field (right figure) as a function of $T_0$.  Blue, green and black points are for $g=5$, $g=10$ and $g=15$ respectively. 
The small dots are the eigenvalue before the noise-correction applied: $\lambda_+(\xi)$ assuming a white noise fluctuation (cyan, dark green and gray for $g=5$, $10$ and $15$ respectively) as a functions of $T_0$.   \label{eigen_v}}
\end{figure}
\begin{figure}[h!]
\begin{center}
\includegraphics[height=5cm,clip]{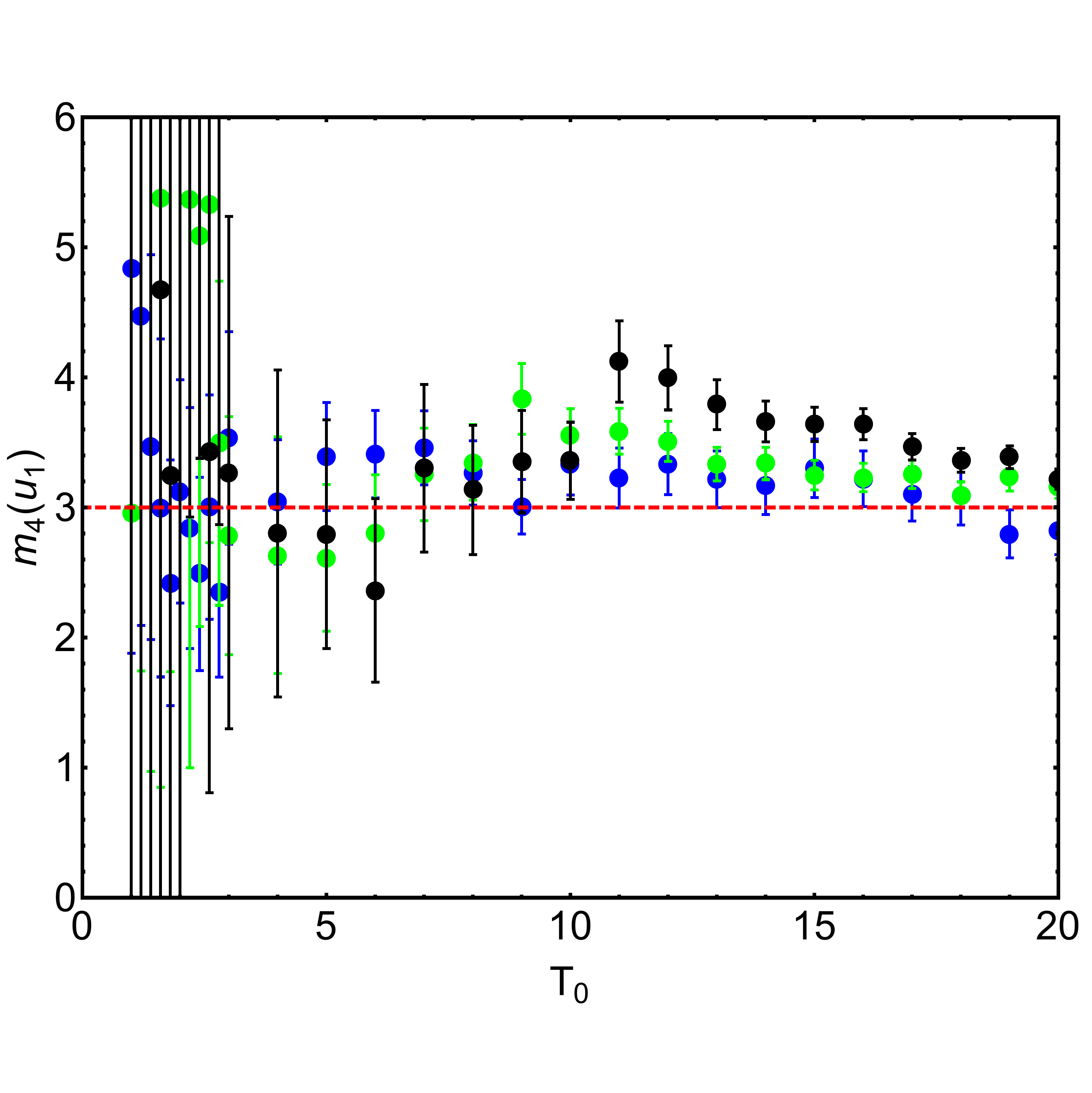}  %velo_moments_3_2_vx.nb
\includegraphics[height=5cm,clip]{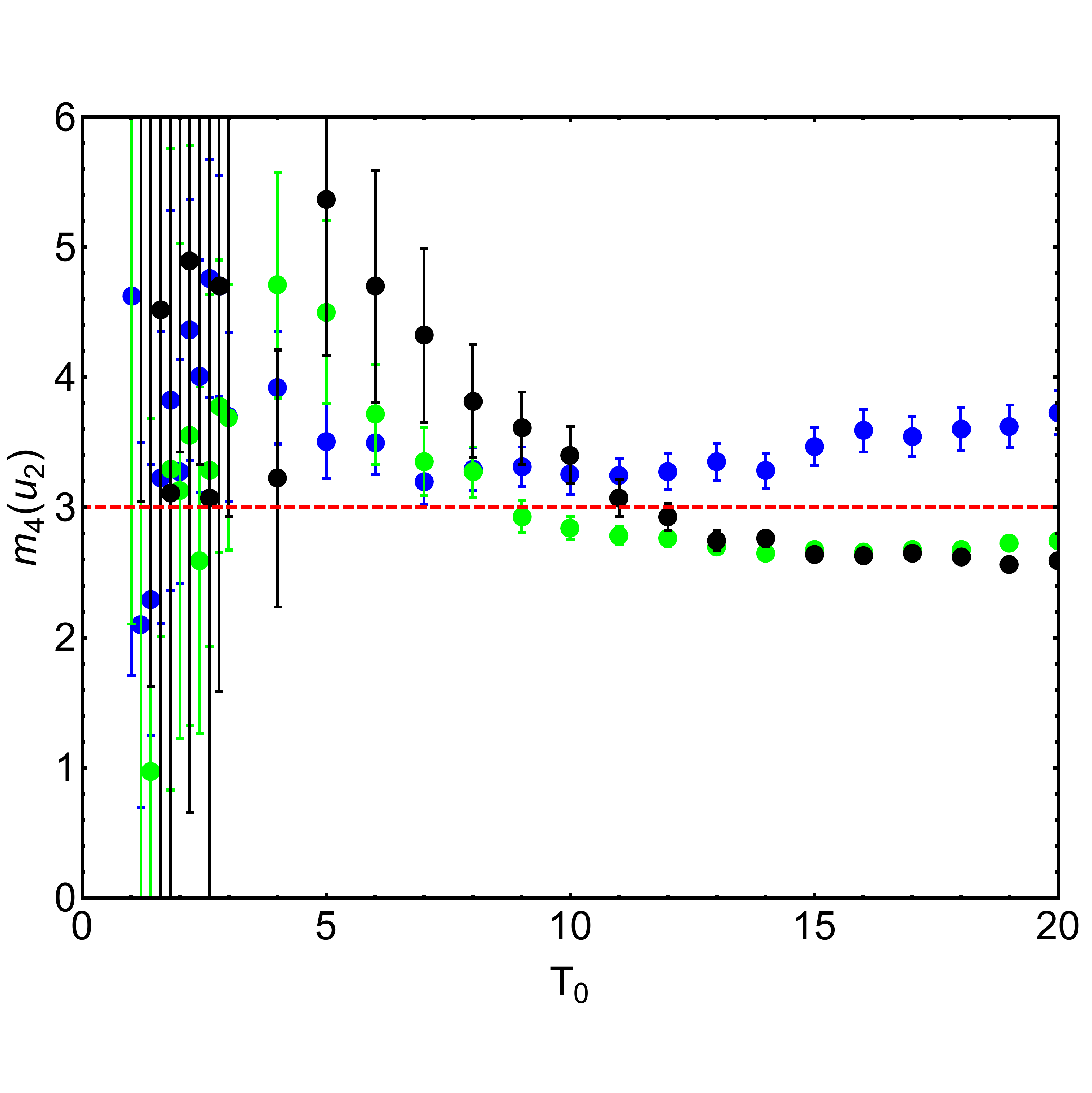}  %velo_moments_3_2_vy.nb 
\end{center}
\kern -1.cm
\caption{Fourth moment of  $u_1^{(s)}$-field (left figure) and  for $u_2^{(s)}$-field (right figure) as a function of $T_0$.  Blue, green and black points are for $g=5$, $g=10$ and $g=15$ respectively. The red dotted line shows the value for a Gaussian distributed data.
  \label{fourth_v}}
\end{figure}

Another global magnitude that capture some of topological structure of a scaled field, $v^{(s)}(x,y)$, is its $n$-th momenta:
 \begin{equation}
 m_n(v^{(s)};\xi)=\frac{1}{N_C}\sum_{(x,y)}v^{(s)}(x,y;\xi)^n
 \end{equation}
Following the same steps as above we can obtain its noise-corrected form and its error bar:
\begin{equation}
m_n(v^{(s)})=m_n(v^{(s)};\xi)-\Lambda_n\pm 3 \epsilon_n
\end{equation}
where
\begin{eqnarray}
\Lambda_n&=&-m_n(v^{(s)};\xi)\frac{n}{2}\left(1-\frac{1}{N_C}\right)\beta(0)+m_n(v^{(s)};\xi)\frac{n(n+2)}{2N_C}\beta(2)+m_{n-1}(v^{(s)};\xi)\frac{n^2}{N_C}\beta(1)\nonumber\\
&+&m_{n-2}(v^{(s)};\xi)\frac{n(n-1)}{2N_C}\beta(0)+\frac{n(n-1)}{2}\left(1-\frac{2}{N_C}\right)\beta(n-2)-\frac{n^2}{N_C}\beta(n)\nonumber\\
\epsilon_n^2&=&\frac{n^2}{N_C}\biggl[m_n(v^{(s)};\xi)^2\beta(2)-2m_n(v^{(s)};\xi)\beta(n)+2m_n(v^{(s)};\xi)m_{n-1}(v^{(s)};\xi)\beta(1)\nonumber\\
&+&\beta(2n-2)-2m_{n-1}(v^{(s)};\xi)\beta(n-1)+
m_{n-1}(v^{(s)};\xi)^2\beta(0)
\biggr]
\end{eqnarray}
where $\beta(n)$ is defined above. We show in figure \ref{fourth_v} the behavior of $m_4$. We observe that the data is noisy but it is compatible with  a constant behavior o to the existence of an asymptotic one for $T_0>13$ in all cases. From the $u_2^{(s)}$ data we see that for the $g=15$ case the constant regime appears for $T_0>14$. Moreover, $m_4(u_1^{(s)})$ shows that  $g=5$ and $g=10$ values have similar behavior while $g=15$ seems to tend to the same value. However, $m_4(u_2^{(s)})$ indicates a clear $g$-dependence at least between $g=5$ and the other $g$-values. We tried to analyze the $m_6$ cases but the associated error bars obscure any useful analysis.

Finally we have measured the euclidean distance between any two given configurations $v^{(s)}(x,y;T_0,g;\xi)$ with a fixed $g$ value but at temperatures $T_0$ and $T_0'$:
\begin{equation}
D(v^{(s)};T_0,T_0',g,\xi)=\left[\frac{1}{N_C}\sum_{(x,y)}(v^{(s)}(x,y;T_0,g;\xi)-v^{(s)}(x,y;T_0',g;\xi))^2\right]^{1/2} \label{distance}
\end{equation} 
We plot the set of pair distances for the scaled fields $u_1^{(s)}(x,y;\xi)$ and $u_2^{(s)}(x,y;\xi)$ for a given $g$ values in figures \ref{distancevx} and \ref{distancevy} (pink surfaces in top figures). We observe that all of surfaces tend to an apparently constant value clearly different from zero. Nevertheless, for any given $T_0>13$ it seems that the distance is constant (non-zero) for all other $T_0'>T_0$. Again, the effect is mainly due to the data noise. We compute the noise correction and we find:
\begin{equation}
D(v^{(s)};T_0,T_0',g)^2=D(v^{(s)};T_0,T_0',g,\xi)^2-\Lambda\pm 3\Sigma
\end{equation}
where
\begin{eqnarray}
\Lambda&=& \frac{2}{N_C}\sum_{(x,y)}(v^{(s)}(x,y;T_0,g;\xi)-v^{(s)}(x,y;T_0',g;\xi))(\alpha_2(x,y;T_0,g)-\alpha_2(x,y;T_0';g))\nonumber\\
&+&\frac{1}{N_C}\sum_{(x,y)}(\alpha_1(x,y;T_0,g)+\alpha_1(x,y;T_0';g))\nonumber\\
\Sigma^2&=&\frac{4}{N_C^2}\sum_{(x,y)}(v^{(s)}(x,y;T_0,g;\xi)-v^{(s)}(x,y;T_0',g;\xi))^2(\alpha_1(x,y;T_0,g)+\alpha_1(x,y;T_0';g))
\end{eqnarray}
and
\begin{eqnarray}
\alpha_1(x,y;T_0,g)&=&(1-\frac{2}{N_C})\frac{\sigma(x,y)^2}{\sigma^2}+\frac{1}{N_C}\beta(2)v^{(s)}(x,y)^2+\frac{\beta(0)}{N_C}-\frac{2}{N_C}v^{(s)}(x,y)^2\frac{\sigma(x,y)^2}{\sigma^2}+\frac{2\beta(1)}{N_C}v^{(s)}(x,y)\nonumber\\
\alpha_2(x,y;T_0,g)&=&\frac{1}{2}\left(\frac{3}{N_C}\beta(2)-(1-\frac{1}{N_C})\beta(0)\right)v^{(s)}(x,y)-\frac{1}{N_C}v^{(s)}(x,y)\frac{\sigma(x,y)^2}{\sigma^2}+\frac{\beta(1)}{N_C}
\end{eqnarray}
where $\beta(n)$, $\sigma(x,y)$, $\sigma$, ..., are computed for a given configuration with $T_0$ and $g$ values following the above definitions.
We expect $D$ to be zero or near to zero. Let us define $D_\pm(v^{(s)};T_0,T_0',g;\xi)=D(v^{(s)};T_0,T_0',g;\xi)\pm\sqrt{\Lambda}$. We expect that 
$D_-(v^{(s)};T_0,T_0',g,\xi)=D_-(v^{(s)};T_0,T_0',g)\pm 3\Sigma_-$ and $D_+(v^{(s)};T_0,T_0',g,\xi)=D_+(v^{(s)};T_0,T_0',g)\pm 3\Sigma_+$. Then, substituting into the $D^2$ expression we find that the non fluctuating parts follow:  $D^2=D_-D_+$ and the error parts: $\Sigma_-D_++\Sigma_+D_-\simeq \Sigma$. Assuming that $D_-$ is going to be almost zero we can approximate $\Sigma_-\simeq\Sigma/D_+$. Finally we can write
\begin{equation}
D_-(v^{(s)};T_0,T_0',g)=D(v^{(s)};T_0,T_0',g;\xi)-\sqrt{\Lambda}\pm 3\Sigma/(D(v^{(s)};T_0,T_0',g;\xi)+\sqrt{\Lambda})\label{distance2}
\end{equation}
We show in figures \ref{distancevx} and \ref{distancevy} the marginal distance $D_-$ for $u_1^{(s)}$ and $u_2^{(s)}$ respectively. We see how the bare distance $D(v^{(s)};T_0,T_0',g;\xi)$ (pink surfaces) have some dependence on $T_0$ and their limiting values are far from zero. Once we introduce the noise correction $\sqrt{\Lambda}$ the marginal distance $D_-$ tend to zero for $T_0>13$ which is coherent with the results obtained from other topological parameters. Moreover, we have an estimate of the error bars that we show at the bottom of the figures. Observe the quality of the convergence to zero of $D_-$for $g=10$ and $15$. $g=5$ is much noisier but still it is consistent with the zero marginal distance value.

\begin{figure}[h!]
\begin{center}
\includegraphics[height=5cm,clip]{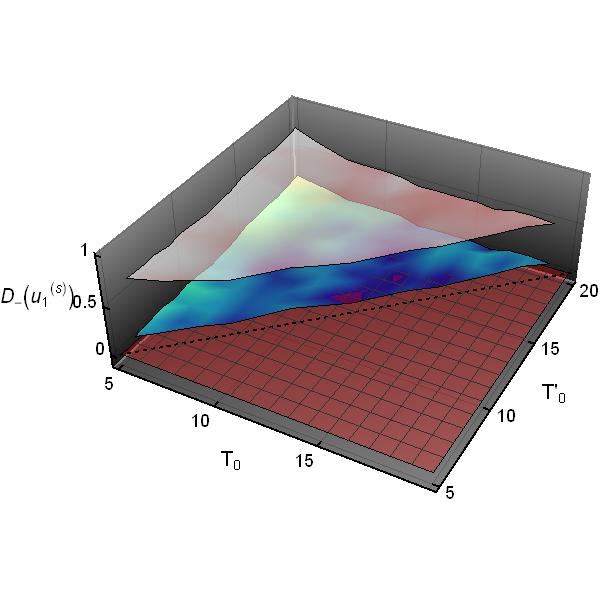}   %velo_vx_E5new.nb
\includegraphics[height=5cm,clip]{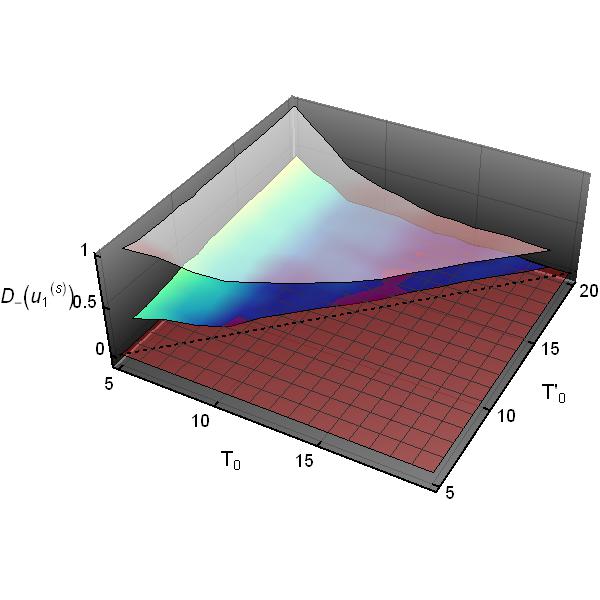}%velo_vx_E10new.nb
\includegraphics[height=5cm,clip]{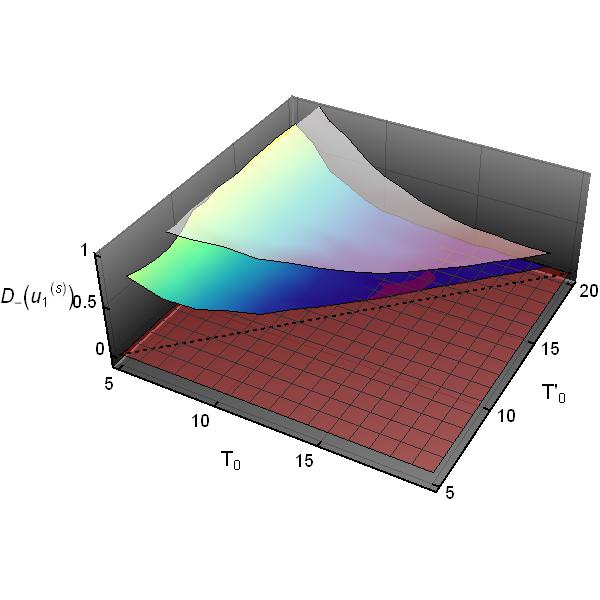}%velo_vx_E15new.nb
\newline\vglue -1cm
\includegraphics[height=4.5cm,clip]{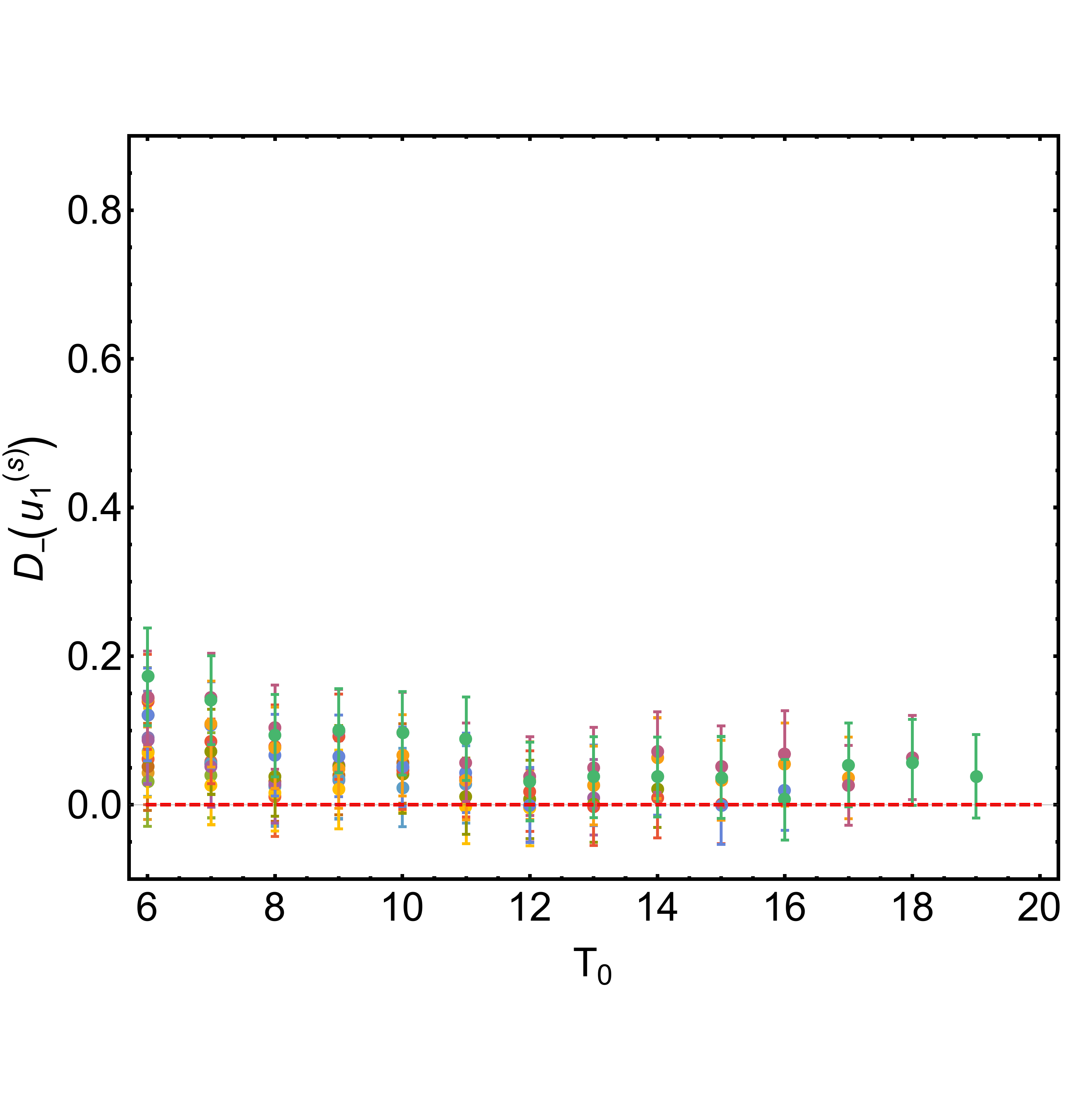}   %velo_vx_E5new.nb
\includegraphics[height=4.5cm,clip]{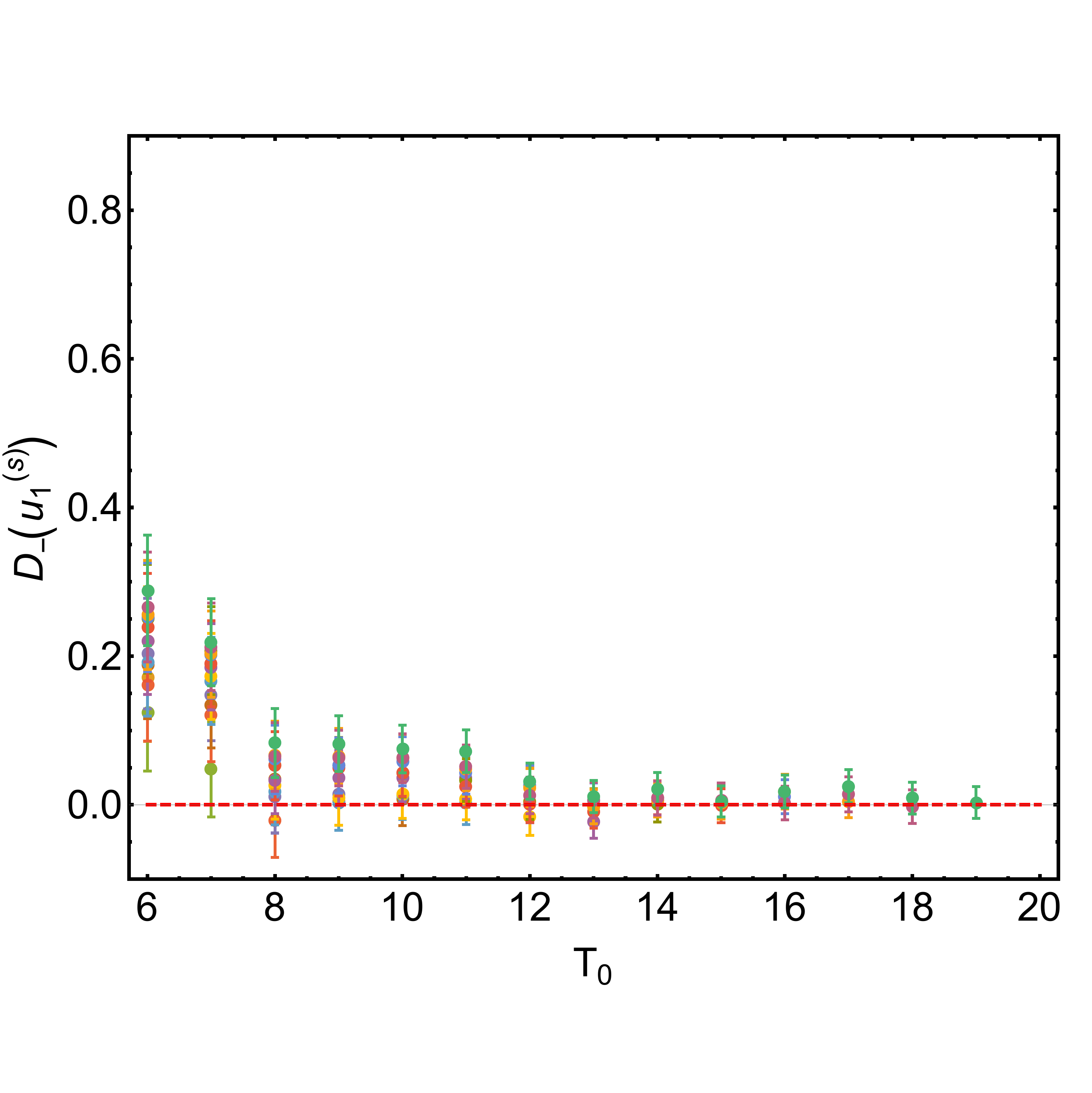}%velo_vx_E10new.nb
\includegraphics[height=4.5cm,clip]{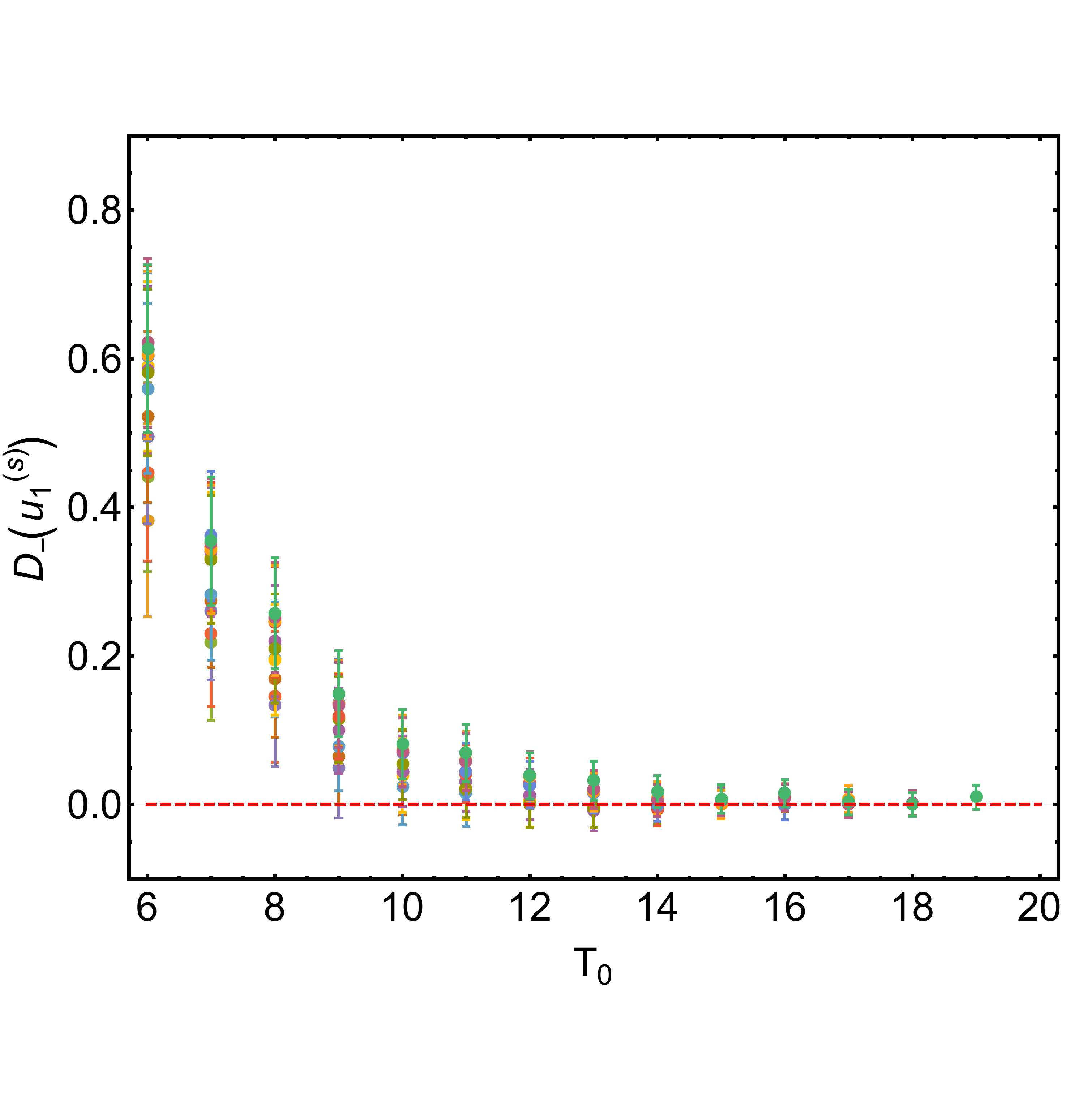}%velo_vx_E15new.nb
\end{center}
\kern -1.cm
\caption{Marginal distance $D_-$ (see text) between hydrodynamic velocity field scaled configurations $u_1^{(s)}$ defined by eq. \ref{distance2} for $g=5$ (figures left), $g=10$ (figures center) and $g=15$ (figures right). for a given $g$ value we only plot pairs of scaled configurations with $T_0$ and $T_0'$ such that $T_0<T_0'$. Top figures: Pink surface are the bare distances $D(u_1^{(s)};\xi)$. Yellow-green surfaces are the distances  after applying the noise correction term $\sqrt{\Lambda}$ (see text). Bottom figures:  Same as top but including error bars. Each color correspond to a given $T_0'$ value. \label{distancevx}}
\end{figure}
\begin{figure}[h!]
\begin{center}
\includegraphics[height=5cm,clip]{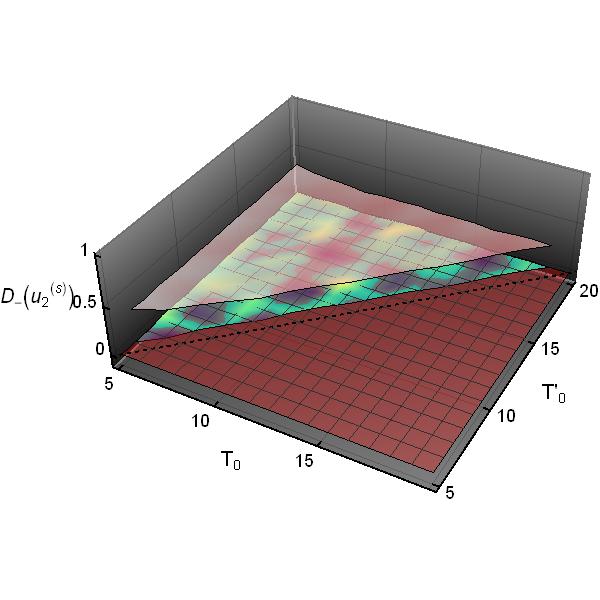}   %velo_vy_E5new.nb
\includegraphics[height=5cm,clip]{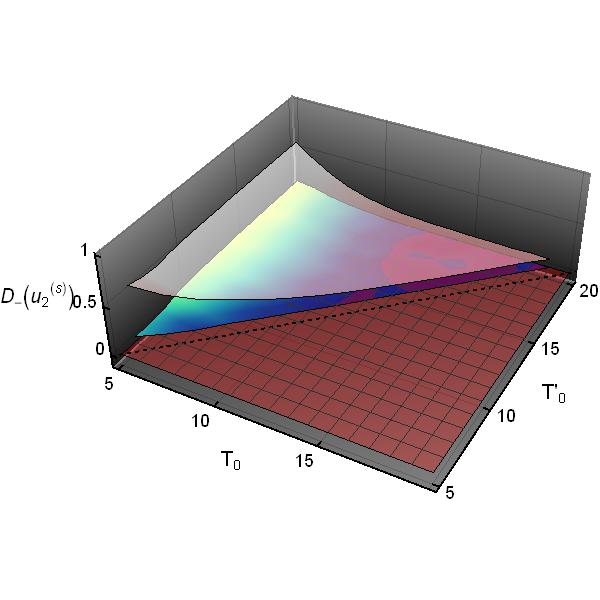}%velo_vy_E10new.nb
\includegraphics[height=5cm,clip]{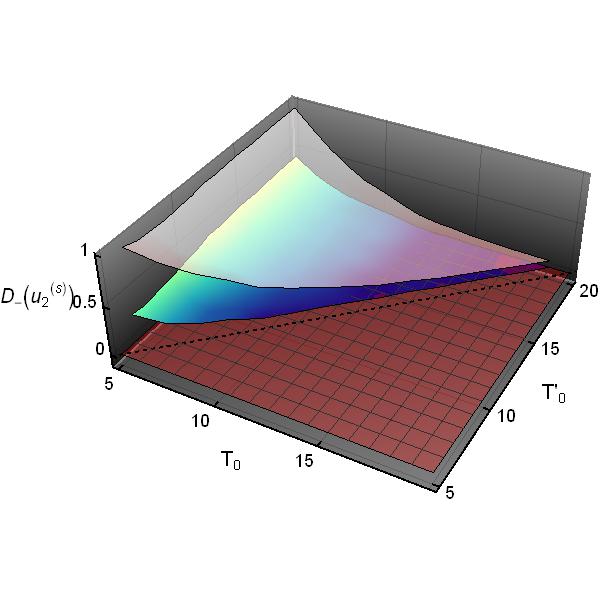}%velo_vy_E15new.nb
\newline\vglue -1cm
\includegraphics[height=4.5cm,clip]{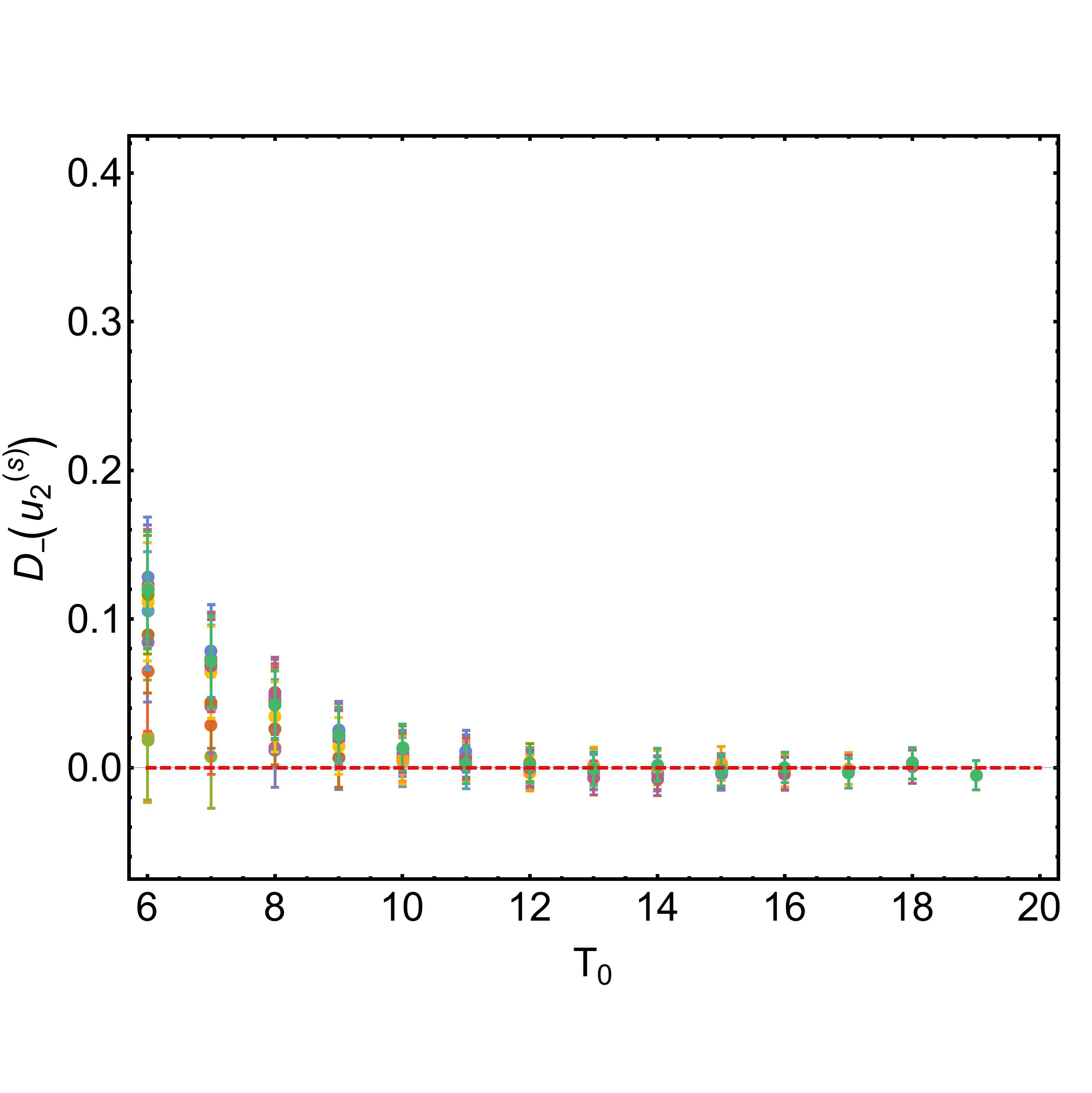}   %velo_vy_E5new.nb
\includegraphics[height=4.5cm,clip]{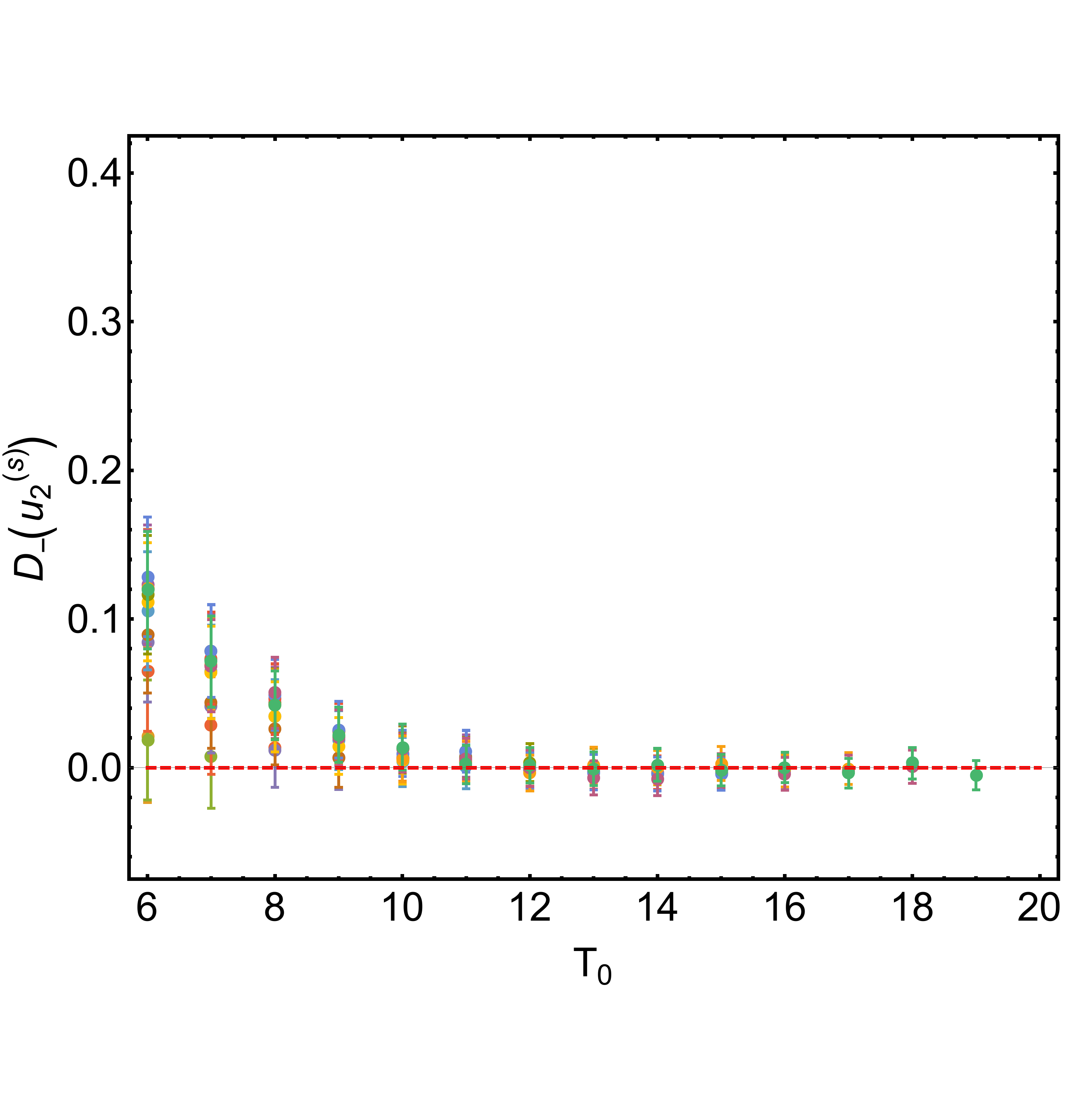}%velo_vy_E10new.nb
\includegraphics[height=4.5cm,clip]{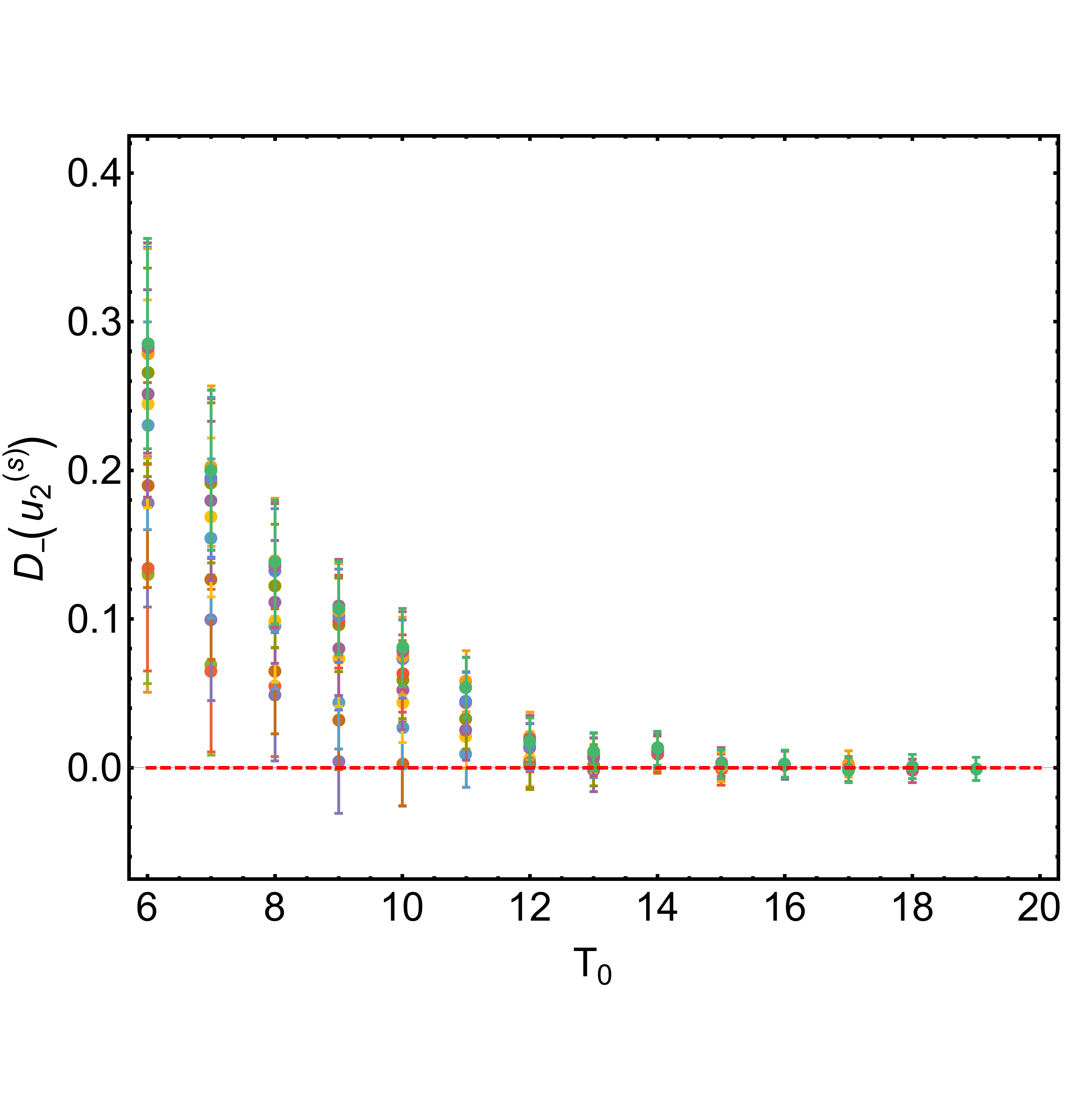}%velo_vy_E15new.nb
\end{center}
\kern -1.cm
\caption{Marginal distance $D_-$ (see text) between hydrodynamic velocity field scaled configurations $u_2^{(s)}$ defined by eq. \ref{distance2} for $g=5$ (figures left), $g=10$ (figures center) and $g=15$ (figures right). for a given $g$ value we only plot pairs of scaled configurations with $T_0$ and $T_0'$ such that $T_0<T_0'$. Top figures: Pink surface are the bare distances  $D(u_2^{(s)};\xi)$. Yellow-green surfaces are the distances after applying the noise correction term $\sqrt{\Lambda}$ (see text). Bottom figures:  Same as top but including error bars. Each color correspond to a given $T_0'$ value. \label{distancevy}}
\end{figure}
\begin{figure}[h!]
\begin{center}
\includegraphics[height=5cm,clip]{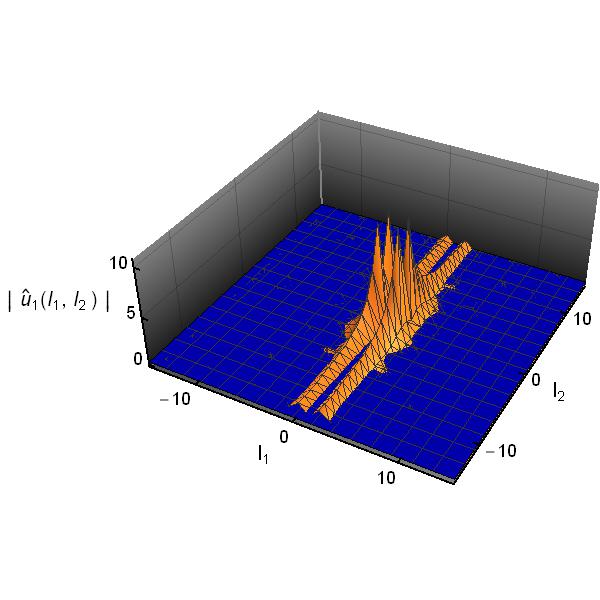}   %velo_vx_E5new.nb
\includegraphics[height=5cm,clip]{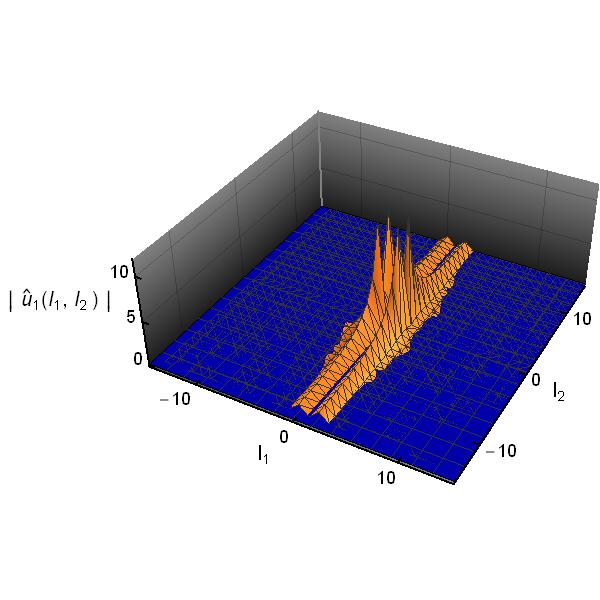}%velo_vx_E10new.nb
\includegraphics[height=5cm,clip]{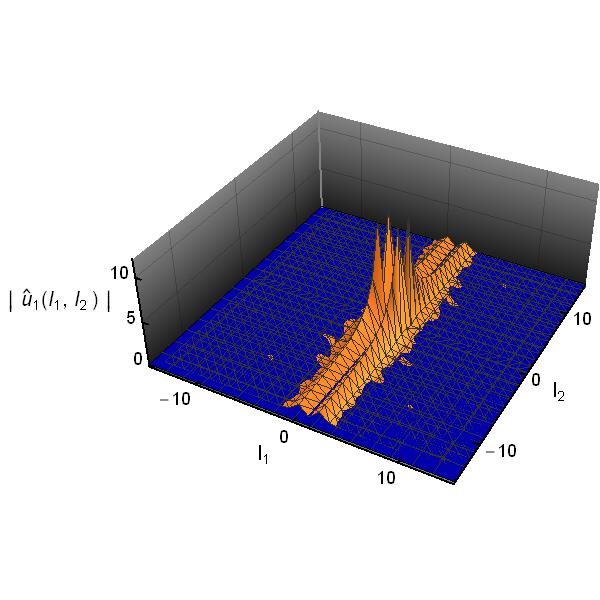}%velo_vx_E15new.nb
\newline\vglue -1cm
\includegraphics[height=5cm,clip]{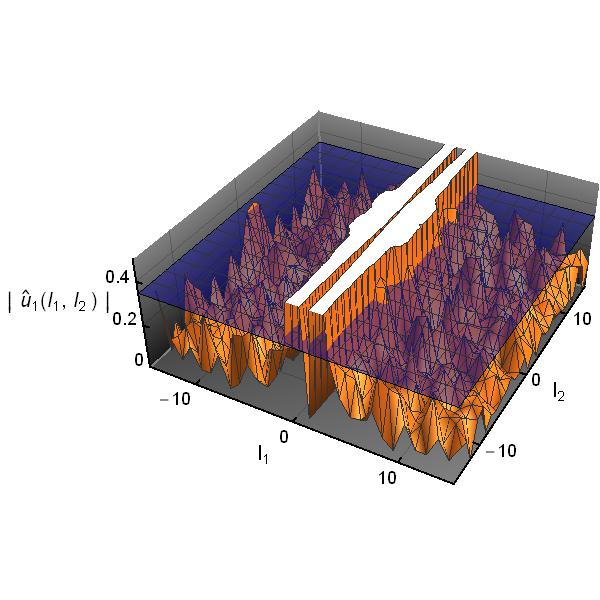}   %velo_vx_E5new.nb
\includegraphics[height=5cm,clip]{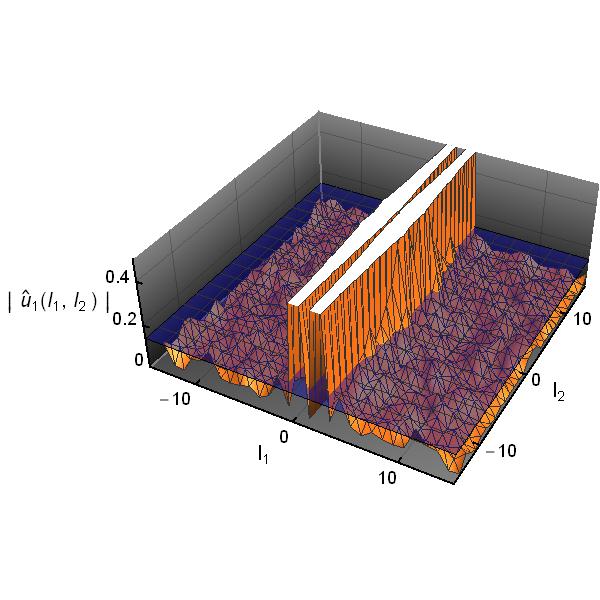}%velo_vx_E10new.nb
\includegraphics[height=5cm,clip]{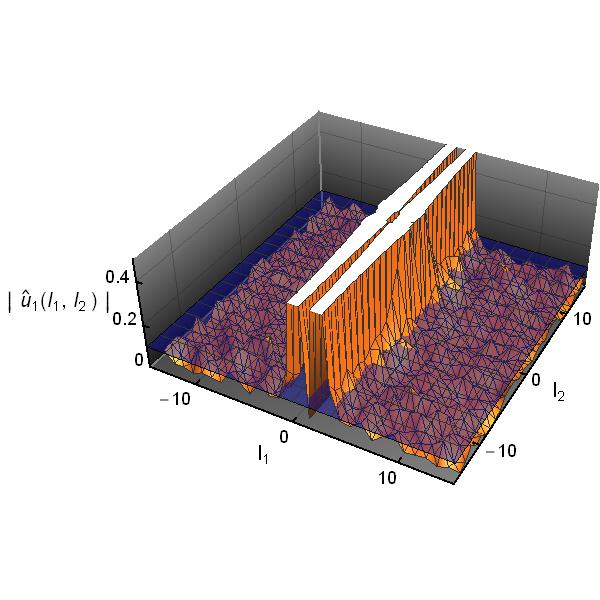}%velo_vx_E15new.nb
\end{center}
\kern -1.cm
\caption{Modulus of the Discrete Fourier Transform obtained by averaging the scaled configurations, $u_1^{(s)}(x,y)$, from $T_0=14,\ldots, 20$ for $g=5$ (left figures), $g=10$ (center figures) and $g=15$ (right figures).  Points below the blue surfaces are discarded and only points above them are used to the subsequent Inverse Fourier Transform to get a smoothed field. Top figures show the modes used in the Discrete Inverse Fourier Transform and bottom ones the detailed behavior of the discarded noisy modes.
 \label{Fouriervx}}
\end{figure}

\begin{figure}[h!]
\begin{center}
\includegraphics[height=5cm,clip]{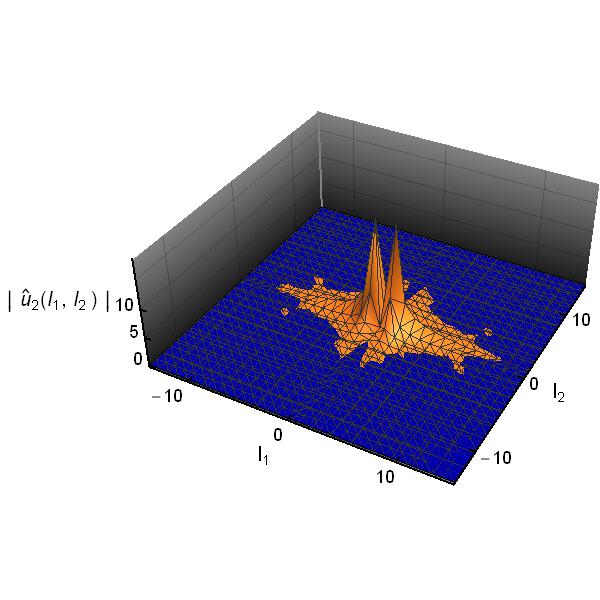}   %velo_vy_E5new.nb
\includegraphics[height=5cm,clip]{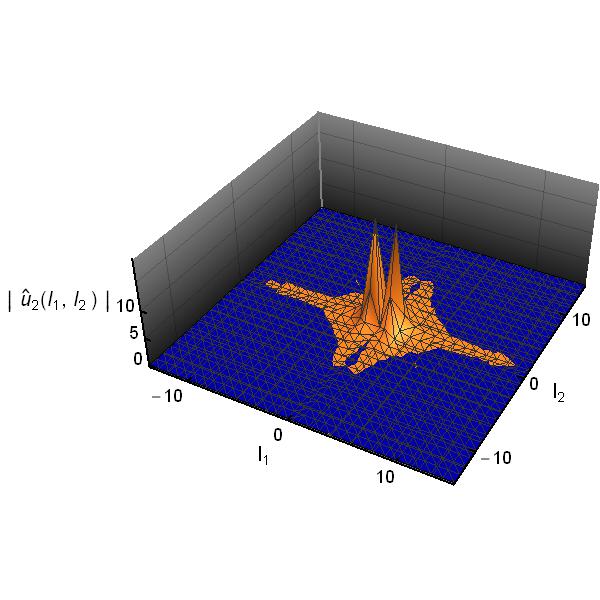}%velo_vy_E10new.nb
\includegraphics[height=5cm,clip]{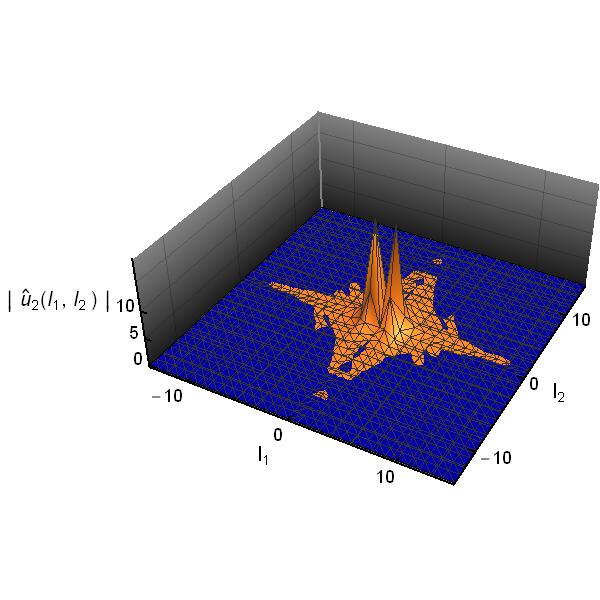}%velo_vy_E15new.nb
\newline\vglue -1cm
\includegraphics[height=5cm,clip]{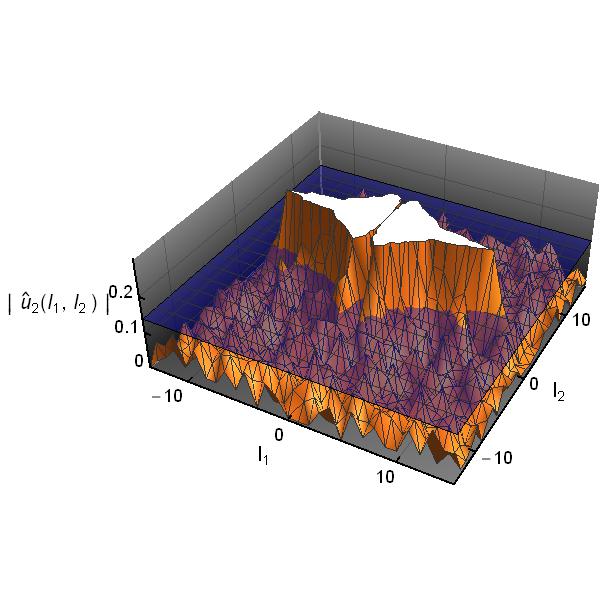}   %velo_vy_E5new.nb
\includegraphics[height=5cm,clip]{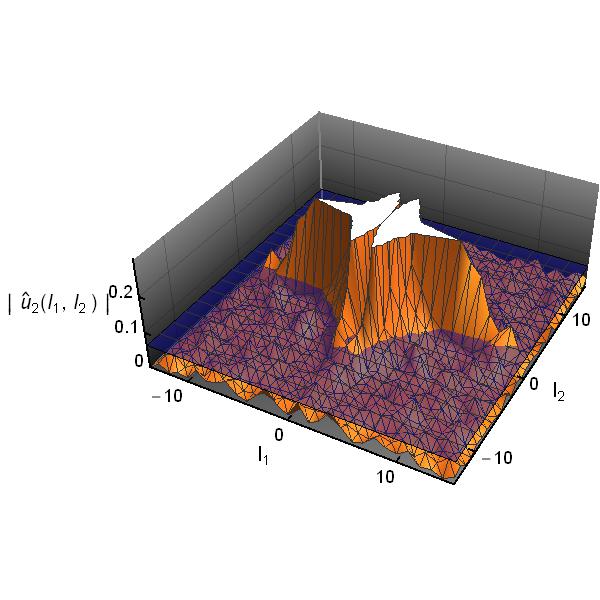}%velo_vy_E10new.nb
\includegraphics[height=5cm,clip]{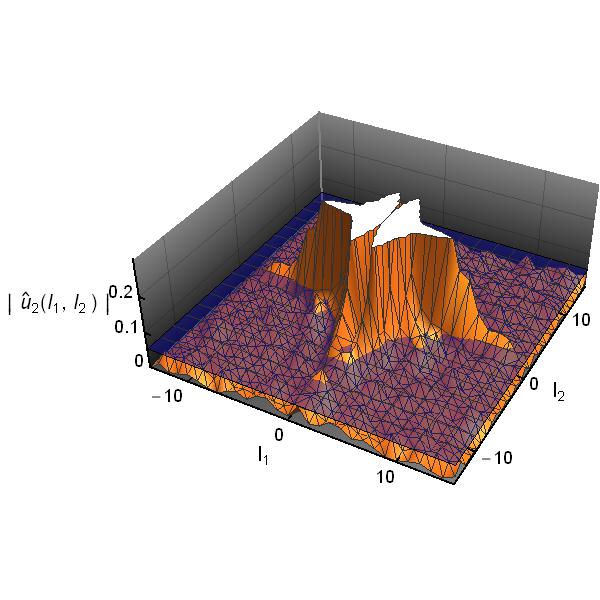}%velo_vy_E15new.nb
\end{center}
\kern -1.cm
\caption{Modulus of the Discrete Fourier Transform obtained by averaging the scaled configurations, $u_2^{(s)}(x,y)$, from $T_0=14,\ldots, 20$ for $g=5$ (left figures), $g=10$ (center figures) and $g=15$ (right figures).  Points below the blue surfaces are discarded and only points above them are used to the subsequent Inverse Fourier Transform to get a smoothed field. Top figures show the modes used in the Discrete Inverse Fourier Transform and bottom ones the detailed behavior of the discarded noisy modes.
 \label{Fouriervy}}
\end{figure}
\begin{table}[h!]
\begin{center}
\resizebox*{!}{3cm}{ 
\begin{tabular}{|c|c|c|}
\hline
$g$&$\vert \hat u_1^{(s)}\vert$&$\vert \hat u_2^{(s)}\vert$\\ \hline
\hline
5&0.35&0.14\\ \hline
10&0.14&0.065\\ \hline
15&0.10&0.055\\ \hline
\end{tabular}}
\end{center}
\caption{Cut-off values. The modes of the Fourier Transform of the field with modulus less than the corresponding cut-off value are discarded. \label{cut}}
\end{table}

\begin{figure}[h!]
\begin{center}
\includegraphics[height=5cm,clip]{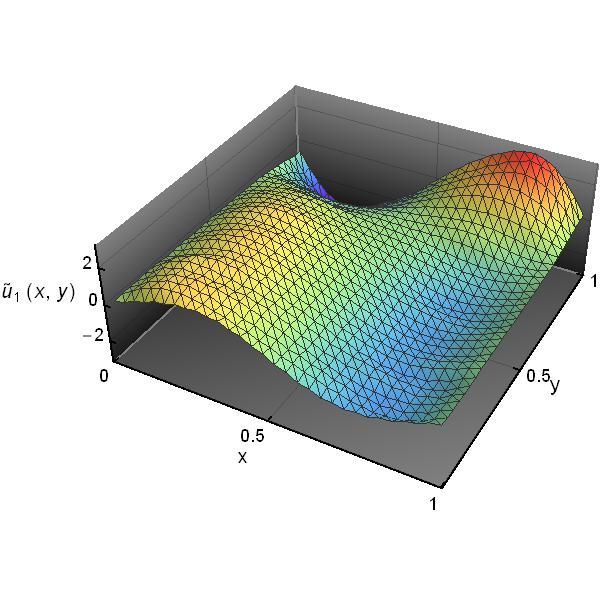}   %velo_vx_E5new.nb
\includegraphics[height=5cm,clip]{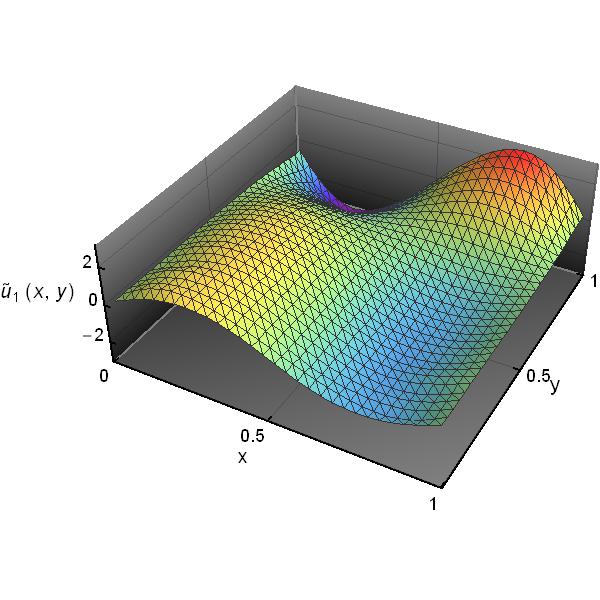}%velo_vx_E10new.nb
\includegraphics[height=5cm,clip]{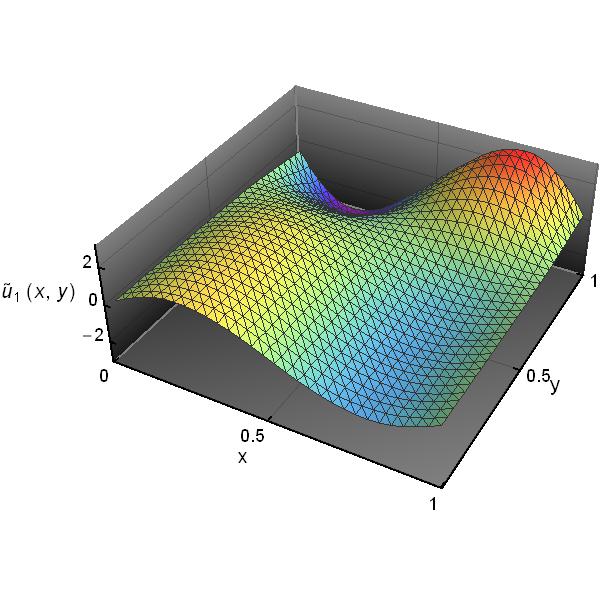}%velo_vx_E15new.nb
\newline\vglue -1cm
\includegraphics[height=5cm,clip]{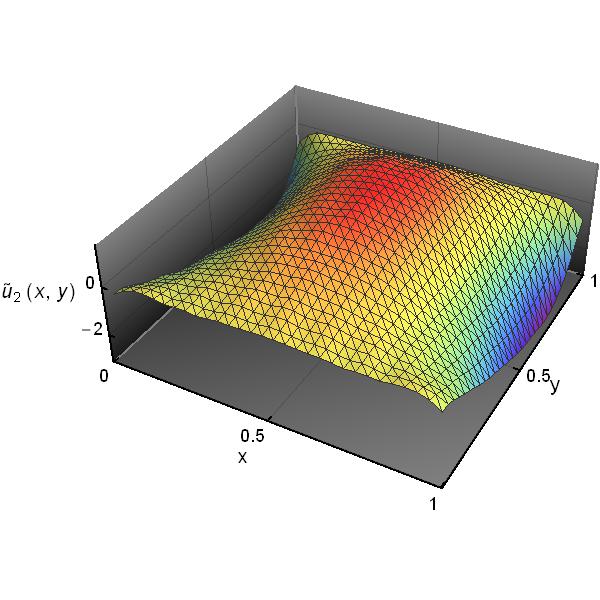}   %velo_vy_E5new.nb
\includegraphics[height=5cm,clip]{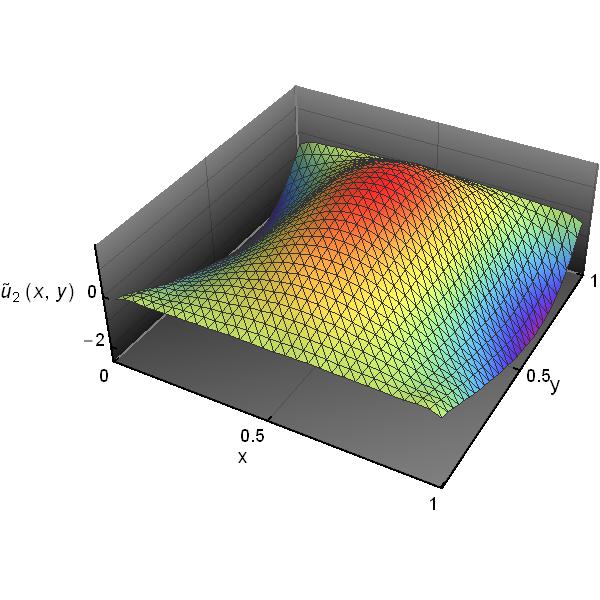}%velo_vy_E10new.nb
\includegraphics[height=5cm,clip]{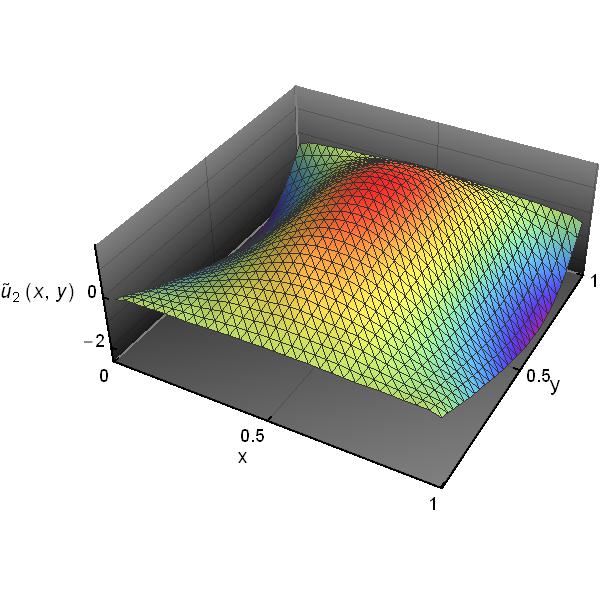}%velo_vy_E15new.nb
\end{center}
\kern -1.cm
\caption{Universal fields $\tilde u_1(x,y)$ (top figures) and $\tilde u_2(x,y)$ (bottom figures) for $g=5$, $10$ and $15$ from left to right. 
 \label{scaledv}}
\end{figure}
\begin{figure}[h!]
\begin{center}
\includegraphics[height=5cm,clip]{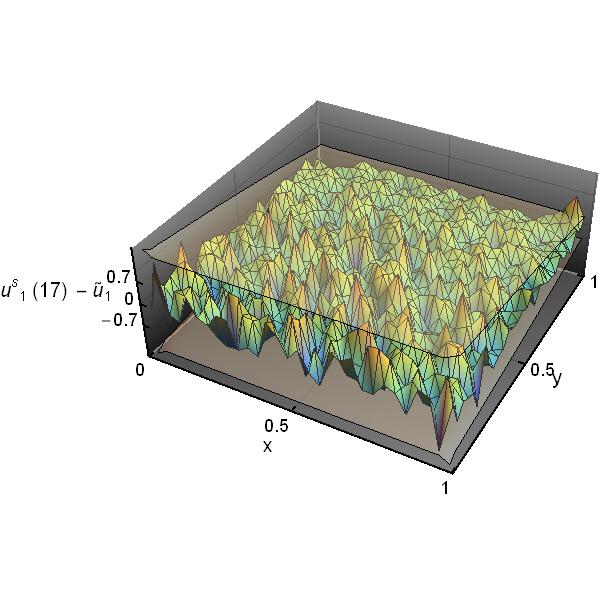}   %velo_vx_E5new.nb
\includegraphics[height=5cm,clip]{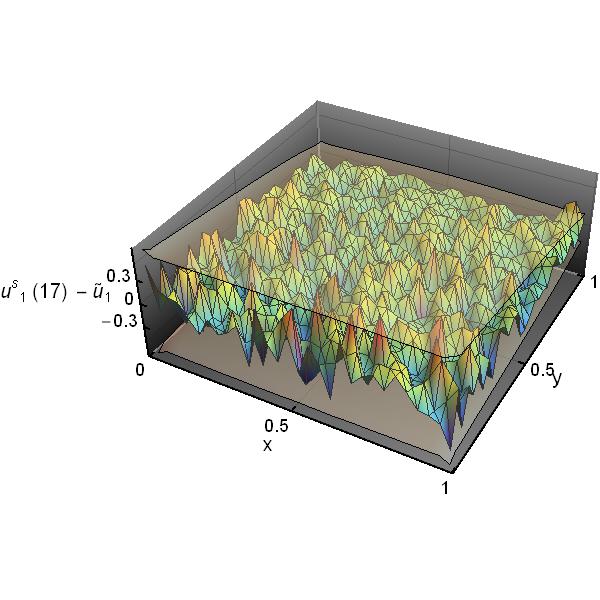}%velo_vx_E10new.nb
\includegraphics[height=5cm,clip]{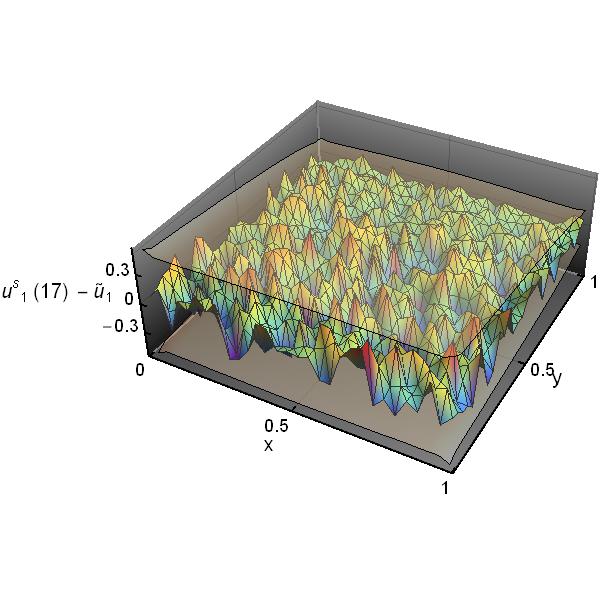}%velo_vx_E15new.nb
\newline\vglue -1cm
\includegraphics[height=5cm,clip]{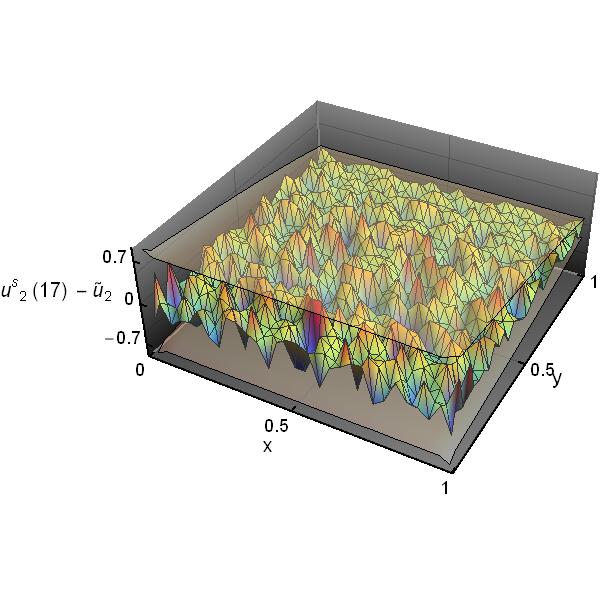}   %velo_vy_E5new.nb
\includegraphics[height=5cm,clip]{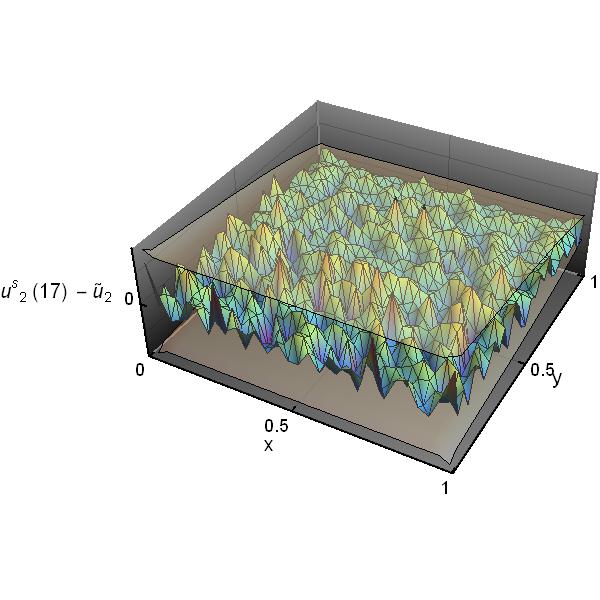}%velo_vy_E10new.nb
\includegraphics[height=5cm,clip]{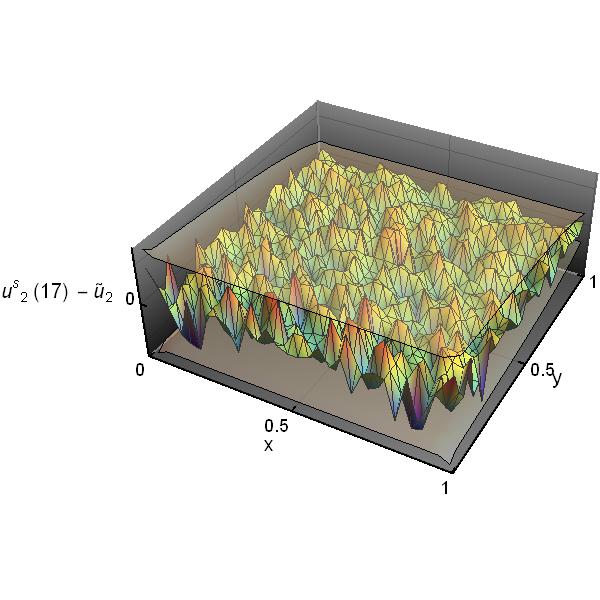}%velo_vy_E15new.nb
\end{center}
\kern -1.cm
\caption{Difference between the scaled field $u_{1,2}^{(s)}(x,y)$ for $T_0=17$ and the corresponding universal field $\tilde u_{1,2}(x,y)$  (top figures and bottom figures respectively) for $g=5$, $10$ and $15$ from left to right. The gray surfaces are the data error bars of the scaled velocity fields.
 \label{fluctuv}}
\end{figure}
\begin{figure}[h!]
\begin{center}
\includegraphics[height=5cm,clip]{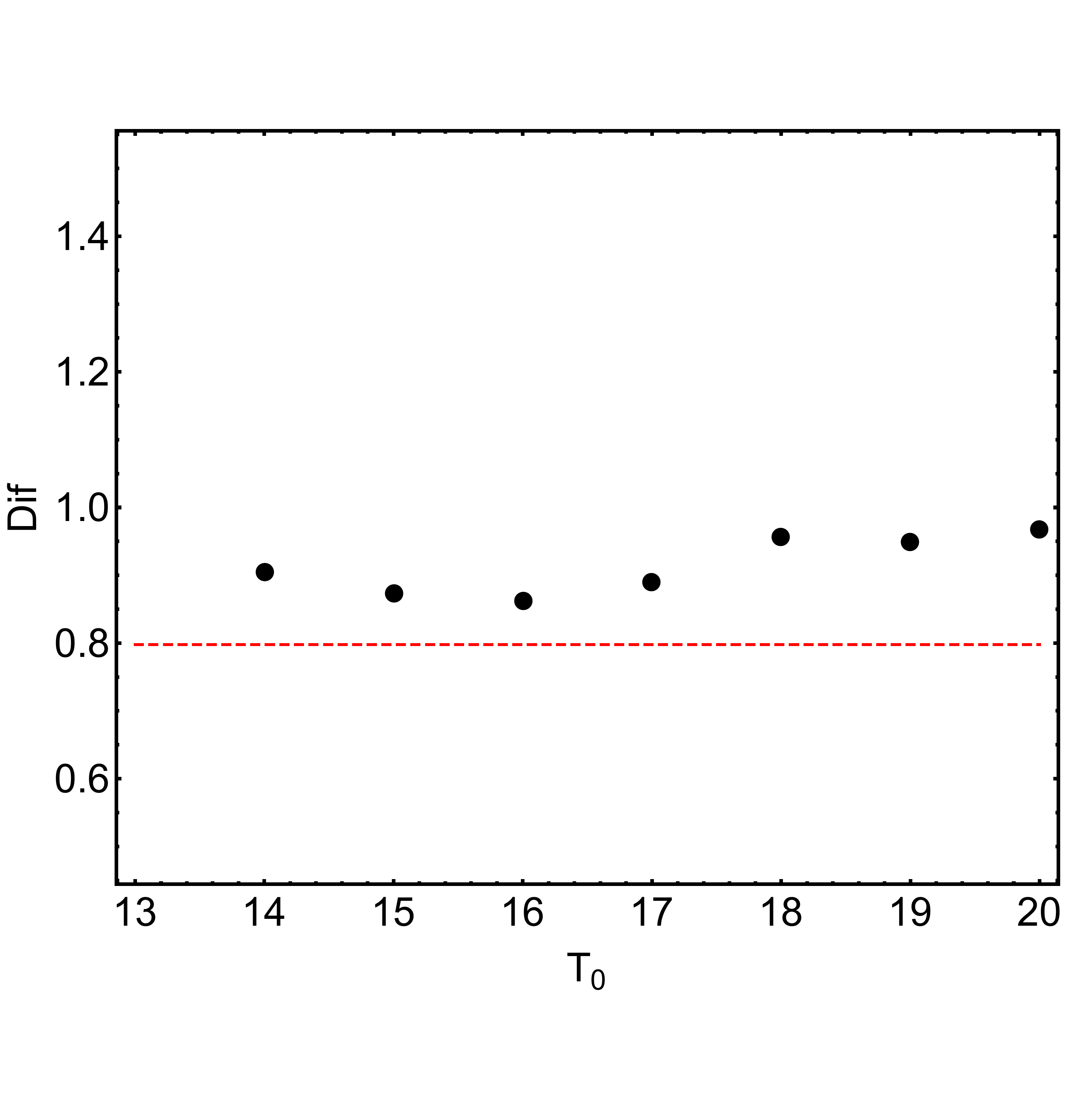}   %velo_vx_E5new.nb
\includegraphics[height=5cm,clip]{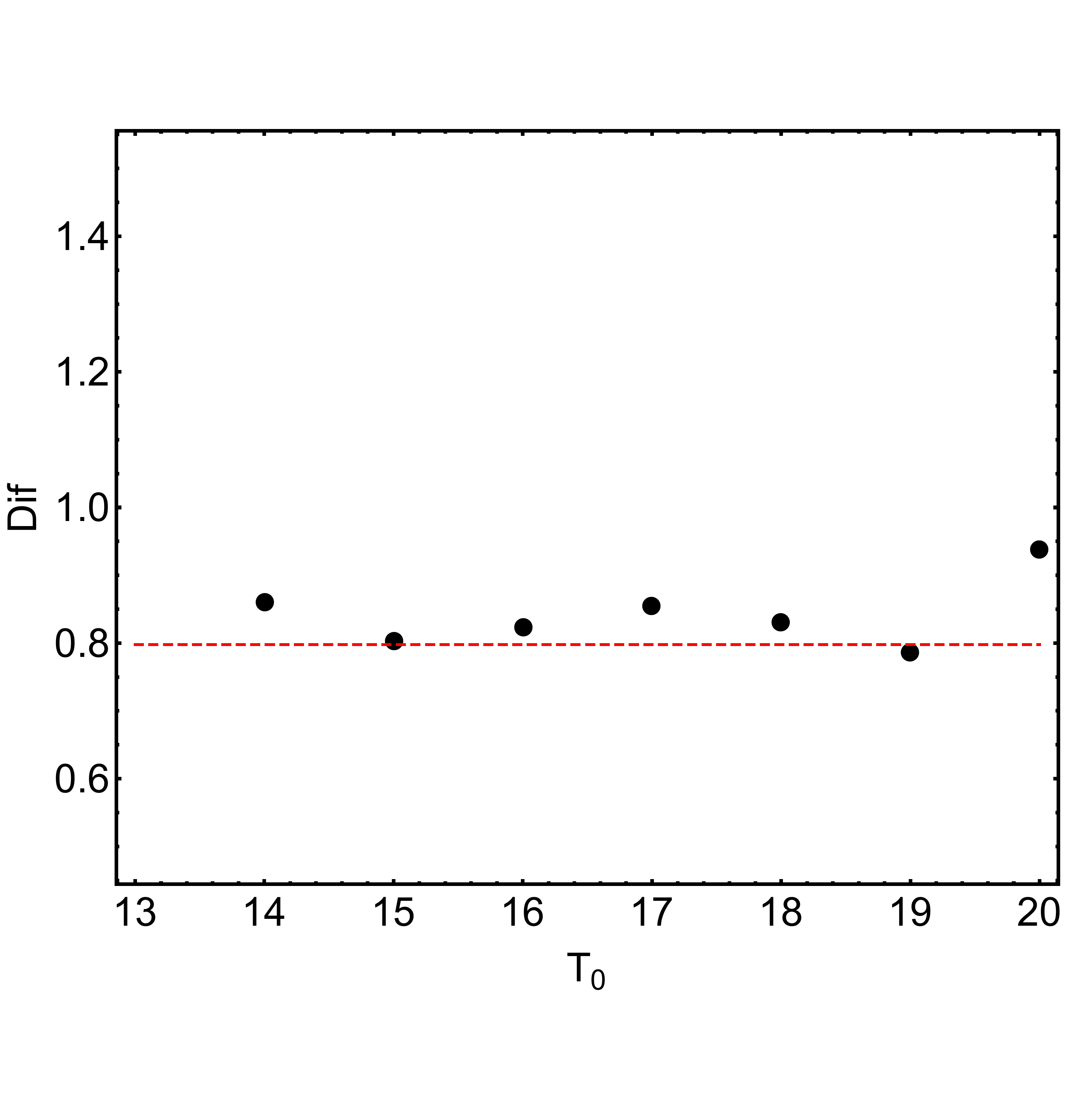}%velo_vx_E10new.nb
\includegraphics[height=5cm,clip]{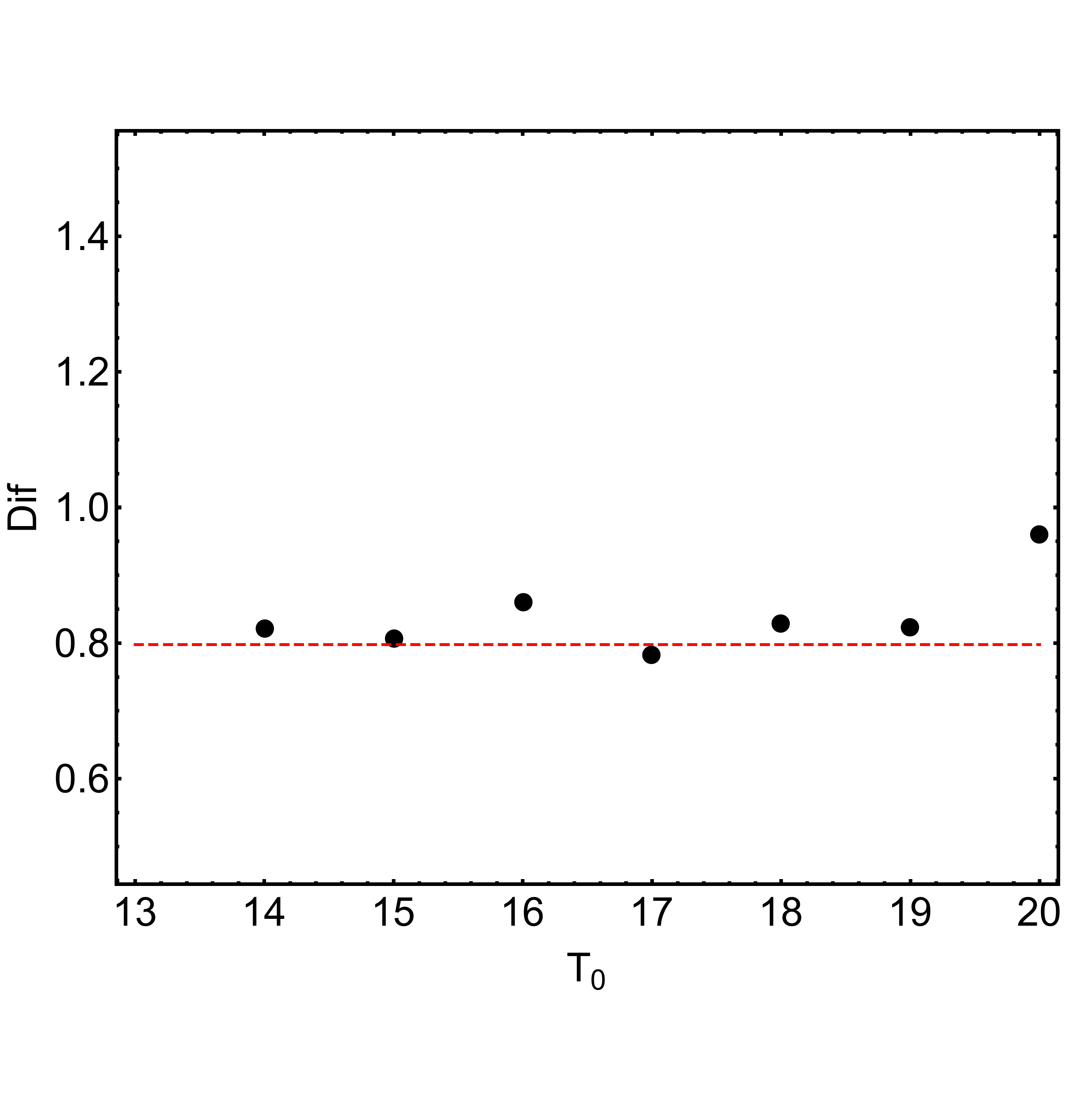}%velo_vx_E15new.nb
\newline\vglue -1cm
\includegraphics[height=5cm,clip]{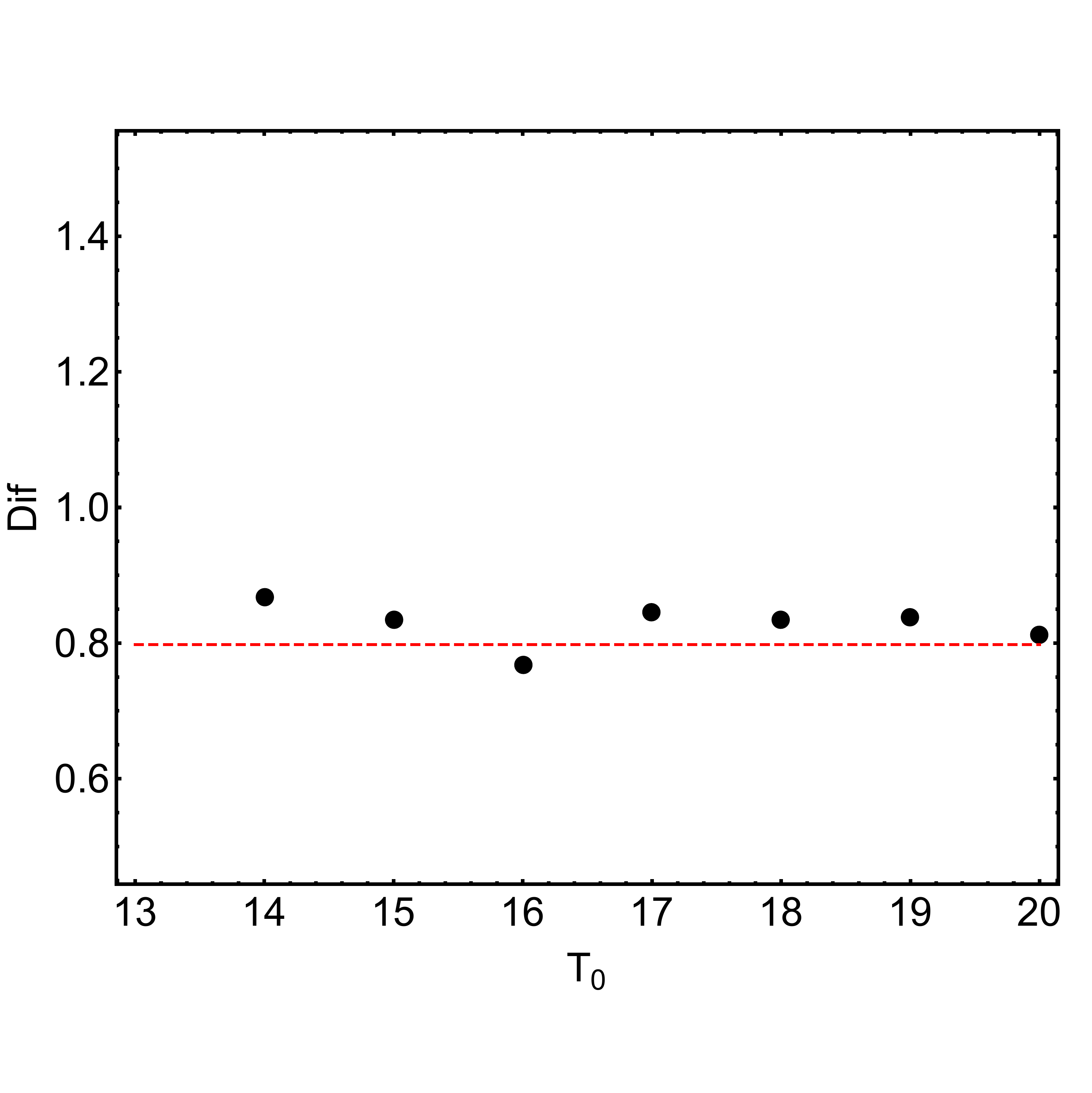}   %velo_vy_E5new.nb
\includegraphics[height=5cm,clip]{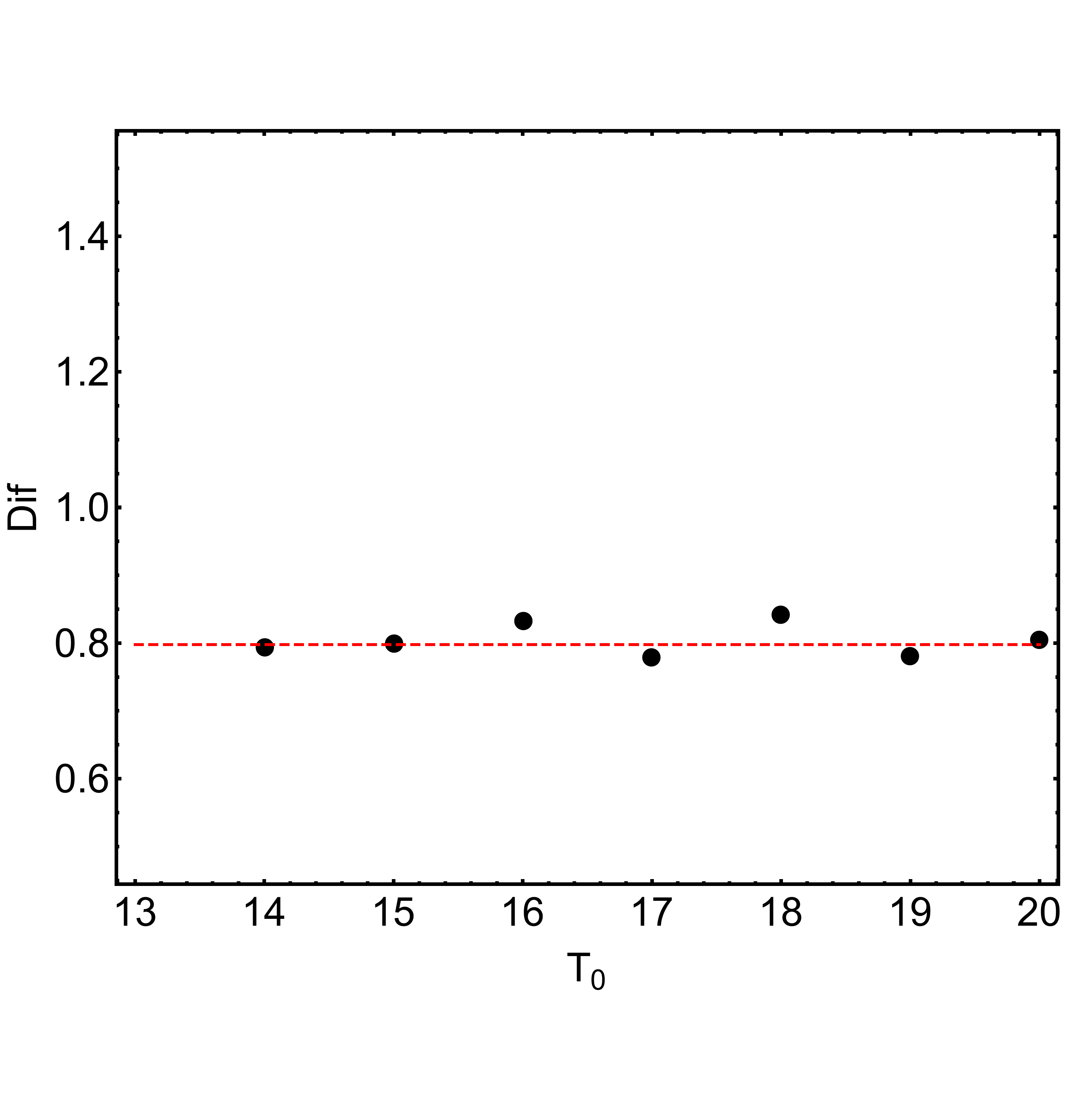}%velo_vy_E10new.nb
\includegraphics[height=5cm,clip]{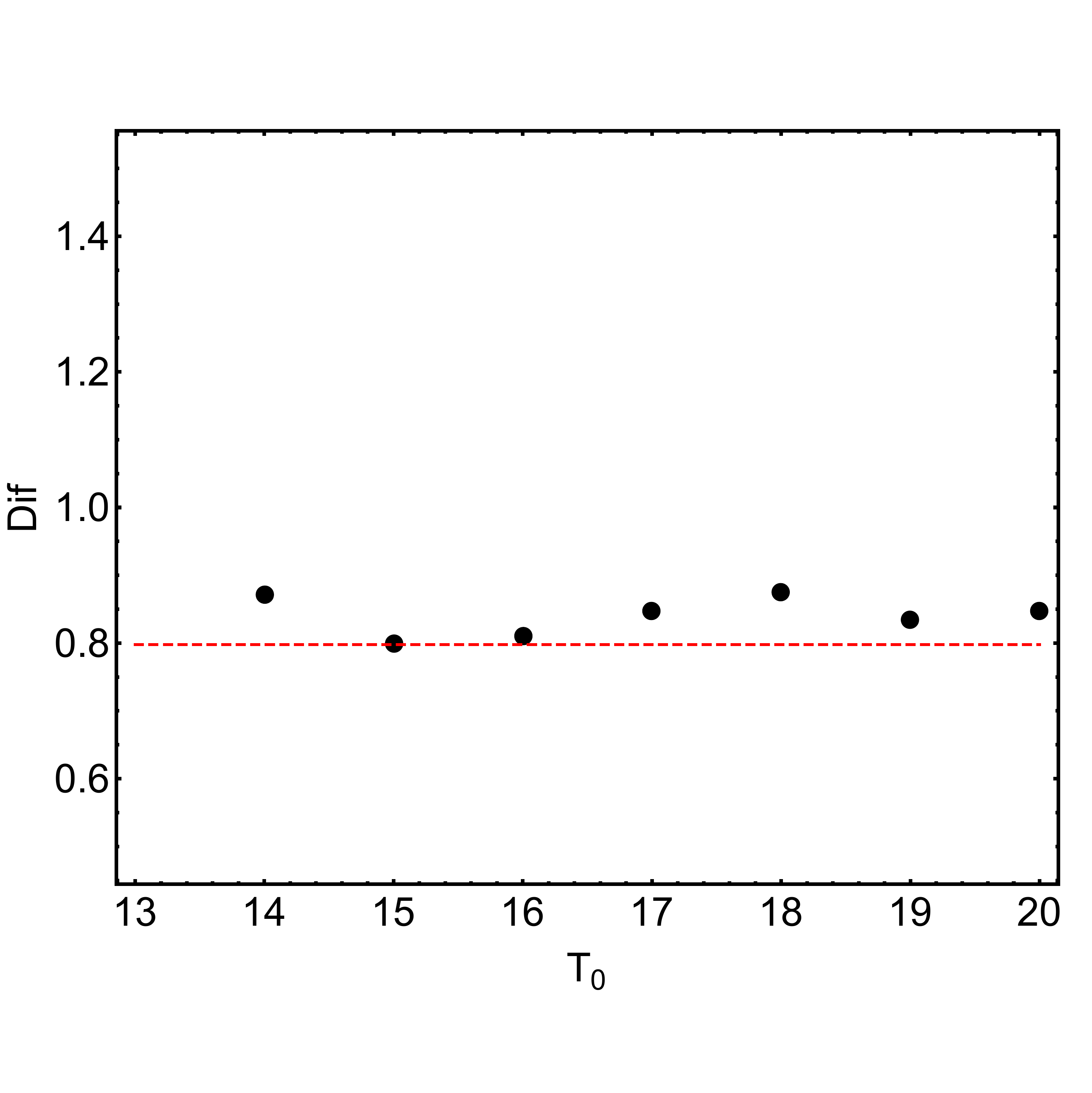}%velo_vy_E15new.nb
\end{center}
\kern -1.cm
\caption{Averaged ratio between the difference between the scaled field $u_{1,2}^{(s)}(x,y)$  and the corresponding universal field $\tilde u_{1,2}(x,y)$ with respect its variance normalized with the profile variance (top figures and bottom figures respectively) for $g=5$, $10$ and $15$ from left to right. 
 \label{fluctuv2}}
\end{figure}

\begin{figure}[h!]
\begin{center}
\includegraphics[height=5cm,clip]{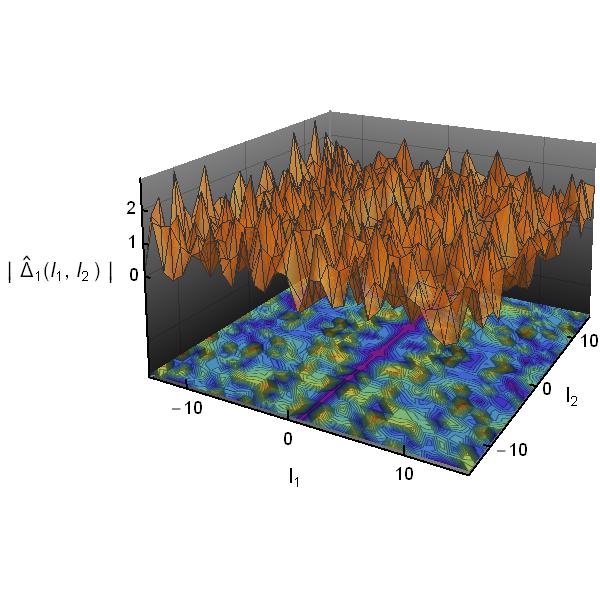}   %velo_vx_E5new.nb
\includegraphics[height=5cm,clip]{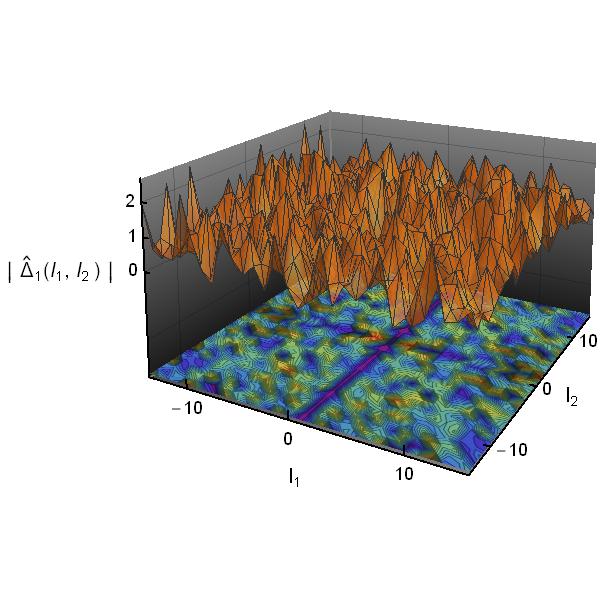}%velo_vx_E10new.nb
\includegraphics[height=5cm,clip]{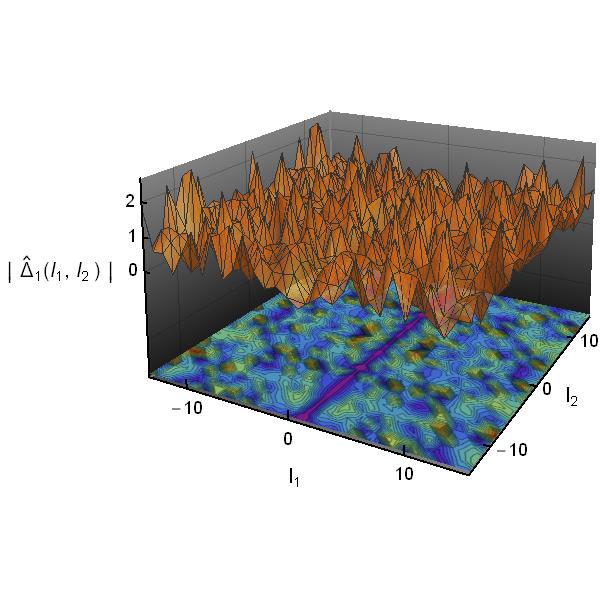}%velo_vx_E15new.nb
\newline\vglue -1cm
\includegraphics[height=5cm,clip]{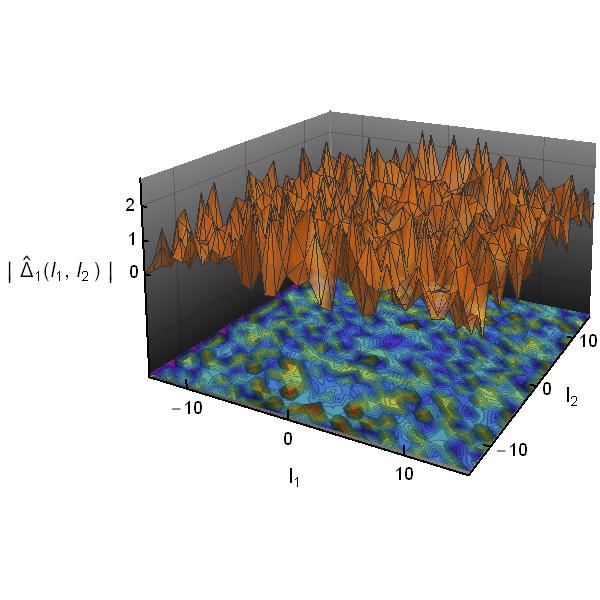}   %velo_vy_E5new.nb
\includegraphics[height=5cm,clip]{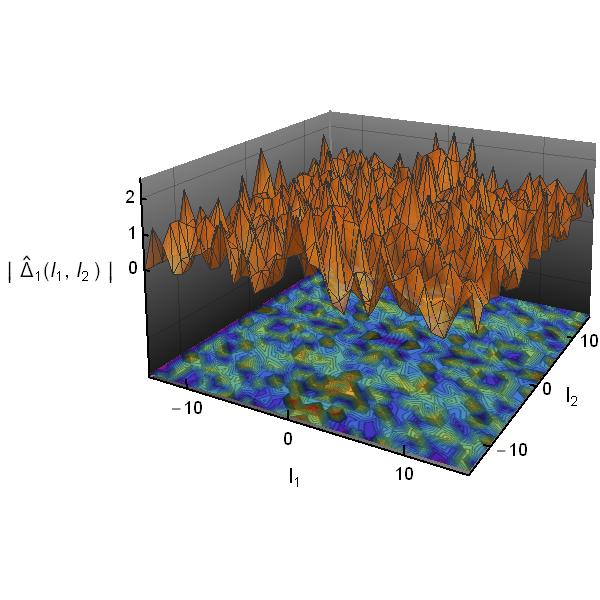}%velo_vy_E10new.nb
\includegraphics[height=5cm,clip]{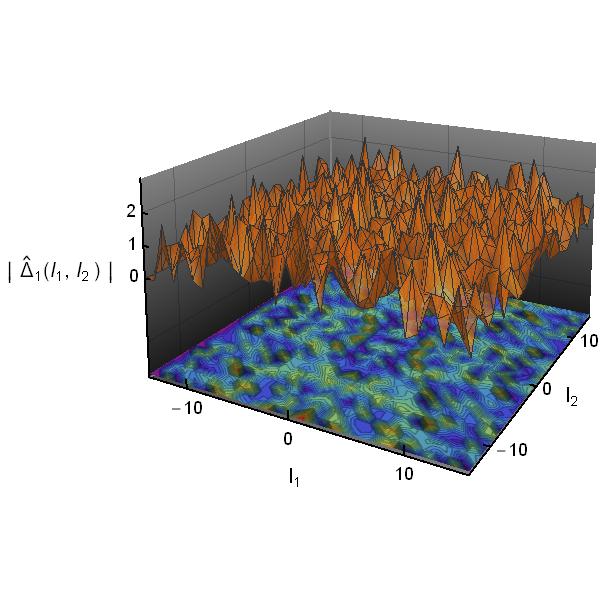}%velo_vy_E15new.nb
\end{center}
\kern -1.cm
\caption{Modulus of the Fourier Transform of the difference  between the scaled field $u_{1,2}^{(s)}(x,y)$ for $T_0=17$ and the corresponding universal field $\tilde u_{1,2}(x,y)$  (top figures and bottom figures respectively) for $g=5$, $10$ and $15$ from left to right. 
 \label{fluctuvf}}
\end{figure}

\item {\it Obtaining the universal fields:}

It is clear now  that the rescaled fields seems to be equal (except for fluctuations) when $T_0>13$. To obtain the universal scaled field $\tilde u_{1,2}(x,y)$ we follow two steps: (i) we average (for a given $g$-value) all the noise-corrected scaled fields $u_{1,2}^{(s)}(x,y)$ with $T_0>13$ and (ii) we Fourier Transform the averaged field, we disregard modes with fluctuating small amplitude and we inverse Fourier Transform the remaining modes. All of these just to get a smooth field without fluctuating parts. 

In figures \ref{Fouriervx} and \ref{Fouriervy} we show the modulus of the Fourier Transform for $u_1$ and $u_2$ fields respectively. We observe how the dominant modes have $\simeq 10$  intensity meanwhile the discarded modes have intensities less than $0.4$. In the table \ref{cut} we show the cut-off values used in each case. In general we choose such values as the minimum one in which we have a compact and continuum set of modes.  In Figure \ref{scaledv} we show the final smoothed universal fields $\tilde u_1(x,y)$ and $\tilde u_2(x,y)$ for $g=5$, $10$ and $15$. One observes at a glance that the universal fields for different $g$-values are very similar. When subtracting we see systematic small differences that have spatial structure. However, such differences are small and of the order of the error bars. At this point and with our set of simulations we cannot conclude about the existence of a $g$-independent universal profile.

The final check is to see if the difference between any $T_0>13$ scaled configuration and the universal field gives place to a random fluctuating field with amplitudes smaller than the error bars of the scaled field. The error for a point $(x,y)$ of any given rescaled configuration have the form of $3\sigma(x,y)/\sigma$ assuming that the central limit theorem holds. Figure \ref{fluctuv} shows, as an example, the difference between $u_{1,2}^{(s)}(x,y;T_0=17)$ and the corresponding $\tilde u_{1,2}(x,y)$. We observe how practically all the fluctuating data is between the error bar interval. Moreover, there is a spatial dependence that is similar for the differences and the error bars. From the point of view of the data errors we cannot do better than this. However, we can try to see if there is any systematic spatial regularity in the differences. 
We define the difference field:
\begin{equation}
w_{d}(x,y;T_0,\xi)=\sigma(T_0)\frac{u_{1,2}^{(s)}(x,y;T_0;\xi)-\tilde u_{1,2}(x,y)}{\sigma(x,y;T_0)}
\end{equation}
In this form, if $u_{1,2}^{(s)}$ follows exactly the noise ansatz: $u_{1,2}^{(s)}(x,y)=\tilde u_{1,2}(x,y)+\sigma(x,y)\xi(x,y)/\sigma$, then $w_{d}(x,y;T_0;\xi)=\xi(x,y)$, that is, we should see a pure white noise. In particular we defined the average of its absolute value:
\begin{equation}
\text{Dif}=\frac{1}{N_C}\sum_{(x,y)}\vert w_{d}(x,y;T_0,\xi) \vert
\end{equation}
In case that $w_d$ was a pure white noise the value of $\text{Diff}$ would be $2/\sqrt{2\pi}\simeq 0.798..$.
We see in figure \ref{fluctuv2} this magnitude as a function of $T_0$ ($\ge 13$). We observe how this magnitude goes around the white noise value. The small deviations observed could be due to the cutoff used in which we could discard some weak structural behavior in modes that are superimposed to the noise and/or some small system correlations. Nevertheless, the white noise average behavior seem to us quite remarkable because it implies that our analysis is capturing the principal parts of the system behavior. Finally, we show in figure \ref{fluctuvf} the Fourier Transform of $w_d$ for $T_0=17$ and $g=5$, $10$ and $15$ just to see if there is any clear spatial structure. We do not see any clear regularity that would mean the existence of a systematic deviation between the universal field and the scaled one. 
 All of this support the idea that our universal field is a good average for each $T_0>13$ velocity field.
\end{enumerate}

We may conclude, after this extent statistical analysis, that the hydrodynamic velocity field components have an universal scaling property, that is, we can write them as:
\begin{equation}
u_{1,2}(x,y)= \sigma(u_{1,2}(T_0,g))\tilde u_{1,2}(x,y;g)
\end{equation}
That is, the $T_0$ dependence can be explicitely extracted from the fields. As we commented above, we cannot exclude the possibility of having $\tilde u_{1,2}$ independent on $g$. Another questions that remains also unsolved is: if we could improve data statistics  for the $T_0<14$ configurations in such a way that their errors were smaller than the typical profile variation, would we still find the same scaling behavior? Is then, the scaling behavior typical of all convective states? or  is it just an asymptotic like property? Those are open questions that are beyond the computational effort presented in this paper.

\begin{itemize}

\item{\it Temperature field:} 

We define local temperature as the kinetic energy measured with respect the local center of mass reference frame. 
In other words, the local temperature at cell $(n,l)$ is:
\begin{equation}
T(n,l)=K(n,l) -\frac{1}{2}u(n,l)^2
\end{equation}
where $K(n,l)$ is the averaged one particle kinetic energy at box $(n,l)$:
\begin{equation}
K(n,l)=\frac{1}{2N(n,l)M}\sum_{t=1}^M\sum_{i:r_i(t)\in B(n,l)}v_i(t)^2 \quad,\quad N(n,l)=\frac{1}{M}\sum_{t=1}^M\sum_{i:r_i(t)\in B(n,l)}1\label{temp}
\end{equation}
and $u(n,l)$ is the modulus of the hydrodynamic velocity studied in the previous section. We see that the temperature definition is composed of two terms that they are independently measured. We consider that the error is the sum of the errors:
\begin{equation}
\epsilon(T)=\epsilon(K)+\vert u_1\vert \epsilon(u_1)+\vert u_2\vert \epsilon(u_2)
\end{equation}

\begin{figure}[h!]
\begin{center}
\includegraphics[height=6cm,clip]{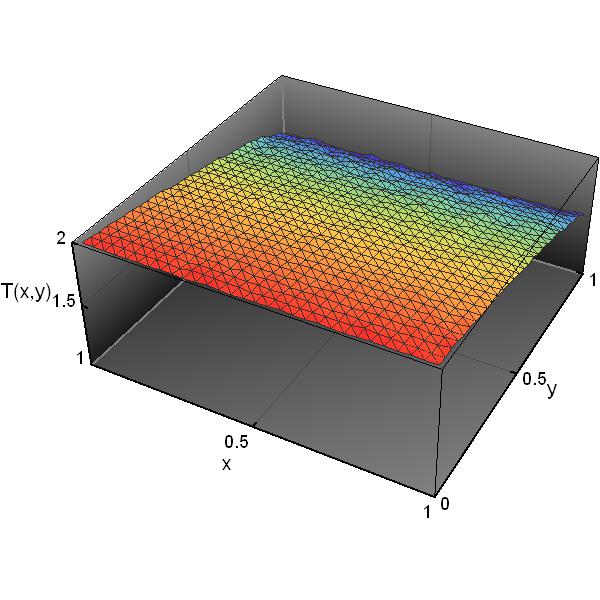}  %temp_profile_1.nb
\includegraphics[height=6cm,clip]{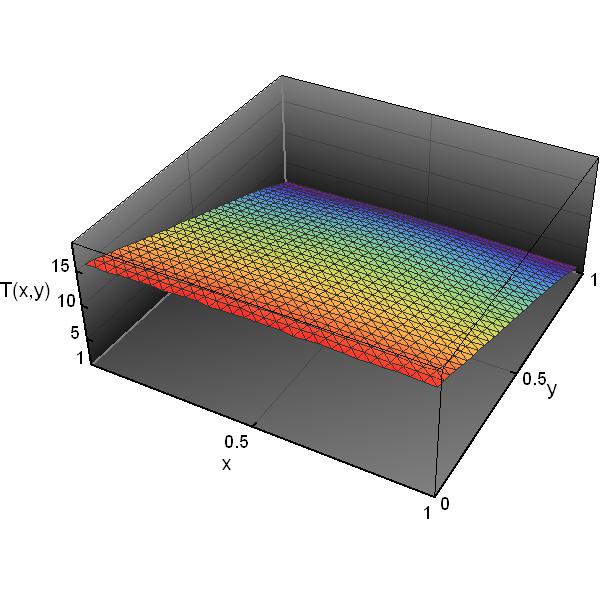}       %temp_profile.nb
\includegraphics[height=5cm,clip]{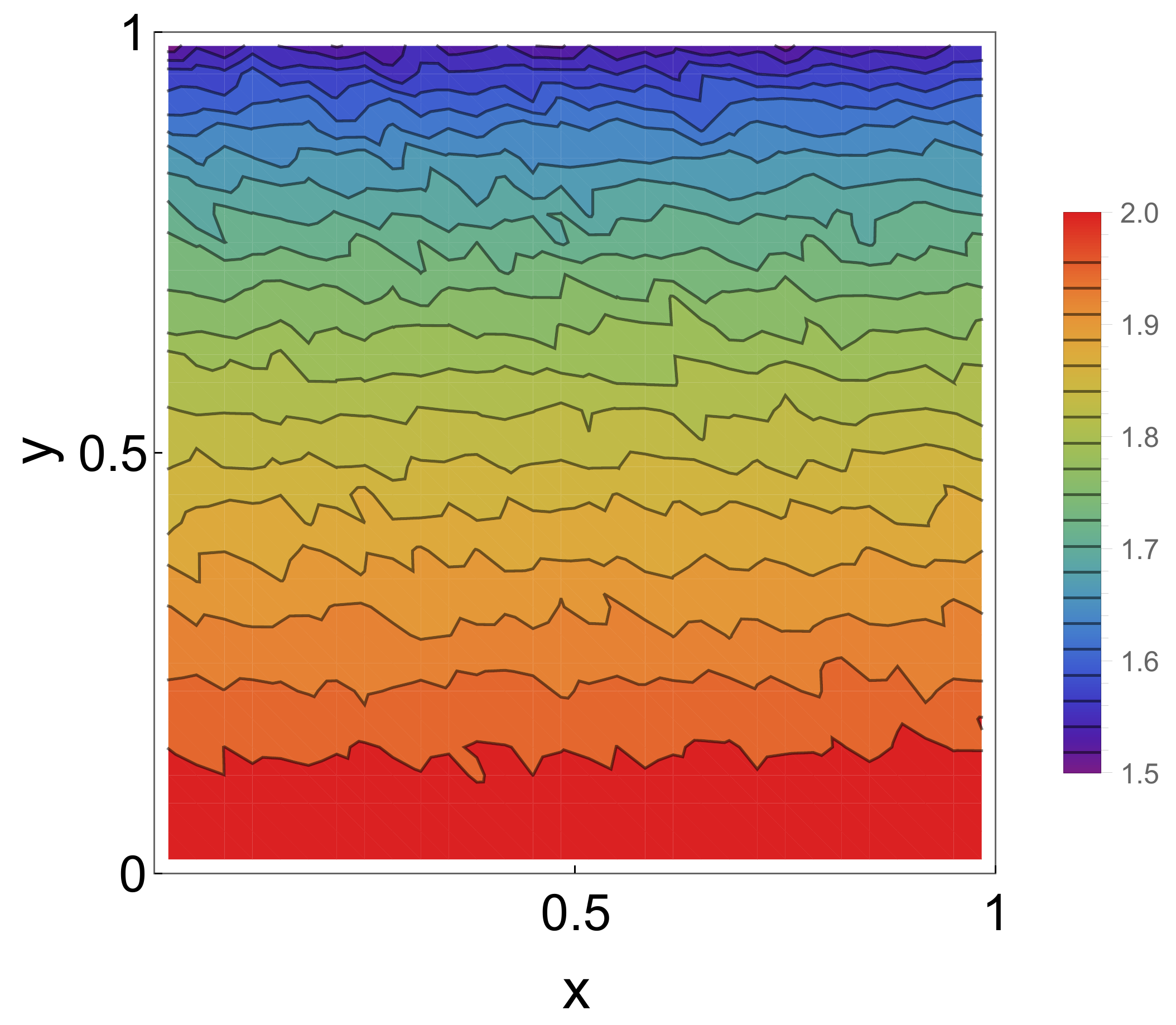}  %temp_profile_1.nb
\includegraphics[height=5cm,clip]{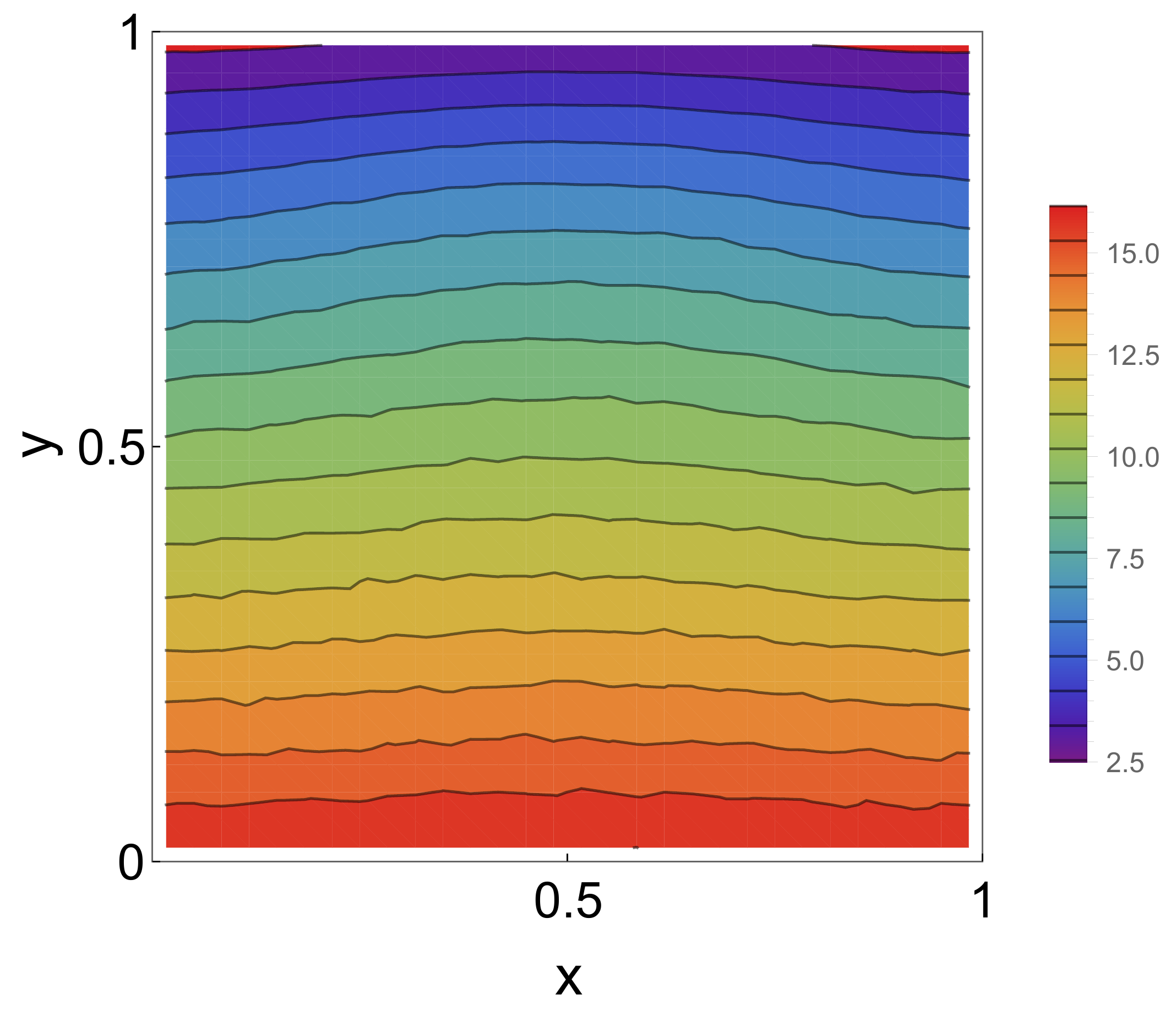}        %temp_profile.nb
\end{center}
\kern -0.5cm
\caption{Temperature field $T(x,y)$  defined in eq. \ref{temp} for $T_0=2$ and $T_0=18$ for left and right columns respectively ($g=10$). Below each of the 3D graphs there are  the corresponding contour plots to show the existence (or not) of a nontrivial spatial structure on $x$ direction. \label{temp0}}
\end{figure}

\begin{figure}[h!]
\begin{center}
\includegraphics[height=4cm,clip]{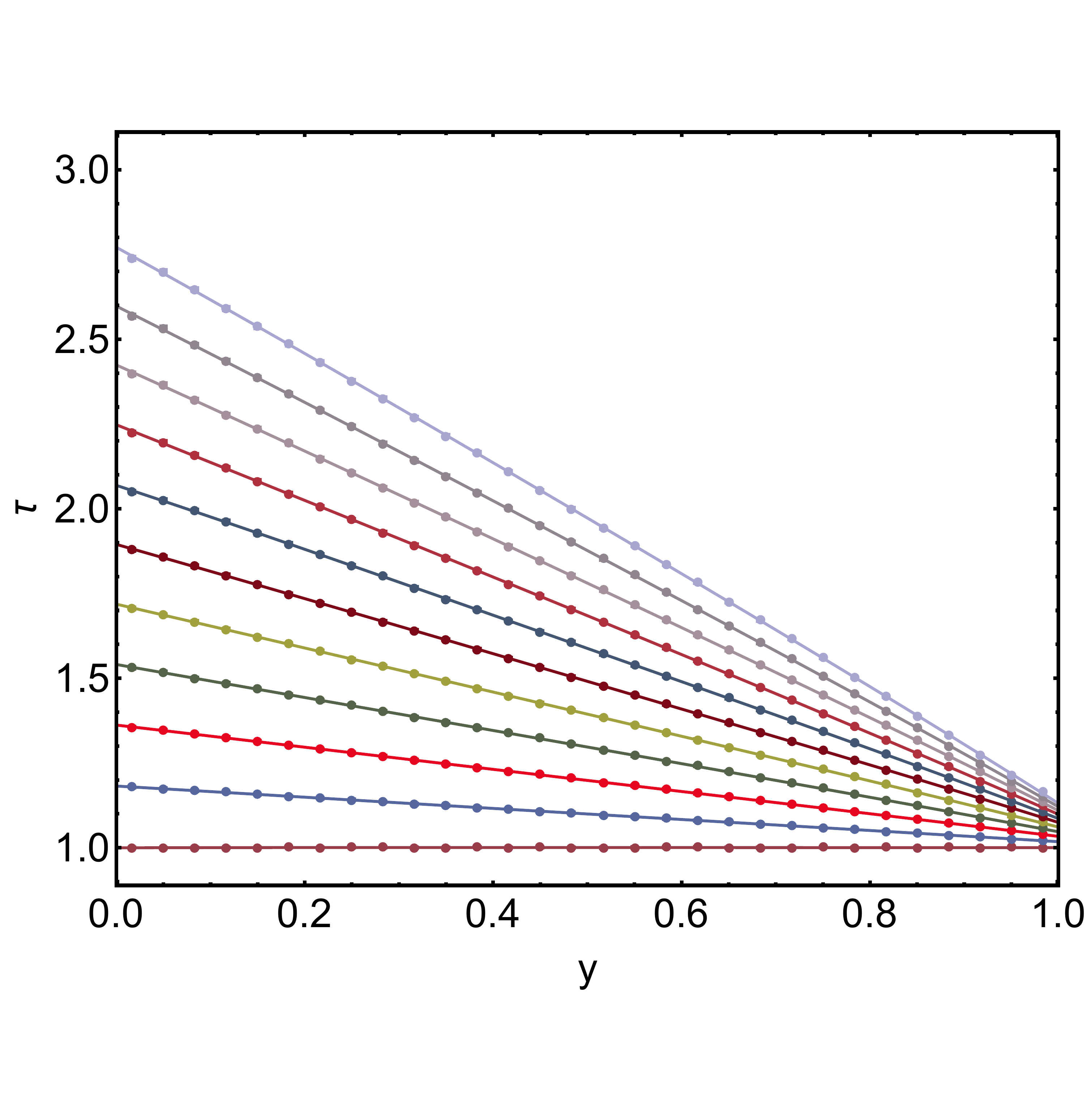}        %tem_profile_y_E0.nb
\includegraphics[height=4cm,clip]{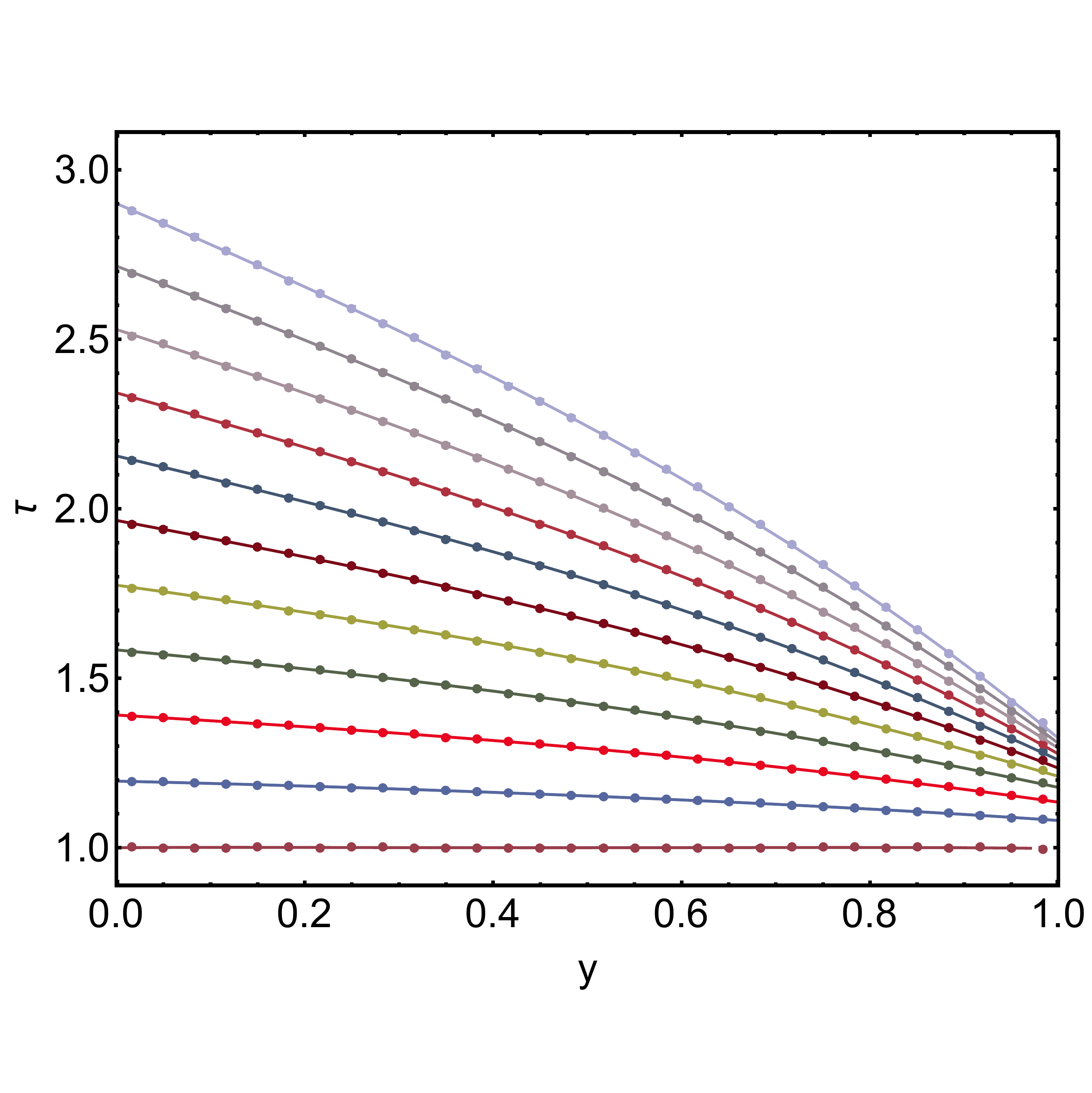}        %tem_profile_y_E5.nb
\includegraphics[height=4cm,clip]{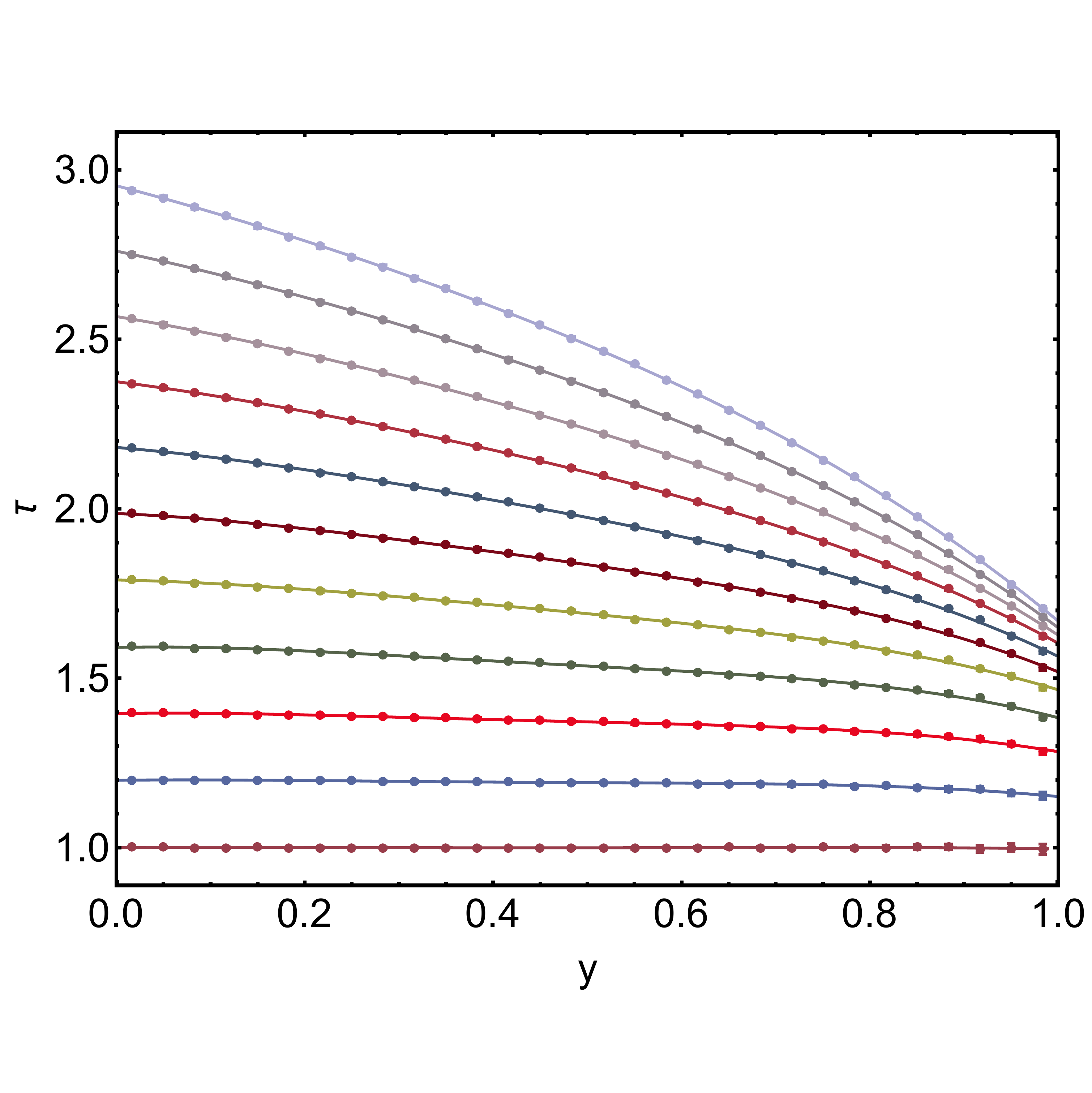}       %tem_profile_y_E10.nb
\includegraphics[height=4cm,clip]{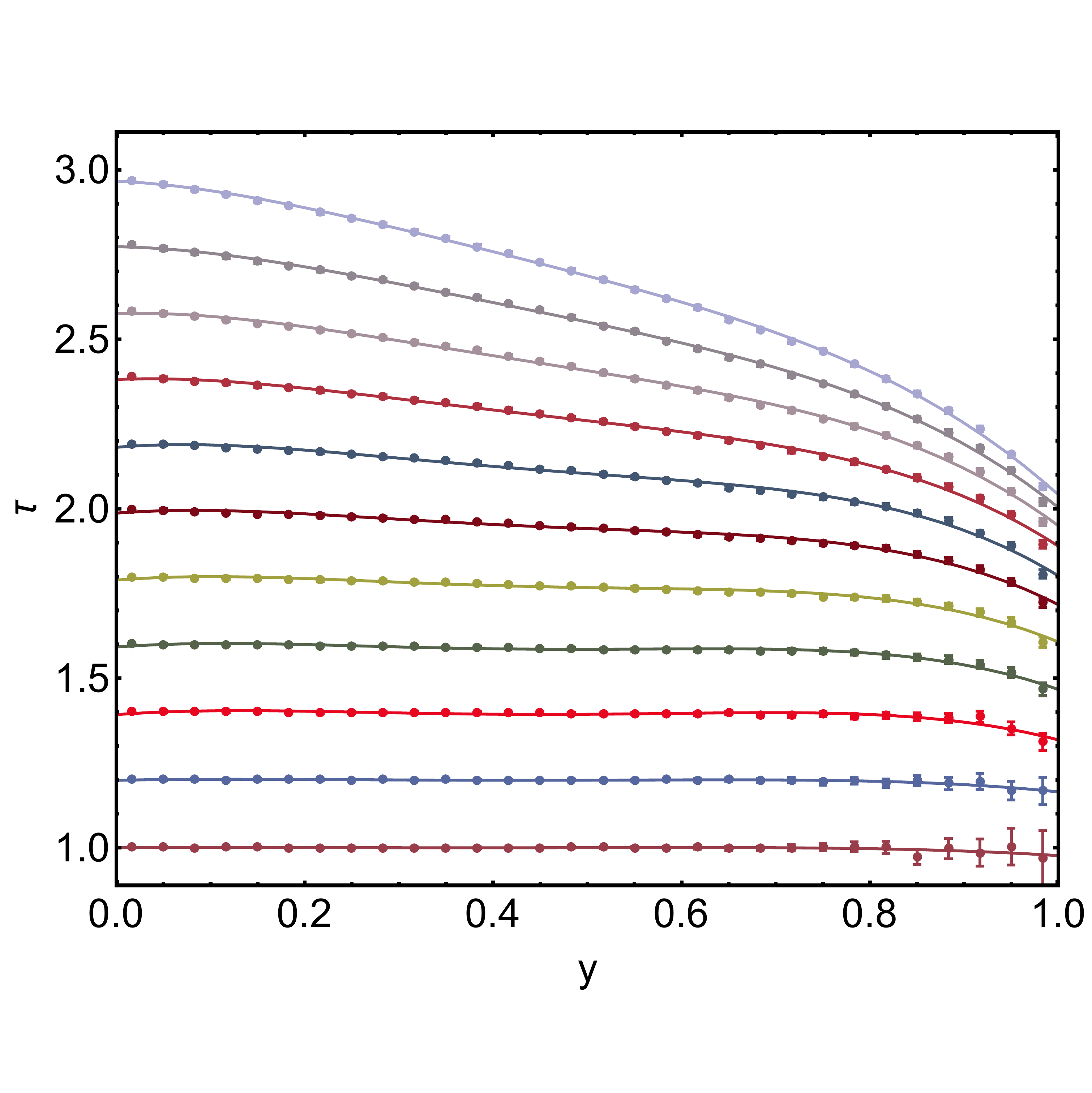}       %tem_profile_y_E15.nb
\includegraphics[height=4cm,clip]{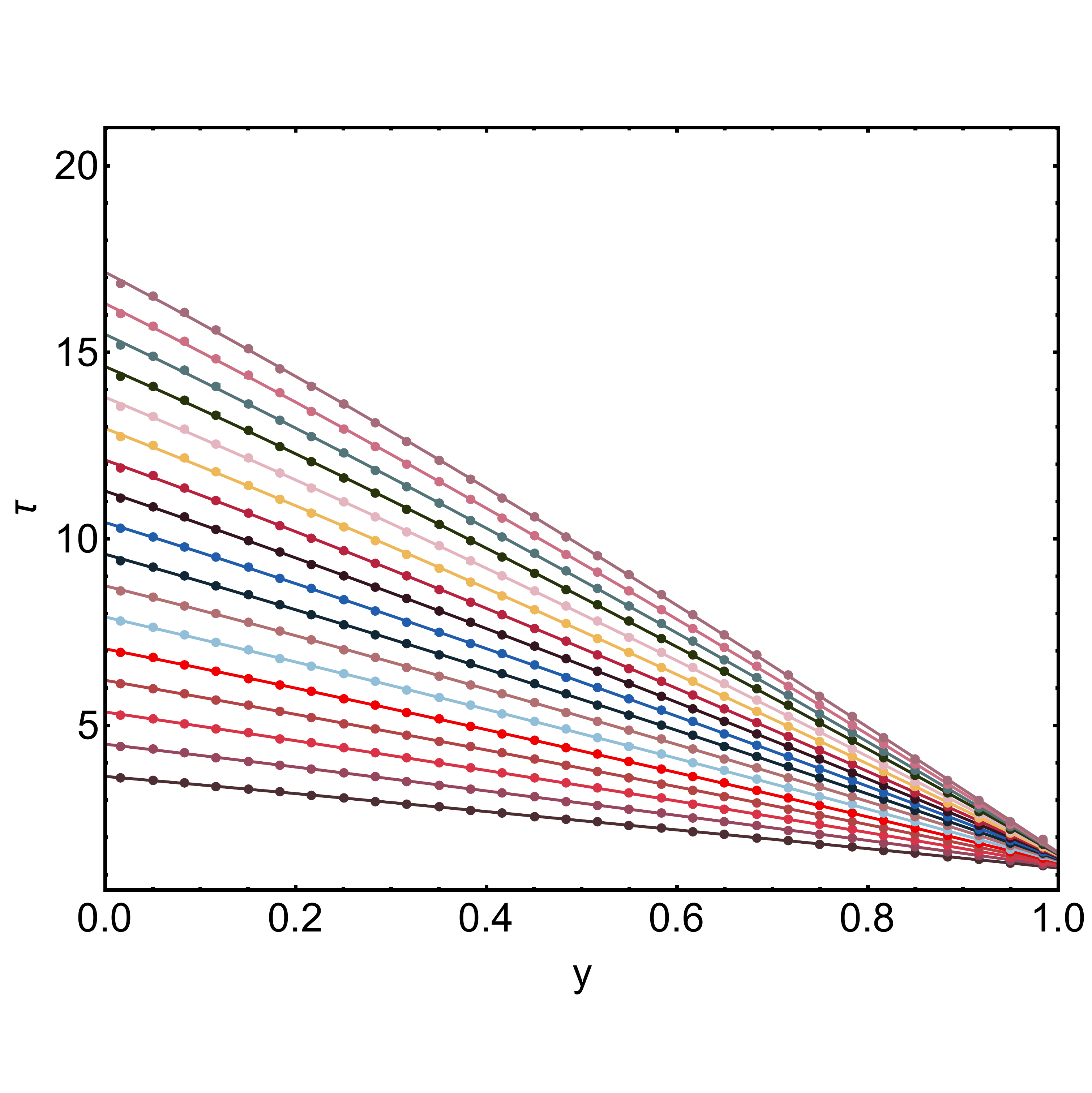}        %tem_profile_y_E0.nb
\includegraphics[height=4cm,clip]{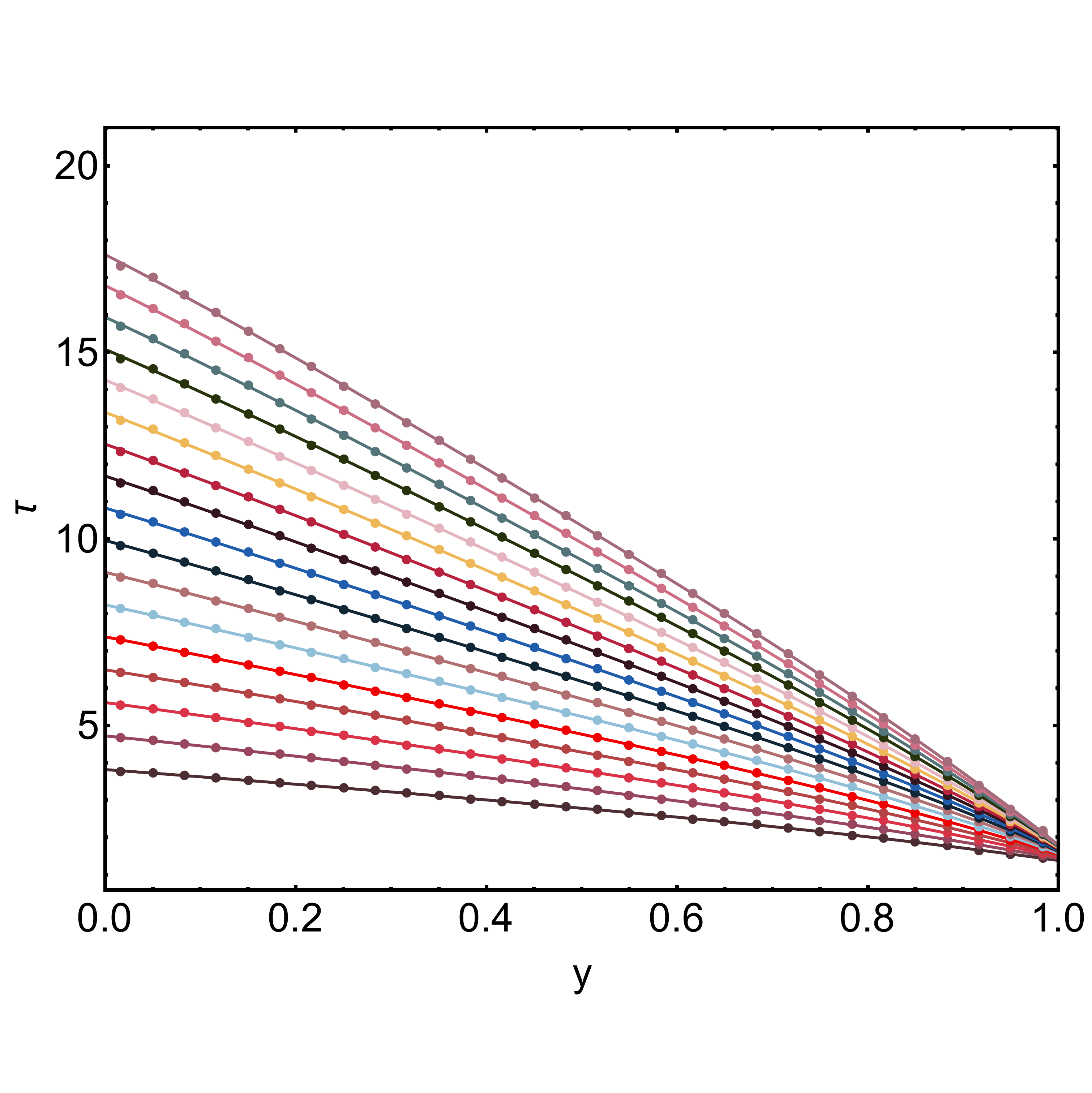}        %tem_profile_y_E5.nb
\includegraphics[height=4cm,clip]{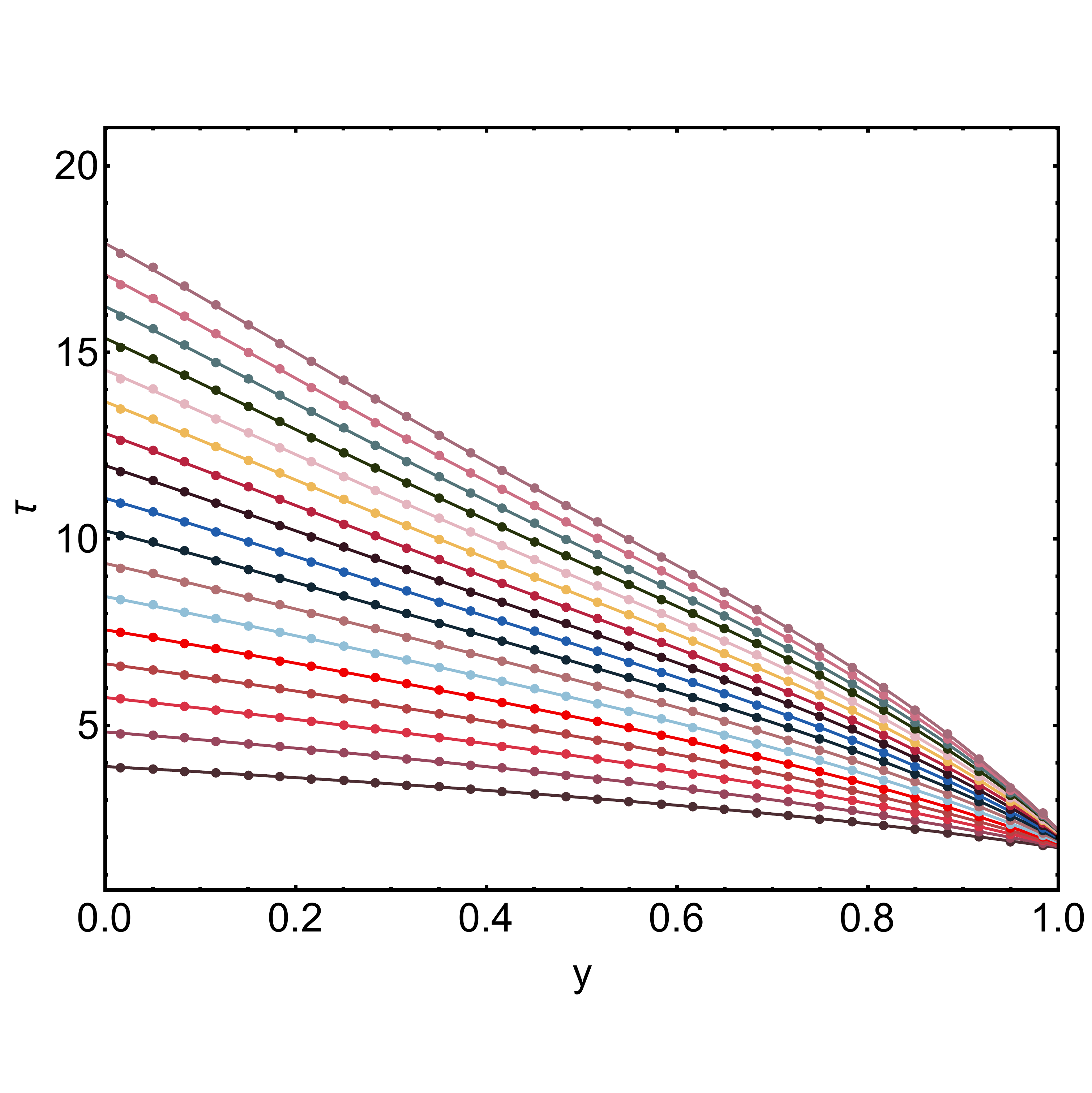}       %tem_profile_y_E10.nb
\includegraphics[height=4cm,clip]{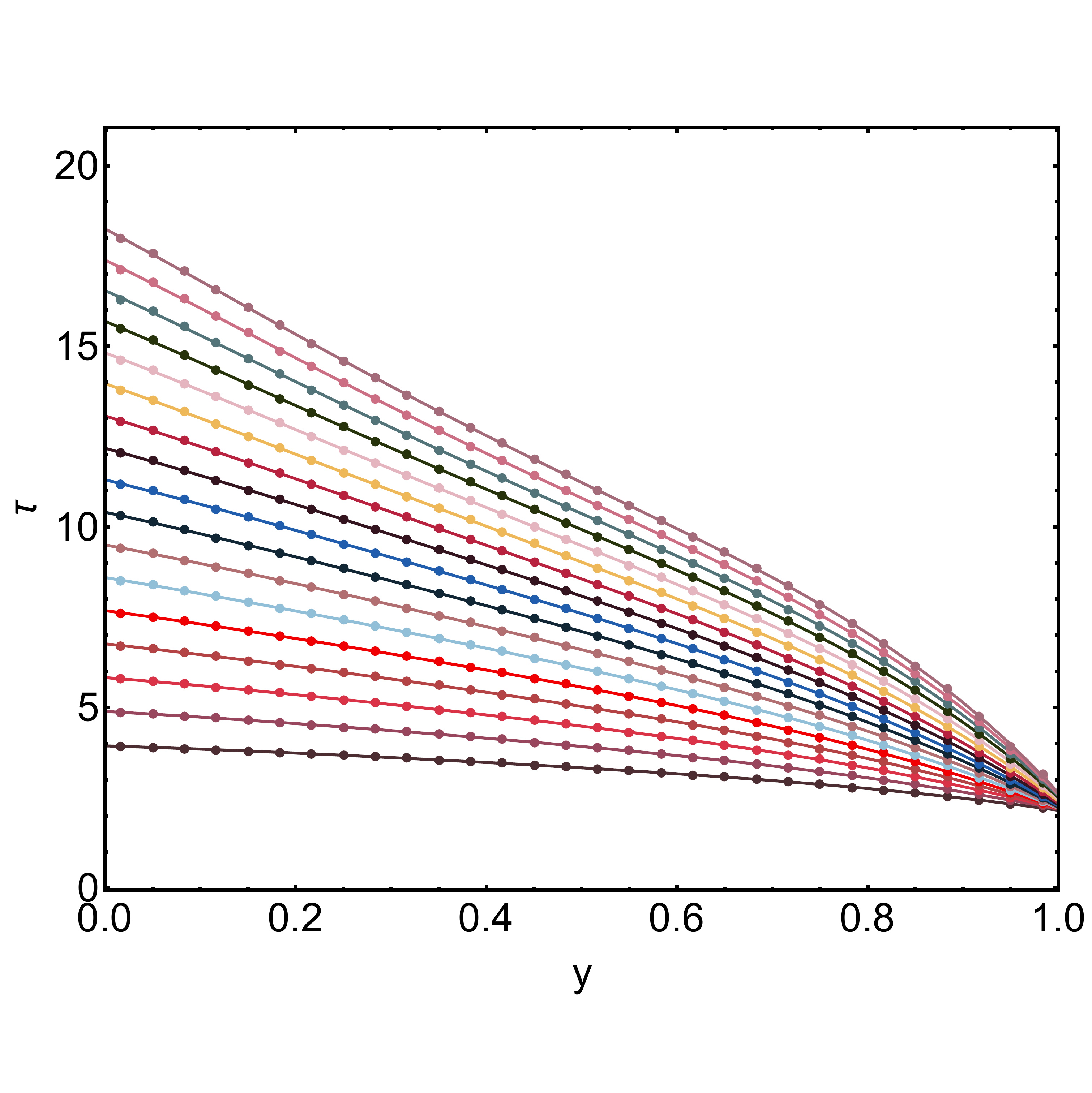}        %tem_profile_y_E15.nb
\end{center}
\kern -0.5cm
\caption{Averaged temperature profiles $\tau(y)$  defined in eq. \ref{temp2} for $g=0$, $5$, $10$ and $15$ from left to right columns respectively. Upper figures: $T_0=1, 1.2, 1.4, 1.6, 1.8, 2.0, 2.2, 2.4, 2.6, 2.8, 3.0$ from bottom to top curves. Bottom figures: $T_0=4, 5, 6, 7, 8, 9, 10, 11, 12, 13, 14, 15, 16, 17, 18, 19, 20$ from bottom to top curves. Error bars are included. Lines are phenomenological fits (see main text) \label{temp1}}
\end{figure}
\begin{figure}[h!]
\begin{center}
\includegraphics[height=4cm,clip]{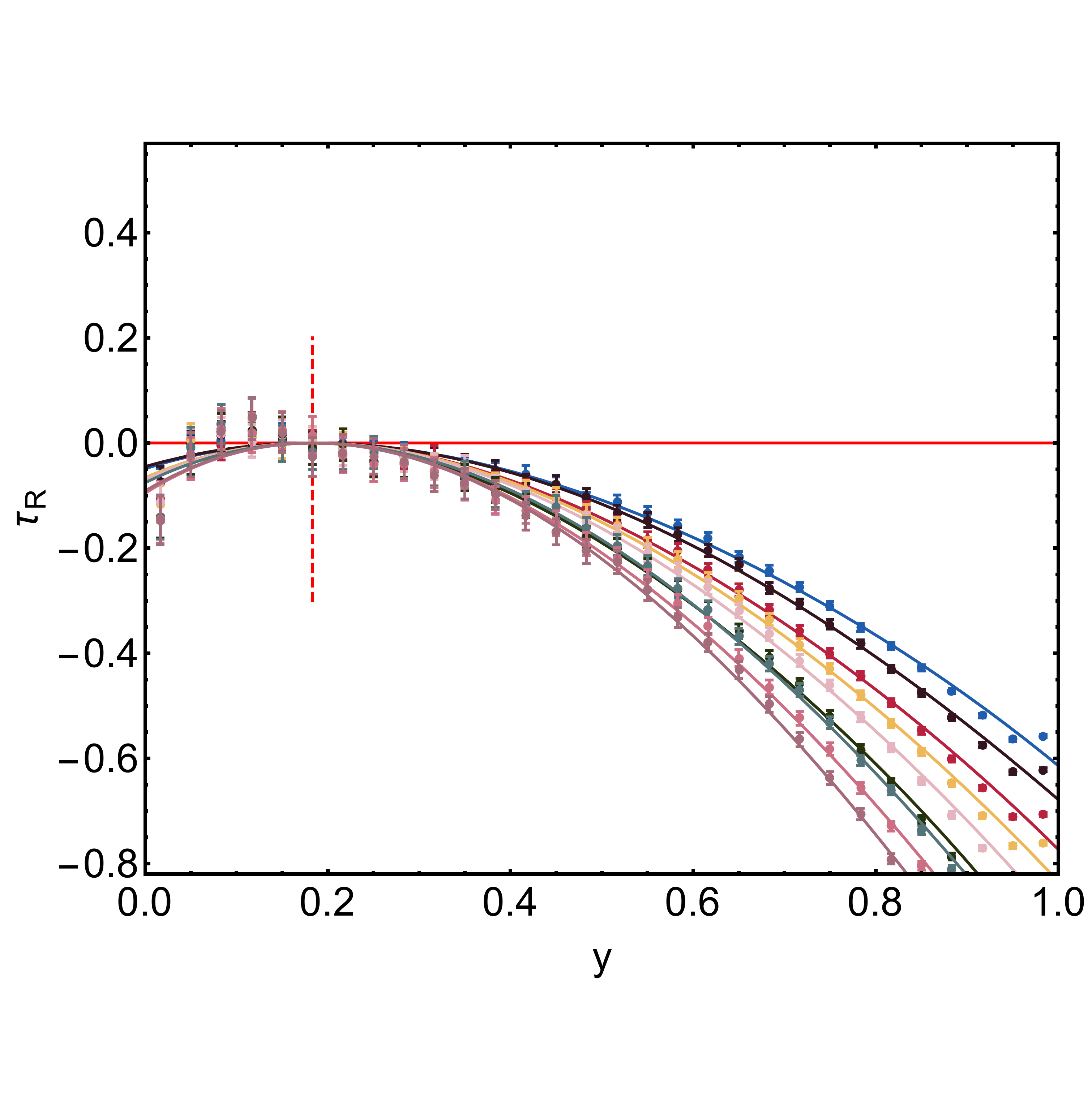}     %tem_profile_y_E0.nb
\includegraphics[height=4cm,clip]{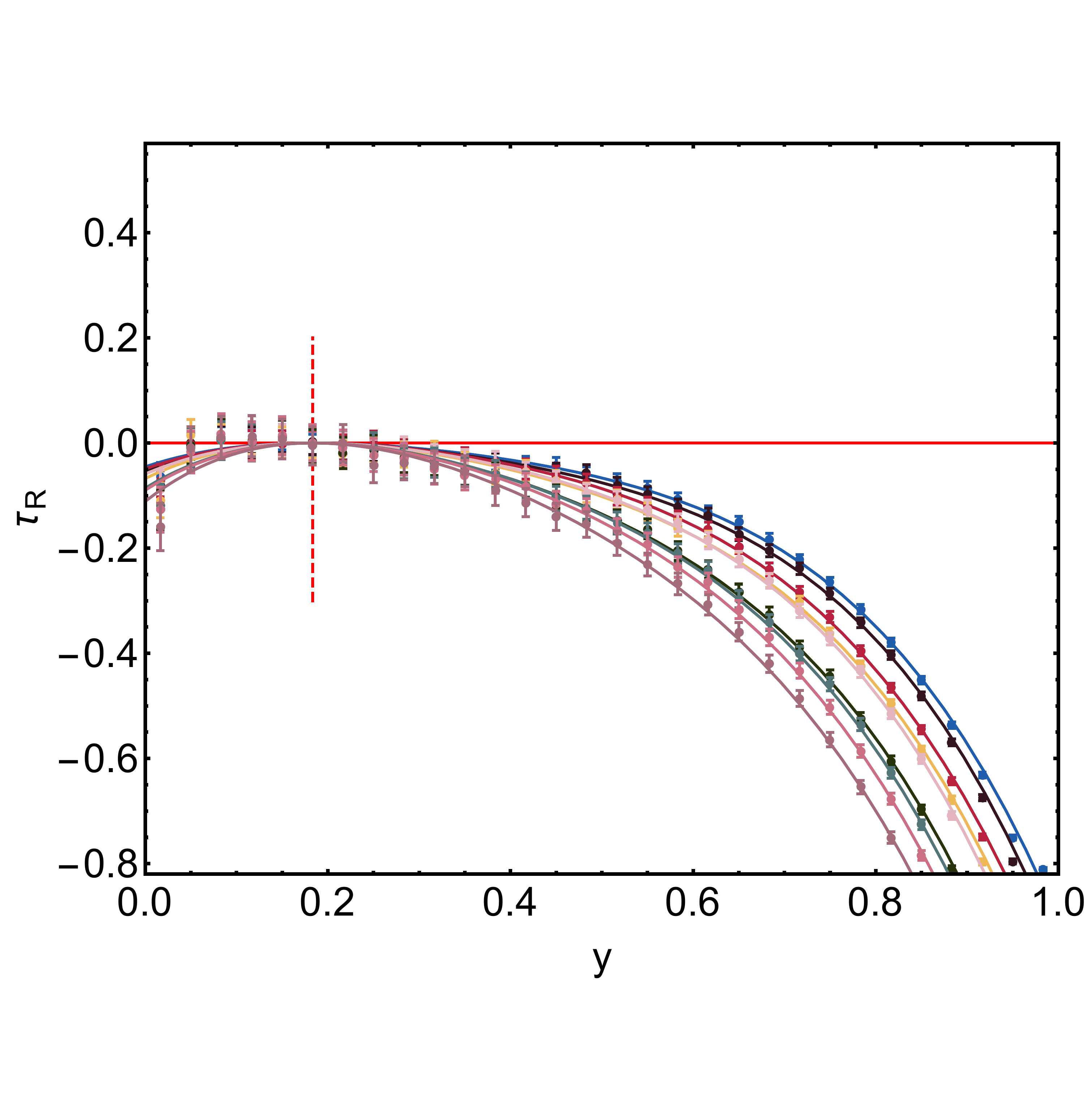}     %tem_profile_y_E5.nb
\includegraphics[height=4cm,clip]{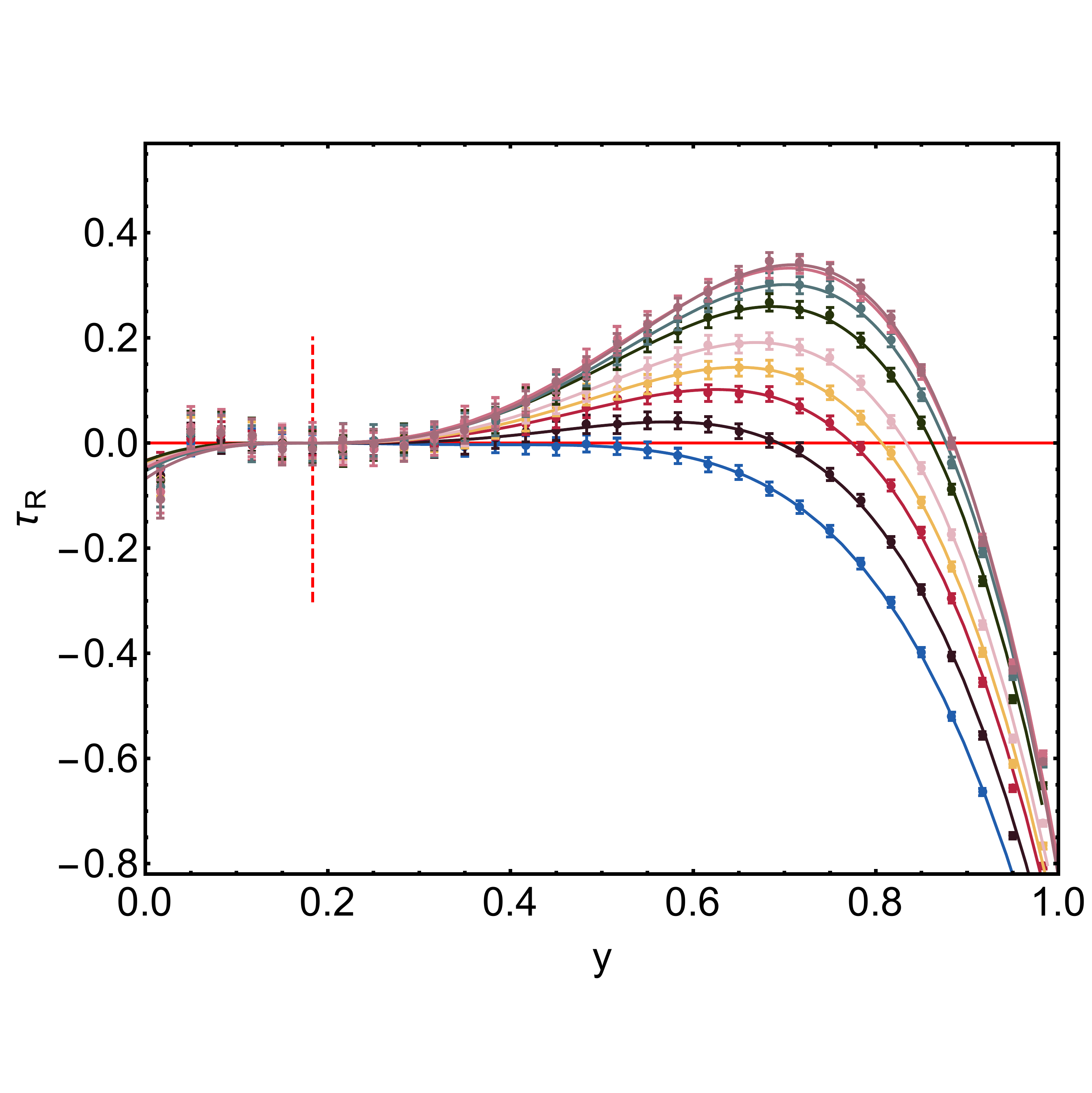}       %tem_profile_y_E10.nb
\includegraphics[height=4cm,clip]{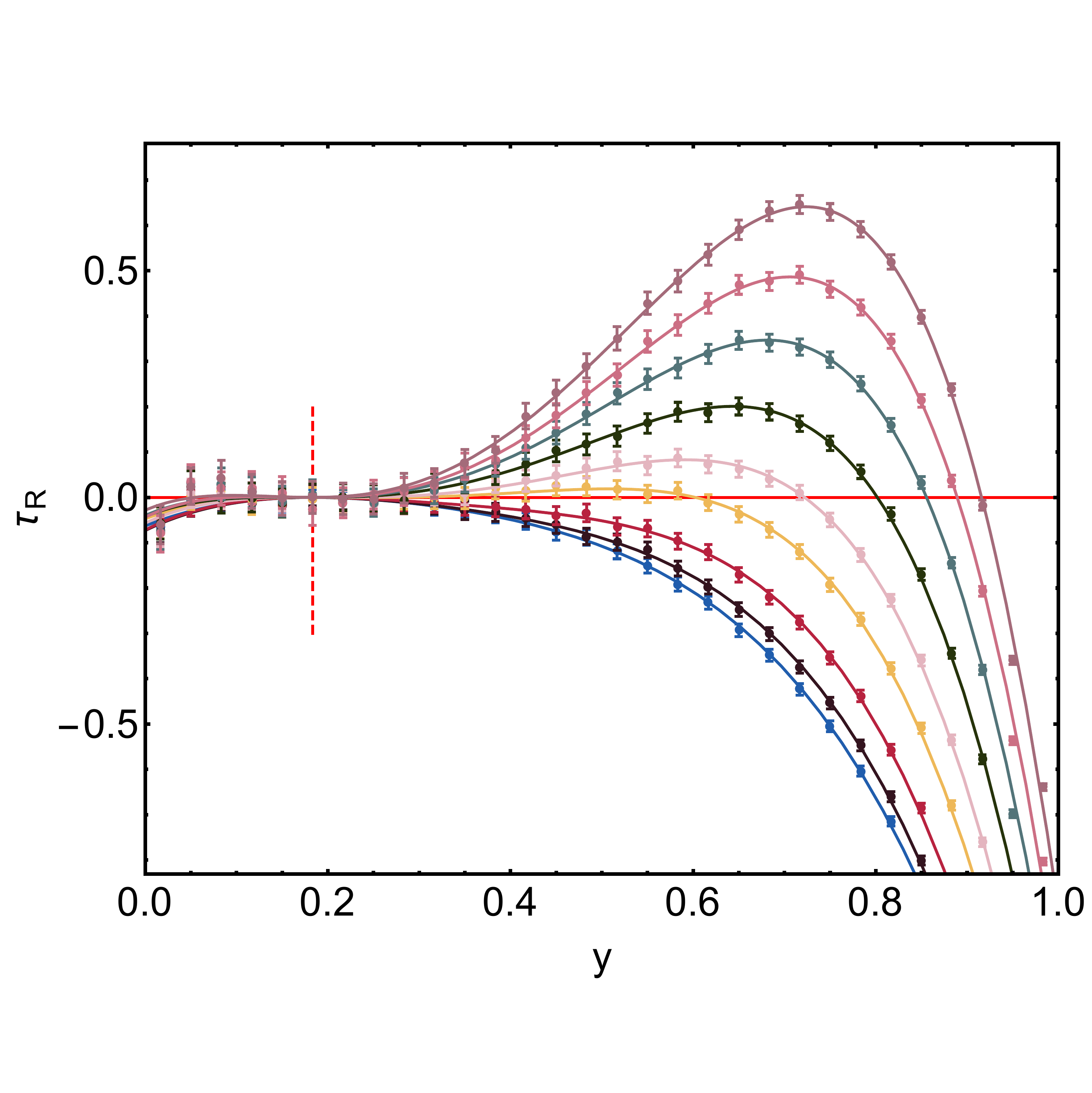}       %tem_profile_y_E15.nb
\end{center}
\kern -0.5cm
\caption{$\tau_R=\tau(y)-\tau(0.2)-\tau'(0.2)(y-0.2)$  for $g=0$, $5$, $10$ and $15$ from left to right respectively . $T_0=13, 14, 15, 16, 17, 18, 19, 20$ from top to bottom curves in left figure and from bottom to top in the others. Error bars are included. Dots are data points and lines are the fitted polynomials minus the tangent at $y=0.11/60\simeq 0.183$ (see main text) \label{temp2}}
\end{figure}
We show in figure \ref{temp0} two typical cases: one without convection ($T_0=2$ and $g=10$) and the other with convection ($T_0=18$ and $g=10$). Observe that both figures present some general features: (1) The profile is nonlinear on the vertical direction $y$, (2) there is no spatial structure on the $x$ direction when we are at a non-convective state and there is a small but systematic $x$ non-constant profile when the system is in a convective state and (3) there is a  thermal gap on the boundaries, that is, the extrapolated profile for $y=0$ and $y=1$ doesn't match the thermal bath temperatures.

In order to characterize the temperature profile we first study the $x$-averaged profiles:
\begin{equation}
\tau(y)=\frac{1}{30}\sum_{n=1}^{30} T(n,l)\quad ,\quad y=\frac{(l-1)}{30}+\frac{1}{60}\label{temp21}
\end{equation}

We see in figure \ref{temp1} the averaged temperature profiles $\tau$. We manage to fit all the data points with a very simple fourth order polynomial that permits us to refine the analysis. First, we observe for $g=10$ and $T_0>12$ that the profile has a small change of curvature at the middle of the profile. The same happens for $g=15$ and $T_0>15$. In order to characterize these changes, we subtract to the data the tangent to the profile at $y=11/60$ (which is the coordinate of the 6th cell) which is computed using the fitted curve. If  the difference is positive means that the slope grows (is negative) and the curve is concave. In contrast, if the difference is negative the curve is convex.
We plot such differences on figures \ref{temp2}. We can see that for $g=0$ and $5$ and any value of $T_0$ the profile is convex. Moreover, the convexity grows with $T_0$. In contrast, for  $g=10$  and $T_0>12$ the $\tau(y)$ profile has a changing convexity: it is almost flat near the hot thermal bath and it is concave for a while and again convex near the cold thermal bath. Also, this changing behavior is reinforced for large values of $T_0$. The same happens for $g=15$ with a concave behavior stronger than $g=10$ but appearing for $T_0>15$. From the phenomenological fit we can get the inflection point by doing the second derivative and looking the zeros of the resulting polynomial. In figure \ref{temp3} it is shown the inflection points $x^*$ as a function of $T_0$ for $g=10$ and $15$. They grow with $T_0$ but all of them are around $x\simeq 0.5$, far from the boundaries. All of this indicate a real strong nonlinear behavior of the $y$-temperature profile. The increment of $g$ emphasizes the nonlinear character of $\tau(y)$. 
\begin{figure}[h!]
\begin{center}
\includegraphics[height=6cm,clip]{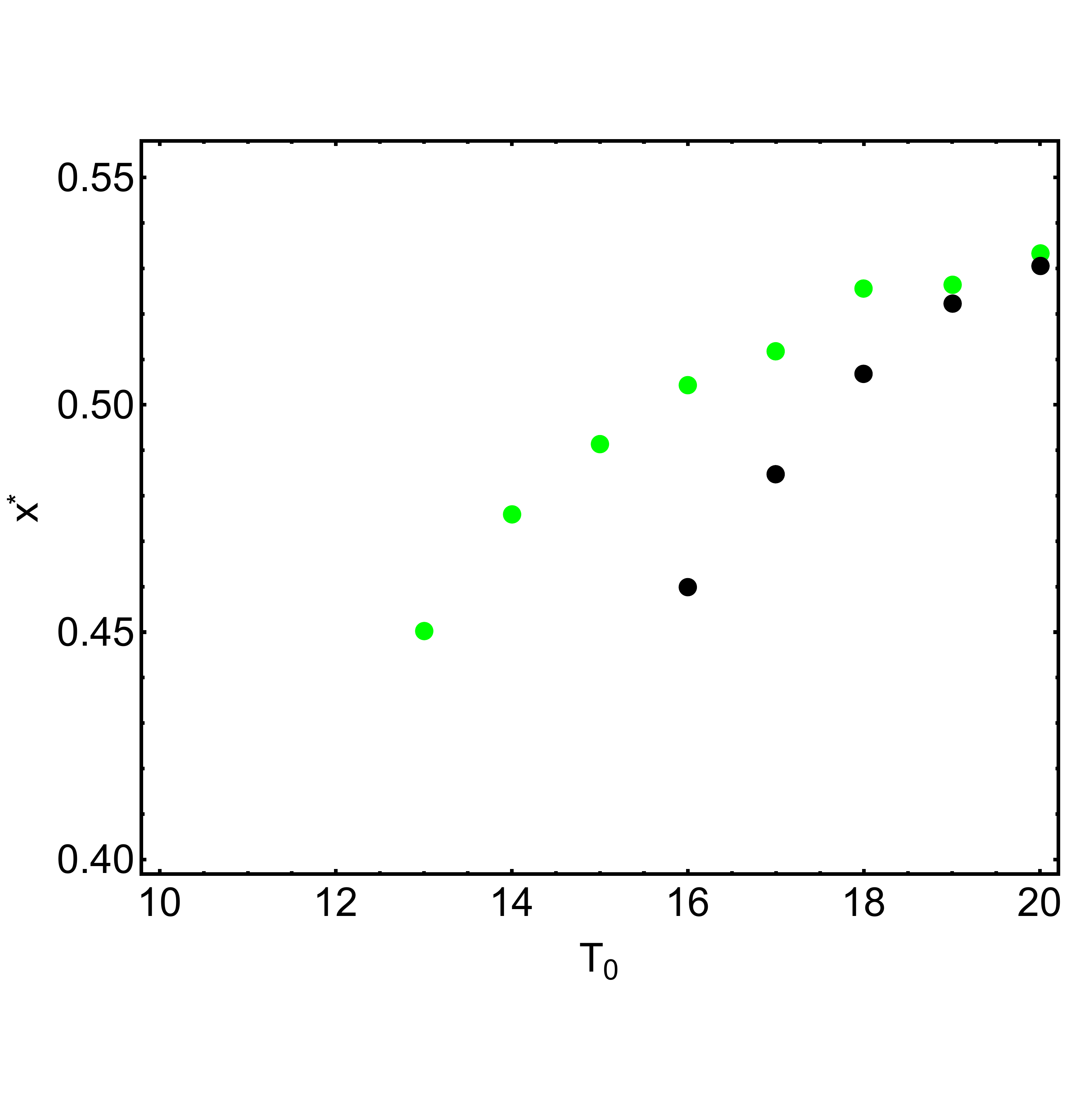}       %tem_profile_y_E15.nb
\end{center}
\kern -1.cm
\caption{Inflection points of $\tau(y)$ for $g=10$ and $15$ (green and black dots respectively)  obtained by doing the second derivative of the fits to the data and the looking for its roots.   \label{temp3}}
\end{figure}
\begin{figure}[h!]
\begin{center}
\includegraphics[height=4cm,clip]{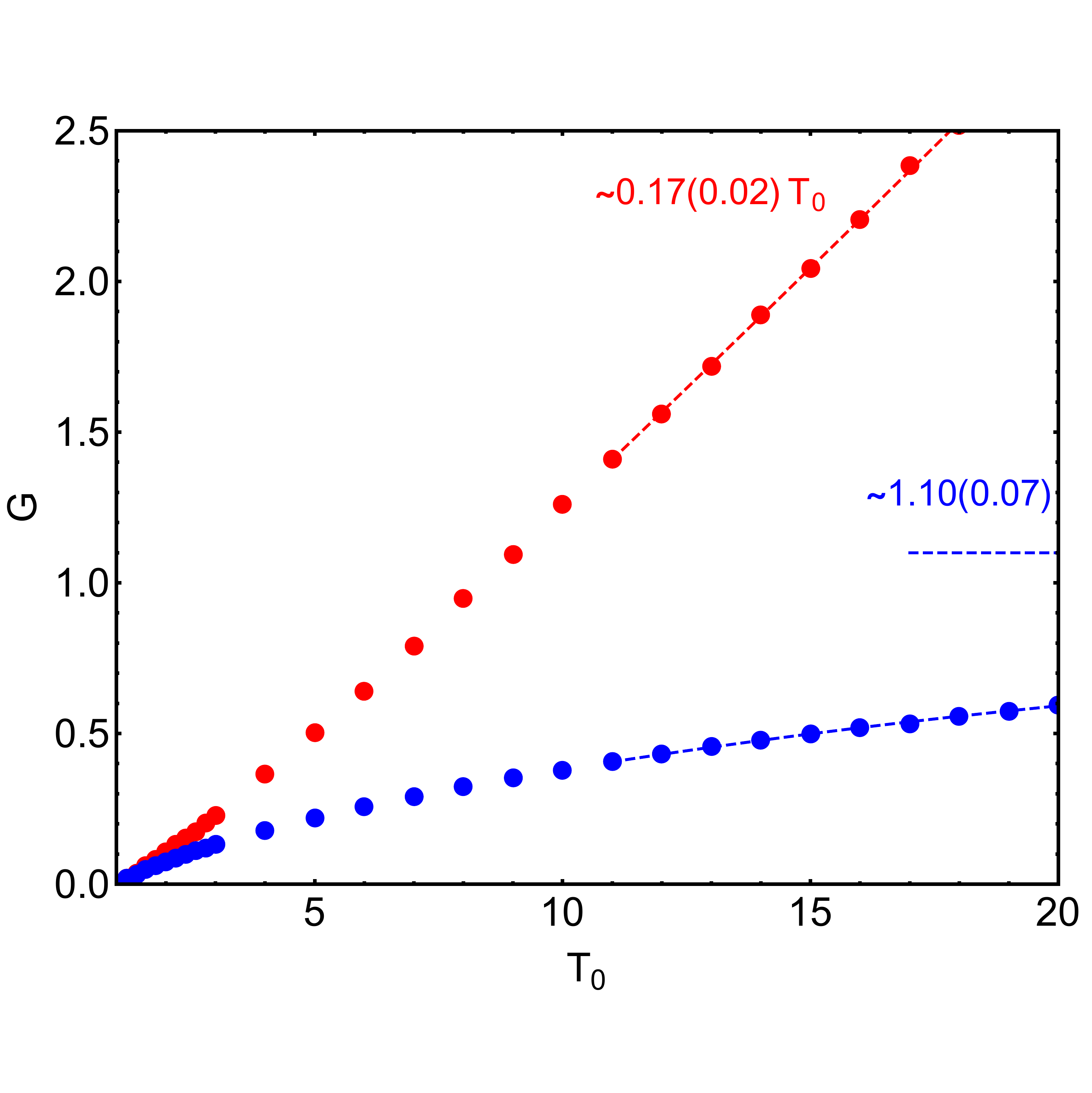}  %tem_profile_y_E0.nb
\includegraphics[height=4cm,clip]{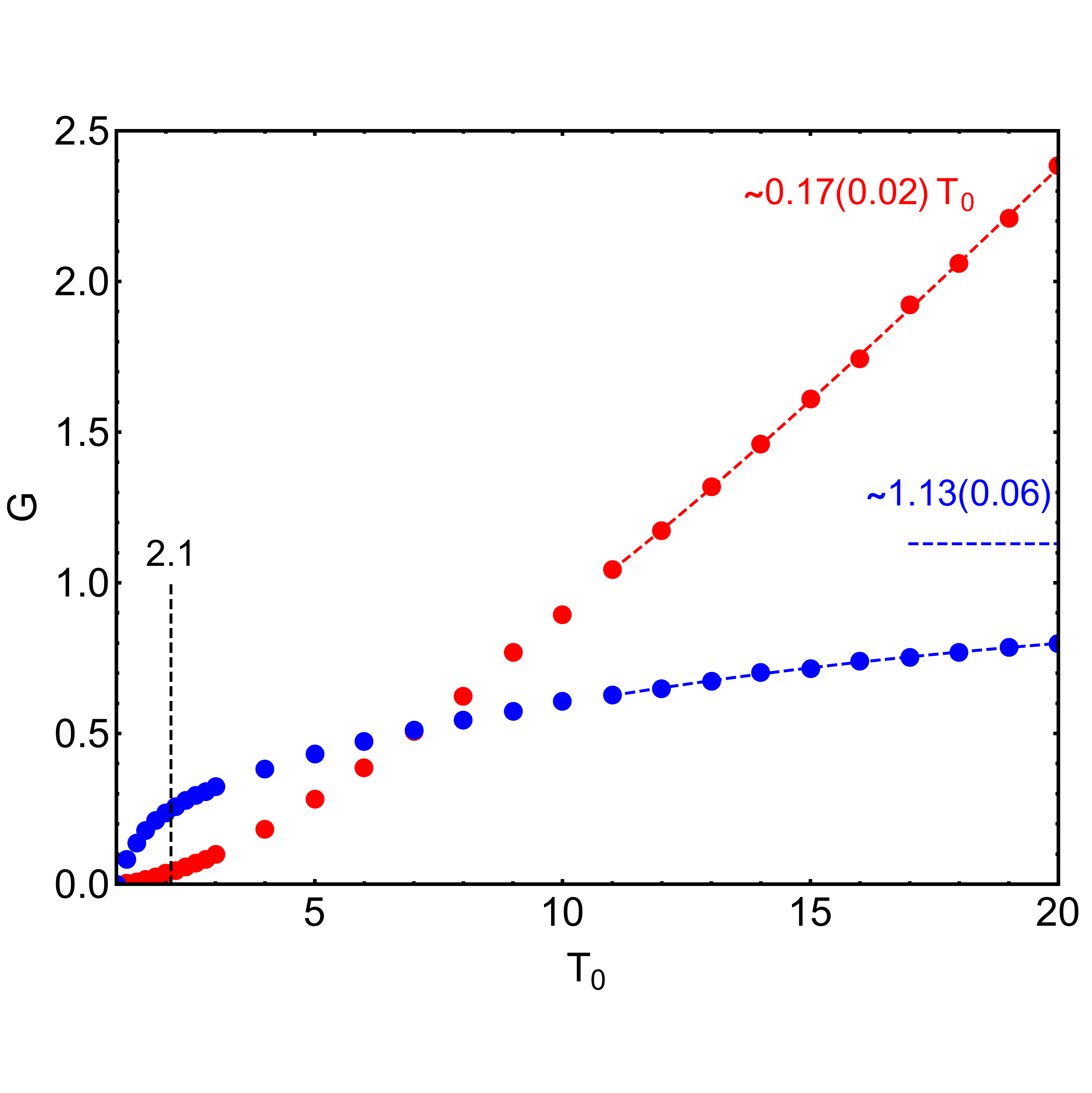}  %tem_profile_y_E5.nb
\includegraphics[height=4cm,clip]{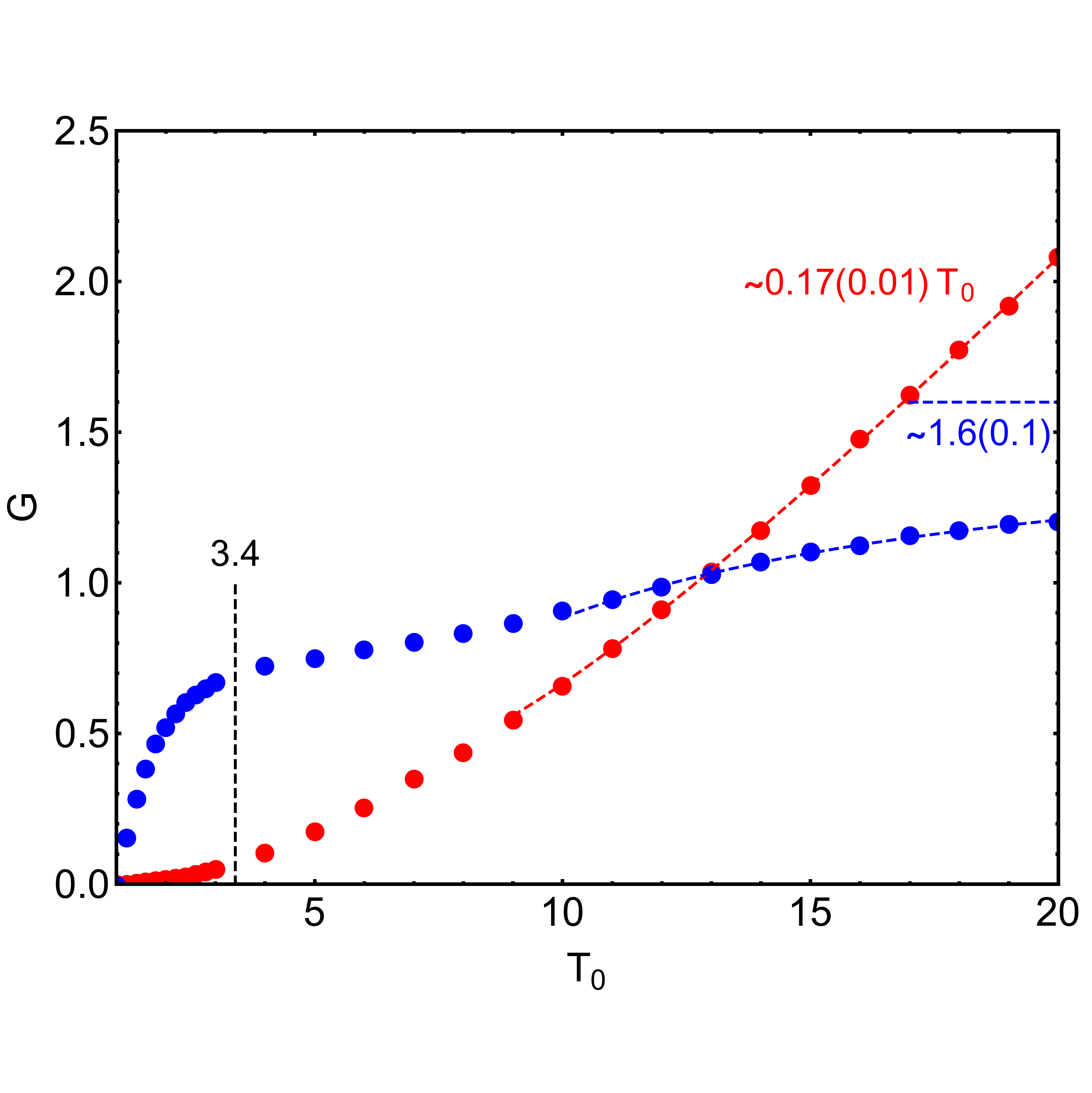}  %tem_profile_y_E10.nb
\includegraphics[height=4cm,clip]{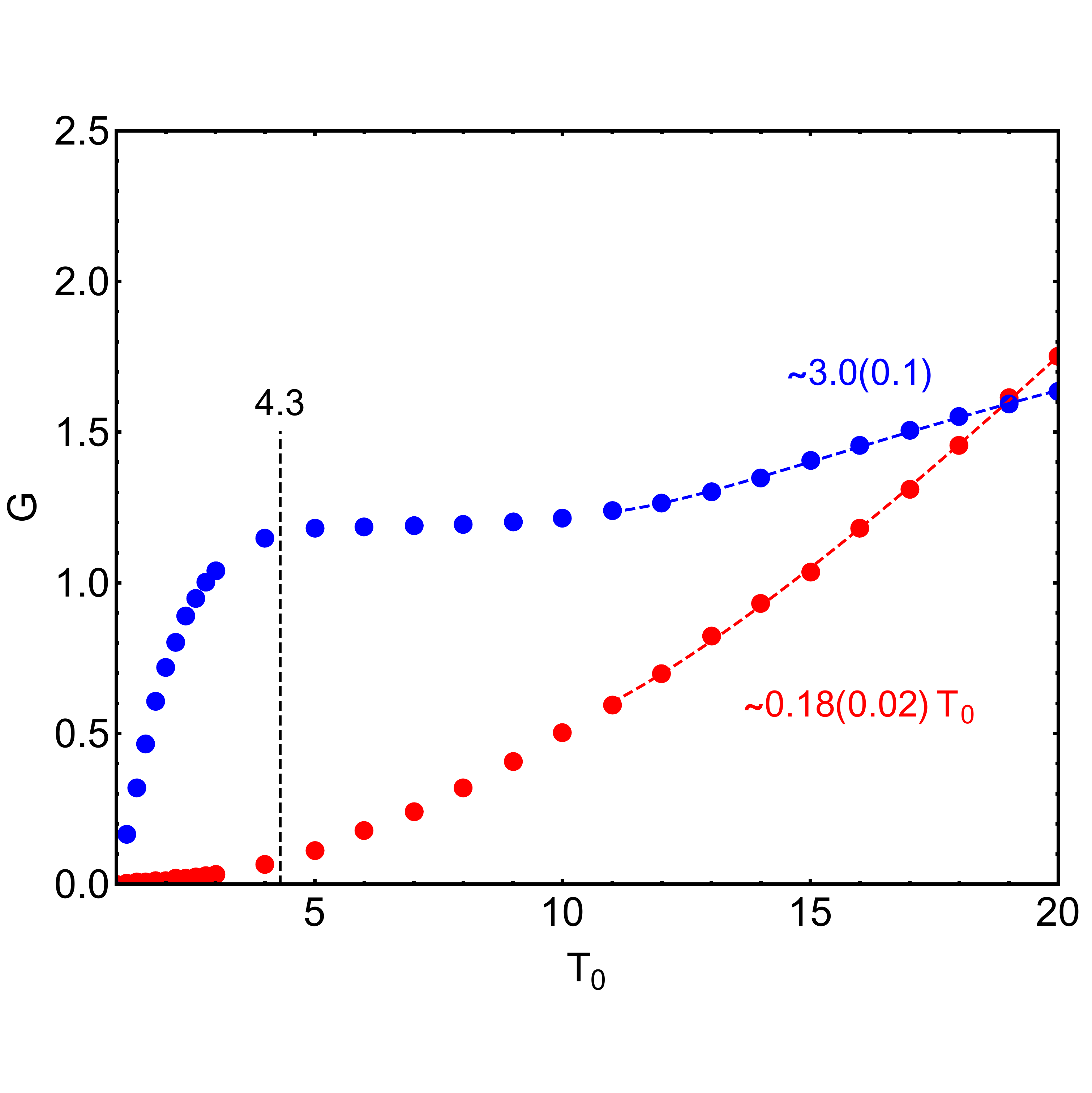}  %tem_profile_y_E15.nb
\end{center}
\kern -0.5cm
\caption{Measured thermal gap, $G$ as a function of $T_0$ for the hot reservoir (red dots) and cold reservoir (blue dots) for $g=0$, $5$,  $10$  and $15$ from left to right respectively. Dotted lines  are fits to the data points they cover to predict large $T_0$ behaviors. The Gaps for the cold heat reservoir tend to a limiting value and the ones for the hot reservoirs grow linearly with $T_0$ \label{temp4}}
\end{figure}
\begin{figure}[h!]
\begin{center}
\includegraphics[height=6cm,clip]{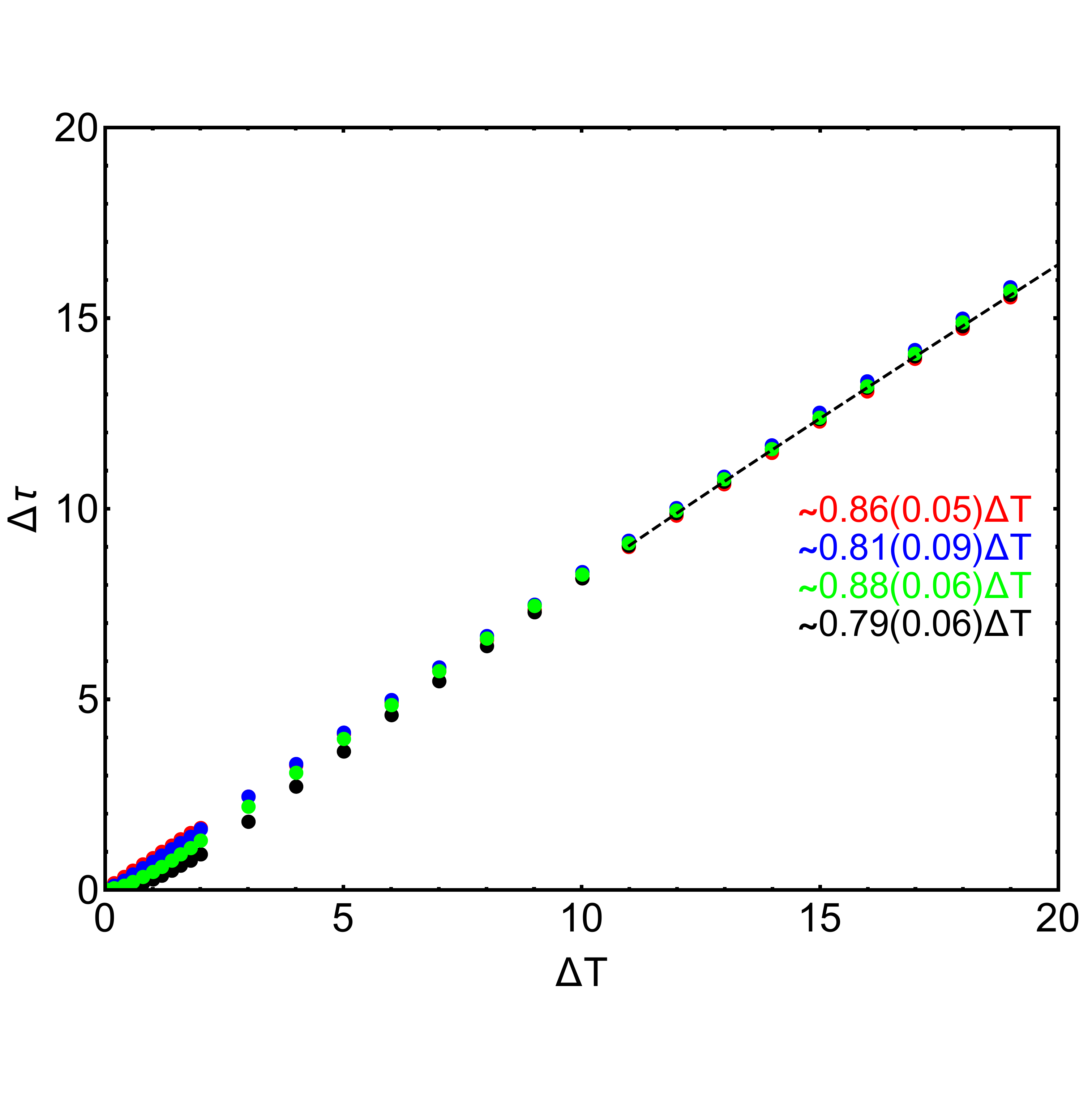}  %tem_profile_y_E15.nb
\end{center}
\kern -1.cm
\caption{Measured effective temperature gradient, $\Delta\tau$ as a function of $\Delta T=T_0-1$  for $g=0$ (red dots), $5$ (blue dots) , $10$ (green dots) and $15$ (black dots). Dotted lines  are fits to the data points  to predict large $T_0$ behaviors. \label{temp5}}
\end{figure}

Finally we have studied the behavior of the thermal gaps at the boundaries. We see that the profiles do not extend to the thermal bath values, that is, $\tau(y=0)\neq T_0$ and $\tau(y=1)\neq 1$. This is a well known phenomena called {\it thermal resistance} that appear when we there are currents crossing a singular boundary that, in our case is due to the use of thermal baths that act to the particles when they hit the walls. Typically it should disappear at the thermodynamic limit. We define the thermal gap as
\begin{equation}
G=\vert \tau(y=0,1)-T_{0,1}\vert
\end{equation}
In figure \ref{temp4} we show how the gaps $G$ behave with $T_0$ for $g=0$, $5$, $10$ and $15$.  The thermal gap for the hot reservoir grows monotonically with $T_0$. For large values of $T_0$ it seems to growth linearly. In this region, we have done fits to the data of the form $G=a_0T_0+a_1+a_2/T_0+a_3/T_0^2+\ldots$ and we got that  $G\simeq 0.17(0.02)T_0$, $0.17(0.01)T_0$, $0.17(0.01)T_0$ and $0.18(0.02)T_0$ for $g=0$, $5$, $10$ and $15$ respectively. We cannot discard a $g$-independent behavior of $G\simeq 0.17T_0$. By other hand, the thermal gap behavior for the cold reservoir has a richer structure (maybe due to the fits used and/or the extrapolated data goodness). For $g=0$ and $5$ it seems to monotonically grow up to a limiting value which is computed by using a fit of the form $G=a_1+a_2/T_0+a_3/T_0^2+\ldots$ to the $T_0>9$ points: $a_1=1.10(0.07)$ and $a_1=1.13(0.06)$ for $g=0$ and $5$ respectively. For $g=10$ and $15$ the gap grows fast up to the critical temperature $T_c$, then stays more or less constant and it smoothly grows again up to a limiting value of $a_1=1.6(0.1)$ and $3.0(0.1)$ for $g=10$ and $15$ respectively. We cannot clearly relate the constant region with the $T_{c,2}$ critical temperature defined above. 

However, independently of the different cold and hot gap behaviors, the effective thermal gradient, $\Delta\tau=\tau(y=0)-\tau(y=1)$, present a very smooth behavior as a function of the imposed external gradient, $\Delta T=T_0-1$. We see in figure \ref{temp5} its behavior with $T_0$ and for the three $g$-values. Except for a small region near the equilibrium, the effective gradient is proportional to $\Delta T$: $\Delta\tau\simeq b\Delta T$, with $b=0.86(0.05)$, $0.81(0.09)$, $0.88(0.06)$ and $0.79(0.06)$ for $g=0$, $5$, $10$ and $15$ respectively. We cannot discard the existence of a common value for all $g$'s. It looks like the system re adapts itself to such effective gradient once we fix the values of the external thermal baths. This could be a characteristic of hard disk systems where the properties of the system are invariant under a global rescale of the bath temperatures. That's why we can fix $T_1=1$ without any loss of generality.

Once we have study the x-averaged behavior of the temperature field, we think it is natural to subtract this $x$-average, $\tau(y)$, to the field in order to study the spatial structure of the deviations that appear in the convective state. Again, we think that this field may also have the scaling behavior that we discovered in the hydrodynamic velocity field. Therefore we define the scaled excess of temperature, $T^{(s)}(x,y)$, as:
\begin{equation}
T^{(s)}(x,y)=\frac{T(x,y)-\tau(y)}{\sigma(T)}
\end{equation}
where
\begin{equation}
\sigma(T)^2=\frac{1}{N_C}\sum_{(x,y)}\left[T(x,y)-\tau(y)\right]^2 \quad,\quad\tau(y)=\frac{1}{L_C}\sum_x T(x,y)
\end{equation}
where $L_C=30$ in our case.

\begin{figure}[h!]
\begin{center}
\includegraphics[height=5cm,clip]{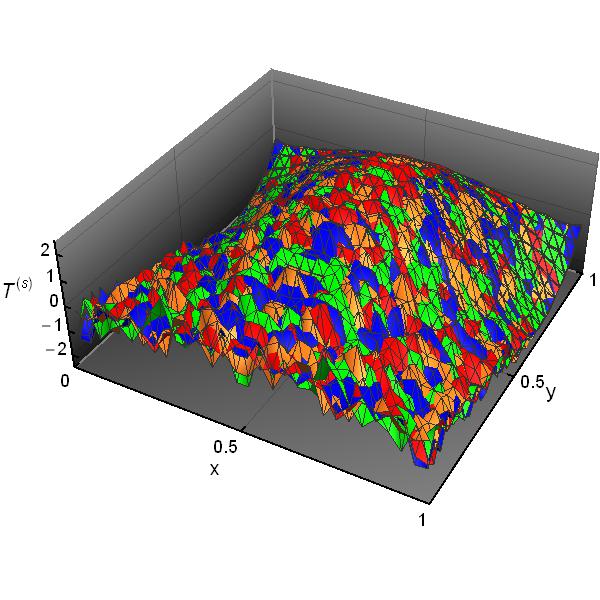}   %temp_field_show.nb
\includegraphics[height=5cm,clip]{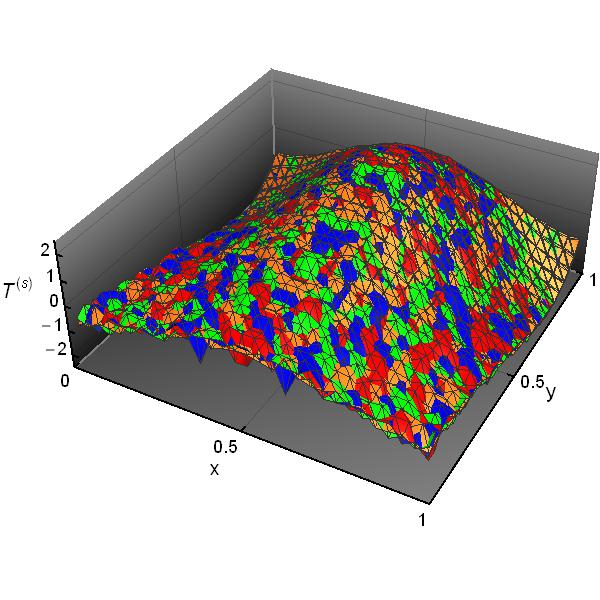}   
\includegraphics[height=5cm,clip]{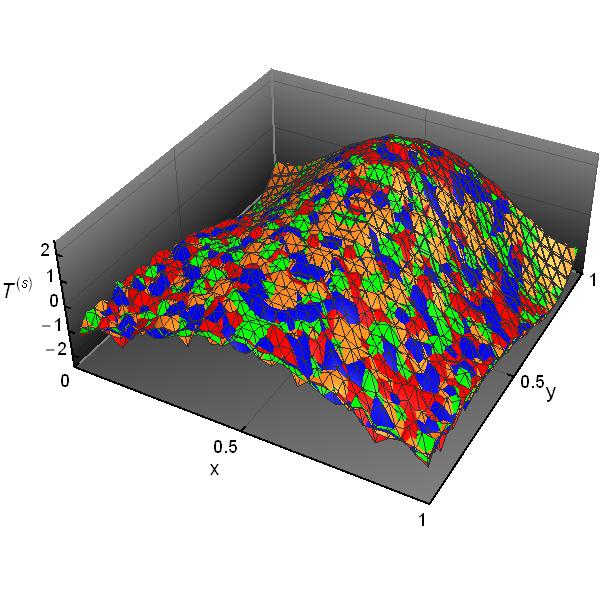}   
\end{center}
\caption{Superposition of the scaled excess of temperature fields, $T^{(s)}$ with $T_0=17$, $18$, $19$ and $20$ (red, blue, green and orange colors) and for different $g$ values (from left to right: $g=5$, $10$ and $15$).    \label{tempfieldsuper}}
\end{figure}

We have plotted in figure \ref{tempfieldsuper} the superposition of the scaled excess of temperature fields for $T_0=17$, $18$, $19$ and $20$ and different $g$ values. We observe that, again, it seems that there are common surfaces for each $g$ value. It is also clear that there are the fields depend on the $g$-value. between $g$ values. We can follow then the same items we did for the hydrodynamic velocity field. Let us go through them.
\begin{figure}[h!]
\begin{center}
\includegraphics[height=6cm,clip]{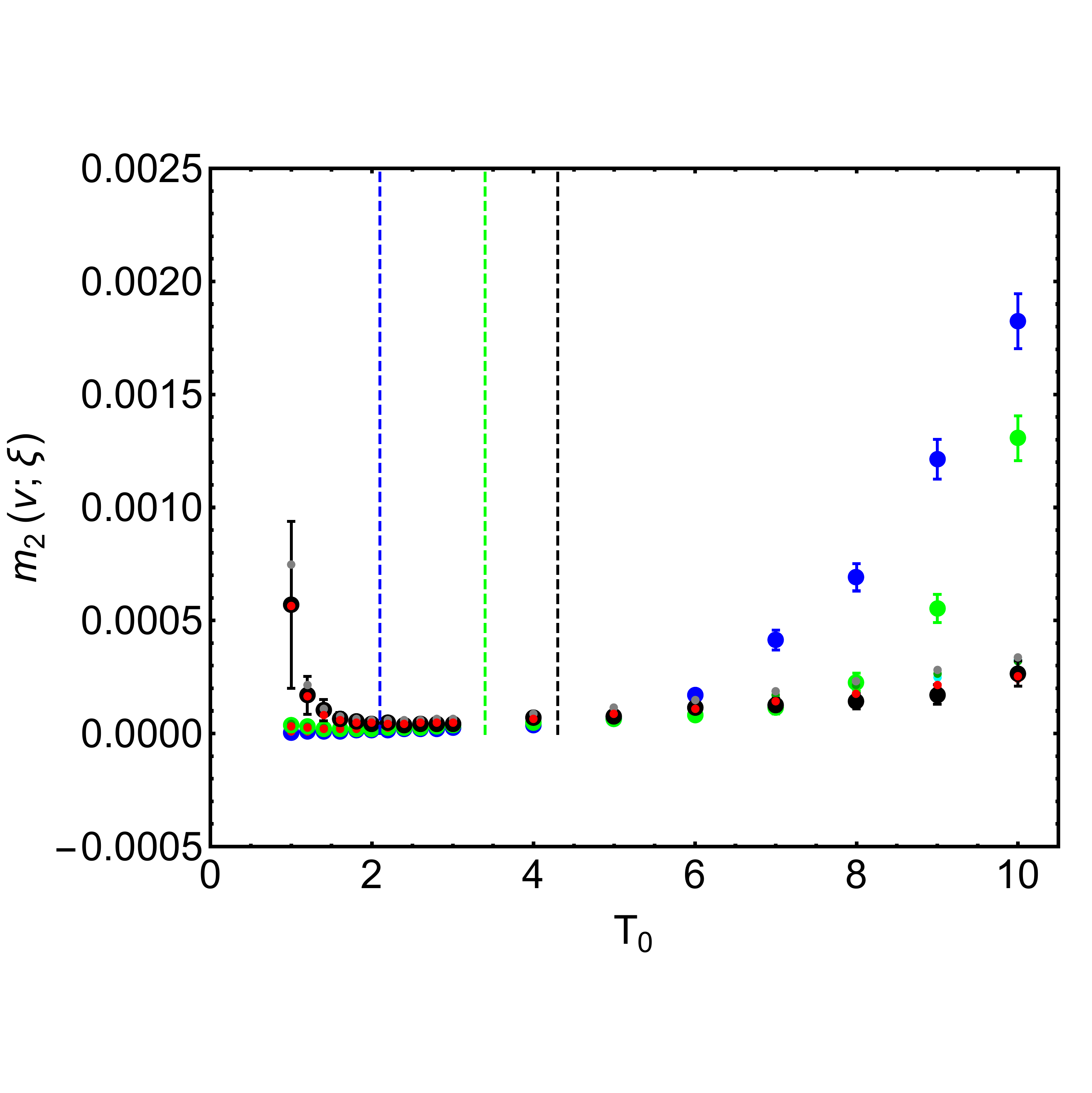}   %temp_moments.nb
\includegraphics[height=6cm,clip]{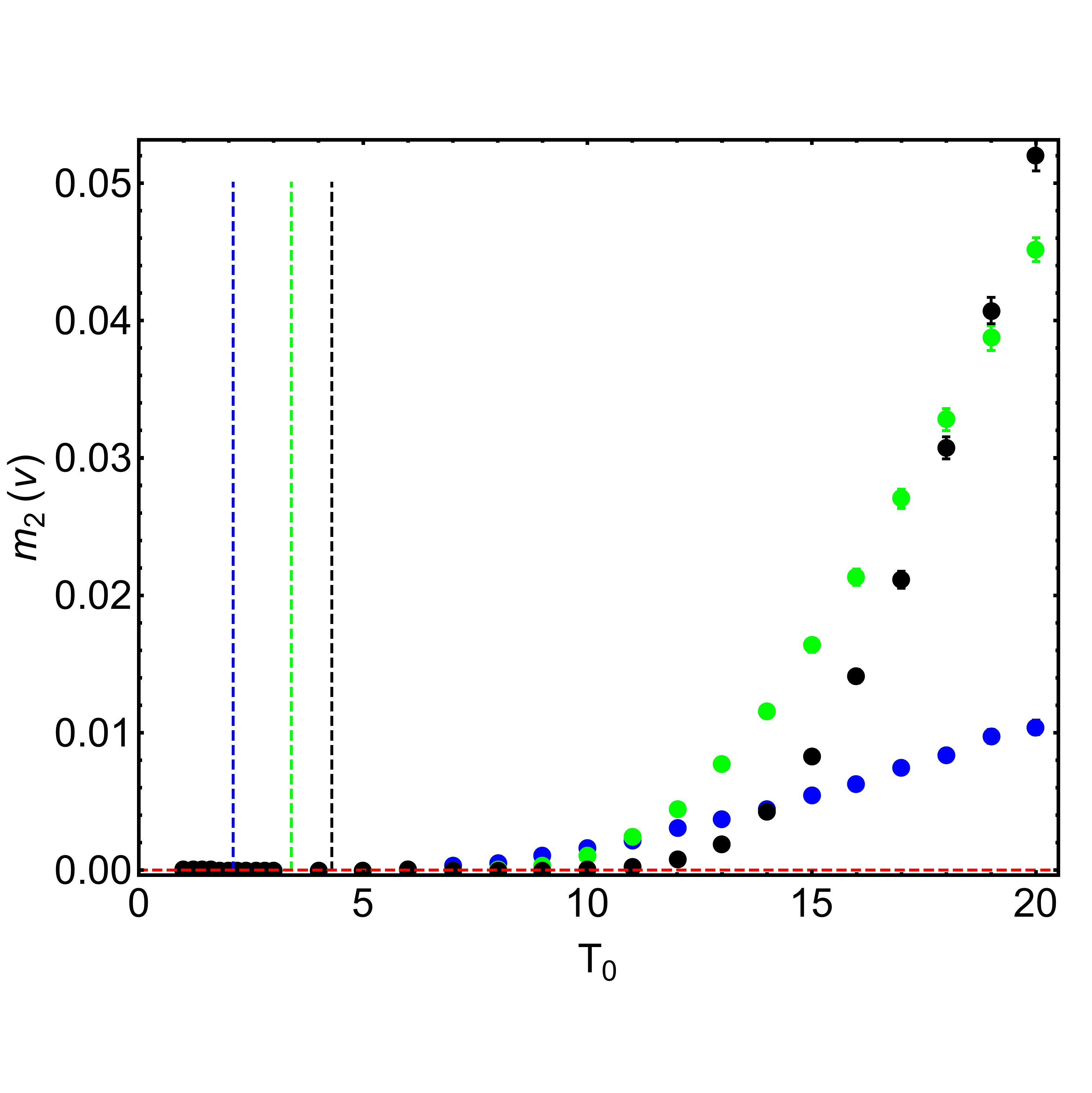}   
\end{center}
\kern -1cm
\caption{Left: Second moment of the measured excess of temperature field: $v(x,y;\xi)$ for $T_0\in[1,10]$ and $g=5$ (blue dots), $g=10$ (green dots) and $g=15$ (black dots). Small gray, dark green and cyan are the values for the $\langle A_2(\xi)\rangle$ term. The red small dots are the $\langle A_2(\xi)\rangle$ values assuming that the $T$ errors are multiplied by $f=0.87$. Right: $m_2(v)=m_2(v;\xi)-\langle A_2(\xi)\rangle$ computed with the errors corrected with the $f$ factor.
   \label{temmom}}
\end{figure}
\begin{enumerate}
\item {\it Preparation of scaled excess field:}
Let $T(x,y;\xi)$ and $\sigma(x,y)$ the averaged measured field abd its standard deviation respectively where we already have symmetrized the data with respect the $x=1/2$ axis. As before we assume that we get a value that is a particular realization of an inherent Gaussian white noise and therefore:
\begin{equation}
T(x,y;\xi)=T(x,y)+\sigma(x,y)\xi(x,y)
\end{equation}
where $\xi(x,y)$ is an uncorrelated Gaussian random variable with zero mean and unit variance. The $y$-profile is then given by:
\begin{equation}
T(y;\xi)=T(y)+\frac{1}{L_C}\sum_x \sigma(x,y)\xi(x,y)\quad,\quad T(y)=\frac{1}{L_C}\sum_x T(x,y)
\end{equation}
and the excess of the $v$-field is defined by:
\begin{equation}
v(x,y;\xi)=T(x,y;\xi)-T(y;\xi)=v(x,y)+\sigma(x,y)\xi(x,y)-\frac{1}{L_C}\sum_x \sigma(x,y)\xi(x,y)
\end{equation}
where $v(x,y)=T(x,y)-T(y)$. Observe that there is a difference with respect the hydrodynamic velocity field case: we subtracted there the overall field average and here just the $y$-averaged profile. This makes the computation rather different: for instance $\sum_x v(x,y;\xi)=0$ and $\sum_x v(x,y)=0$. The second momenta is now given by:
\begin{equation}
m_2(v;\xi)=\frac{1}{N_C}\sum_{(x,y)}v(x,y;\xi)^2=m_2(v)+A_1(v;\xi)+A_2(v;\xi)\quad,\quad m_2(v)=\frac{1}{N_C}\sum_{(x,y)}v(x,y)^2
\end{equation}
where
\begin{eqnarray}
A_1&=&\frac{2}{N_C}\sum_{(x,y)}v(x,y)\sigma(x,y)\xi(x,y)\nonumber\\
A_2&=&\frac{1}{N_C}\sum_{(x,y)}\sigma(x,y)^2\xi(x,y)^2-\frac{1}{L_C}\sum_{y}\left[\frac{1}{L_C}\sum_{x'}\sigma(x',y)\xi(x',y)\right]^2\nonumber
\end{eqnarray}
following the techniques already explained we find that
\begin{equation}
m_2(v)=m_2(v;\xi)-\langle A_2(\xi)\rangle \pm 3 \sqrt{\langle A_1(\xi)^2\rangle}
\end{equation}
where $\langle\cdot\rangle$ the the average over the uncorrelated white noise field. We show in figure \ref{temmom} the behaviors of $m_2(v;\xi)$ and $m_2(v)$.
First, we observe how the $m_2(v;\xi)$ have a queue for small $T_0$ values that it is due to the noisy data behavior. The computed $\langle A_2(\xi)\rangle$ term is sensibly larger that the measured value. This effect could be due to our assumption that the $\sigma(x,y)$ we have obtained is not the intensity of a pure white noise, that is, there are correlations effects not taked into account in the error theory we have developed. Another related possibility is that we have build  $\sigma(x,y)$ as the straight superposition of the $K$, $u_1$ and $u_2$ errors assuming that they were mutually independent. We know that there are some correlations between the fluctuations in the total kinetic energy of a cell and the fluctuations of the average values of the cell velocity. That is, we are overestimating the error in $T$ by definition.  With the set of data we have obtained we cannot make a precise and independent computation of the $T$'s fluctuation to study its behavior. In any case we observe that multiplying the $\sigma(x,y)$ by a factor $f=0.87$ we manage to fit the observed queue with the computed $A_2$ correction in all cases. We are going to use this effective factor in the error definition from now on. For instance, in figure \ref{temmom} (right) we show $m_2(v)$ as a function of $T_0$ and  we already included this error correction. In this case, observe that the values of $m_2(v)$ are very small for $T_0<13$ in contrast with the behavior of the hydrodynamic velocity field. Moreover, we see there that $m_2(v)$ grows and it doesn't seem to reach any finite limit. 

Finally, the measured scaled excess field is given by
\begin{equation}
T^{(s)}(x,y;\xi)=\frac{T(x,y;\xi)-T(y;\xi)}{\sqrt{m_2(T(x,y;\xi)-T(y;\xi))}}=\frac{v(x,y;\xi)}{\sqrt{m_2(v;\xi)}}
\end{equation} 

It can be expanded with respect of $\sigma(x,y)$ which is assumed to be small:
\begin{equation}
T^{(s)}(x,y;\xi)=T^{(s)}(x,y)+\alpha_1(x,y;\xi)+\alpha_2(x,y;\xi)+O(\sigma^3)
\end{equation}
where
\begin{eqnarray}
\alpha_1(x,y;\xi)&=&\tilde\sigma(x,y)\xi(x,y)-\frac{1}{L_C}\sum_{x'}\tilde\sigma(x',y)\xi(x',y)-\frac{1}{2}\bar A_1 T^{(s)}(x,y)\nonumber\\
\alpha_2(x,y;\xi)&=&-\frac{1}{2}\bar A_1\tilde\sigma(x,y)\xi(x,y)+\frac{1}{2}\bar A_1 \frac{1}{L_C}\sum_{x'}\tilde\sigma(x',y)\xi(x',y)-\frac{1}{2}\bar A_2 T^{(s)}(x,y)+\frac{3}{8}\bar A_1^2 T^{(s)}(x,y)
\end{eqnarray}
where $\tilde\sigma(x,y)=\sigma(x,y)/\sigma(v)$, $\bar A_i=A_i/\sigma(v)^2$ ($i=1,2$) and $\sigma(v)=m_2(v)^{1/2}$.

\item {\it Analysis of the scaled configuration:}
The Inertial Tensor of the scaled field is diagonal by construction and it doesn't depend on any $T^{(s)}$ property. By other hand we have studied the $4$'th momenta of it: 
\begin{equation}
m_4(T^{(s)};\xi)=\frac{1}{N_C}\sum_{(x,y)}T^{(s)}(x,y;\xi)^4
\end{equation}
Its corrected noise form as a function of the observed one is given by:
\begin{equation}
m_4(T^{(s)})=m_4(T^{(s)};\xi)-\Lambda \pm 3 \epsilon
\end{equation}
with
\begin{eqnarray}
\Lambda&=&\frac{4}{N_C}\sum_{(x,y)}T^{(s)}(x,y)^3\langle\alpha_2(x,y;\xi)\rangle+\frac{6}{N_C}\sum_{(x,y)}T^{(s)}(x,y)^2\langle\alpha_1^2\rangle\nonumber\\
\epsilon&=&\frac{16}{N_C^2}\sum_{(x,y)}\sum_{(x',y')}T^{(s)}(x,y)^3T^{(s)}(x',y')^3\langle\alpha_1(x,y;\xi)\alpha_1(x',y';\xi)\rangle
\end{eqnarray}

\begin{figure}[h!]
\begin{center}
\includegraphics[height=6cm,clip]{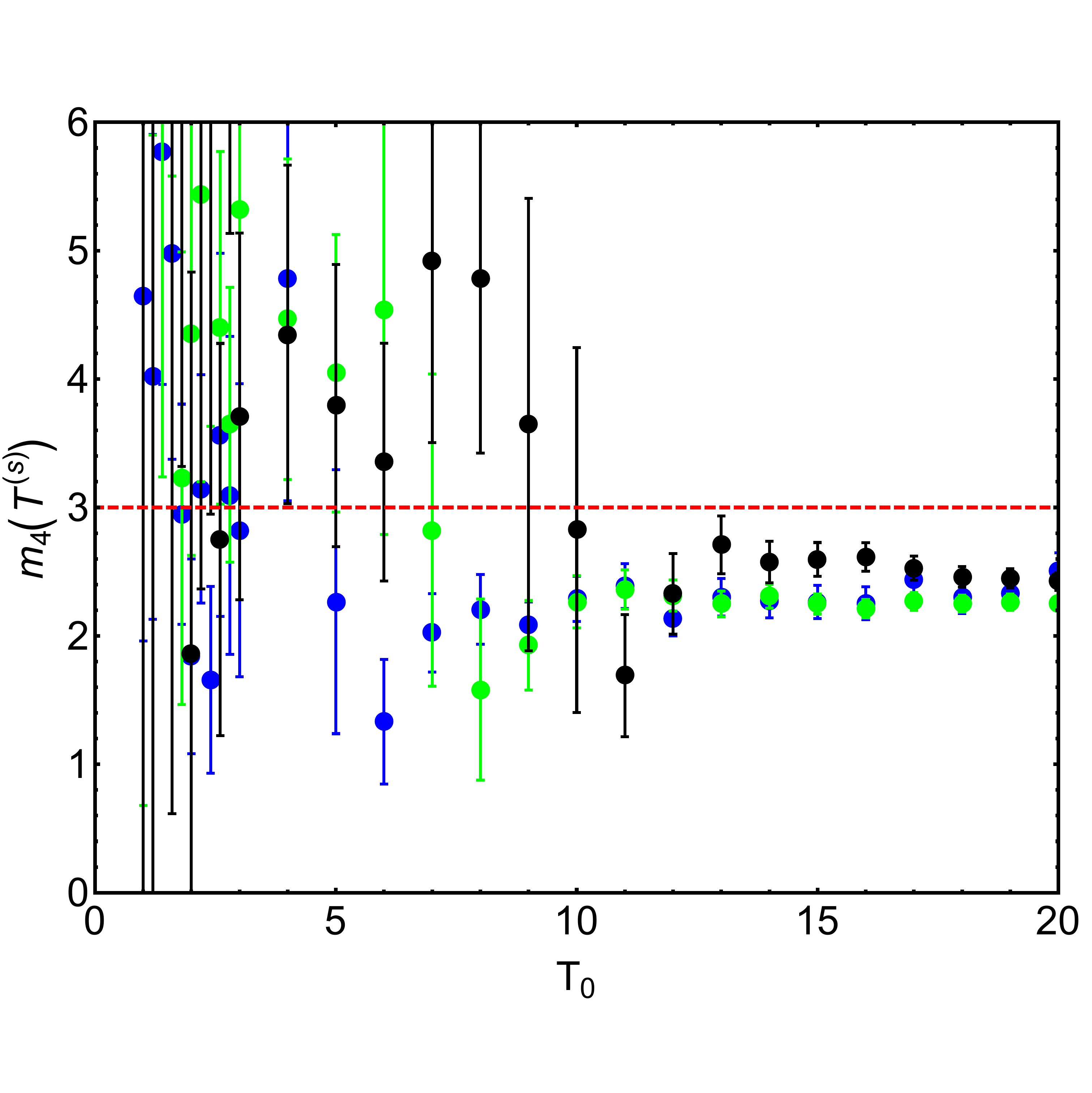}   %temp_moments.nb
\end{center}
\kern -1cm
\caption{Fourth moment of the scaled excess of temperature field,  $m_4(T^{(s)})$, as a function of $T_0$. 
   \label{temmom4}}
\end{figure}

In figure \ref{temmom4} we observe again how the $m_4$'th moment tends for each $g$ to a constant value independent on $T_0$. The $g$ dependence of such constants is small.

We have also studied the averaged euclidean distance between any two configurations $T^{(s)}(x,y;T_0,g;\xi)$ with a fixed $g$ value and different $T_0$ temperatures. As in the case of the hydrodynamic velocity field, we can write:
\begin{equation}
D(T^{(s)};T_0,T_0',g)^2=D_-(T^{(s)};T_0,T_0',g)D_+(T^{(s)};T_0,T_0',g)
\end{equation}
where
\begin{eqnarray}
D_-(T^{(s)};T_0,T_0',g)&=&D(T^{(s)};T_0,T_0',g;\xi)- \Lambda \pm 3\epsilon(D_-)\nonumber\\
D_+(T^{(s)};T_0,T_0',g)&=&D(T^{(s)};T_0,T_0',g;\xi)+ \Lambda \pm 3\epsilon(D_+)\nonumber
\end{eqnarray}
\begin{eqnarray}
\Lambda^2&=&\frac{1}{N_C}\sum_{(x,y)}\biggl[\langle \alpha_1(x,y;\xi;T_0)^2\rangle+ \langle \alpha_1(x,y;\xi;T_0')^2\rangle\nonumber \\
&+&2(T^{(s)}(x,y;T_0)-T^{(s)}(x,y;T_0'))(\langle \alpha_2(x,y;\xi;T_0)\rangle-\langle \alpha_2(x,y;\xi;T_0')\rangle)\biggr]
\end{eqnarray}
and
\begin{equation}
\epsilon(D_-)=\frac{\sqrt{\langle a(\xi)^2\rangle}}{\vert D(T^{(s)};T_0,T_0',g;\xi)- \Lambda\vert+D(T^{(s)};T_0,T_0',g;\xi)+ \Lambda }
\end{equation}
with
\begin{equation}
a(\xi)=\frac{2}{N_C}\sum_{(x,y)}\left(T^{(s)}(x,y;T_0)- T^{(s)}(x,y;T_0')\right)(\alpha_1(x,y;\xi;T_0)-\alpha_1(x,y;\xi;T_0'))
\end{equation}
\end{enumerate}
We plot in figure \ref{distanceT} the mutual marginal distance, $D_-$, between configurations. We observe how the raw distance tend to a non-zero constant that, once we include the noise corrections, is zero for $T_0>11$ in all cases. Let us remark that we have included the correcting factor in the $T$'s errors. Without including it, the corrected distances would be constant but negative. That is consistent with our observation that we overestimated the error in $T$.

\begin{figure}[h!]
\begin{center}
\includegraphics[height=5cm,clip]{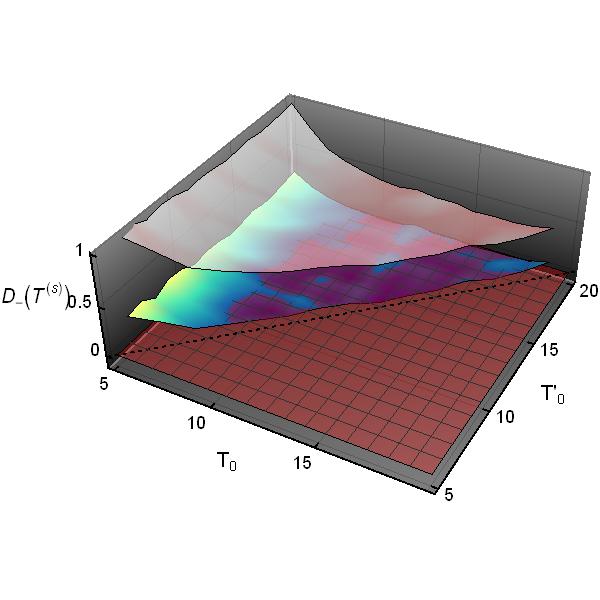}   %temp_E5new.nb
\includegraphics[height=5cm,clip]{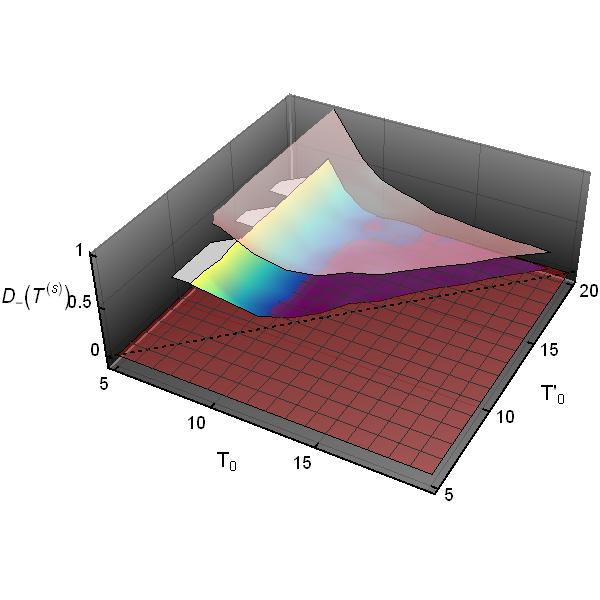}%temp_E10new.nb
\includegraphics[height=5cm,clip]{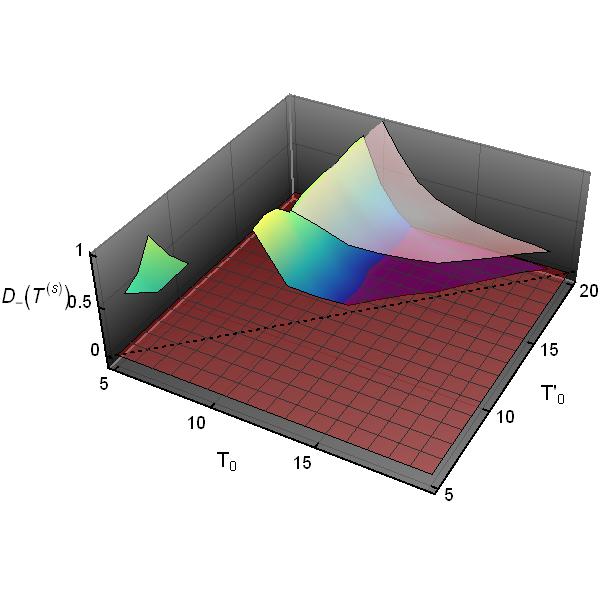}%temp_E15new.nb
\newline\vglue -1cm
\includegraphics[height=4.5cm,clip]{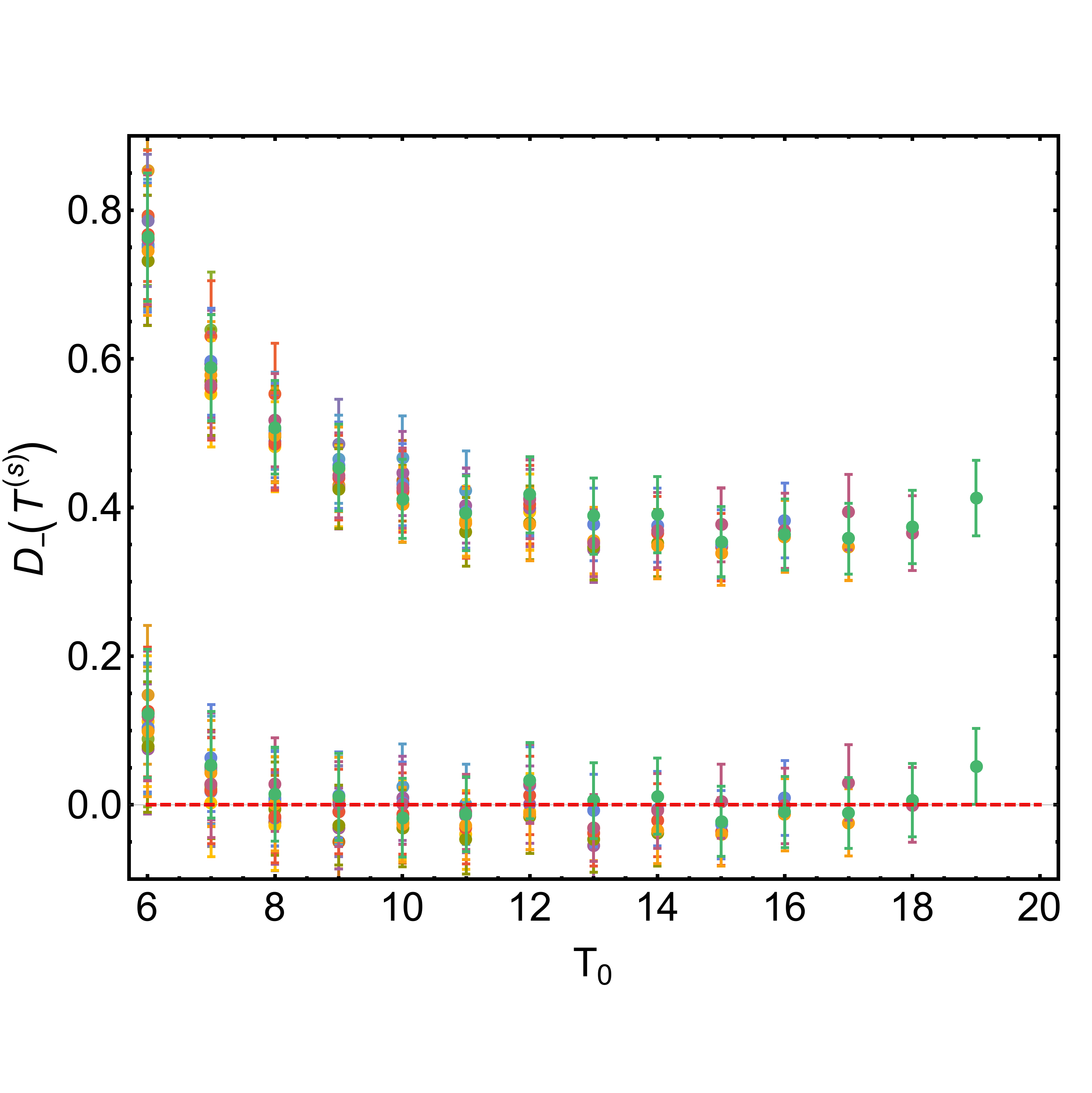}   %temp_E5new.nb
\includegraphics[height=4.5cm,clip]{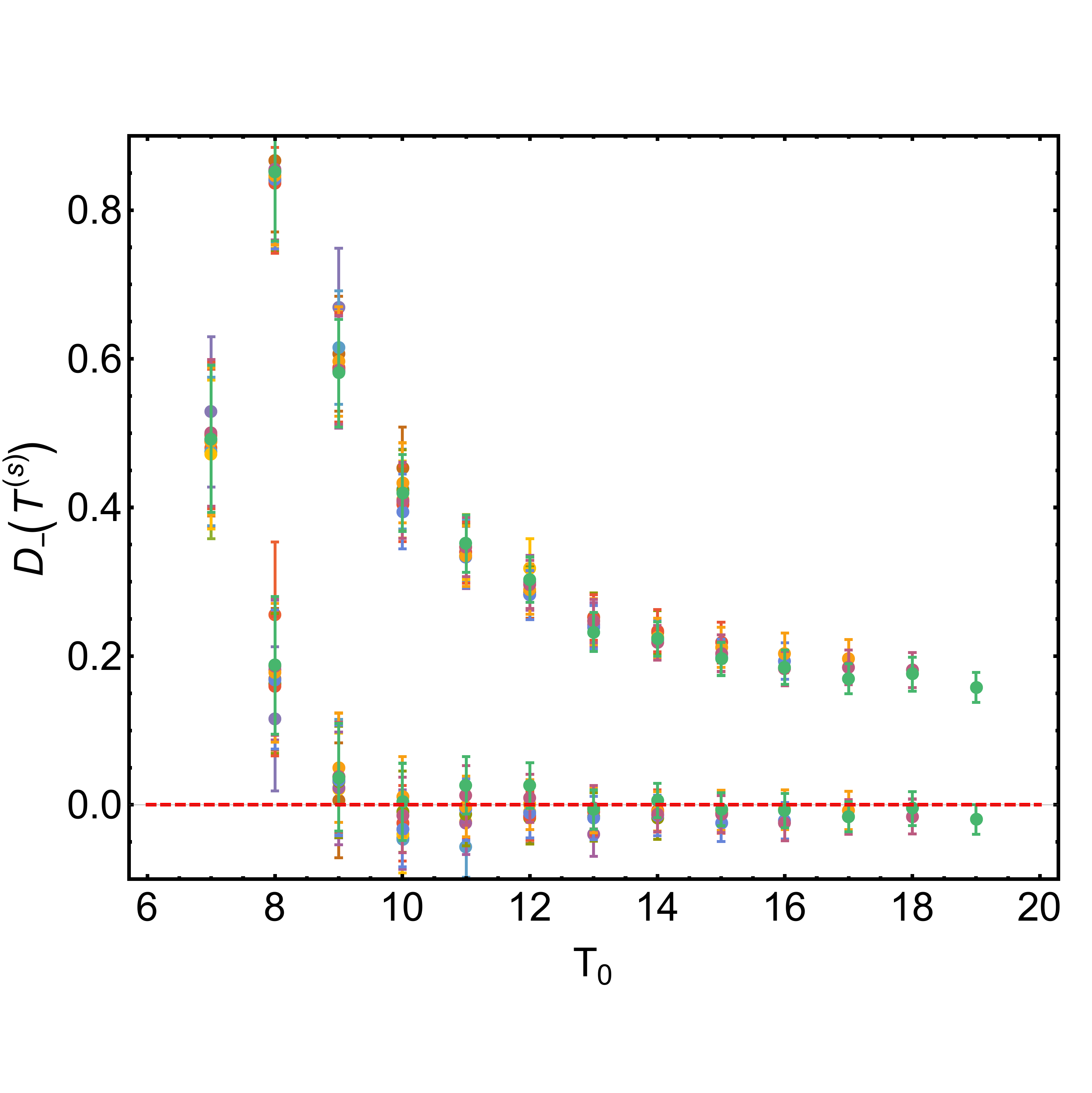}%temp_E10new.nb
\includegraphics[height=4.5cm,clip]{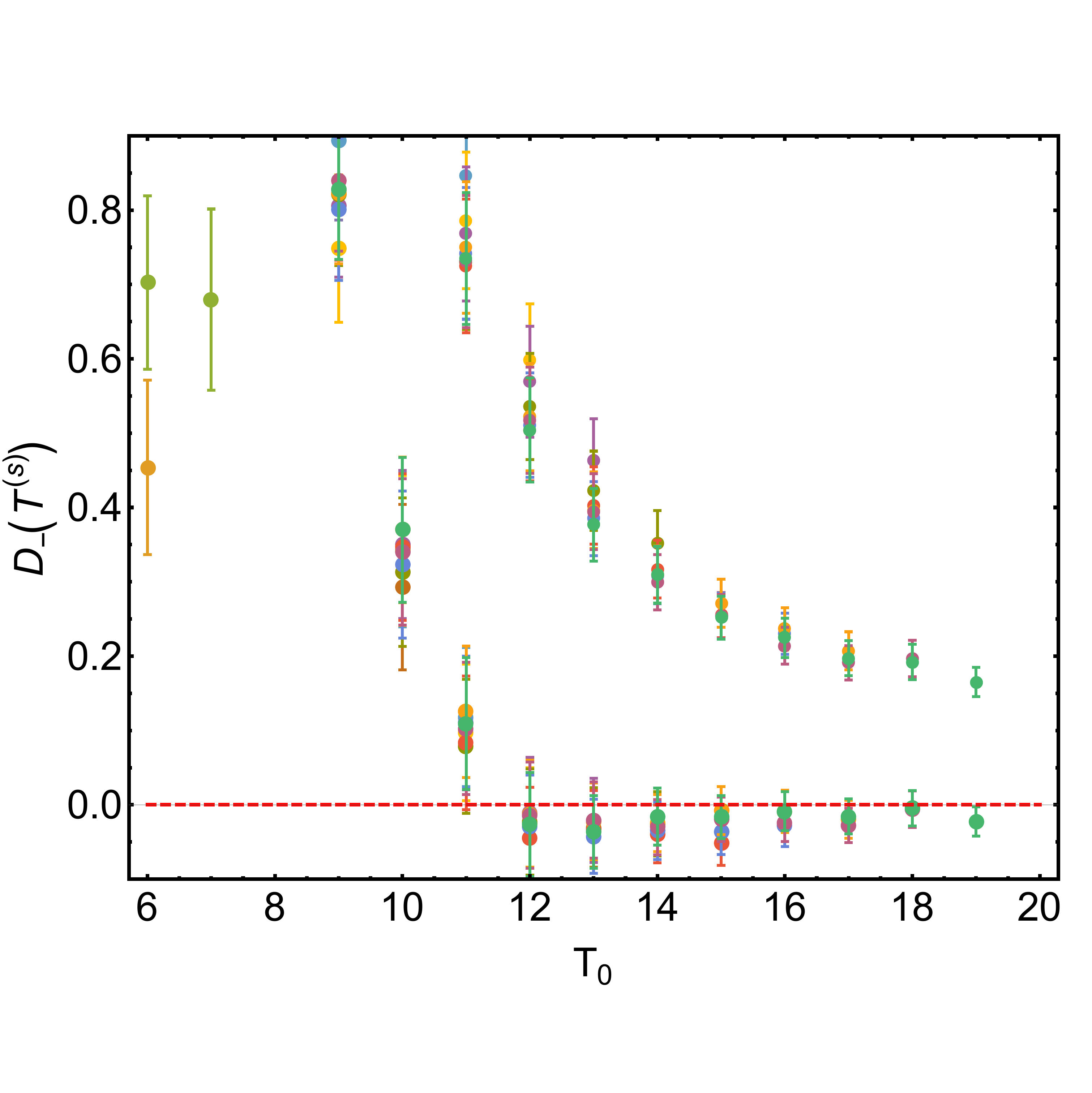}%temp_E15new.nb
\end{center}
\kern -1.cm
\caption{Marginal distance $D_-$ (see text) between temperature field scaled configurations $T^{(s)}$ for $g=5$ (figures left), $g=10$ (figures center) and $g=15$ (figures right). for a given $g$ value. We only plot pairs of scaled configurations with $T_0$ and $T_0'$ such that $T_0<T_0'$. Top figures: Pink surface are the bare distances $D(T^{(s)};\xi)$. Yellow-green surfaces are the distances  after applying the noise correction term $\Lambda$ (see text). Bottom figures:  Same as top but including error bars. Each color corresponds to a given $T_0'$ value. \label{distanceT}}
\end{figure}

\item {\it The universal fields:} At this point we follow the same path we did with the hydrodynamic velocity field. First, we averaged the scaled excess temperature fields with $T_0=14$, $\ldots 20$ just to have better statistics. Then we do a discrete Fourier Transform of the resulting field (see figure \ref{FourierT}) discarding the noisy wave numbers (we show in table \ref{cut2} the limiting intensity below which the modes are neglected) and, finally, we do the inverse Fourier Transform obtaining the universal scaled excess of temperature field (see figure \ref{universalT}).

 Finally we have also compared the universal scaled field with the one with $T_0=17$. We again see in figure \ref{fluctuT} how such differences are constrained to the $3\sigma(x,y)$ error bars of the measured field. Morever, we again see the average of the absolute value of $w_d$-like field (as we defined it in the hydrodynamic velocity section) follows remarkably well the white noise value. (see figure \ref{fluctuT2}). The Fourier Transform of the $w_d$ field is shown in figure \ref{fluctuTf}. We observe in this case a very weak structure around the $(0,0)$ mode. It could be just our effective error definition that it does not capture correctly the real spatial correlations of the noise.

\begin{figure}[h!]
\begin{center}
\includegraphics[height=5cm,clip]{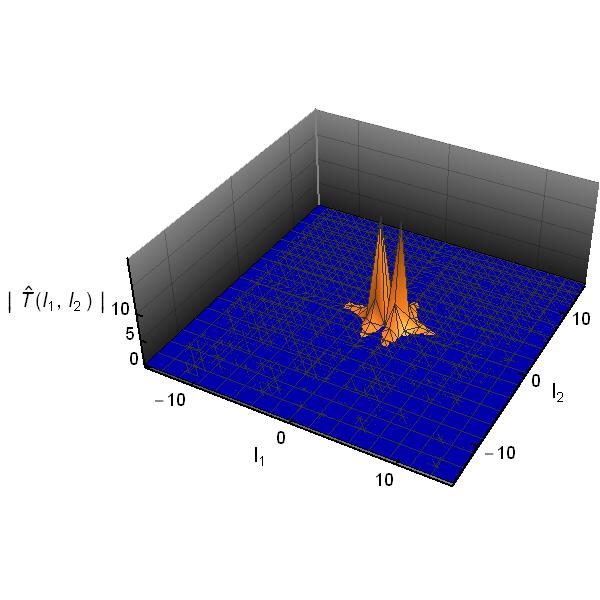}   %temp_E5new.nb
\includegraphics[height=5cm,clip]{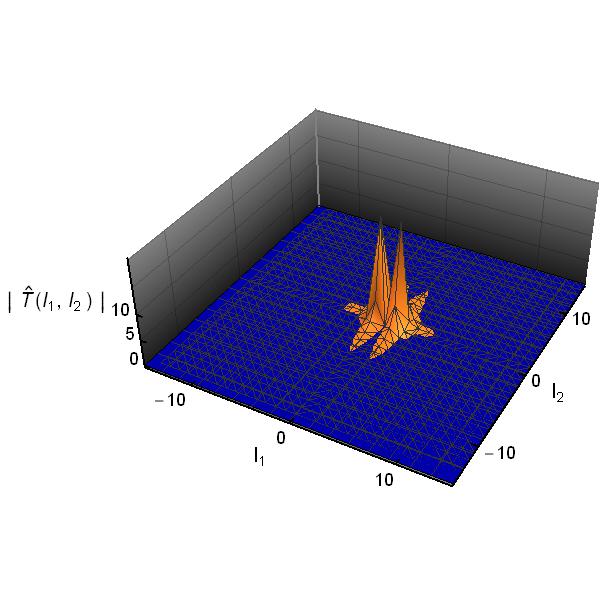}%temp_E10new.nb
\includegraphics[height=5cm,clip]{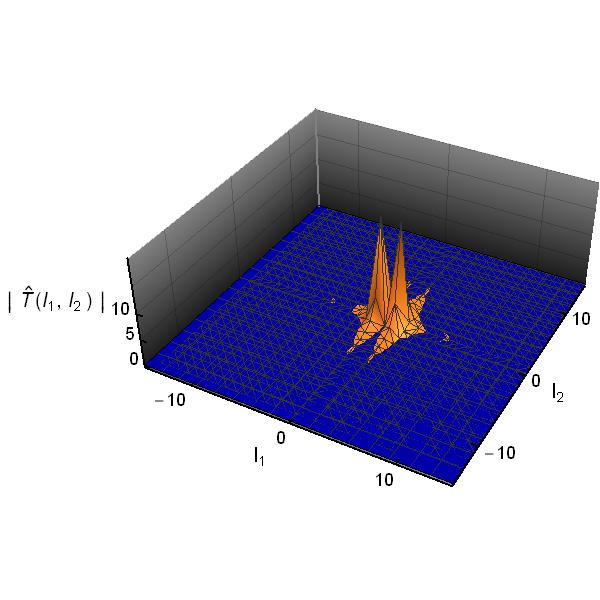}%temp_E15new.nb
\newline\vglue -1cm
\includegraphics[height=5cm,clip]{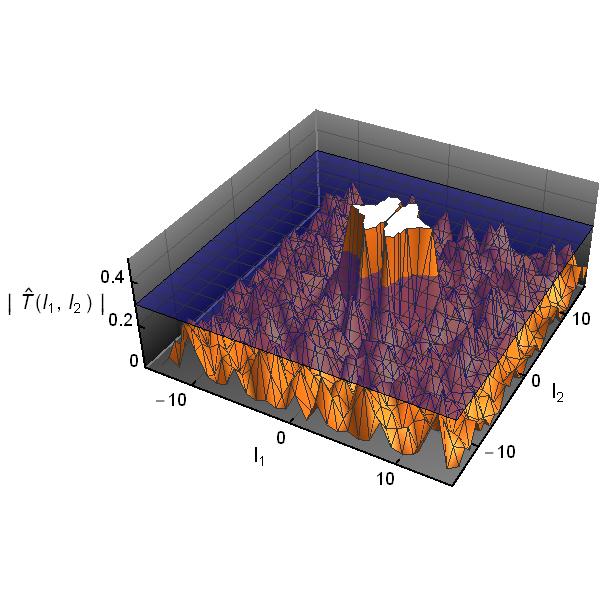}   %temp_E5new.nb
\includegraphics[height=5cm,clip]{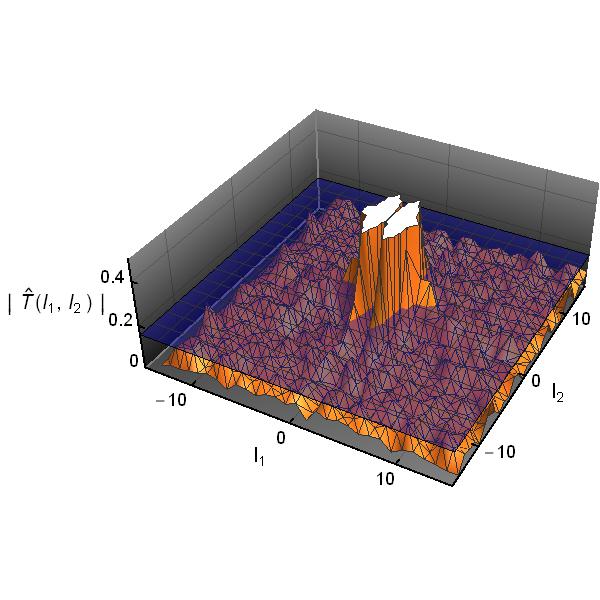}%temp_E10new.nb
\includegraphics[height=5cm,clip]{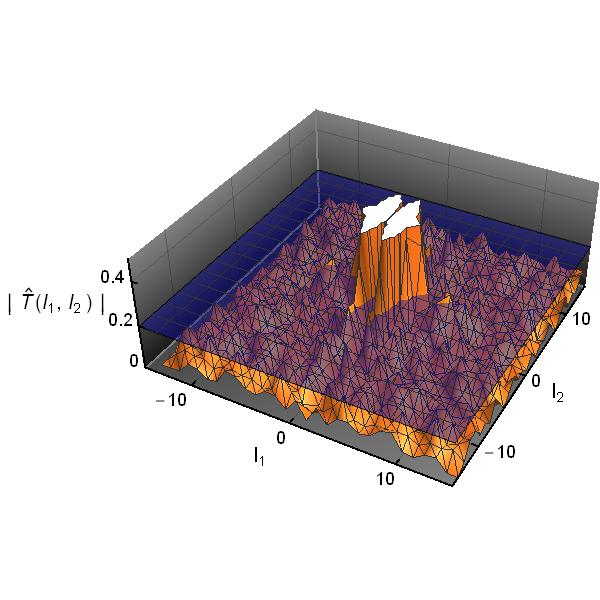}%temp_E15new.nb
\end{center}
\kern -1.cm
\caption{Modulus of the Discrete Fourier Transform obtained by averaging the scaled configurations, $T^{(s)}(x,y)$, from $T_0=14,\ldots, 20$ for $g=5$ (left figures), $g=10$ (center figures) and $g=15$ (right figures).  Points below the blue surfaces are discarded and only points above them are used to the subsequent Inverse Fourier Transform to get a smoothed field. Top figures show the modes used in the Discrete Inverse Fourier Transform and bottom ones the detailed behavior of the discarded noisy modes.
 \label{FourierT}}
\end{figure}

\begin{table}[h!]
\begin{center}
\resizebox*{!}{3cm}{ 
\begin{tabular}{|c|c|}
\hline
$g$&$\vert \hat T^{(s)}\vert$\\ \hline
\hline
5&0.3\\ \hline
10&0.16\\ \hline
15&0.20\\ \hline
\end{tabular}}
\end{center}
\caption{Cut-off values. The modes of the Fourier Transform of the field with modulus less than the corresponding cut-off value are discarded. \label{cut2}}
\end{table}

\begin{figure}[h!]
\begin{center}
\includegraphics[height=5cm,clip]{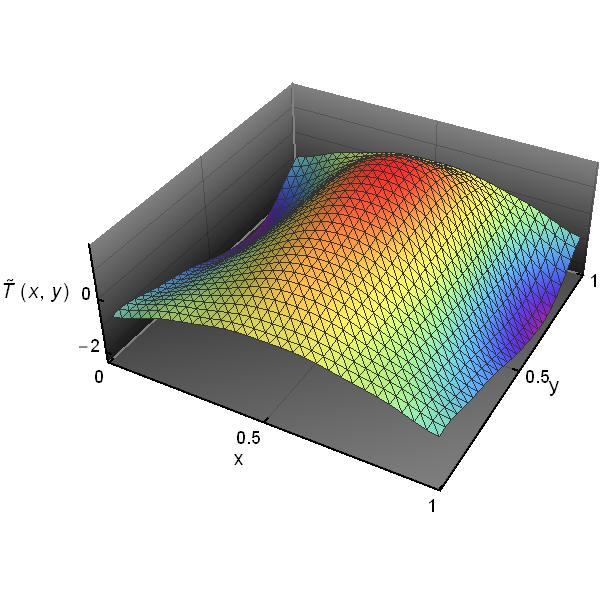}   %temp_E5new.nb
\includegraphics[height=5cm,clip]{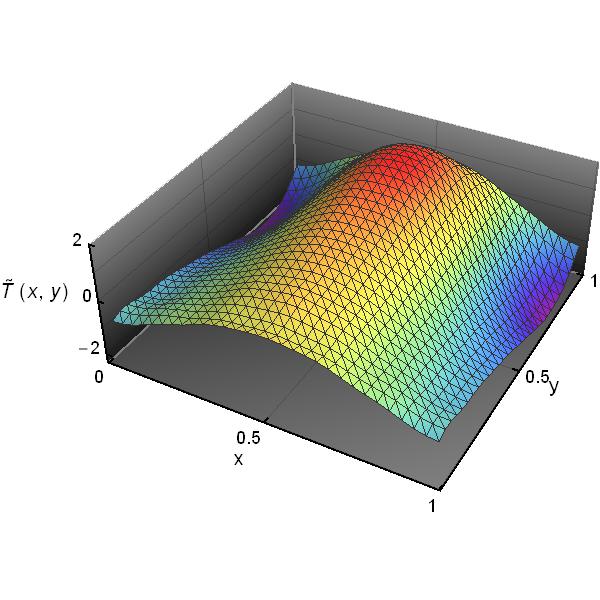}%temp_E10new.nb
\includegraphics[height=5cm,clip]{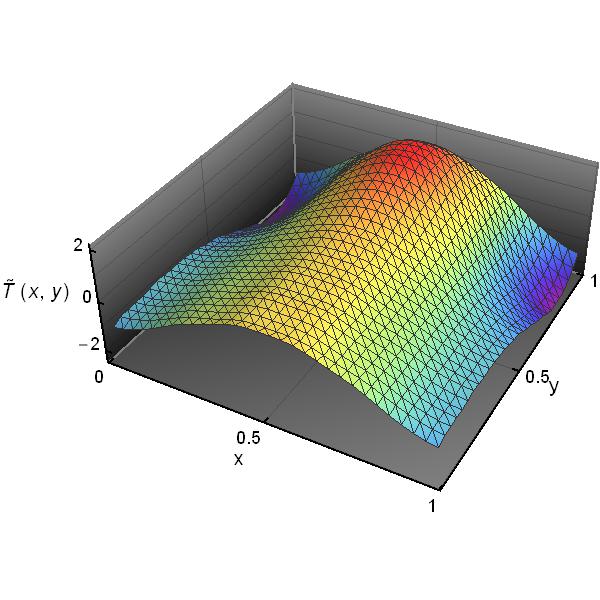}%temp_E15new.nb
\end{center}
\kern -1.cm
\caption{Universal fields $\tilde T(x,y)$   for $g=5$, $10$ and $15$ from left to right. 
 \label{universalT}}
\end{figure}
\begin{figure}[h!]
\begin{center}
\includegraphics[height=5cm,clip]{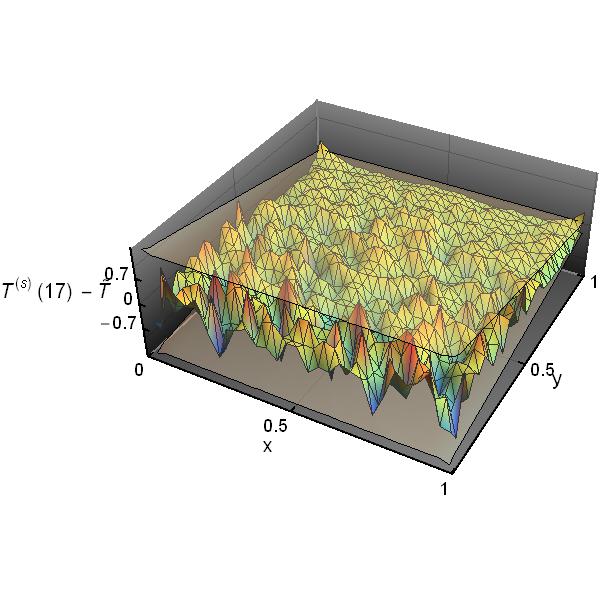}   %temp_E5new.nb
\includegraphics[height=5cm,clip]{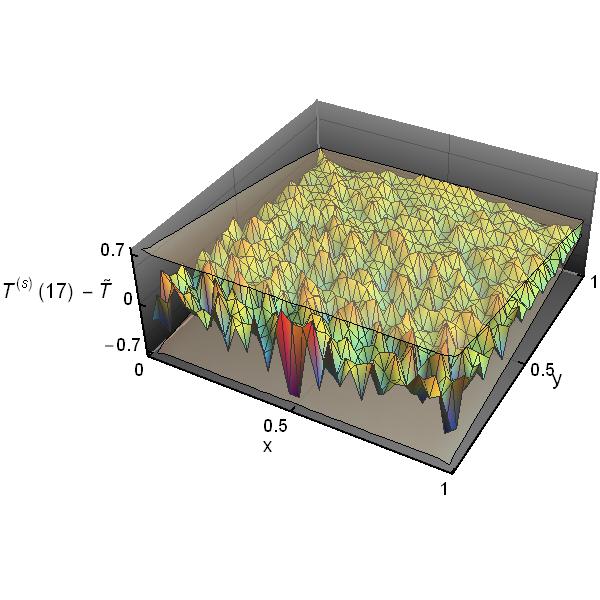}%temp_E10new.nb
\includegraphics[height=5cm,clip]{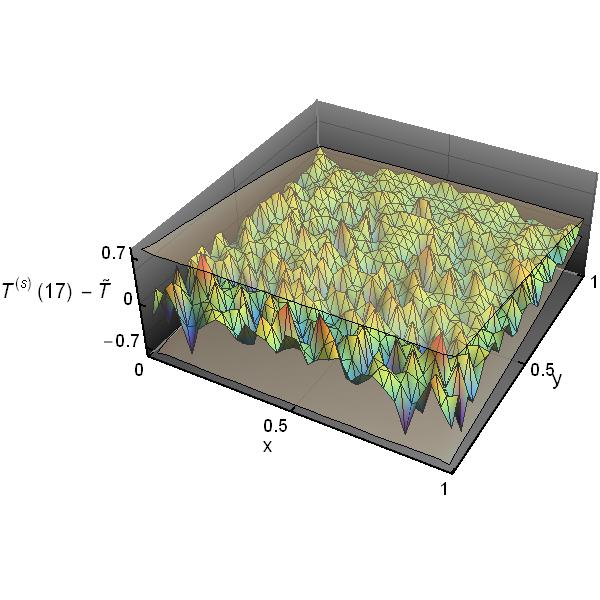}%temp_E15new.nb
\end{center}
\kern -1.cm
\caption{Difference between the scaled field $T^{(s)}(x,y)$ for $T_0=17$ and the corresponding universal field $\tilde T(x,y)$  for $g=5$, $10$ and $15$ from left to right. The gray surfaces are the data error bars of the scaled temperature fields.
 \label{fluctuT}}
\end{figure}
\begin{figure}[h!]
\begin{center}
\includegraphics[height=5cm,clip]{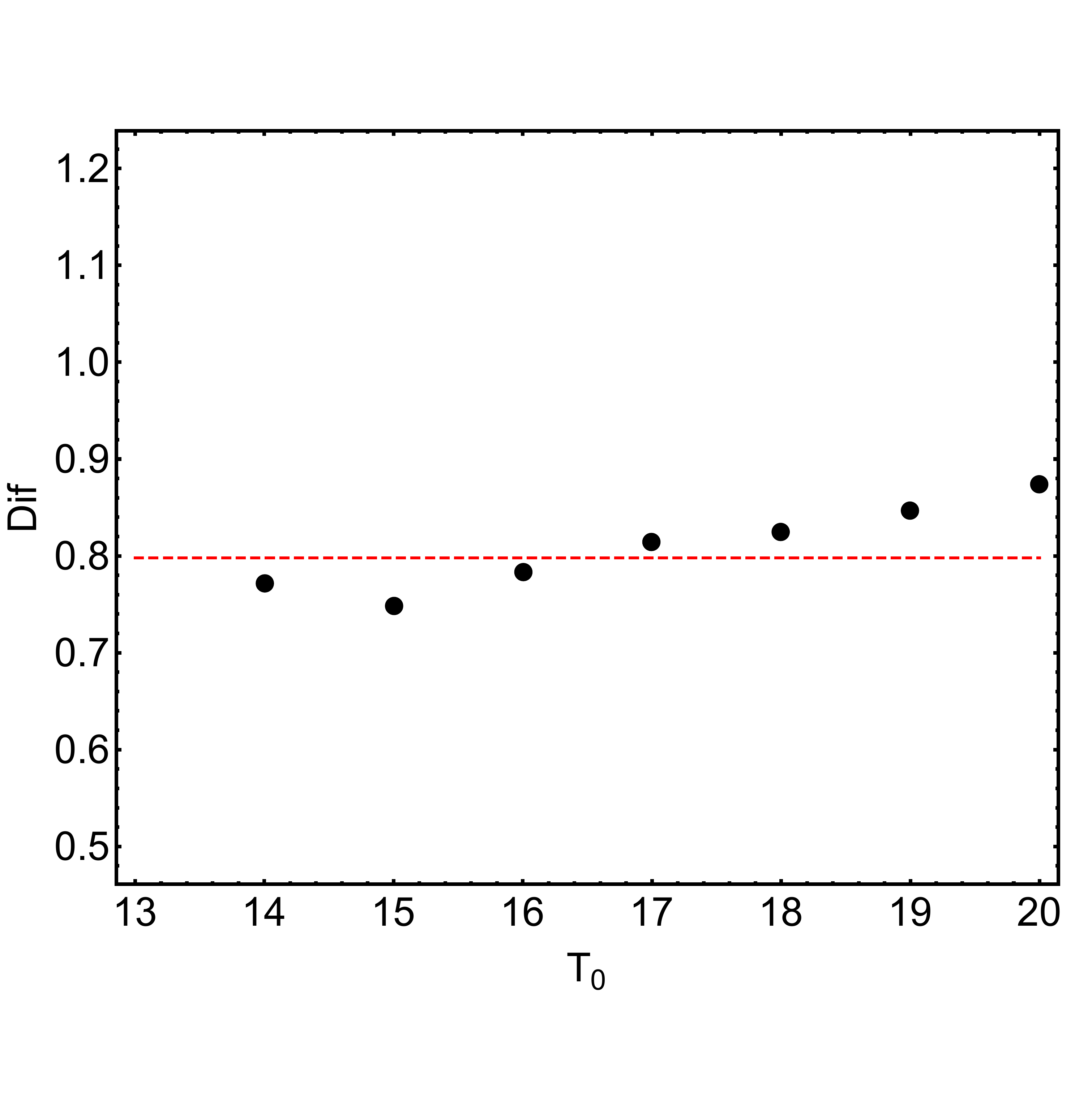}   %temp_E5new.nb
\includegraphics[height=5cm,clip]{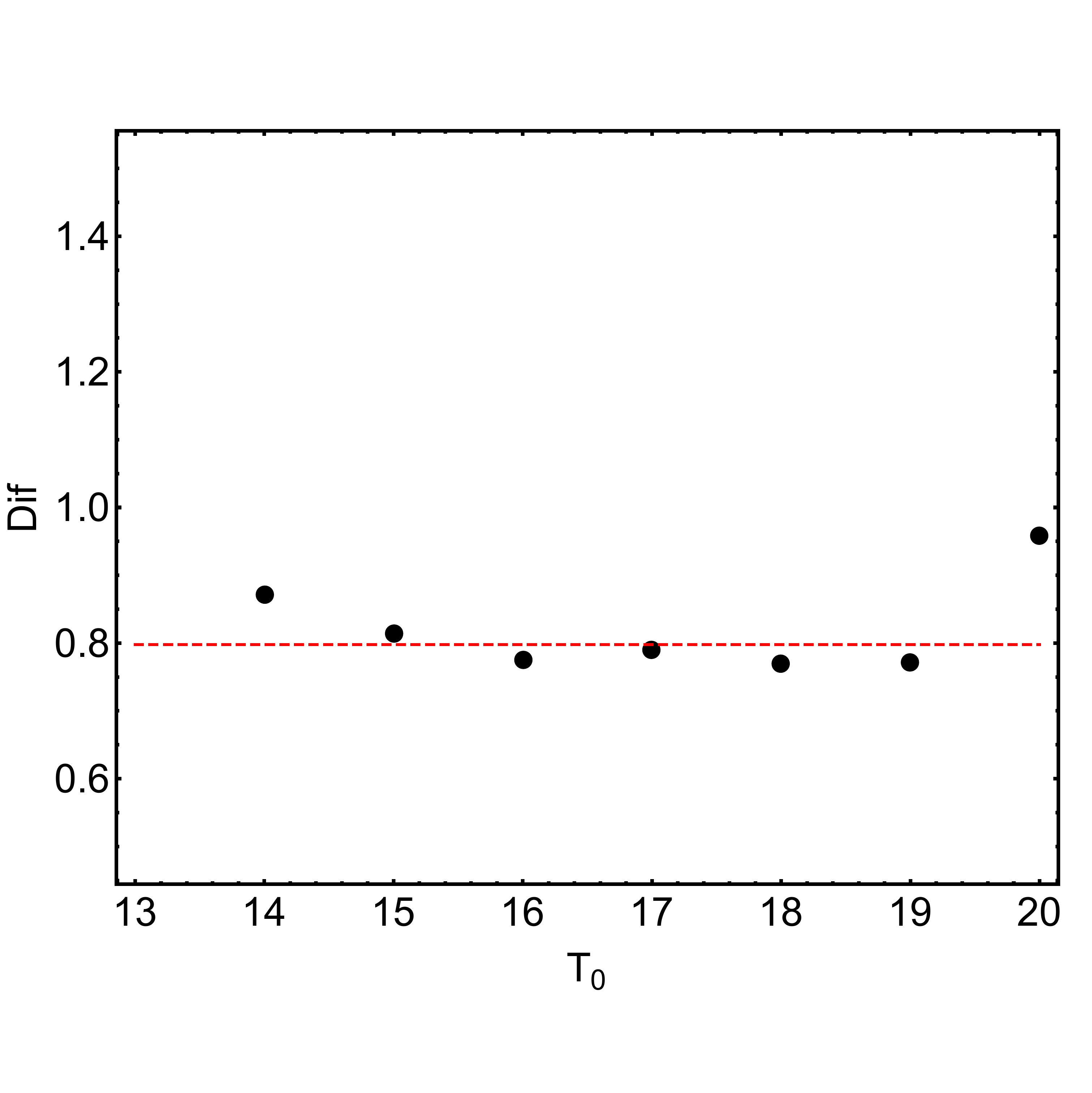}%temp_E10new.nb
\includegraphics[height=5cm,clip]{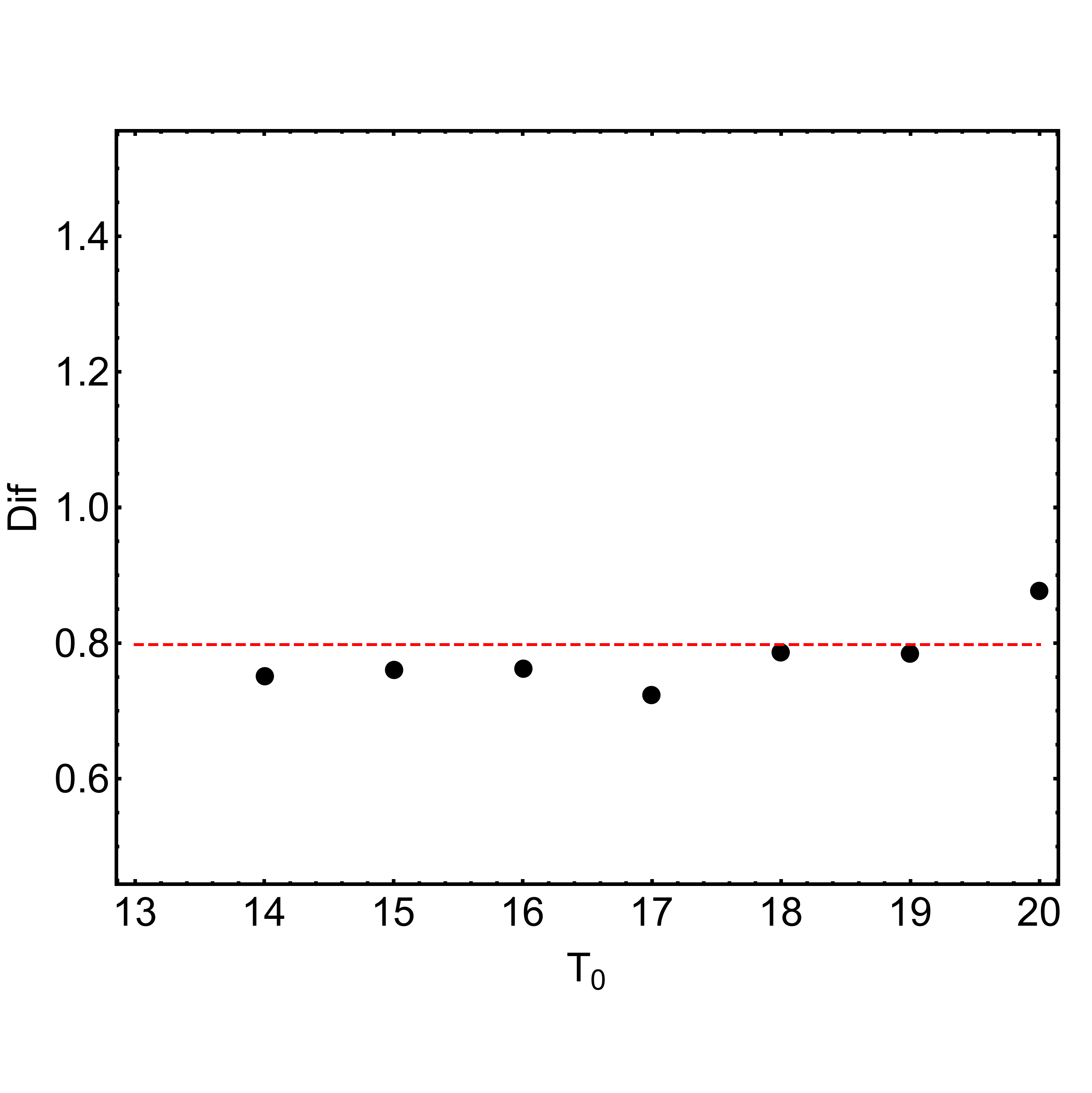}%temp_E15new.nb
\end{center}
\kern -1.cm
\caption{Averaged ratio between the difference between the scaled field $T^{(s)}(x,y)$  and the corresponding universal field $\tilde T(x,y)$ with respect its variance normalized with the profile variance for $g=5$, $10$ and $15$ from left to right. 
 \label{fluctuT2}}
\end{figure}

\begin{figure}[h!]
\begin{center}
\includegraphics[height=5cm,clip]{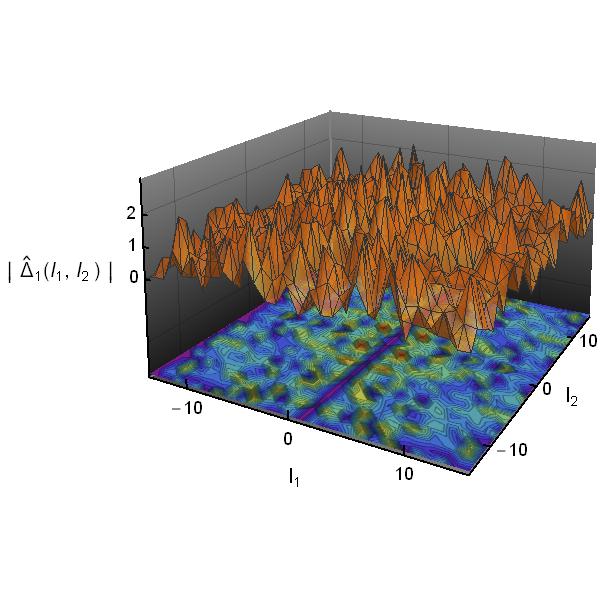}   %temp_E5new.nb
\includegraphics[height=5cm,clip]{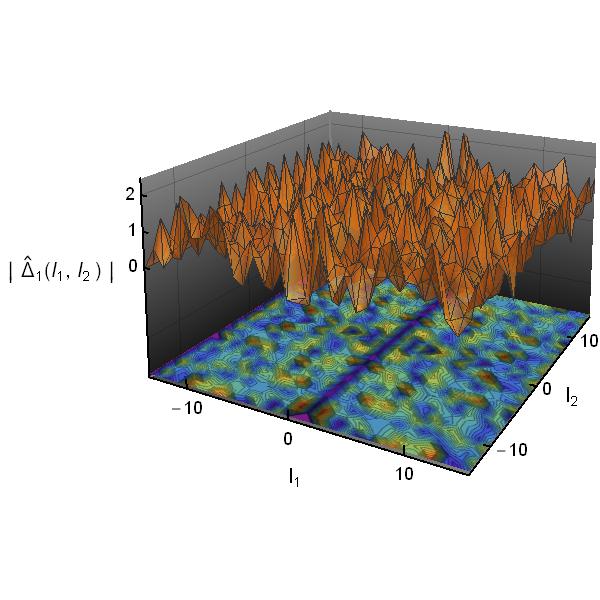} %temp_E10new.nb
\includegraphics[height=5cm,clip]{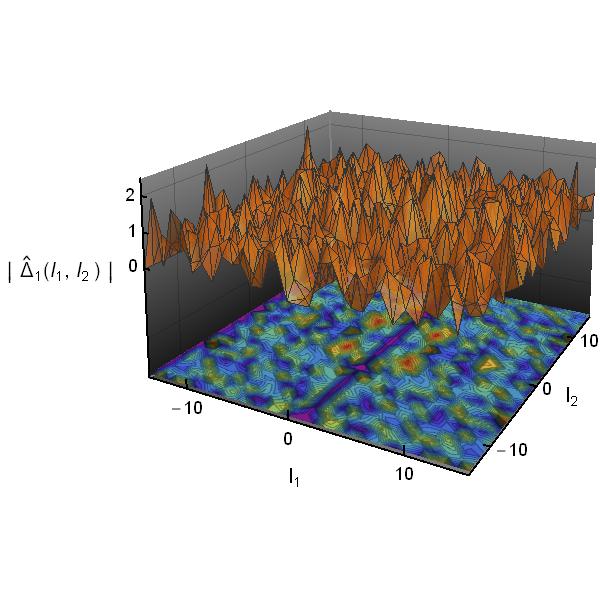} %temp_E15new.nb
\end{center}
\kern -1.cm
\caption{Modulus of the Fourier Transform of the difference  between the scaled field $T^{(s)}(x,y)$ for $T_0=17$ and the corresponding universal field $\tilde T(x,y)$  for $g=5$, $10$ and $15$ from left to right. 
 \label{fluctuTf}}
\end{figure}

We can conclude that there exists an universal field $\tilde T(x,y)$ in such a way that
\begin{equation}
T(x,y;T_0,g)= T(y;T_0,g)+\sigma(T(x,y)-T(y);T_0,g))\tilde T(x,y;g)
\end{equation}
for the configurations with $T_0\in[14,20]$.

\item{\it Density profiles:}

\begin{figure}[h!]
\begin{center}
\includegraphics[height=6cm,clip]{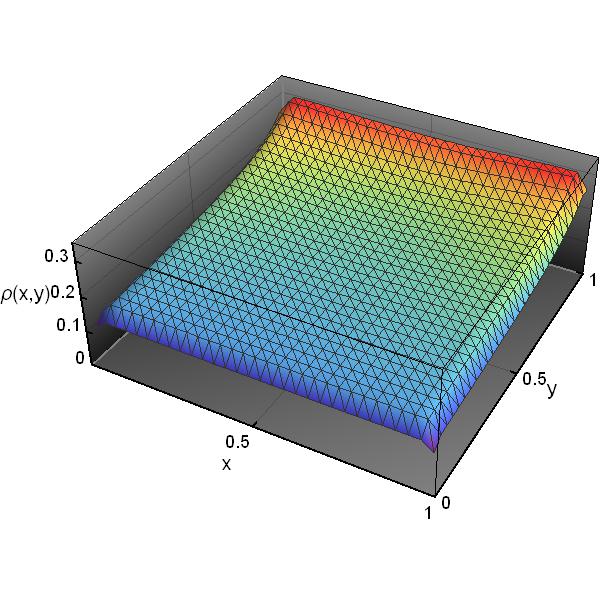}  %ro_profile.nb
\includegraphics[height=6cm,clip]{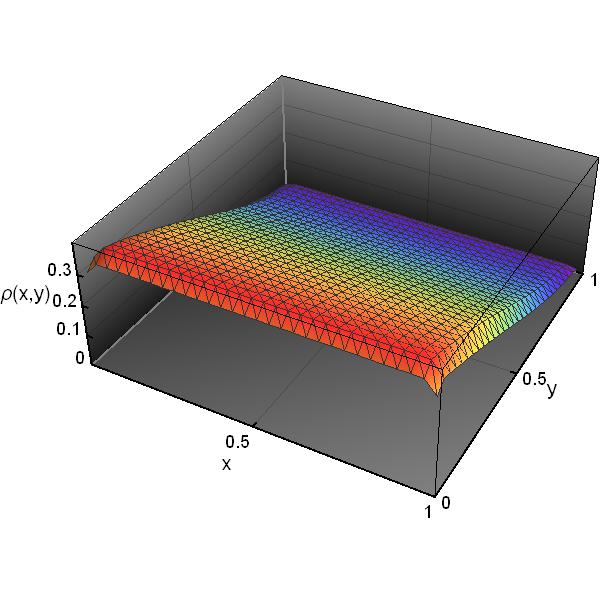}       %ro_profile_1.nb
\includegraphics[height=5cm,clip]{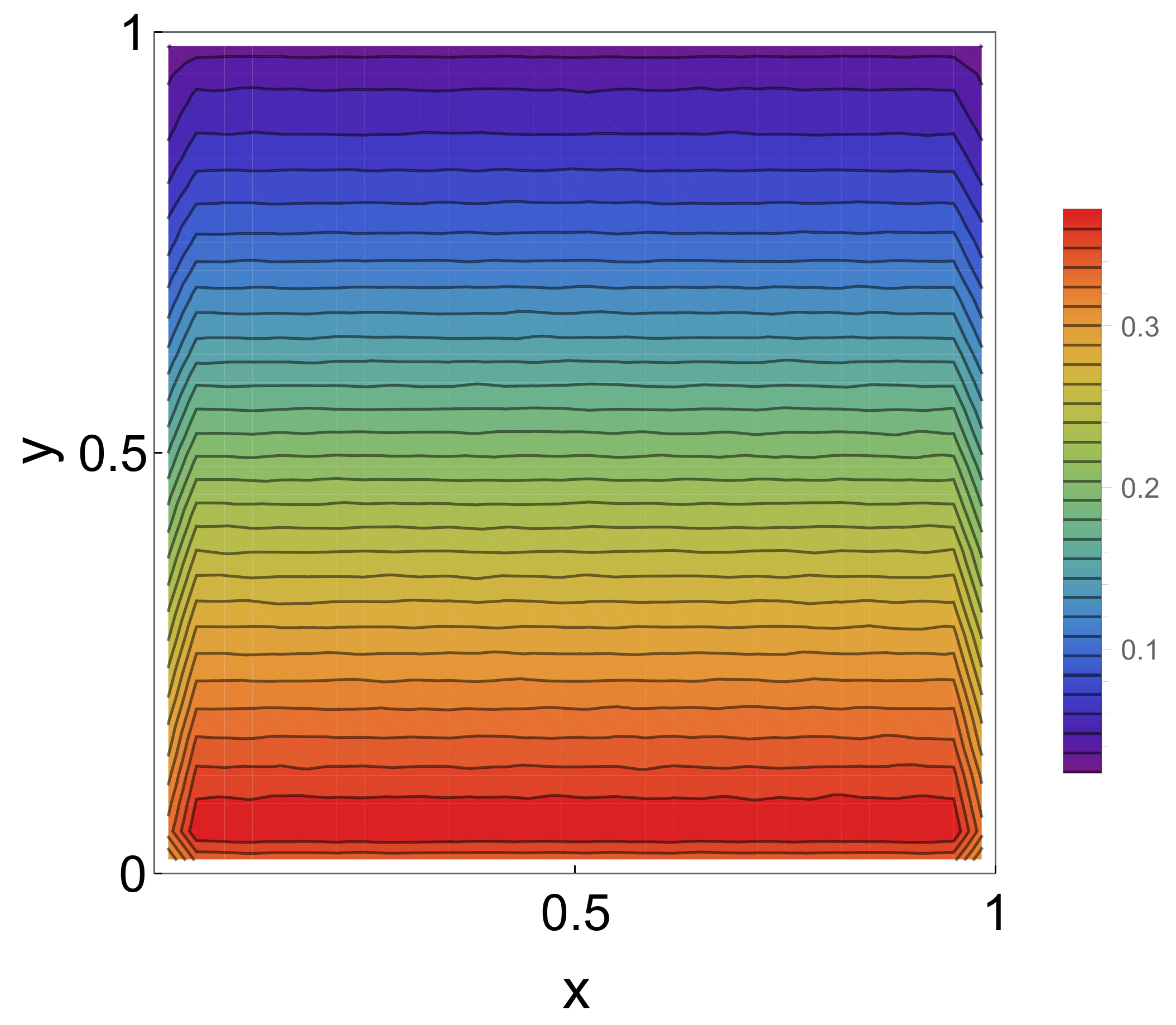}  %ro_profile.nb
\includegraphics[height=5cm,clip]{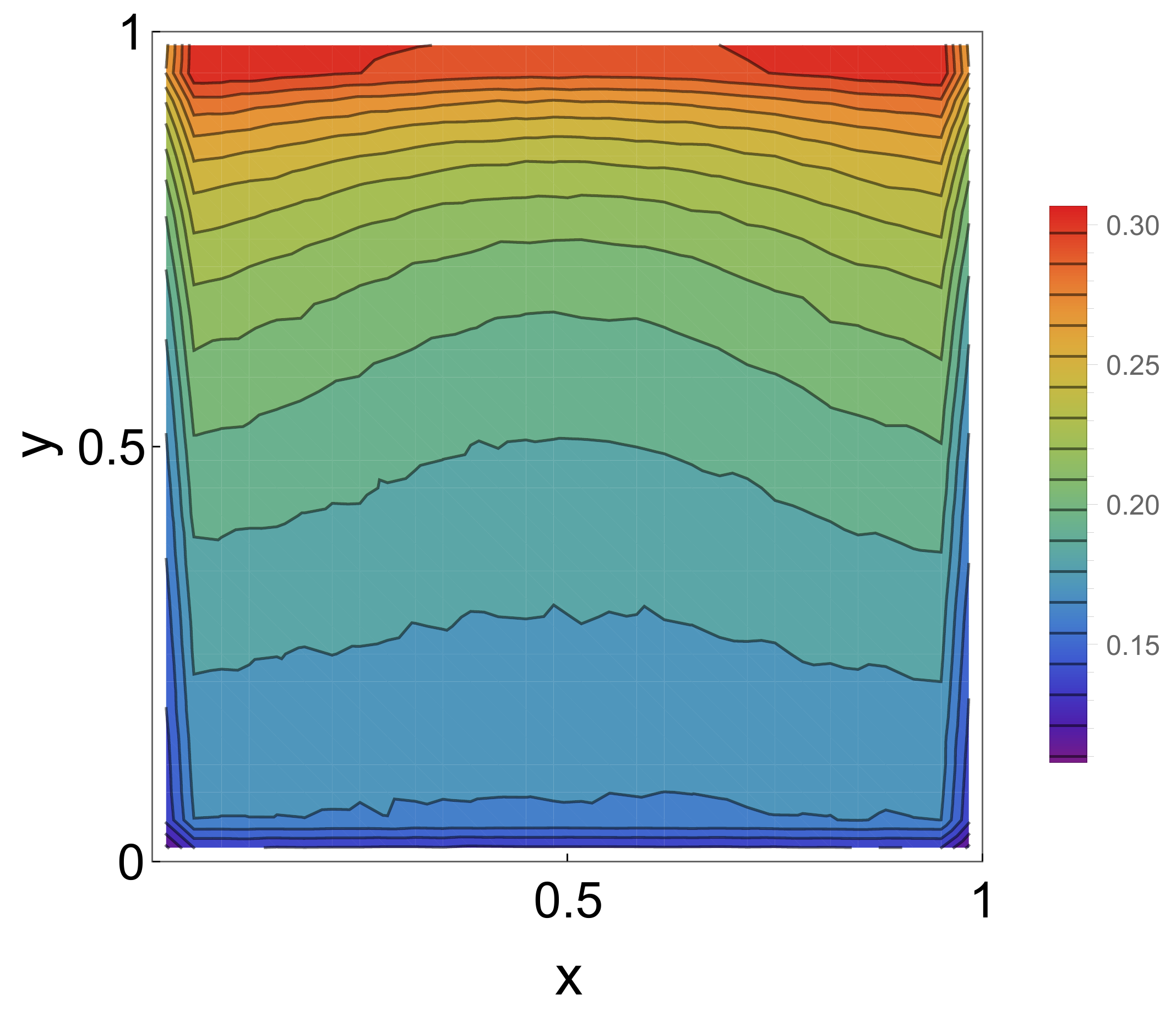}        %ro_profile_1.nb
\end{center}
\kern -0.5cm
\caption{Density field $\rho(x,y)$  defined in eq. \ref{ro} for $T_0=2$ and $T_0=18$ for left and right columns respectively ($g=10$). Below each of the 3D graphs there are  the corresponding contour plots to show the existence (or not) of a nontrivial spatial structure on $x$ direction. \label{ro0}}
\end{figure}

The local density is defined by eq. (\ref{ro}). In figure \ref{ro0} we present two typical measured density  fields for $g=10$: $T_0=2$ (left figures) and $T_0=18$ (right figures).  For $T_0=2$ we are at a non-convective state and we observe that a barometric type of density profile is present: due to the effect of the driven external field the particles tend to accumulate at the bottom of the system. We also see that (as in the temperature case) $\rho(x,y)$ does not present structure on the $x$-direction. However, for $T_0=18$ the density profile is inverted and now the low density region is at the bottom of the cage and the high density is at the top of it. Moreover, there is a clear spatial structure on the $x$-direction. We can see in both cases the boundary effects we already commented and that it results in a decrease of their value near the walls.

It seems reasonable to try to characterize the density marginal $\omega(y)$ defined by the average of its $x$ values:
\begin{equation}
\omega(y)=\frac{1}{N_A}\sum_{x\in A}\rho(x,y)\label{roy}
\end{equation}
where $A$ is the set of cells we use for the averaging and $N_A$ is the total number of cells. We discard the two first columns and rows near the boundaries. In our analysis we choose $A=\{(2n-1)/60 \vert n=3,\ldots,28\}$  and $N_A=26$.
\begin{figure}[h!]
\begin{center}
\includegraphics[height=4cm,clip]{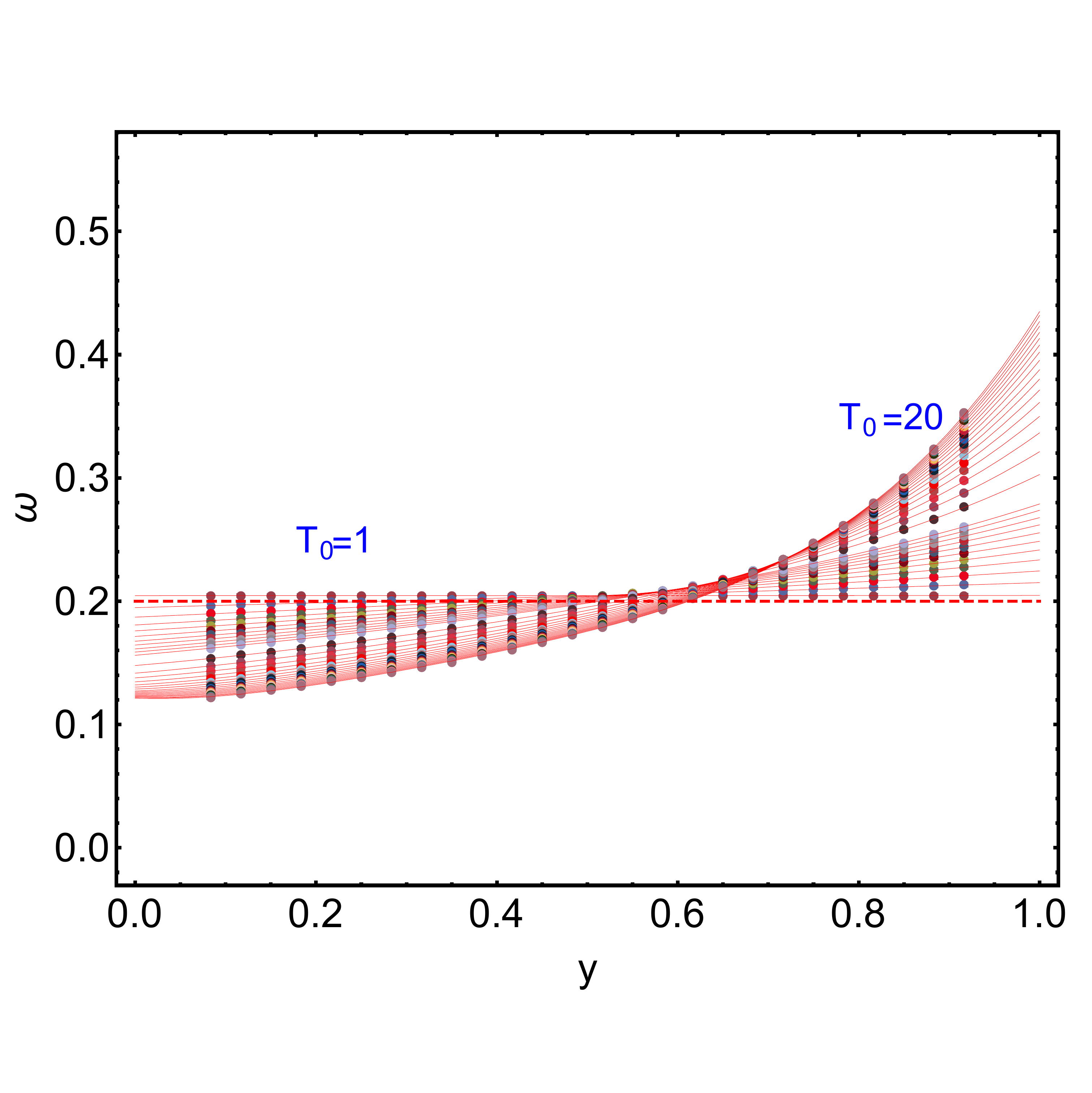}  %ro_profile_y_E0.nb
\includegraphics[height=4cm,clip]{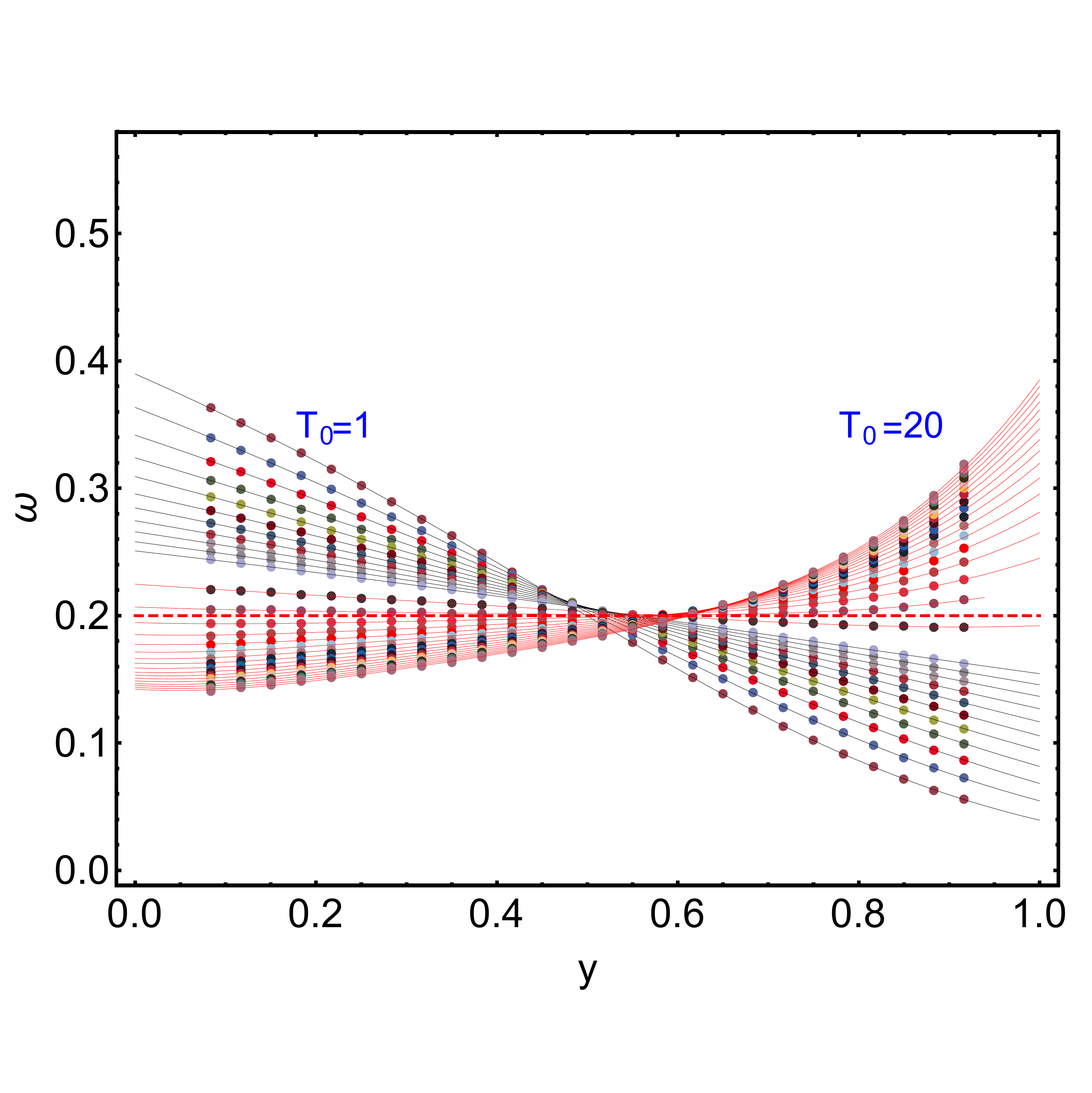}  %ro_profile_y_E5.nb
\includegraphics[height=4cm,clip]{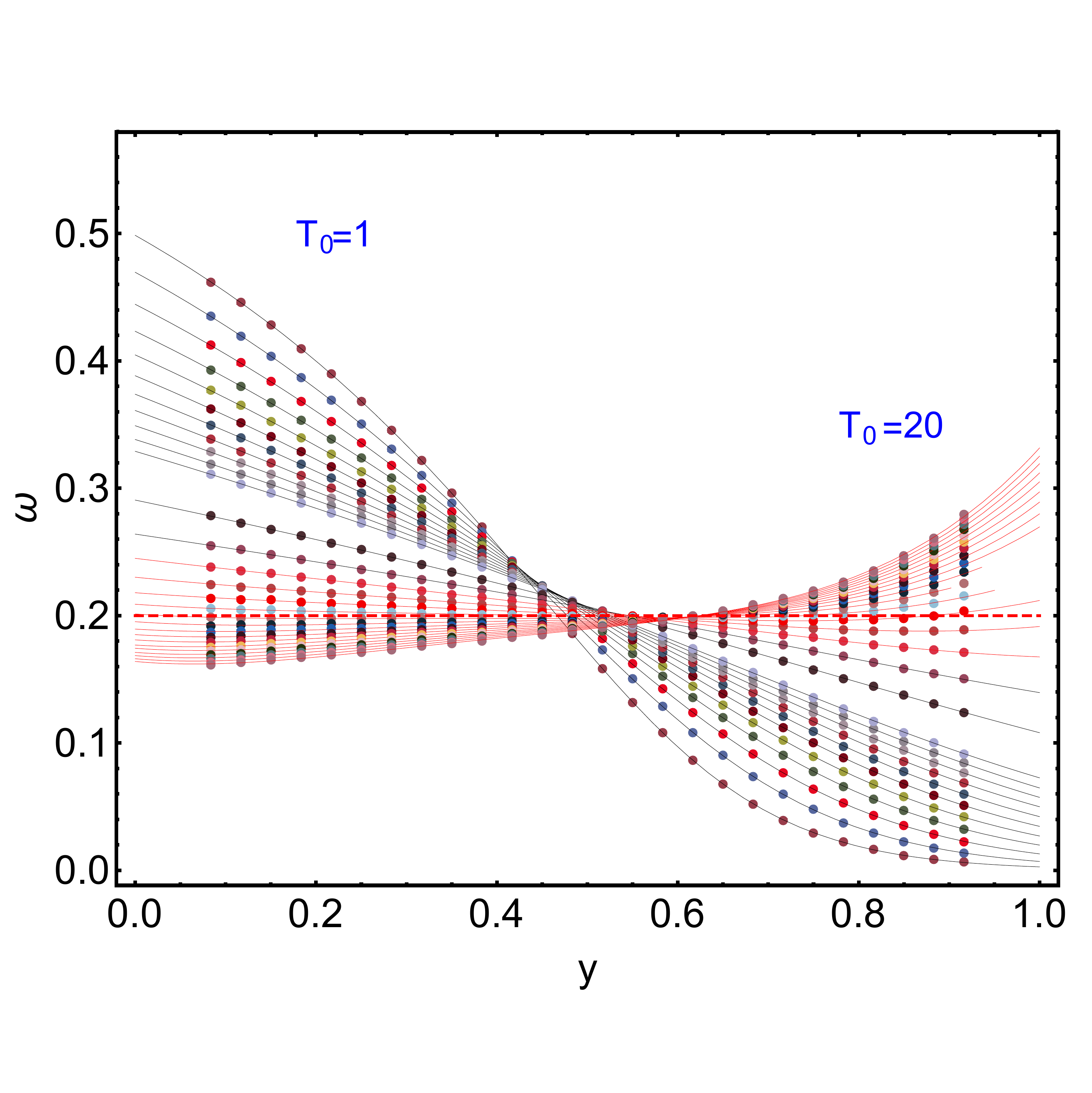}  %ro_profile_y_E10.nb
\includegraphics[height=4cm,clip]{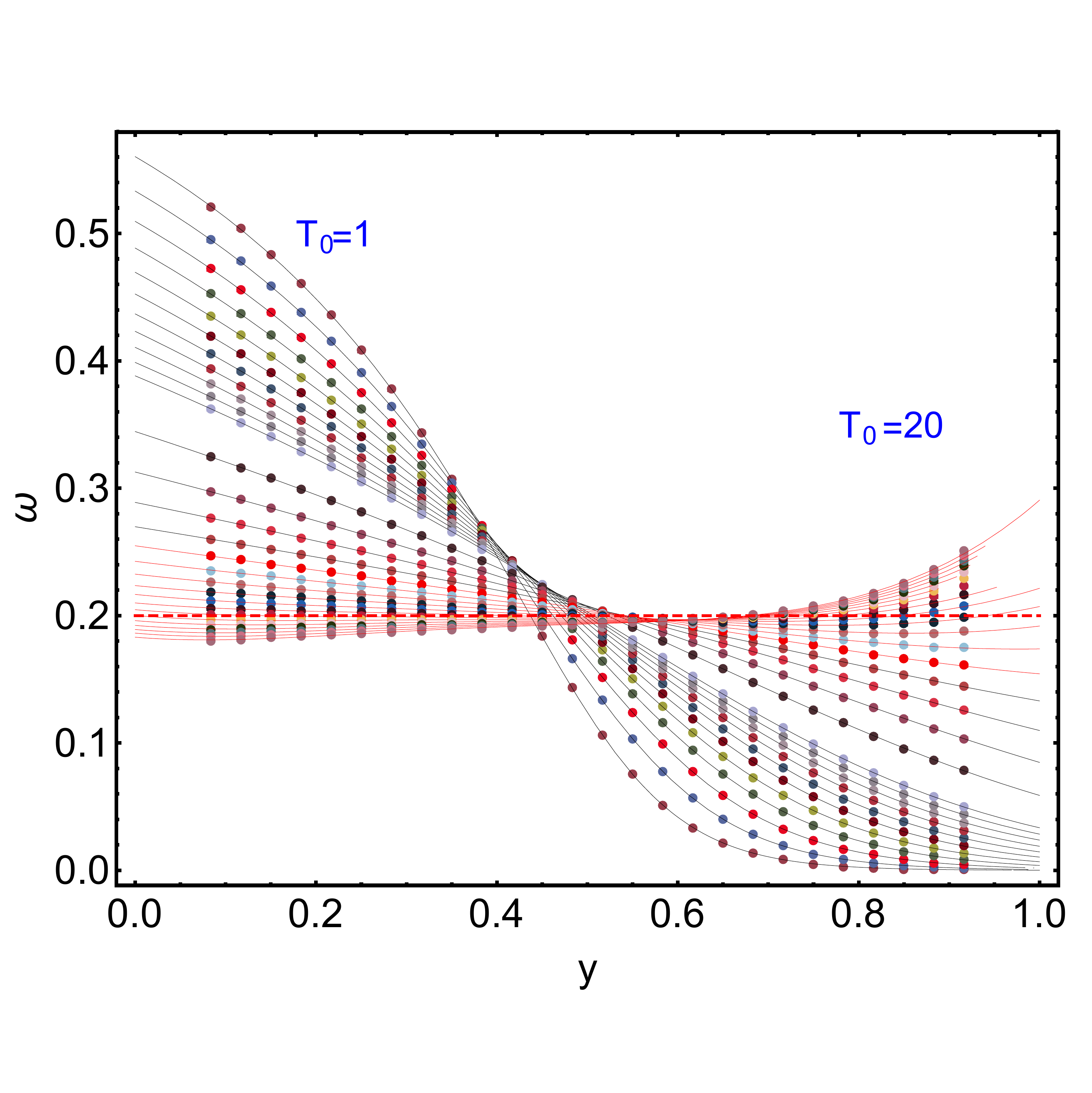}       %ro_profile_y_E15.nb
\end{center}
\kern -0.5cm
\caption{Average y-density profiles $\omega(y)$  defined in eq. \ref{roy} for $g=0$, $5$, $10$ and $15$ from left to right. At each figure we plot the points with $T_0=1, 1.2, 1.4, 1.6, 1.8, 2.0, 2.2, 2.4, 2.6, 2.8, 3.0, 4, 5, 6, 7, 8, 9, 10, 11, 12, 13, 14, 15, 16, 17, 18, 19, 20$ from top to bottom left. Error bars are included. Red thin lines are phenomenological fits of fourth order polynomials: $w=a_0+a_1 y+\ldots+a_4 y^4$. Black thin lines are equilibrium-like density profiles fits (see text). Dotted red line is the fixed global density used during the simulations \label{ro1}}
\end{figure}

We observe some interesting properties (see figure \ref{ro1}): (1) For $T_0=1$ the density profile presents a typical barometric equilibrium density distribution we already studied.  When the external field is on the bottom density is larger than the top density. (2) The barometric-like profile tends to disappear as we increase the value of $T_0$. (3) There is an inversion of the density for large values of $T_0$: the large density region is at the top of the cage. (4) For $g=0$ the density is always a monotonous increasing function for $T_0>1$.
(5) We do not see relation between being at the non-convective or convective state and the profile type. In fact, for $g=5$ and $T_0>4$ the profiles are monotonous increasing function of $y$ and for $T_0<4$ are decreasing functions of $y$, far from the appearance of the convective state which occurs at $T_c\simeq 2.1$ in this case. Similar thing occurs for $g=10$ and $g=15$. However, there is a clear relation with $T_{c,2}$: $w(y)$ is a monotonous decreasing function for $T_0<T_{c,2}$ and it looses such monotonicity for $T_0>T_{c,2}$ appearing first some increasing parts (positive slopes) for $y\simeq 1$ when $T_0$ has values near $T_{c,2}$ and $w$ being a monotonous increasing function for large enough values of $T_0$. Again it seems that the {\it strong convecting state} has the properties that a fluid in a convective state should have: low density fluid is at the bottom of the container and it tends to go up and the cold high density fluid is at the top and it tends to go down. The unclear issue is to fit this image for the region $T_c<T_0<T_{c,2}$ where we see convection and there is no density inversion.
(6) It is curious that for $T_0<T_{c,2}$ the best fitting function is the equilibrium like one with varying constants: $y=a(0)+a(1)\log w+a(2)\log(1-w)+a(3)/(1-w)+a(4)/(1-w)^2$. For $T_0>T_{c,2}$ the best fit is a fourth order polynomial.

\begin{figure}[h!]
\begin{center}
\includegraphics[height=5cm,clip]{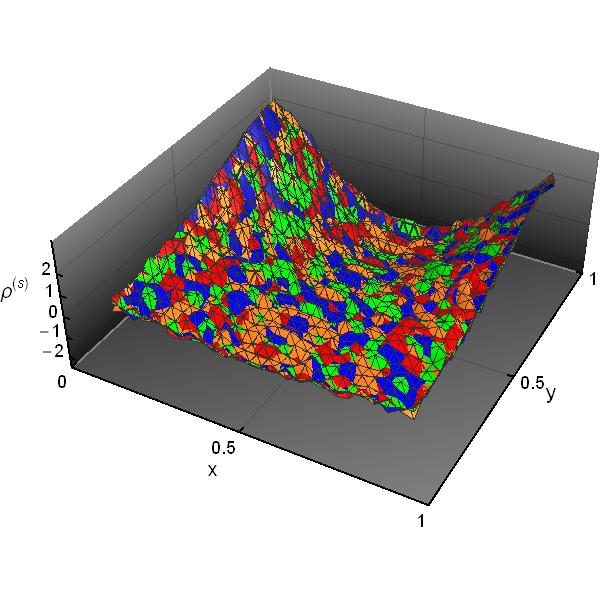}   %ro_field_show.nb
\includegraphics[height=5cm,clip]{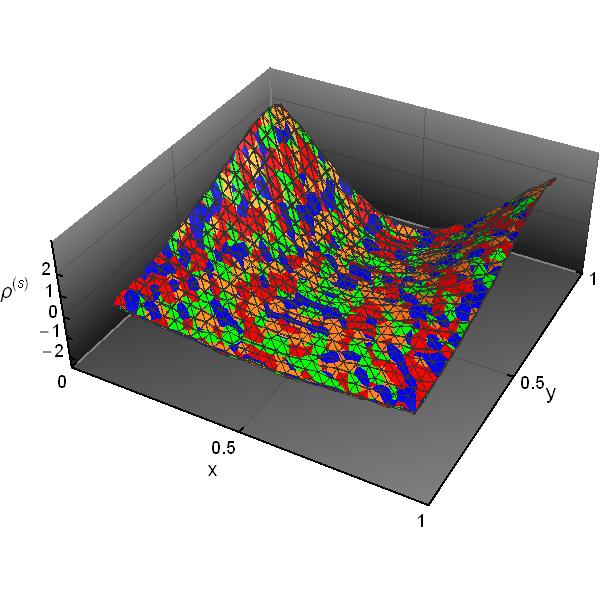}   
\includegraphics[height=5cm,clip]{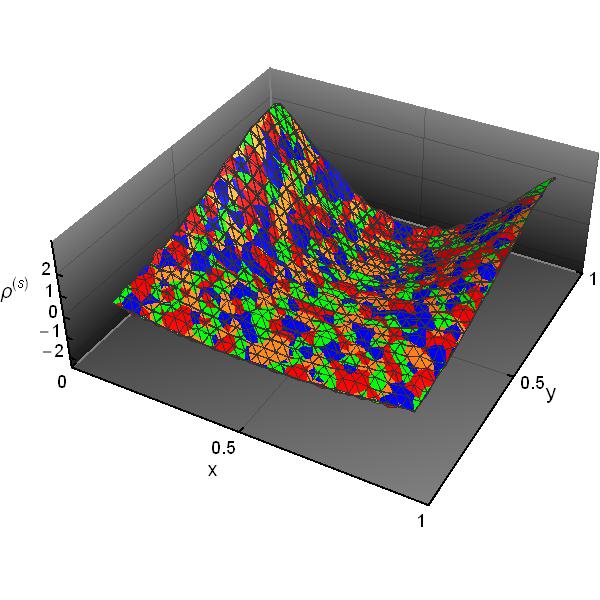}   
\end{center}
\caption{Superposition of the scaled excess of density fields, $\rho^{(s)}$ with $T_0=17$, $18$, $19$ and $20$ (red, blue, green and orange colors) and for different $g$ values (from left to right: $g=5$, $10$ and $15$).    \label{rofieldsuper}}
\end{figure}

\begin{figure}[h!]
\begin{center}
\includegraphics[height=6cm,clip]{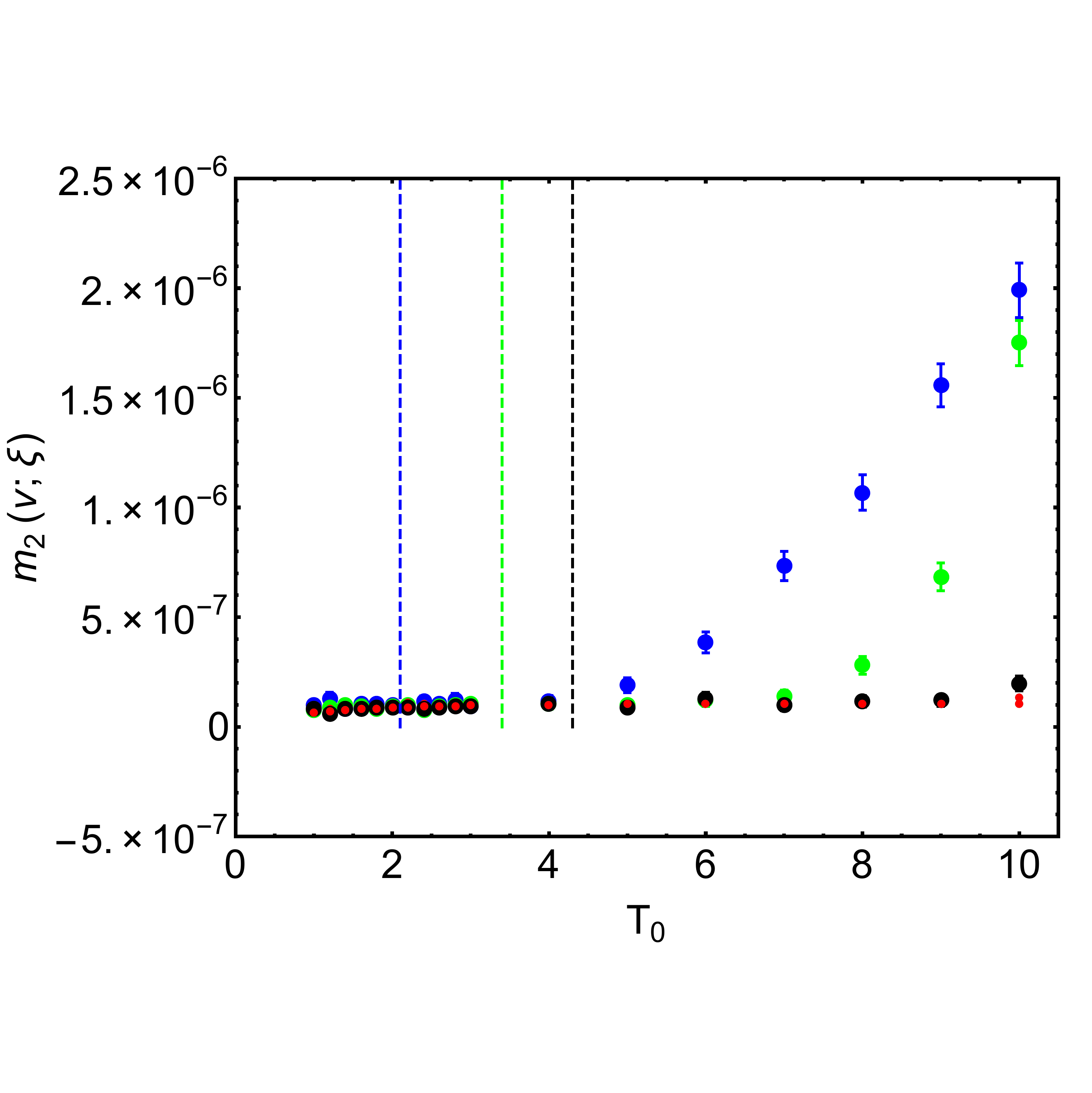}   %ro_moments.nb
\includegraphics[height=6cm,clip]{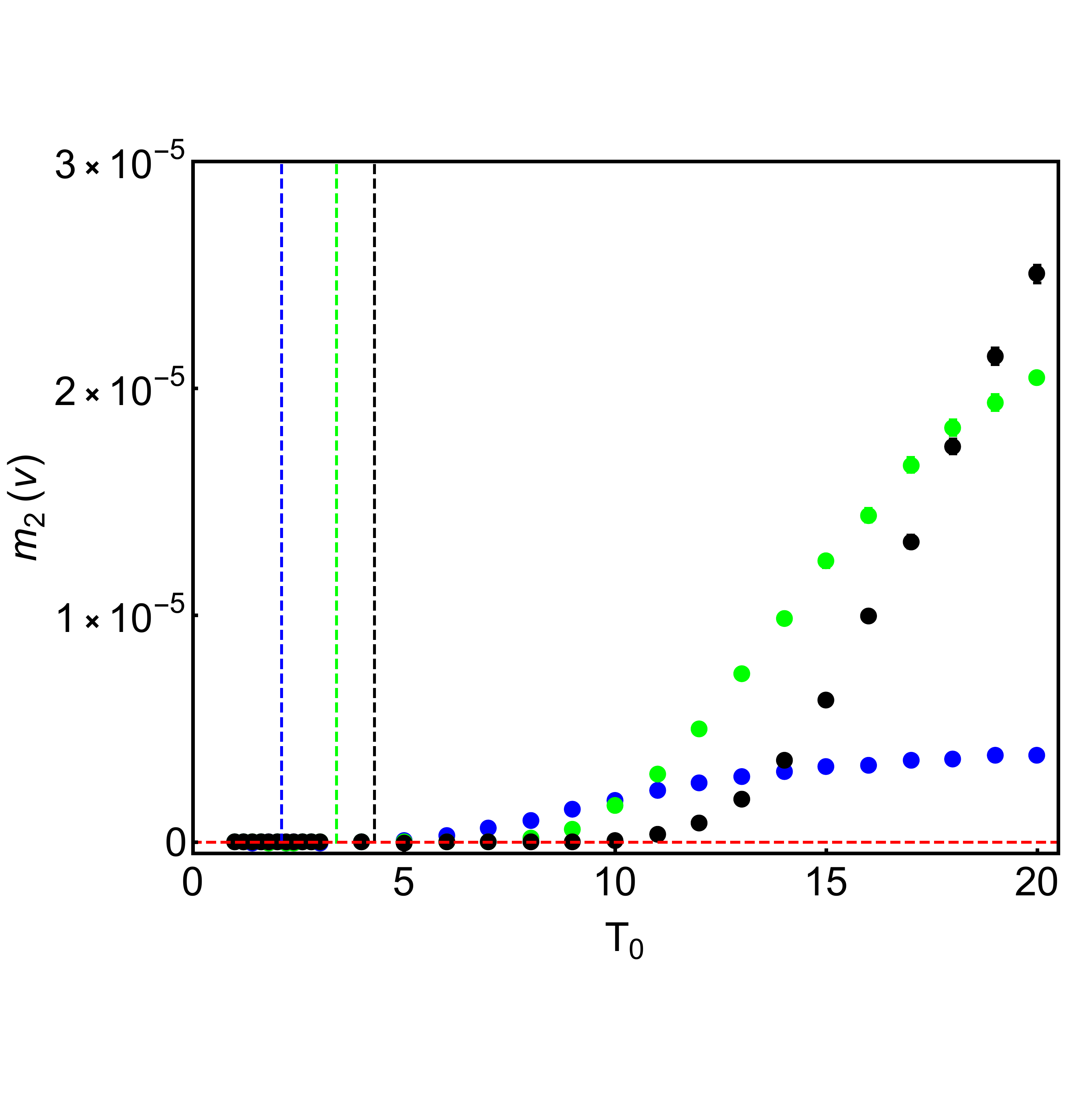}   
\end{center}
\kern -1cm
\caption{Left: Second moment of the measured excess of density field: $v(x,y;\xi)=\rho(x,y;\xi)-w(y;\xi)$ for $T_0\in[1,10]$ and $g=5$ (blue dots), $g=10$ (green dots) and $g=15$ (black dots).  The red small dots are the $\langle A_2(\xi)\rangle$.  Right: $m_2(v)=m_2(v;\xi)-\langle A_2(\xi)\rangle$.
   \label{romom}}
\end{figure}
\begin{figure}[h!]
\begin{center}
\includegraphics[height=6cm,clip]{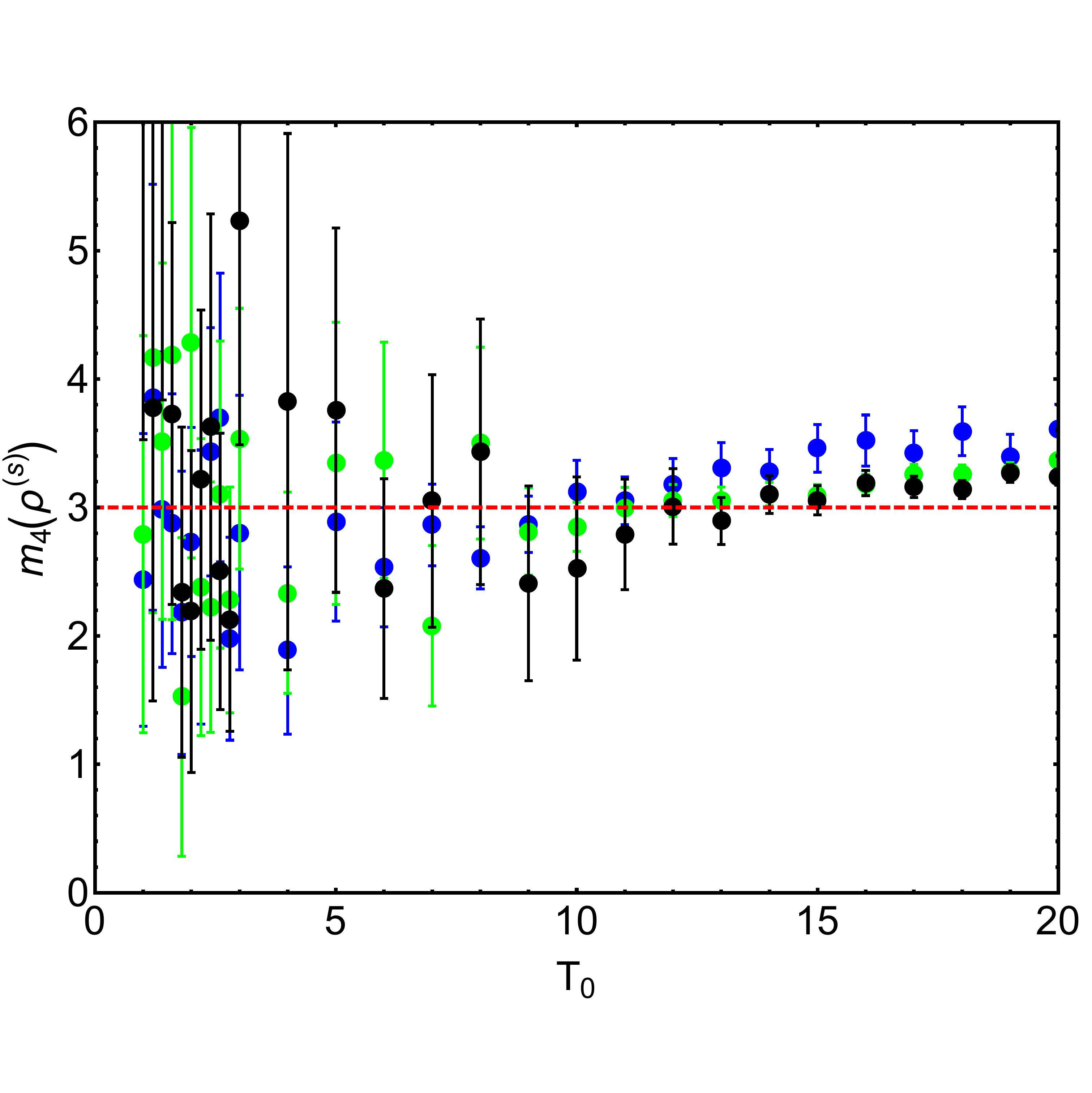}   %ro_moments.nb
\end{center}
\kern -1cm
\caption{Fourth moment of the scaled excess of density field,  $m_4(\rho^{(s)})$, as a function of $T_0$. 
   \label{romom4}}
\end{figure}

\begin{figure}[h!]
\begin{center}
\includegraphics[height=5cm,clip]{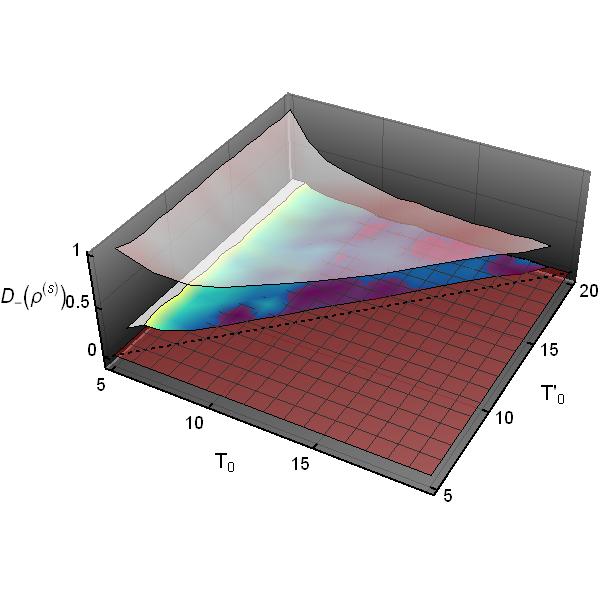}   %ro_E5new.nb
\includegraphics[height=5cm,clip]{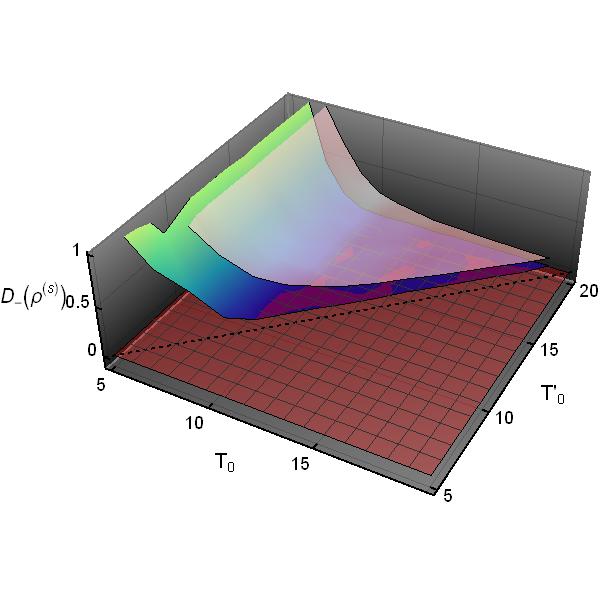}%ro_E10new.nb
\includegraphics[height=5cm,clip]{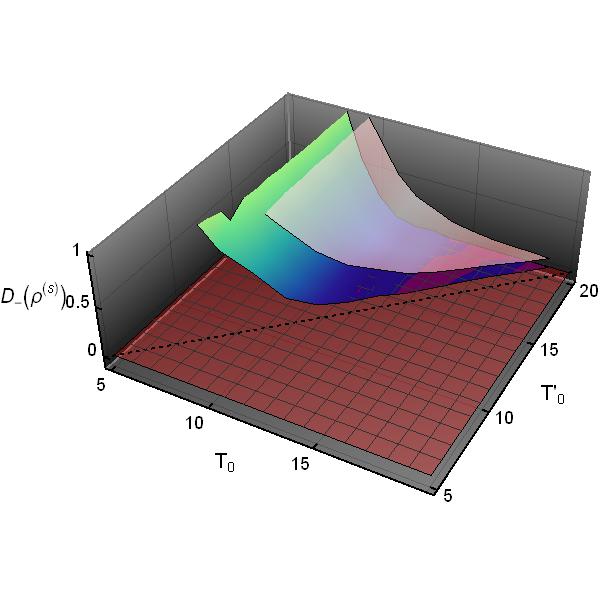}%ro_E15new.nb
\newline\vglue -1cm
\includegraphics[height=4.5cm,clip]{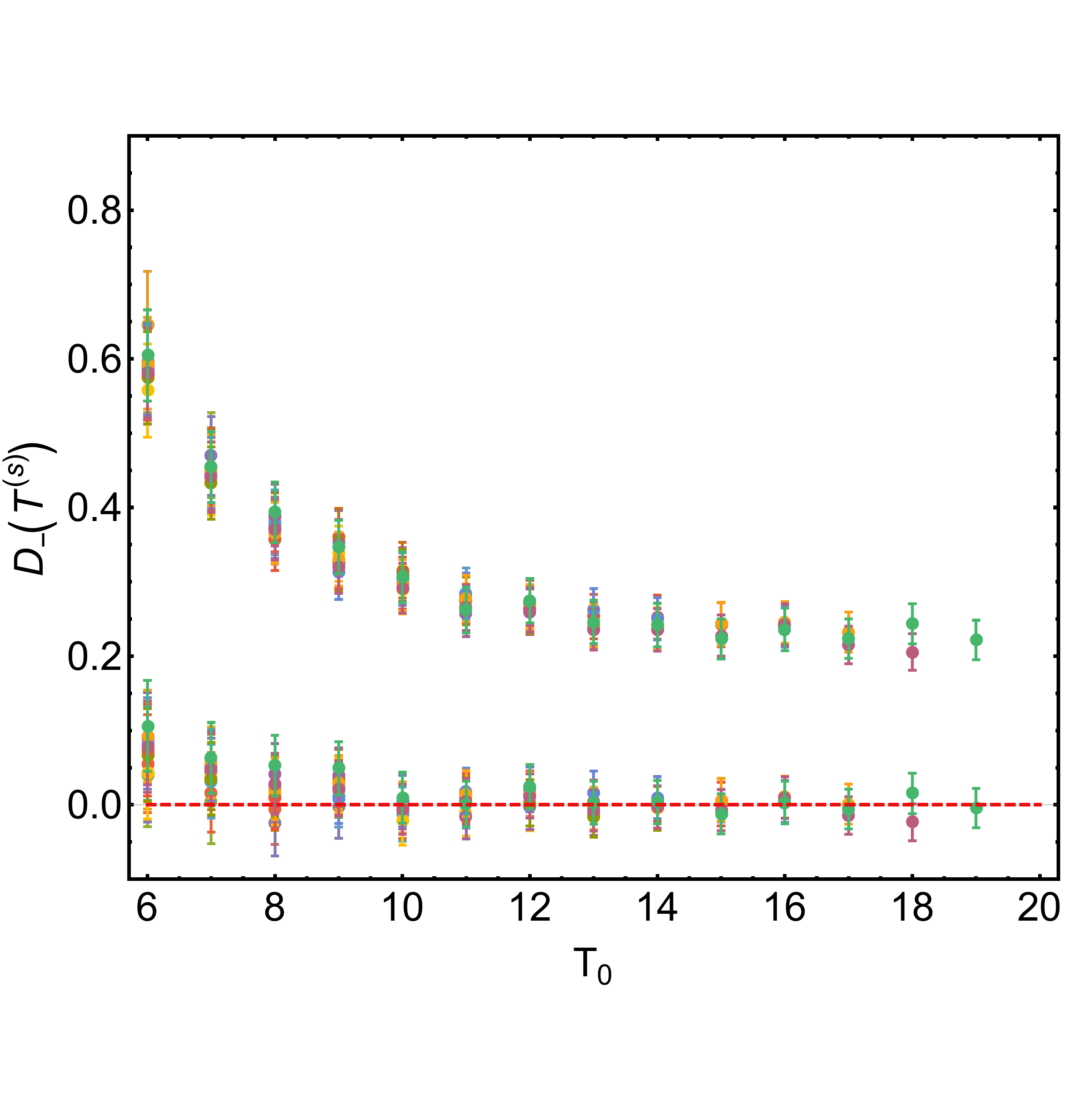}   %ro_E5new.nb
\includegraphics[height=4.5cm,clip]{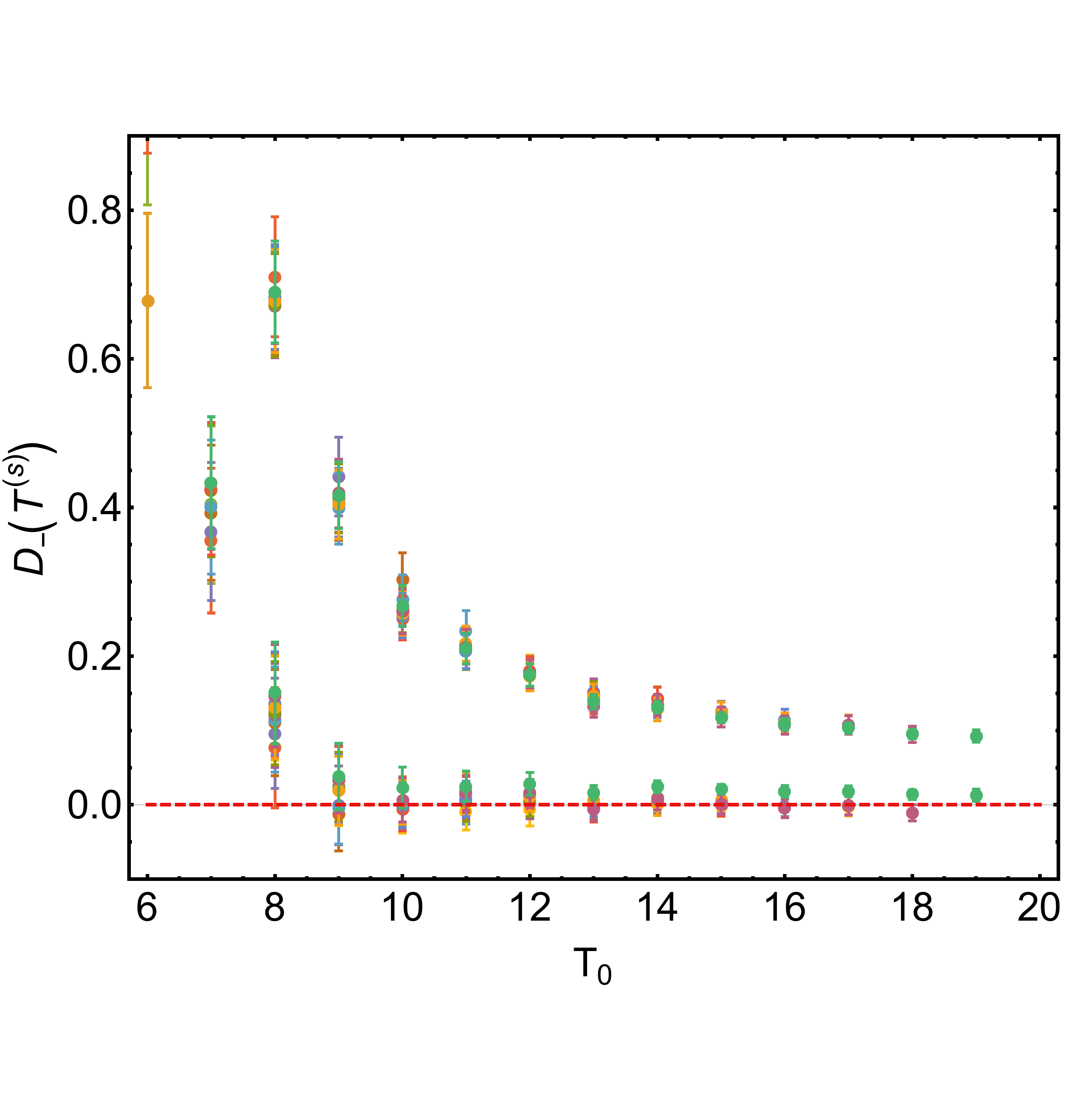}%ro_E10new.nb
\includegraphics[height=4.5cm,clip]{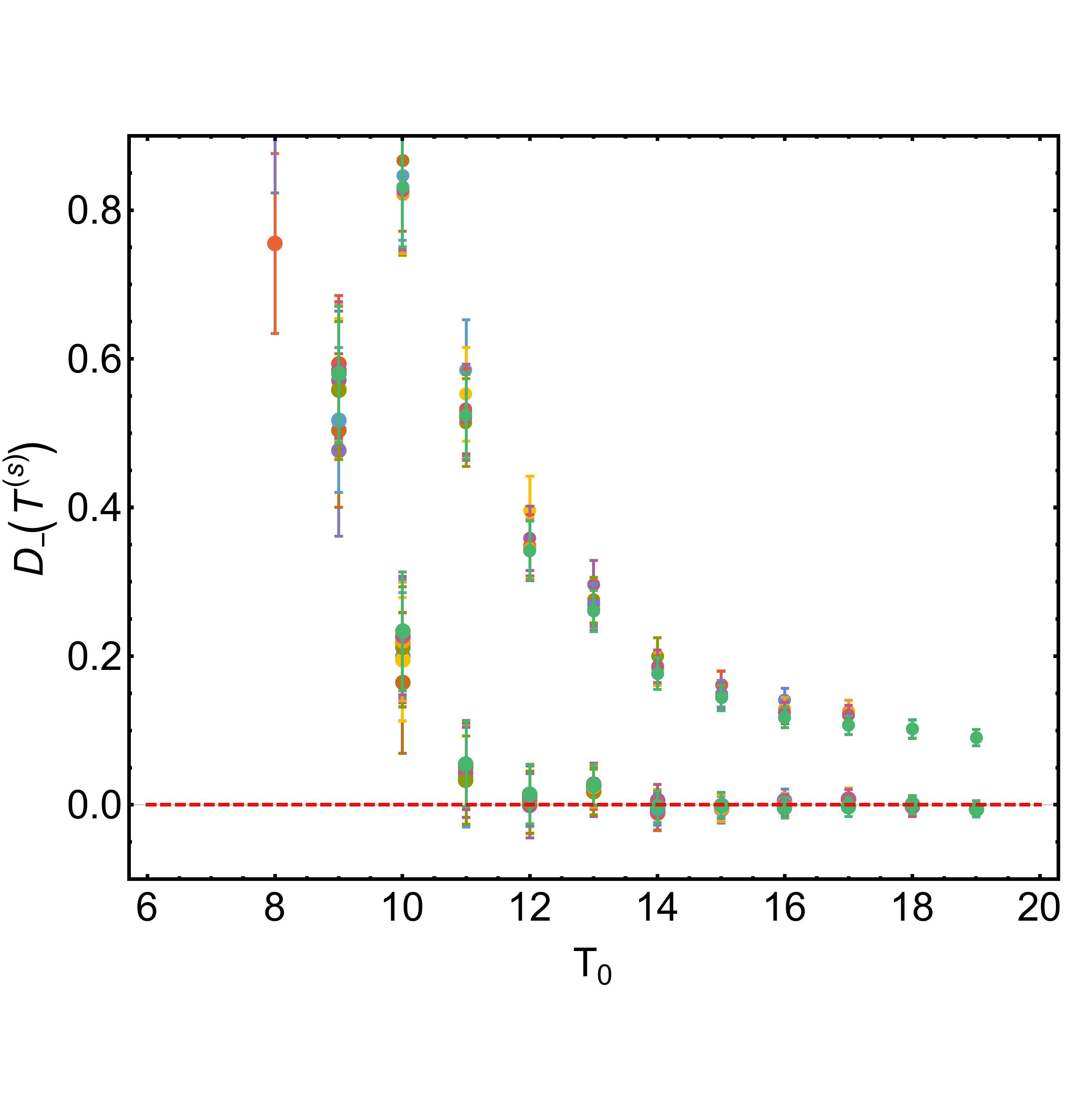}%ro_E15new.nb
\end{center}
\kern -1.cm
\caption{Marginal distance $D_-$ (see text) between the density field scaled configurations $\rho^{(s)}$ for $g=5$ (figures left), $g=10$ (figures center) and $g=15$ (figures right). for a given $g$ value. We only plot pairs of scaled configurations with $T_0$ and $T_0'$ such that $T_0<T_0'$. Top figures: Pink surface are the bare distances $D(\rho^{(s)};\xi)$. Yellow-green surfaces are the distances  after applying the noise correction term $\Lambda$ (see text). Bottom figures:  Same as top but including error bars. Each color corresponds to a given $T_0'$ value. \label{distancero}}
\end{figure}

\begin{figure}[h!]
\begin{center}
\includegraphics[height=5cm,clip]{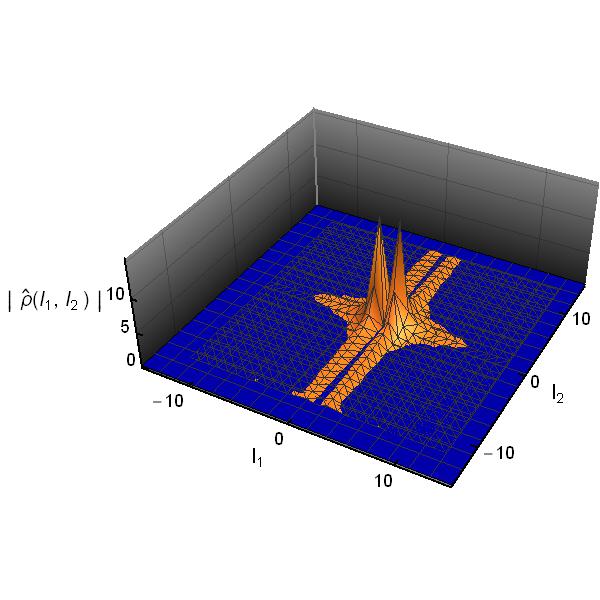}   %ro_E5new.nb
\includegraphics[height=5cm,clip]{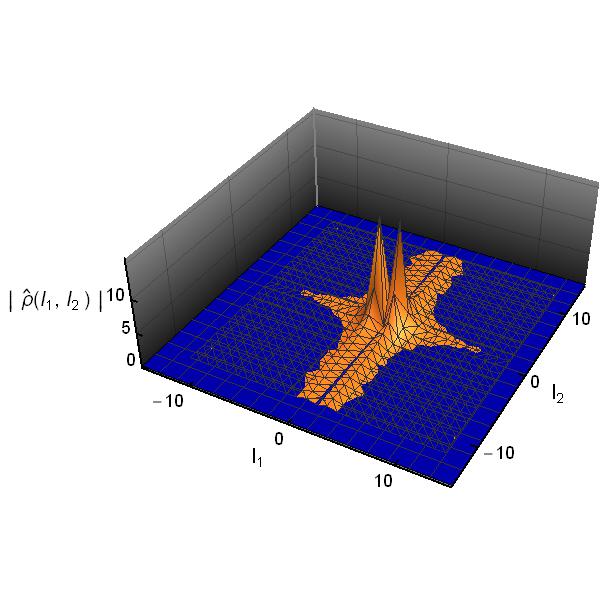}%ro_E10new.nb
\includegraphics[height=5cm,clip]{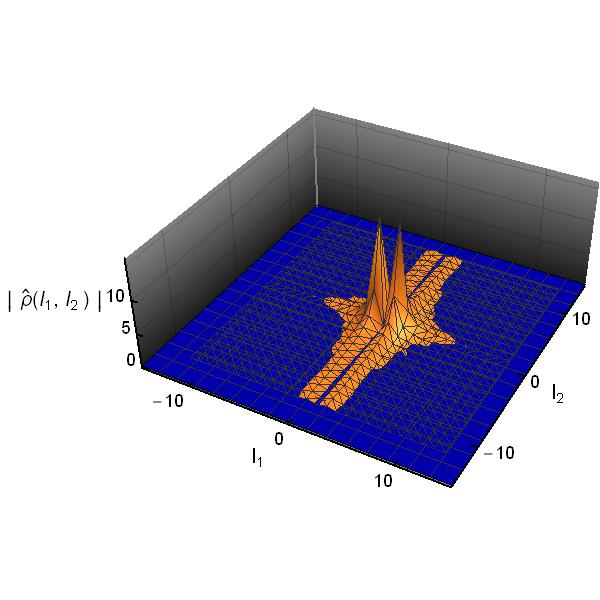}%ro_E15new.nb
\newline\vglue -1cm
\includegraphics[height=5cm,clip]{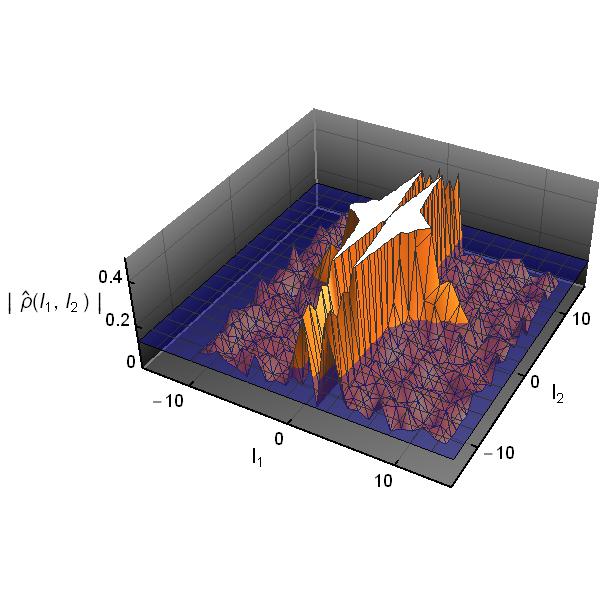}   %ro_E5new.nb
\includegraphics[height=5cm,clip]{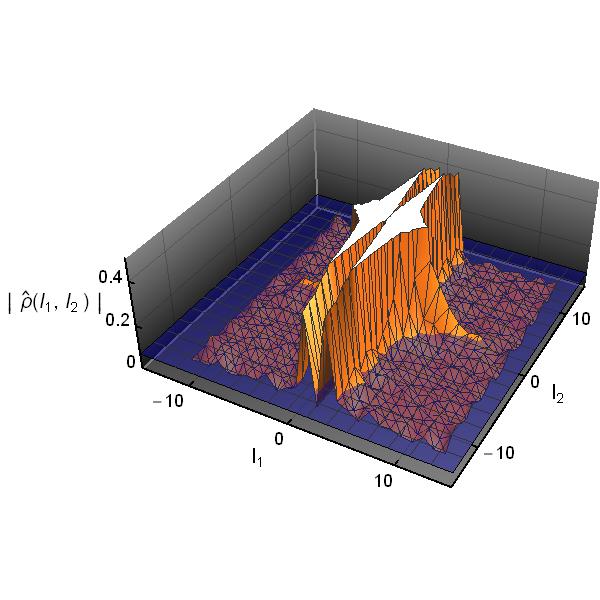}%ro_E10new.nb
\includegraphics[height=5cm,clip]{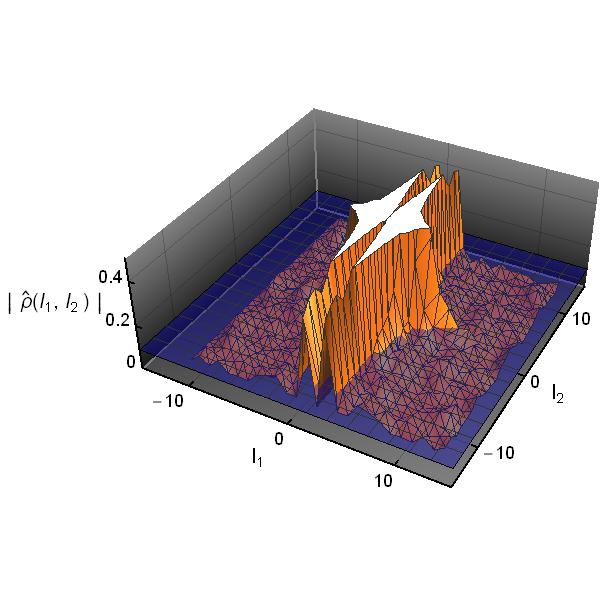}%ro_E15new.nb
\end{center}
\kern -1.cm
\caption{Modulus of the Discrete Fourier Transform obtained by averaging the scaled configurations, $\rho^{(s)}(x,y)$, from $T_0=14,\ldots, 20$ for $g=5$ (left figures), $g=10$ (center figures) and $g=15$ (right figures).  Points below the blue surfaces are discarded and only points above them are used to the subsequent Inverse Fourier Transform to get a smoothed field. Top figures show the modes used in the Discrete Inverse Fourier Transform and bottom ones the detailed behavior of the discarded noisy modes.
 \label{Fourierro}}
\end{figure}

\begin{table}[h!]
\begin{center}
\resizebox*{!}{3cm}{ 
\begin{tabular}{|c|c|}
\hline
$g$&$\vert \hat \rho^{(s)}\vert$\\ \hline
\hline
5&0.13\\ \hline
10&0.07\\ \hline
15&0.10\\ \hline
\end{tabular}}
\end{center}
\caption{Cut-off values for the $\rho^{(s)}$ averaged. The modes of the Fourier Transform of the field with modulus less than the corresponding cut-off value are discarded. \label{cut2ro}}
\end{table}

\begin{figure}[h!]
\begin{center}
\includegraphics[height=5cm,clip]{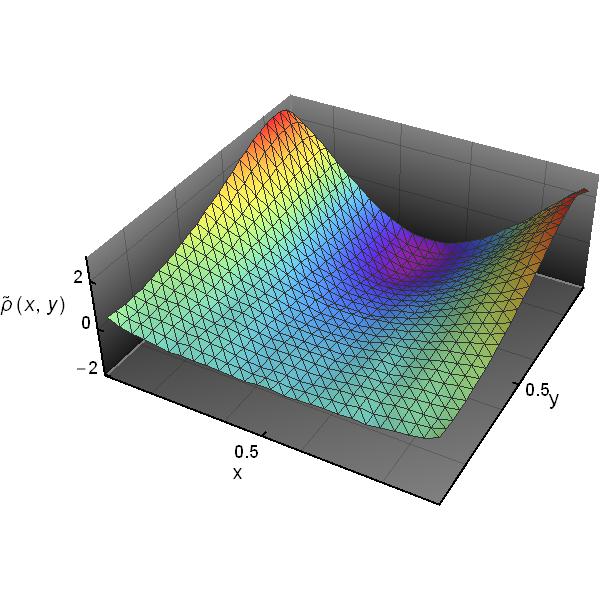}   %ro_E5new.nb
\includegraphics[height=5cm,clip]{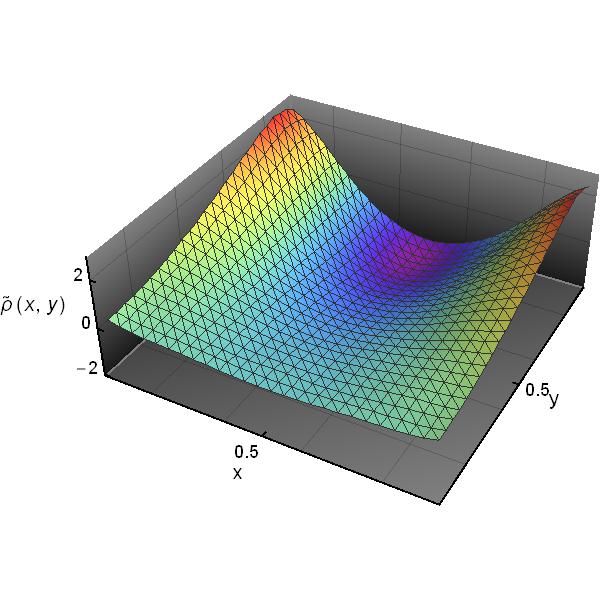}  %ro_E10new.nb
\includegraphics[height=5cm,clip]{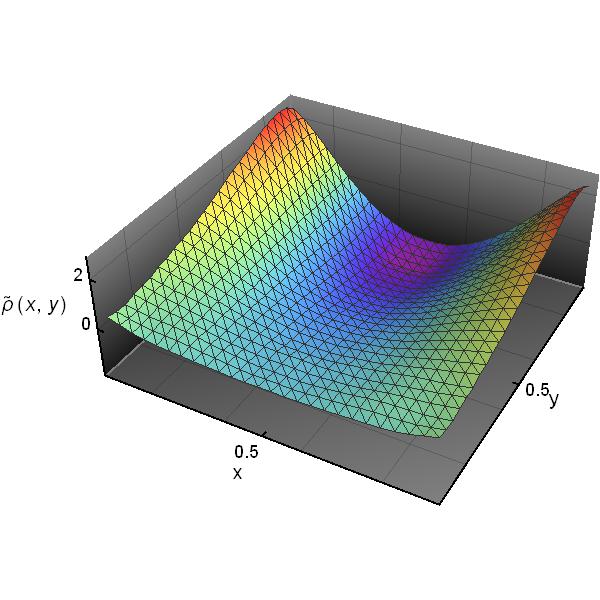}  %ro_E15new.nb
\end{center}
\kern -1.cm
\caption{Universal fields $\tilde\rho(x,y)$   for $g=5$, $10$ and $15$ from left to right. 
 \label{universalro}}
\end{figure}
\begin{figure}[h!]
\begin{center}
\includegraphics[height=5cm,clip]{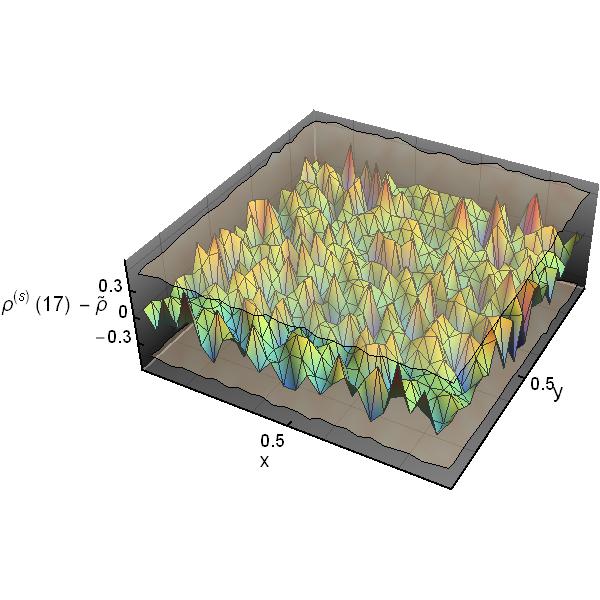}   %ro_E5new.nb
\includegraphics[height=5cm,clip]{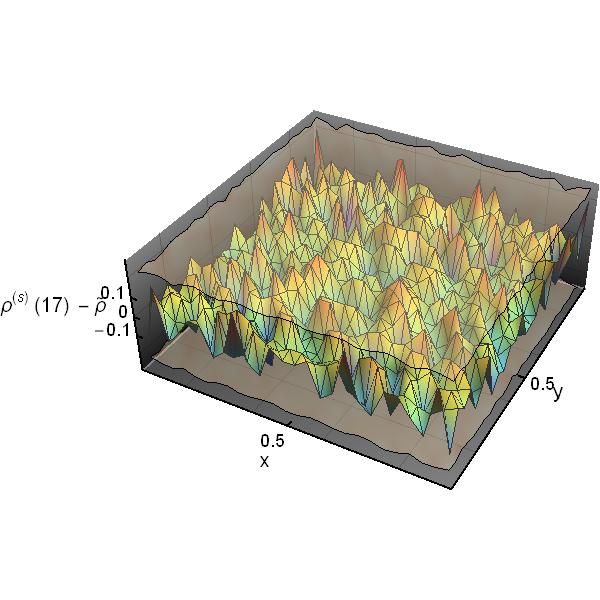} %ro_E10new.nb
\includegraphics[height=5cm,clip]{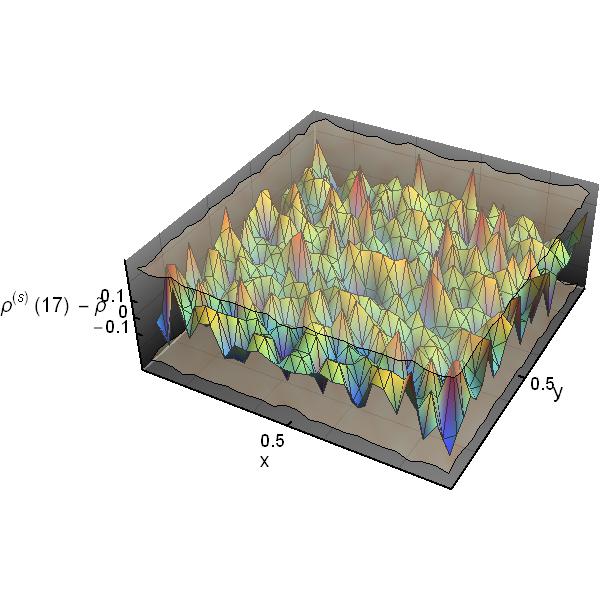} %ro_E15new.nb
\end{center}
\kern -1.cm
\caption{Difference between the scaled field $\rho^{(s)}(x,y)$ for $T_0=17$ and the corresponding universal field $\tilde\rho(x,y)$  for $g=5$, $10$ and $15$ from left to right. The gray surfaces are the data error bars of the scaled density fields.
 \label{flucturo}}
\end{figure}
\begin{figure}[h!]
\begin{center}
\includegraphics[height=5cm,clip]{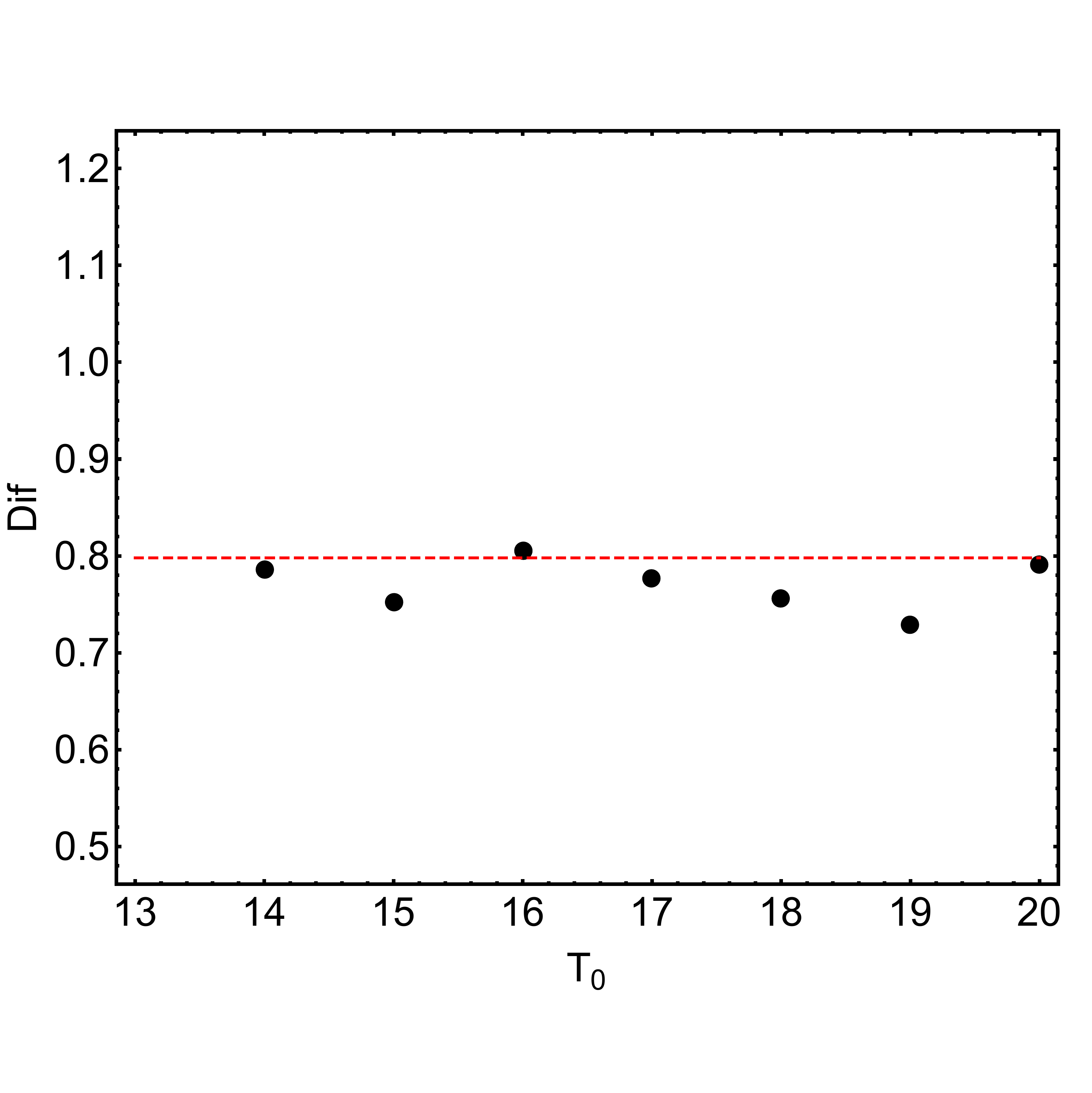}   %ro_E5new.nb
\includegraphics[height=5cm,clip]{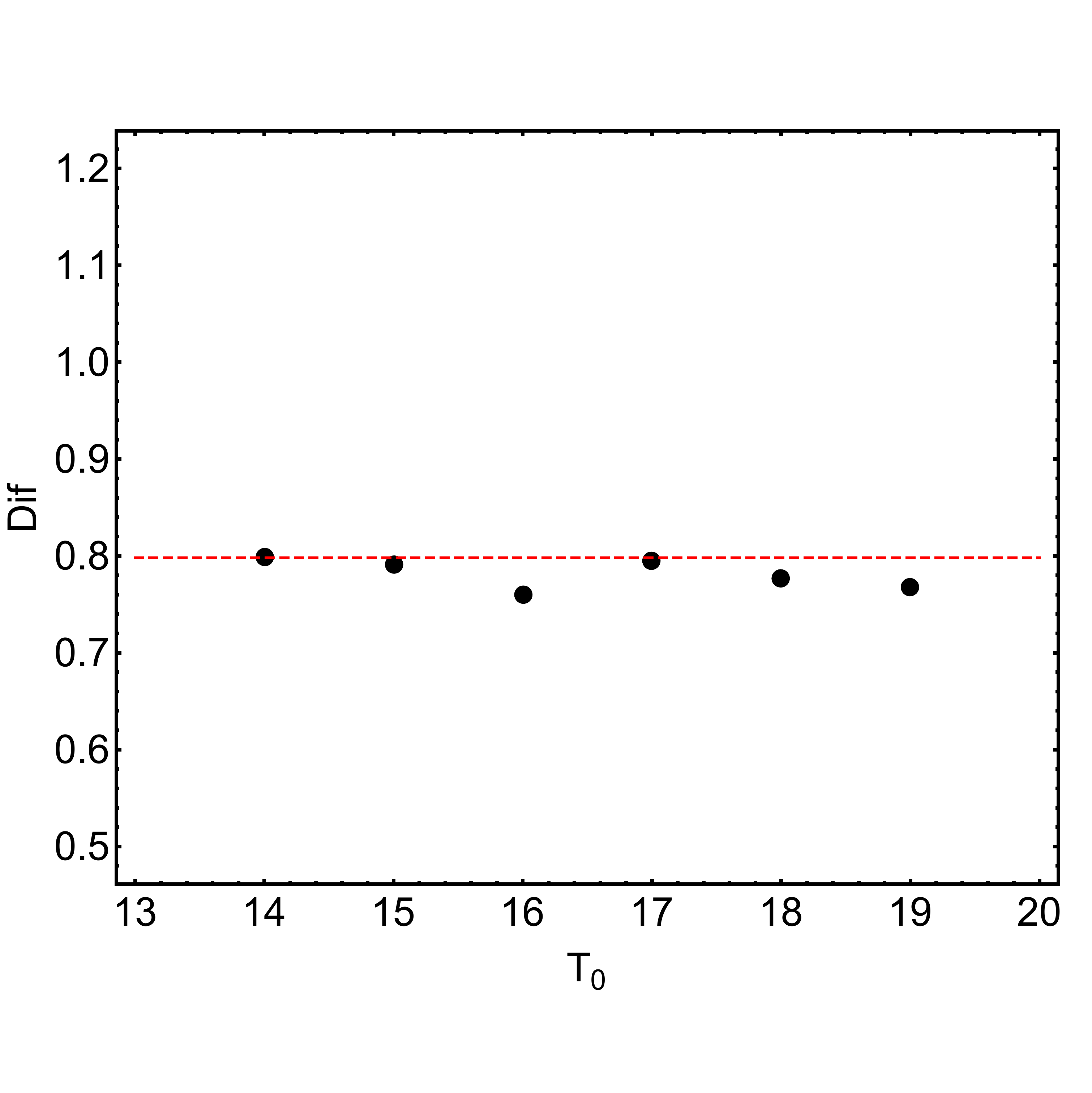} %ro_E10new.nb
\includegraphics[height=5cm,clip]{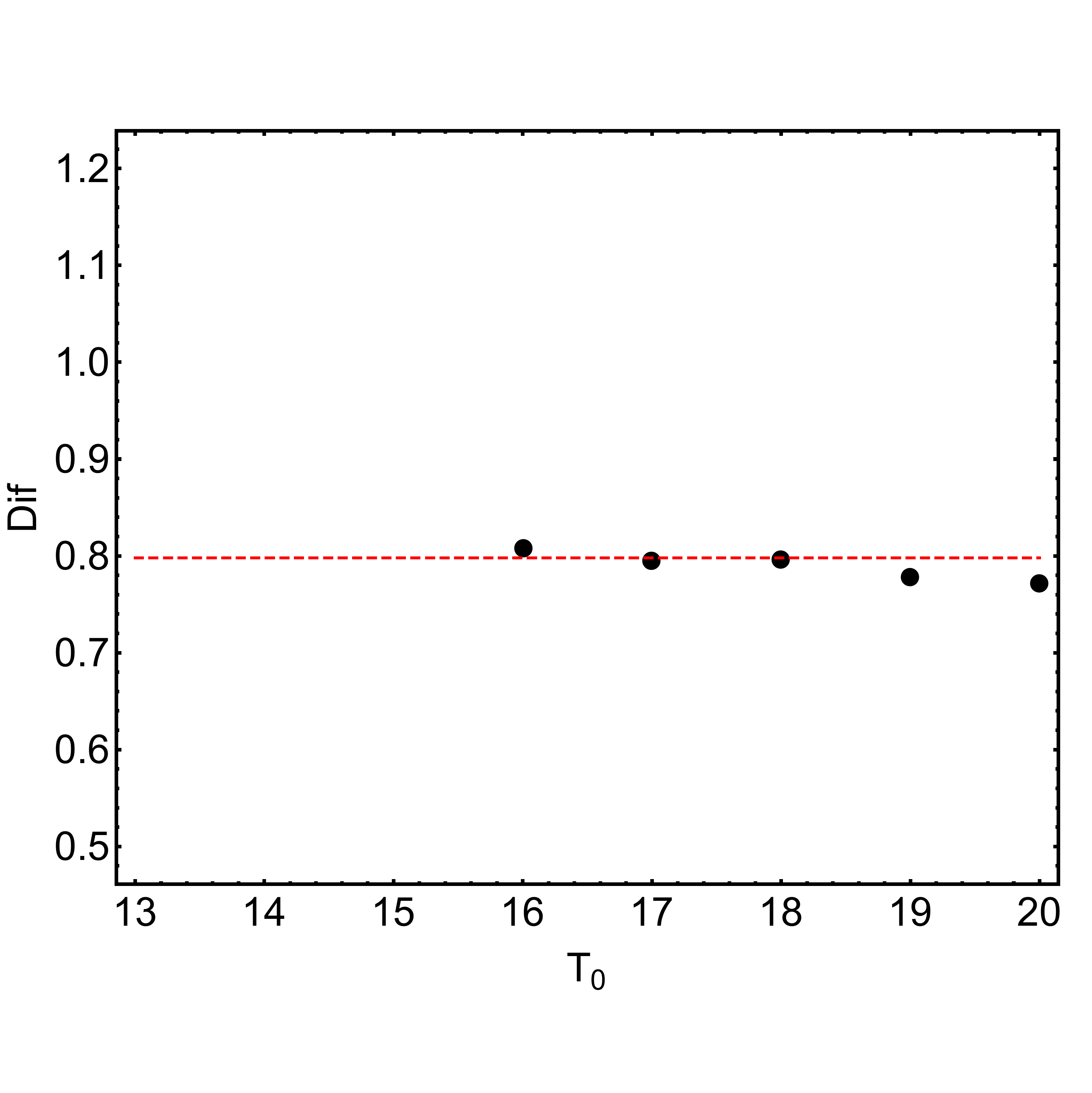} %ro_E15new.nb
\end{center}
\kern -1.cm
\caption{Averaged ratio between the difference between the scaled field $\rho^{(s)}(x,y)$  and the corresponding universal field $\tilde\rho(x,y)$ with respect its variance normalized with the profile variance for $g=5$, $10$ and $15$ from left to right. 
 \label{flucturo2}}
\end{figure}

\begin{figure}[h!]
\begin{center}
\includegraphics[height=5cm,clip]{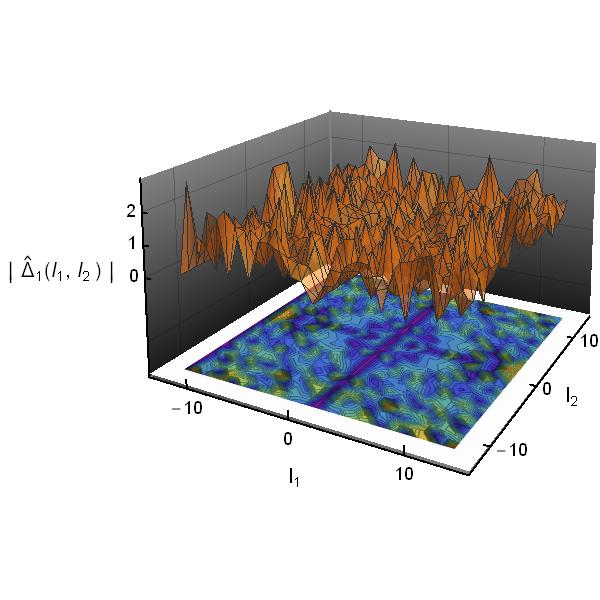}   %ro_E5new.nb
\includegraphics[height=5cm,clip]{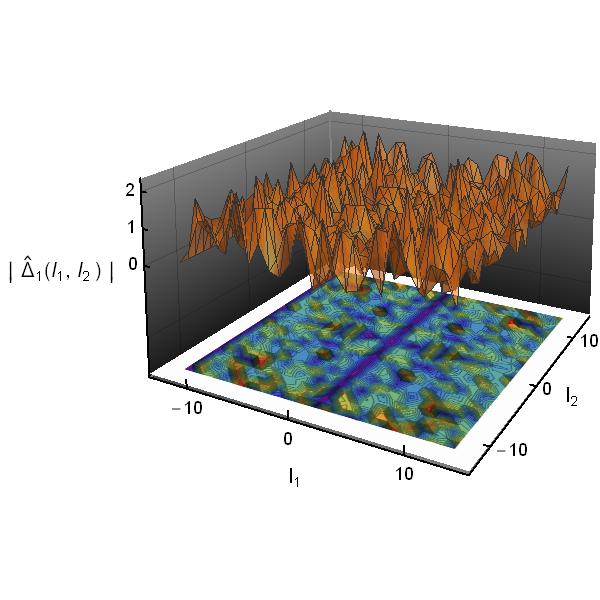} %ro_E10new.nb
\includegraphics[height=5cm,clip]{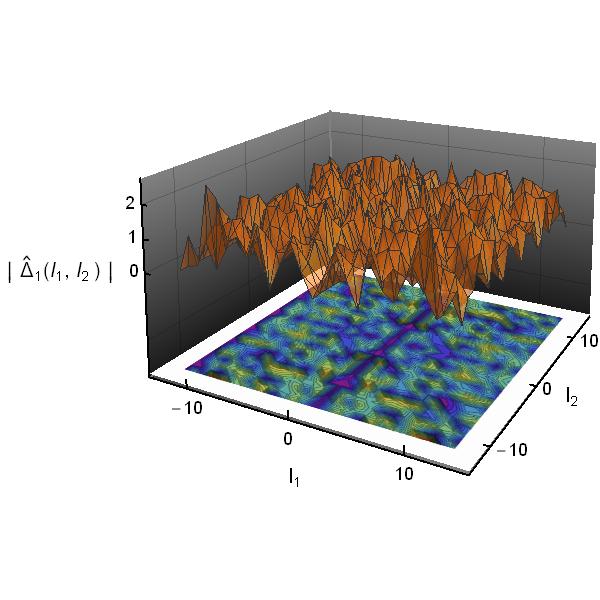} %ro_E15new.nb
\end{center}
\kern -1.cm
\caption{Modulus of the Fourier Transform of the difference  between the scaled field $\rho^{(s)}(x,y)$ for $T_0=17$ and the corresponding universal field $\tilde\rho(x,y)$  for $g=5$, $10$ and $15$ from left to right. 
 \label{flucturo3}}
\end{figure}

Again we can study now the scaled excess density field: :
\begin{equation}
\rho^{(s)}(x,y)=\frac{\rho(x,y)-w(y)}{\sigma(\rho)}
\end{equation}
where
\begin{equation}
\sigma(\rho)^2=\frac{1}{N_C}\sum_{(x,y)}\left[\rho(x,y)-w(y)\right]^2 \quad,\quad w(y)=\frac{1}{L_C}\sum_x \rho(x,y)
\end{equation}
and $N_C=L_C^2$ and $L_C=26$  in our case and the sums exclude the two nearest boundary rows and columns.
At this point the analysis follows the same steps we did with the temperature field. We show in figures \ref{rofieldsuper} the superposition of the scaled density fields $\rho^{(s)}$ for $T_0=17$, $18$, $19$ and $20$ we see again that all of them seem to follow the same surface without any systematic deviation. In figures \ref{romom} we present the behavior of $\sigma(\rho)^2$. We observe  that here there is not a systematic tail on $\sigma(\rho)^2$ for small values of $T_0$. Moreover it seems that the values are non-zero and small.  The correction we compute assuming a random white noise, $\langle A_2\rangle$, (small dots with red color in the figure) is exactly the part to be corrected.
Finally, in figure \ref{romom4} we show how the fourth momenta of $\rho^{(s)}$ tends to a constant for large values of $T_0$. We see in figure \ref{distancero} how the mutual marginal distance tend to zero once we subtract the noise effect. We average the  scaled configurations with $T_0\in [14,20]$ values for $g=5$ and $g=10$ and with $T_0\in [16,20]$  for $g=15$ (because we detected in this case some small but systematic deviations when we included the ones with $T_0=14$ and $15$). 
In order to get rid of the noise effects, we do the Fourier Transform  of the averaged configurations (figure \ref{Fourierro}) and we discard the modes below the cutoff showed in Table \ref{cut2ro}. After doing the inverse Fourier Transform to the remaining modes we get the universal scaled configuration for the excess of the density field $\tilde\rho(x,y)$ (figure \ref{universalro}). Finally, the differences between these universal configurations and the scaled ones behave like  a ramdom white noise that is bounded by our measured error (figures \ref{flucturo}, \ref{flucturo2} and \ref{flucturo3} and see comments in the temperature field section).

We conclude this analysis by remarking that we have observed that there exists  an universal field $\tilde\rho(x,y)$ such that:
\begin{equation}
\rho(x,y:T_0,g)=w(y;T_0,g)+\sigma(\rho(T_0,g)-w(T_0,g))\tilde\rho(x,y;g) 
\end{equation}
for $T_0$'s large enough. 
The universal profiles seem to be equal. In fact we see that the difference between different $g$-values is, at most of $5\%$ and there is a small but systematic spatial deviations on the differences. However we cannot discard from the numerics the possible existence of a common universal profile independently of the value of $g$. 

{\it NOTE:} We have discarded two rows and columns near the boundary to study the scaling of the excess density field. That changes the $\sigma(\rho(T_0,g)-w(T_0,g))$ with respect to the one we would get if we used all the points in the system and it maybe could change the universal character of the scaled density field.  Let us find the changes that we get when we change the domain to compute the scaled field.
Let us assume that in a given square domain $D\times D$ we have a field that can be written as
\begin{equation}
f(x,y;T_0)=f(y;T_0)+\sigma(T_0)\tilde f(x,y)
\end{equation}
where
\begin{equation}
f(y;T_0)=\frac{1}{\vert D\vert}\int_{D}dx f(x,y;T_0)\quad,\quad \sigma(T_0)^2=\frac{1}{\vert D\vert^2}\int_{D}dx\int_{D}dy \left[f(x,y;T_0)-f(y;T_0)\right]^2
\end{equation}
Let us now consider that we compute the scaled field in a different domain $D'\times D'$:
\begin{equation}
f^{(s)}(x,y;T_0,D')=\frac{f(x,y;T_0)-f(y;T_0,D')}{\sigma(T_0,D')}
\end{equation} 
where now
\begin{equation}
f(y;T_0,D')=\frac{1}{\vert D'\vert}\int_{D'}dx f(x,y;T_0)\quad,\quad \sigma(T_0,D')^2=\frac{1}{\vert D'\vert^2}\int_{D'}dx\int_{D'}dy \left[f(x,y;T_0)-f(y;T_0,D')\right]^2
\end{equation}
Is it $f^{(s)}(x,y;T_0,D')$ universal?, that is, $f^{(s)}(x,y;T_0,D')=\tilde f(x,y;D')$? It is a matter of simple algebra to see that:
\begin{eqnarray}
f(y;T_0,D')&=&f(y;T_0)+\frac{\sigma(T_0)}{\vert D'\vert}\int_{D'}dx\tilde f(x,y)\nonumber\\
\sigma(T_0,D')&=&\sigma(T_0)\Sigma(D')
\end{eqnarray}
where
\begin{equation}
\Sigma(D')^2=\frac{1}{\vert D'\vert^2}\int_{D'}dx\int_{D'}dy\left[\tilde f(x,y)-\frac{1}{\vert D'\vert}\int_{D'}dx'\tilde f(x',y)  \right]^2
\end{equation}
We can substitute these relations into the $f^{(s)}(x,y;T_0,D')$ definition and we get
\begin{equation}
f^{(s)}(x,y;T_0,D')=\frac{1}{\Sigma(D')}\left[\tilde f(x,y)-\frac{1}{\vert D'\vert}\int_{D'}dx'\tilde f(x',y) \right]\equiv\tilde f(x,y;D')
\end{equation}
that is, the scaled field in the new domain is still universal (it doesn't depend on $T_0$). However we pay the price of having a different scaled universal function because, in any case, it depends on the domain in which we define the scale.

\begin{figure}[h!]
\begin{center}
\includegraphics[height=6cm,clip]{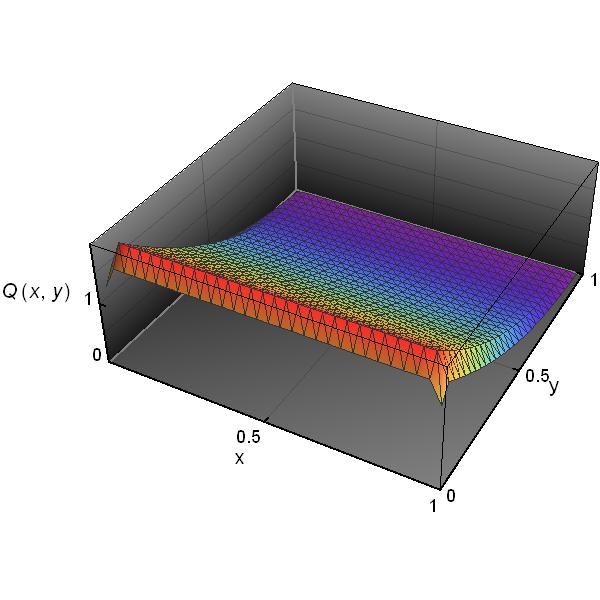}  %press_profile_0.nb
\includegraphics[height=6cm,clip]{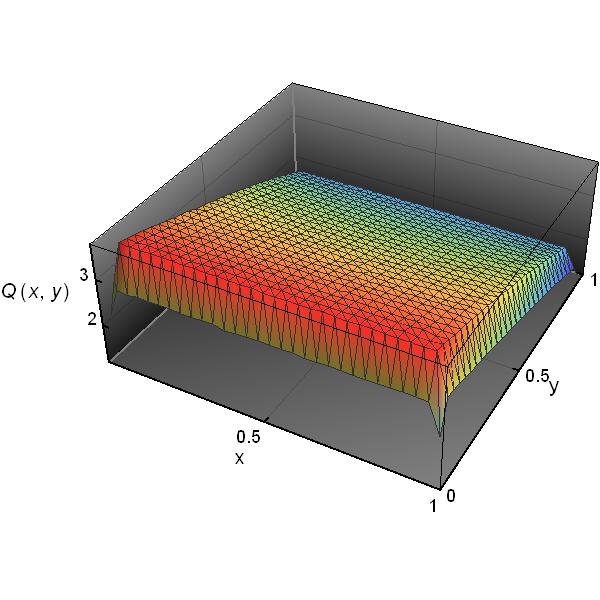}       %press_profile.nb
\includegraphics[height=5cm,clip]{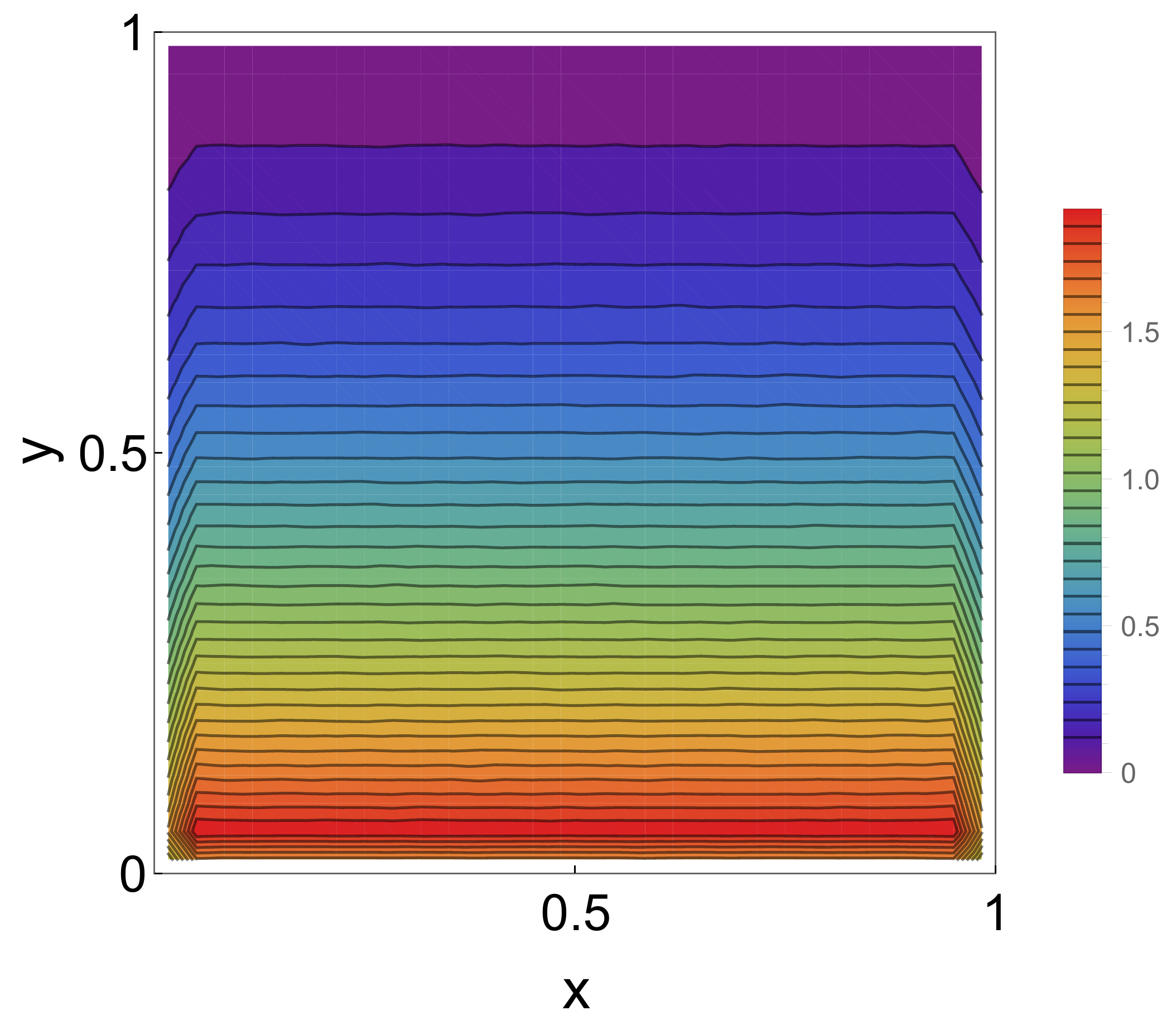}  %press_profile_0.nb
\includegraphics[height=5cm,clip]{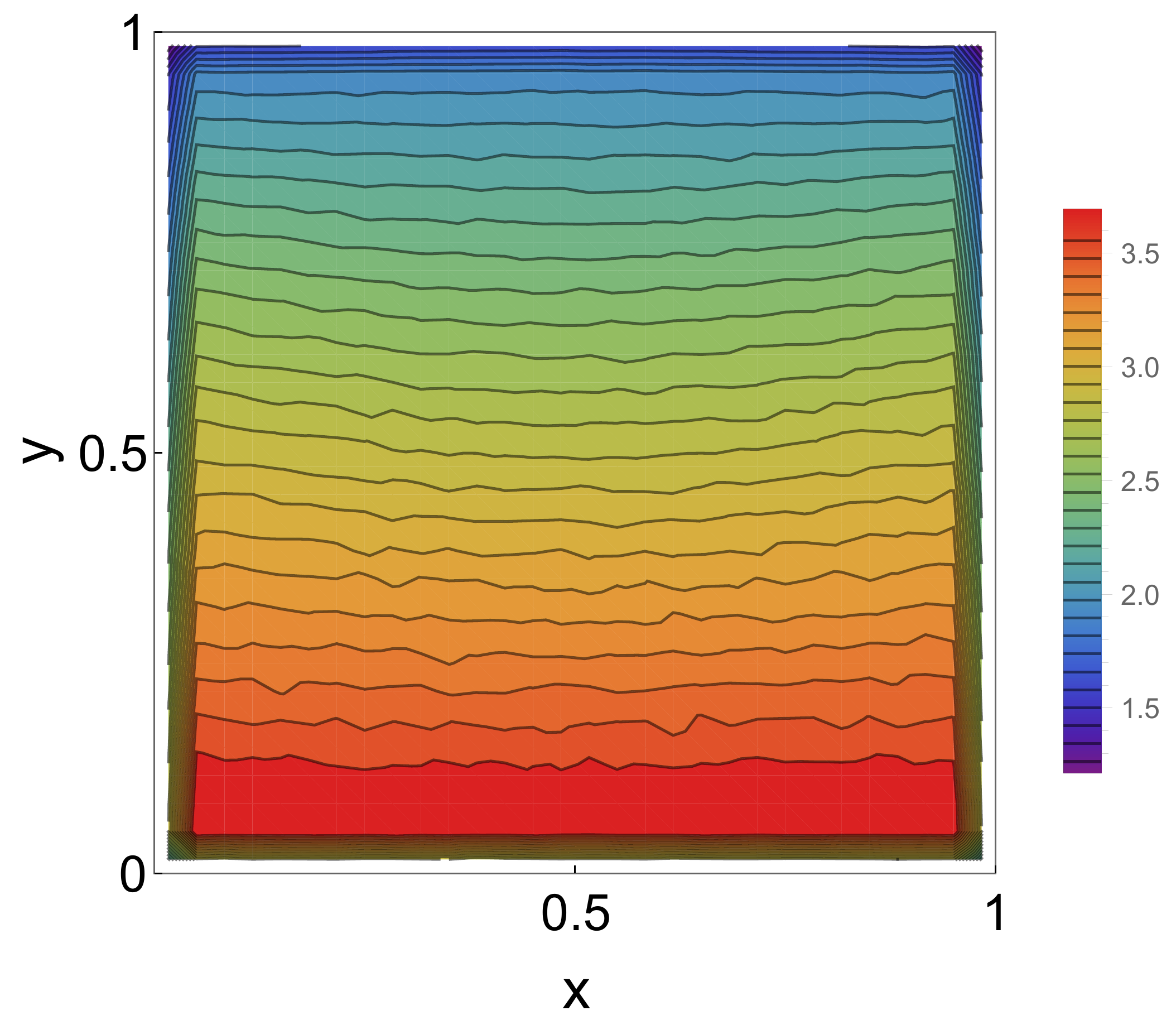}        %press_profile.nb
\end{center}
\kern -0.5cm
\caption{Pressure field $Q(x,y)$  defined in eq. \ref{pre11} for $T_0=2$ and $T_0=18$ for left and right columns respectively ($g=10$). Below each of the 3D graphs there are  the corresponding contour plots to show the existence (or not) of a nontrivial spatial structure on $x$ direction. \label{pre1}}
\end{figure}

\item{\it Pressure profiles:} 
We use the virial theorem to compute the pressure field $P(x,y)$ even in the nonequlibrium stationary state:
 \begin{equation}
 P(x,y)=\frac{\rho(x,y) e_c(x,y)}{\pi r^2}+\lim_{\tau\rightarrow\infty}\frac{1}{2 \Delta^2 \tau}\sum_{n:t_n\in[0,\tau]}r_{ij}\cdot p_{ij}
 \label{pre11}
 \end{equation}
where $\Delta=1/30$ is the side length of a cell, $e_c(x,y)=\langle E_c(x,y)\rangle/\langle N(x,y)\rangle$ is the average total kinetic energy in the cell and the sum runs over all particle-particle collisions that occur at the cell $(x,y)$ at the time interval $[0,\tau]$ assuming that at time zero the system is at the stationary state. From the derivation of this expression we can call $P(x,y)$ to be the local {\it mechanical pressure} that, as we will see, it is non trivially  related with the {\it thermodynamic pressure} when there is a nonzero hydrodynamic velocity field. From a computational point of view we already checked this expression for the equilibrium case $T_0=T_1=1$ and we showed the goodness of our simulation when we compared the data obtained with the exact result (assuming the Henderson's EOS).  
\begin{figure}[h!]
\begin{center}
\includegraphics[height=4.cm,clip]{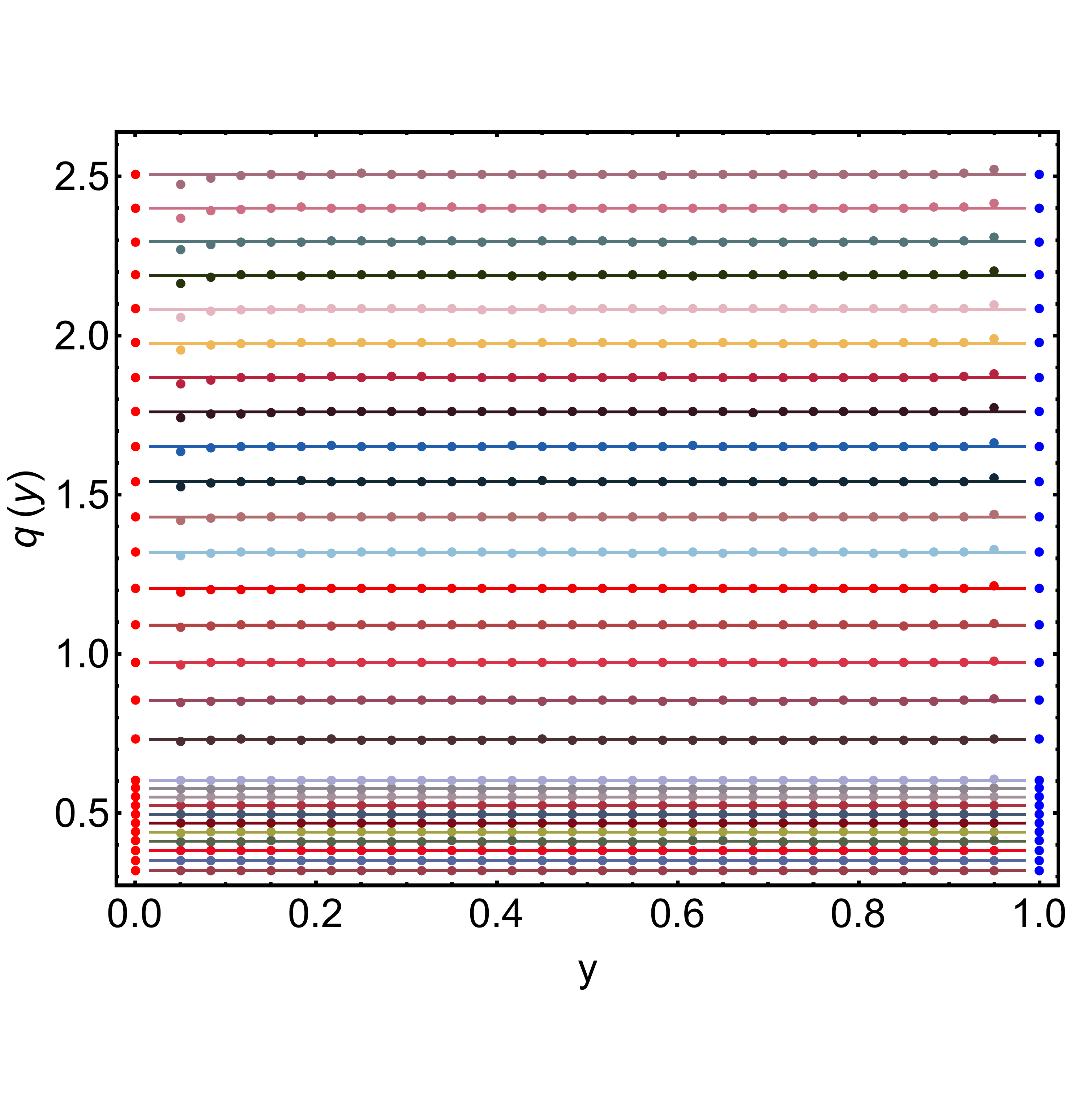}  %pre_profile_y_E0.nb
\includegraphics[height=4.cm,clip]{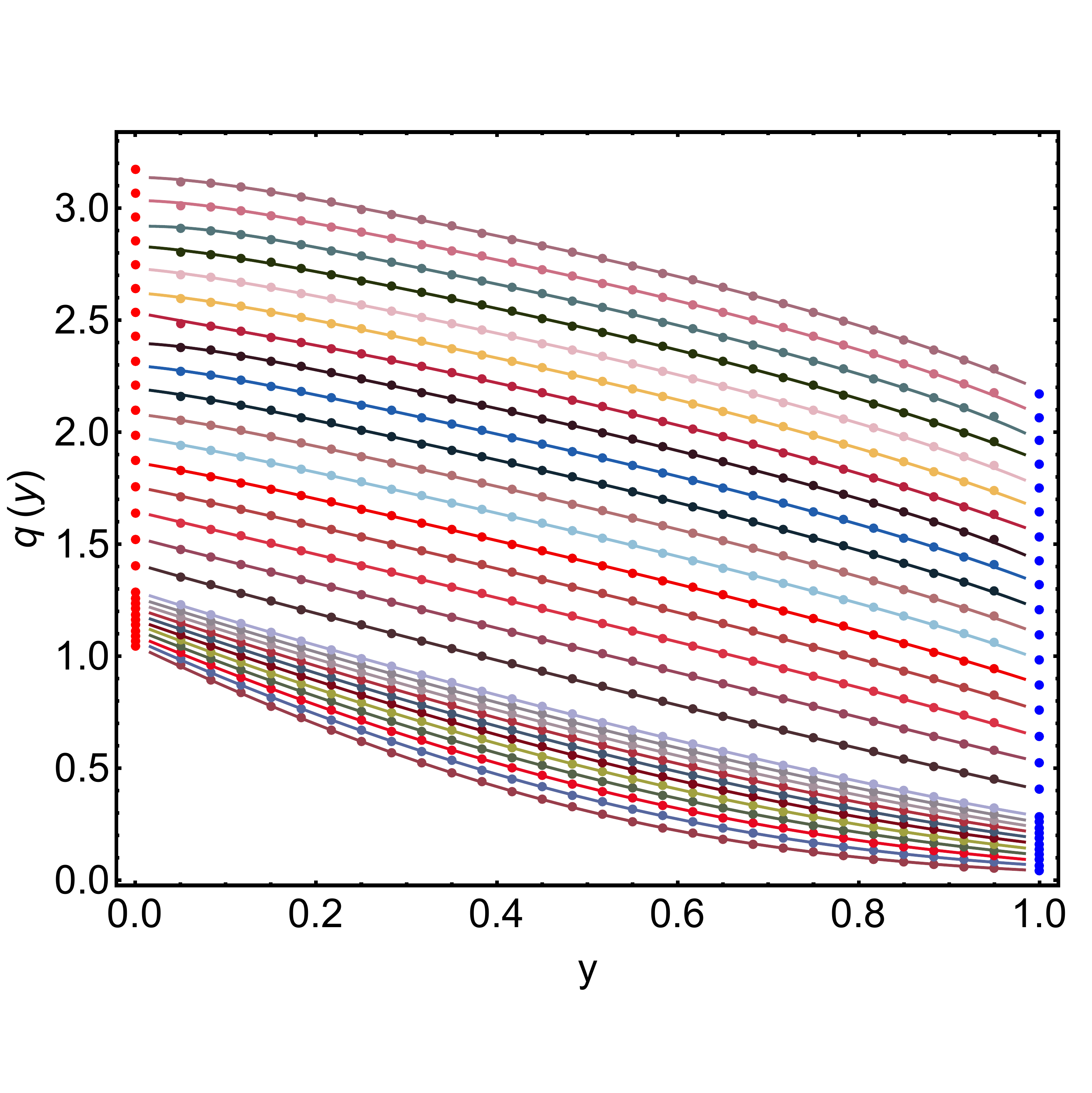}       %pre_profile_y_E5.nb
\includegraphics[height=4.cm,clip]{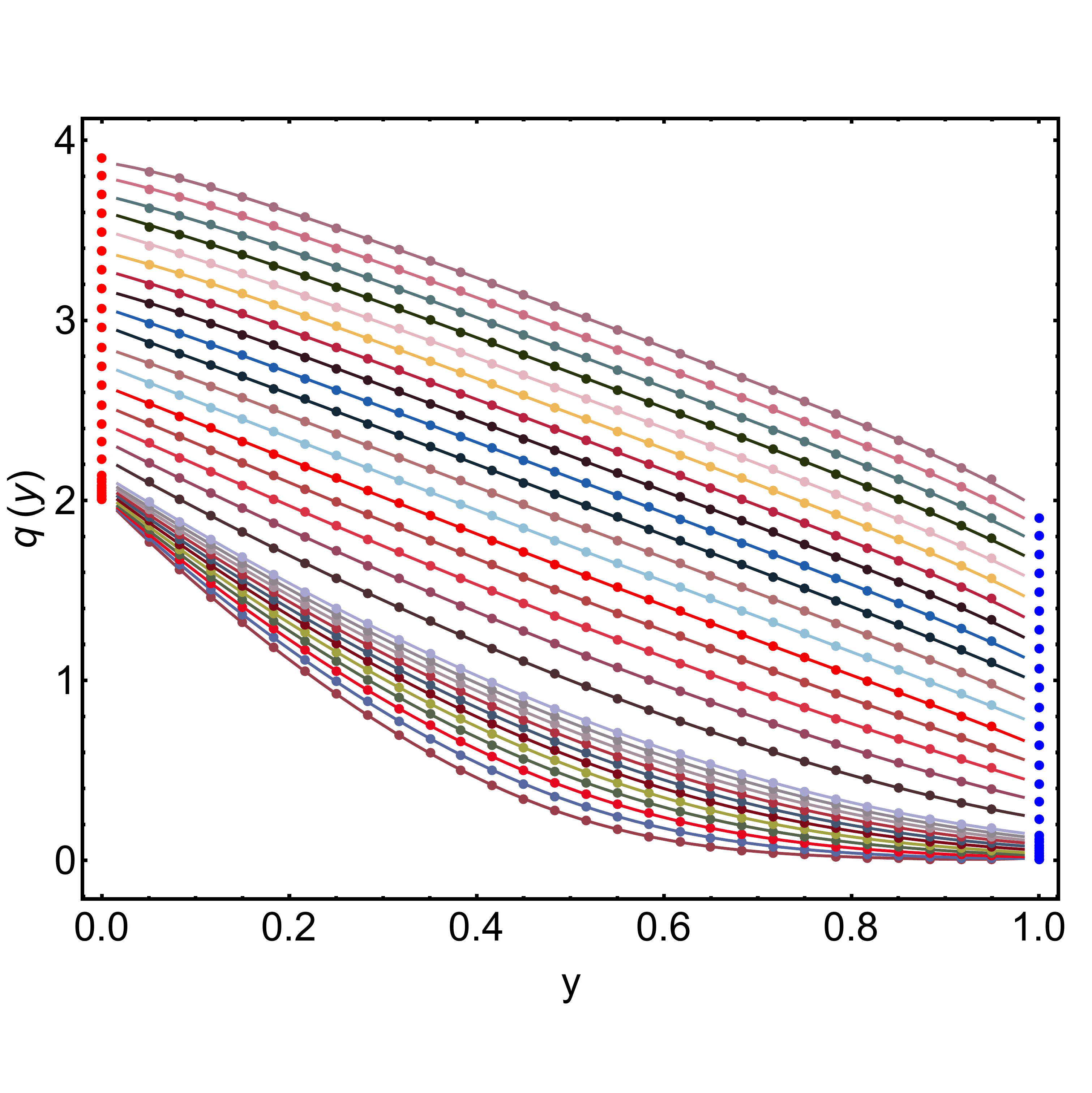}  %pre_profile_y_E10.nb
\includegraphics[height=4.cm,clip]{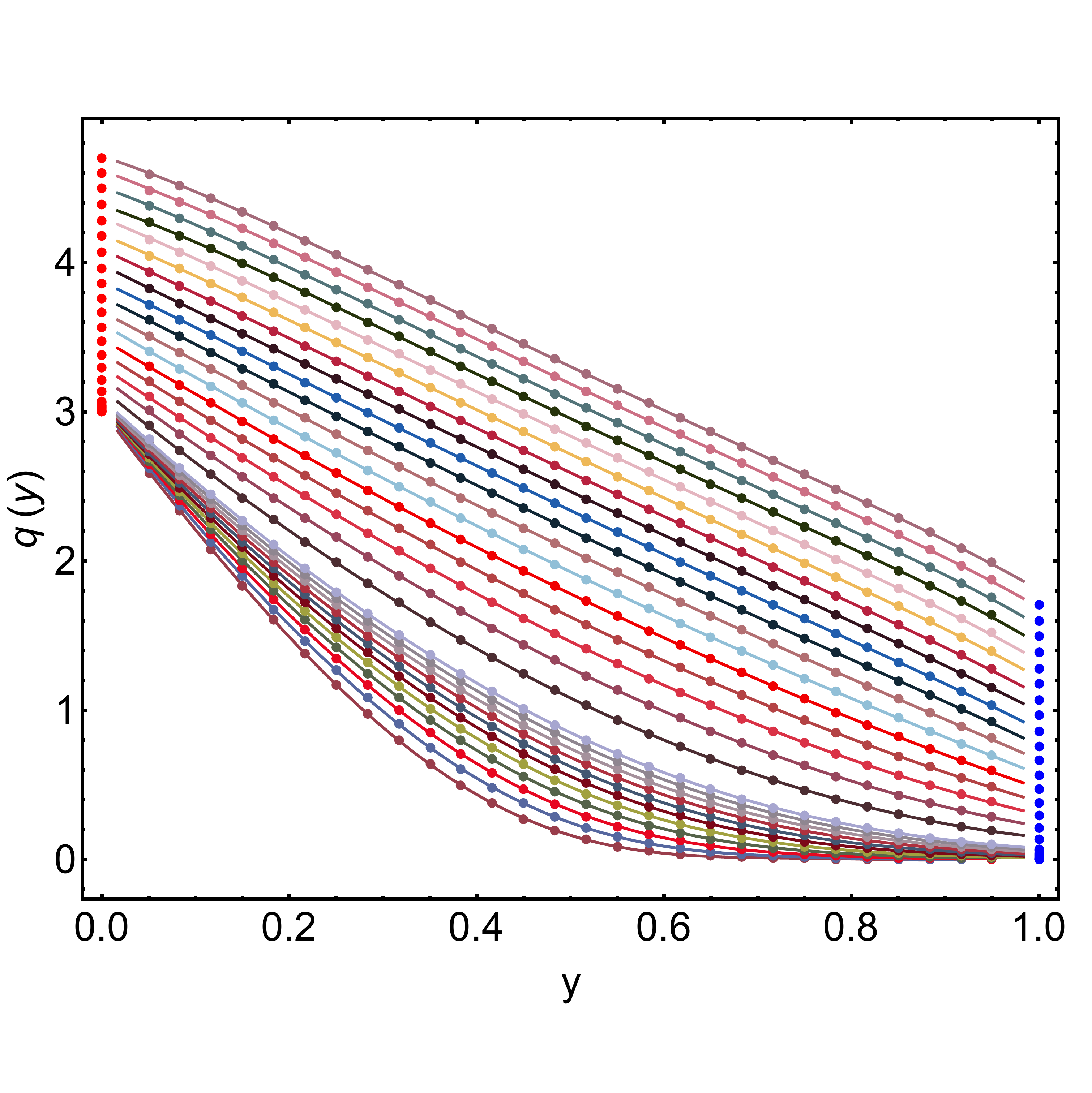}  %pre_profile_y_E15.nb
\includegraphics[height=4.cm,clip]{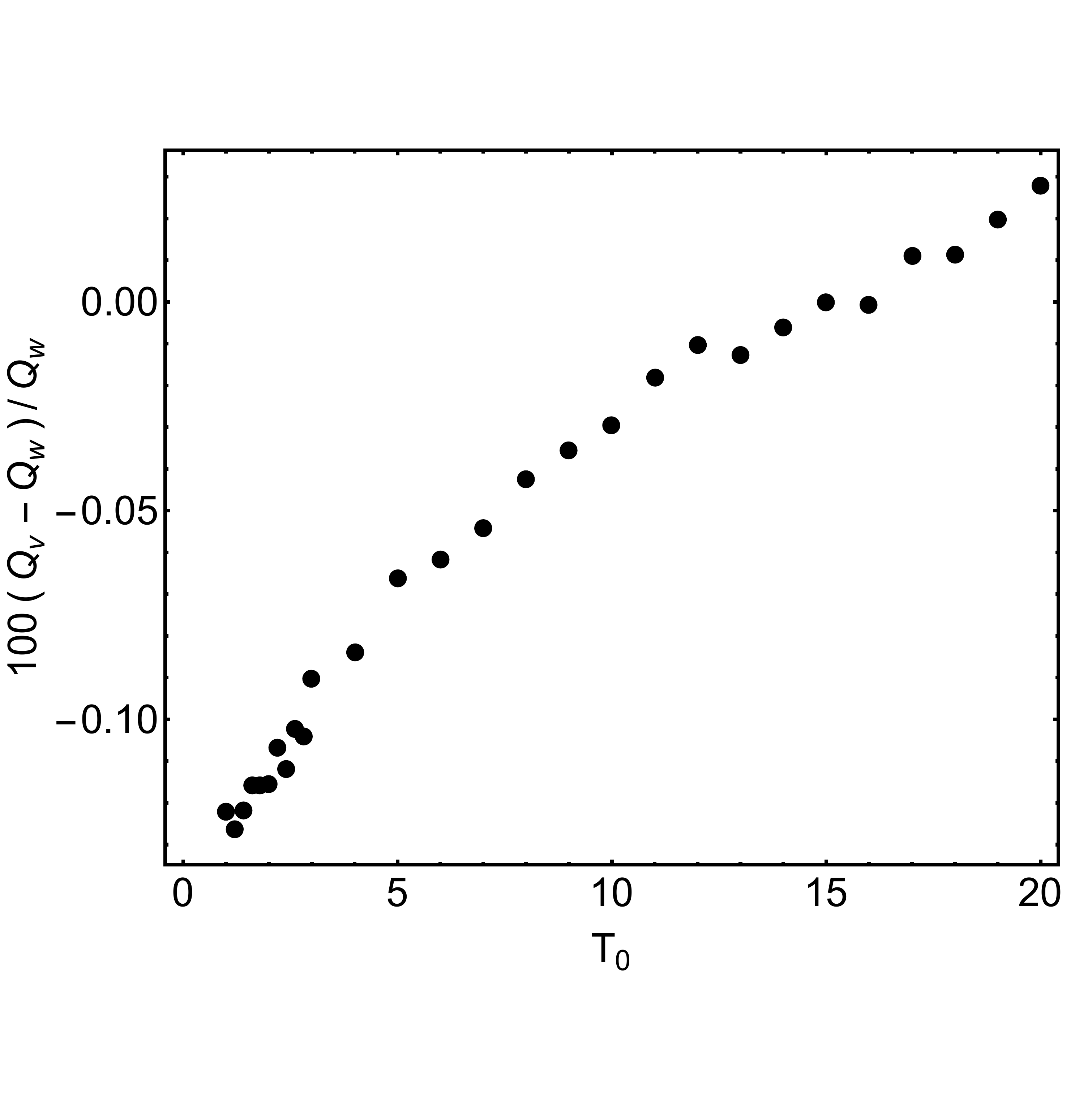}  %pre_profile_y_E0.nb
\includegraphics[height=4.cm,clip]{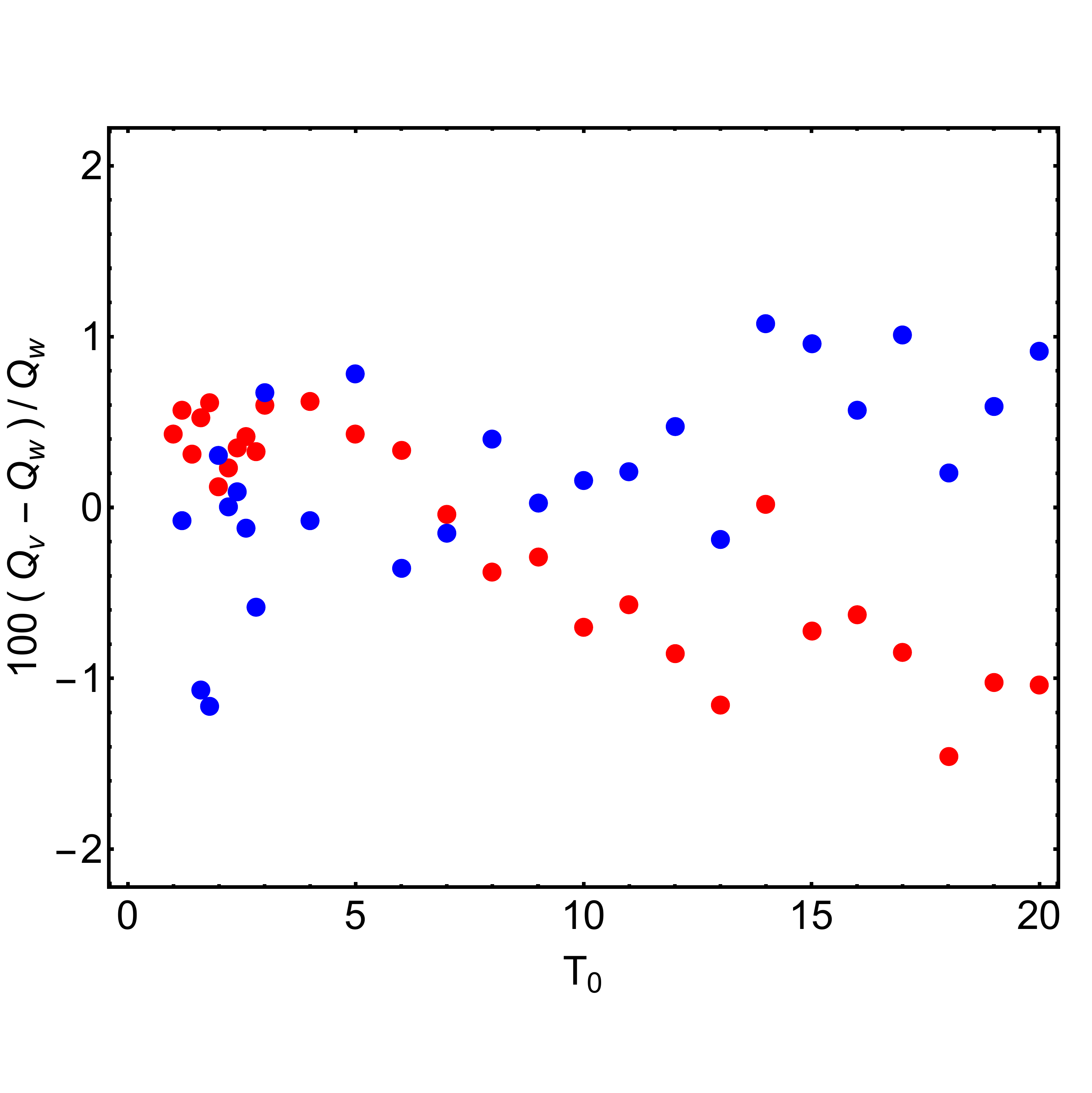}       %pre_profile_y_E5.nb
\includegraphics[height=4.cm,clip]{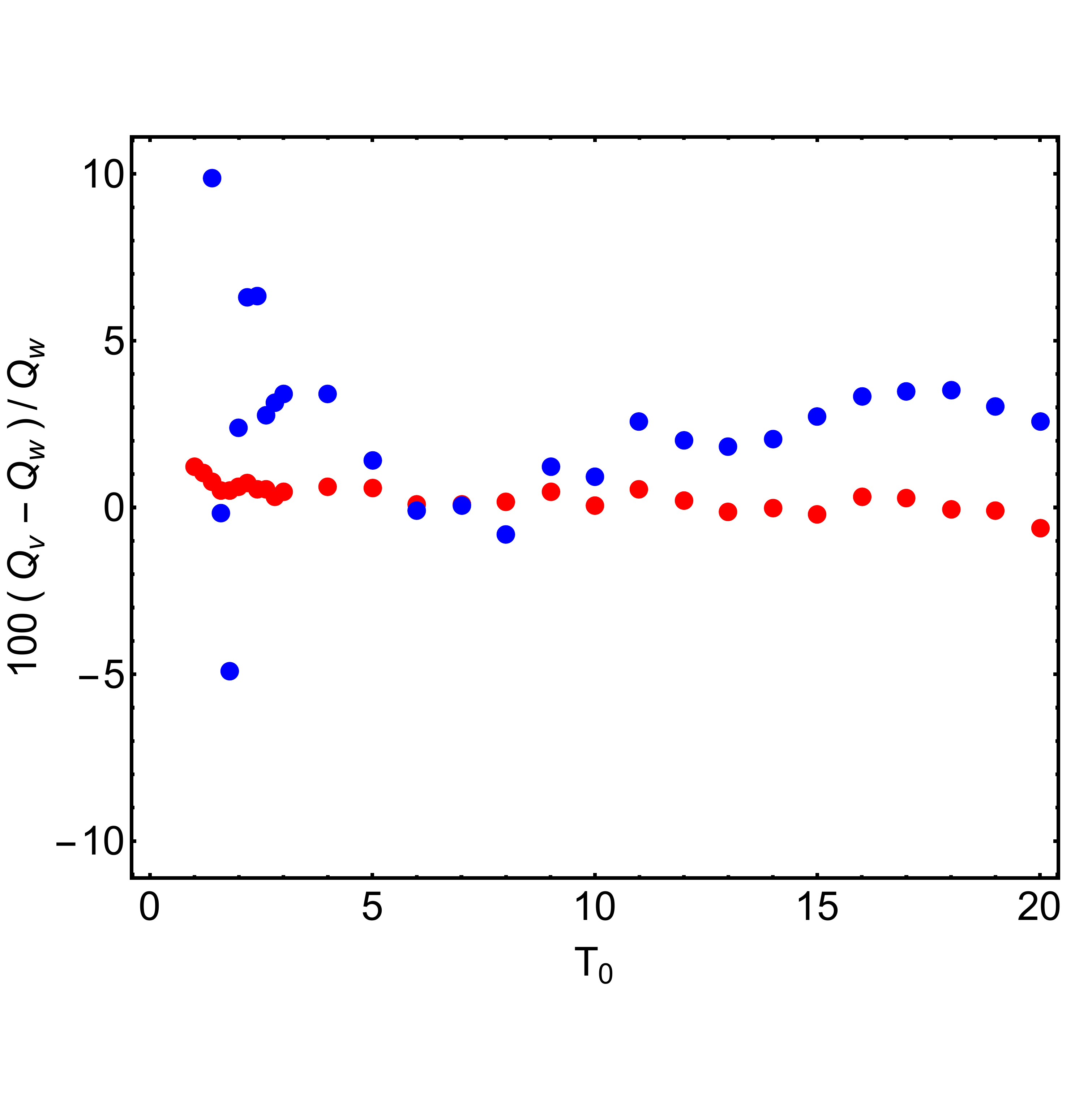}  %pre_profile_y_E10.nb
\includegraphics[height=4.cm,clip]{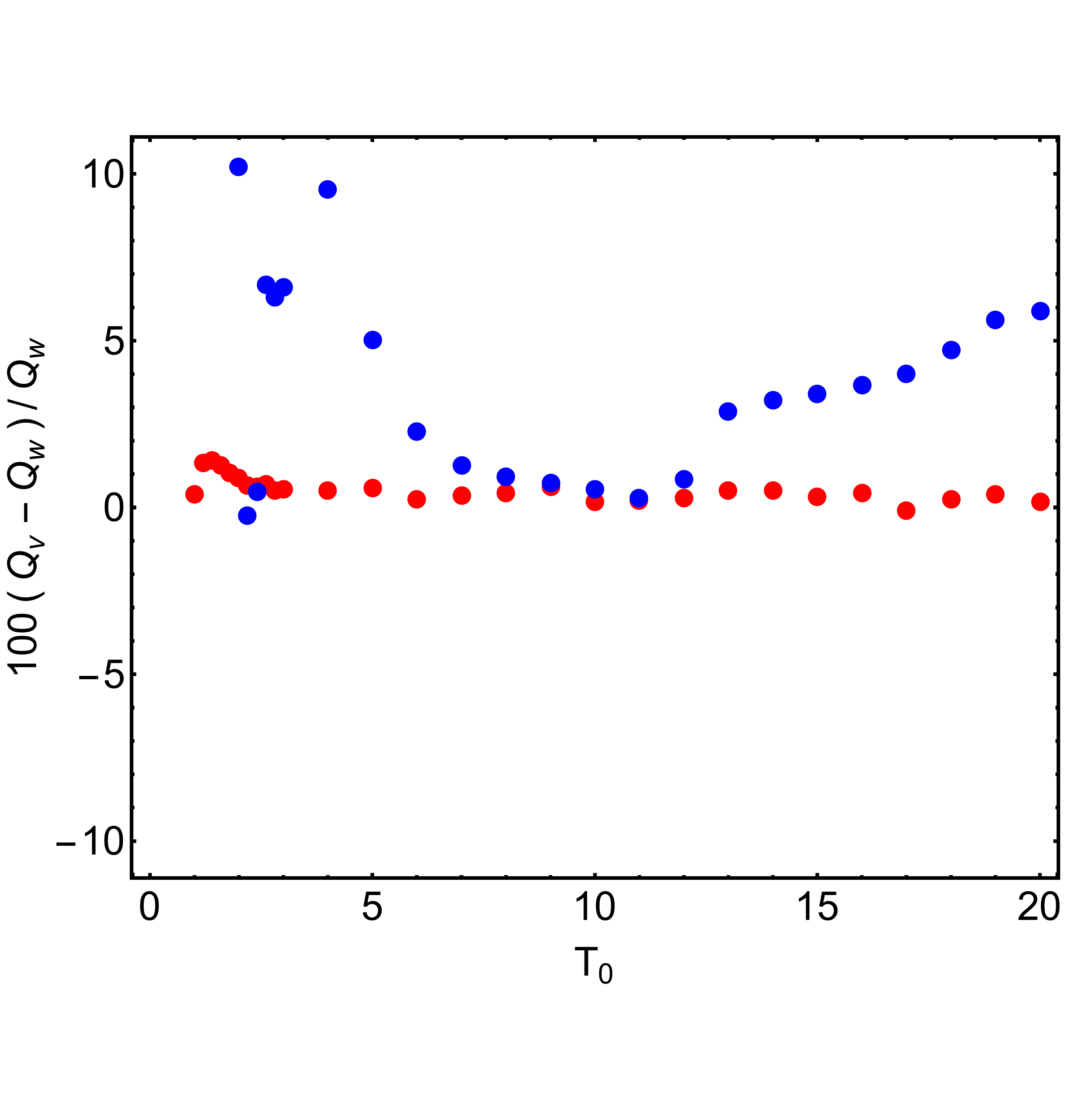}  %pre_profile_y_E15.nb
\end{center}
\kern -0.5cm
\caption{Top row: Average y-pressure profiles $q(y)$  defined in eq. \ref{pressy} for $g=0$ $5$, $10$ and $15$ (from left to right). At each figure we plot the points with $T_0=1, 1.2, 1.4, 1.6, 1.8, 2.0, 2.2, 2.4, 2.6, 2.8, 3.0, 4, 5, 6, 7, 8, 9, 10, 11, 12, 13, 14, 15, 16, 17, 18, 19, 20$ from bottom to top.  Lines are phenomenological fits of sixth order polynomials: $w=a_0+a_1 y+\ldots+a_6 y^6$. Red points at $y=0$ and blue points at $y=1$ are the measured pressure at the boundaries (see section {\it Global magnitudes: The Pressure}). Bottom row: Relative error between the extrapolated virial pressure at $y_0=0,1$ ($Q_v=q(y_0)$) and the pressure measure at the wall ($Q_w$) (red dots for $y=0$ and  blue dots for $y=1$) as a function of $T_0$ for $g=0$, $5$, $10$ and $15$ (from left to right). Notice that for the $g=0$ case there is no difference between $y=0$ and $y=1$ relative errors \label{pre2}}
\end{figure}
At figure \ref{pre1} we present two typical measured pressure fields for $g=10$: $T_0=2$ (left figures) and $T_0=18$ (right figures).  For $T_0=2$ we are at a non-convective state and we observe that the pressure decreases with $y$ and it has some structure on $x$ when we are at the convective state (right figure in \ref{pre1}).

Again we first study the marginal $q(y)$ defined by the average of its $x$ values:
\begin{equation}
q(y)=\frac{1}{N_A}\sum_{x\in A}Q(x,y)\label{pressy}
\end{equation}
where $A$ is the set of cells we use for the averaging and $N_A$ is the total number of cells. We discard only the two rows near the thermal walls. However, we do not discard the columns near the vertical boundaries in order to compare with the measured pressure on the top and bottom boundaries (see section {\it Global Magnitudes: the pressure}). In our analysis we choose $A=\{(2n-1)/60 \vert n=1,\ldots,30\}$  and $N_A=30$. 

 In figure \ref{pre2}  we see: (1) the pressure is a decreasing monotone function on $y$,  (2) there is a smooth change of curvature of the pressure profiles with the temperature gradient from concave to convex at around $T_0=T_{c,2}$. However we cannot precise the transition temperature because the second derivatives of the polynomial fitted functions present some wavy behavior that obscure the convexity property, (3) we see in eq. \ref{eqpress0} that the effect of $g$ into the equilibrium pressure profiles is through a trivial rescaling of the $y$ coordinate by $g$: $\bar y=gy$. However, for any nonequlibrium state this doesn't happen and the pressure profiles have a nontrivial dependence on $g$ and (4) the naive extrapolation of the fitted $6$th order polynomial to the extreme points $y=0,1$ ($Q_v$) coincide with the measured values of the pressure at the boundaries ($Q_w$). In fact, for the $g=0$ case the relative error is, at most, of about $0.15\%$. For the $g=5$, $10$ and $15$ cases the relative errors for the hot boundary ($y=0$) is about $1\%$. For the cold boundary ($y=1$) the relative errors fluctuate much more as we see in figure \ref{pre2}. That is due to the fitting function we are using that doesn't represent the fast decay that is present for small values of $T_0$. We already had some similar troubles in the study of the density field for such values. There we used the equilibrium form for the fitting function in order to get very good fits.   

\begin{figure}[h!]
\begin{center}
\includegraphics[height=5cm,clip]{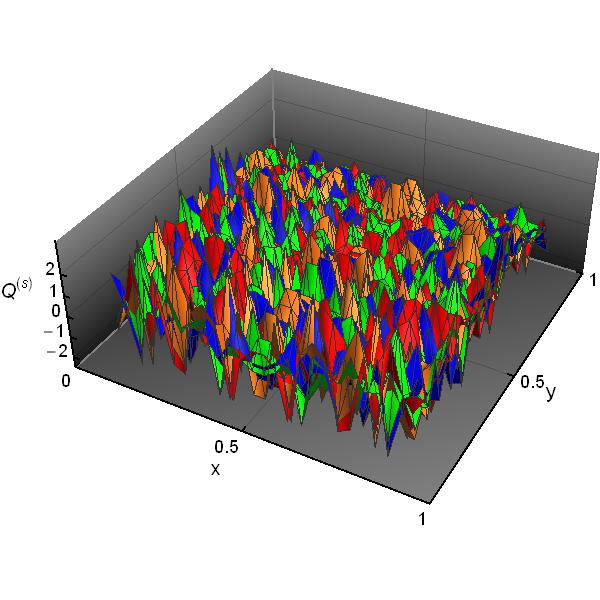}   %pre_field_show.nb
\includegraphics[height=5cm,clip]{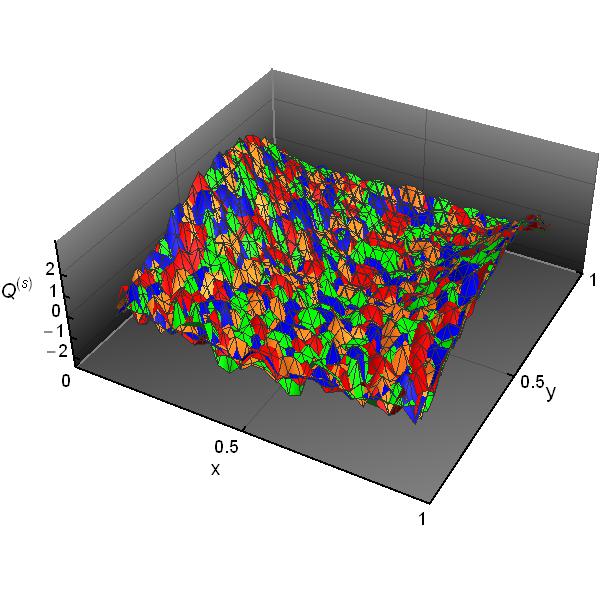}   
\includegraphics[height=5cm,clip]{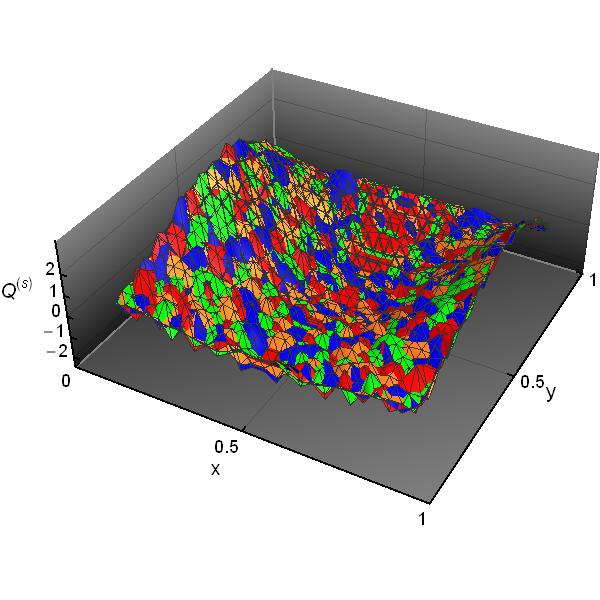}   
\end{center}
\caption{Superposition of the scaled excess of pressure fields, $\rho^{(s)}$ with $T_0=17$, $18$, $19$ and $20$ (red, blue, green and orange colors) and for different $g$ values (from left to right: $g=5$, $10$ and $15$).    \label{prefieldsuper}}
\end{figure}

\begin{figure}[h!]
\begin{center}
\includegraphics[height=6cm,clip]{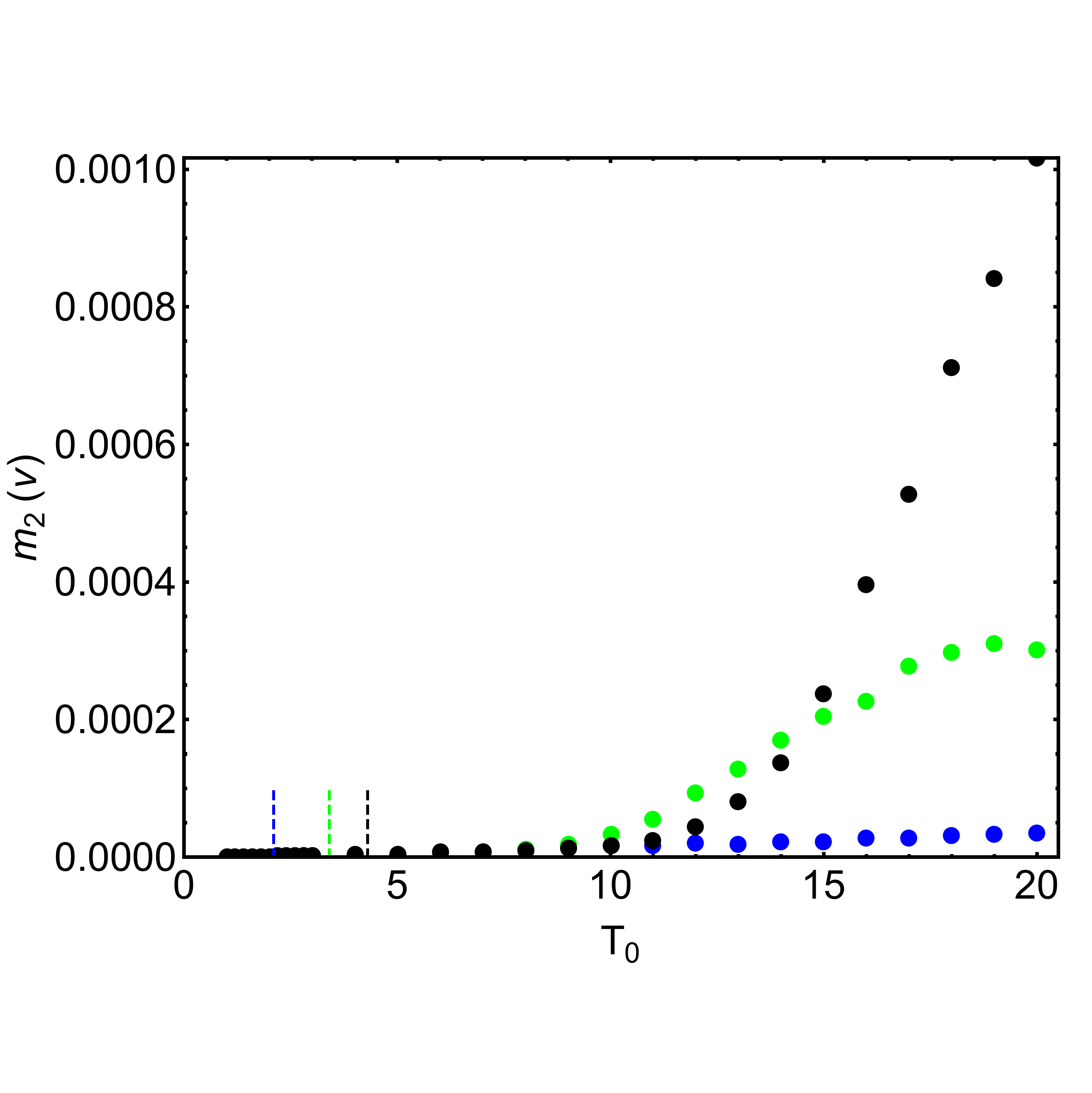}   %pre_moments.nb
\includegraphics[height=6cm,clip]{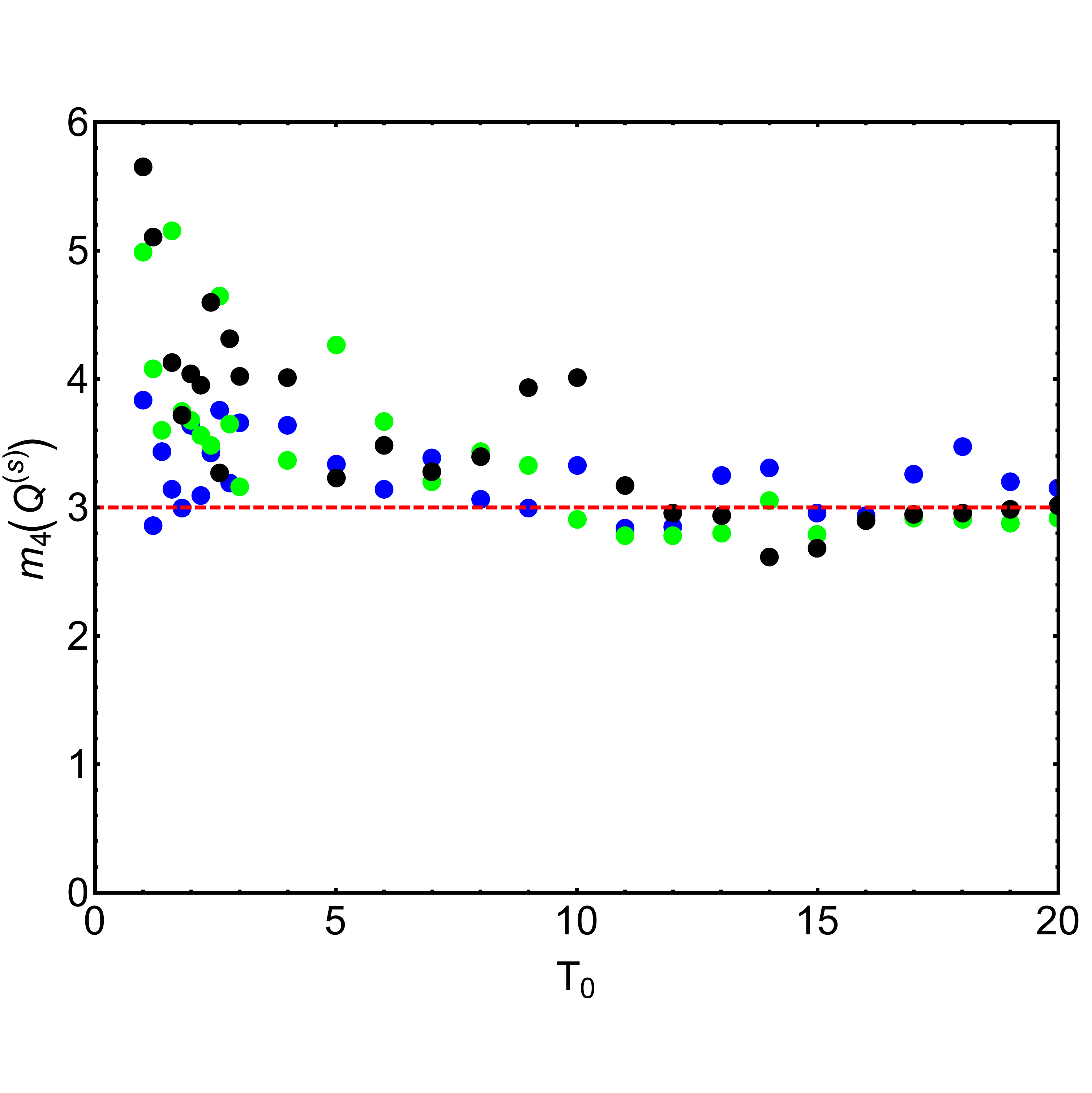}   
\end{center}
\kern -1cm
\caption{Left: Second moment of the measured excess of density field: $v(x,y;\xi)=Q(x,y;\xi)-q(y;\xi)$ for $T_0\in[1,10]$ and $g=5$ (blue dots), $g=10$ (green dots) and $g=15$ (black dots). Right: Fourth moment of the scaled excess of density field,  $m_4(Q^{(s)})$, as a function of $T_0$. 
   \label{premom}}
\end{figure}
\begin{figure}[h!]
\begin{center}
\includegraphics[height=5cm,clip]{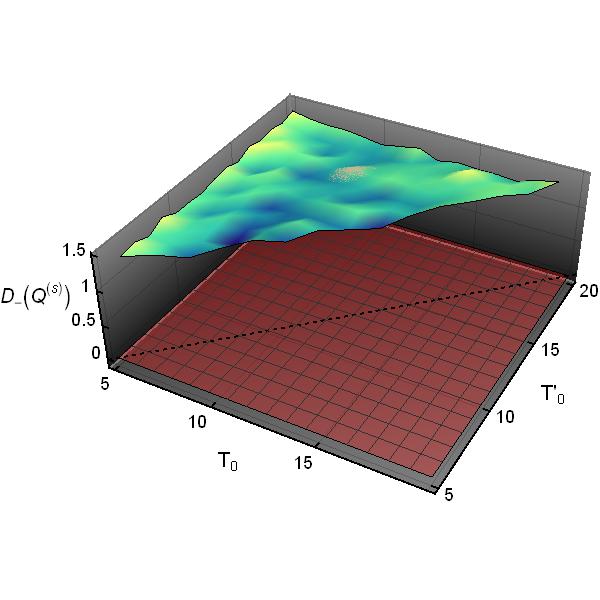}   %pre_E5new.nb
\includegraphics[height=5cm,clip]{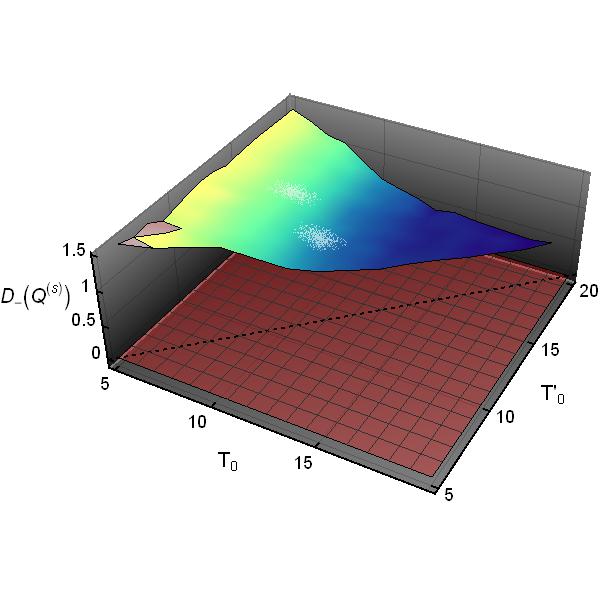}%pre_E10new.nb
\includegraphics[height=5cm,clip]{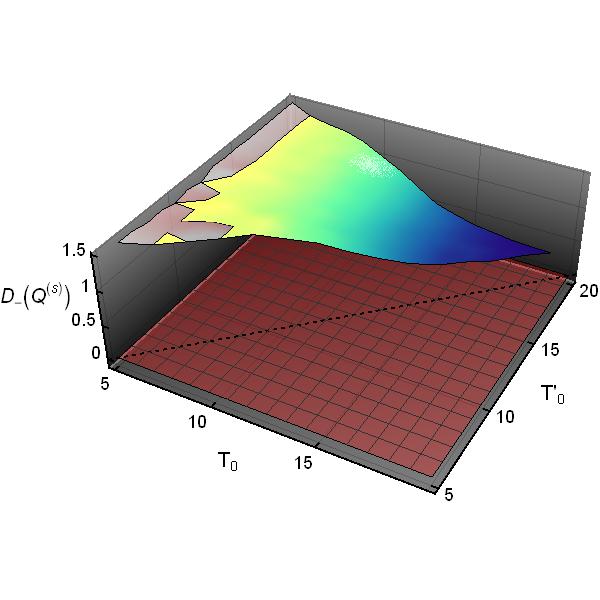}%pre_E15new.nb
\newline\vglue -1cm
\includegraphics[height=4.5cm,clip]{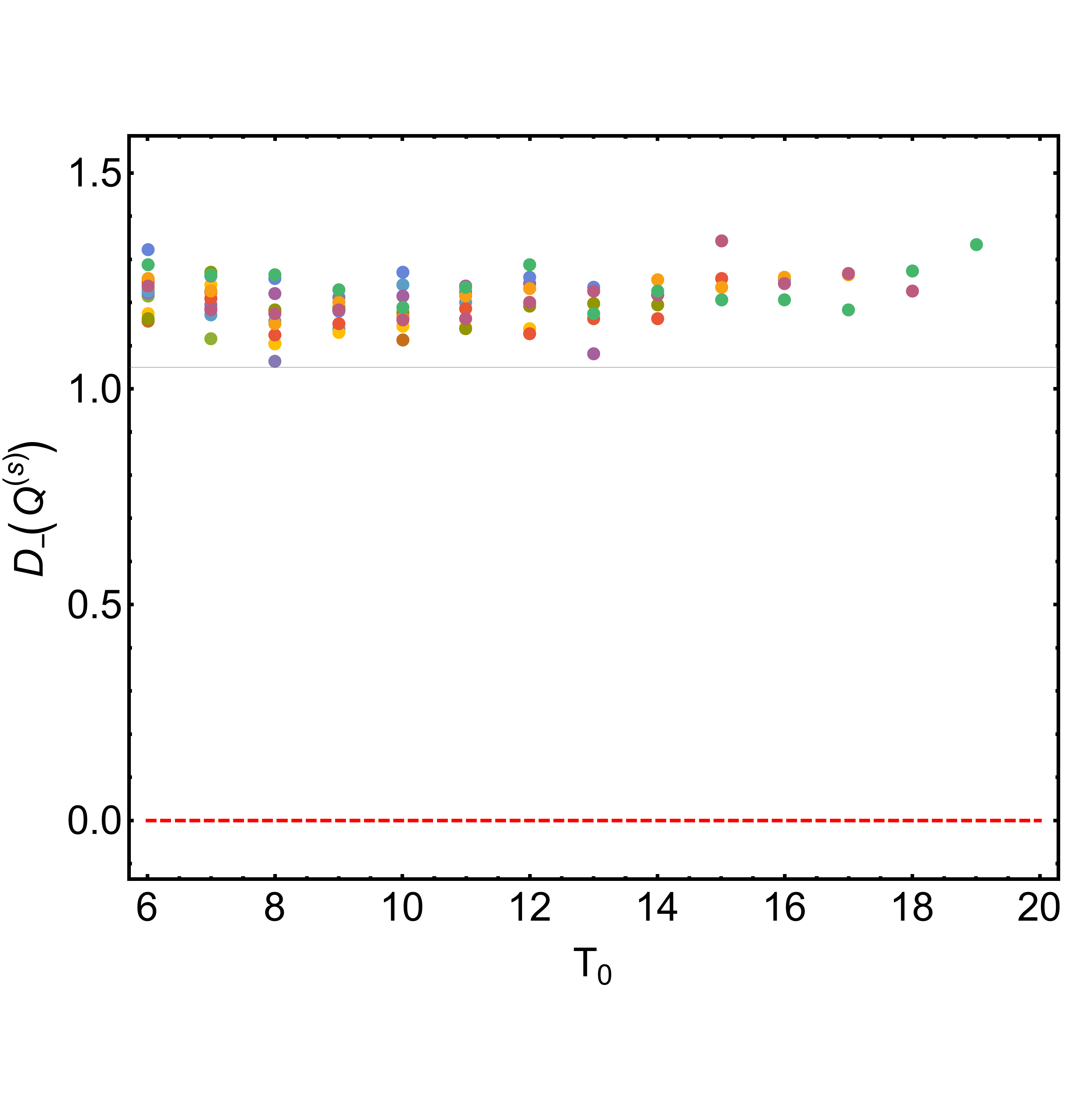}   %pre_E5new.nb
\includegraphics[height=4.5cm,clip]{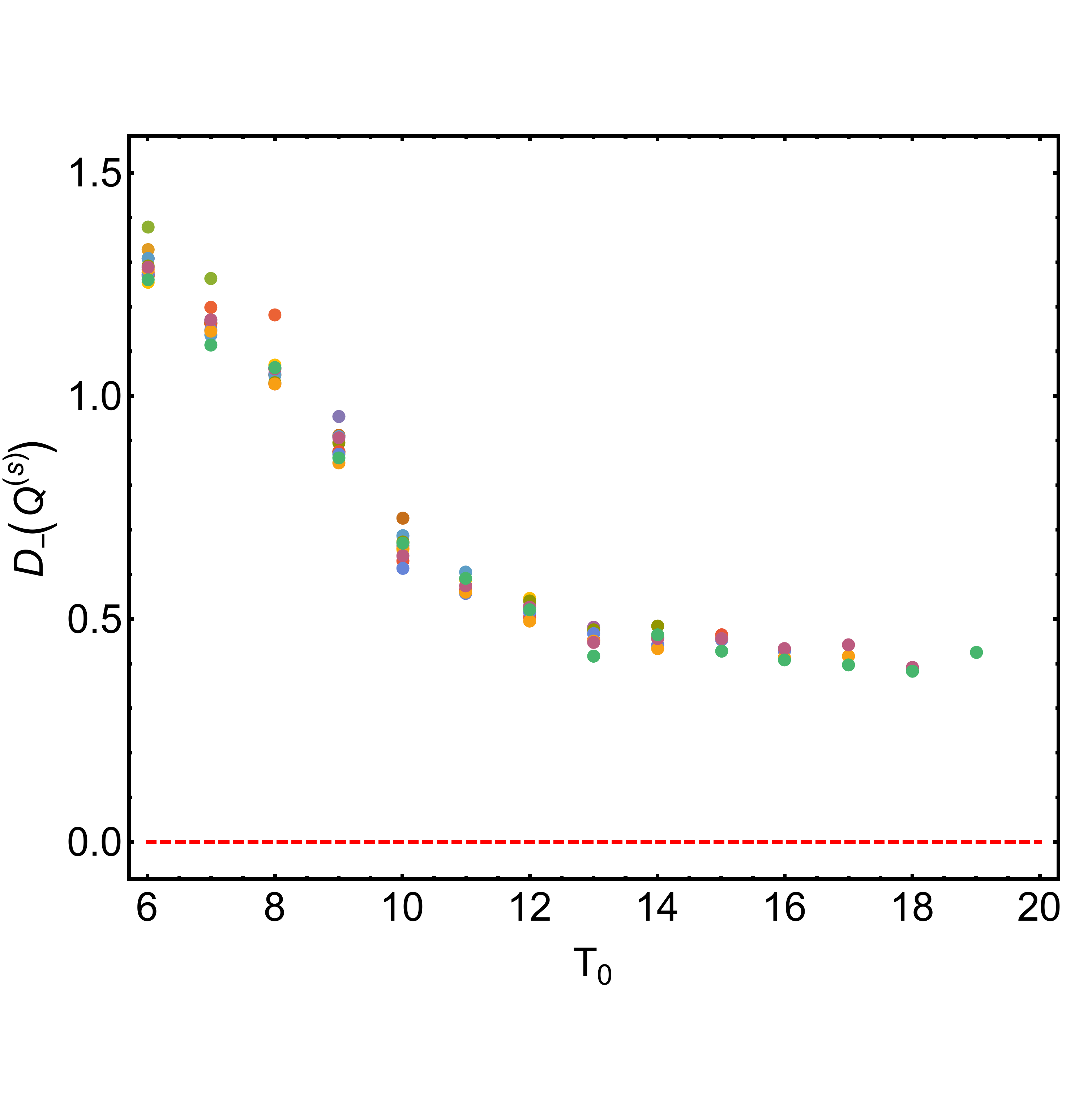}%pre_E10new.nb
\includegraphics[height=4.5cm,clip]{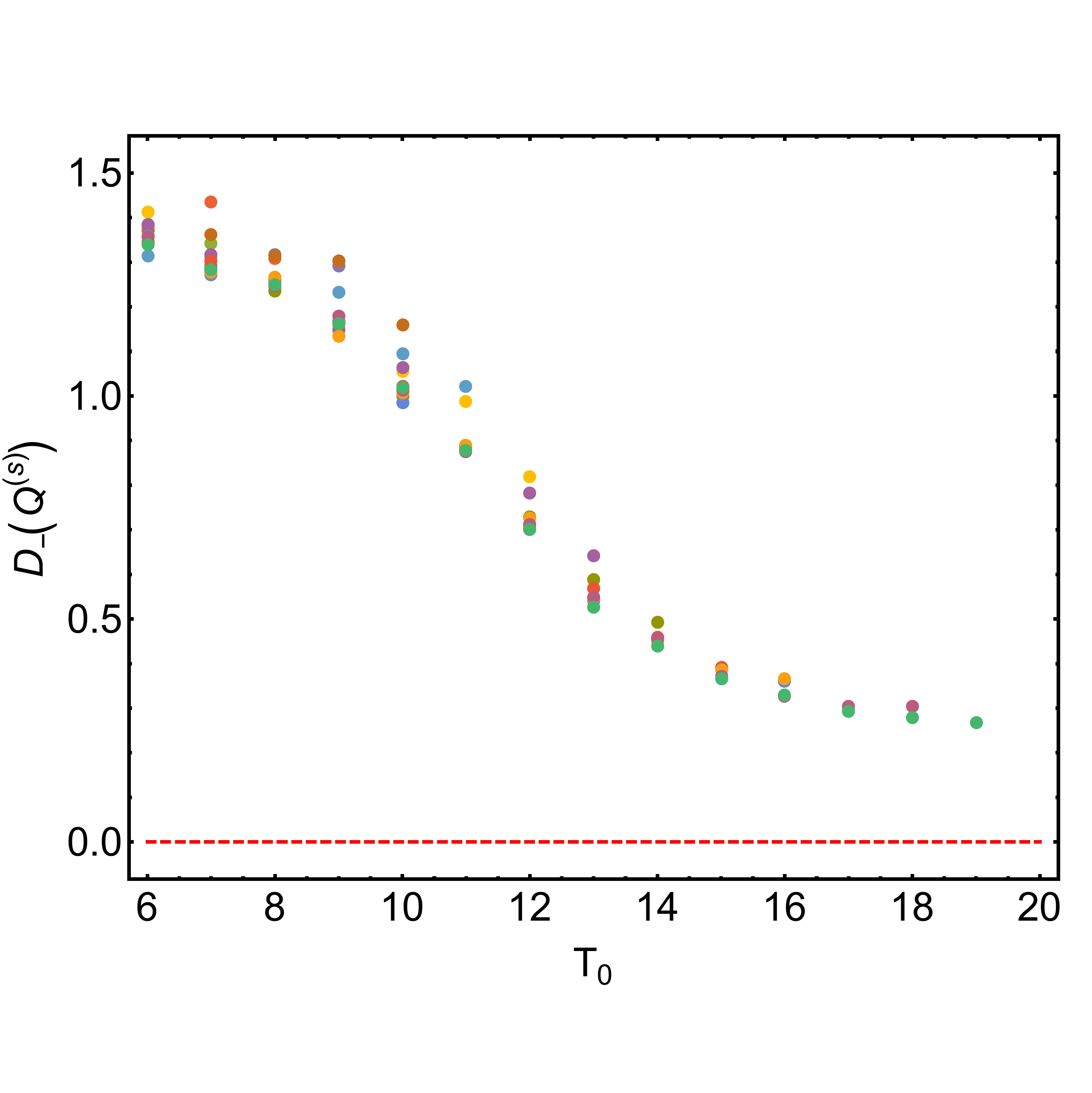}%pre_E15new.nb
\end{center}
\kern -1.cm
\caption{Marginal distance $D_-$ (see text) between the pressure field scaled configurations $Q^{(s)}$ for $g=5$ (figures left), $g=10$ (figures center) and $g=15$ (figures right). for a given $g$ value. We only plot pairs of scaled configurations with $T_0$ and $T_0'$ such that $T_0<T_0'$. Top figures: Pink surface are the bare distances $D(Q^{(s)};\xi)$. Bottom figures:  Same as top.  Each color correspong to a given $T_0'$ value. \label{distanceQ}}
\end{figure}

\begin{figure}[h!]
\begin{center}
\includegraphics[height=5cm,clip]{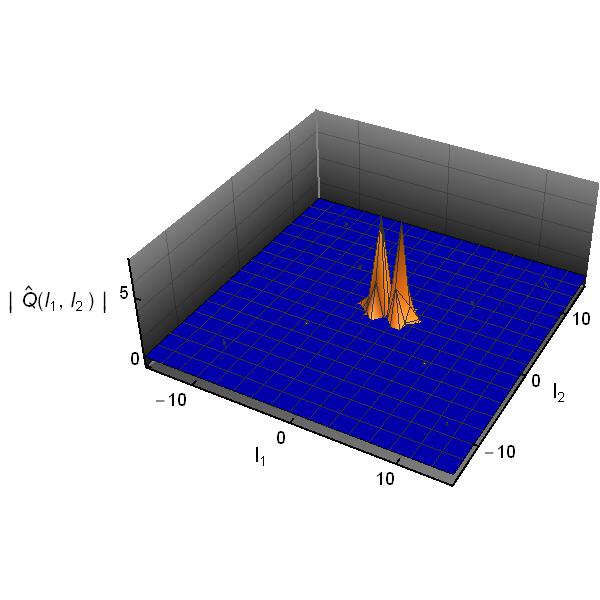}   %pre_E5new.nb
\includegraphics[height=5cm,clip]{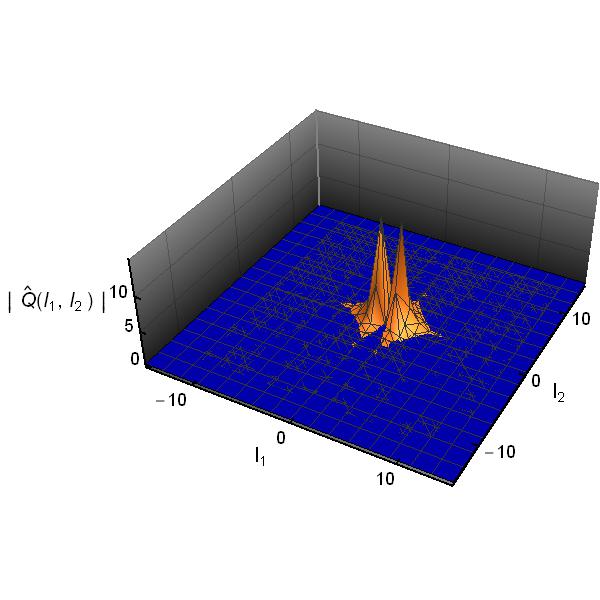} %pre_E10new.nb
\includegraphics[height=5cm,clip]{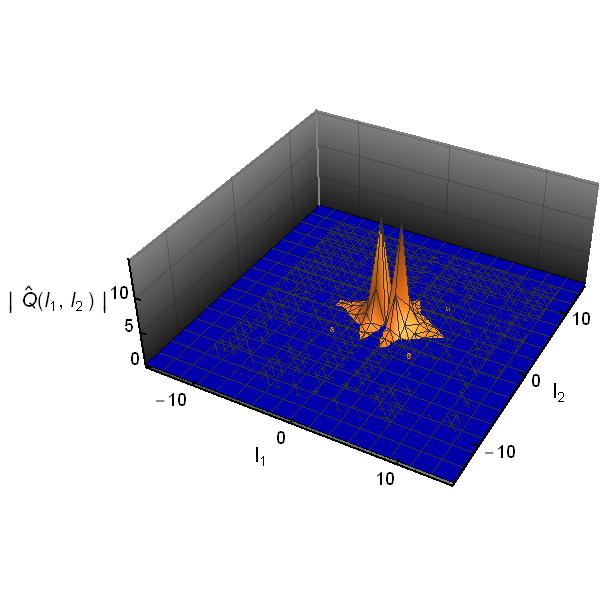} %pre_E15new.nb
\newline\vglue -1cm
\includegraphics[height=5cm,clip]{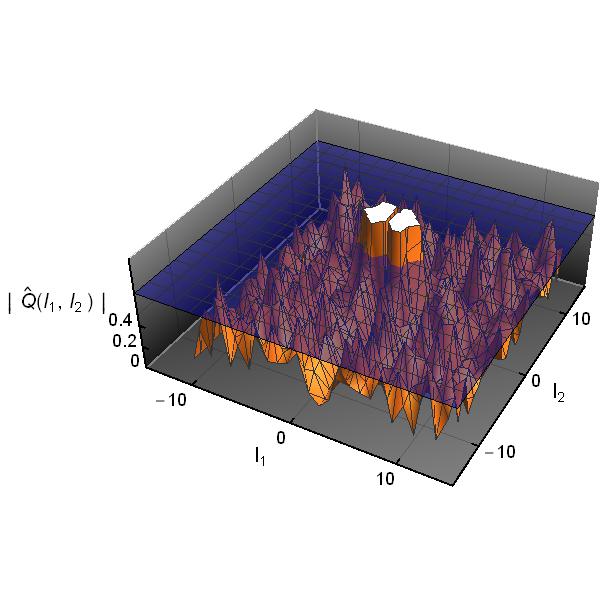}   %pre_E5new.nb
\includegraphics[height=5cm,clip]{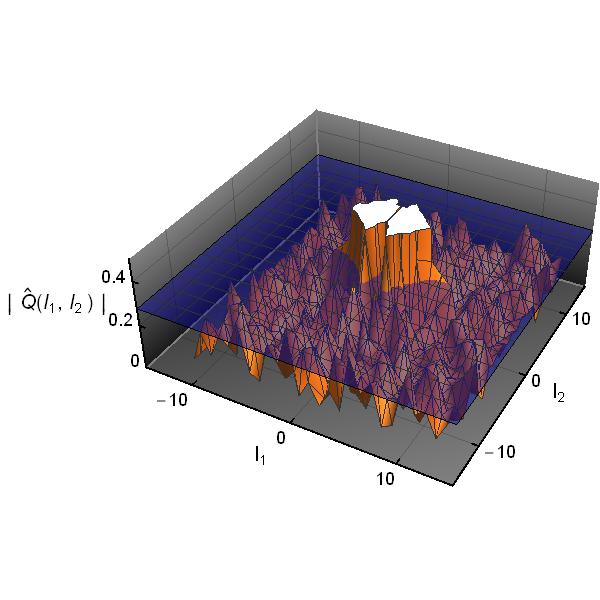} %pre_E10new.nb
\includegraphics[height=5cm,clip]{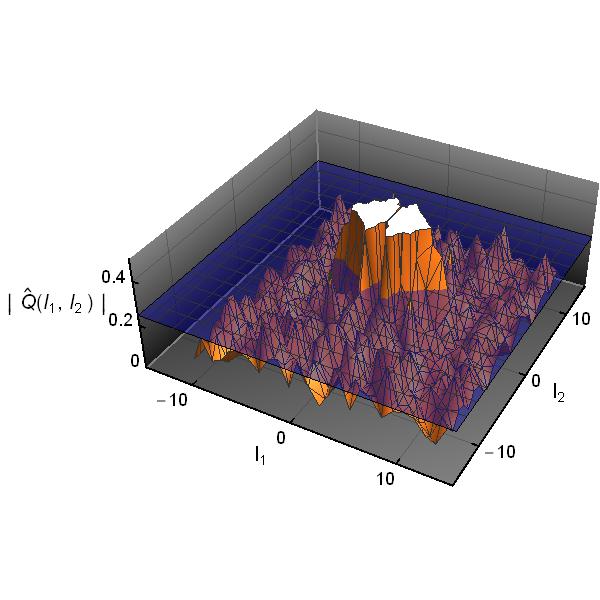} %pre_E15new.nb
\end{center}
\kern -1.cm
\caption{Modulus of the Discrete Fourier Transform obtained by averaging the scaled configurations, $Q^{(s)}(x,y)$, from $T_0=14,\ldots, 20$ for $g=5$ (left figures), $g=10$ (center figures) and $g=15$ (right figures).  Points below the blue surfaces are discarded and only points above them are used to the subsequent Inverse Fourier Transform to get a smoothed field. Top figures show the modes used in the Discrete Inverse Fourier Transform and bottom ones the detailed behavior of the discarded noisy modes.
 \label{FourierQ}}
\end{figure}

\begin{table}[h!]
\begin{center}
\resizebox*{!}{3cm}{ 
\begin{tabular}{|c|c|}
\hline
$g$&$\vert \hat Q^{(s)}\vert$\\ \hline
\hline
5&0.7\\ \hline
10&0.28\\ \hline
15&0.25\\ \hline
\end{tabular}}
\end{center}
\caption{Cut-off values for the $Q^{(s)}$ averaged. The modes of the Fourier Transform of the field with modulus less than the corresponding cut-off value are discarded. \label{cut2Q}}
\end{table}

\begin{figure}[h!]
\begin{center}
\includegraphics[height=5cm,clip]{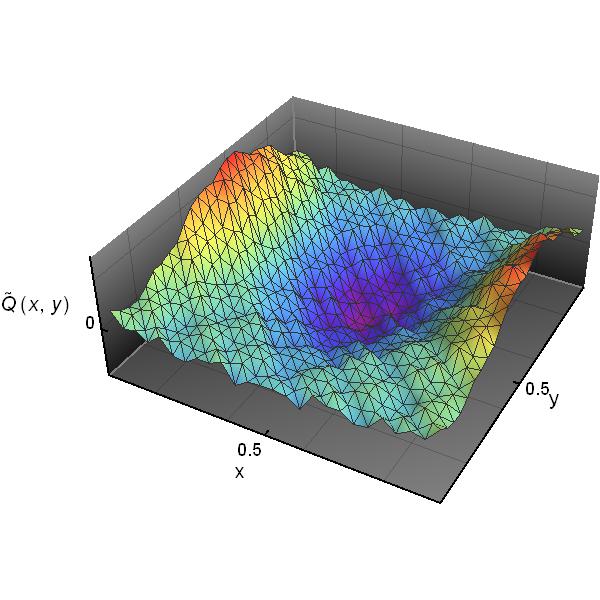}   %pre_E5new.nb
\includegraphics[height=5cm,clip]{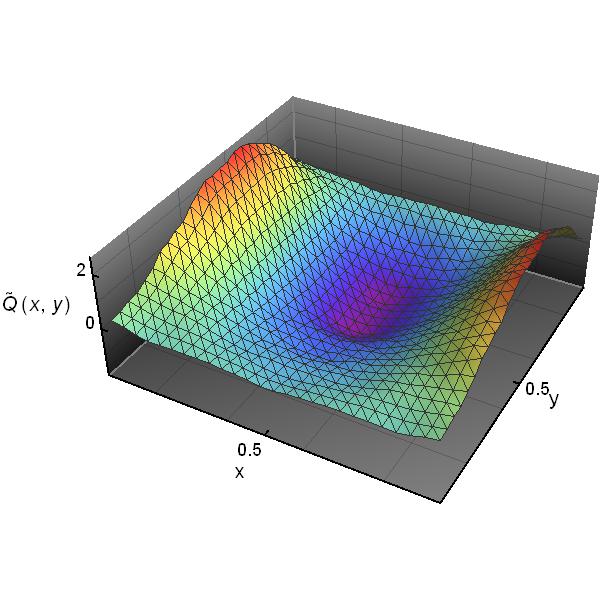}  %pre_E10new.nb
\includegraphics[height=5cm,clip]{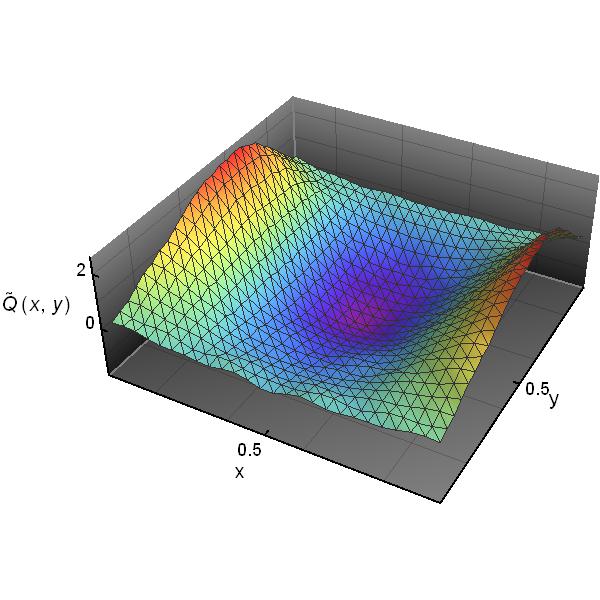}  %pre_E15new.nb
\end{center}
\kern -1.cm
\caption{Universal fields $\tilde Q(x,y)$   for $g=5$, $10$ and $15$ from left to right. 
 \label{universalQ}}
\end{figure}
\begin{figure}[h!]
\begin{center}
\includegraphics[height=5cm,clip]{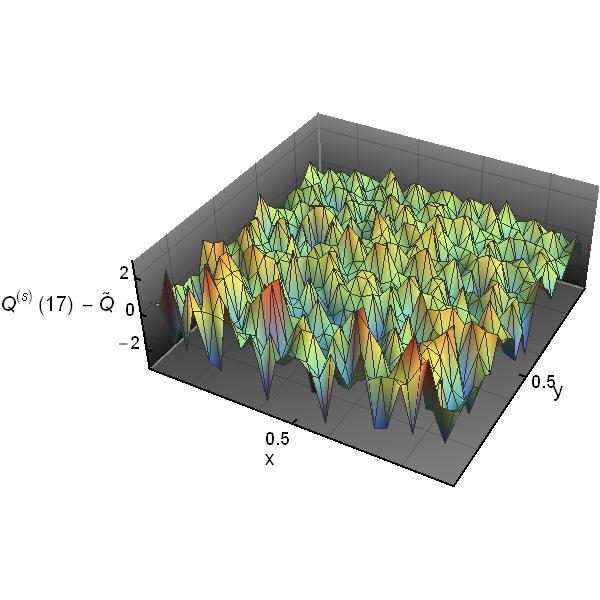}   %pre_E5new.nb
\includegraphics[height=5cm,clip]{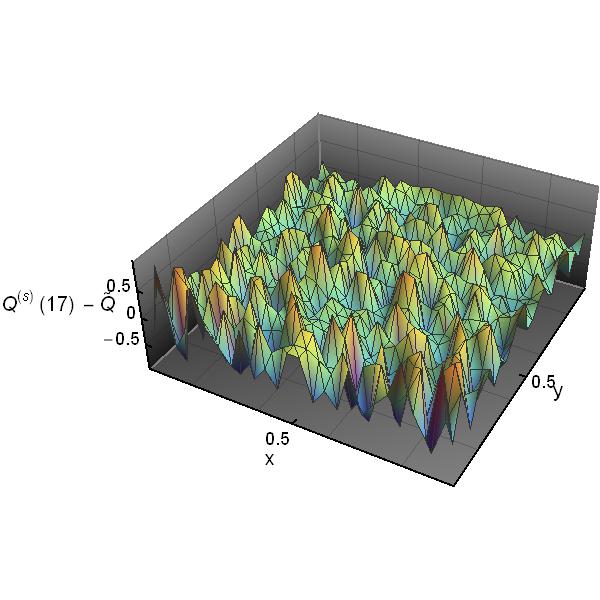} %pre_E10new.nb
\includegraphics[height=5cm,clip]{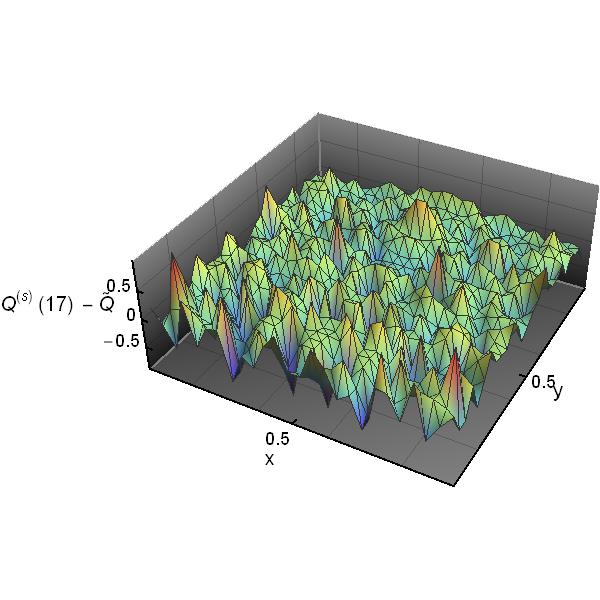} %pre_E15new.nb
\end{center}
\kern -1.cm
\caption{Difference between the scaled field $Q^{(s)}(x,y)$ for $T_0=17$ and the corresponding universal field $\tilde Q(x,y)$  for $g=5$, $10$ and $15$ from left to right. 
 \label{fluctuQ}}
\end{figure}

We define the scaled excess pressure field:
\begin{equation}
Q^{(s)}(x,y)=\frac{Q(x,y)-q(y)}{\sigma(Q)}
\end{equation}
where
\begin{equation}
\sigma(Q)^2=\frac{1}{N_C}\sum_{(x,y)}\left[Q(x,y)-q(y)\right]^2 \quad,\quad q(y)=\frac{1}{L_C}\sum_x Q(x,y)
\end{equation}
and $N_C=L_C^2$ and $L_C=26$  in our case and the sums exclude the two nearest boundary rows and columns.

We show in figures \ref{prefieldsuper} the superposition of the scaled pressure fields $Q^{(s)}$ for $T_0=17$, $18$, $19$ and $20$. We see how the $g=5$ is so noisy that we cannot discriminate, at a glance, a regular surface. For $g=10$ and $15$ we see again that all of them seem to follow the same surface without any systematic deviation. In figures \ref{premom} we present the behavior of $\sigma(Q)^2$ and  the fourth momenta of $Q^{(s)}$.
Due to computational problems we didn't get the errors in the local virial pressure and we are unable to get the corrections to the field due to the noise of the data. In any case we see that their behavior is very similar to the ones of the temperature or the density fields. In particular the fourth momenta tend to a constant value independent of $T_0$. The rest of the analysis follows the same path of the other already analyzed magnitudes.

 We see in figure \ref{distanceQ} how the mutual marginal distance tend to a constant. The constant is larger in the $g=5$ case due to the intrinsic fluctuations. We average the  scaled configurations with $T_0\in [14,20]$ values for all $g$ values. 
In order to get rid of the noise effects, we do the Fourier Transform  of the averaged configurations (figure \ref{FourierQ}) and we discard the modes below the cutoff showed in Table \ref{cut2Q}. After doing the inverse Fourier Transform to the remaining modes we get the universal scaled configuration for the excess of the pressure field $\tilde Q(x,y)$ (figure \ref{universalQ}). Finally, the differences between these universal configurations are shown in figure \ref{fluctuQ}.  .

We again observe that, for the configurations with $T_0\in[14,20]$ exists an universal field $\tilde Q(x,y)$ such that 
\begin{equation}
Q(x,y;T_0,g)= q(y;T_0,g)+\sigma(Q(T_0,g)-q(T_0,g))\tilde Q(x,y;g)
\end{equation}

\section{Connecting with hydrodynamics}

Once we have studied global and local magnitudes from the microscopic model, we would like to compare the observed system behavior with the one we would obtain from stationary Navier-Stokes equations applied to a hard disk system. 
 We already explained that NS equations are based on local conservation laws and in two relevant assumptions:
\begin{itemize}
\item {\it Local equilibrium hypothesis:} It is assumed that thermodynamics holds locally. That is, we can define local thermodynamic quantities like density, temperature, pressure, ... and they are assumed to be related via the equilibrium thermodynamic relations. For instance by the equation of state.
\item {\it The constitutive relations:} The local energy current is assumed just to be proportional to the temperature gradient (Fourier law) and the pressure tensor has only linear dependence on the hydrodynamic velocity gradients (newtonian fluid). Moreover, there are transport coefficient functions like the thermal conductivity, the shear viscosity and the second viscosity that they are assumed to depend only on the local values of density and temperature.
\end{itemize}
Some questions may arise at this point: Up to what extend the computer simulation is going to give us results to be compared with hydrodynamics?  Is local equilibrium hypothesis valid always or only for (relatively) small external temperature gradients? and the Fourier law or the newtonian fluid hypothesis? Also there are other interesting questions to discuss, for instance What is the relation between the mechanical pressure and the thermodynamic pressure in the convective regime?, or
Is it correct to use the Boussinesq approximation to study the non-convective-convective transition? We'll intent to answer some of these questions. 

\begin{itemize}
\item{\bf 0. The Problem of the numerical derivatives}

We have measured in our computer simulation a set of observables  in  each of the $30\times 30$ virtual cells in which  we have divided our system. The obtained averaged values are, in most cases, smooth enough to look like a spatial regular field. In order to compare our results with the NS-equations (without solving them) we need to do the derivatives to such computed {\it discrete fields}. There are many strategies to do derivatives of a discrete set of variables, for instance discrete local schemes or data polynomial interpolations before doing analytical derivatives. However, if we want to get the goodness or consistency of our results compared with the NS equations, we need to control the propagation of the errors on the derivatives. They  are coming from two sources: the intrinsic error bars in each measured datum and the chosen derivative scheme. After playing with several possibilities we decided use the following strategy. Let $(a(n,l),\sigma(n,l))$ the computed value of the observable and its statistical standard deviation at cell $(n,l)$ then: 
\begin{itemize}
\item (a) We create a new data set by perturbing $a(n,l)$ by a gaussian random number with zero mean and $\sigma(n,l)$ standard deviation: $\tilde a(n,l)=a(n,l)+\sigma(n,l)\xi(n,l)$ with $\xi(n,l)\in N(0,1)$.
\item (b)We fit the data to a seventh order polynomial on each variable. That is,  the fitted function to the data, $A(x,y)$, is of the form:
\begin{equation}
A(x,y)=\sum_{n=1}^{N_x}\sum_{l=1}^{N_y}c(n,l)(x-\frac{1}{2})^n(y-\frac{1}{2})^l
\end{equation}
where typically $N_x=N_y=7$.
We can include boundary terms if necessary. For instance, if we know that the observable should be zero at $x=0,1$ we multiply the polynomial by $x(1-x)$.  
\item (c) We do an analytic derivative to the fitted function $A(x,y)$. 
\item (d) We compute the values of the derivative and its square at the $(x,y)$ points corresponding to the center of the virtual cells. 
\item (e) We iterate steps (a)-(d) $1000$ times and average at each cell point the derivative and its square. 
\item (f) the local derivative is the average value and its dispersion is the standard deviation of the thousand data values obtained. 
\end{itemize}
We show in figures \ref{partialx} and \ref{partialy} the results of this strategy by plotting the partial derivatives of the density, $\rho$, pressure, $Q$, temperature $T$ and the hydrodynamic velocity components $u_1$ and $u_2$ for $g=15$ and $T_0=20$. Observe how the derivatives present a larger error interval on the boundaries where there are no constraints which it is due to the variability of the fitted function for values on the boundary (remind that we take the center of the box, $x=1/2$ and $y=1/2$, as the seed of the polynomial expansion). It is clear that the error is reasonably small  and it gives us some hope into obtain some reasonable info about the NS-equations behavior.
\begin{figure}[h!]
\begin{center}
\includegraphics[height=5cm,clip]{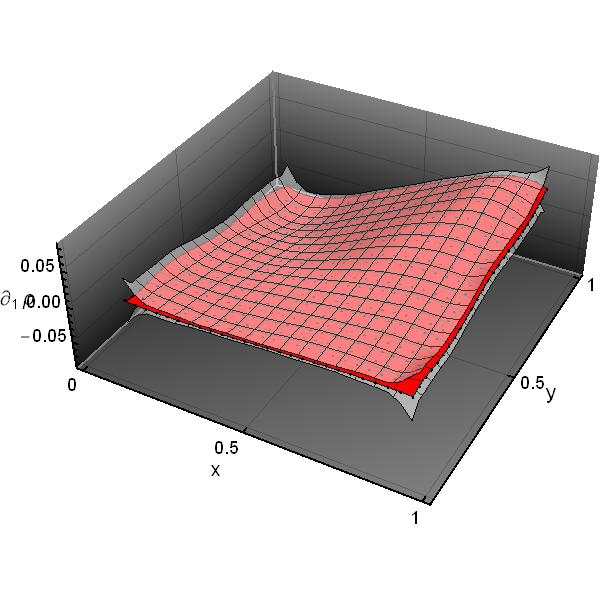}   %derivada_den.nb
\includegraphics[height=5cm,clip]{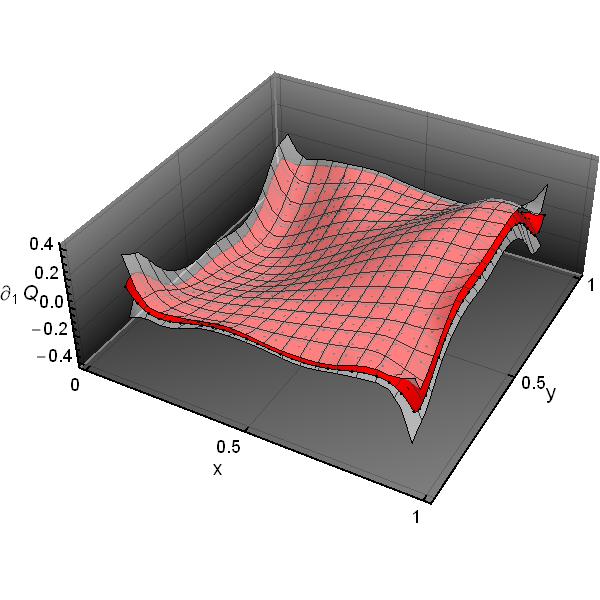}  %derivada_P.nb
\includegraphics[height=5cm,clip]{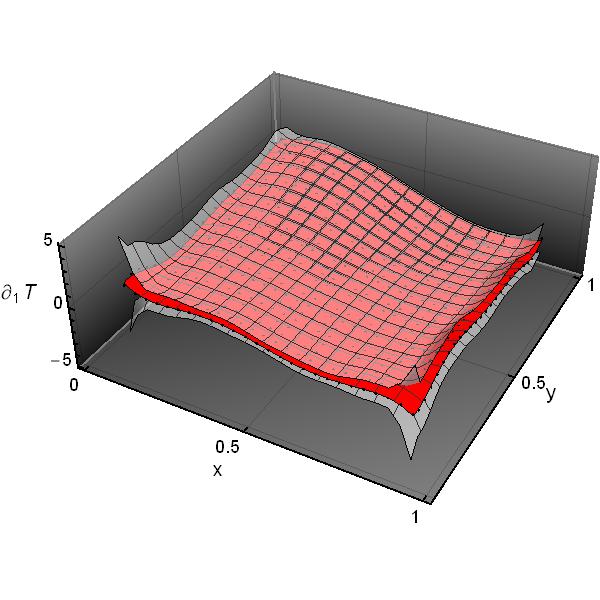}  %derivada_T.nb
\includegraphics[height=5cm,clip]{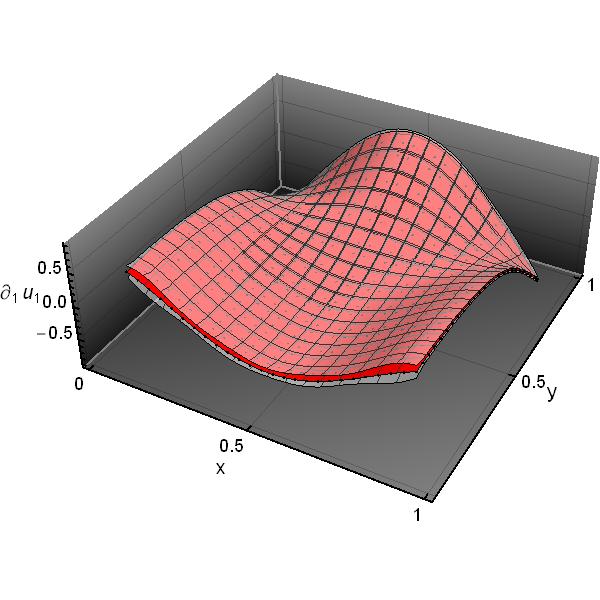}  %derivada_u1.nb
\includegraphics[height=5cm,clip]{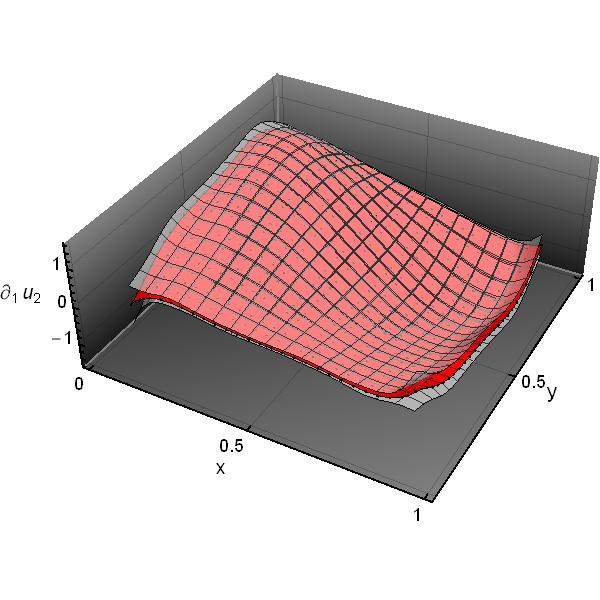}  %derivadau2.nb
\end{center}
\kern -1.cm
\caption{Partial derivatives respect $x$ for the fields, $\rho$, $Q$, $T$, $u_1$ and $u_2$ for $g=15$ and $T_0=20$ from left to right. White surfaces limit the error interval. Red surface is the averaged derivative values and black dots are the derivative of the data fit.
 \label{partialx}}
\end{figure}
\begin{figure}[h!]
\begin{center}
\includegraphics[height=5cm,clip]{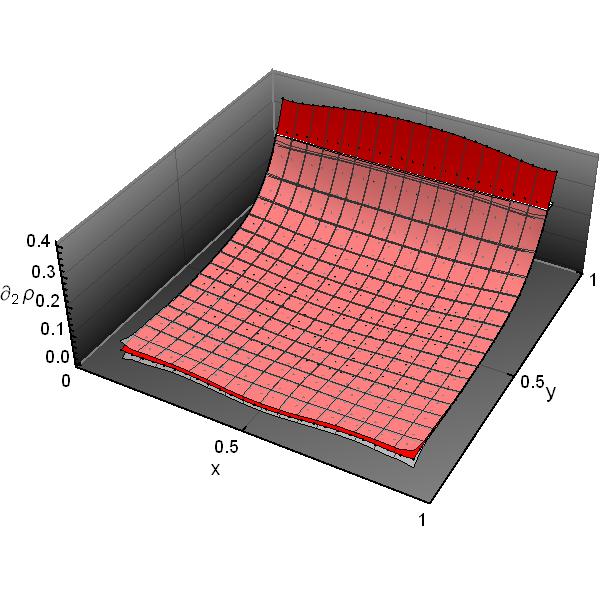}   %derivada_den.nb
\includegraphics[height=5cm,clip]{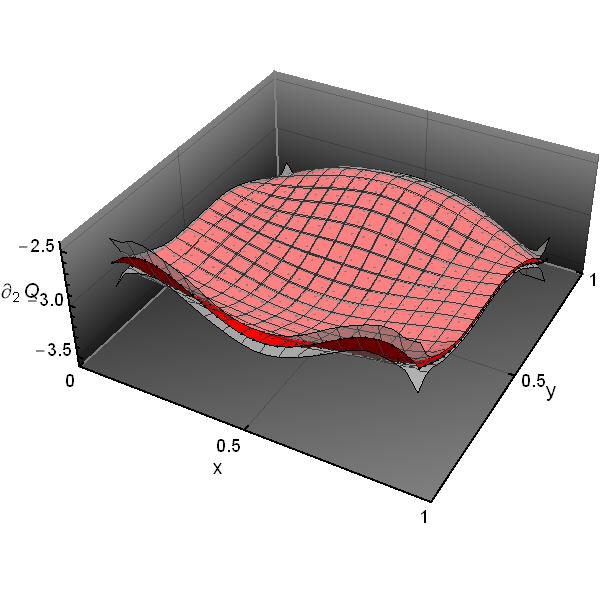}  %derivada_P.nb
\includegraphics[height=5cm,clip]{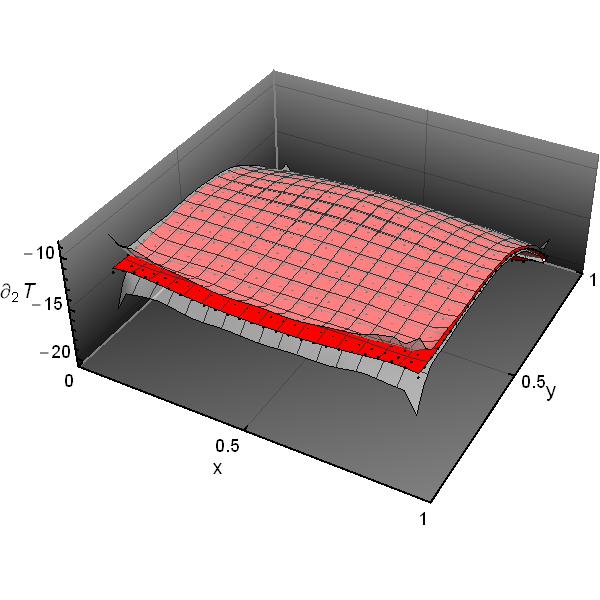}  %derivada_T.nb
\includegraphics[height=5cm,clip]{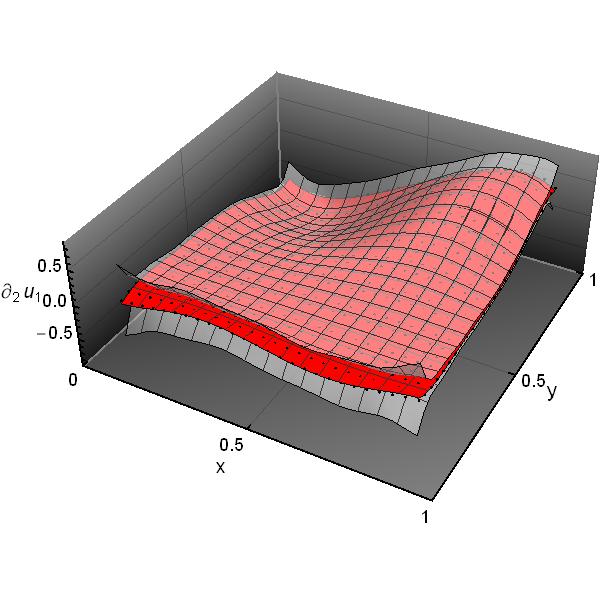}  %derivada_u1.nb
\includegraphics[height=5cm,clip]{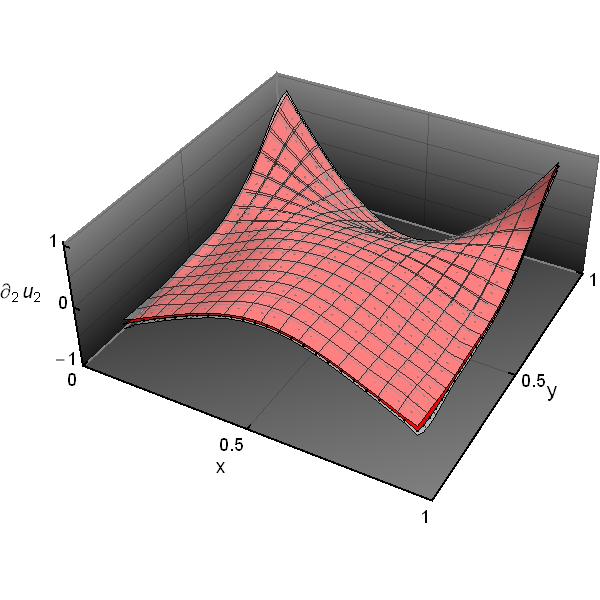}  %derivadau2.nb
\end{center}
\kern -1.cm
\caption{Partial derivatives respect $y$ for the fields, $\rho$, $Q$, $T$, $u_1$ and $u_2$ for $g=15$ and $T_0=20$ from left to right. White surfaces limit the error interval. Red surface is the averaged derivative values and black dots are the derivative of the data fit.
 \label{partialy}}
\end{figure}

\item{\bf 1. The continuity equation}

The continuity equation express the hydrodynamic mass conservation:
\begin{equation}
\partial_1(\rho u_1)+\partial_2(\rho u_2)=0
\end{equation}
We may compute each of the derivatives of our discrete fields $\rho u_1$ and $\rho u_2$ obtained in the computer simulation and check that its sum gives zero. Obviously, from our polynomial fitting method (5 or 7th order polynomial) we cannot expect a perfect matching (that would only occur if we used tha exact analytic solution). However we should see that the answer is contained inside the computed error bars. In order to convince the reader that the continuity equation holds let us use two complementary points of view. First we show in figure \ref{cont1} the numerically obtained $\partial_1(\rho u_1)$  and $\partial_2(\rho u_2)$ fields, and its corresponding sum for the case with $T_0=20$ and $g=5$, $10$ and $15$. We observe that in all cases the sum of each component is compatible with the zero value. The relative error is smaller as we increase the $g$ value. For $g=15$, the variability of the derivatives is of order $0.4$ and the error of the sum is bounded between $\pm 0.03$, that is a $0.06$ interval of variation. That is, we are seeing a true cancellation when summing up the derivatives of two different functions more beyond the error noise.

The second strategy is to plot $\partial_1(\rho u_1)$ vs $\partial_2(\rho u_2)$ for each virtual cell point where the derivative is computed. If continuity equation holds the data should follow the line $y=-x$ for any value of $T_0$ and $g$. We show in figure \ref{cont2} the data for $T_0=14$, $17$ and $20$ for $g=5$, $10$ and $15$. We observe how, beyond error bars, the data follows the expected line. We can conclude that the numerical analysis permits us to show that the basic hydrodynamic continuity equation is correct This is the minimum requirement we had to pass to continue, with some confidence, the analysis of the NS hydrodynamic equations from our computer simulation.
\begin{figure}[h!]
\begin{center}
\includegraphics[height=5cm,clip]{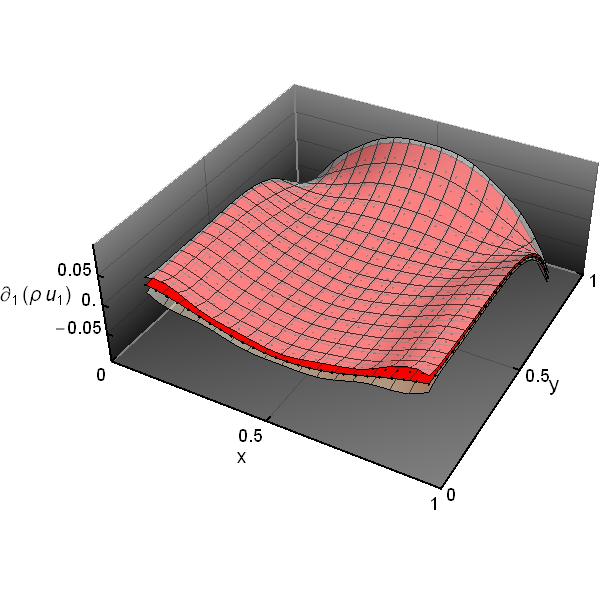}   %derivadas_1_g5.nb
\includegraphics[height=5cm,clip]{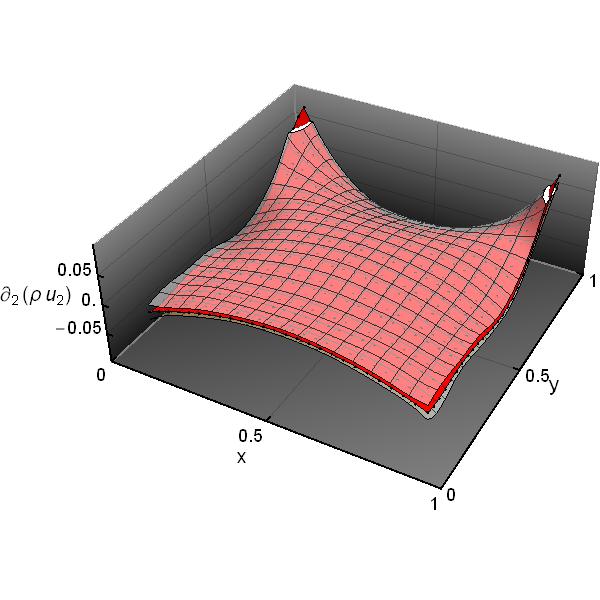}  %derivadas_1_g5.nb
\includegraphics[height=5cm,clip]{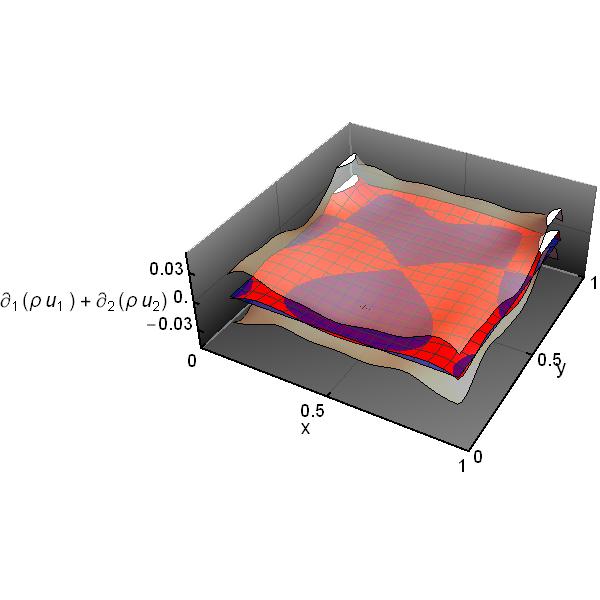}  %derivadas_1_g5.nb
\includegraphics[height=5cm,clip]{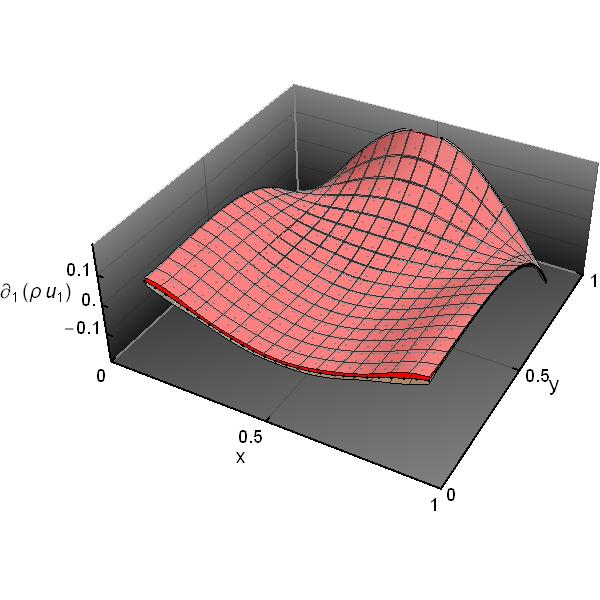}   %derivadas_1_g10.nb
\includegraphics[height=5cm,clip]{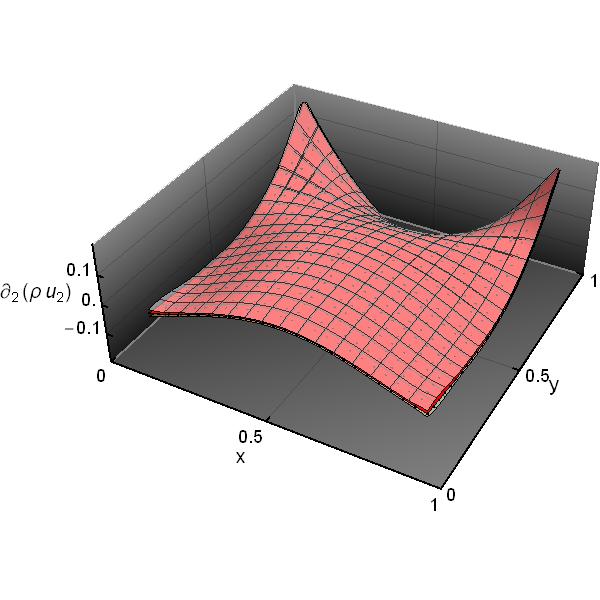}  %derivadas_1_g10.nb
\includegraphics[height=5cm,clip]{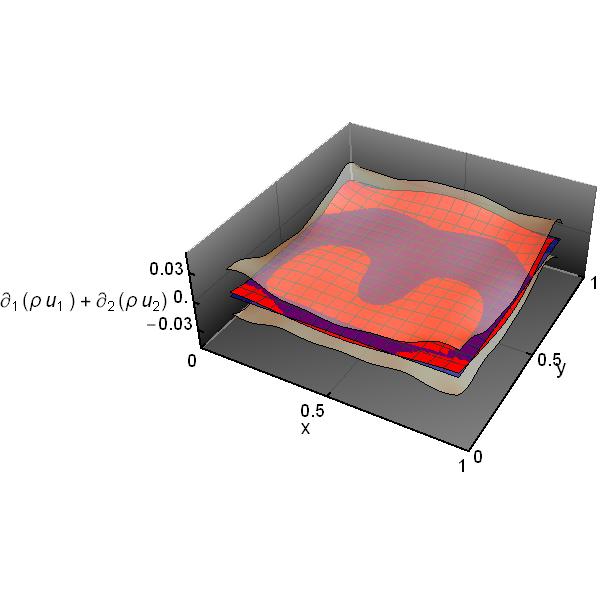}  %derivadas_1_g10.nb
\includegraphics[height=5cm,clip]{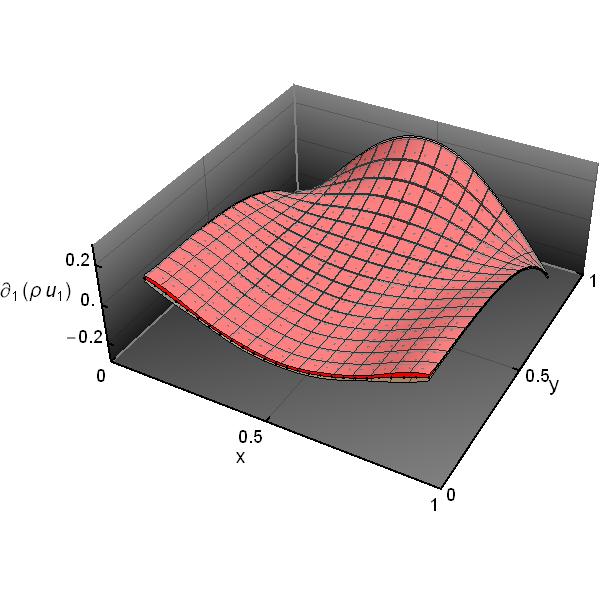}   %derivadas_1_g15.nb
\includegraphics[height=5cm,clip]{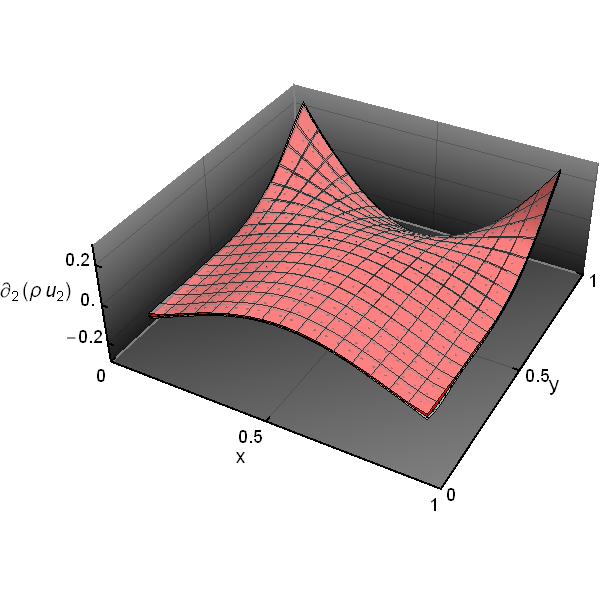}  %derivadas_1_g15.nb
\includegraphics[height=5cm,clip]{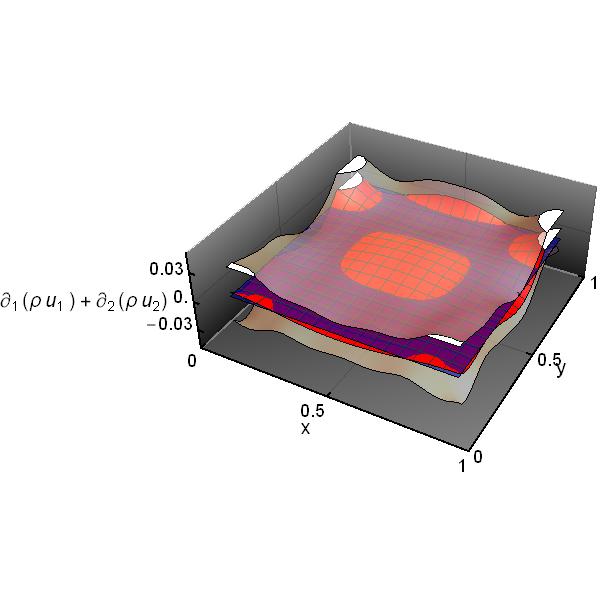}  %derivadas_1_g15.nb
\end{center}
\kern -1.cm
\caption{Numerically computed $\partial_1(\rho u_1)$ (first column), $\partial_2(\rho u_2)$ (second column) and its sum (third column) for $T_0=20$ and $g=5$, $g=10$ and $g=15$ from top to bottom rows. White surfaces limit the error interval. Red surface is the averaged derivative values and black dots are the derivative of the data fit. Blue surface is the reference zero.
 \label{cont1}}
\end{figure}
\begin{figure}[h!]
\begin{center}
\includegraphics[height=5cm,clip]{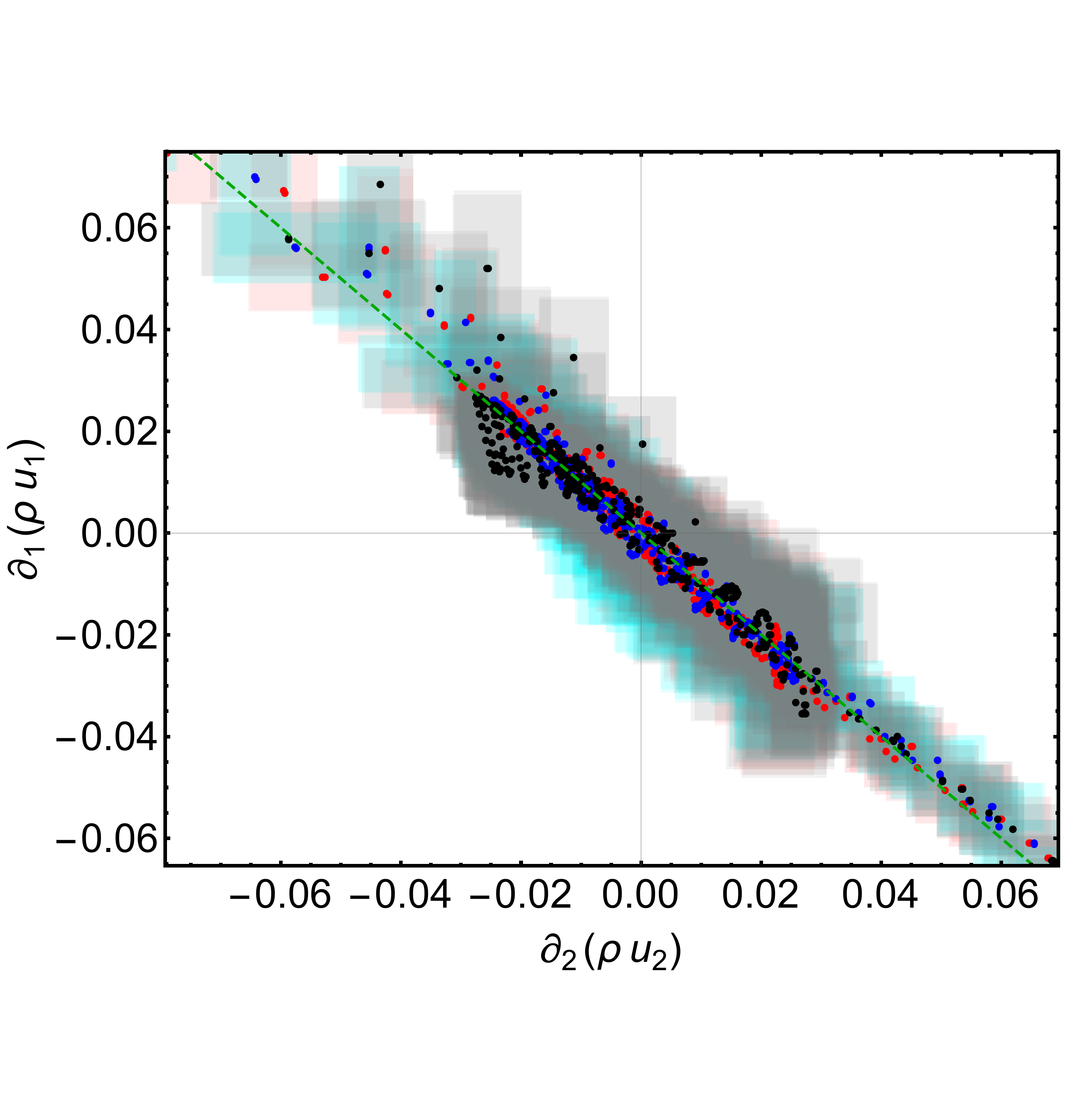}   %derivadas_2_g5.nb
\includegraphics[height=5cm,clip]{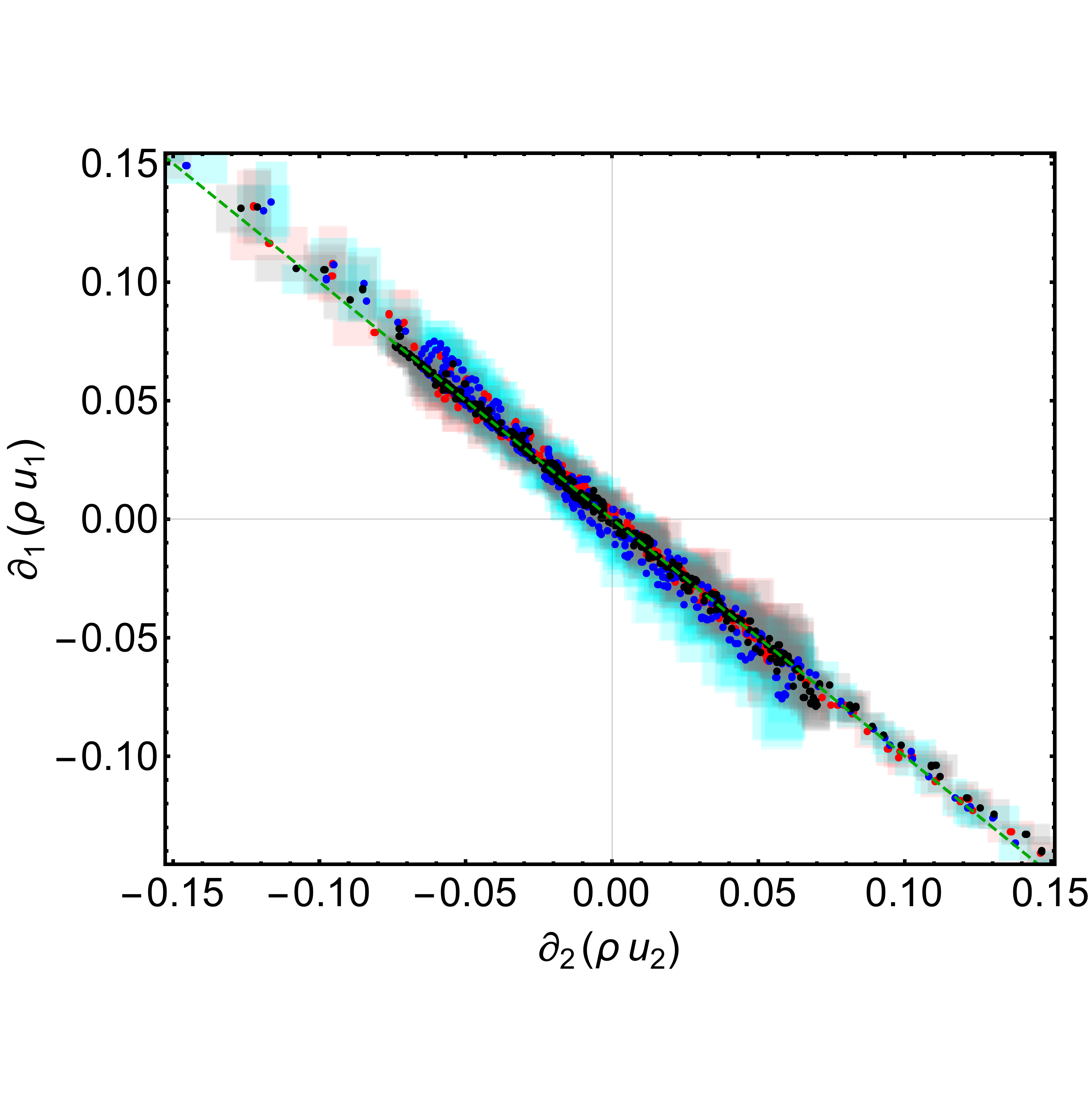}  %derivadas_2_g10.nb
\includegraphics[height=5cm,clip]{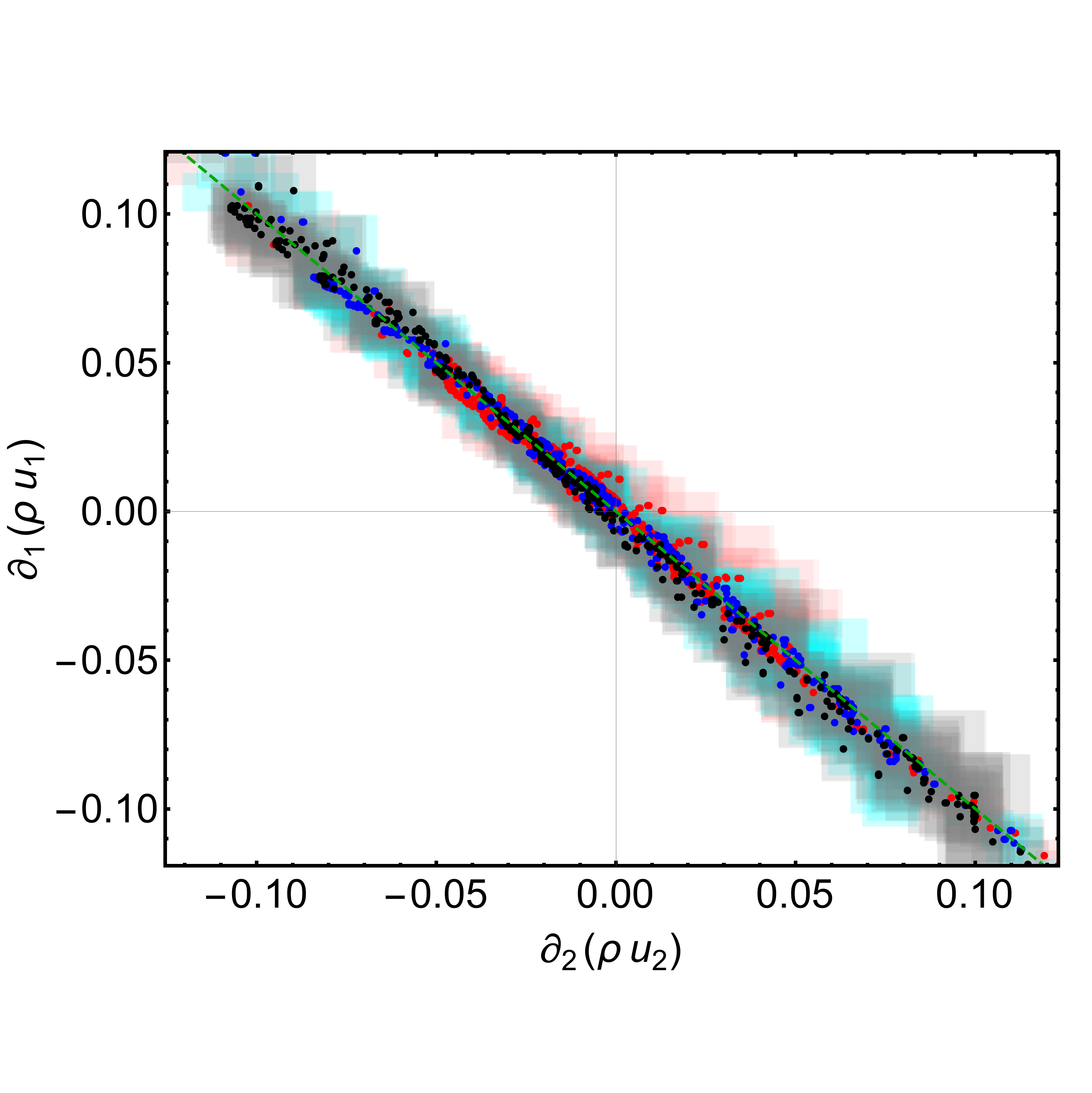}  %derivadas_2_g15.nb
\end{center}
\kern -1.cm
\caption{Numerically computed $\partial_1(\rho u_1)$ versus $\partial_2(\rho u_2)$ at the virtual cell points for $T_0=14$ (Red dots), $17$ (Blue dots) and $20$ (Black dots) for  $g=5$, $10$ and $15$ from left to right. Squares represent the error area. The dashed line is the expected theoretical result if continuity equation holds.
 \label{cont2}}
\end{figure}

\item{\bf 2. Checking the Local Equilibrium hypothesis}

In order to close the conservative equations in Navier-Stokes we need to relate the local temperatures, pressures and densities assuming that equilibrium  applies at each macroscopic spatial point. This is a well assumed property that is called {\it local equilibrium hypothesis}. Let us emphasize that it doesn't mean that the fluid is at equilibrium locally. That is, the local probability measures is not a Gibbs measure because there are currents of energy, momentum and matter that breaks the symmetry properties of the Hamiltonian. Local equilibrium just means that the equation of state of equilibrium applies at each macroscopic point. This could be so, probably because the nonequilbrium measure distribution shape near the most probable configurations is similar to the corresponding Gibbs distribution but the tails deviate from it in a highly nontrivial way.

We know that the equilibrium equation of state (EOS) for hard disks is unknown. However, there are a set of analytic expressions that approximate the measured results with relative errors smaller than $1\%$ (for instance the Henderson EOS). These expressions may help us to check, up to some extend, if local equilibrium holds for the hard disk system. In any case, we have an important advantage in the hard disk case. We know from the equilibrium Gibbs probability measure that the EOS has the structure:
 \begin{equation}
 Q=T F(\rho)=T\rho H(\rho) \label{eos}
 \end{equation}
because the interaction potential has only two values: zero (when disks move freely) or infinity (when disks are in contact).
If local equilibrium applies,  the values of density, temperature and  pressure measured at {\it each cell} should follow {\it the same} relation given by eq.(\ref{eos}) and it should be independent on the external gradient $T_0$ or the value of the external driving field $g$. This is a very exigent assumption. In figure \ref{eos1} we plot the $75712$ points of $(\rho(x,y),T(x,y),Q(x,y))$ that we have obtained in the set of all simulations shown in this paper. We have discarded the points near the boundaries (the nearest two rows and columns) to eliminate the exclusion surface effects due the existence of the rigid boundaries. We observe in the figures the variety of profiles in the $(\rho,T,Q)$ space and how all follow (apparently) the same surface in such space. 
\begin{figure}[h!]
\begin{center}
\includegraphics[height=7cm,clip]{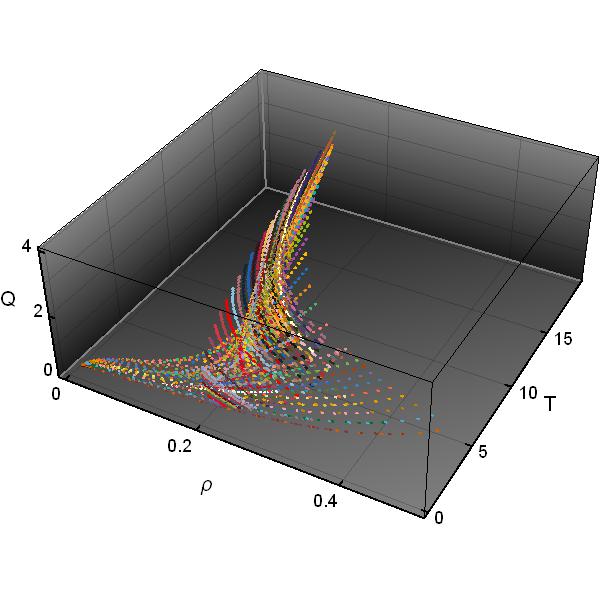}  %eos.nb
\end{center}
\kern -1.cm
\caption{Local equilibrium: each point has the coordinate $(\rho(x,y),T(x,y),Q(x,y))$ that are the values of density, Temperature and Pressure measured at each cell $(x,y)$  (excluding the sites near the boundaries) for all numerical experiments done in this paper ($T_0\in[1,20]$ and $g=0, 5, 10, 15$). There are $75712$ data points.
\label{eos1}}
\end{figure}
\begin{figure}[h!]
\begin{center}
\includegraphics[height=7cm,clip]{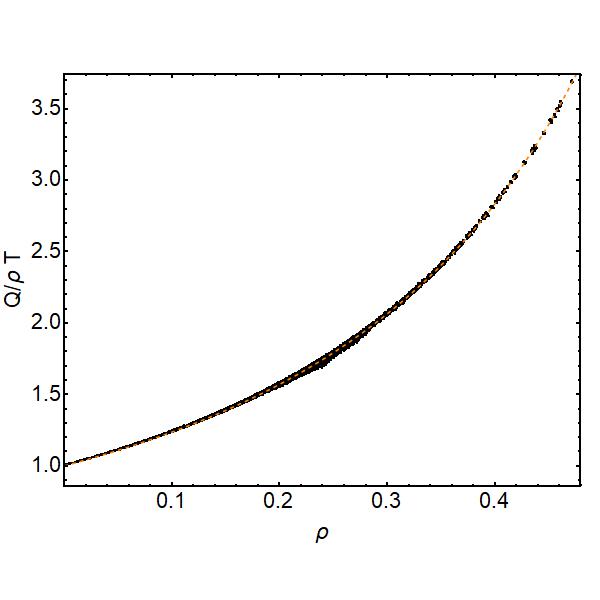}  %eos.nb
\end{center}
\kern -1.cm
\caption{Local equilibrium: each point has the coordinate $(\rho(x,y),Q(x,y)/\rho(x,y)T(x,y))$ that are the values of density, Temperature and Pressure measured at each cell $(x,y)$  (excluding the sites near the boundaries) for all numerical experiments done in this paper ($T_0\in[1,20]$ and $g=0, 5, 10, 15$). There are $75712$ data points.
\label{eos2}}
\end{figure}

We can be more precise. If we plot $(\rho(x,y),Q(x,y)/\rho(x,y)T(x,y))$, the EOS for hard disks, eq.(\ref{eos}), predicts that all points should follow exactly the same function: $H(\rho)$. 
In figure \ref{eos2} we show the   $75712$ data points in such graph. The dotted line is the function proposed by Henderson that is assumed to be correct (up to $1\%$ of relative error):
\begin{equation}
H_H(\rho)=\frac{1+\rho^2/8}{(1-\rho)^2}\label{hender}
\end{equation}

\begin{figure}[h!]
\begin{center}
\includegraphics[height=6cm,clip]{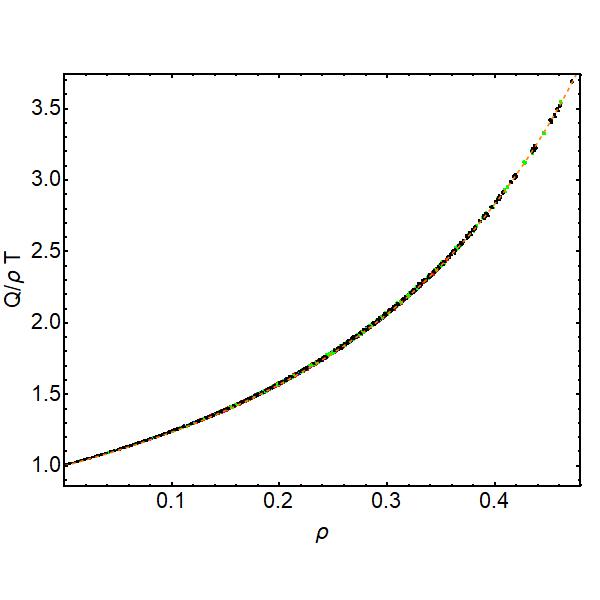}  %eos.nb
\includegraphics[height=6cm,clip]{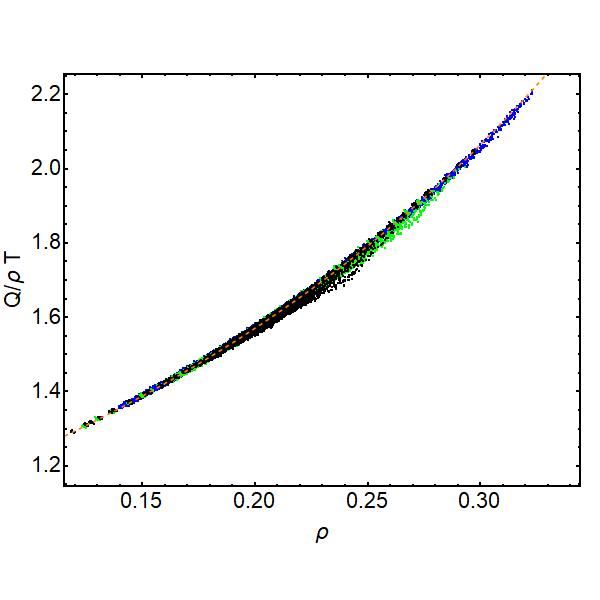}  %eos.nb
\end{center}
\kern -1.cm
\caption{Local equilibrium: each point has the coordinate $(\rho(x,y),Q(x,y)/\rho(x,y)T(x,y))$ that are the values of density, Temperature and Pressure measured at each cell $(x,y)$  (excluding the sites near the boundaries). Left: data points of system at non-convective regime (18928 points). Right: data points of systems at the convective regime (56784 points).  $g=0, 5, 10, 15$ are the red, blue, green and black dots respectively.  The dotted line is the Henderson's proposal. 
\label{eos3}}
\end{figure}
\begin{figure}[h!]
\begin{center}
\includegraphics[height=6cm,clip]{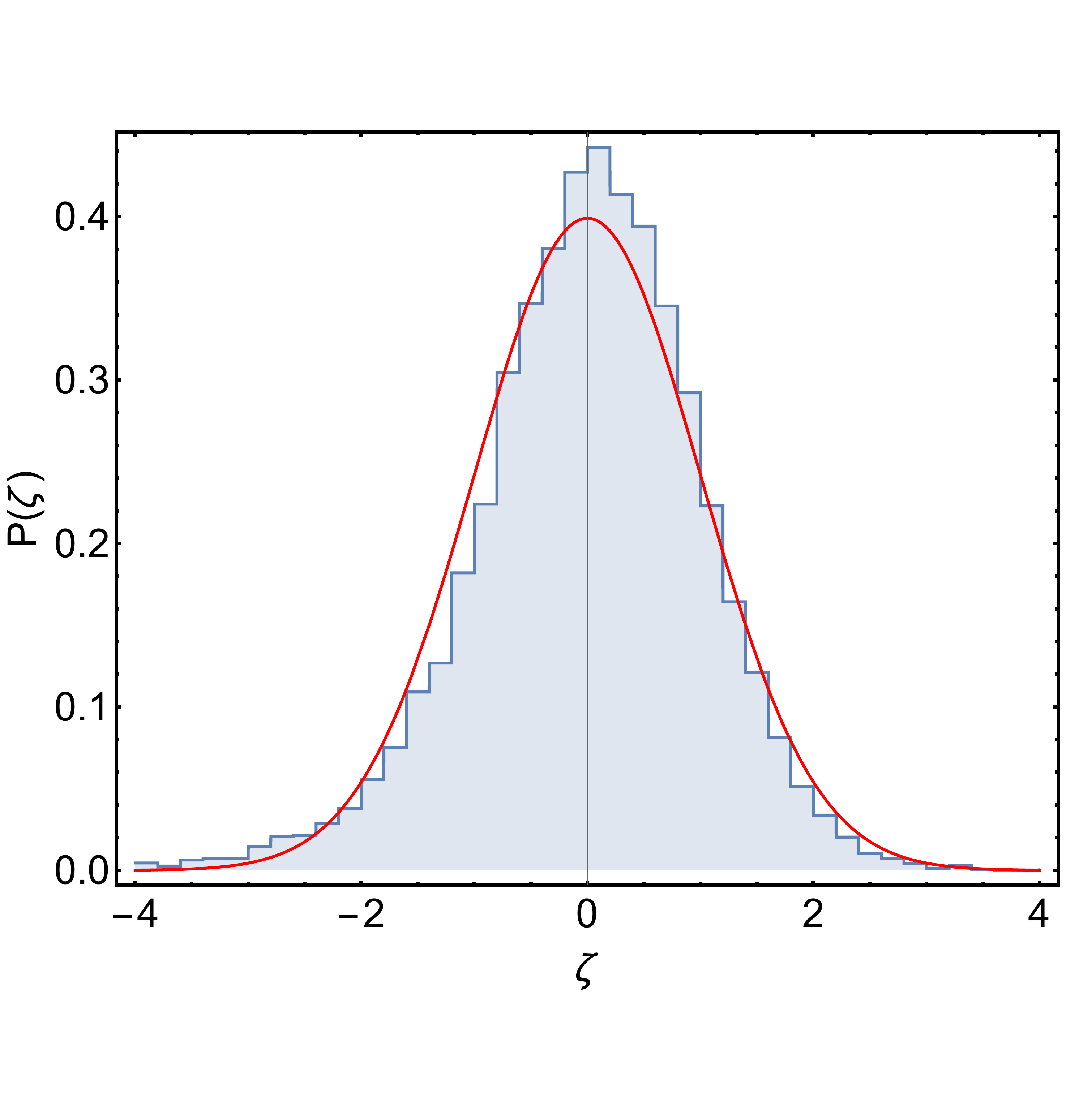}  %eos.nb
\end{center}
\kern -1.cm
\caption{Local equilibrium: Distribution of the differences for systems at the non-convective regime $\xi=(\Delta H-\langle\Delta H\rangle)/\sigma(\Delta H)$ where $\Delta H = H-H_H(\rho)$ and $H=\pi r^2 P/\rho T$ and $P$, $\rho$ and $T$ are the measured pressure, density and temperature respectively.  Red solid line is the normalized gaussian with zero average and unit standard deviation.
\label{eos5}}
\end{figure}
\begin{figure}[h!]
\begin{center}
\includegraphics[height=6cm,clip]{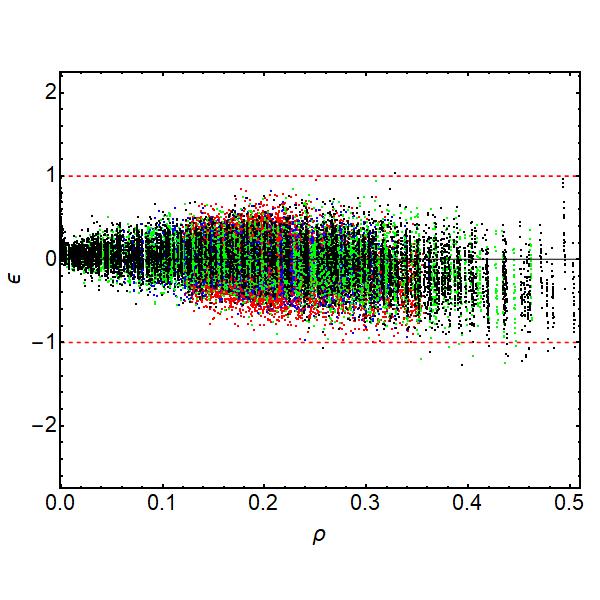}  %eos.nb
\includegraphics[height=6cm,clip]{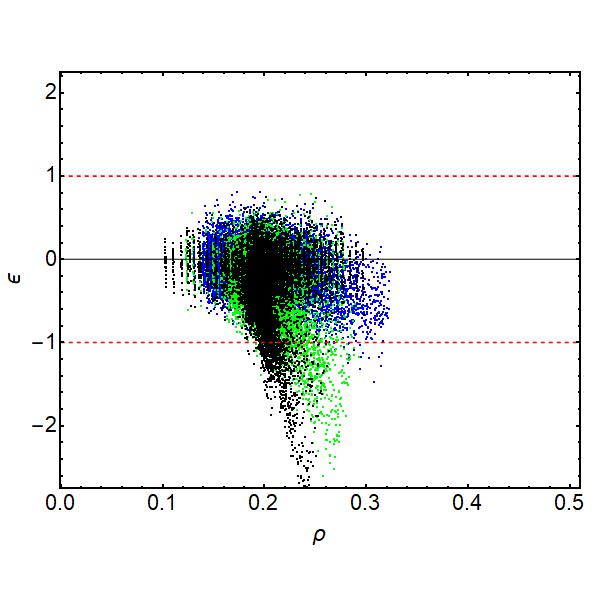}  %eos.nb
\end{center}
\kern -1.cm
\caption{Local equilibrium: The data relative error with respect Henderson's EOS proposal as a function of $\rho$. $\epsilon=100(Q/(\rho T H_H(\rho))-1)$. It is accepted that Henderson's EOS is correct under a $1\%$ relative error with the real EOS. Dashed lines shows such limits. Left: data points of systems at non-convective regime. Right: data points of systems at the convective regime. 
\label{eos4}}
\end{figure}

We observe how the complete set of data follows the same curve that coincides apparently with the Henderson's proposal. However, we appreciated two small but systematic deviations: 
\begin{itemize}
\item (1) The points corresponding to the non-convective regime deviate, in average, from the Henderson's EOS: $\langle\Delta H\rangle=\sum_{i=1}^{N_D}(H_i-H_H(\rho_i))/N_D=-0.00098$ where $i$ runs over all data cells pertaining to configurations at non-convective states, the total number of such data is $N_D=18928$, $H_i=\pi r^2 P_i/\rho_iT_i$, and $P_i$, $\rho_i$ and $T_i$ are the measured pressure, density and temperature at each cell $i$. The data mean standard deviation of such differences is $\sigma(\Delta H)=0.00402297...$ and the statistical error of the average is $3\sigma(\Delta H)/\sqrt{N_D}=0.0000877$ which is almost $10$ times smaller that the observed average. That is, we are seeing some deviation from the Henderson's EOS that is small but systematic. Moreover, if we normalize the differences: $\xi=(\Delta H-\langle\Delta H\rangle)/\sigma(\Delta H)$ we obtain a non-Gaussian behavior (see figure \ref{eos5}). This implies that $H_H(\rho)$ is not the real EOS (as we knew already). It is surprising to us how our data detects such fact in a statistic way. Nevertheless the data set is in the $1\%$ relative error taking as reference the Henderson's EOS as it was expected (see figure \ref{eos4} left).
\item (b) Some points in the range $\rho\in[0.2,0.3]$ deviated (more than an usual fluctuation) from the rest. After some analysis we found that part of the data corresponding to the convective regime (see figure \ref{eos3} right) are the one that deviates systematically from the scaling behavior. In order to study such observed deviations we show in figure \ref{eos4}  the relative differences between the data points with the Henderson's proposal. 
We observe how, for all the data pertaining to states in non-convecting states, the points are distributed around zero and they do not cross the limiting errors lines of $1\%$. However, the data from convective states are distributed in a non symmetric manner around zero and there is a set of points deviating far from the limiting error lines. We think that this is not casual and there should be some reason beyond fluctuating data and/or errors. That is, it seems that the existence of a nonzero hydrodynamic velocity field affects to the existence of a unique $H(\rho)$ function. In fact, the observed deviation grows systematically with $g$ which is correlated with the fact that the modulus of the observed velocity field also grows with $g$.

 \end{itemize}

In order to understand the apparent violation of the local equilibrium from the data corresponding  to systems in the convective regime, we should remind some definitions.
In Navier-Stokes equations, the pressure appears as a part of the stress tensor:
\begin{equation}
\tau_{ij}=-P\delta_{ij}+\eta(\partial_iu_j+\partial_ju_i)+\eta'\delta_{ij}\sum_k\partial_ku_k
\end{equation}
where $P$ is the Pressure that appears in the EOS (also called {\it Thermodynamic Pressure}). One can also define  the {\it mean normal stress}:
\begin{equation}
\bar P=-\frac{1}{d}\text{Tr}(\tau)
\end{equation}
where $d$ is the system dimension and it is written is this way to guarantee its invariance with respect any change of coordinates. Let us recall that the stress tensor is defined from the vector small  forcing ($\delta f$) acting over a small surface ($n\delta S$):
\begin{equation}
\delta f_i=\sum_j\tau_{ij}n_j\delta S
\end{equation}
For a fluid at rest ($u=0$) we find that $\delta f=-P n\delta S$. That is, the mean normal stress coincides with the Thermodynamic Pressure which is the force applied by unit surface and perpendicular to it. When the fluid is moving, $\bar P$ and $P$ differ and they are related by:
\begin{equation}
\bar P=P-(\eta'+\frac{2}{d}\eta)\nabla\cdot u\label{eos6}
\end{equation}
where $\eta'+\frac{2}{d}\eta=\xi$ is known as the {\it bulk viscosity}.
$\bar P$ is often called {\it Mechanical Pressure} because it measures the total averaging forcing at a given fluid point. Let us mention that typically it is assumed that the bulk viscosity is a function of the local variables $\rho(x)$ and $T(x)$. However there is also the possibility that $\xi$ being a function of $\bar P$, $T$ and $\rho$ implying that $\bar P$ could be a nonlinear function of $\nabla\cdot u$ (see for instance \cite{Raja2}).

Observe that when the fluid is at rest or the fluid is incompressible ($\nabla\cdot u=0$) both magnitudes coincide $\bar P=P$. Stokes in 1845 assumed that, in general, the fluids have $\xi=0$ (Stokes assumption). This assumption is still under discussion but it seems that the actual theoretical and experimental evidence is against it (see for instance \cite{Gad,Buresti,Raja}).

At this point we should think about what we have really measured by using the virial pressure expression (\ref{pre0}). Looking with care its derivation we can conclude that the virial pressure fields $P(x,y)$ analyzed above are in fact the {\it Mechanical Pressure fields} $\bar P(x,y)$ because we are measuring the TOTAL external forcing applied to a given region. It this is so, the observed deviations to the expected EOS from the data pertaining to convecting states should be due to the use in the EOS of $\bar P$ instead to the correct $P$ fields. Therefore, we would like to get the thermodynamic pressure from our data by using eq. (\ref{eos6}) and afterwards to check if the systematic observed deviations disappear. Obviously the first question to be solved is to see how $\bar P$ depends on the local $\rho$, $T$ and $\nabla\cdot u$. In particular we would like to check if it is just a linear function on $\nabla\cdot u$.

Let us first write eq. (\ref{eos6}) for hard disk systems. We assume the general form:
\begin{equation}
\bar P(x,y)=\frac{\rho(x,y) T(x,y)H(\rho(x,y))}{\pi r^2}-\xi(\rho(x,y),T(x,y),\bar P(x,y))\nabla\cdot u\label{dim1}
\end{equation}
where $H(\rho)=\pi r^2 P/\rho T$ (see eq.(\ref{eos})). Dimensional arguments imply that the bulk viscosity should be of the form
\begin{equation}
\xi(\rho,T,\bar P)=\frac{\sqrt{T}}{r} \bar E_B\left(\rho,\frac{\pi r^2\bar P}{T}\right)\label{dim2}
\end{equation}
where the arguments of the function $\bar E_B$ are dimensionless (see for instance the expression in Enskog's approximation). Substituting into eq.(\ref{dim1}) we find:
\begin{equation}
\bar H(\rho)=H(\rho)- \tilde E_B\left(\rho,\bar H\right) \frac{\nabla\cdot u}{\sqrt{T}}\label{dim3}
\end{equation}
where $\bar H=\pi r^2\bar P/\rho T$ and $\tilde E_B=\pi r\bar E_B/\rho$. That is, $\bar H$ is solution of an implicit equation and it has the general form of: 
\begin{equation}
\bar H=G\left(\rho, \frac{\nabla\cdot u}{\sqrt{T}}\right)
\end{equation}
 We know that for $u=0$, $\bar H=H$ and then $G(\rho,0)= H(\rho)$. Then it is convenient to write:
\begin{equation}
\bar H\left(\rho,\frac{\nabla\cdot u}{\sqrt{T}}\right)=H(\rho)+G_1\left(\rho,\frac{\nabla\cdot u}{\sqrt{T}}\right)  \label{barH}
\end{equation}
where we have assumed that $G(x,y)$ is an analytic function. 

Observe that if $G_1(x,y)=g(x)y$ we are in the case where the bulk viscosity only depends on the local temperature and local density ($\xi=\xi(\rho,T)=\sqrt{T}\bar E_B(\rho)/r$) and $g(\rho)=-\pi r \bar E_B(\rho)/\rho$. In a general case, $G_1(x,y)= g_1(x)y+O(y^2)$ when $y\rightarrow 0$. We expect also that for very low density the system tends to the ideal gas behavior and then $\bar H\simeq 1+O(\rho)$. But $H\simeq 1+O(\rho)$ and then $G_1$ should have the form: $G_1(x,y)=x\bar G_1(x,y)$ with $\bar G_1$ being an analytic function on both arguments.

Then, by studying eq.(\ref{barH}) using our data for $\bar H$, $\rho$, $\nabla\cdot u$ and $T$, we can try to fit $H$ and $G_1$ functions. However we deal with several numerical problems: (a) We see that $\bar H$ does not differ too much from $H$ and the data obtained for $\bar H$ fluctuates around $H$, therefore it is going to be very difficult to obtain a clear picture of the $G_1$ function and (b) the numerical derivatives contain some rough structure and errors that causes a noisy result that obscures the system behavior. Therefore, we can probably discard or confirm tendencies from a statistical analysis of a set of structured cloud of points.

\begin{figure}[h!]
\begin{center}
\includegraphics[height=5cm,clip]{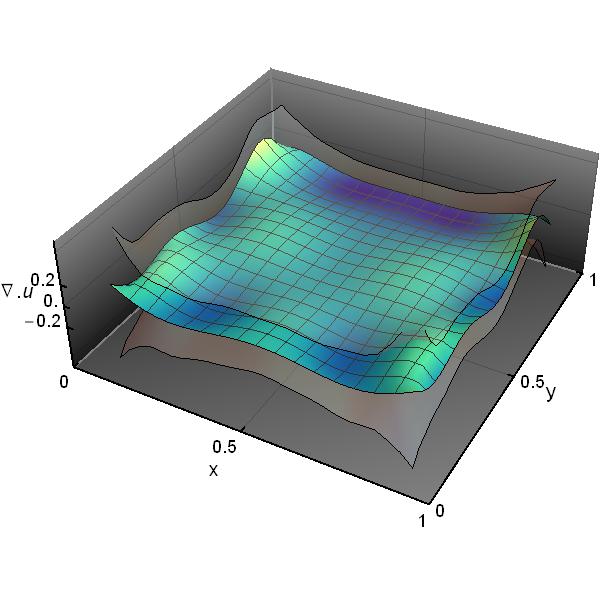}  %divergence_g15.nb
\includegraphics[height=5cm,clip]{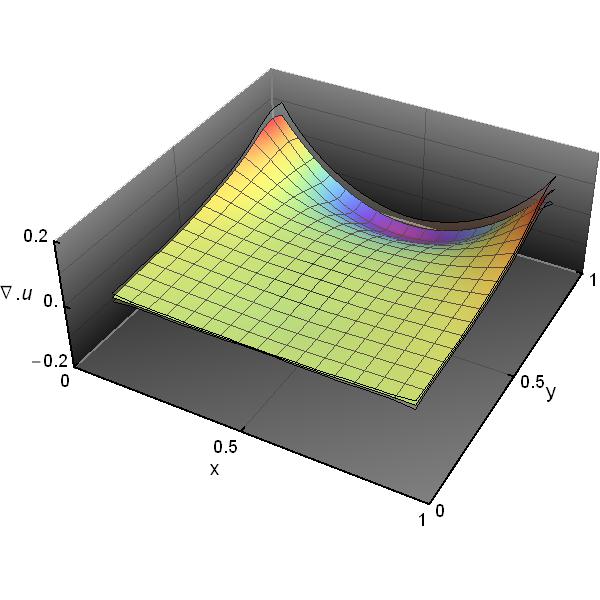} 
\includegraphics[height=5cm,clip]{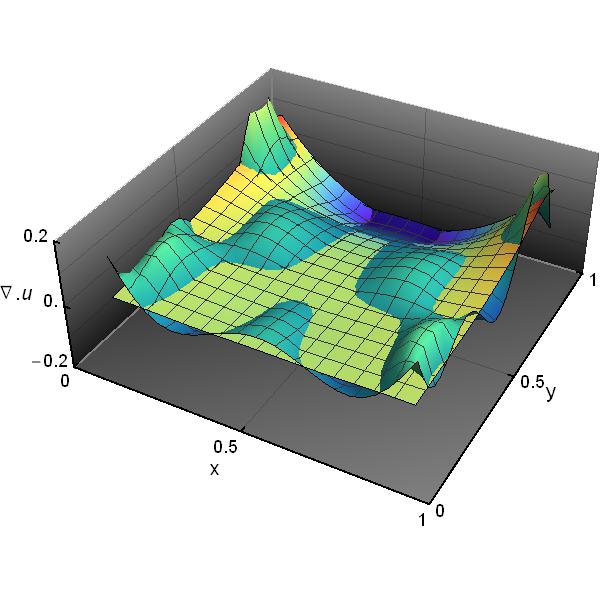} 
\end{center}
\kern -1.cm
\caption{Divergence of the velocity field for configurations at $T_0=17$ and $g=15$. Left:  the divergence is obtained by the direct computation of the velocity derivatives, $\nabla\cdot u=\partial_1 u_1+\partial_2 u_2$. Center: the divergence of the velocity is computed using the continuity equation, $\nabla\cdot u=-u\cdot\nabla\rho/\rho$, that is, through the computation of the density field derivatives. White surfaces is the error bar interval. Right: comparison of both results (without error bars)
\label{div1}}
\end{figure}

We already know that, in general, numerical derivatives contain an extra noise due to the necessity of doing some fit and/or interpolation to the discrete data field. When computing the divergence of the velocity field we have two ways of getting it: (1) obtaining the derivatives  $\partial_1u_1$ and $\partial_2u_2$ (using the method above explained) and summing up both fields or (2) by assuming that the continuity equation is correct we obtain from it the relation $\nabla\cdot u=-u\cdot\nabla\rho/\rho$, then we just need to compute the derivatives of the $\rho$ field. Each method depend on the numerical computation of different field derivatives. We show $\nabla\cdot u$ in figure \ref{div1} computed by the two methods for $T_0=17$ and $g=15$. Observe that the direct method (1)  give us a wavy field that seems to follow, in average, the one we obtain by the method (2) which is much smoother and its error bars are smaller.  That is due to the less fluctuating behavior of the density field compared with the hydrodynamic velocity field that also affects to the error propagation when doing the derivatives. It is clear that the second method is the one we should use in the EOS analysis.

\begin{figure}[h!]
\begin{center}
\includegraphics[height=6cm,clip]{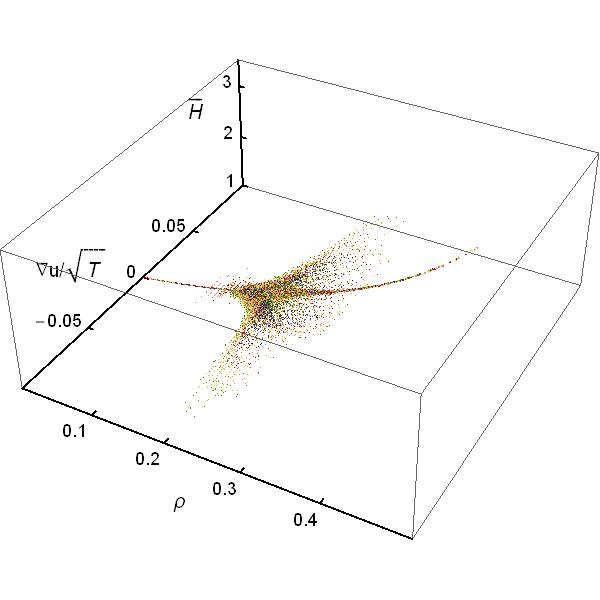}  %mechpress.nb
\end{center}
\kern -1.cm
\caption{Reduced mechanical pressure $\bar H=\pi r^2 \bar P/\rho T$ versus $\rho$ and $\nabla\cdot u/\sqrt{T}$ for the complete set of configurations ($T_0\in[1,20]$ and $g=0,5,10,15$). There are $75712$ data points in the figure. A given color is data from a configuration with a fix value of $g$ and $T_0$ ($26\times 26=676$ data points where we excluded the two columns and rows nearest to  the four boundaries). 
\label{diver1}}
\end{figure}

Once we solved the problem of how we measure the $\nabla\cdot u$, we can continue the analysis of the mechanical pressure. First we plot in figure left in (\ref{diver1}) the measured local pressure $\bar H=\bar Q/\rho T=\pi r^2\bar P/\rho T$ as a function of local $(\rho,\nabla\cdot u/\sqrt{T})$ for all the cells (discarding  two rows and columns nearest to the four boundaries) from stationary configurations we have obtained in this numerical experiment ($T_0\in[1,20]$ and $g=0,5,10$ and $15$). Let us point out that we did the full analysis separately for the different $g$ values and we did not see any clear dependence on $g$, thats why we join all data sets. There are $75712$ data points in total. Observe in the figure how the data from non-convective configurations  accumulate into a curve with $\nabla\cdot u=0$. The rest of the data extends over the space following (apparently) a smooth surface.

We have to fit a regular surface of the form given by eq.(\ref{barH}) to all these points without any theoretical help to build a test function. All we know is: (a) $H(\rho)$ is the real EOS for hard disks and that the Henderson's EOS is a good approximation to the real one but there are other well known EOS approximations having good behaviors on different density ranges (see for instance the Chapter 3 in \cite{Mulero}) and (b) the function $G_1$ is assumed to be analytic and it is of the form $G_1(x,y)= xyG_2(x,y)$ due to physics constraints (see above).

Therefore, after many trials, we have chosen to fit functions of the form:
\begin{equation}
\bar H(x,y)=\frac{1}{(1-x)^2}\left[1+\sum_{n=1}^{M_1}c_2(n)x^{n+1} \right]+\sum_{n=1}^{M_2}\sum_{l=1}^{M_3} c_1(n+M_2(l-1))x^{n}y^{l} \label{trial1}
\end{equation} 
where $M=M_1+M_2*M_3$ is the total number of parameters we use to fit the surface. Let's point out that many  coefficients that measure the goodness of a fit improve with the total number of parameters. Therefore if we want to compare different approaches we need to fix $M$ to do not introduce spurious improvement in the fit.
Let us show explicitly the analysis over three particular structures with $M=8$ and $M=10$ as typical examples that confirms our final answer. We have tested many other $M$'s that give us the same final results. These trial functions are for M=8: $A(M_1=3,M_2=1,M_3=5)$ that permits a nonlinear behavior in $\nabla\cdot u$, $B(8,0,0)$ that assumes that the Stoke's assumption holds and tries to fit only the EOS, and $C(3,5,1)$ that assumes that the bulk viscosity depends only on the local values of $\rho$ and $T$. For $M=10$ the corresponding cases are $A(4,2,3)$, $B(10,0,0)$ and $C(4,6,1)$.

In table \ref{table1} we show the obtained parameter values with their standard deviations.
\begin{table}
\begin{center}
 \begin{tabular}{| c || c | c | c ||c | c | c |}
 \hline
  &\multicolumn{3}{|c||}{M=8}& \multicolumn{3}{|c|}{M=10}\\ \hline
{\bf c's}&{\bf A(3,1,5)}&{\bf B(8,0,0)}&{\bf C(3,5,1)}&{\bf A(4,2,3)}&{\bf B(10,0,0)}&{\bf C(4,6,1)}\\
 \hline\hline
$c_1(1)$&-0.27(0.01)&-&117(7)&-2.33(0.08)&-&-47.9(23.2)\\ \hline
$c_1(2)$&-40.8(0.4)&-&-2200(122)&9.9(0.3)&-&1731(541)\\ \hline
$c_1(3)$&-25(9)&-&14862(794)&-78.4(1.9)&-&-21715(4954)\\ \hline
$c_1(4)$&5050(94)&-&-43038(2277)&225.8(7.7)&-&123556(22331)\\ \hline
$c_1(5)$&28887(1462)&-&45421(2423)&-717(36)&-&-326800(49579)\\ \hline
$c_1(6)$&-&-&-&2733(149)&-&326899(43399)\\ \hline\hline
$c_2(1)$&0.101(0.002)&-1.4(0.1)&0.074(0.002)&-0.073(0.005)&1.92(0.38)&-0.038(0.006)\\ \hline
$c_2(2)$&-0.052(0.009)&53(4)&0.06(0.01)&1.63(0.05)&-77.4(15.3)&1.16(0.05)\\ \hline
$c_2(3)$&0.19(0.01)&-674(41)&0.008(0.02)&-4.91(0.14)&1412(260)&-3.3(0.2)\\ \hline
$c_2(4)$&-&4290(245)&-&4.90(0.14)&-13903(2452)&3.3(0.2)\\ \hline
$c_2(5)$&-&-15244(849)&-&-&81012(14177)&-\\ \hline
$c_2(6)$&-&30851(1702)&-&-&-291883(52385)&-\\ \hline
$c_2(7)$&-&-33351(1835)&-&-&656146(124241)&-\\ \hline
$c_2(8)$&-&14986(823)&-&-&-893959(183061)&-\\ \hline
$c_2(9)$&-&-&-&-&673126(152545)&-\\ \hline
$c_2(10)$&-&-&-&-&-213897(54927)&-\\ \hline
 \hline 
  \end{tabular}
\end{center}
\caption{Parameter values obtained when fitting the data to eq.(\ref{trial1}) to cases $M=8$ and $10$ with $A(M_1=3,M_2=1,M_3=5)$, $B(8,0,0)$ and $C(3,5,1)$ and $A(4,2,3)$, $B(10,0,0)$ and $C(4,6,1)$ respectively.}\label{table1}
\end{table}

We should develope a criteria for a given $M$ value to elect which one, A, B or C, is the best fit. We focus on the distribution of the differences between the data and each of the fitted surfaces:
$\Delta H_i=\bar H_i-\bar H(\rho_i,\nabla\cdot u_i/\sqrt{T_i}$ for $i=1,\ldots, N_D=75712$. 
 In particular we have computed the statistics of the differences that includes the distribution of values and the first four central moments:
 \begin{eqnarray}
 \langle\Delta H\rangle&=&\frac{1}{N_D}\sum_{i=1}^{N_D}\Delta H_i \nonumber\\
 \sigma(\Delta H)^2&=&\frac{1}{N_D}\sum_{i=1}^{N_D}(\Delta H_i-\langle\Delta H\rangle)^2\nonumber\\
\epsilon(\Delta H)&=&3\sigma(\Delta H)/\sqrt{N_D}\nonumber\\
 s(\Delta H)&=&\frac{1}{N_D\sigma(\Delta H)^3}\sum_{i=1}^{N_D}(\Delta H_i-\langle\Delta H\rangle)^3\nonumber\\
 \kappa(\Delta H)&=&\frac{1}{N_D\sigma(\Delta H)^4}\sum_{i=1}^{N_D}(\Delta H_i-\langle\Delta H\rangle)^4-3
  \label{parameters}
  \end{eqnarray}
 $\epsilon(\Delta H)$ is a measure of the statistical error of $\langle\Delta H\rangle$.
 Moreover, we have also studied the corresponding spatial point distribution by computing first the {\it center of mass}:
  \begin{equation}
 R=\frac{1}{N_D}\sum_{i=1}^{N_D}r_i\quad\quad r_i \equiv (\Delta H_i,\rho_i,\frac{\nabla\cdot u_i}{\sqrt{T_i}})
 \end{equation}
and then obtaining the inertial tensor with respect the center of mass: 
\begin{equation}
I_{\alpha,\beta}=\frac{1}{N_D}\sum_{i=1}^{N_D}\left[{r'}_i^2\delta_{\alpha,\beta}-{r'}_{i,\alpha}{r'}_{i,\beta}\right]
\end{equation}
where $r'_i=r_i-R$. From the inertial tensor we can get the 
principal axes of inertia that are defined by its eigenvectors which give us some averaged idea of the spatial point distribution.

In order to interpret all these parameters let us think what would happened to the data if the fitted function $\bar H$ was the exact one. If we neglect the mutual correlations from the data (they come from fields whose values at each cell are not independent from the other cells), then we assume that the points $\Delta H_i$ should be distributed, in this case, evenly around the plane $(x,y,0)$. We do not expect a total homogeneous distribution in space because the sampling is, by construction, not homogeneous (it depends on temperature gradients, external fields and the resulting fields have non-linear and non-uniform values). Moreover, we also assume that the probability distribution of $\Delta H$ should be a Gaussian and/or a near Gaussian distribution with large kurtosis that would reflect the correlation of the  data.  In fact, it is simple to show that the moments of the distribution $\xi=(\Delta H-\langle\Delta H\rangle)/\sigma(\Delta H)$ are proportional to the moments of the differences between the fitted function and the correct one (assuming that the data fluctuates with a Gaussian distribution around the correct values of the function). That is, we would get $s=0$ and $\kappa=0$ if the fitted function was {\it exactly} the correct one (except for errors due to the finite size of the sampling).

That is the picture we have in mind. Therefore, a given fit would be better than other if: (a) $\langle\Delta H\rangle=R_3$ (the third coordinate of the center of mass) is nearest to zero, (b) $s(\Delta H)$ and $\kappa(\Delta H)$ have the smaller possible value and (c) one of the principal axes of inertia is nearest to the vector $(0,0,1)$ because it means that the points are more evenly distributed in space around the plane $(x,y,0)$.
A final element  is that we can compute the error of the center of mass third coordinate $R_3=\langle\Delta H\rangle$: $\epsilon(\Delta H)=3\sigma(\Delta H)/\sqrt{N_D}$. The fit is good if the error bars around the value of $R_3$ {\it contains the zero value}: $0\in[R_3-\epsilon(\Delta H),R_3+\epsilon(\Delta H)]$. If that doesn't happen the fit is considered wrong.

In figure \ref{diver2} we show, for the case $M=8$, the point spatial distributions with the center of mass and the principal axes of inertia and the probability distribution of $\xi=(\Delta H-\langle\Delta H\rangle)/\sigma(\Delta H)$ that it is compared with a Gaussian $N(0,1)$.
We show in Table \ref{table2} the values obtained for the above parameters for the three reference fits A,B and C and $M=8$ and $10$.  We highlighted the fit with the best value.
\begin{figure}[h!]
\begin{center}
\includegraphics[height=6cm,clip]{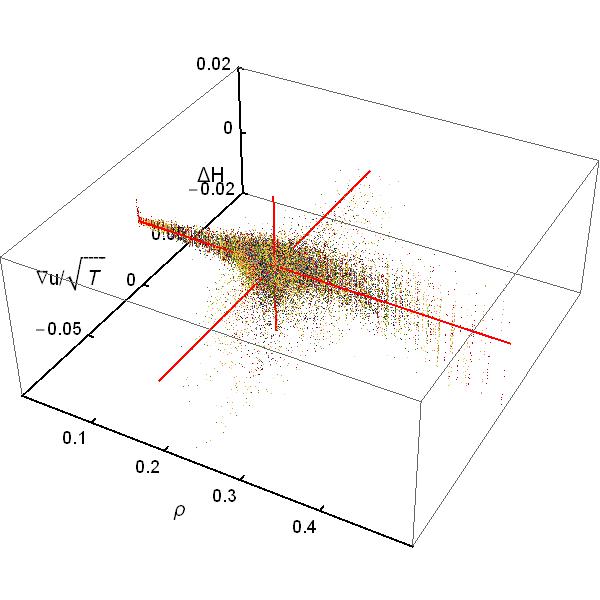}  %mechpress.nb
\includegraphics[height=6cm,clip]{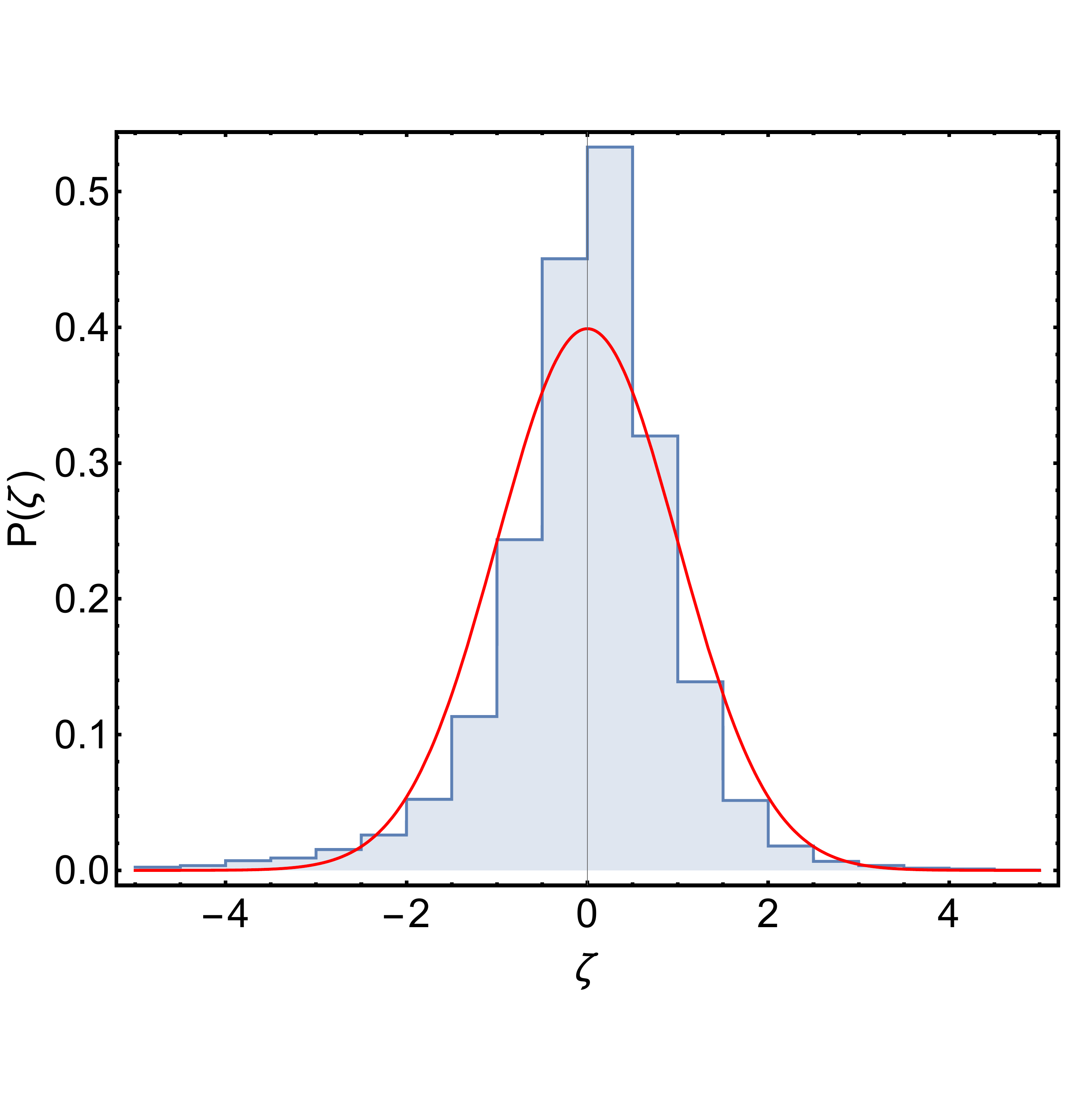}
\includegraphics[height=6cm,clip]{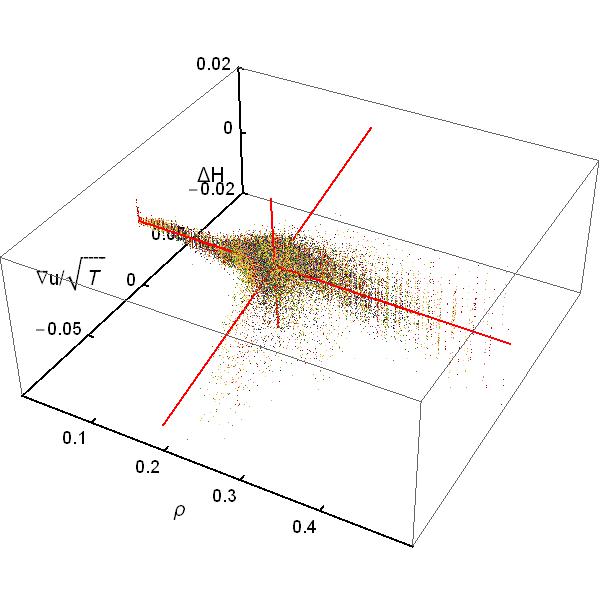}
\includegraphics[height=6cm,clip]{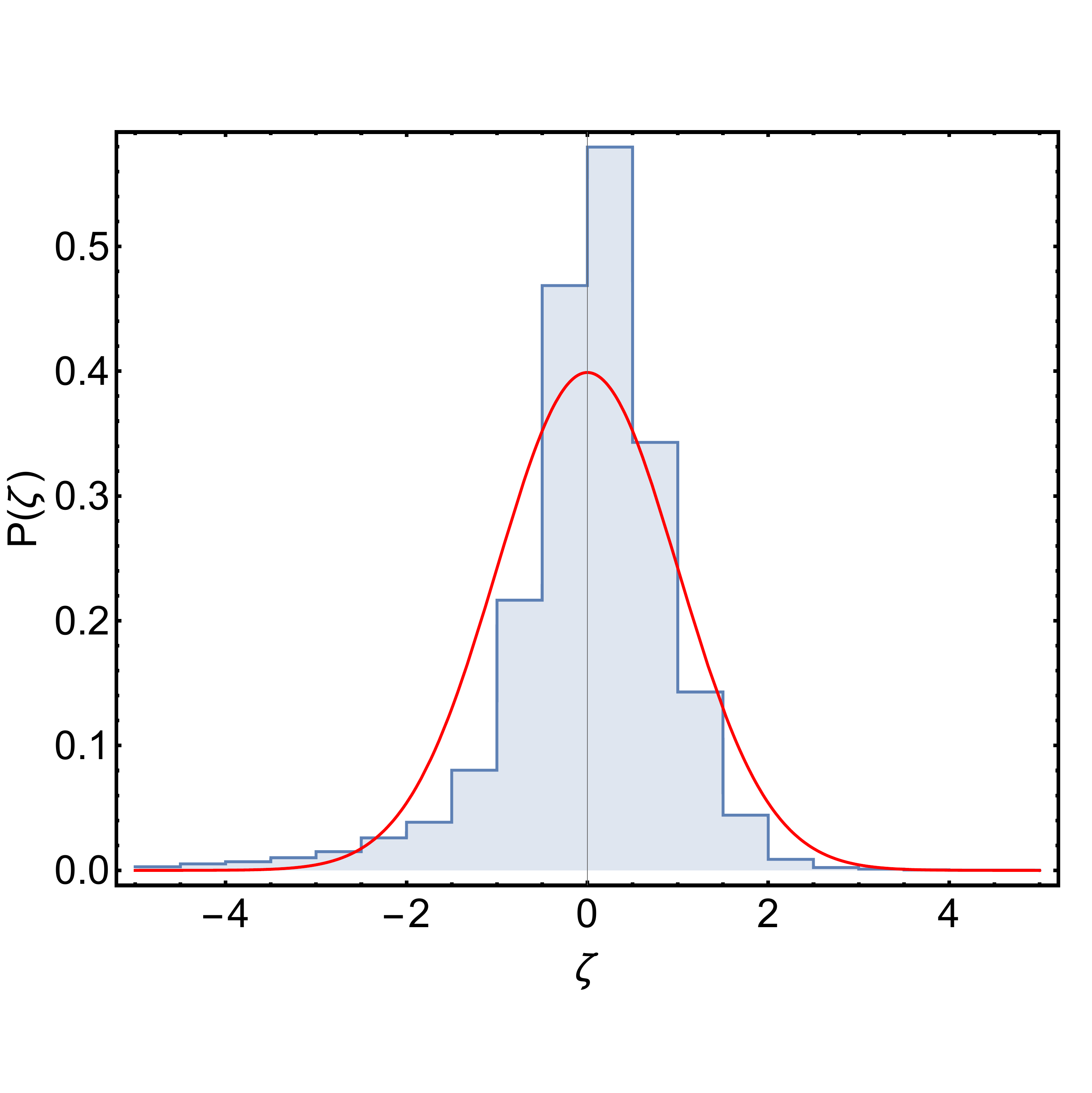}
\includegraphics[height=6cm,clip]{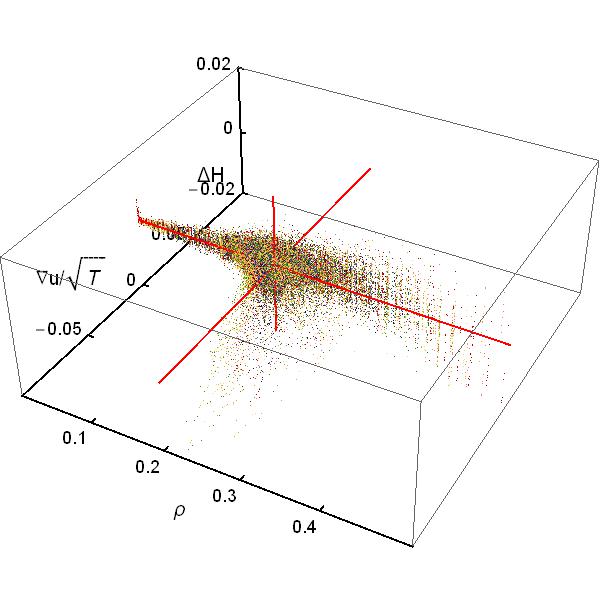}
\includegraphics[height=6cm,clip]{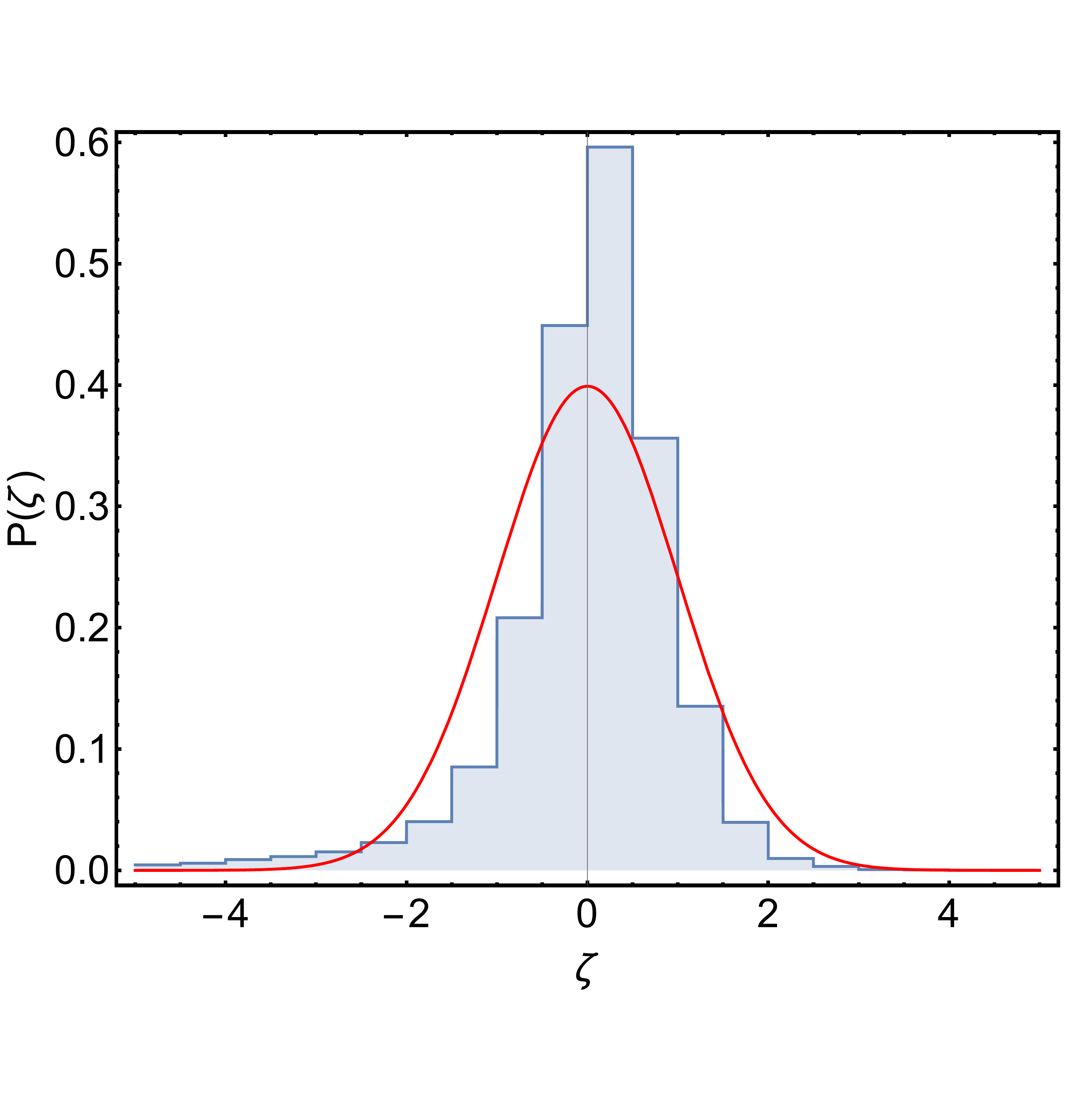}
\end{center}
\kern -1.cm
\caption{Left column: The set of points $\{(\Delta H_i,\rho_i,\nabla\cdot u_i/\sqrt{T_i})\}$ for $i=1,\ldots, N_D$ for the three fits (from top to bottom) A, B and C for $M=8$.  Red lines are the principal axes of inertia for each case. They cross at the center of mass (see table \ref{table2}. Right column: Probability distribution of $\xi=(\Delta H-\langle\Delta H\rangle)/\sigma(\Delta H)$ for the three fits (from top to bottom) A,B and C for $M=8$. The red lines are Gaussian distributions $N(0,1)$.  
\label{diver2}}
\end{figure}

\begin{table}
\begin{center}
 \begin{tabular}{| c || c | c | c || c | c | c |}
 \hline
  \hline
  &\multicolumn{3}{|c||}{M=8}& \multicolumn{3}{|c|}{M=10}\\
 \hline
&{\bf A(3,1,5)}&{\bf B(8,0,0)}&{\bf C(3,5,1)}&{\bf A(4,2,3)}&{\bf B(10,0,0)}&{\bf C(4,6,1)}\\
 \hline\hline
 $\langle\Delta H\rangle=R_3$&$\mathbf{3.12 \times10^{-7}}$& $1.93\times 10^{-5}$&$4.73 \times 10^{-5}$&
 $9.60 \times10^{-5}$& $\mathbf{7.04\times 10^{-6}}$&$1.07 \times 10^{-4}$\\ \hline
$\sigma(\Delta H)$&0.00401831&0.004612&0.00456807&0.0040069&0.004608&0.004552\\ \hline
$\epsilon(\Delta H)$&$4.38 \times10^{-5}$&$5.03 \times 10^{-5}$&$4.98 \times 10^{-5}$&$4.37 \times10^{-5}$&$5.02 \times 10^{-5}$&$4.96 \times 10^{-5}$\\ \hline
\hline
$s(\Delta H)$&{\bf -0.198521}&-2.01974&-1.30117&{\bf -0.683441}&-1.97379&-1.54625\\ \hline
$\kappa(\Delta H)$&11.3585&12.0316&{\bf 10.9428}&{\bf 7.62006}&11.8523&9.63263\\ \hline\hline
$v_3$&0.999976&0.992871&{\bf 0.999998}&{\bf 0.999994}&0.992949&0.999993\\ \hline
 \hline 
  \end{tabular}
\end{center}
\caption{Parameter values obtained when fitting the data to eq.(\ref{trial1}) to cases $M=8$ and $A(M_1=3,M_2=1,M_3=5)$, $B(8,0,0)$ and $C(3,5,1)$ and $M=10$ with $A(4,2,3)$, $B(10,0,0)$ and $C(4,6,1)$. In bold face there are the best fit for a given parameter. $\langle\Delta H\rangle$: average value of the differences $\Delta H_i$. $\sigma(\Delta H)$: standard deviation of the set of differences. $\epsilon(\Delta H)=3\sigma(\Delta H)/\sqrt{N_D}$: estimated error of $\langle\Delta H\rangle$. $s$ and $\kappa$: third central moment and kurtosis (respectively) of the distribution of $\xi=(\Delta H-\langle\Delta H\rangle)/\sigma(\Delta H)$.
 $v_3$: third coordinate of the vector defining the principal axe of inertia nearest to $(0,0,1)$. }\label{table2}
\end{table}

 There are some comments to make the best fit's final choice:
\begin{itemize}
\item[(a)] The fit B (Stokes assumption) is the worse one always. None of the parameters in table \ref{table2} does better than the fits A or C for any $M$ value. We can conclude that the Stokes assumption is not followed by our system. 
\item[(b)] The fit C (Mechanical pressure proportional to $\nabla\cdot u$) is,in general, worse than the fit A. For instance, for the $M=8$ case we observe that the $R_3$ value is so large that the zero value is barely included in the error interval of $\epsilon(\Delta H)$ and for the $M=10$ case the zero is out of the error interval meaning that the fit is useless.  Moreover the skewness of the data with fit C is always larger  the one's with A. C does better in the kurtosis value for $M=8$ but no with $M=10$ (but it just reflects the existence of data correlation that we cannot evaluate its real magnitude). C also does better that A in the  the third component of the principal inertial axis for $M=8$  but this is maybe due to a better description by the fitted function of the density dependence of $G_1$. This is confirmed when we use $M=10$ where we see that  $v_3$ for A is larger than the corresponding one with the C fit.  
\item[(c)] We have observed that the overall behavior of fits type A, B or C is conserved when we do other fits with different total number of parameters. 
\end{itemize}
We can conclude that, from a statistical point of view, the fit type A (for most of the $M$ values checked)  is  the most consistent with the data. That is, from our analysis we can clearly discard the Stokes assumption in our system and that the mechanical pressure is a nonlinear function on $\nabla\cdot u$.  Therefore, we are quite convinced  that 
the bulk viscosity is a function of the local density, $\rho$, local temperature $T$ and the local mechanical pressure $\bar P$.

\begin{figure}[h!]
\begin{center}
\includegraphics[height=5cm,clip]{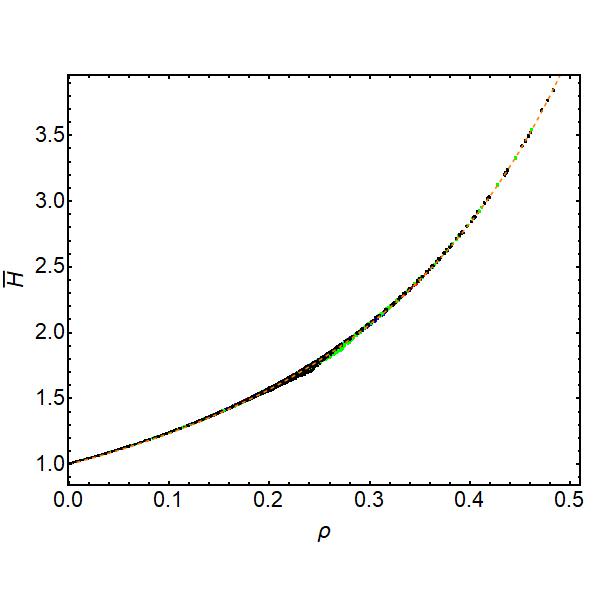}  %mechpress.nb
\includegraphics[height=5cm,clip]{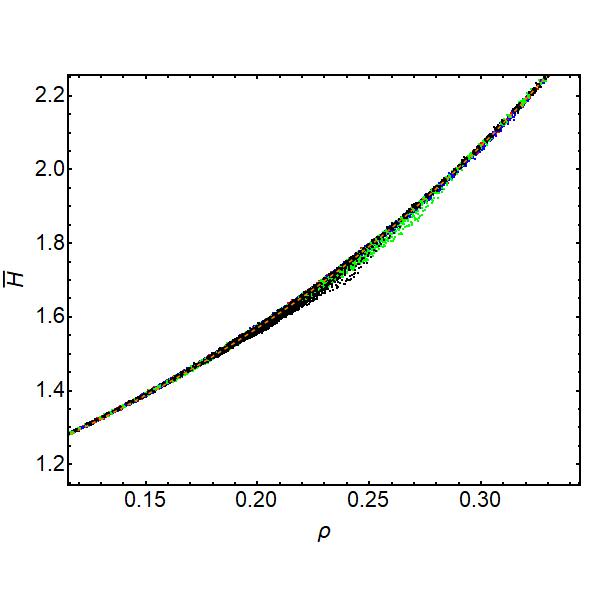}
\includegraphics[height=5cm,clip]{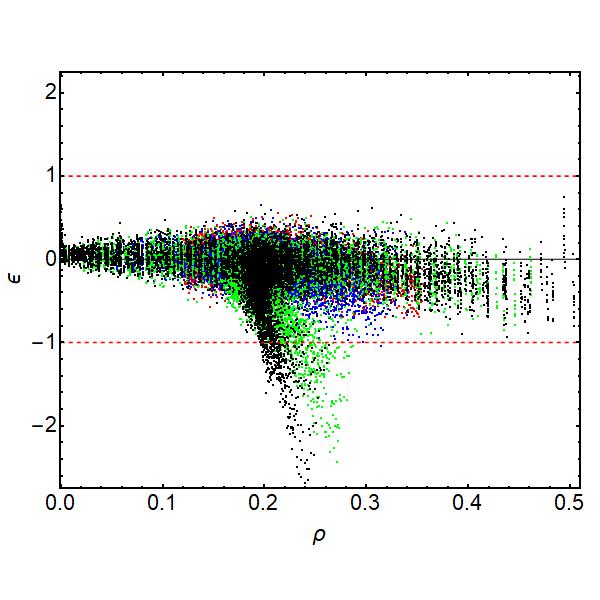}
\includegraphics[height=5cm,clip]{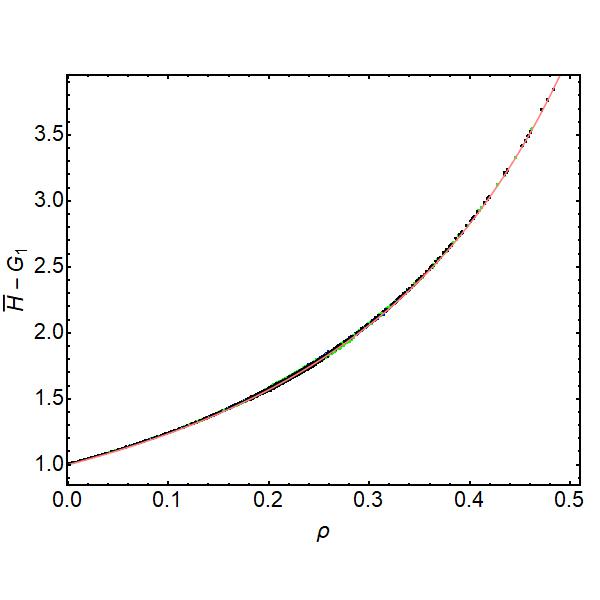}
\includegraphics[height=5cm,clip]{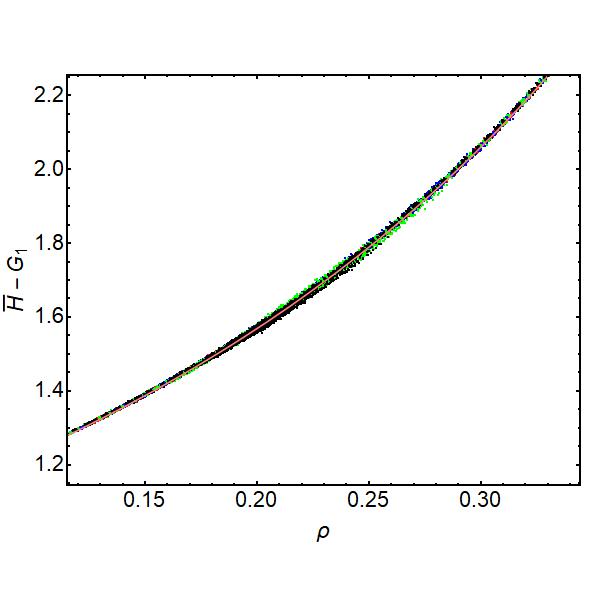}
\includegraphics[height=5cm,clip]{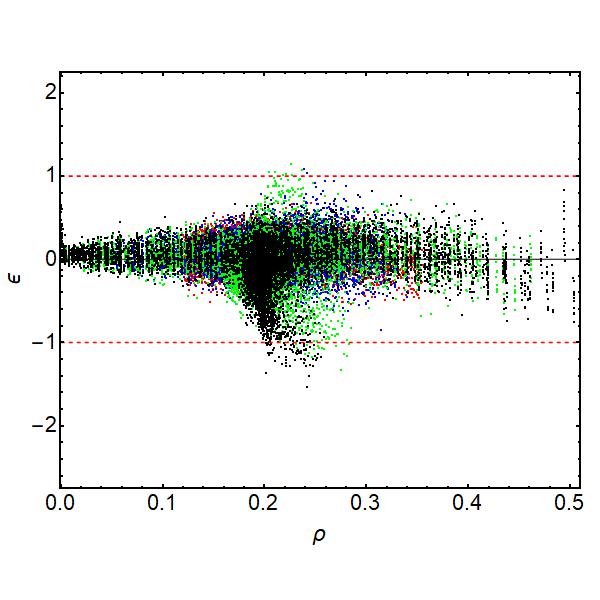}
\includegraphics[height=5cm,clip]{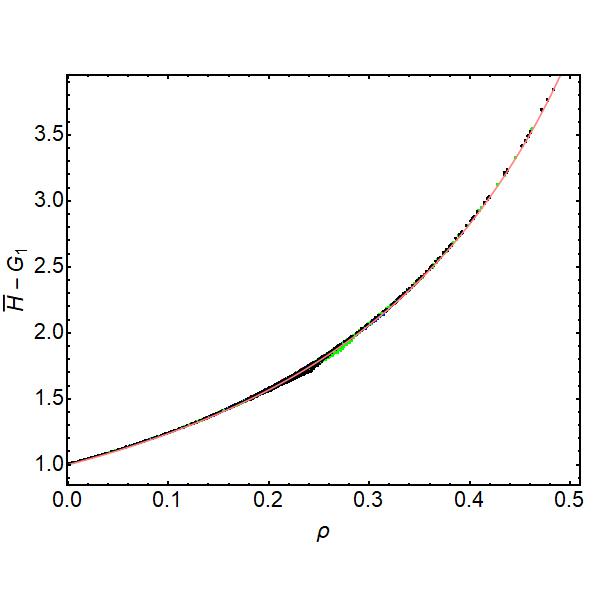}
\includegraphics[height=5cm,clip]{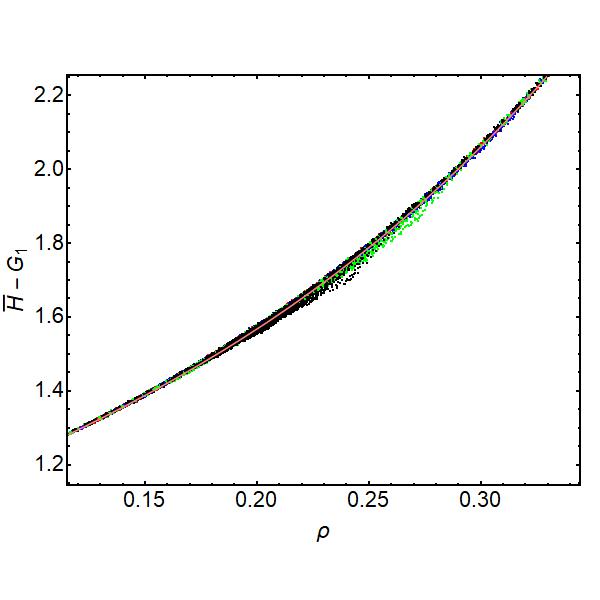}
\includegraphics[height=5cm,clip]{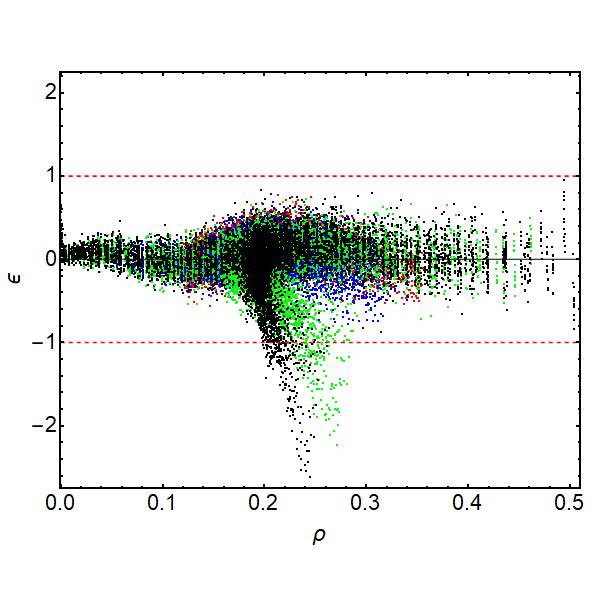}
\includegraphics[height=5cm,clip]{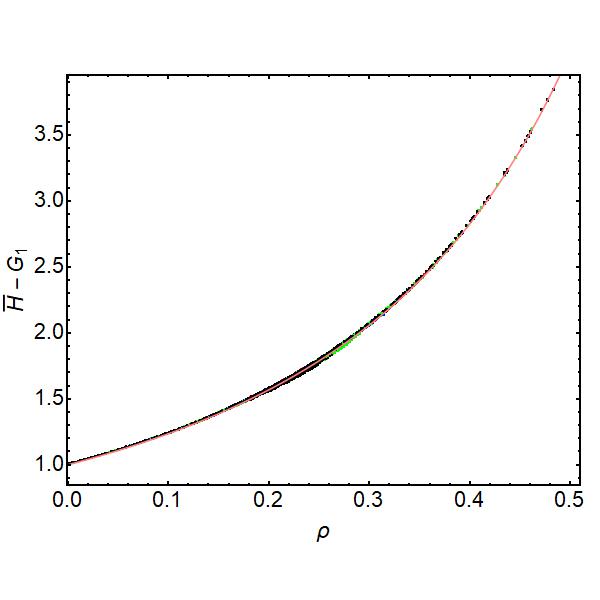}
\includegraphics[height=5cm,clip]{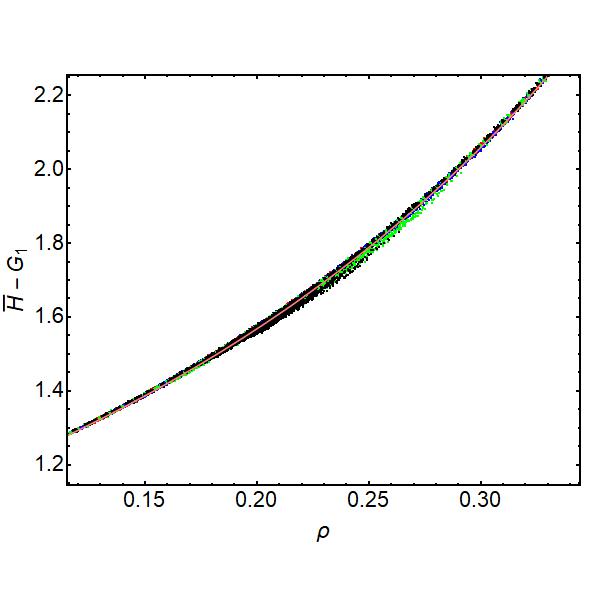}
\includegraphics[height=5cm,clip]{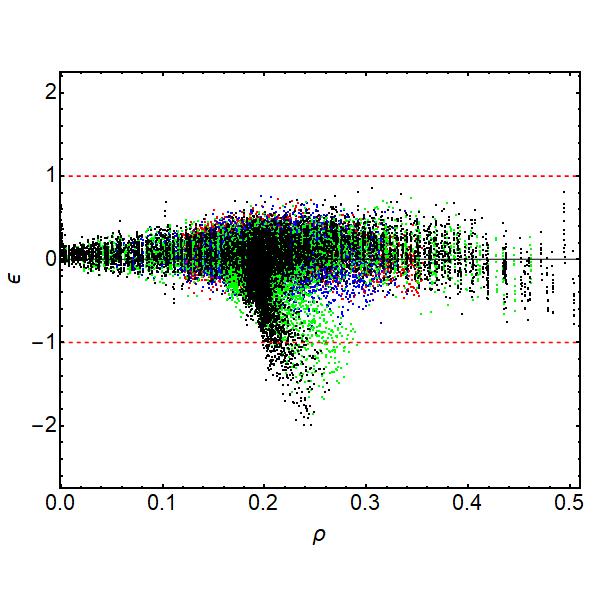}
\end{center}
\kern -1.cm
\caption{First Row: (left) $\bar H$ vs density, $75712$ data points. Dashed line is $H$ corresponding to the Henderson's EOS. Solid line is the $H$ corresponding to the EOS obtained by fit A and $M=8$ (see text). (center) Focus or $\bar H$ vs. $\rho$ to show data deviations to the regular behavior fixed by Henderson's EOS (or A fit EOS). (right) Relative error $\epsilon_i=100(\bar H_i/\bar H_H(\rho_i)-1)$. Horizontal dashed lines are the expected Henderson's interval of precision. 
Second Row: (left) Corrected points using the $G_1$ function obtained by fit A and $M=8$ (see text), $H_i=\bar H_i-G_1(\rho_i,\nabla\cdot u_i)/\sqrt{T_i}$. Solid line is the A fit EOS. (center) as center in first row. (right) Relative error of data with respect A fit EOS.
Third and Four Rows: Same as second row using the B and C-fits respectivelly with $M=8$.
\label{diver3}}
\end{figure}

In figure \ref{diver3} we show the row data $(\rho_i,\bar H_i)$ (first row) and the corrected mechanical pressure using the $G_1$ function obtained with the A,B and C-fits, $(\rho_i,\bar H_i-G_1(\rho_i,\nabla\cdot u_i/\sqrt{T_i})$ in the second. third and fourth rows respectively. We appreciate with some detail at the center figures at each row how the data is corrected and they scale in a more homogeneous way for the fit A compared with the original set of data. The correction is worse for fits B and C.

\begin{figure}[h!]
\begin{center}
\includegraphics[height=6cm,clip]{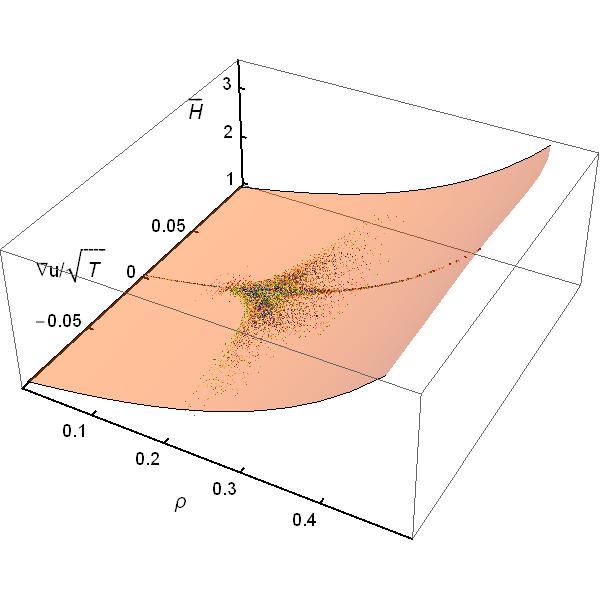}  %mechpress.nb
\end{center}
\kern -1.cm
\caption{Reduced mechanical pressure $\bar H=\pi r^2 \bar P/\rho T$ versus $\rho$ and $\nabla\cdot u/\sqrt{T}$ for the complete set of configurations ($T_0\in[1,20]$ and $g=0,5,10$ and $15$). There are $75712$ data points in the figure. A given color is data from a configuration with a fix value of $g$ and $T_0$ ($26\times 26=676$ data points where we excluded the two columns and rows nearest to  the four boundaries). The orange surface is the A fit to the data (see text).
\label{diver4}}
\end{figure}

We show in figure \ref{diver4} the A fit overlapped to the data. The fitted function can be written:
\begin{equation}
\bar H=H_A(\rho)-\frac{\pi r^2}{\rho T}\xi(\rho,\frac{\nabla\cdot u}{\sqrt{T}})\nabla\cdot u
\end{equation}
with
\begin{equation}
H_A(\rho)=\frac{1-0.0726 \rho^2 +1.6265\rho^3 -4.9088 \rho^4+4.9040\rho^5}{(1 - \rho)^2}
\end{equation}
that is the A-fit EOS and
\begin{equation}
\xi(\rho,\frac{\nabla\cdot u}{\sqrt{T}})=\frac{\sqrt{T}}{r}\rho^2\left[91.08-386.40\rho+78.45\tilde y-225.82\rho\tilde y+18.38\tilde y^2-70.02\rho\tilde y^2\right]
\end{equation}
where  $\tilde y=\nabla\cdot u /(\pi r\sqrt{T})$ anr $\pi r= 0.02562..$ in our simulations. We can compare with the result at the Enskog approximation. There the leading term is $49.72$ that is almost half of our result.  We do not pretend that these fitted functions have the correct mathematical structure. Nevertheless, there are reasonable smooth fits and they are compatible with the already known approximations (for instance with the hard disk EOS) and, more important, there are {\it necessary} to rescale the data to obtain a better distribution of them around the fitted functions.  After all these analysis we may conclude that: (1) we detect the existence of the  mechanical pressure concept and its different behavior at the convective regime with respect the thermodynamic pressure, (2) Assuming that the average stress tensor is the mechanical pressure that we have measured, we conclude that the Stokes relation cannot be correct (in this case), (3) the bulk viscosity seems to be a function of the local density, temperature and the mechanical pressure and then the mechanical pressure cannot be a linear function of the divergence of the hydrodynamic velocity vector field.
%aqui
\item{\bf 3. Checking the stationary Navier-Stokes equations for non-convective states}

We already know that the stationary Navier Stokes equations for non-convective states are:
\begin{eqnarray}
\frac{dQ}{dy}&=&-g\rho\\
\sqrt{T}K(\rho)\frac{dT}{dy}&=&-J
\end{eqnarray}
where it is assumed the functions only depend on the $y$-coordinate. We should add the local equilibrium hypothesis to these equations and the temperature boundary conditions. Our aim is to check that the data is compatible with these two equations. Let us study first the equation for the pressure.

We can integrate the pressure equation:
\begin{equation}
Q(y)-Q(y_0)=-g\int_{y_0}^y d\bar y\rho(\bar y) \label{inte}
\end{equation}
If we want to get a precise check, it is convenient to take into account the differences between the measured variables in a cell and the above continuum expression.  First, we can define the pressure value over a one-dimensional cell, $\bar Q(n)$:
\begin{equation}
\bar Q(n)=\frac{1}{N_C}\sum_{m=3}^{N_C-2} Q(m,n)
\end{equation}
where $Q(m,n)$ is the pressure measured in the cell $(m,n)$ and we have discarded the columns near the vertical boundaries.
Assuming the existence of the continuum functions $Q(y)$ and $\rho(y)$ we can write
\begin{equation}
\bar Q(n)=\frac{1}{\Delta}\int_{\tilde y(n)-\Delta/2}^{\tilde y(n)+\Delta/2} dy \,Q(y)\label{disc}
\end{equation}
where $\tilde y(n)=(n-1/2)\Delta$. 
Substituting equation (\ref{inte}) into eq. (\ref{disc}) we get:
\begin{equation}
\bar Q(n)=Q(y_0)-\frac{g}{\Delta} \int_{\tilde y(n)-\Delta/2}^{\tilde y(n)+\Delta/2} dy \int_{y_0}^yd\bar y\rho(\bar y)
\end{equation}

The integration variables of the double integral can be exchanged with the corresponding change of the integral limits. In this way we can do one of the integrals.The result, when $y_0<\tilde y(n)-\Delta/2$, is:
\begin{equation}
\bar Q(n)=Q(y_0)-g\int_{y_0}^{\tilde y(n)-\Delta/2}dy\rho(y)-\frac{g}{\Delta}\int_{\tilde y(m,n)-\Delta/2}^{\tilde y(m,n)+\Delta/2} dy\rho(y)\left(\tilde y(n)+\frac{\Delta}{2}-y\right)
\end{equation} 
We now choose $y_0=\tilde y(n_0)-\Delta/2=(n_0-1)\Delta$ (the bottom of the cell $n_0$) and we can write
\begin{equation}
\bar Q(n)=Q(y_0)-g\Delta\sum_{l=n_0}^{n-1}\bar\rho(l)-\frac{g}{\Delta}\int_{(n-1)\Delta}^{n\Delta}dy\rho(y)\left(n\Delta-y\right)
\end{equation}
where $\bar\rho(l)$ is the average density at a cell at height $l$. The $\rho(y)$ inside the integral can be Taylor expanded around the $\tilde y(n)$ coordinate and we integrate order by order in the expansion. Afterwards, we obtain:
\begin{equation}
\bar Q(n)=Q(y_0)-g\Delta\sum_{l=n_0}^{n-1}\bar\rho(l)-\frac{g}{\Delta}\sum_{k=0}^{\infty}\frac{(-1)^k\Delta^{k+1}}{k! 2^{k+2}}\left(\frac{1-(-1)^k}{k+2}+\frac{1+(-1)^k}{k+1}\right)\frac{d^k\rho}{dy^k}\biggr\vert_{y=\tilde y(n)}
\end{equation}
To find $Q(y_0)$ as a function of $\tilde Q(n_0)$ it is enough to substitute $n=n_0$ in the last expression. After its substitution we get:
\begin{eqnarray}
\bar Q(n)-\bar Q(n_0)&=&-g\Delta\sum_{l=n_0}^{n-1}\bar\rho(l)\nonumber\\
&-&\frac{g}{\Delta}\sum_{k=0}^{\infty}\frac{(-1)^k\Delta^{k+1}}{k! 2^{k+2}}\left(\frac{1-(-1)^k}{k+2}+\frac{1+(-1)^k}{k+1}\right)\left[\frac{d^k\rho}{dy^k}\biggr\vert_{y=\tilde y(n)}-\frac{d^k\rho}{dy^k}\biggr\vert_{y=\tilde y(l_0)}\right]\label{pres31}
\end{eqnarray}
Finally, the derivatives at a central point can be expressed as variations with cell functions $\bar\rho$ as it was explained in the introduction. Then we get the cell version of the barometric equation:
\begin{eqnarray}
\bar Q(n)-\bar Q(n_0)&=&-g\Delta\sum_{l=n_0}^{n-1}\bar\rho(l)-\frac{g\Delta}{2}(\bar\rho(n)-\bar\rho(n_0))\nonumber\\
&+&\frac{g\Delta}{24}\left[\bar\rho(n+1)-\bar\rho(n-1)-\bar\rho(n_0+1)+\bar\rho(n_0-1)\right]+O(\Delta^3)\equiv -D(n,n_0) \label{densi}
\end{eqnarray}
We use this last expression to check the barometric equation. We measure by one hand
$\Delta Q(n,n_0)\equiv \bar Q(n_0)-\bar Q(n)$ and by the other hand $D(n,n_0)$ defined in eq. (\ref{densi}). We choose the values $n_0=4,\ldots,N_C-4$ and $n=n_0+1,\ldots,N_C-3$ to not to use the data from the boundary cells.
\begin{figure}[h!]
\begin{center}
\includegraphics[height=5cm,clip]{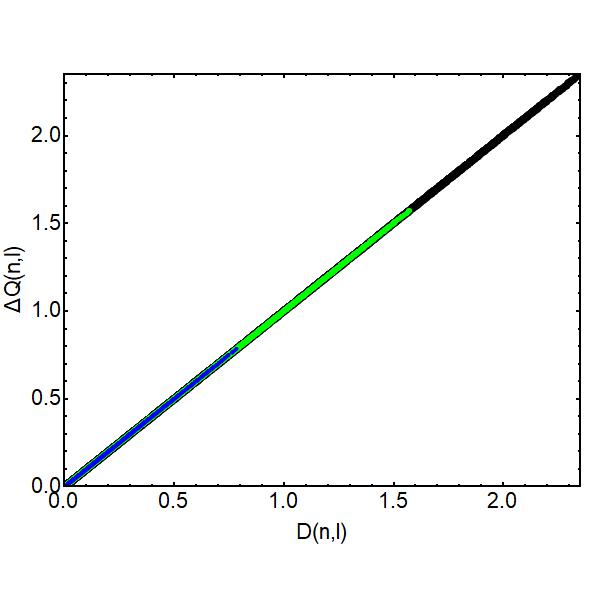} %press
\includegraphics[height=5cm,clip]{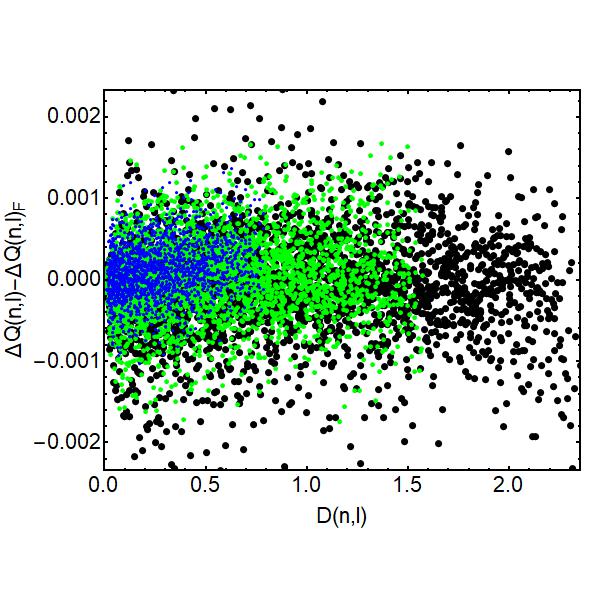}
\includegraphics[height=5cm,clip]{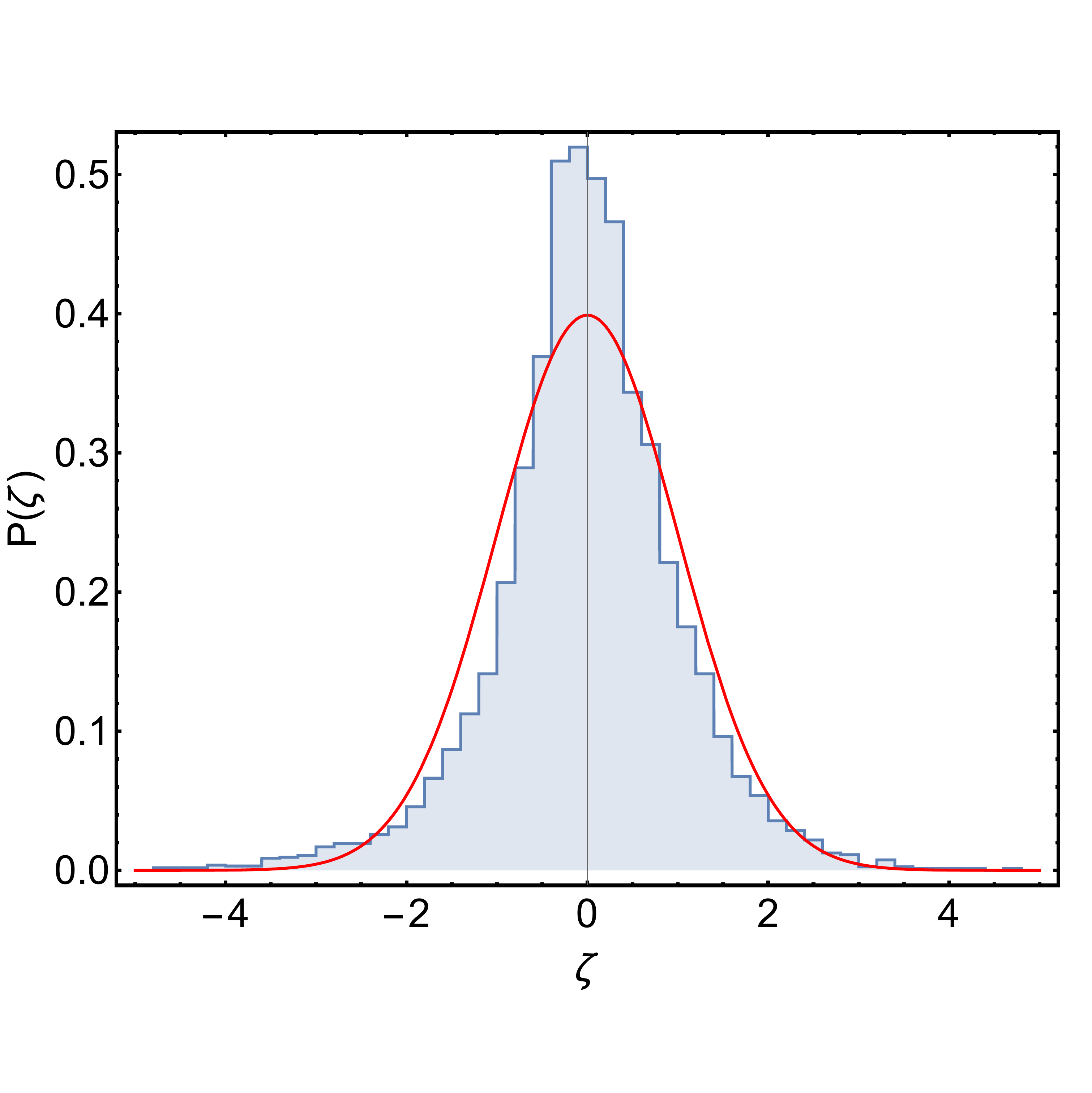}
\includegraphics[height=5cm,clip]{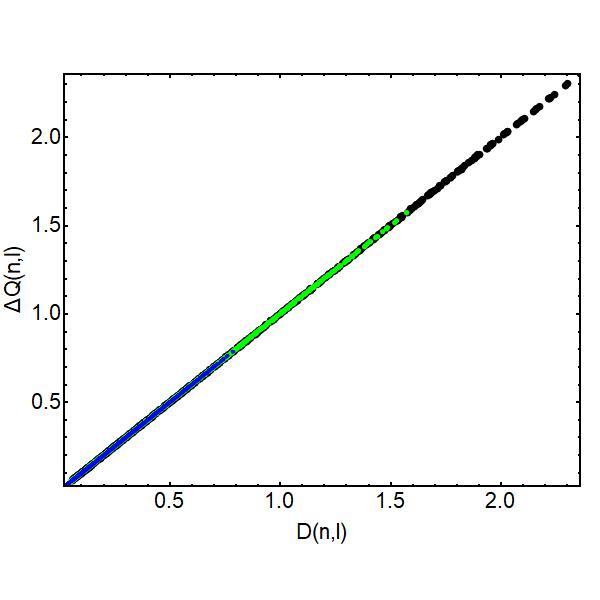}%press_2
\includegraphics[height=5cm,clip]{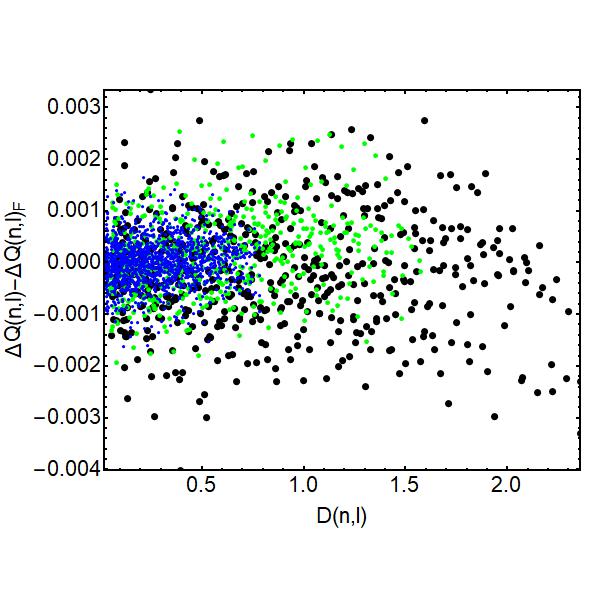}
\includegraphics[height=5cm,clip]{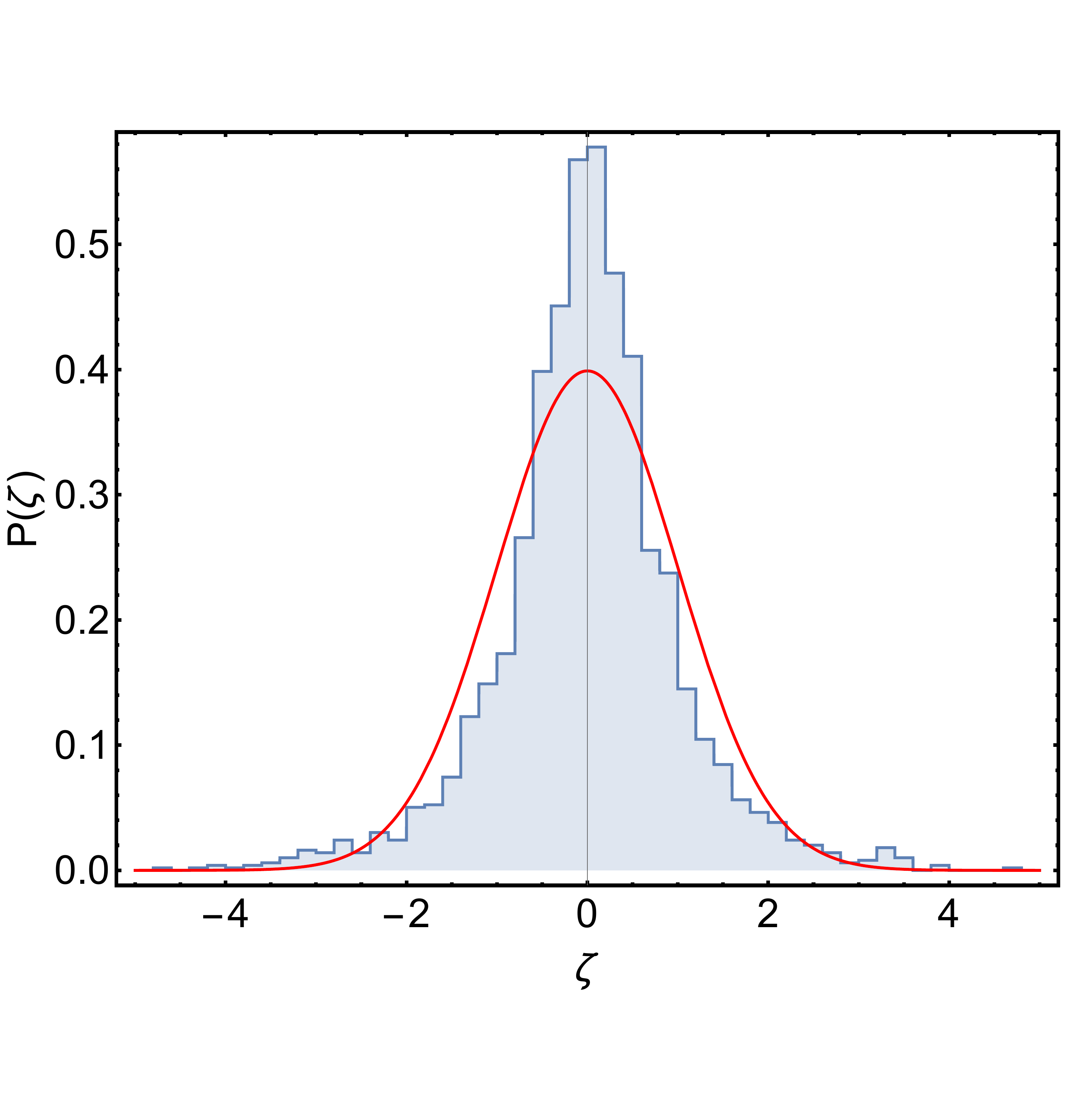}
\includegraphics[height=5cm,clip]{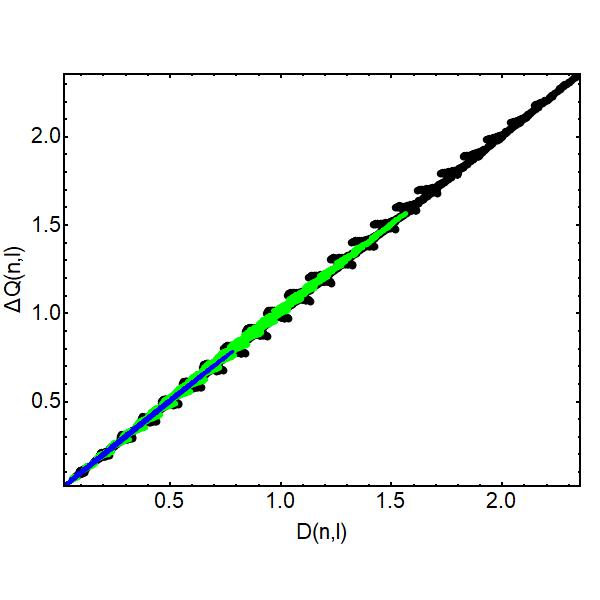}%press_3
\includegraphics[height=5cm,clip]{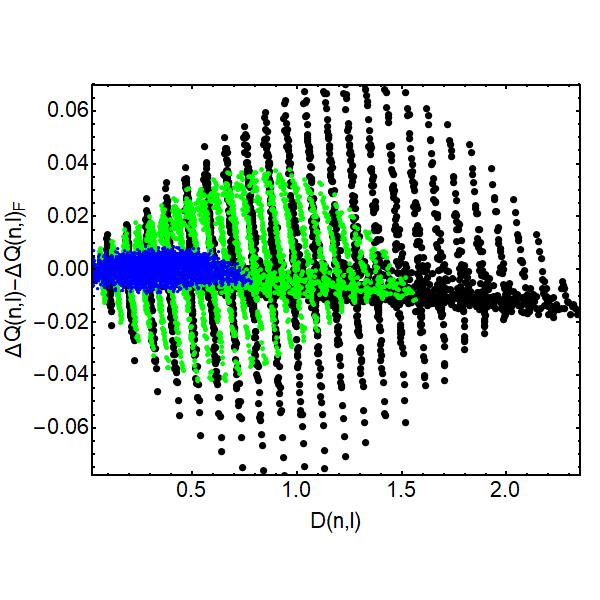}
\includegraphics[height=5cm,clip]{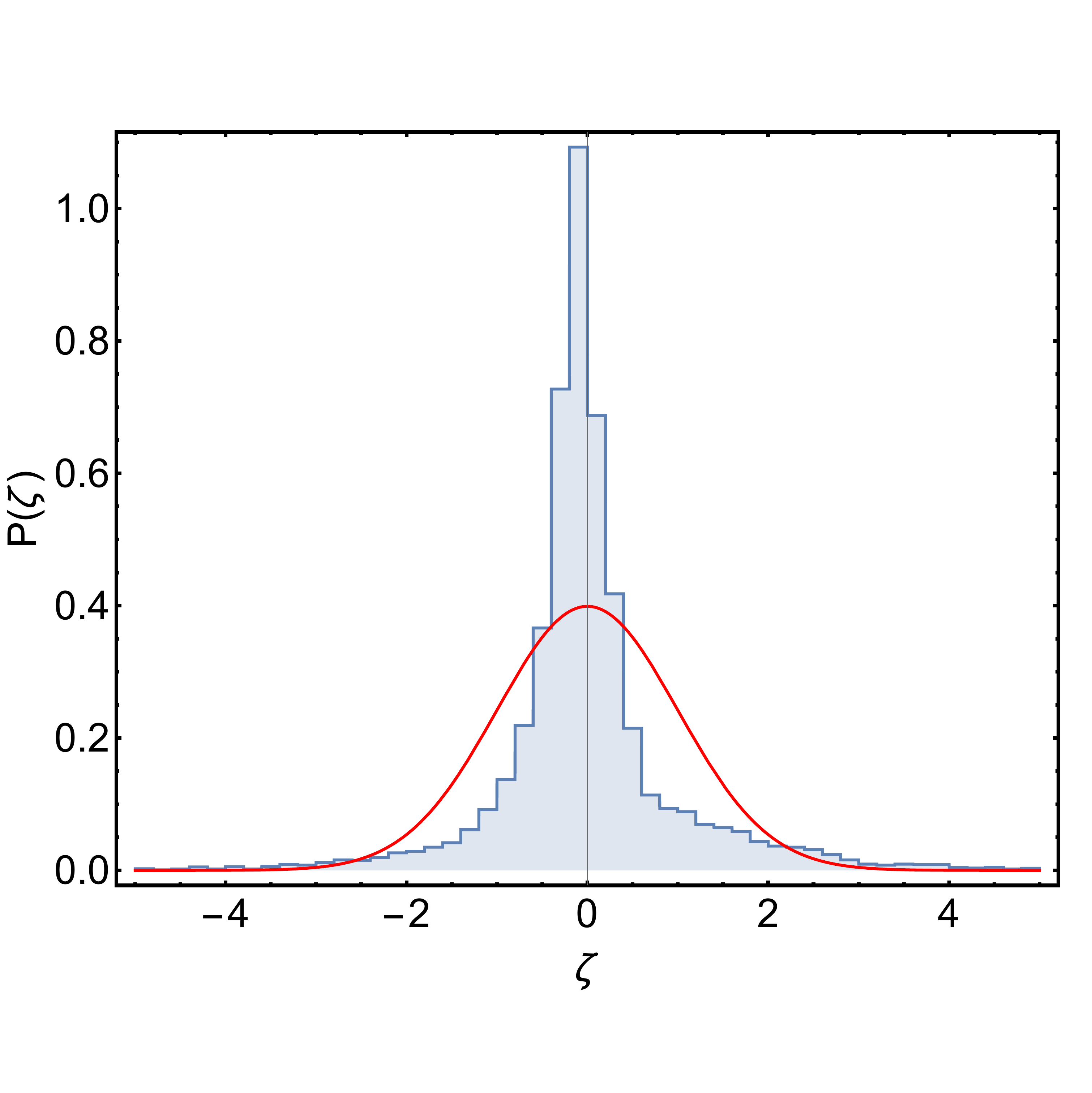}
\end{center}
\kern -1.cm
\caption{Check of the barometric equation: $dQ/dy=-g\rho$. First column: $\Delta Q(n,n_0)$ vs $D(n,n_0)$ for $n_0=4,\ldots,N_C-4$ and $n=n_0+1,\ldots,N_C-3$. Second column: Deviations of the data with respect their linear fit.Third column: Distribution of the deviations of the data with respect their linear fit.  First row: Data corresponding to $T<T_c$. Second row: Data with $T_c<T<T_{c,2}$. Third row: Data with $T>T_{c,2}$. All of them for $g=5$ (blue dots), $g=10$ (green dots) and $g=15$ (black dots).
\label{pres3}}
\end{figure}
\begin{table}
\begin{center}
 \begin{tabular}{| c | c || c | c || c | c | c | c |}
 \hline
&{\bf Ndat}&{$\mathbf a_0$}&{$\mathbf a_1$}&{$\mathbf m_1$}&{$\mathbf m_2$}&{$\mathbf m_3$}&{$\mathbf \kappa$}\\
 \hline\hline
$T<T_c$&$8004$&$-1.32423\times10^{-6}$&$1.00068$&$2.89343\times10^{-16}$&$2.5312\times10^{-7}$&$-3.04246\times10^{-11}$&1.89561\\ \hline
$T_c<T<T_{c,2}$&$2484$&$51.5847\times 10^{-6}$&$1.00151$&$2.98171\times10^{-17}$&$5.1005\times10^{-7}$&$-6.75406\times10^{-11}$&$2.56314$\\ \hline
$T>T_{c,2}$&$12696$&$898.071\times 10^{-6}$&$1.00852$&$4.70725\times10^{-17}$&$0.000163174$&$9.265\times10^{-7}$&$5.85464$\\ \hline
 \hline 
  \end{tabular}
\end{center}
\caption{Check of the barometric equation: Linear fits parameters $a_0$ and $a_1$,  central moments $m_1$, $m_2$ and $m_3$ and kurtosis of the data deviation from the fit. (see text)}\label{table7}
\end{table}
We study the barometric equation in three temperature regions for all the $g$'s we have simulated: (I) $T<T_c(g)$ (non-convecting states), (II) $T_c(g)<T<T_{c,2}(g)$ (weak convecting states) and (III) $T>T_c(g)$ (fully convecting states). Obviously we expect only good agreement for data in region (I).In figure \ref{pres3} we show  the results obtained.
We see in the first row how the $8004$ data of region (I) behave. $\Delta Q(n,l)$ follows a clean straight line for all distance differences and $g$'s studied. We do a linear fit to the data 
$\Delta Q(n,l)_F=a_0+a_1 D(n,l)$ whose results are in table \ref{table7}. In order to see small deviations we study the remainders: $\Delta Q(n,l)-\Delta Q(n,l)_F$ and their statistical distribution and central moments. We observe that the data follows a deformed Gaussian distribution with a high kurtosis value. That may indicate that the data is somehow correlated. That's could be because the data differences are build over all pairs of a given set of pressures and densities. In region (II) we cannot see any obvious deviation to the barometric equation.
Finally, we see in region (III) how the data presents an obvious structure indicating the they do not follow the barometric equation. The data structure grows with the $g$ value in accordance with the strong convection behavior.
It is somehow curious how in regions (II) and (III) the linear fit still is very good. That is, in average the data seems to follow the barometric equation with some small dispersion even though there are convecting structures on the system. That is consitent with the fact that the overall difference between the pressures on top and bottom of the system always follow the barometric relation (see above).

Finally, we can check if the Fourier's Law  is consistent with our data:
\begin{equation}
\sqrt{T}K(\rho)\frac{dT}{dy}=-J
\end{equation}
in order to do that we should show that exists a function $K(\rho)$ that is independent on $g$, $T_0$ and $\bar\rho$) and  has the form:
\begin{equation}
K(\bar\rho)=-J\sqrt{T(y)}\left[\frac{dT}{dy}\right]^{-1}\biggr\vert_{y=\rho^{-1}(\bar\rho)} \label{conduc}
\end{equation}
In order to get $K$ we should compute the local spatial derivatives of the $T(y)$ profiles at a given cell, its local density and the heat current passing through the system. The main problem is to get a precise value of the derivative of a function build with only $30$ points. The measured profiles are quite smooth and with relative small error bars then the appropriate strategy is to fit a function to the data and doing its analytic derivative. There are several problems however: (1)  Any function we use to interpolate the data to the data minimize the distance to the data points but, typically, they do not cross through the points. That is, even though we had a set of discrete data points obtained from a given function, the fitted one will cross the real function by above and below it. Therefore, the derivatives at a given point have a systematic and rather large dispersion due to such undulations. This effect increases if, like in our case, the data have error bars, and (2) for small gradients and/or large $g$ values, the temperature profiles are very flat and the heat currents very small. That is, the ratio current/derivative is in this case very sensitive to the errors in the data and in the fit. Typically, we will discard these profiles.   

We observe that the temperature profiles have a well defined convexity: $d^2T/dy^2<0$. Therefore we have used a fitting function with such property:
\begin{equation}
p(y)=-\sum_{n=1}^s\sum_{l=1}^s \frac{a(n)a(l)}{(n+l+2)(n+l+1)}y^{n+l+2}+a(s+1)y+a(s+2)
\end{equation}
with $a(n)$, $n=1,\ldots , s+2$ being the free parameters to be fitted.

In figure \ref{fourier1} we show the fits for the values of $T_0$ being at the non-convective regime for $g=0$, $5$, $10$ and $15$. The fits are reasonable good with regression coefficients near one. Observe that some data for low gradients present the commented smaller values of $R$ due to the flat profile problem we already commented.
In figure \ref{fourier2} we show the values of $K$ obtained by doing the derivative of $T(y)$ of the fitted function at the point $y$ for states at the non-convecting regime (left figure) and convecting one (right figure) for most of the gradients and all the $g$'s we have simulated. We do not show the cases having large dispersions due to having near flat profiles and/or fits with low values of the regression coefficient.  We observe that the data scale with still some dispersion. Moreover we also see the already observed deviation from Enskog result. 

Finally we checked the possibility that the $y$ profile for $T_0$ values at the convective regime, might follow the same Fourier's Law and/or some effective one. In figure \ref{fourier2} (right)  we show that there is no trace of any scaling. 

\begin{figure}[h!]
\begin{center}
\includegraphics[height=4.cm,clip]{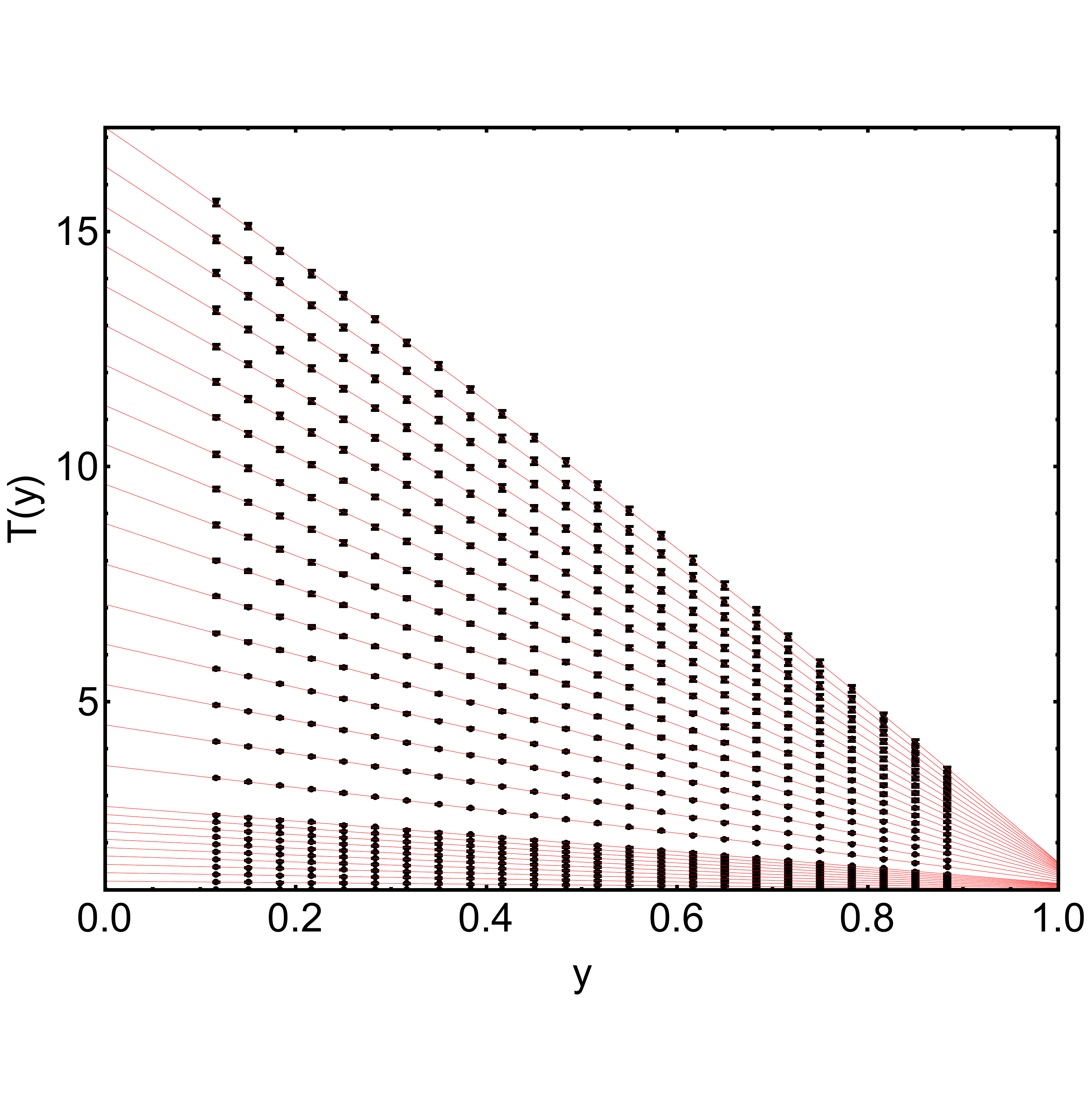}  %fourier5.nb
\includegraphics[height=4.cm,clip]{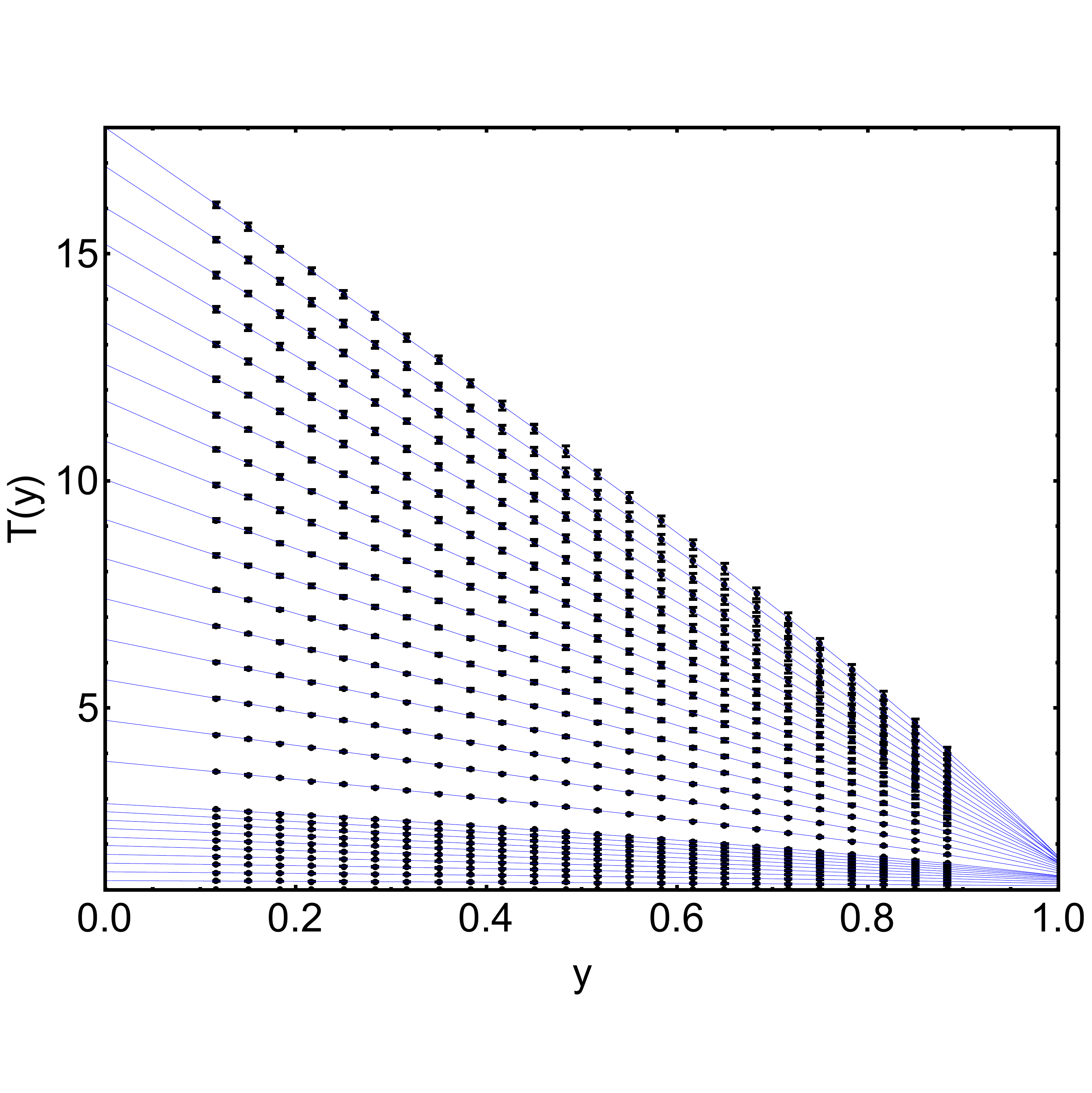}
\includegraphics[height=4.cm,clip]{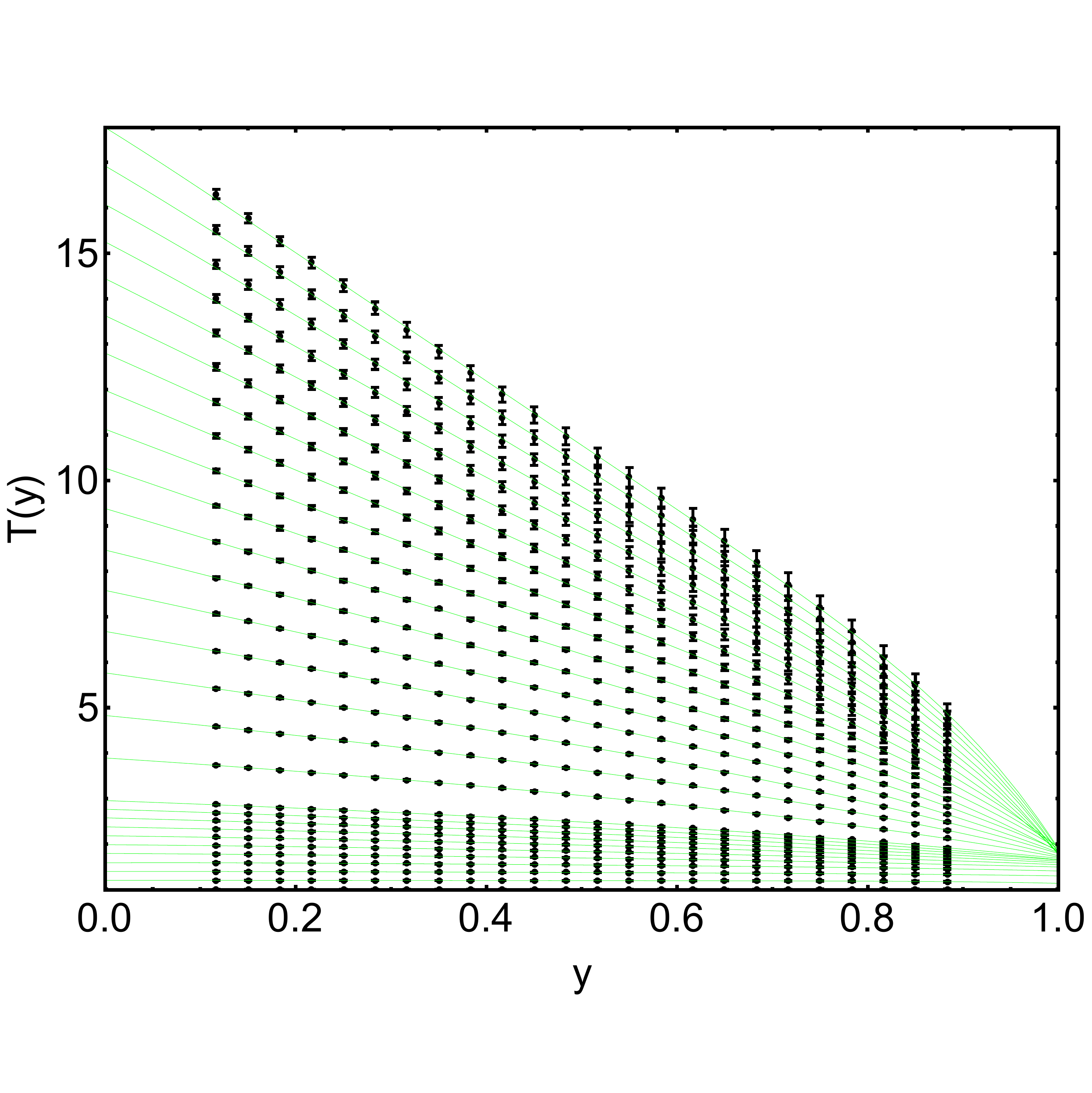}  
\includegraphics[height=4.cm,clip]{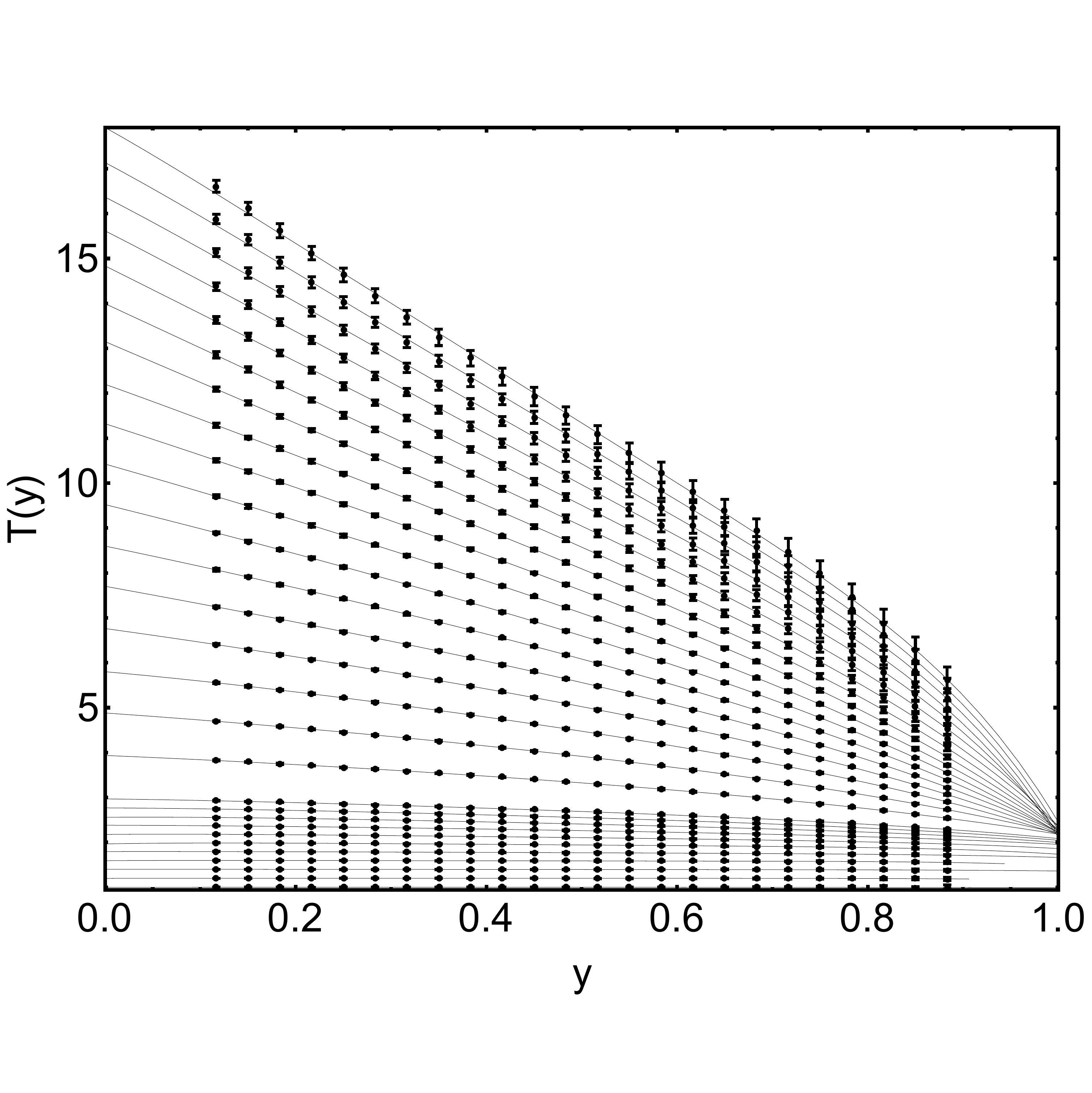}
\includegraphics[height=5cm,clip]{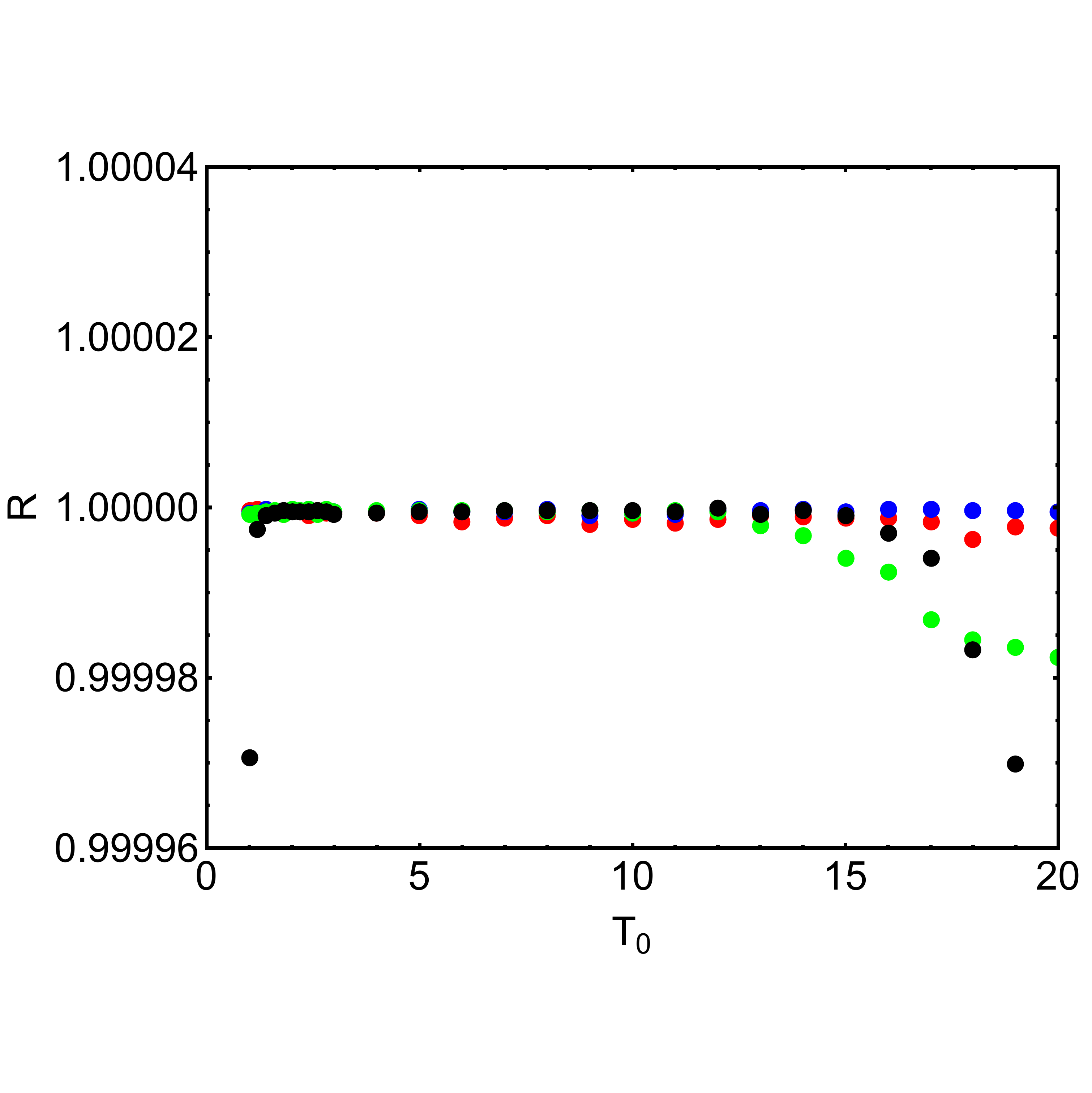}
\end{center}
\kern -1.cm
\caption{Temperature profiles and their fits  (excluding some of the points near the boundaries) for $g=0$, $5$, $10$ and $15$ (from left to right). Bottom: regression coefficients of the fits vs $T_0$ for $g=0$ (red dots), $g=5$ (blue dots), $g=10$ (green dots) and $g=15$ (black dots) 
\label{fourier1}}
\end{figure}

\begin{figure}[h!]
\begin{center}
\includegraphics[height=6cm,clip]{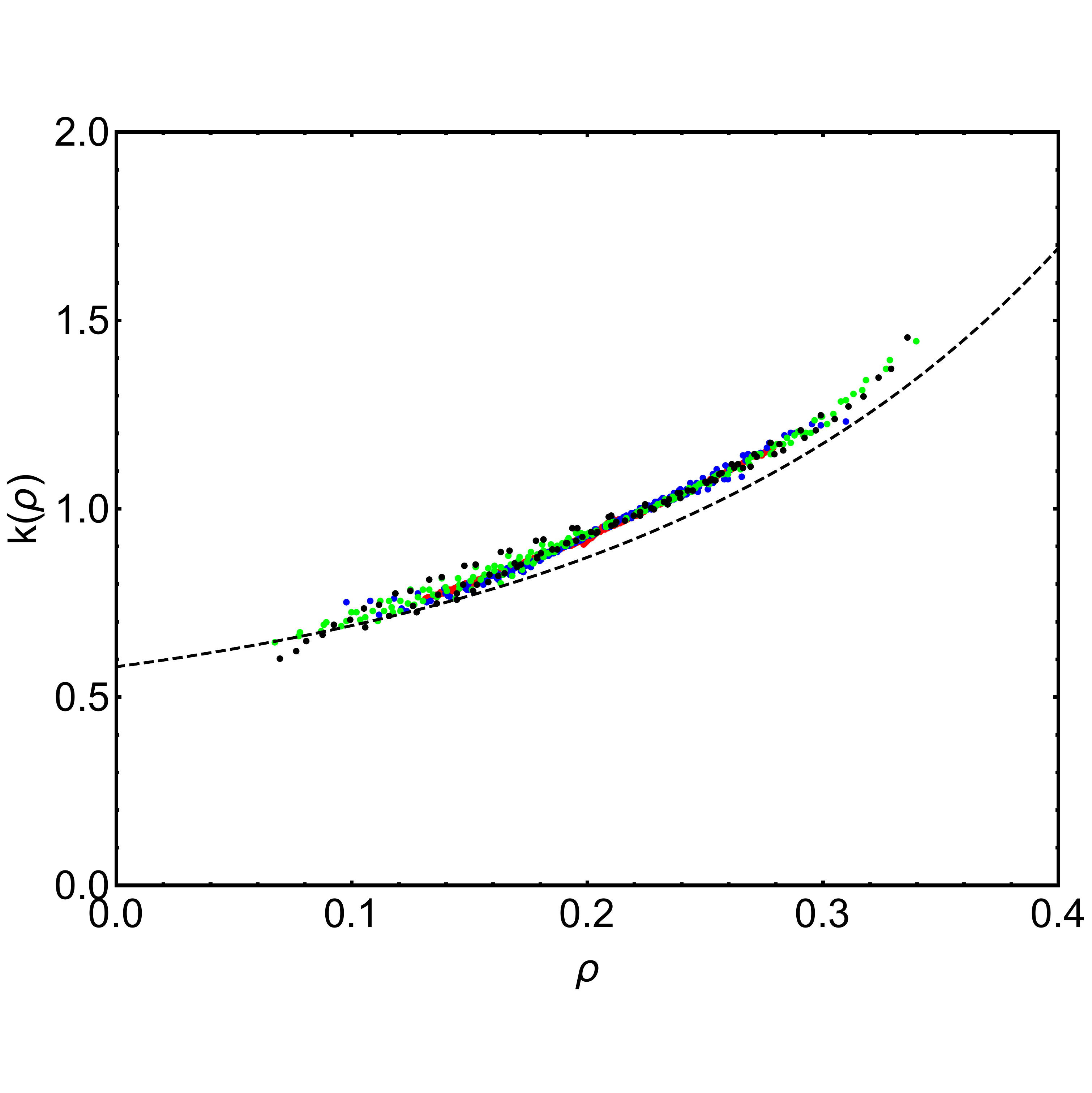}  %fourier5.nb
\includegraphics[height=6cm,clip]{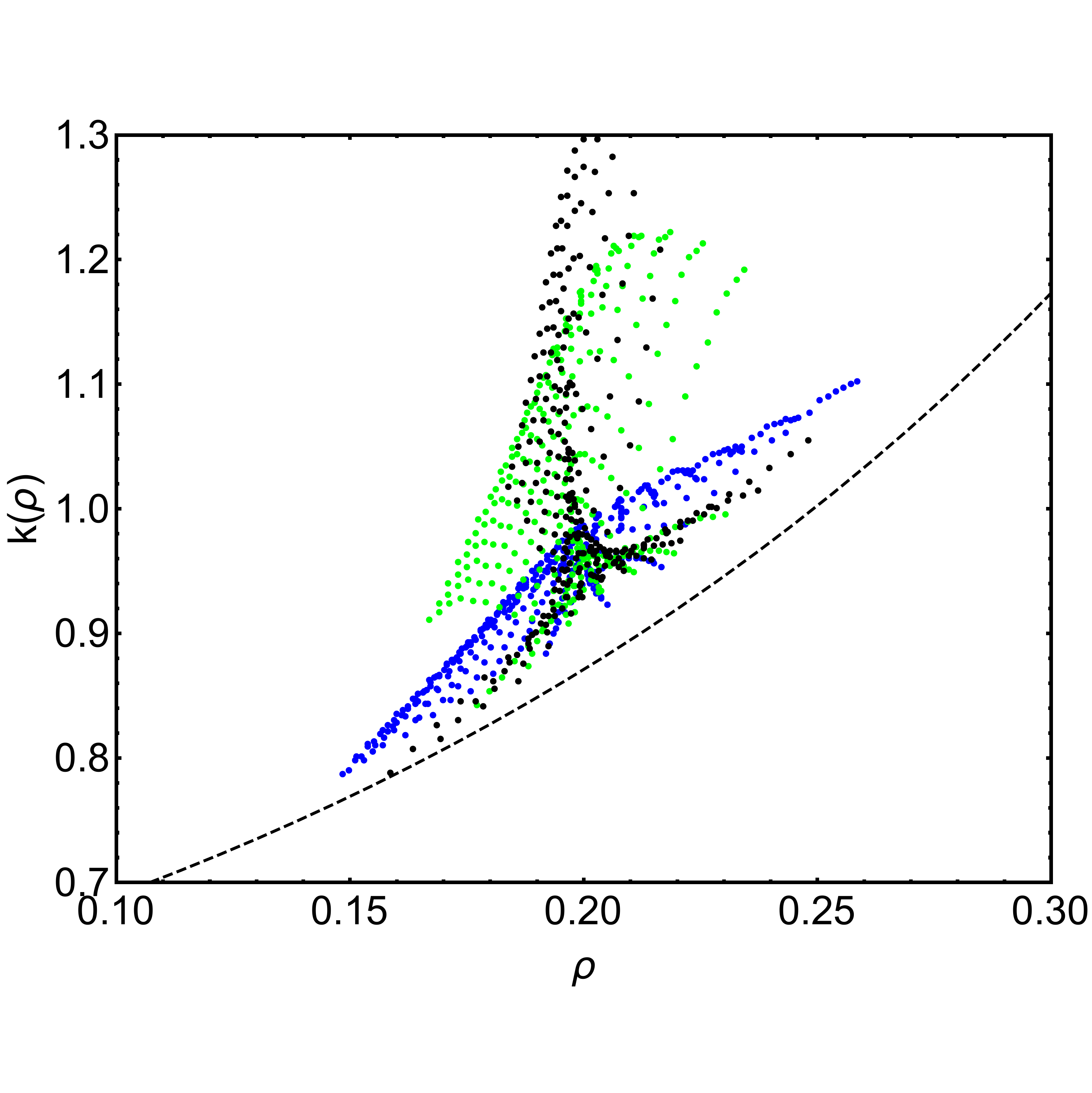}  
\end{center}
\kern -1.cm
\caption{$K(\rho)$ defined by eq. (\ref{conduc}) vs. $\rho$. Each dot is the corresponding $K$ value for a given cell-$y$. Left figure: non-convecting regime. Right figure: convecting regime for $g=0$ (red dots) $g=5$ (blue dots), $g=10$ (green dots) and $g=15$ (black dots). Each dot represent a given temperature gradient and the conductivity obtained in a virtual cell.  Dotted line is the Enskog prediction.
\label{fourier2}}
\end{figure}

\item{\bf 4. Some comments about the boundary conditions}

In the computer simulation the boundary conditions choosed (reflecting vertical walls at $x=0,1$ and thermal walls at $y=0,1$) condition the particle behavior and the fields measured at the stationary state. In particular we observe:
\begin{equation}
u_1(0,y)=0 \quad u_1(1,y)=0\quad u_2(x,0)=0 \quad u_2(x,1)=0
\end{equation}
We also showed above that the difference of pressures measured at the top and bottom of the system follow the barometric formula with high precision (see figure \ref{press}):
\begin{equation}
P(1)-P(0)=-g\rho
\end{equation}
where $\rho$ is the system average density. We can use this information to know the kind of boundary conditions we should use on the Navier-Stokes equations to obtain this property. We know that the momentum equations in NS can be writen: 
\begin{eqnarray}
\partial_1P&=&\partial_1(\tau_{11}'-\rho u_1^2)+\partial_2(\tau_{12}'-\rho u_1u_2)\nonumber\\
\partial_2P&=&\partial_1(\tau_{12}'-\rho u_1u_2)+\partial_2(\tau_{22}'-\rho u_2^2)-\rho g
\end{eqnarray}
We can integrate both equations on $x$ and $y$ over all the domain. The left hand sides are:
\begin{eqnarray}
\int_0^1dx\int_0^1dy\partial_1 P(x,y)&=&\int_0^1dy\left[P(1,y)-P(0,y)\right]=0\nonumber\\
\int_0^1dx\int_0^1dy\partial_2 P(x,y)&=&\int_0^1dx\left[P(x,1)-P(x,0)\right]\equiv P(1)-P(0)
\end{eqnarray} 
The first average is zero by the boundary conditions symmetry. The complete set of pressure equations are, after integration,
\begin{eqnarray}
0&=&\int_0^1dy\left[\tau_{11}'(1,y)-\tau_{11}'(0,y)\right]+\int_0^1 dx\left[\tau_{12}'(x,1)-\tau_{12}'(x,0) \right]\nonumber\\
P(1)-P(0)&=&-g\rho+ \int_0^1dy\left[\tau_{12}'(1,y)-\tau_{12}'(0,y)\right]+\int_0^1 dx\left[\tau_{22}'(x,1)-\tau_{22}'(x,0) \right]
\end{eqnarray}
Therefore, the conditions to get the observed barometric formula are:
\begin{eqnarray}
\int_0^1dy\left[\tau_{11}'(1,y)-\tau_{11}'(0,y)\right]+\int_0^1 dx\left[\tau_{12}'(x,1)-\tau_{12}'(x,0) \right]&=&0\nonumber\\
 \int_0^1dy\left[\tau_{12}'(1,y)-\tau_{12}'(0,y)\right]+\int_0^1 dx\left[\tau_{22}'(x,1)-\tau_{22}'(x,0) \right]&=&0
\end{eqnarray}
In hydrodynamics  one can use many different boundary conditions for the hydrodynamic velocities. For instance, the {\it no-slip} boundary conditions consider $u(x,y)=(0,0) \forall (x,y)\in\partial\Lambda$ that is incompatible with the results we have obtained in our simulation. On the other hand, the {\it stress-free} boundary conditions are 
\begin{equation}
u(x,y)\cdot n=0 \quad\quad,\quad \tau_{12}'(x,y)=0  \quad\quad\forall (x,y)\in\partial\Lambda
\end{equation}
where $n$ is a unitary perpendicular vector to the boundary. Observe that this boundary conditions are compatible with our observations if
\begin{eqnarray}
\int_0^1dy\tau_{11}'(1,y)&=&\int_0^1dy\tau_{11}'(0,y)\nonumber\\
\int_0^1 dx\tau_{22}'(x,1)&=&\int_0^1 dx\tau_{22}'(x,0)\label{extra}
\end{eqnarray}
In conclusion, our computer simulation is compatible with the well known in hydrodynamics {\it slip boundary conditions} if we add the properties given by equations (\ref{extra}).

\end{itemize}

\section{Appendix I: Lyapunov Central Limit Theorem}

Let $X_1, X_2, \dots, X_N$ independent random variables such that 
\begin{equation}
\langle X_i\rangle=\bar X_i \quad, \quad \langle (X_i-\bar X_i)^2\rangle=\sigma_i < \infty
\end{equation}
If for any $\delta >0$ the following condition holds (Lyapunov condition):
\begin{equation}
\lim_{N\rightarrow\infty}\frac{1}{s_N^{2+\delta}}\sum_{i=1}^N\langle\vert X_i-\bar X_i\vert^{2+\delta} \rangle=0
\end{equation}
where
\begin{equation}
s_N^2=\sum_{i=1}^N\sigma_i^2
\end{equation}
Then 
\begin{equation}
\lim_{N\rightarrow\infty}\frac{1}{s_N}\sum_{i=1}^N(X_i-\bar X_i) \overset{d}{\rightarrow}  N(0,1)
\end{equation}
where $d$ means a convergence on distribution and $N(0,1)$ is a gaussian distribution with $0$ average and $\sigma=1$. We can formally rewrite the later property:
\begin{equation}
\frac{1}{N}\sum_{i=1}^NX_i \underset{N\rightarrow\infty}{\simeq}\frac{1}{N}\sum_{i=1}^N\bar X_i+\frac{\xi}{N}\sqrt{\sum_{i=1}^N\sigma_i^2}
\end{equation}
where $\xi$ is a gaussian random variable with $N(0,1)$ distribution.

As an example  let us assume that the $X_i$ random variables are gaussian distributed with $N(\bar X_i,\sigma_i)$ where
$\sigma_i\in [\sigma_0,\sigma_1]$. In this case the Lyapunov condition holds. Let us show it.

First we compute  the average:
\begin{equation}
\langle\vert X_i-\bar X_i\vert^{2+\delta} \rangle=\frac{1}{\sqrt{2\pi}\sigma_i}\int_{-\infty}^{\infty}dx\,\vert x-\bar X_i\vert^{2+\delta} \exp\left[-\frac{(x-\bar X_i)^2}{2\sigma_i^2}\right]=C(\delta)\sigma_i^{2+\delta}
\end{equation}
where $C(\delta)=\Gamma((3+\delta)/2)2^{(3+\delta)/2}/\sqrt{2\pi}$ is a positive constant.
 Then, the Lyapunov condition can be written:
 \begin{equation}
 \frac{\sum_{i=1}^N \sigma_i^{2+\delta}}{\left(\sum_{i=1}^N \sigma_i^2\right)^{1+\delta/2}}\underset{N\rightarrow\infty}{\rightarrow} 0
 \end{equation}
 For $\delta=2$ we know that
  \begin{equation}
  \frac{\sum_{i=1}^N \sigma_i^{4}}{\left(\sum_{i=1}^N \sigma_i^2\right)^{2}} <\frac{1}{1+(N-1)\sigma_0^4/\sigma_1^4}
  \end{equation}
  and it goes to zero when $N\rightarrow\infty$ whenever $\sigma_0>0$ and $\sigma_1<\infty$.
  
\section{Acknowledgements}

We want to acknowledge the support from  National Science Foundation [grant DMR1104500] and AFOSR [grant F49620-01-0154] and FIS2017-84256-P Spanish Ministry MINECO. 

\end{itemize}


%merlin.mbs apsrev4-1.bst 2010-07-25 4.21a (PWD, AO, DPC) hacked
%Control: key (0)
%Control: author (72) initials jnrlst
%Control: editor formatted (1) identically to author
%Control: production of article title (-1) disabled
%Control: page (0) single
%Control: year (1) truncated
%Control: production of eprint (0) enabled
\begin{thebibliography}{0}%
\makeatletter
\providecommand \@ifxundefined [1]{%
 \@ifx{#1\undefined}
}%
\providecommand \@ifnum [1]{%
 \ifnum #1\expandafter \@firstoftwo
 \else \expandafter \@secondoftwo
 \fi
}%
\providecommand \@ifx [1]{%
 \ifx #1\expandafter \@firstoftwo
 \else \expandafter \@secondoftwo
 \fi
}%
\providecommand \natexlab [1]{#1}%
\providecommand \enquote  [1]{``#1''}%
\providecommand \bibnamefont  [1]{#1}%
\providecommand \bibfnamefont [1]{#1}%
\providecommand \citenamefont [1]{#1}%
\providecommand \href@noop [0]{\@secondoftwo}%
\providecommand \href [0]{\begingroup \@sanitize@url \@href}%
\providecommand \@href[1]{\@@startlink{#1}\@@href}%
\providecommand \@@href[1]{\endgroup#1\@@endlink}%
\providecommand \@sanitize@url [0]{\catcode `\\12\catcode `\$12\catcode
  `\&12\catcode `\#12\catcode `\^12\catcode `\_12\catcode `\%12\relax}%
\providecommand \@@startlink[1]{}%
\providecommand \@@endlink[0]{}%
\providecommand \url  [0]{\begingroup\@sanitize@url \@url }%
\providecommand \@url [1]{\endgroup\@href {#1}{\urlprefix }}%
\providecommand \urlprefix  [0]{URL }%
\providecommand \Eprint [0]{\href }%
\providecommand \doibase [0]{http://dx.doi.org/}%
\providecommand \selectlanguage [0]{\@gobble}%
\providecommand \bibinfo  [0]{\@secondoftwo}%
\providecommand \bibfield  [0]{\@secondoftwo}%
\providecommand \translation [1]{[#1]}%
\providecommand \BibitemOpen [0]{}%
\providecommand \bibitemStop [0]{}%
\providecommand \bibitemNoStop [0]{.\EOS\space}%
\providecommand \EOS [0]{\spacefactor3000\relax}%
\providecommand \BibitemShut  [1]{\csname bibitem#1\endcsname}%
\let\auto@bib@innerbib\@empty
%</preamble>
\end{thebibliography}%


%merlin.mbs apsrev4-1.bst 2010-07-25 4.21a (PWD, AO, DPC) hacked
%Control: key (0)
%Control: author (72) initials jnrlst
%Control: editor formatted (1) identically to author
%Control: production of article title (-1) disabled
%Control: page (0) single
%Control: year (1) truncated
%Control: production of eprint (0) enabled
%


\newpage
\begin{thebibliography}{99}
\bibitem{Alder62} Alder, B.J. and Wainwright, T.E., {\it Phase Transitions in Elastic Disks}, Physical Review, {\bf 127},  359-361 (1962).
\bibitem{Batchelor} Batchelor, G.K., {\it An introduction to fluid dynamics}, Cambridge University Press (2000).
\bibitem{Boden} Bodenschatz, E., Pesch, W. and Ahlers, G., {\it Recent developments in Rayleigh-B{\'e}nard convection}, Annu. Rev. Fluid Mech. {\bf 32}, 709-778 (2000).
\bibitem{Bormann} Bormann, Andreas S., {\it The onset of convection in the Rayleigh-B{\'e}nard problem for compressible fluids}, Continuum Mechanics and Thermodynamics, {\bf 13}, 9--23 (2001). {\it Numerical Linear Stability Analysis for Compressible Fluids} in {\it Analysis and Numerics for Conservation Laws}, Warnecke, Gerald (ed.), pp. 93--105,  Springer Berlin Heidelberg (2005).
\bibitem{Buresti} Buresti, G., {\it A note on Stokes' hypothesis}, Acta Mechanica, {\bf 226}, 3555-3559 (2015).
\bibitem{Chandra} Chandrasekhar, S., {\it Hydrodynamic and Hydromagnetic stability}, Dover, New York (1981).
\bibitem{Cordero95} Cordero, P., Mar{\'\i}n, M. and Risso, D., {\it Efficient Simulations of Microscopic Fluids: Algorithm and Experiments.}, Chaos, Solitons \& Fractals,  {\bf 6},  95-104, (1995). 
\bibitem{delPozo} del Pozo, J., Garrido, P.L. and Hurtado P.I., {\it Probing local equilibrium in nonequilibrium fluids}, Physical Review E, {\bf 92}, 022117 (2015); {\it Scaling laws and bulk-boundary decoupling in heat flow}, Physical Review E, {\bf 91}, 032116 (2015). 
\bibitem{delPozo2} del Pozo, J., Garrido, P.L. and Hurtado P.I. in preparation.
\bibitem{Engel} Engel, Michael and Anderson, Joshua A. and Glotzer, Sharon C. and Isobe, Masaharu and Bernard, Etienne P. and Krauth, Werner, {\it Hard-disk equation of state: First-order liquid-hexatic transition in two dimensions with three simulation methods}, Phys. Rev E, {\bf 87}, 042134, (2013).
\bibitem{Esposito} Esposito, R., Lebowitz, J.L. and Marra, R., {\it On the derivation of hydrodynamics from the Boltzmann equation}, Physics of Fluids, {\bf 11}, 2354-2366 (1999).
\bibitem{Gad} Gad-el-Hak, M., {\it Stokes's hypothesis for a Newtonian, Isotropic Fluid},  Journal of Fluids Engineering {\bf 117}, 3-5 (1995).
\bibitem{Gallavotti} Gallavotti, G., {\it Foundations of Fluid Dynamics}, Springer, Berlin (2002).
\bibitem{Gass71} Gass, D.M., {\it  Enskog Theory for a Rigid Disk Fluid}, Journal of Chemical Physics {\bf 54}, 1898-1902 (1971).
\bibitem{Hen75} Henderson, D., {\it A Simple Equation of State for Hard Discs}, Mol. Phys. {\bf 30}, 971 (1975).
\bibitem{Lappa} Lappa, M. {\it Thermal Convection: Patterns, Evolution and Stability}, Wiley 2010. 
\bibitem{Manela} Manela, A. and Frankel, I., {\it On the Rayleigh-B{\'e}nard problem: dominant compressibility effects}, J. Fluid Mech. {\bf 565}, 461-475 (2006).
\bibitem{Mar87} Mareschal, M. and Kestemont, E., {\it Order and Fluctuations in Nonequilibrium Molecular Dynamics Simulations of Two-Dimensional Fluids}, Journal of Statistical Physics {\bf 48}, 1187-1201 (1987).
\bibitem{Mar88} Mareschal, M., Malek Mansour, M., Puhl, A. and Kestermont, E., {\it Molecular Dynamics versus Hydrodynamics in a two dymensional Rayleigh-B{\'e}nard System}, Phys. Rev. Lett. {\bf 61} 2550-2553 (1988).
\bibitem{MartinLof} Martin-Lof, A., {\it Statistical Mechanics and the Foundations of Thermodynamics}, Lecture Notes in Physics {\bf 101}, Springer-Verlag, Berlin (1979).
\bibitem{Mulero} Mulero, A. (Ed.), {\it Theory and Simulation of Hard-Sphere Fluids and Related Systems},
Lect. Notes Phys. {\bf 753} (Springer, Berlin Heidelberg 2008).
\bibitem{Mutabazi} Mutabazi, I.,  Wesfreid, J.E. and Guyon, E. (eds.) {\it Dynamics of Spatio-Temporal Cellular Structures: Henri Benard Centenary Review}, in Springer Tracts in Modern Physics {\bf 207} (Springer, 2006). 
\bibitem{Puhl} Puhl, A., Malek Mansour, M. and Mareschal, M., {\it Quatitative comparison of molecular dynamics with hydrodynamics in Rayleigh.B{\'e}nard convection}, Physical Review A, {\bf 40}, 1999-2012 (1989).
\bibitem{Raja} Rajagopal, K.R., {\it A new development and interpretation of the Navier Stokes fluid which
reveals why the ‘‘Stokes assumption’’ is inapt}, International Journal of Non-Linear Mechanics {\bf 50}, 141–151  (2013).
\bibitem{Raja2} Rajagopal, K.R., {\it Remarks on the notion of “pressure”}, International Journal of Non-Linear Mechanics {\bf 71}, 165–172  (2015). 
\bibitem{Rapaport_1986} Rapaport, D.C., {\it Eddy formation in obstructed fluid flow: a molecular dynamics study}, Physical Review Letters, {\bf 57}, 695-698 (1986).
\bibitem{Rapaport_1988} Rapaport, D.C., {\it Molecular Dynamics study of Rayleigh-Bénard Convection}, Physical Review Letters, {\bf 60}, 2480-2483 (1988).
\bibitem{Rapaport_1992} Rapaport, D.C., {\it Unpredictable convection in a small box: Molecular dynamics experiments}, Physical Review A, {\bf 46}, 1971-1984 (1992).
\bibitem{Rapa09} Rappaport, D.C., {\it The Event-Driven Approach to N-Body Simulation
}, Progress of Theoretical Physics Supplement, {\bf 178}, 5-14 (2009).
\bibitem{Risso} Risso, D. and Cordero, P., {\it Empirical determination of the onset of convection for a hard disk system}, in {\it Instabilities
and Nonequilibrium Structures IV}, Tirapegui, E. and Zeller, W. (eds.). Kluwer Academic Publishers (1993).
\bibitem{Roman} Rom{\'a}n, F.L., White, J.A.  and Velasco, S., {\it Fluctuations in an equilibrium hard-disk fluid: Explicit size effects}. Journal of  Chemical Physics {\bf 107} (12), 4635-4641 (1997).
\bibitem{Stalling} Stalling, D. and  Hege, H.C., {\it Fast and Resolution Independent Line Integral Convolution}, in Proceeding
SIGGRAPH '95 Proceedings of the 22nd annual conference on Computer graphics and interactive techniques,  Pages 249-256, (1995) ISBN:0-89791-701-4. Cabral, B. and Leedom, L.  {\it Imaging vector fields using line integral convolution}. Computer Graphics, {\bf 27}, 263–272 (1993).
\bibitem{Wain71} Wainwright, T.E., Alder, B.J., and Gass, D.M., {\it Decay of Time Correlations in Two Dimensions},  Physical Review A, {\bf 4}, 233-237 (1971).
\end{thebibliography}
\end{document}